\documentclass[aps,prl,onecolumn,showpacs,groupedaddress,amssymb]{revtex4}
\usepackage{graphics}
\usepackage{graphicx}
\usepackage{multibib}
\newcites{ltex}{For some reason it is needed by multibib}

\begin{document}
\title{ENERGY MEDIATED BY INTERFERENCE OF PARTICLES (Parts I-IV):\\ The Way to Unified Classical  and Quantum Fields and Interactions\\
\vspace{0.2cm}
{\it {Part I. Introduction to Unified Classical and Quantum Fields}}}

\author{S. V. Kukhlevsky}
\affiliation{Department of Physics, Faculty of Natural Sciences, University of Pecs,
Ifjusag u. 6, H-7624 Pecs, Hungary}


\begin{abstract}
It is known that the pure additive or subtractive interference of two waves does induce the positive or negative cross-correlation energy that would contradict the conservation of energy. In the classical wave physics, the problem can be explained simply by a fact that the pure additive or subtractive interference never occurs in the real experiments. The alternately additive and subtractive interference of waves in the experiments changes only the local energy (intensity) of waves, while the total wave energy remains constant. The problem arises again if the conditions of the pure constructive or destructive interference have been created, for instance, in a subwavelength set-up or a resonator. In quantum physics, the Copenhagen-Dirac interpretation of quantum interference gets over the difficulty associated with the cross-correlation energy by assuming that interference between two different particles of matter never occurs. In the present study, the interference-induced energy, which does not exist from the point of view of both the quantum mechanics and particle field theory, is not ignored. In both the classical and quantum fields, the two-times increase of the wave amplitude does increase the wave energy in four times, and the wave with zero amplitude has zero energy. The problem of nonconservation of the energy is overcame by taking into account the fact that the creation of the conditions of pure additive or subtractive interference requires "additional" energy. In Part I, the Hamiltonians describing the cross-correlation energy in the basic classical and quantum fields are derived. The conditions of pure constructive or destructive interference are found. The influence of the cross-correlation energy on the basic physical properties of boson and fermion fields is demonstrated. The energy, mass, charge, and momentum of the interfering fields are calculated. It is shown that the gradient of the cross-correlation energy induces attractive or repulsive forces. These forces could be attributed to the all-known classical and quantum fields, for instance, to the gravitational and Coulomb fields. To this end, the model shows a key role of the cross-correlation energy in several basic physical phenomena, such as the Bose-Einstein condensation, super-radiation, superfluidity, superconductivity, and supermagnetism.
\end{abstract}


\maketitle

%
\maketitle
\section{1. Introduction}
The additive and subtractive interference of waves is a basic phenomenon in the classical wave physics. If two waves having the same amplitudes are made to interfere only destructively, then the waves and the energy associated with these waves do vanish. The pure constructive interference of two waves would produce a wave with the energy four times larger the single wave. The increase or annihilation of the wave energy by the positive or negative cross-correlation term inducing by the interference would contradict the conservation of energy. The cross-correlation (interference) term that increases or decreases the energy logically to call the cross-correlation energy. In the classical wave physics, the problem connected with the energy nonconservation by the cross-correlation is usually explained by mentioning a fact that the pure additive or subtractive interference never happens in the real experiments. The alternately constructive and destructive interference of waves in the experiments redistributes spatially the local energy (intensity) of waves, while the total energy of waves remains constant. The problem of energy nonconservation appears again if the amplitude and phase conditions of the pure additive or subtractive interference have been created in the appropriate physical system, for instance, in a subwavelength set-up or inside a resonator cavity. In such a case, the interference-induced positive or negative cross-correlation energy, which is the "nonexistent energy" from the point of view of both the classical and quantum physics, violates the energy conservation. 

In the past, the problem associated with the cross-correlation energy has been {\it{considered}} in quantum mechanics under the study of interference of particles of matter in the context of the double-slit experiment (see, textbooks). If the particle waves interfere in such an experiment as particles, then on the basis of the exclusively additive nature of particles, the density of interfering particles is expected to be equal to the sum of those before the interference. However, it was found experimentally that the interference changes the local density of particles. This can be explained only by the alternately additive and subtractive interference (cross-correlation) of the particle waves. According to the classical wave physics, the pure constructive or destructive interference could result into violation of the energy conservation in the interfering waves. To overcome the problem Paul Dirac has suggested that interference between two different particles (photons~\cite{Planck,Einst1}) never occurs~\cite{Dir1}: {\it{"... Suppose we have a beam of light consisting of a large number of photons split up into two components of equal intensity. On the assumption that the beam is connected with the probable number of photons in it, we should have half the total number going into each component. If the two components are now made to interfere, we should require a photon in one component to be able to interfere with one in the other. Sometimes these two photons would have to annihilate one another and other times they would have to produce four photons. This would contradict the conservation of energy. The new theory, which connects the wave-function with probabilities for one photon gets over the difficulty by making each photon go partly into each of the two components. Each photon then interferes only with itself. Interference between two different photons never occurs. ..."}} \cite{Dir2}. In the Dirac model of the "interference-less" particles, the main reason of exclusion of the interference between the "self-interfering" particles is the requirement of conservation of the number and energy of particles. According to the Copenhagen-Dirac model of quantum interference, the probability amplitude attributed to the wave-function of a particle in the Copenhagen interpretation of quantum mechanics must be used to calculate the likelihood of a single particle (photon) to be in a particular single-particle Fock state. Another way in which probability can be applied to the behavior of particles is calculation of the probable number of photons in a particular state. Although the modern physics does use mainly the Dirac ({\it{traditional}}) interpretation based on the Copenhagen interpretation of quantum mechanics, the former model was also applied, for instance, for coherent states and statistical mixtures of such, as thermal light. The model allows existence of the interference between two different states (wave-functions). Therefore, formally, the energy conservation can be violated by the cross-correlation of states (particles). 

In the classical wave physics, the cross-correlation term (energy) plays a key role in description of the pure constructive or destructive interference. The Copenhagen-Dirac ({\it{canonical}}) interpretation of the double-slit quantum experiment in terms of the "interference-less" particles, namely the self-interfering particles that never interfere with each other, has solved the problem of the energy nonconservation, but at the price of exclusion of the cross-correlation energy from quantum mechanics and particle field theory. The Copenhagen-Dirac {\it{postulate}}, which is the viable, if non-intuitive, opinion, has been suggested well before the modern understanding of quantum mechanics. Nevertheless, {\it{the huge progress of the quantum mechanics and quantum field theory based on the concept (postulate) of the "interference-less" (self-interfering) particles completely justified the Copenhagen-Dirac model of quantum interference}}. Although the problem associated with violation of the energy conservation by the cross-correlation energy has been solved by Dirac, the several principal questions remain unexplained up to now. For instance, in the quantum mechanics and quantum electrodynamics, the wave-functions ${\psi _n(\mathbf{r},t)}$ of the $N$ particles are not additive [$\psi (\mathbf{r},t) {\neq}\sum_{n=1}^N{\psi _n(\mathbf{r},t)}$] due to non-additivity of the probability amplitudes associated with the wave-functions of particles in the Copenhagen interpretation of quantum mechanics. In other words, the waves of probabilities associated with the wave-functions of particles are not additive because the particles never interfere with each other. In the classical electrodynamics, however, the electromagnetic field of the $N$ electromagnetic fields (waves) is considered as a sum ({\it{superposition}}) of the interfering electric [${\mathbf{E}_n(\mathbf{r},t)}$] and magnetic [${\mathbf{H}_n(\mathbf{r},t)}$] fields or the respective derivatives of the electromagnetic four-potential [${{A}_{\mu n}(\mathbf{r},t)}$] fields: $\mathbf{E}(\mathbf{r},t)=\sum_{n=1}^N{\mathbf{E}_n(\mathbf{r},t)}$, $\mathbf{H}(\mathbf{r},t)=\sum_{n=1}^N{\mathbf{H}_n(\mathbf{r},t)}$, and ${A}_{\mu}(\mathbf{r},t)=\sum_{n=1}^N{{A}_{\mu n}(\mathbf{r},t)}$. That contradicts the {\it{general}} correspondence principle that the quantum and classical treatments must be in agreement not only  for a very large number of particles (photons). The electric, magnetic and vector-potential fields of a particle (photon) in quantum electrodynamics are not attributed to the probability amplitude of the wave of probability associated with the wave-function of the photon. Strictly speaking, the generally accepted wave-function of a photon {\it{does not exist}} up to now. Furthermore, the traditional interpretation of quantum interference {\it{forbids existence}} of the cross-correlation energy (each particle  interferes only with itself), while the classical wave physics does use this energy in description of the pure additive or subtractive interference. Thus the field energy calculated by using the canonical interpretation of the quantum interference would {\it{never approach}} the classical value for the pure constructive or destructive interference. In addition, the particle energy ${\varepsilon}=\hbar\omega + (1/2)\hbar\omega$ calculated in the quantum field theory based on the Copenhagen-Dirac postulate is {\it{different}} from the Planck-Einstein particle energy ${\varepsilon}=\hbar\omega$~\cite{Planck,Einst1}. The quantum energy $(1/2)\hbar\omega$, which is usually identified as the vacuum energy, does {\it{contradict}} both the Einstein theory and classical physics of empty space. In particle field theory based on the Copenhagen-Dirac postulate, the force acting upon a particle is seen as the action of the respective gauge field that is present at the particle location. The force is attributed to the exchange of the field virtual particles. Such a physical picture {\it{does not compare well}} with the Einstein theory of general relativity, where the gravitational interaction between two objects (particles) is not viewed as a force, but rather, objects moving freely in gravitational fields travel under their own inertia in straight lines through curve spacetime~\cite{Einst2}. In spite of the Dirac solution of the problem of the nonconservation of energy under quantum interference, the nonconservation of the number of particles and energy in quantum field theories based on the Copenhagen-Dirac postulate does remain up to now one of the key problems. The founders of quantum mechanics and particle field theory Max Born, Werner Heisenberg, Erwin Schr{\"o}dinger, Pascual Jordan, and Paul Dirac already in the 1930s demonstrated that in perturbative calculations many integrals are divergent resulting in {\it{nonconservation}} of the number of particles and energy in quantum fields. Since then the different powerful renormalization procedures have been developed to avoid the divergences (for instance, see Refs. \cite{Tomo,Beth,Schw,Feyn}), but they {\it{did not provide}} the general solution of the problem. One could mention also the problem associated with the superluminal signalling in the {\it{canonical}} quantum mechanics (the Einstein-Podolsky-Rosen paradox~\cite{Ein}) that follows from the J.S. Bell inequalities~\cite{Bell}, but {\it{contradicts}} the Einstein special relativity. Finally, the Copenhagen {\it{interpretation}} ({\it{phylosophy}}) of the de Broglie wave associated with a particle as the wave of probability presents a more or less intuitively transparent background for the physical interpretation of quantum mechanics. Unfortunately, up to now, it is {\it{not completely clear}} how to interpret physically the wave (field) of operators in particle field theory. In any physical interpretation of the wave of operators, the operator would be rather a pure mathematical object than a real physical matter. In addition, the Copenhagen interpretation (philosophy) of the de Broglie wave using the pure mathematical object (probability), strictly speaking, {\it{did not solve}} the problem of physical interpretation of quantum mechanics. The probability or the wave of probability is  also rather a pure mathematical object than a real physical substance. In this regard, one could mention the well-known "ironic interpretation" of the {\it{canonical}} quantum mechanics attributing to Richard Feynman, "{\it{Shut up and calculate!}}". The similar problems do exist also in the quantum field theory based on the Copenhagen-Dirac postulate of "interference-less", self-interfering particles. With no solution of the aforementioned problems, especially of the problem of nonconservation of the number of particles and energy in quantum fields, it appears that an {\it{incompatibility}} exists between the quantum theories and the non-quantum physics (including the Einstein theory of relativity). For the detailed descriptions of classical wave physics, canonical quantum mechanics and particle field theory see the canonical articles and traditional textbooks~\cite{Land,Jack,Bere,Itz,Ryde,Pes,Grei,Wein} and the references therein. 

In the present study (Part I and II), the interference between particles and the interference-induced positive and negative cross-correlation energies, which do not exist from the point of view of the canonical quantum mechanics and particle field theory, are not ignored. In both the classical and quantum fields of the present model, the two-times increase of the wave amplitude does increase the wave energy in four times, and the wave with zero amplitude has zero energy. The problem of nonconservation of the energy and number of particles by the cross-correlation is overcame by taking into consideration the fact that creation of the conditions of the pure additive or subtractive interference of the waves (fields) requires "additional" energy that must be added or subtracted from the physical system. Then the "additional" energy does provide conservation of the total energy of the system. In Part I, the Hamiltonians describing the energy mediated by the cross-correlation (interference) in the basic classical and quantum fields are derived. The conditions of pure constructive or destructive interference are found by using these Hamiltonians. The influence of the interference-induced  cross-correlation energy on the energy, mass, charge, and momentum of the interfering fields is demonstrated. The calculated particle energy ${\varepsilon}=\hbar\omega=(  {\mathbf  p}^2  c ^2 + m^2c^4)^{1/2}$ is equal to the Planck-Einstein energy of the particle, but is different from the value ${\varepsilon}=\hbar\omega + (1/2)\hbar\omega$ of traditional quantum field theory. The calculated vacuum energy, in agreement with the Einstein and classical physics of the empty space, is equal to zero. In the present model, the unit-wave associated with a boson or fermion particle, unlike the wave of probability or operators in quantum mechanics or particle field theory, is a real, finite unit-wave (unit-field) of the physical matter, whose gradient ("curvature") can be changed spatially and/or temporally. The present model {\it{suggests the very simple physical interpretations}} of quantum phenomena. For instance, the physical interpretation of the position-momentum uncertainty of a particle (unit-field) in the present model is different from the some (non-generally accepted) interpretations of quantum mechanics, which are based on the "compression" of the pure mathematical object (probability) associated with a particle by the material boundaries of macroscopic objects. In the present model, the physical mechanism behind the position-momentum uncertainty is attributed simply to the increase of the gradient (spatial "curvature") of a real,  material unit-field (particle) by interaction and interference of the unit-field of the matter (mass-energy) with the boundaries of macroscopic objects. In order to account for the well-known discrepancies between measurements based on the mass of the visible matter in astronomy and cosmology and definitions of the mass made through dynamical or general relativistic means, the present model does not need in a hypothesis of existence of the "dark" mass. In the present model, the "dark" cosmological energy-mass as well as the well-known spiral cosmological structures are attributed to the cross-correlation energy-mass of the moving cosmological objects. According to the present model, the Bell superluminal signals (the Einstein-Podolsky-Rosen paradox) and the well-known astronomical objects, which propagate with velocities greater than the velocity of light, involve no physics incompatible with the theory of special relativity. The superluminal velocities, if they do really exist, are attributed rather to the physical properties of the unit-field like material mediums than the empty space (vacuum) of the Einstein special relativity. It is also shown that the positive or negative gradient of the cross-correlation energy does mediate the attractive or repulsive forces, respectively. These forces could be connected with the all-known classical and quantum fields, for instance, with the gravitational and Coulomb fields. In the present model, the physical mechanism behind the Bose-Einstein statistics, Fermi-Dirac statistics and Pauli exclusion principle is attributed to the attractive and repulsive forces associating with the subtractive and additive interference (cross-correlation) of the real unit-waves (bosons or fermions) of the matter. To this end, the model shows a key role of the interference-induced cross-correlation energy in several basic coherent quantum phenomena, such as the Bose-Einstein condensation, super-radiation, Bosenova effect, superfluidity, superconductivity, supermagnetism, and quantum anomalous and fractional Hall effects.

Part I of the present study is organized as follows. Section (1) provides an introduction to the problem associated with violation of the energy conservation by the cross-correlation energy in the classical and quantum physics under the pure additive or subtractive interference. The solution of the problem by the Copenhagen-Dirac interpretation of quantum interference, which forbids both the existence of interference between particles and the interference induced cross-correlation energy, is discussed. {\it{Motivation}} of the introduction of the interference between particles (bodies) and the respective cross-correlation energy into classical physics, quantum mechanics and particle field theory is presented. In Sec. (2), the energy conservation in the classical and quantum fields under the interference of fields is considered in the context of Noether's theorem~\cite{Noeth,Bye}. The Dirac interpretation of violation of the energy conservation under quantum interference is reanalyzed by taking into account the energy spent on the creation of the conditions of pure additive or subtractive interference. The Hamiltonians describing the energy induced by the cross-correlation (interference) in the basic classical and quantum fields are derived. The influence of the cross-correlation energy on the basic physical properties of boson and fermion fields is demonstrated. The energy, mass, charge, and momentum of the interfering fields are calculated. It is shown that the gradient of the cross-correlation energy induces the forces that could be attributed to all known fields, for instance, to the gravitational or Coulomb fields [Sec. (3)]. The conditions of pure constructive or destructive interference of classical and quantum fields are found by using the derived Hamiltonians in Sec. (4). The first-order and higher order coherence (interference) and the respective cross-correlation energies are discussed. In Sec. (5), the influence of the cross-correlation energy on the basic coherent properties of boson and fermion fields is demonstrated. The section shows a key role of the cross-correlation energy in several basic coherent physical phenomena, such as Bose-Einstein condensation, super-radiation, Bosenova effect, superfluidity, superconductivity, supermagnetism, and quantum anomalous and fractional Hall effects. The traditional approach of insertion of the interference and cross-correlation energy of particles into the many-particle models of the canonical quantum mechanics and particle field theory, which strictly forbid the existence of the interference (cross-correlation) between particles, is illustrated in Sec. 6 by analyzing the traditional consideration of a many-particle quantum system. The summary and conclusions are presented in Sec. (7). In order to make the interference between particles and the respective cross-correlation energy understandable {\it{not only to experts}} in the research field, the model concepts are {\it{analyzed and reanalyzed}} in many {\it{philosophic, mathematical and physical details}} in the context of each section of Secs. (1)-(7).

\section{2. Cross-correlation energy in boson and fermion fields}

The energy conservation law states that the field energy can be converted from one form to another, but it cannot be created or destroyed. The energy conservation is a mathematical consequence of the shift symmetry of time. More abstractly, energy is a generator of continuous time-shift symmetry of the physical system under study. When a physical system has a time-shift symmetry, Noether's theorem~\cite{Noeth} implies the existence of a conserved current. The thing that "flows" in the current is the energy, the energy is the generator of the symmetry group~\cite{Noeth,Bye}. {\it{In the modern physics, the quantum fields are modelled by using simultaneously the energy conservation law and the Copenhagen-Dirac postulate of "interference-less", self-interfering particles}}~\cite{Land,Jack,Bere,Itz,Ryde,Pes,Grei,Wein}. According to the traditional quantum field theory based on the postulate (interpretation) there is only one global (fundamental) spatially-infinite field of the particles of a specific (particular) kind, which must satisfy both the energy conservation and the "interference-less" properties of the  self-interfering particles. The particle is created or destroyed by interaction of the field with the fundamental (global) field of another kind. For an example, photons of the electromagnetic field are created or destroyed by interaction of the infinite electromagnetic field with the infinite field of electrons. When the fields satisfy the shift symmetry of time, the energy and number of particles are conserved. The interaction of fields results just into redistribution of the energy and number of particles in the space and time. The phenomenon of redistribution of the field energy (intensity) by the interaction of quantum fields is quit similar to the redistribution of the field energy (intensity) under the ordinary interference of classical fields. Therefore, the interaction of fields can be considered, at least formally, as the interference of fields. Under the canonical quantum interference, however, each particle interferes only with itself. The interference (cross-correlation) between two different particles never occurs. {\it{Thus in the modern physics, which is based on the Copenhagen-Dirac postulate of "interference-less", self-interfering particles, the interaction of particles is not attributed to the interference and the respective cross-correlation energy of classical or quantum fields}}. In addition, the boundaries located infinitely far from any spacetime point in the traditional quantum field theory do affect the field energy in this point instantaneously. In other words, movement of the boundaries located infinitely far from the space point does modulate the field in this point with zero time-delay, which is associated with infinite velocity of the signal propagation. If an ensemble of the interacting fields does not satisfy the shift symmetry of time, then the system does not conserve the total energy and number of particles. 

In contrast to the traditional approach, the present model suggests existence of an arbitrary number of the finite fields of particular (specific) kinds, for instance, the beams of photons or electrons. The spatially and temporally finite (quantum or classical) fields can interfere (cross-correlate) with each other and/or with the finite fields of another kind. For instance, the electron fields can interfere with each other, as well as with the photon fields. The interference between particles does increase or decrease not only the local density of particles, but at the pure additive or subtractive interference creates or destroys the total field energy and number of particles by the cross-correlation. The phenomenon of redistribution of the field energy (intensity) by the interaction of classical or quantum fields is considered as the interference of the classical or quantum fields. In the present model, the classical and quantum fields of elementary particles are considered as a superposition of the unit-fields associated with these particles. The material unit-field (unit-wave) associated with a particle, in contrast to the non-material wave of probability or operators in quantum mechanics or particle field theory, is a real, finite unit-wave (unit-field) of the mass-energy. Under the interference, each particle (unit-wave) can interfere with itself ({\it{quantum interference}}). However, unlike in the canonical quantum models, the unit-fields (particles) could interfere (interact) also with each other ({\it{classical interference}}). Thus the attractive or repulsive forces associated with interaction of the particles are associated simultaneously with the classical and quantum interference (cross-correlations) of the interfering (interacting) classical or quantum unit-fields. In other words, the positive or negative gradient of the field cross-correlation energy does induce the attractive or repulsive forces that redistribute the field energy (intensity). For both the classical and quantum waves (fields), the two-times increase of the wave amplitude does increase the wave energy in four times and the wave with zero amplitude has zero energy. In the frame of such an approach, the quantum and classical treatments of the fields are in agreement with each other for an arbitrary number of particles. It should be stressed that the physical system of the interfering (interacting) fields includes not only the interfering fields, but also the environment (material boundaries) that could provide the pure constructive or destructive interference. In other words, the fields and unit-fields (particles) do interfere and interact not only with each other, but also with the particles of boundaries of the physical system. Although the field energy is changed under the pure constructive or destructive interference of the fields (waves), the total energy of the physical system is conserved. The pure additive or subtractive interference is provided by the "additional" energy, which should be added or subtracted from the physical system of interfering fields before the interference of fields with each other. In such a case, the physical system of the interfering (interacting) fields does not obey the shift symmetry of time. Therefore, the energy nonconservation associated with the cross-correlation energy does not contradict the Noether theorem. The energy conservation of the total physical system is provided rather by the exchange of the field energy with the environment (usually, particles of material boundaries) than by the shift symmetry of time. In contrast to the pure additive or subtractive interference, the creation of conditions of the normal (not pure constructive or destructive) interference does not require the additional energy. The normal interference is a reversible process. The energy conservation of the total physical system under the normal interference is provided by the shift symmetry of time in the system. Although the cross-correlation (interference) of fields is relevant to the uncertainty of momentum and energy of a particle, the uncertainty associated with the cross-correlation should not be confused with the uncertainty attributing to the uncertainty principle~\cite{Heis1,Heis2}. In quantum mechanics, the non-commutation of the momentum operator with the coordinate operator results into the Heisenberg position-momentum uncertainty, which means it is impossible to determine simultaneously both the position and momentum of a particle with any great degree of accuracy or certainty. The lack of commutation of the time derivative operator with the time operator itself mathematically results into an uncertainty principle for time and energy: the longer the period of time, the more precisely energy can be defined. The problem will be analyzed later [see, Sec. (4.2.)] in details by using the Hamiltonians that include the cross-correlation energy.

\subsection{2.1. Boson scalar fields}
\subsubsection{2.1.1. Boson scalar fields without the cross-correlation}
The interference-induced cross-correlation energy of classical and quantum electromagnetic fields has been recently introduced and preliminary investigated in Refs.~\cite{Kukh1}(a-c). With the objective of deriving the Hamiltonians that take into account the interference, cross-correlation and cross-correlation energy of particles in other boson and fermion fields under classical or quantum interference, let me begin the analysis with the conventional consideration~\cite{Land,Jack,Bere,Itz,Ryde,Pes,Grei,Wein} of an infinite boson scalar (spin $s$ = 0) field $\psi (\mathbf{r},t)$, which is described by the Lagrangian density ${\cal L}=({\partial}_{\mu}{\psi}^*)({\partial}^{\mu}{\psi})-m^2{\psi}^*{\psi}$. The dynamics of the field is determined by the Euler-Lagrange equation of motion, which yields the well-known Klein-Gordon-Fock equation for the complex field $\psi$. The present study follows the theoretical formulation and notations of Refs.~\cite{Land} and \cite{Bere}, which use the natural units $\hbar=c=1$. Due to the spacetime shift symmetry, the field $\psi$ obeys the conservation laws for the energy and momentum. The Lagrangian density is invariant under the $U(1)$ local gauge transformation ($\psi \rightarrow \psi '=e^{i\alpha } \psi $) given rise to the conserved current. The continuity equation for the 4-vector of the Noether current density $j_\mu = m(\psi ^* (\partial_\mu \psi) -(\partial _\mu \psi ^*)\psi  )$ yields the conservation of the field charge (mass) $Q={\int}j_0d^3x$. The canonical, non-quantum Hamiltonian of the
system is given by
\begin{eqnarray} \label{eq1}
{\cal H}= {\int_{0}^{\infty}}hd^3x,
\end{eqnarray}
where
\begin{eqnarray}\label{eq2}
{{h}}= \frac {\partial {\psi ^*}} {\partial t} \frac {\partial {\psi}} {\partial t} +
\nabla {\psi ^*} \cdot \nabla {\psi} + m^2 {\psi^*} {\psi}  
\end{eqnarray}
is the Hamiltonian density. The second quantization of the field is performed by the multimode expansion of the fields $\psi$ and $\psi ^*$
and the subsequent replacement of the fields by the respective multimode field operators $\hat \psi$ and $\hat \psi ^{\dagger }$:
\begin{eqnarray} \label{eq3}
\hat \psi=\sum_{\mathbf{k}}{(2V{\varepsilon}_{\mathbf{k}})^{-1/2}}({\hat a}_{\mathbf{k}}e^{-i{\mathbf{k}
{\mathbf{r}}}}+{\hat b}_{\mathbf{k}}^{\dagger}e^{i{\mathbf{k}{\mathbf{r}}}}),
\end{eqnarray}
\begin{eqnarray} \label{eq4}
\hat \psi ^{\dagger }
=\sum_{\mathbf{k}}{(2V{\varepsilon}_{\mathbf{k}})^{-1/2}}({\hat
a}_{\mathbf{k}}^{\dagger} e^{i{\mathbf{k}{\mathbf{r}}}}+{\hat
b}_{\mathbf{k}}e^{-i{\mathbf{k}{\mathbf{r}}}}),
\end{eqnarray}
where ${ \varepsilon}_{ \mathbf {k} } =\omega _{ \mathbf {k}}= ({ \mathbf {k} }^2 + m^2)^{1/2}$ is the Planck-Einstein particle energy; 
${ \hat a^{\dagger}}_{ \mathbf {k} }$, $ { \hat a }_{ \mathbf {k}}$, 
${ \hat b^{ \dagger }}_{ \mathbf {k} }$, and ${ \hat b}_{\mathbf {k} }$ are
respectively the creation and destruction operators for the particles (bosons) and antiparticles (antibosons) inside the infinite volume $V$. The operators satisfy the canonical commutation relations $[ \hat a _{k_n},\hat a^{ \dagger }_{k_m}]$ = $[ \hat b_{k_n}, \hat b^{\dagger }_{k_m}]$ = $\delta _{nm}$ for the bosons and antibosons; the other operator pairs commute. Here, $\delta _{nm}$ is the Kronecker symbol. Equations (\ref{eq1})-(\ref{eq4}) yielded the canonical quantum Hamiltonian
\begin{eqnarray} \label{eq5}
{\hat {\cal H}}=\sum_{\mathbf{k}}{\varepsilon}_{\mathbf{k}}(\mathbf{\hat {\cal N}}_{\mathbf{k}}+
\mathbf{\hat {\bar {\cal  N}}}_{\mathbf{k}}+1/2+1/2)
\end{eqnarray}
corresponding the quantum field energy $\varepsilon = \langle {\hat {\cal H}} \rangle$, where $\mathbf{\hat {\cal N}}_{\mathbf{k}}={{\hat a}_{\mathbf{k}}^{\dagger }}{{\hat a}_{\mathbf{k}}}$
and $\mathbf{\hat {\bar {\cal  N}}}_{\mathbf{k}}={{\hat b}_{\mathbf{k}}^{\dagger }}{{\hat b}_{\mathbf{k}}}$
are the number operators for the particles and antiparticles, respectively. According to the traditional correspondence principle, the quantum and classical treatments must be in agreement for a very large number of particles. The quantum field energy $\varepsilon = \langle {\hat {\cal H}} \rangle$ is equal to the classical field energy $\varepsilon = {{\cal H}}$ for the very large number of particles: $\varepsilon = \langle {\hat {\cal H}} \rangle ={{\cal H}}$ if $\langle \mathbf{\hat {\cal N}}_{\mathbf{k}} \rangle  >>1/2$ and $\langle \mathbf{\hat {\bar {\cal  N}}}_{\mathbf{k}} \rangle  >>1/2$.

The above-presented, canonical model of boson fields may be summarized and {\it{interpreted}} as follows. In the model, unlike in canonical quantum mechanics, position is not an observable, and thus, one does not need the concept of a position-space probability density. The resonator-like boundaries are located infinitely far from any spacetime point. The wave-functions (resonator modes) $\psi_{ \mathbf {k} } (\mathbf{r})$ of the quantum mechanical system, in the non-operator or operator form, do not have a probabilistic interpretation in position space. The quantum mechanical system is considered as an infinite quantum field inside the infinite resonator. The field elementary degrees of freedom are the occupation numbers, and each occupation number is indexed by a vector number ${ \mathbf {k} }$, indicating which of the single-particle states it refers to. The properties of this global field are explored by defining creation and annihilation quantum operators with the canonical commutation relations imposed, which add and subtract particles. The field is considered as a set of degrees of freedom indexed by position, where the second quantization indexes the infinite field by enumerating the single-particle quantum states. Mathematically, the second quantization of the global (infinite) non-quantum field is performed by replacing the field by the respective field operator. In other words, the real multimode field is replaced by the multimode field of operators. The classical (\ref{eq1}) and quantum (\ref{eq5}) Hamiltonians are invariant under the $U(1)$ local gauge transformation ($\psi _{ \mathbf {k}} \rightarrow \psi ' _{ \mathbf {k}}=e^{i\alpha _ { \mathbf {k}}} \psi _{ \mathbf {k}}$). The $U(1)$ gauge symmetry of the classical Hamiltonian (\ref{eq1}), which is provided by the orthogonality of wave-functions (the resonator modes are orthogonal to each other in the Hilbert space), gives rise to the interference-less behavior of the wave-functions and the independence of the Hamiltonian on the phases $\alpha _ { \mathbf {k}}$. Each boson or antiboson interferes (interacts) only with itself due to the canonical commutation relations $[ \hat a _{k_n},\hat a^{ \dagger }_{k_m}]$ = $[ \hat b_{k_n}, \hat b^{\dagger }_{k_m}]$ = $\delta _{nm}$ (the other operator pairs commute), which have been constructed to satisfy the $U(1)$ gauge symmetry of the quantum Hamiltonian (\ref{eq5}). {\it{Although the strict interference-less behaviour of the n-th and m-th particles could be attributed to the commutation relations and/or the orthogonality of the wave-functions (the n-th and m-th resonator modes), the Hamiltonian symmetry also could be considered as a reason providing the interference-less behaviour of the particles.}} In order to control the absence of both the interference between particles and the respective cross-correlation energy, the field Hamiltonians have been engineered by using the $U(1)$ gauge symmetry. In other words, the interference, cross-correlation and interaction between two different bosons or antibosons never occurs, in agreement with the Copenhagen-Dirac (canonical) interpretation of quantum interference, due to the $U(1)$ gauge symmetry of the mathematically constructed  Hamiltonians. In accord with the {\it{canonical quantum interference}}, a particle (wave-function, resonator mode) interferes only with itself. Thus, for instance, the boson and antiboson could not annihilate each other due to the gauge symmetry. Both the boson and antiboson have the positive energy-mass ${\varepsilon}_{ \mathbf {k} }=({\mathbf{k}}^2+m^2)^{1/2}$, or more precisely, ${\varepsilon}_{ \mathbf {k} }=({\mathbf{k}}^2+m^2)^{1/2} (1+1/2)$ [see, Eq. (\ref{eq5})]. Notice, the positive energy-mass ${\varepsilon}_{ \mathbf {k} }= ({\mathbf{k}}^2+m^2)^{1/2}$  of the antiboson is different from the negative energy-mass ${\varepsilon}_{ \mathbf {k} }=- ({\mathbf{k}}^2+m^2)^{1/2}$ of the Dirac antiparticle. Although both the antiparticle with positive energy-mass and the antiparticle having negative energy-mass do satisfy the Einstein energy-mass relation ${\varepsilon}_{ \mathbf {k} }^2= {\mathbf{k}}^2+m^2$, the only negative energy-mass of an antiparticle provides physical annihilation of the positive energy-mass particle in the traditional quantum field models. Indeed, in the Dirac model of electrons and positrons, which is based on the Copenhagen-Dirac postulate, the antiparticle (positron) corresponds to the negative-energy solution of the Dirac equation. The physical interpretation of the negative energy (mass) of antiparticles  remains up to now one of the unsolved problems of traditional quantum field theories.   

In the above-presented model of the boson field theory, one should not confuse the wave (field) of operators with the wave of probability associated with the wave-function of a boson in the canonical interpretation of quantum mechanics. The fields corresponding the field operators  (\ref{eq3}) and (\ref{eq4}) are not attributed to the probability amplitude of the wave of probability associated with the wave-function of the boson or antiboson. Unlike in quantum mechanics, the generally accepted wave-function of a boson (antiboson) associated with the probability amplitude does not exist up to now. The properties of bosons in the boson field theory may be attributed rather to the properties of a wave of operators than the wave of probability of quantum mechanics. The Copenhagen (traditional) interpretation (philosophy) of the de Broglie wave associated with a boson (antiboson) as the wave of probability presents a more or less intuitively transparent background for the physical interpretation of quantum mechanics. Unfortunately, the physical interpretation of the wave (field) of operators in the boson field theory has not been presented up to now. Indeed, in any physical interpretation, an operator should be considered rather as a pure mathematical object than a real physical substance. It could be mentioned again in this regard that the canonical interpretation of the de Broglie wave using the pure mathematical object (probability) does not solve really the problem of physical interpretation of quantum mechanics. The wave of probability is also rather a pure mathematical object than a real physical matter. Notice, the dynamics of the above-considered infinite field of scalar bosons is determined by the Euler-Lagrange equation of motion in the Klein-Gordon-Fock form. Unlike in the traditional interpretation of the Klein-Gordon-Fock infinite field of bosons by the quantum field theory, the Klein-Gordon-Fock equation was originally formulated and interpreted by the authors as a {\it{single-particle equation}} analogous to the Schr{\"o}dinger equation. In the traditional quantum field theory, the resonator-like boundaries located infinitely far from any spacetime point of the infinite resonator do affect the global field and the field energy at this point instantaneously. Movement of the boundaries located infinitely far from a space point does change the resonator modes and modulate the field at this point with zero time-delay, which is associated with infinite velocity of the signal propagation. The infinite velocities of the signals in the resonator, which has never been interpreted physically, is incompatible with the Einstein theory of special relativity. Also note that the numbers of particles 1 and 1/2 in the expression ${\varepsilon}_{ \mathbf {k} }=({\mathbf{k}}^2+m^2)^{1/2} (1+1/2)$ [Eq. (\ref{eq5})] could be associated respectively with the whole particle (boson or anti-boson) and the half of the particle. Although the half-particles  have never been observed experimentally, the energy ${\varepsilon}_{ \mathbf {k} }=({\mathbf{k}}^2+m^2)^{1/2} (1/2)$ of such half-particles in the quantum field theory is usually interpreted as the vacuum energy.

\subsubsection{2.1.2. Boson scalar non-quantum and quantum fields with the cross-correlation }

Let me now suggest the existence of an arbitrary number $N$ of the finite boson scalar fields ${\psi _n(\mathbf{r},t)}$. These spatially and temporally finite fields (beams), unlike in the quantum mechanics and particle field theory based on the traditional interpretation of quantum interference, could interfere with each other. The model should include not only the increase or decrease of the local density of the energy and number of particles by the interference (cross-correlation) of the interfering (interacting) fields, but at the pure additive or subtractive interference should take into account the creation or destruction of the energy and number of particles. The dynamics of the field $\psi (\mathbf{r},t) =\sum_{n=1}^N{\psi _n(\mathbf{r},t)}$ of the interfering fields ${\psi _n(\mathbf{r},t)}$ is determined by the Euler-Lagrange equation of motion with the initial and boundary conditions imposed. The second quantization of these "non-quantum" fields, the replacement of the fields by the fields of operators, will be presented in the next subsection. For the sake of simplicity, the formulas in the present study are presented for the discrete number of fields. One can easily rewrite the formulas for the continuous spectra by integrating the number density. The generalization of the canonical non-quantum Hamiltonian (\ref{eq1}) for the superposition $\psi (\mathbf{r},t) =\sum_{n=1}^N{\psi _n(\mathbf{r},t)}$ of the interfering fields $\psi _n(\mathbf{r},t)$ yields 
\begin{eqnarray} \label{eq6}
{\cal H}={\int_{0}^{\infty}}hd^3x=\sum_{n=1}^N{\cal H}_{nn}+\sum_{n\neq m}^{N^2-N}{\cal H}_{nm},
\end{eqnarray}
where
\begin{eqnarray} \label{eq7}
{\cal H}_{nm}={\int_{0}^{\infty}}h_{nm}d^3x,
\end{eqnarray}
and 
\begin{eqnarray} \label{eq8}
{h}_{nm}= \frac {\partial {\psi ^*_n}}{\partial t} \frac {\partial {\psi_m}}{\partial t} +
\nabla {\psi ^*_n}\cdot \nabla {\psi_m} + m_n{\psi^*_n}m_m{\psi_m}.   
\end{eqnarray}
It could be mentioned that such a generalization, in fact, is the direct generalization of the Einstein famous energy-mass relation ${\varepsilon}^{2} ={\mathbf{k}}^2+m^2$ for the particles (bodies), which interfere with each other. For more details see Part II of the present study. Notice, the Hamiltonian density $h$ is not invariant under the $U(1)$ local gauge transformation ($\psi _n \rightarrow \psi ' _n=e^{i\alpha _n} \psi _n$) given rise to the interference between the non-quantum finite fields ${\psi _n(\mathbf{r},t)} \equiv \phi _n (\mathbf{r},t) e^{i\alpha _n} $ and ${\psi _m(\mathbf{r},t)} \equiv \phi _m(\mathbf{r},t) e^{i\alpha _m} $ and the dependence of the Hamiltonian on the phases $\alpha _n$ and $\alpha _m$. In the present model of {\it{unified}} fields and interactions, the classical and quantum treatments must be in agreement not only for a very large number of particles. This {\it{general}} correspondence principle could be considered as a {\it{generalization}} of the {\it{traditional}} correspondence principle of quantum physics. In contrast to the canonical model considered in Sec. 2.1.1. and in agreement with the {\it{general}} correspondence principle, the field energy in the present model is determined by the relation $\varepsilon = \langle {\hat {\cal H}} \rangle = {{\cal H}}$ for the classical and quantum fields with {\it{an arbitrary number of particles}}. The term ${\cal H}_{nn}$ is associated with the canonical energy of the single field ${\psi _n(\mathbf{r},t)}$. In the terms of the present model, the Hamiltonian ${\cal H}_{nn}$ describes the correlation (interference) of the finite field ${\psi _n(\mathbf{r},t)}$ with itself. In the limit of the infinite field ${\psi _n(\mathbf{r},t)}$, the self-correlation energy ${\cal H}_{nn}$ could be interpreted as the Hamiltonian (self-energy) of the single mode of the global (infinite) field, which is used in particle field theory before the second quantization of this (non-quantum) field. The cross-correlation energy ${\cal H}_{nm}$, which does not exist in quantum mechanics and particle field theory~\cite{Land,Jack,Bere,Itz,Ryde,Pes,Grei,Wein}, is attributed to the interference (cross-correlation) of the finite fields ${\psi _n(\mathbf{r},t)}$ with ${\psi _m(\mathbf{r},t)}$. It will be shown later [see, Sec. (6)] that the cross-correlation energy ${\cal H}_{nm}$ in the limit of the infinite fields ${\psi _n(\mathbf{r},t)}$ and ${\psi _m(\mathbf{r},t)}$ can be interpreted (remember that the cross-correlation energy associated with the ordinary interference does not exist in quantum mechanics and particle field theory) as the energy of interaction of the resonator modes of the infinite (global) field in particle field theory. The Hamiltonian (\ref{eq6}) describes the interference of the boson fields, ${\psi _n(\mathbf{r},t)}\equiv {\psi^b _{n}(\mathbf{r},t)}$. If the field ${\psi _n(\mathbf{r},t)}$ is composed from the boson field ${\psi^b _{n}(\mathbf{r},t)}$ and the antiboson field ${\psi ^{ab}_{n}(\mathbf{r},t)}$, then the Hamiltonian is described by Eqs. (\ref{eq6})-(\ref{eq8}), where ${\psi _n(\mathbf{r},t)}={\psi^b _{n}(\mathbf{r},t)}+{\psi ^{ab}_{n}(\mathbf{r},t)}$ and ${\psi^* _n(\mathbf{r},t)}={\psi^{*b} _{n}(\mathbf{r},t)}+{\psi ^{*ab}_{n}(\mathbf{r},t)}$. In the case of ${\psi^b _{n}(\mathbf{r},t)}=-{\psi ^{ab}_{n}(\mathbf{r},t)}$ and ${\psi^{*b} _{n}(\mathbf{r},t)}=-{\psi ^{*ab}_{n}(\mathbf{r},t)}$, the superposition of the fields results into their annihilation. It should be noted that the energy conservation of the total physical system under the annihilation is provided rather by the exchange of the energy with the environment (boundaries or other fields) than by the shift symmetry of time.

We now consider the Hamiltonian (6) of the interfering fields ${\psi _{n}(\mathbf{r},t)}$ and ${\psi _{m}(\mathbf{r},t)}$ in more details. At the time moment $t$ the finite fields ${\psi _{n}(\mathbf{r},t)}$ and ${\psi_{m}(\mathbf{r},t)}$ occupy respectively the finite volumes $V_n $ and $V_m $. The cross-correlation energy ${\cal H}_{nm}$ vanishes, ${\cal H}_{nm}={\cal H}_{nm} \delta _{nm}$, if the fields ${\psi _n (\mathbf {r},t)}$ and ${\psi_m (\mathbf {r},t)}$ do not overlap (interfere) with each other. Here, $\delta _{nm}$ is the Kronecker delta. Notice, the infinite fields occupying the infinite volume are overlapped in the whole space. In such a case, the field energy always includes both the canonical energies ${\cal H}_{nn}$ of the single fields ${\psi _{n}(\mathbf{r},t)}$ and the cross-correlation energies ${\cal H}_{nm}$ associated with the ordinary interference (cross-correlation) between the fields ${\psi _{n}(\mathbf{r},t)}$ and ${\psi_{m}(\mathbf{r},t)}$. In contrast to the infinite fields, the ordinary interference of the finite fields (beams) requires the creation of the interference conditions. In order to interfere, the finite fields ${\psi _n (\mathbf{r},t)}$ and ${\psi_m (\mathbf{r},t)}$ or their parts have to occupy (share) the same volume ${V_{nm} } = {V_n} = {V_m}$ at the same time moment  $ t_{nm} = {t_n} = { t_m}$. The total energy $\cal H$ at the moment $ t_{nm}$ is calculated by integrating the Hamiltonian density $h$ in the infinite volume. The Hamiltonian contains the canonical energy of the single fields ${\psi _n (\mathbf{r},t)}$ occupying the volumes $V_n$ and the cross-correlation energy of the fields overlapped in the volume ${V_{nm}}$ at the time $ t_{nm}$:
\begin{eqnarray} \label{eq9}
{\cal H}= \sum_{n=1}^N{\int_{V_n}}{h}_{nn}d^3x+\sum_{n\neq{m}}^{N^2-N}{\int_{V_{nm}}}{h}_{nm}d^3x.
\end{eqnarray}
According to Eq. (\ref{eq9}), the canonical {\it{self-energies}} ${\cal H}_{nn}$ of the fields ${\psi _{n}(\mathbf{r},t)}$ are permanent (constant) in the time, while the cross-correlation energy ${\cal H}_{nm}$ depends on the time moment of the overlapping (interference) of the finite fields ${\psi _{n}(\mathbf{r},t)}$ with ${\psi_{m}(\mathbf{r},t)}$: ${\cal H}(t)=\sum_{n=1}^N{\cal H}_{nn}+\sum_{{n\neq m}}^{N^2-N}{\cal H}_{nm}(t)$. The difference $\Delta {\cal H}(t)={\cal H}(t)-\sum_{n=1}^N{\cal H}_{nn}=\sum_{{n\neq m}}^{N^2-N}{\cal H}_{nm}(t)$ between the generalized energy (\ref{eq6}) and the canonical energy $\sum_{n=1}^N{\cal H}_{nn}$ logically to call the {\it{cross-correlation energy or the defect of energy-mass}}. Notice, the boundaries located infinitely far from the finite field, unlike in the case of an infinite field of particles in quantum field theory, do not affect the field and energy. In other words, movement of the boundaries located infinitely far in the transverse or longitudinal direction from the finite field (beam) could not modulate this field. One should not confuse vanishing the cross-correlation energy ${\cal H}_{nm}$ by non-overlapping the finite fields ${\psi _{n}(\mathbf{r},t)}$ and ${\psi_{m}(\mathbf{r},t)}$ with the case of the "orthonormal" fields, which satisfy the well-known condition ${\int_{ V_{nm}}}{\psi^*_n(\mathbf{r})}{\psi_m(\mathbf{r})}  d^3x = \delta _{nm} $ of the orthonormality in the Hilbert space. The orthonormal fields do overlap, but the cross-correlation integrals (\ref{eq7}) are equal to zero due to the orthonormality, ${\cal H}_{nm}={\cal H}_{nm}\delta _{nm}$. Among the orthonormal fields that satisfy the Euler-Lagrange equation of motion  with the initial and boundary conditions imposed one could mention the transverse and longitudinal (stationary or transient~\cite{Kukh1}(d)) modes, the Fourier decomposition terms (time harmonic plane-waves with different spatial and/or temporal frequencies), and other eigensolutions. 

The above-considered boson fields can exist in the form of the waves, as well as the static (time-independent) fields. Let me first consider the wave-like fields. The configuration and dynamics of the wave-like field are determined by the Euler-Lagrange equation of motion with the initial and boundary conditions imposed. The simplest solution of the equation for the finite wave ${\psi _{n}(\mathbf{r},t)}$, which is located infinitely far from the boundaries, is the time-harmonic (stationary) wave
\begin{eqnarray} \label{eq10}
\psi_n (\mathbf{r},t)={a}_ne^{-i{\varepsilon}_nt}e^{i({\mathbf{k}_n{\mathbf{r}}}-\alpha _n)}.
\end{eqnarray}
Notice, ${\varepsilon}_n = {\omega }_n$ in the natural units. Other solutions of the Euler-Lagrange equation are considered in Part II of the present study. In the initial time moment $t= t_{n}=0$, the field ${\psi _{n}(\mathbf{r},t)}$ occupies the finite volume ${V_n}$. The spatial distribution of the field in this volume is given by  
\begin{eqnarray} \label{eq11}
\psi_n (\mathbf{r})=a_n e^{i({\mathbf{k}_n{\mathbf{r}}}-\alpha _n)},
\end{eqnarray}
where ${\alpha _n}$ is the wave phase. At the time moment $t= t_{nm}\neq 0$, the Hamiltonian ${\cal H}=\sum_{n=1}^N{\cal H}_{nn}+\sum_{{n\neq m}}^{N^2-N}{\cal H}_{nm}(t)$ of the superposition $\psi (\mathbf{r},t) =\sum_{n=1}^N{\psi _n(\mathbf{r},t)}$ of the cross-correlating fields ${\psi _{n}(\mathbf{r},t)}$  is given by Eq. (\ref{eq9}):
\begin{eqnarray} \label{eq12}
{\cal H}=\sum_{n=1}^N{\int_{V_n}}[\varepsilon_n^2+\mathbf{k}_n^2+m_n^2]{a_n^2}d^3x+\sum_{n\neq{m}}^{N^2-N}{\int_{V_{nm}}}[{\varepsilon_n}{\varepsilon_m}+{\mathbf{k}_n}{\mathbf{k}_m }+{m_n}{m_m}]{a_n{a_m}}e^{-i({\Delta \mathbf{k}_{nm}{\mathbf{r}}}-{\Delta \varepsilon_{nm}{t}}-{\Delta \alpha _{nm}})}d^3x,
\end{eqnarray}
where $\Delta \mathbf{k}_{nm}=\mathbf{k}_n-\mathbf{k}_m$, ${\Delta \varepsilon_{nm}}=\varepsilon_{n} - \varepsilon_{m}$ and $\Delta \alpha _{nm}=\alpha _{n}-\alpha _m$. The first term of the right-hand side of Eq. (\ref{eq12}) does not take into account the  cross-correlation phenomenon associated with the ordinary interference of the fields ${\psi _{n}(\mathbf{r},t)}$. The term is associated with the field superposition $\psi (\mathbf{r},t) =\sum_{n=1}^N{\psi _n(\mathbf{r},t)}$, where each field ${\psi _{n}(\mathbf{r},t)}$ correlates (interferes) only with itself. The total energy of the $N$ correlation-free fields ${\psi _{n}(\mathbf{r},t)}$ is permanent [${\cal H}\neq {\cal H}(t)$] and its value does not depend on the interference conditions: 
\begin{eqnarray} \label{eq13}
{\cal H}=\sum_{n=1}^N{\cal H}_{nn}= \sum_{n=1}^N[\varepsilon_n^2+\mathbf{k}_n^2+m_n^2]{a_n^2}V_n. 
\end{eqnarray}
The Hamiltonian    
\begin{eqnarray} \label{eq14}
{\cal H}_{nn}=[\varepsilon_n^2+\mathbf{k}_n^2+m_n^2]{a_n^2}V_n 
\end{eqnarray}
corresponds to the canonical Hamiltonian of the single field (interference-less resonator mode) $\psi_{\mathbf{k}_n} (\mathbf{r},t)$ in the traditional particle field theory (Sec. 2.1.1.). The second term of the right-hand side of Eq. (\ref{eq12}) plays a key role in description of the increase or decrease of the local density of the field energy under the interference. The positive or negative cross-correlation energy $\sum_{{n\neq m}}^{N^2-N}{\cal H}_{nm}(t)$ is responsible for the energy associated with the interference (cross-correlation) between the waves $\psi_{\mathbf{k}_n} (\mathbf{r},t)$ and $\psi_{\mathbf{k}_m} (\mathbf{r},t)$ that can change the total energy of the interfering fields. If the waves interfere only destructively in the common volume ${V_{nm}}$, then the interference of waves completely destroys the wave energy. The interference increases the wave energy if the waves interfere only constructively. The energy of interfering waves can be increased or completely destroyed in an ensemble of the coherent waves by the appropriate modification of the wave phases $\alpha _{n}$. The pure additive or subtractive interference is provided by the "additional" energy, which should be added or subtracted from the total physical system before the interference. In such a case, the physical system does not obey the shift symmetry of time. Therefore, the energy nonconservation associated with the cross-correlation energy does not contradict the Noether theorem. The energy conservation of the total physical system is provided rather by the exchange of the energy with the environment (material boundaries or external fields) than by the shift symmetry of time. {\it{The physical mechanism}} behind the cross-correlation energy in the above-considered non-quantum boson fields is simply the phenomenon of classical (ordinary) interference.

Let me now suggest existence of a basic unit of the boson field, an indivisible unit-field ${\psi _{0}(\mathbf{r},t)}$ of the matter (mass-energy), which could be associated with a boson (particle). The Hamiltonian of the field $\psi_{n} (\mathbf{r},t) =\sum_{n=1}^N{\psi _{0n}(\mathbf{r},t)}$, which is the superposition of indivisible unit-fields ${\psi _{0n}(\mathbf{r},t)}$, is given by Eqs. (\ref{eq6}) and (\ref{eq7}) with the Hamiltonian cross-correlation density ${h}_{nm}= \frac {\partial {\psi ^*_{0n}}}{\partial t} \frac {\partial {\psi_{0m}}}{\partial t} + \nabla {\psi ^*_{0n}}\cdot \nabla {\psi_{0m}} + m_n{\psi^*_{0n}}m_m{\psi_{0m}}$. The concrete parameters (shape, spatial and temporal distributions and phase) of the indivisible unit-field ${\psi _{0n}(\mathbf{r},t)}$, which depend on the experimental conditions, are determined by the Euler-Lagrange equation of motion with the initial and boundary conditions imposed. According to the {\it{general}} correspondence principle, the energy ${\cal H}_{11}={\int_{0}^{\infty}}h_{11}d^3x$ of a boson unit-field ${\psi _{0}(\mathbf{r},t)}$ should be equal to the Planck-Einstein particle energy ${\varepsilon}_0=\omega _0=({\mathbf{k}}_0^2+m_0^2)^{1/2}$ independently from the concrete form of this unit-field: ${\cal H}_{11}={\int_{0}^{\infty}}[{\frac {\partial {\psi ^*_0}}{\partial t} \frac {\partial {\psi_0}}{\partial t} +
\nabla {\psi ^*_0}\cdot \nabla {\psi_0} + m_0{\psi^*_0}m_0{\psi_0}}]d^3x=({\mathbf{k}}_0^2+m_0^2)^{1/2}$. In the case of a free boson describing by the time-harmonic unit-field (unit-wave) located infinitely far from the boundaries [see, Eqs. (\ref{eq10}) and (\ref{eq14})], the correspondence principle yields the boson energy    
\begin{eqnarray} \label{eq15}
{\cal H}_{11}=[\varepsilon_0^2+\mathbf{k}_0^2+m_0^2]{a_0^2}V_0=[{\mathbf{k}}_0^2+m_0^2]^{1/2}, 
\end{eqnarray}
which may be called the self-correlation energy, self-interference energy or simply self-energy. Then the boson unit-wave (de Broglie's wave) of the mass-energy associated with the particle is given by 
\begin{eqnarray} \label{eq16}
\psi_0 (\mathbf{r},t)=a_0e^{-i{\varepsilon}_0t}e^{i({\mathbf{k}_0{\mathbf{r}}}-\alpha )},
\end{eqnarray}
where the amplitude $a_0=(2V_0 [\mathbf{k}_0^2+m_0^2]^{1/2})^{-1/2}=(2V_0 {\varepsilon}_0)^{-1/2}$ [see, Eqs. (\ref{eq3}) and (\ref{eq4}) for comparison of the field amplitudes]. Remember, ${\varepsilon}_0 = {\omega }_0$ in the natural units. It could be noted that other solutions of the Euler-Lagrange equation determining the unit-field configuration ${\psi _{0}(\mathbf{r},t)}$ are considered in Part II of the present study. In the case of the superposition $\psi (\mathbf{r},t) =\sum_{n=1}^N{\psi _{0n}(\mathbf{r},t)}$  of the correlation-free unit-waves $\psi_{0n} (\mathbf{r},t)=a_{0}e^{i{\varepsilon}_0t}e^{-i({\mathbf{k}_{0n}{\mathbf{r}}}-\alpha _n)}$, the Hamiltonian ${\cal H}=\sum_{n=1}^N{\cal H}_{nn}$ is given by Eq. (\ref{eq13}):
\begin{eqnarray} \label{eq17}
{\cal H}=N{{\varepsilon}_0}, 
\end{eqnarray}
where ${\cal H}\neq {\cal H}(t)$. The number of the interference-free unit-waves, which is a positive integer, is then given by 
\begin{eqnarray} \label{eq18}
{N}={{\varepsilon}_0}^{-1}{\cal H},  
\end{eqnarray}
where $N\neq N(t)$. For the  $N$ cross-correlating unit-waves ${\psi _{0n}(\mathbf{r},t)}$ describing by the Hamiltonian ${\cal H}={\cal H}(t)=\sum_{n=1}^N{\cal H}_{nn}+\sum_{{n\neq m}}^{N^2-N}{\cal H}_{nm}(t)$, Eq. (\ref{eq12}) yields
\begin{eqnarray} \label{eq19}
{ \cal H }= \sum_{n=1}^N \varepsilon_{0n} + \sum_{n \neq{m} }^{N^2-N} { \int_{V_{nm} }} 
[ {\varepsilon}_{0n} {\varepsilon}_{0m} + { \mathbf{k}_{0n} } { \mathbf{k}_{0m} } + m_{0n} m_{0m} ] a_{0n} a_{0m} e^{{-i( \Delta \mathbf{k}_{0nm} {\mathbf{r}}-\Delta \varepsilon_{0nm} {t}-\Delta \alpha _{nm}})}d^3x.
\end{eqnarray}
The expression (\ref{eq19}) could be considered as the direct {\it{generalization}} of the Planck-Einstein energy  ${\varepsilon}_0 = {\omega }_0 = ({\mathbf {k}}_0^2 + m_0^2)^{1/2}$ for the $N$ interfering particles (bodies). The generalized energy of the body composed from the $N$ interfering bodies (particles) is given by the energy ${\cal H} \neq N ({\mathbf {k}}_0^2 + m_0^2)^{1/2}$. The difference $\Delta {\cal H}(t) = \sum_{n > {m}}^{N^2-N} {\int_{V_{nm} }} [{ \varepsilon}_{0n} {\varepsilon}_{0m} + { {\mathbf {k}}_{0n} } { {\mathbf{k}}_{0m} } + m_{0n} m_{0m} ] a_{0n} a_{0m}2 cos( \Delta {\mathbf {k}}_{0nm} {\mathbf {r}} - \Delta \varepsilon_{0nm}t - \Delta \alpha _{nm})d^3x $ between the generalized energy (\ref{eq19}) and the Planck-Einstein self-correlation energy $ \sum_{n=1}^N \varepsilon_{0n}$ is the cross-correlation energy (defect of energy-mass) of the interfering particles (unit-fields). Notice, the energy-mass defect (cross-correlation energy-mass), which depends on the unit-field phases $\alpha _{n}$ and $\alpha _{m}$, can be positive or negative. The sign depends on the interference conditions determining by the values $\Delta {\mathbf {k}}_{0nm} {\mathbf {r}} - \Delta \varepsilon_{0nm}t - \Delta \alpha _{nm}$. In the case of a single particle ($N=1$), the generalized energy (\ref{eq19}) is equal to the Planck-Einstein energy ${ \cal H }={\varepsilon}_0$ of the particle. The energy (\ref{eq19}) of a body composed from the $N$ incoherent unit-waves (particles) has the conventional value ${ \cal H }=N{\varepsilon}_0$. In other words, the conventional (Planck-Einstein) energy-mass is the pure incoherent energy-mass, while the generalized energy (\ref{eq19}) includes the additional energy, namely the coherent cross-correlation energy-mass (energy-mass defect). That is a difference between the conventional and generalized energy-masses. It should be stressed that the {\it{physical mechanism}} behind the cross-correlation energy-mass in Eq. (\ref{eq19}) is the interference between particles (unit-fields). In the case of $V_{0n} = V_{0m} = {V_{0}}$, $m_{0n} = m_{0m} = m_{0}$, ${\mathbf {k}}_{0n} = {\mathbf {k}}_{0m} = {\mathbf {k}}_{0}$ and ${\varepsilon}_{0n} = {\varepsilon}_{0m} = {\varepsilon}_{0}$, Eq. (\ref{eq19}) is given by 
\begin{eqnarray} \label{eq20}
{\cal H}={\cal N}{{\varepsilon}_0}, 
\end{eqnarray}
where 
\begin{eqnarray} \label{eq21}
{\cal N}=N+\sum_{n\neq{m}}^{N^2-N}e^{i\Delta \alpha _{mn}}
\end{eqnarray}
is the effective number  ${\cal N}$ of the interfering unit-waves (particles), which depends on the interference conditions determining by the phases differences $\Delta \alpha _{mn}$; the effective number ${\cal N}$ can be a non-integer, for instance, fractional number bigger or smaller than one. The respective defect energy (mass) is given by $\Delta {\cal H} = {{\varepsilon}_0}\sum_{n\neq{m}}^{N^2-N}e^{i\Delta \alpha _{mn}}$. One can easily demonstrate that the energy of the cross-correlating unit-waves 
\begin{eqnarray} \label{eq22}
0\leq{ \cal H }\leq N^2{{\varepsilon}_0},
\end{eqnarray}
and the respective effective number of the unit-waves     
\begin{eqnarray} \label{eq23}
0\leq{{\cal N}}\leq N^2. 
\end{eqnarray}
For instance, Eqs. (\ref{eq20}) and (\ref{eq21}) yielded the energy ${\cal H}=N^2{{\varepsilon}_0}$ and the effective number ${\cal N}=N^2$ for the pure additive interference ($\Delta \alpha _{nm}=0$). The pure subtractive interference results into annihilation of the unit-waves and energy. The annihilation takes place if the pairs of unit-waves satisfy the phase condition $\alpha _n-\alpha _m=\pm \pi$. The annihilation ${\psi^b _{0n}(\mathbf{r},t)}+{\psi ^{ab}_{0n}(\mathbf{r},t)}=0$ of the pairs of unit-waves can be represented as the pure subtractive interference of the boson unit-fields 
\begin{eqnarray} \label{eq24}
\psi_{0n}^b (\mathbf{r},t)={a}_0e^{-i{\varepsilon}_0t}e^{i({\mathbf{k}_{0}{\mathbf{r}}}-{\alpha _n})} 
\end{eqnarray}
with the antiboson unit-fields
\begin{eqnarray} \label{eq25}
\psi_{0n}^{ab} (\mathbf{r},t)={a}_0e^{-i{\varepsilon}_0t} e^{i({\mathbf{k}_0{\mathbf{r}}}-{\alpha _n} )} e^{\pm i\pi} .
\end{eqnarray}
In the case of $N_b=N_{ab}$, the pure subtractive interference of the boson field 
\begin{eqnarray} \label{eq26}
\psi^b(\mathbf{r},t) =\sum_{n=1}^{N_b}{\psi _{0n}^b}(\mathbf{r},t) 
\end{eqnarray}
with the antiboson field 
\begin{eqnarray} \label{eq27}
\psi ^{ab}(\mathbf{r},t) =\sum_{n=1}^{N_{ab}} \psi _{0n}^{ab}(\mathbf{r},t),
\end{eqnarray}
results into annihilation of the fields and energy. It should be stressed again that the energy conservation of the total physical system under the annihilation is provided rather by the exchange of the energy with the environment (for instance, boundaries or other fields) than by the shift symmetry of time. The unit-waves do annihilate only partially if $\alpha _n-\alpha _m=\pm \pi  \pm \delta $, where $|\delta| <\pi $. In order to provide the annihilation of a boson unit-wave with an antiboson unit-wave for the arbitrary wave-number (momentum) vectors and phases, the waves should satisfy the annihilation condition 
\begin{eqnarray} \label{eq28}
{\psi^b _{0}(\mathbf{r},t;{\mathbf{k}}_{0}^b,\alpha _b )}+{\psi ^{ab}_{0}(\mathbf{r},t;{\mathbf{k}}_{0}^{ab}, \alpha _{ab})}=0
\end{eqnarray}
independently from the values of the wave momentum (wave-number) vectors and phases. In such a case, the antiboson unit-field is the true antiparticle unit-field, which cannot be obtained by the phase shift of the boson field. In the case of ${\mathbf{k}}_{0}^b={\mathbf{k}}_{0}^{ab}=0$ and $\alpha _b= \alpha _{ab} = 0$, however, the massive ($m_0\neq 0$) fields only satisfy the condition (\ref{eq28}). Therefore, the mass-less fields do not have the true antiparticle unit-fields (antiparticles). According to Eqs. (\ref{eq6}) and (\ref{eq28}), the energy ${\cal H}=0$ of the annihilated boson-antiboson pair can be presented as the sum ${\cal H}_{11}^{b-b}+{\cal H}_{22}^{ab-ab}+(-{\cal H}_{12}^{b-ab})+(-{\cal H}_{21}^{ab-b})=0$ of the positive boson-boson self-correlation energy ${\cal H}_{11}^{b-b}$, the positive antiboson-antiboson self-correlation energy ${\cal H}_{22}^{ab-ab}$, the negative boson-antiboson cross-correlation energy $(-{\cal H}_{12}^{b-ab})$ and the negative antiboson-boson cross-correlation energy $(-{\cal H}_{21}^{ab-b})$. That means the energies of both the boson and antiboson unit-fields (bosons and antibosons) are always positive. The negative signs of the boson-antiboson and antiboson-boson cross-correlation energies do reflect just a fact of the destructive interference (cross-correlation) of the boson and antiboson unit-waves. Notice, the positive sign of the energy-mass ${\varepsilon}_0=({\mathbf{k}}_0^2+m_0^2)^{1/2}$ of the antiboson is different from the negative sign of the energy-mass ${\varepsilon}_0=-({\mathbf{k}}_0^2+m_0^2)^{1/2}$ of the Dirac antiparticle satisfying the Einstein condition ${\varepsilon}_0^2= [- ({\mathbf{k}}_0^2+m_0^2)^{1/2}]^2$. 

The above-described model can be applied for other physical parameters of the cross-correlating unit-waves (fields) ${\psi _{0n}}(\mathbf{r},t)$. For instance, the momentum $\mathbf{P}$ of the field $\psi(\mathbf{r},t ) =\sum_{n=1}^{N}{\psi _{0n}}(\mathbf{r},t)$ is given by 
\begin{eqnarray} \label{eq29}
\mathbf{P}={\cal P}^i \mathbf{e}_i,
\end{eqnarray}
where 
\begin{eqnarray} \label{eq30}
{\cal P}^i=\sum_{n=1}^N{{\cal P}}_{nn}^i+\sum_{n\neq{m}}^{N^2-N}{{\cal P}}_{nm}^i,
\end{eqnarray}
\begin{eqnarray} \label{eq31}
{{\cal P}}_{nm}^i={\int_{0}^{\infty}} p_{nm}^id^3x,
\end{eqnarray}
and 
\begin{eqnarray} \label{eq32}
p_{nm}^i= -\frac {\partial {\psi ^*_{0n}}}{\partial t} \frac {\partial {\psi_{0m}}}{\partial x^i}-\frac {\partial {\psi _{0n}}}{\partial t} \frac {\partial {\psi ^*_{0m}}}{\partial x^i}.    
\end{eqnarray}
Here, $p_{nm}^i$ is the momentum density, which takes into account the cross-correlation of the unit-fields ${\psi _{0n}}(\mathbf{r},t)$. The equation (\ref{eq30}) yields
\begin{eqnarray} \label{eq33}
{\cal P}^i=\sum_{n=1}^N{{k}_{0n}^i} +\sum_{n\neq{m}}^{N^2-N}{\int_{V_{nm}}}[{\varepsilon}_{0n}{{k}_{0m}^i}+{\varepsilon}_{0m}{{k}_{0n}^i}]a_{0n}{a_{0m}}e^{{-i( \Delta \mathbf{k}_{0nm} {\mathbf{r}}-\Delta \varepsilon_{0nm} {t}-\Delta \alpha _{nm}})}d^3x,
\end{eqnarray}
where $a_{0n}=(2V_{0n} {\varepsilon}_{0n})^{-1/2}$ and $a_{0m}=(2V_{0m} {\varepsilon}_{0m})^{-1/2}$. If the values $V_{0n} = V_{0m} = {V_{0}}$, $m_{0n} = m_{0m} = m_{0}$ and ${\mathbf {k}}_{0n} = {\mathbf {k}}_{0m} = {\mathbf {k}}_{0}$ (respectively, ${\varepsilon}_{0n} = {\varepsilon}_{0m} = {\varepsilon}_{0}$), then Eq. (\ref{eq29}) is given by 
\begin{eqnarray} \label{eq34}
\mathbf{P}={\cal N}{\mathbf{k}_{0} }, 
\end{eqnarray}
where $\cal N$ is the effective number (\ref{eq21}) of the unit-waves. Notice, the angular momentum $\mathbf{L}$ of the interfering unit-fields can be found by the similar generalization of the angular momentum $\mathbf{L}=\mathbf{r}\times \mathbf{P}$ or the relativistic angular momentum tensor $L^{nm}= \sum (x^np^m-x^mp^n)$. The generalized  rest mass ${\cal M}$ ($0\leq{{\cal M}}\leq N^2m_0$) of the system, which is determined by Eq. (\ref{eq20}) with ${\mathbf{k}}_{0}\rightarrow 0$, is given by 
\begin{eqnarray} \label{eq35}
{\cal M}={\cal N}m_0,
\end{eqnarray}
where $m_0=\varepsilon_0$. Indeed, for ${\varepsilon}_0=({\mathbf{k}}_0^2+m_0^2)^{1/2}$ and ${\mathbf{k}}_{0} \rightarrow 0$, the energy ${\varepsilon}_0 \rightarrow  m_0$. The expression (\ref{eq35}) could be considered as a {\it{generalization}} of the Einstein energy-mass relation ${\varepsilon}=mc^2$ (${\varepsilon}=m$ in the natural units) for the $N$ interfering particles (bodies). The generalized rest mass of the body composed from the $N$ interfering bodies (particles) is given by the mass ${ \cal M }\neq Nm_0$. The difference $\Delta {\cal M}=({\cal N}-N)m_0$ between the generalized rest mass (\ref{eq35}) and the conventional (Newton-Einstein) rest mass $Nm_0$ logically to call the cross-correlation rest mass or the rest-mass defect. In the case of a single particle ($N=1$), the generalized rest mass (\ref{eq35}) is equal to the Newton-Einstein rest mass  ${\cal M}=m_0$ of the particle. The generalized rest mass (\ref{eq35}) of a body composed from the $N$ incoherent unit-waves (particles) has the conventional value ${\cal M}=Nm_0$. It should be stressed, in this regard, that the cross-correlation (coherent) mass does not exist in the Newton and Einstein mechanics. In astrophysics and cosmology, the invisible mass is usually called the "dark" mass. In order to account for the well-known discrepancies between measurements based on the mass of the visible matter in astronomy and cosmology and definitions of the mass made through dynamical or general relativistic means, the present model does not need in hypothesizing the existence of "dark" mass. In the present model, the "dark" cosmological mass as well as the well-known spiral cosmological structures are associated with the cross-correlation (coherent) energy-mass of the moving cosmological objects. Note that the generalized mass ${\cal M}=0$ of an annihilated boson-antiboson pair can be presented as the sum of the positive boson-boson self-correlation mass ${\cal M}_{11}^{b-b}=m^b_0=m_0$, the positive antiboson-antiboson self-correlation mass ${\cal M}_{22}^{ab-ab}=m^{ab}_0=m_0$, the negative boson-antiboson cross-correlation mass $(-{\cal M}_{12}^{b-ab})=-(m^b_0)^{1/2}(m^{ab}_0)^{1/2}=-m_0$ and the negative antiboson-boson cross-correlation mass $(-{\cal M}_{21}^{ab-b})=-(m^{ab}_0)^{1/2}(m^{b}_0)^{1/2}=-m_0$:
\begin{eqnarray} \label{eq36}
{\cal M}={\cal M}_{11}^{b-b}+{\cal M}_{22}^{ab-ab}+(-{\cal M}_{12}^{b-ab})+(-{\cal M}_{21}^{ab-b})=0.
\end{eqnarray}
That means the masses of both the boson and antiboson unit-fields (bosons and antibosons) are always positive. The negative signs of the boson-antiboson and antiboson-boson cross-correlation masses do reflect just a fact of the destructive interference (cross-correlation) of the boson and antiboson unit-waves [see, the comments to Eq. (\ref{eq28})]. It should be stressed again that the positive sign of the rest mass (energy ${\varepsilon}_0=m_0$) of the antibosons in the present model is different from the negative sign of the rest mass (energy ${\varepsilon}_0=-(m_0^2)^{1/2}$) of the Dirac antiparticle. If the unit-charge $q_0$ is associated with the mass $m_0$ of the unit-wave ${\psi _{0n}}(\mathbf{r},t)$ [see comments to Eq. (\ref{eq120}), Eq. (\ref{eq121}) and Figs. (5)-(8)], then the effective charge ${\cal Q}$ of the cross-correlating unit-fields is given by 
\begin{eqnarray} \label{eq37}
{\cal Q}={\cal N}{q}_0,
\end{eqnarray}
where $0\leq{{\cal Q}}\leq N^2 q_0$. Similarly, if the magnetic moment ${\vec {\mu }}_0$ is associated with the mass $m_0$ of the unit-wave ${\psi _{0n}}(\mathbf{r})$, then the effective magnetic moment $\vec {M}$ of the cross-correlating unit-fields is given by 
\begin{eqnarray}
{ \vec {M}}={\cal N}{\vec {\mu }}_0
\end{eqnarray}
with $0\leq{{\vec {M}}}\leq N^2 {\vec {m}}_0$. In the case of the pure subtractive interference of the boson field (\ref{eq26}) with the antiboson field (\ref{eq27}), the effective charge is given by the sum of the positive charge of the boson field with the negative charge of the antiboson field: 
\begin{eqnarray} \label{eq38}
{\cal Q}={\cal N}_b q_0+{\cal N}_{ab}(-q_0).
\end{eqnarray}
Although the charge and the charge-associated parameters of the particle (boson or antiboson) have been formally introduced [see, Eqs. (\ref{eq37})-(\ref{eq38})] into the scalar boson model, the existence or nonexistence of the boson and antiboson charges is determined rather by the empirical data than the model. In other words, the introduction or non-introduction of the charges into the model should correspond to the experimental observations of the nature of concrete field. 

The basic properties of the interfering particles (unit-waves) of the above-presented model can be summarized as follows. {\it{The unit-fields of the present model are not the point particles and waves (particle-wave dualism) of quantum mechanics or the point particles of quantum field theory. The unit-fields have simultaneously the properties of the point particles and waves of quantum mechanics, the point particles of quantum field theory, as well as the properties of the 3-dimensional (in space) waves of theory of classical fields.}} The Hamiltonian of the superposition of indivisible material unit-waves ${\psi _{0n}(\mathbf{r},t)}$ is given by Eqs. (\ref{eq6}) and (\ref{eq7}), with the Hamiltonian cross-correlation density ${h}_{nm}= \frac {\partial {\psi ^*_{0n}}}{\partial t} \frac {\partial {\psi_{0m}}}{\partial t} + \nabla {\psi ^*_{0n}}\cdot \nabla {\psi_{0m}} + m_n{\psi^*_{0n}}m_m{\psi_{0m}}$. The exact form of the indivisible unit-field ${\psi _{0n}(\mathbf{r},t)}$, which depends on the experimental conditions, is determined by the Euler-Lagrange equation of motion with the initial and boundary conditions imposed. In the case of the boson fields composed from the time-harmonic finite unit-fields located infinitely far from the boundaries, the above-derived number of unit-waves, magnetic moment, mass and charge of the waves are equal to the respective values of the traditional particle field theory if the unit-fields are free from the cross-correlation associated with the ordinary interference phenomenon. However, the calculated energy ${\varepsilon}_0=\omega_0=(  {\mathbf k}_0^2  + m_0^2)^{1/2}$ and momentum ${\mathbf  p}_0={\mathbf  k}_0$ of the unit-wave (particle), which are equal to the Planck-Einstein energy and momentum of the particle, are different from the canonical values ${\varepsilon}_0=\omega_0 + (1/2)\omega_0$ and ${\mathbf  p}_0={\mathbf  k}_0+(1/2){\mathbf  k}_0$ of traditional particle field theory. The positive energy-mass ${\varepsilon}_0=({\mathbf{k}}_0^2+m_0^2)^{1/2}$ of antibosons in the present model is different from the negative energy-mass ${\varepsilon}_0=- ({\mathbf{k}}_0^2+m_0^2)^{1/2}$ of the Dirac antiparticles satisfying the energy-mass relation ${\varepsilon}_0^2= [- ({\mathbf{k}}_0^2+m_0^2)^{1/2}]^2$. The energy and momentum of the empty vacuum, in agreement with the Einstein and classical physics energies of the empty space, are equal to zero. In the present model, the dynamics of the finite fields is determined by the Euler-Lagrange equation of motion with the initial and boundary conditions imposed. The physical parameters of the cross-correlating fields have been derived for the relativistic fields with the unit-field energy ${\varepsilon}_0=( \mathbf{k}_0^2 + m_0^2 )^{1/2}$. One can easily demonstrate that the physical parameters of the non-relativistic ($ \mathbf{k}_0 \rightarrow 0$) fields are described by Eqs. (\ref{eq6})-(\ref{eq38}), where the unit-field energy is given by $ \varepsilon_0 \approx ( \mathbf{k}_0^2 / 2m_0) + m_0$. The absolute minimum of energy of the unit-field in both the relativistic and non-relativistic cases is given by ${\varepsilon}_0^{min} = m_0$. Notice, the aforementioned transition from the relativistic cross-correlating boson unit-fields to non-relativistic ones is somewhat similar to the transition from the relativistic correlation-free spinor (electron) infinite field describing by the Dirac equation of motion with the resonator-like boundary conditions to the non-relativistic correlation-free finite waves of probabilities (finite wave-functions), whose dynamics is determined by the Schr{\"o}dinger equation with the quantum mechanical boundary conditions imposed. In this regard, one could mention that the Klein-Gordon-Fock equation describing dynamics of the infinite boson field was {\it{originally formulated and interpreted by the authors}} as a single-particle equation similar to the Schr{\"o}dinger equation. It should be {\it{stressed again}} that the physical system of the interfering boson-fields contains not only the interfering fields, but also the boundaries that may provide the pure constructive or destructive interference. The energy of interfering fields is changed under the pure constructive or destructive interference, however, the total energy of the physical system is conserved. The external energy, which is added or subtracted from the interfering fields before the interference, insures the pure additive or subtractive interference. Thus the physical system of interfering fields does not exhibit the shift symmetry of time. The energy conservation of the physical system is provided rather by the irreversible exchange of the energy with the environment (boundaries or other fields) than by the shift symmetry of time. Therefore, the energy nonconservation associated with the cross-correlation energy is in agreement witht the Noether theorem. In contrast to the pure constructive or destructive interference, the creation of conditions of the normal (not pure constructive or destructive) interference of the boson fields does not require the additional energy from the environment. Indeed, the energy conservation of the total physical system under the normal interference of the  boson fields is provided by the shift symmetry of time in such a system. 

The above-described model of the boson unit-fields suggests the {\it{simple interpretations and explanations}} of the quantum phenomena associated with the bosons. Indeed, the Copenhagen interpretation (philosophy) of the de Broglie wave using the pure mathematical object (probability) does not solve really the problem of physical interpretation of the quantum mechanics of bosons. In any physical interpretation, the wave of probability associated with the boson would be rather a pure mathematical object than a real material substance. In the boson field theory, up to now, it is not completely clear how to interpret physically the wave (field) of boson operators. The boson operator is also rather a pure mathematical object than a real physical matter. The real finite-wave or real finite unit-wave of the boson matter (mass-energy) described in Sec. 2.1.2 is not the wave of probability of quantum mechanics or the wave of operators of the canonical particle field theory. The division of the "non-quantum" field $\psi$ of the boson matter (mass-energy) into the indivisible "non-quantum" unit-fields $ \psi_{0n} $ of the boson matter (mass-energy) [see, Eqs. (\ref{eq15})-(\ref{eq38})], in fact, is the {\it{second quantization of the real, material boson field without the use of the non-material fields of operators or probabilities}}. The second quantization performed by replacing the "non-quantum" unit-fields of the boson matter (mass-energy) by the non-material unit-fields of boson operators is presented in the next section. 

Finally, it should be noted that the unified description of the interfering elementary particles (unit-fields) is given in Part II of the present study, where the Hamiltonian ${\cal H}_{11}$ of the massive (the rest mass $m_0\neq 0$) particle is associated with the unit-field energy squared (${\cal H}_{11}={\varepsilon}_0^{2} ={\mathbf{k}}_0^2+m_0^2$). Respectively, the Hamiltonian ${\cal H}$ of the massive field composed from the $N$ particles (unit-fields) is attributed to the total field-energy squared (${\cal H}={\varepsilon}^{2}=\sum_{n=1}^N{\cal H}_{nn}+\sum_{{n\neq m}}^{N^2-N}{\cal H}_{nm}$). In such a case, the physical parameters of the interfering massive unit-fields are described by the above-presented equations with $0\leq{ \varepsilon  }\leq N{{\varepsilon}_0}$ and $0\leq{{\cal N}}\leq N$. In Part I of the present study, the Hamiltonian ${\cal H}_{11}$ of the particle is associated with the unit-field energy ${\cal H}_{11}={\varepsilon}_0 =[{\mathbf{k}}_0^2+m_0^2]^{1/2}$, and the Hamiltonian ${\cal H}$ of the field composed from the $N$ particles having the non-zero rest mass is attributed to the total-energy ${\cal H}={\varepsilon}=\sum_{n=1}^N{\cal H}_{nn}+\sum_{{n\neq m}}^{N^2-N}{\cal H}_{nm}$. The energy of the cross-correlating massive unit-fields is then given by $0\leq{ \cal H }\leq N^2{{\varepsilon}_0}$ with the  respective effective number $0\leq{{\cal N}}\leq N^2$ of the unit-fields. 

\subsubsection{2.1.3. Boson scalar quantum-fields of operators with the cross-correlation}

According to the traditional field theory~\cite{Dir2,Land,Jack,Bere,Itz,Ryde,Pes,Grei,Wein}, the second quantization of the non-quantum fields $\psi$ and $\psi ^*$, which are free from the interference, cross-correlation and cross-correlation energy, is performed by replacing the fields by the respective field operators $\hat \psi$ and  $\hat \psi ^{\dagger }$ [see, Sec. (2.1.1.)]. I would apply this approach to the cross-correlating "non-quantum" fields described in the previous section. 
The operator ${\hat \psi}_{0} $ of the unit-field
\begin{eqnarray} \label{eq39}
\psi_0 (\mathbf{r},t)=a_0e^{-i{\varepsilon}_0t}e^{i({\mathbf{k}_0{\mathbf{r}}}-\alpha )}
\end{eqnarray}
is given by 
\begin{eqnarray} \label{eq40}
{\hat \psi}_{0}= {\hat a}a_0e^{i({\mathbf{k}_{0}{\mathbf{r}}}-{i\alpha })},
\end{eqnarray}
and the operator ${\hat \psi ^{\dagger }}_0$ of the unit-field
\begin{eqnarray} \label{eq41}
\psi_{0} ^*(\mathbf{r},t)=a_0e^{i{\varepsilon}_0t}e^{-i({\mathbf{k}_{0}{\mathbf{r}}}-\alpha )}
\end{eqnarray}
is given by 
\begin{eqnarray} \label{eq42}
{\hat \psi ^{\dagger }}_0= {\hat a}^ {\dagger}a_0e^{-i({\mathbf{k}_{0}{\mathbf{r}}}-{i\alpha })},
\end{eqnarray}
where $a_0= {(2V_0{\varepsilon}_{0}})^{-1/2}$ and ${\varepsilon}_0 =[{\mathbf{k}}_0^2+m_0^2]^{1/2}$ [see, also Eqs. (\ref{eq3}), (\ref{eq4}) and (\ref{eq16})]. Notice, the field operators (\ref{eq40}) and (\ref{eq42}) are similar to the respective boson parts of the multimode operators (\ref{eq3}) and (\ref{eq4}). According to the {\it{general}} correspondence principle $\varepsilon = \langle {\hat {\cal H}} \rangle = {{\cal H}}$ for both the classical and quantum fields with an arbitrary number of particles. Therefore the energy ${\varepsilon}_0={\cal H}_{11}={\int_{0}^{\infty}}[{\frac {\partial {\psi ^*_0}}{\partial t} \frac {\partial {\psi_0}}{\partial t} +
\nabla {\psi ^*_0}\cdot \nabla {\psi_0} + m_n{\psi^*_0}m_m{\psi_0}}]d^3x$ of a single particle (a boson unit-field) is given by
\begin{eqnarray} \label{eq43}
{\varepsilon}_0 =  {\cal H}_{11}= {\int_{V_0}}{a_0e^{-i({\mathbf{k}_{0}{\mathbf{r}}}-\alpha )}}[({\hat a}^ {\dagger}{\hat a}+{\hat a}{\hat a}^ {\dagger})/2}][{\varepsilon}_{0}^2+{\mathbf{k}_0^2+m_0^2]a_0e^{i({\mathbf{k}_{0}{\mathbf{r}}}-\alpha )}d^3x,
\end{eqnarray}
or in the conventional (Dirac) notations
\begin{eqnarray} \label{eq44}
\varepsilon_0 = { \cal H }_{11} =  \langle \psi_0 | {\hat {\cal H}}_{11} | \psi_0 \rangle = \langle 1 | {\hat {\cal H}} | 1 \rangle  ,
\end{eqnarray}
where
\begin{eqnarray} \label{eq45}
{\hat {\cal H}}={\varepsilon}_0({\hat a}^ {\dagger}{\hat a}+{\hat a}{\hat a}^{\dagger})/2 
\end{eqnarray}
is the Hamiltonian operator of the quantum unit-field (boson) presented in the form, which does not depend on the order of the operators $\hat a$ and  $\hat a^{\dagger }$. Equation (\ref{eq43}) yields the "commutator like" relation $({\hat a}^ {\dagger}{\hat a}+{\hat a}{\hat a}^ {\dagger})/2=1$. One can use also the equivalent non-symmetric operator 
\begin{eqnarray} \label{eq46}
{\hat {\cal H}}={\varepsilon}_0{\hat a}^{\dagger}{\hat a}, 
\end{eqnarray}
or
\begin{eqnarray} \label{eq47}
{\hat {\cal H}}={\varepsilon}_0{\hat a}{\hat a}^{\dagger}, 
\end{eqnarray}
with the "commutator like" relation ${\hat a}^{\dagger}{\hat a}={\hat a}{\hat a}^{\dagger}=1$. Note that the energy of the quantum unit-field (boson) describing by the operator equations (\ref{eq43})-(\ref{eq45}) can be considered, in terms of the virtual processes (transitions) of the traditional particle field theory, as the product of the virtual processes of the simultaneous creation and destruction of the unit-field (boson) at the every time moment $t$. If a quantum unit-field (boson) is absent ($\psi_{0} (\mathbf{r})={\hat \psi_{0}} (\mathbf{r}) = 0\cdot \psi_{0} (\mathbf{r}) =0\cdot {\hat \psi_{0}} (\mathbf{r}) =0$) in the volume $V_0$, then the energy of the empty space associated with the absence of the boson in this volume is equal to zero: 
\begin{eqnarray} \label{eq48}
 {\cal H}_{00}= {\int_{V_0}}{[0\cdot a_0]e^{-i({\mathbf{k}_{0}{\mathbf{r}}}-\alpha )}}[({\hat a}^ {\dagger}{\hat a}+{\hat a}{\hat a}^ {\dagger})/2}][{\varepsilon}_{0}^2+{\mathbf{k}_0^2+m_0^2][0\cdot a_0]e^{i({\mathbf{k}_{0}{\mathbf{r}}}-\alpha )}d^3x=0 
\end{eqnarray}
or in the Dirac notations
\begin{eqnarray} \label{eq49}
{\cal H}_{00}=\langle 0\cdot  \psi_0 |{\hat {\cal H}}_{00} | 0\cdot  \psi_0 \rangle = 0 \cdot \langle \psi_0 |{\hat {\cal H}}_{00} | \psi_0 \rangle = \langle 0  |{\hat {\cal H}} | 0 \rangle =0.  
\end{eqnarray}
That also means the unit-wave (unit-field) with zero amplitude has zero energy. It should be stressed that the quantum particle energy $ {\cal H}_{11}$ calculated by using the operator equation (\ref{eq43}) is equal to the Planck-Einstein energy ${\varepsilon}_0=\omega _0=({\mathbf{k}}_0^2+m_0^2)^{1/2}$ of a single boson, as well as to the energy of the "non-quantum" unit-field describing by the non-operator equation (\ref{eq15}). Remember that the division of the "non-quantum" field $\psi$ into the indivisible "non-quantum" unit-fields $ \psi_{0n} $ [see, Eqs. (\ref{eq15})-(\ref{eq38})], in fact, is the {\it{second quantization of the field without the use of the field operators}}. Therefore, the energy of the "non-quantum" unit-field describing by the non-operator equation (\ref{eq15}), in fact, is the quantum energy of the quantum unit-field (particle) in the field of operators. If the unit-field (boson) is absent in the volume $V_0$, then the vacuum quantum energy (\ref{eq48}), the Planck-Einstein energy of the empty volume, and the "non-quantum" unit-field energy (\ref{eq15}) of the empty space (zero-energy of a zero-amplitude unit-wave) are equal to zero. In order to obtain the non-zero vacuum energy in the present model, one can impose the additional condition for the operator pairs, namely the canonical commutation relation $[{\hat a},{\hat a^{\dagger }}]=1$ of quantum mechanics and particle field theory. The use of this commutation relation in Eqs. (\ref{eq43}) and (\ref{eq48}) yields  
\begin{eqnarray} \label{eq50}
\langle 1 |{\hat {\cal H}} | 1 \rangle = {\varepsilon}_0+ (1/2) {\varepsilon}_0
\end{eqnarray}
and 
\begin{eqnarray} \label{eq51}
\langle 0 |{\hat {\cal H}} | 0 \rangle = 0, 
\end{eqnarray}
respectively. The quantum energy (\ref{eq50}), which is equal to the canonical quantum energy of a single boson in p{article field theory~\cite{Dir1,Dir2,Land,Jack,Bere,Itz,Ryde,Pes,Grei,Wein}, is different from the Planck-Einstein particle energy ${\varepsilon}_0=\omega _0=({\mathbf{k}}_0^2+m_0^2)^{1/2}$ and the particle energy (\ref{eq15}) of the "non-quantum" unit-field (particle). The vacuum energy (\ref{eq51}) could be in artificial agreement with the canonical quantum value $\langle 0 |{\hat {\cal H}} | 0 \rangle = {\varepsilon}_0/2$ if the zero energy ${\cal H}_{00} =0$ determining by Eq. (\ref{eq48}) is assumed to be equal to ${\varepsilon}_0/2$. Mathematically, that means the redefinition ($0\cdot a \neq 0$) of the canonical zero. Then the particle energy (\ref{eq50}) can be presented in the form $\langle 1 |{\hat {\cal H}} | 1 \rangle = {\varepsilon}_0+ (1/2) {\varepsilon}_0=\langle 1 |{\hat {\cal H}} | 1 \rangle + \langle 0 |{\hat {\cal H}} | 0 \rangle$, which corresponds to the particle energy of traditional quantum field theories. In such a case, the half of the boson energy (mass $m_0$) remains in the volume $V_0$ when a boson leaves this volume. In other words, the half of the boson remains in the volume. Furthermore, the infinite energy (mass) of the half-particles is accumulated in the volume $V_0$ if the infinite number of bosons leaves this volume. Although the canonical values $\langle 1 |{\hat {\cal H}} | 1 \rangle = {\varepsilon}_0+ (1/2) {\varepsilon}_0 $ and $\langle 0 |{\hat {\cal H}}| 0 \rangle = (1/2) {\varepsilon}_0 $ of the particle field theory~\cite{Dir1,Dir2,Land,Jack,Bere,Itz,Ryde,Pes,Grei,Wein} can be obtained by the above-described artificial increase of the energies (\ref{eq43}) and (\ref{eq48}) on the value ${\varepsilon}_0/2$, {\it{I will not use this procedure}}. In the present model, the particle quantum energy (\ref{eq43}) is {\it{equal}} to the Planck-Einstein particle energy ${\varepsilon}_0=\omega _0=({\mathbf{k}}_0^2+m_0^2)^{1/2}$, as well as to the particle energy (15) of the "non-quantum" unit-field (particle), but is different from the {\it{canonical value}} of quantum field theory. The vacuum energy ${\cal H}_{00}=\langle 0 |{\hat {\cal H}}_{00} | 0 \rangle=0$ determining by Eq. (\ref{eq48}), which is the quantum energy of an empty volume, is in agreement with the Planck-Einstein vacuum energy (Fig.~1). One should not confuse here the vacuum energy of the Einstein spacial relativity and Newton mechanics having the "straight" geometry of the spacetime field with that of the spacetime "curve" geometry of the Einstein general relativity. The energies of the "straight" space ({\it{absolute vacuum}}) of the Einstein spacial relativity and Newton mechanics are equal to {\it{zero}}, while the energy of the Einstein "curve" space ({\it{non-absolute vacuum}})  has the {\it{non-zero}} value. Here, one should not confuse also the absolute minimum ${\varepsilon}_0^{min}=m_0 $ of the energy of the unit-field (particle) with the minimum of the canonical vacuum energy $\langle 0 |{\hat {\cal H}}| 0 \rangle = m_0/2$ connected with the particle. It could be mentioned, in this regard, that the problem of negative energies of the antiparticles of traditional quantum field theories will be discussed in the context of the present model at the end of Sec. (2.1.3).
\begin{figure}
\begin{center}
\includegraphics[keepaspectratio, width=0.5\columnwidth]{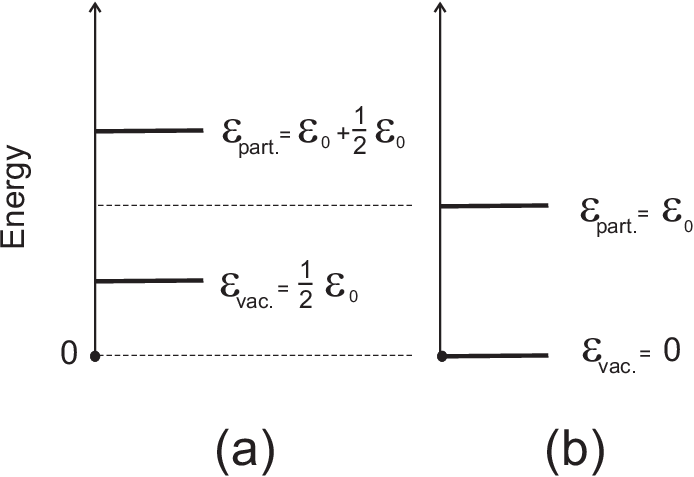}
\end{center}
\caption{(a) The particle (${\varepsilon}_{part.}={\cal H}_{11}$) and vacuum (${\varepsilon}_{vac.}={\cal H}_{00}$) energies in the traditional quantum field theory. (b) The particle (${\varepsilon}_{part.}={\cal H}_{11}$) and vacuum (${\varepsilon}_{vac.}={\cal H}_{00}$) energies in the Planck and Einstein theories and the present model. Here, ${\varepsilon}_0=\omega _0=({\mathbf{k}}_0^2+m_0^2)^{1/2}$ is the Planck-Einstein particle energy.}
\label{fig:Fig1}
\end{figure}

We now consider the Hamiltonian operator of the superposition of quantum operator unit-fields. The operators of the fields 
\begin{eqnarray} \label{eq52}
\psi (\mathbf{r},t) =\sum_{n=1}^N{\psi _{0n}(\mathbf{r},t)}
\end{eqnarray}
and 
\begin{eqnarray} \label{eq53}
\psi ^*(\mathbf{r},t) =\sum_{n=1}^N{\psi_{0n}^*(\mathbf{r},t)}, 
\end{eqnarray}
which are the superpositions of the respective unit-fields $ \psi_{0n} (\mathbf{r},t)=a_{0n}e^{-i{\varepsilon}_0t}e^{i({\mathbf{k}_{0n}{\mathbf{r}}}-\alpha _n)}$ and $ \psi_{0n} (\mathbf{r})^*=a_{0n}e^{i{\varepsilon}_0t}e^{-i({\mathbf{k}_{0n}{\mathbf{r}}}-\alpha _n)}$, are respectively given by the operators
\begin{eqnarray} \label{eq54}
{\hat \psi}= \sum_{n=1}^N{\hat a}_na_{0n}e^{i({\mathbf{k}_{0n}{\mathbf{r}}}-{\alpha _n})}
\end{eqnarray}
and
\begin{eqnarray} \label{eq55}
{\hat \psi ^{\dagger }} =\sum_{n=1}^N{\hat a}_n^ {\dagger}a_{0n}e^{-i({\mathbf{k}_{0n}{\mathbf{r}}}-{\alpha _n})},
\end{eqnarray}
where
$a_{0n}={(2V_{0n}{\varepsilon}_{0}})^{-1/2}$. In the case of the superposition $\psi (\mathbf{r},t) =\sum_{n=1}^N{\psi _{0n}(\mathbf{r},t)}$  of the unit-waves $\psi_{0n} (\mathbf{r},t)=a_{0n}e^{-i{\varepsilon}_{0n}t}e^{i({\mathbf{k}_{0n}{\mathbf{r}}}-\alpha _n)}$, which are free from the cross-correlation, the energy ${\cal H}=\sum_{n=1}^N{\cal H}_{nn}+\sum_{{n\neq m}}^N{\cal H}_{nm}=\sum_{n=1}^N{\cal H}_{nn}$  determining by the field operators is given by 
\begin{eqnarray} \label{eq56}
 {\cal H}= \sum_{n=1}^N {\int_{V_{0n}}}{a_{0n}e^{-i( {\mathbf{k}_{0n} {\mathbf{r}}}- \alpha  _n )}}[({\hat a}_n^ {\dagger}{\hat a}_n+{\hat a}_n{\hat a}_n^ {\dagger})/2][\varepsilon _{0n}^2+\mathbf{k}_{0n}^2+m_0^2]a_{0n}e^{i( {\mathbf{k}_{0n}{\mathbf{r}}}- \alpha  _ n)}d^3x={\varepsilon}_0 N,
\end{eqnarray}
or in the conventional (Dirac) notations
\begin{eqnarray} \label{eq57}
{\cal H}= \langle \psi | \sum_{n=1}^N{\hat {\cal H}_{nn}} | \psi \rangle =\langle N |{\hat {\cal H}} | N \rangle = \varepsilon _0 N.
\end{eqnarray}
Here, the Hamiltonian operator
\begin{eqnarray} \label{eq58}
{\hat {\cal H}}={{\varepsilon}_0}{\hat N}
\end{eqnarray}
and the particle number operator
\begin{eqnarray} \label{eq59}
{\hat N} =({\hat a}_n^ {\dagger}{\hat a}_n+{\hat a}_n{\hat a}_n^ {\dagger})/2=({\hat a}^ {\dagger}{\hat a}+{\hat a}{\hat a}^ {\dagger})/2
\end{eqnarray}
are determined by the commutation-like relation 
\begin{eqnarray} \label{eq60}
({\hat a}^ {\dagger}{\hat a}+{\hat a}{\hat a}^ {\dagger})/2=1.
\end{eqnarray}
The number of the correlation-free unit-waves (particles), which is an integer, is given by 
\begin{eqnarray} \label{eq61}
N = \langle N |{\hat N} | N \rangle ={\varepsilon}_0^{-1}{\langle N |{\hat {\cal H}} | N \rangle}.
\end{eqnarray}
The quantum energy (\ref{eq57}) and the quantum number of particles (\ref{eq61})  of the unit-waves (particles), which are free from the cross-correlation, are equal to the respective non-quantum (classical) values determining by Eqs. (\ref{eq17}) and (\ref{eq18}) for an arbitrary number of the unit waves (particles). This result is different from the traditional particle field theory, where the quantum and classical values are in agreement in the case of very big number of particles, only. The quantum Hamiltonian ${\cal H}$ of the  $ N$ cross-correlating unit-waves ${\psi _{0n}(\mathbf{r})}$ determining by the field operators is given by
\begin{eqnarray} \label{eq62}
{\cal H}=\sum_{n=1}^N{\cal H}_{nn}+\sum_{{n\neq m}}^{N^2-N}{\cal H}_{nm},
\end{eqnarray}
where the "self-correlation" energy
\begin{eqnarray} \label{eq63}
\sum_{n=1}^N{\cal H}_{nn}={{\varepsilon}_0}N,
\end{eqnarray}
and the cross-correlation energy
\begin{eqnarray} \label{eq64}
\sum_{n\neq m}^{N^2-N}{\cal H}_{nm}=\sum_{n\neq m}^{N^2-N}{\int_{V_{nm}}}[({\hat a}^{\dagger}_n{\hat a}_m + {\hat a}_n{\hat a}^{\dagger}_m)/2][{\varepsilon}_{0n}{\varepsilon}_{0m}+{\mathbf{k}_{0n}}{\mathbf{k}_{0m} }+{m_{0n}}{m_{0m}}]a_{0n}a_{0m}e^{{-i( \Delta \mathbf{k}_{0nm} {\mathbf{r}}-\Delta \alpha _{nm}})}d^3x,
\end{eqnarray} 
with the commutation-like relation 
\begin{eqnarray} \label{eq65}
({\hat a}_n^ {\dagger}{\hat a}_m+{\hat a}_n{\hat a}_m^ {\dagger})/2=e^{{i\Delta \varepsilon_{0nm} {t}}}.
\end{eqnarray}
Notice, the relation (\ref{eq65}) is obtained by comparison of Eq. (\ref{eq64}) with the time-dependent part of Eq. (\ref{eq19}). In the case of the "non-quantum" unit-fields considered in the previous section, the cross-correlation energy has been attributed to the ordinary interference phenomenon. The cross-correlation energy of the quantum fields describing by the operators could be attributed to the quantum interference phenomenon. Notice, the ordinary and quantum interference are indistinguishable phenomena in the present model. That means that the "non-quantum" and quantum energies are equal to each other for an arbitrary number of the unit-field (particles). In the terms of the virtual processes (transitions) of the traditional particle field theory, the quantum cross-correlation energy can be considered as the product of the virtual exchange of the indistinguishable unit-fields (bosons) [see, Eq. (\ref{eq64})]. In other words, the cross-correlation energy could be attributed to the simultaneous creation and destruction of the unit-field (boson) ${\psi _{0n}(\mathbf{r})}$ and the simultaneous destruction and creation of the unit-field (boson) ${\psi _{0m}(\mathbf{r})}$. The cross-correlation in such a virtual process is associated with the indistinguishableness of identical unit-fields ${\psi _{0n}(\mathbf{r})}$ and ${\psi _{0m}(\mathbf{r})}$ under the interference (interaction). The total energy of the interfering unit-fields [see, Eq. (\ref{eq62})] consists of the self-correlation energy (\ref{eq56}) and the cross-correlation energy (\ref{eq64}), which do correspond to the simultaneous creation/destruction of the unit-fields (bosons) ${\psi _{0n}(\mathbf{r})}$ and the simultaneous destruction/creation of the unit-fields (bosons) ${\psi _{0m}(\mathbf{r})}$ in the self-correlation and cross-correlation (interference) processes. The virtual quantum exchange of the indistinguishable unit-waves (particles) under the non-quantum or quantum interference is quite similar to the exchange of virtual particles for a short time ($\Delta t \leq 1/\Delta \varepsilon $) in the traditional perturbative particle field theory. It should be stressed, however, that the cross-correlation energy (\ref{eq64}) does {\it{not exist}} in canonical quantum mechanics and particle field theory based on the Copenhagen-Dirac postulate of "interference-less", self-interfering particles~\cite{Dir1,Dir2,Land,Jack,Bere,Itz,Ryde,Pes,Grei,Wein}.

In the case of $V_{0n} = V_{0m} = {V_{0}}$, $m_{0n} = m_{0m} = m_{0}$ and ${\mathbf {k}}_{0n} = {\mathbf {k}}_{0m} = {\mathbf {k}}_{0}$, Eqs. (\ref{eq62})-(\ref{eq65}) yielded the equation
\begin{eqnarray} \label{eq66}
{\cal H}= {{\varepsilon}_0}N +\sum_{n\neq{m}}^{N^2-N}{\int_{V_{0}}} [ ({\hat a}^ {\dagger}_n{\hat a}_m+{\hat a}_n{\hat a}^ {\dagger}_m)/2] [{{\varepsilon}_0}^2+{\mathbf{k}_{0}^2}+{m_{0}^2}]a_{0n}a_{0m}e^{i{\Delta \alpha _{mn}}}d^3x,
\end{eqnarray} 
which in the Dirac notations is given by
\begin{eqnarray} \label{eq67}
{\cal H} = {\langle \psi} | \sum_{n=1}^N{\hat {\cal H}}_{nn}+\sum_{{n\neq m}}^{N^2-N}{\hat {\cal H}}_{nm} | \psi \rangle ={\langle N} | {\hat {\cal H}} | N \rangle ={\varepsilon}_0 {\langle N} | {\hat {\cal N}} | N \rangle={{\varepsilon}_0} { \cal N}, 
\end{eqnarray}
where the Hamiltonian operator 
\begin{eqnarray} \label{eq68}
{\hat {\cal H}}={{\varepsilon}_0}{\hat {\cal N}}, 
\end{eqnarray}
and the effective number of the cross-correlating unit-waves (particles), which can be a non-integer, is given by 
\begin{eqnarray} \label{eq69}
{\cal N} = \langle N |{\hat {\cal N}} | N \rangle ={\varepsilon}_0^{-1}{\langle N |{\hat {\cal H}} | N \rangle}
\end{eqnarray}
with the effective number operator
\begin{eqnarray} \label{eq70}
{\hat {\cal N}}={\hat N}+\sum_{n\neq{m}}^{N^2-N}[ ({\hat a}^ {\dagger}_n{\hat a}_m+{\hat a}_n{\hat a}^ {\dagger}_m)/2]e^{i\Delta \alpha  _{mn}}={\hat N}+\sum_{n\neq{m}}^{N^2-N}e^{i\Delta \alpha  _{mn}}.
\end{eqnarray}
Notice, in Eqs. (\ref{eq66})-(\ref{eq70}), I used the relation 
\begin{eqnarray} \label{eq71}
({\hat a}^ {\dagger}_n{\hat a}_m+{\hat a}_n{\hat a}^ {\dagger}_m)/2=({\hat a}^ {\dagger}{\hat a}+{\hat a}{\hat a}^ {\dagger})/2=1,
\end{eqnarray}
which could be attributed to the indistinguishableness of the identical unit-waves (particles) that satisfy the condition $V_{0n} = V_{0m} = {V_{0}}$, $m_{0n} = m_{0m} = m_{0}$ and ${\mathbf {k}}_{0n} = {\mathbf {k}}_{0m} = {\mathbf {k}}_{0}$. One can easily demonstrate that the energy of the cross-correlating unit-waves describing by Eqs. (\ref{eq66})-(\ref{eq71}) is given by
\begin{eqnarray} \label{eq72}
0\leq{ \cal H }\leq N^2{{\varepsilon}_0},
\end{eqnarray}
and the respective effective number of the waves     
\begin{eqnarray} \label{eq73}
0\leq{{\cal N}}\leq N^2. 
\end{eqnarray}
Equations (\ref{eq67}) and (\ref{eq69}) yielded the energy ${\cal H}=N^2{{\varepsilon}_0}$ and the effective number ${\cal N}=N^2$ for the pure additive interference ($\Delta \alpha _{mn}=0$). The pure subtractive interference results into annihilation of the unit-waves and energy. The annihilation takes place if the pairs of unit-wave operators satisfy the phase condition $\alpha _m-\alpha _n=\pm \pi$. The annihilation ${\hat \psi^b _{0n}(\mathbf{r},t)}+{\hat \psi ^{ab}_{0n}(\mathbf{r},t)}=0$ of the boson and antiboson operators ${\hat \psi^b _{0n}(\mathbf{r},t)}$ and ${\hat \psi ^{ab}_{0n}(\mathbf{r},t)}=0$ can be represented as the pure subtractive interference of the boson unit-field operators 
\begin{eqnarray} \label{eq74}
{\hat \psi}_{0n}^b (\mathbf{r},t)={\hat a}a_{0}e^{i({\mathbf{k}_{0}{\mathbf{r}}}-{\alpha _n})} 
\end{eqnarray}{
with the antiboson unit-field operators
\begin{eqnarray} \label{eq75}
{\hat \psi}_{0n}^{ab} (\mathbf{r},t)={\hat b} a_{0}e^{i({\mathbf{k}_0{\mathbf{r}}}-{\alpha _n} )} ,
\end{eqnarray}
where $ {\hat b}={\hat a}e^{\pm i\pi}$. Similarly, the annihilation ${\hat {\psi ^ {\dagger}}^b _{0n}(\mathbf{r},t)}+{\hat {\psi ^ {\dagger} }^{ab}_{0n}(\mathbf{r},t)}=0$ of the boson ${\hat {\psi ^ {\dagger}}^b _{0n}(\mathbf{r},t)}$ and antiboson ${\hat {\psi ^ {\dagger} }^{ab}_{0n}(\mathbf{r},t)}=0$ operators can be presented as the pure subtractive interference of the operators 
\begin{eqnarray} \label{eq76}
{\hat {\psi ^ {\dagger}}}_{0n}^b (\mathbf{r},t)={\hat a ^ {\dagger}}a_{0}e^{-i({\mathbf{k}_{0}{\mathbf{r}}}-{\alpha _n})} 
\end{eqnarray}{
with the operators 
\begin{eqnarray} \label{eq77}
{\hat {\psi ^ {\dagger}}}_{0n}^{ab} (\mathbf{r},t)={\hat b ^ {\dagger}} a_{0}e^{-i({\mathbf{k}_0{\mathbf{r}}}-{\alpha _n} )}
\end{eqnarray}
where $ {\hat b ^ {\dagger}}={\hat a ^ {\dagger}}e^{\pm i\pi}$. For instance, the pure subtractive interference of the boson field operators
\begin{eqnarray} \label{eq78}
{\hat \psi}^b(\mathbf{r},t) =\sum_{n=1}^{N_b}{\hat \psi _{0n}^b}(\mathbf{r},t) 
\end{eqnarray}
with the antiboson field operators
\begin{eqnarray} \label{eq79}
{\hat \psi} ^{ab}(\mathbf{r},t) =\sum_{n=1}^{N_{ab}}\hat \psi _{0n}^{ab}(\mathbf{r},t),
\end{eqnarray}
results into annihilation of the fields and energy if $N_b=N_{ab}$. It should be noted that the unit-wave operators do annihilate partially if $\alpha _n-\alpha _m=\pm \pi  \pm \delta $, where $|\delta| <\pi $. In order to provide the annihilation of a boson unit-wave operator with an antiboson unit-wave operator for the arbitrary momentum (wave-number) vectors and phases, the operators should satisfy the annihilation condition 
\begin{eqnarray} \label{eq80}
{\hat \psi^b _{0}(\mathbf{r},t;{\mathbf{k}}_0^b,\alpha _b)}+{\hat \psi ^{ab}_{0}(\mathbf{r},t;{\mathbf{k}}_0^{ab}, \alpha _{ab})}=0
\end{eqnarray}
independently from the values of the momentums (wave-numbers) and phases. In such a case, the antiboson unit-field operator is the true antiparticle unit-field operator, which cannot be obtained by the phase shift of the boson field operator. In the case of ${\mathbf{k}}_0^b={\mathbf{k}}_0^{ab}=\alpha  _b= \alpha  _{ab}=0$, the massive ($m_0\neq 0$) fields only satisfy the condition (\ref{eq80}). The mass-less fields do not have the true antiparticle unit-fields (antiparticles) and the respective antiparticle operators. Notice, the annihilation condition (\ref{eq80}) applied to the boson-antiboson pair yields the commutation-like relation $({\hat a}^ {\dagger}{\hat a}+{\hat b}{\hat b}^{\dagger}+{\hat a}^ {\dagger}{\hat b}+{\hat b}{\hat a}^ {\dagger})=0$. Notice, the positive energy-mass ${\varepsilon}_0=({\mathbf{k}}_0^2+m_0^2)^{1/2}$ of antibosons in the present model is different from the negative energy-mass of the Dirac antiparticles satisfying the energy-mass relation ${\varepsilon}_0=- ({\mathbf{k}}_0^2+m_0^2)^{1/2}$.

The second quantization based on the replacement of the cross-correlating fields by the respective field operators can be applied for other physical parameters, such as momentum, number of waves, mass and charge. For instance, the momentum $\mathbf{P}$ of the field $\psi(\mathbf{r},t ) =\sum_{n=1}^{N}{\psi _{0n}}(\mathbf{r},t)$ is given by 
\begin{eqnarray} \label{eq81}
\mathbf{P}={\cal P}^i \mathbf{e}_i={\langle \psi} |  {\hat {\cal P}^i} | \psi {\rangle} \mathbf{e}_i  ,
\end{eqnarray}
where 
\begin{eqnarray} \label{eq82}
{\hat {\cal P}^i}=\sum_{n=1}^N{k_{0n}^i} +\sum_{n\neq{m}}^{N^2-N}{\int_{V_{nm}}}[ ({\hat a}^ {\dagger}_n{\hat a}_m+{\hat a}_n{\hat a}^ {\dagger}_m)/2][{\varepsilon}_{0n}{{k}_{0m}^i}+{\varepsilon}_{0m}{{k}_{0n}^i}]a_{0n}a_{0m}e^{{-i( \Delta \mathbf{k}_{0nm} {\mathbf{r}}-\Delta \alpha _{nm}})}d^3x.
\end{eqnarray}
In the case of $V_{0n} = V_{0m} = {V_{0}}$, $m_{0n} = m_{0m} = m_{0}$ and ${\mathbf {k}}_{0n} = {\mathbf {k}}_{0m} = {\mathbf {k}}_{0}$, Eq. (\ref{eq82})  yields 
\begin{eqnarray} \label{eq83}
\mathbf{P}={\cal N}{\mathbf{k}_{0} }={\langle \psi} |  {\hat {\cal N}} | \psi {\rangle}{\mathbf{k}_{0} }, 
\end{eqnarray}
where $\hat {\cal N}$ is the effective number operator (\ref{eq70}) of the waves. Notice, the angular momentum $\mathbf{L}$ and the respective operator of the interfering unit-fields can be found by the similar generalization of the angular momentum $\mathbf{L}=\mathbf{r}\times \mathbf{P}$ or the relativistic angular momentum tensor $L^{nm}= \sum (x^np^m-x^mp^n)$.  The generalized rest mass $\cal M$ of the system according to Eqs. (\ref{eq67}) and (\ref{eq69}) with ${\mathbf{k}}_{0}\rightarrow 0$ (respectively, $\varepsilon_0 \rightarrow  m_0$) is given by 
\begin{eqnarray} \label{eq84}
{\cal M}={\cal N}m_0 = {\langle \psi} |  {\hat {\cal N}} | \psi {\rangle}m_0 
\end{eqnarray}
with $0 \leq {{\cal M}} \leq N^2m_0$. In the case of the pure subtractive interference of the boson field with the antiboson field, the field energy can be formally presented as the absolute value of the sum of the positive energy of the boson unit-field with the negative energy of the antiboson field in the interference process. Respectively, the generalized rest mass $\cal M$ is given by the absolute value of the sum of the positive mass of the boson field with the "negative mass" of the antiboson field: 
\begin{eqnarray} \label{eq85}
{\cal M}=|{\cal N}_bm_0+{\cal N}_{ab}(-m_0)|=|\langle {\hat \psi}^b _0 |  {\hat {\cal N}} | {\hat \psi}^b _0 {\rangle}m_0 -   \langle {\hat \psi}^{ab} _0 | {\hat {\cal N}_{ab}} | {\hat \psi}^{ab} _0 {\rangle}  m_0 |.
\end{eqnarray}
It should be stressed that the mass-energies of particles and antiparticles are positive. The negative sign in Eq.~(\ref{eq85}) reflects only a subtractive character of the interaction (interference). Notice, the positive rest energy-mass ${\varepsilon}_0=(m_0^2)^{1/2}$ of antibosons in the present model is different from the negative rest energy-mass ${\varepsilon}_0=- (m_0^2)^{1/2}$ of the Dirac antiparticles. If the unit-charge $q_0$ is associated with the mass $m_0$ of the unit-wave ${\psi _{0n}}(\mathbf{r},t)$ [see comments to Eqs. (\ref{eq120}) and (\ref{eq121}), and Figs. (5)-(8)], in other words, if the unit wave $\psi _{0n}(\mathbf{r})$ has the unit charge $q_0$, then the effective charge ${\cal Q}$ of the cross-correlating unit-fields is given by
\begin{eqnarray} \label{eq86}
{\cal Q}={\cal N}q_0 = {\langle \psi} |  {\hat {\cal N}} | \psi {\rangle}q_0,
\end{eqnarray}
where $0\leq{{\cal Q}}\leq N^2 q_0$. Similarly, if the magnetic moment ${\vec {\mu }}_0$ is associated with the mass $m_0$ of the unit-wave ${\psi _{0n}}(\mathbf{r})$, then the effective magnetic moment $\vec {M}$ ($0\leq{{ \vec {M}}}\leq N^2{\vec {\mu }}_0$) of the cross-correlating unit-fields is given by
\begin{eqnarray}
{\vec {M}}={\cal N}{\vec {\mu }}_0.
\end{eqnarray}
In the case of the pure subtractive interference of the boson field with the antiboson field he effective charge is given by the sum of the positive charge of the boson field with the negative charge of the antiboson field: 
\begin{eqnarray} \label{eq87}
{\cal Q}={\cal N}_bq_0+{\cal N}_{ab}(-q_0)=\langle {\hat \psi}^b _{0} |  {\hat {\cal N}} | {\hat \psi}^b _{0} {\rangle}q_0 -   \langle {\hat \psi}^{ab} _{0} | {\hat {\cal N}_{ab}} | {\hat \psi}^{ab} _{0} {\rangle}q_0.
\end{eqnarray}
Notice, although the boson parameters associated with the charge have been formally introduced [see, Eqs. (\ref{eq86})-(\ref{eq87})] into the scalar boson model, the existence or nonexistence of the boson/antiboson charges is determined rather by the experiment than the model. The introduction of the antibosons into the model also should correspond to the experimental observations of the concrete field.   

In the context of  {\it{physical interpretations}} of the above-presented model, one should mention that the unit-fields (particles) of the present model are not the point particles of quantum field theory. In the particle field theory, the physical interpretation of the wave (field) of operators as the real physical object is very problematic. An operator is a pure mathematical object, which should not be interpreted as a real physical matter. In the present model, the division of the real "non-quantum" field $\psi$ of the matter (mass-energy) into the indivisible "non-quantum" unit-fields $ \psi_{0n}$ of the matter (mass-energy) [see, Eqs. (\ref{eq15})-(\ref{eq38})], in fact, is {\it{the second quantization of the field without the use of the fields of operators}}. The above-presented model of the boson quantum-fields based on the operator representation of the fields is {\it{equivalent}} to the "non-quantum" model of the real unit-fields of the matter considered in Sec. (2.1.2.). That means the physical parameters describing by Eqs. (\ref{eq39})-(\ref{eq87}) are equal to the respective values derived in Sec. (2.1.2.) without the use of the fields of  operators. In the case of the fields (waves), which are free from the cross-correlation associated with the ordinary interference phenomenon, the above-derived number of unit-waves, magnetic moment, mass and charge of the waves are equal to the respective values of the traditional particle field theory. However, the calculated energy ${\varepsilon}_0=\omega_0=(  {\mathbf  k}_0^2   + m_0^2)^{1/2}$ and momentum ${\mathbf  p}_0={\mathbf  k}_0$ of a unit-wave (particle), which are equal to the Planck-Einstein energy and momentum of the particle, are different from the canonical values ${\varepsilon}_0=\omega_0 + (1/2)\omega_0$ and ${\mathbf  p}_0={\mathbf  k}_0+(1/2){\mathbf  k}_0$ of the traditional particle field theory~\cite{Dir2,Land,Jack,Bere,Itz,Ryde,Pes,Grei,Wein}. The positive energy-mass ${\varepsilon}_0=({\mathbf{k}}_0^2+m_0^2)^{1/2}$ of antibosons in the above-described model is different from the negative energy-mass of the Dirac antiparticles satisfying the energy-mass relation ${\varepsilon}_0=- ({\mathbf{k}}_0^2+m_0^2)^{1/2}$. The energy and momentum of the empty space, which are in agreement with the Planck-Einstein and classical physics of the vacuum, are equal to zero. 

\subsubsection{2.1.4. Boson vector fields}

So far we have considered the interferences between "non-quantum" and quantum scalar ($s$ = 0) boson particles (unit-fields). The cross-correlation model of the interfering boson unit-fields having the spin $s > 0$ would be similar to the model of scalar fields considered in the previous sections. One can easily follow the above-presented cross-correlation model for an arbitrary value of the spin $s$. As an example, let me briefly describe the cross-correlating vector finite-fields composed from the mass-less vector unit-fields (vector bosons) having the spins $s=1$. Such fields correspond to the electromagnetic (EM) finite-fields of the photons. The infinite EM field is described by the canonical Lagrangian density ${\cal L}=(-1/4){{\psi}_{\mu \nu }}{{\psi}^{\mu \nu } }$, where the 4-tensor $\psi ^{\mu \nu }$ is the field tensor $F^{\mu \nu }$; the 4-vector $\psi ^\mu $ is the 4-potential $A^\mu$~\cite{Dir2,Land,Jack,Bere,Itz,Ryde,Pes,Grei,Wein}. According to the present model, the generalization of the canonical Lagrangian (Hamiltonian) and the second quantization should be performed for the superposition of the respective interfering unit-fields. The configuration and dynamics of the unit-field is determined by the Euler-Lagrange equation of motion with the initial and boundary conditions imposed. The simplest wave-like solution  $\psi_{0\mu}$ of the equation for the unit-field, which is located infinitely far from the boundaries, is the time-harmonic plane wave. The operator ${\hat \psi}_{0\mu} $ of the cross-correlating, time-harmonic, plane, vector unit-wave 
\begin{eqnarray} \label{eq88}
{\psi}_{0\mu} (\mathbf{r},t)=a_{0\alpha }{u_{\mu }^{\alpha}}e^{-i{\varepsilon}_0t} e^{i({\mathbf{k}_{0}{\mathbf{r}}}-\alpha )}
\end{eqnarray}
of the physical matter (vector boson, for instance, photon) is given by 
\begin{eqnarray} \label{eq89}
{\hat \psi}_{0\mu}= {\hat a}_{\alpha }{a}_{0\alpha }{u_{\mu }^{\alpha}}e^{i({\mathbf{k}_{0}{\mathbf{r}}}-{\alpha })},
\end{eqnarray}
and the operator ${\hat \psi}_{0\mu}^{\dagger }$ of the unit-wave
\begin{eqnarray} \label{eq90}
{\psi}_{0\mu}^* (\mathbf{r},t)={a}_{0\alpha }{u_{\mu }^{\alpha ^*}}e^{i{\varepsilon}_0t}e^{-i({\mathbf{k}_{0}{\mathbf{r}}}-\alpha )}
\end{eqnarray}
is given by 
\begin{eqnarray} \label{eq91}
{\hat \psi}_{0\mu}^{\dagger }= {\hat a}_{\alpha }^{\dagger }{a}_{0\alpha }{u_{\mu }^{\alpha ^*}}e^{-i({\mathbf{k}_{0}{\mathbf{r}}}-{\alpha })},
\end{eqnarray}
where ${a}_{0\alpha }$ is the wave amplitude. The unit 4-vector of polarization is given by ${u_{\mu }^{\alpha }}$, where the index $\alpha $ corresponds to the two independent polarizations. Then the operators of the finite fields 
\begin{eqnarray} \label{eq92}
{\psi}_{\mu} (\mathbf{r},t) =\sum_{n=1}^N{\psi _{0\mu n}(\mathbf{r},t)}
\end{eqnarray}
and 
\begin{eqnarray} \label{eq93}
{\psi}_{\mu}^* (\mathbf{r},t) =\sum_{n=1}^N{\psi _{0\mu n}^*(\mathbf{r},t)}, 
\end{eqnarray}
which are the superpositions of the respective unit-fields $ \psi_{0\mu n} (\mathbf{r},t)$ and $ \psi_{0\mu n}^* (\mathbf{r},t)$, could be presented respectively as
\begin{eqnarray} \label{eq94}
{\hat \psi}_{\mu}= \sum_{n=1}^N{\hat \psi}_{0\mu n}
\end{eqnarray}
and
\begin{eqnarray} \label{eq95}
{\hat \psi ^{\dagger }}_{\mu} =\sum_{n=1}^N{\hat \psi ^{\dagger }}_{0\mu n} .
\end{eqnarray}
Notice, the superpositions (\ref{eq92}) and (\ref{eq93}) may be reinterpreted as the Fourier decomposition of the field composed from the indivisible, time-harmonic, plane unit-waves (plane-wave photons). One should not confuse here the plane-wave photon with the non-plane unit-wave, for instance, with the spherical-wave photon. The Fourier decomposition could be useful in an analysis of the near-field and sub-wavelength phenomena (see, Sec. 4.2.), where the non-plane unit-wave ({\it{indivisible}} non-plane photon) producing near to the sharp boundaries is presented formally as the Fourier {\it{decomposition}} of the plane-wave photons.  The boson vector fields (EM fields) describing by Eqs. (\ref{eq88})-(\ref{eq95}) do not have the rest mass ($m_0=0$). It is not necessary to repeat all the mathematical procedures of Secs. (2.1.2.) and (2.1.3.) for the mass-less vector unit-fields (photons). One can easily demonstrate that the physical parameters [for instance, energy, momentum and number of the unit-waves (photons)] in the linearly polarized EM fields are described by the respective equations of Secs. (2.1.2.) and (2.1.3.) with the particle rest mass $m_0=0$ and energy ${\varepsilon}_0=k_0$, which correspond to the rest mass and energy of the Planck-Einstein photon. In the case of the fields with the two different polarizations, the summations of the kind 
\begin{eqnarray} 
\sum_{n=1}^N(...)+\sum_{n\neq{m}}^{N^2-N}(...)
\end{eqnarray}
in the equations should be replaced by 
\begin{eqnarray} \label{eq96}
\sum_{n=1;{\alpha }_n}(...)+\sum_{n\neq m;{{\alpha}_n ,{\alpha}_m}}(...){\delta _{{\alpha}_n {\alpha}_m}},
\end{eqnarray}
where the index $\alpha $ corresponds to the two independent polarizations, and $\delta _{{\alpha}_n {\alpha}_m}$ is the Kronecker symbol. For further details see the introduction and preliminary investigation of the cross-correlation and cross-correlation energy of the classical and quantum EM fields presented in Refs.~\cite{Kukh1}(a-c). It could be noted that Eqs. (\ref{eq35}), (\ref{eq36}), (\ref{eq84}), and (\ref{eq85}) describe the "moving mass" of the cross-correlating EM fields that have the rest mass $m_0=0$ in these equations. The fields (\ref{eq92}) - (\ref{eq95}), which have the positive energies, could annihilate each other if $\alpha _m-\alpha _n=\pm \pi$ for the unit-field (photon) pairs. From a point of view of the first photon of the annihilated photon pair the second photon may be considered formally as an anti-photon. In order to provide the annihilation of an EM unit-wave (photon) by another EM unit-wave (anti-photon), the unit-waves should satisfy the annihilation conditions (\ref{eq28}) and (\ref{eq80}) for the arbitrary momentum (wave-number) vectors and phases. In such a case, the anti-photon would be the true antiparticle unit-field, which cannot be obtained by the phase shift of an EM unit-wave (photon). The EM fields do not satisfy the aforementioned annihilation conditions. Therefore the EM fields do not have the true antiparticle unit-fields (anti-photons). That is in agreement with the experimental fact that the mass-less ($m_0=0$) unit-field (photon) has zero charge, $q_0=0$.

\subsubsection{2.1.5. Fermion fields}

It is clear that the cross-correlation model of the "non-quantum" and quantum fields of the interfering fermions (unit-fields with half-integer spin $s$) should be similar to the above-considered boson fields. As an example, let me briefly describe the cross-correlating "non-quantum" and quantum fermion finite-fields with the unit-fields (spinor particles) having the spin $s=1/2$. Such finite fields correspond to the finite fields of spinor particles, fore instance, electrons. The Dirac infinite spinor field, which is free from the cross-correlation (interference), is described by the Lagrangian density ${\cal L}=i\bar \psi \gamma ^{\mu }\partial _{\mu}\psi -m\bar \psi \psi $~\cite{Dir2,Land,Jack,Bere,Itz,Ryde,Pes,Grei,Wein}. According to the present model, the generalization of the canonical Lagrangian (Hamiltonian) and the second quantization should be performed for the superposition of the respective interfering unit-fields (spinors). The configuration and dynamics of the unit-field is determined by the Euler-Lagrange equation of motion, which is indistinguishable from the Dirac equation of motion, with the initial and boundary conditions imposed. The simplest wave-like solution  of the equation for the unit-field, which is located infinitely far from the boundaries, is the time-harmonic plane wave. The operator ${\hat \psi}_{0} $ of the spinor unit-wave 
\begin{eqnarray} \label{eq97}
{\psi}_{0} (\mathbf{r},t)={c}_{0 \sigma  }{u_{\sigma}}e^{-i{\varepsilon}_0t}e^{i({\mathbf{k}_{0}{\mathbf{r}}}-\alpha )}
\end{eqnarray}
of the material substance (spinor particle, for instance, electron) is given by 
\begin{eqnarray} \label{eq98}
{\hat \psi}_{0}= {\hat c}_{\sigma }{c}_{0 \sigma  }{u_{\sigma}}e^{i({\mathbf{k}_{0}{\mathbf{r}}}-\alpha )},
\end{eqnarray}
and the operator ${\hat {\bar \psi }}_{0}$ of the unit-field
\begin{eqnarray} \label{eq99}
{\psi}_{0}^* (\mathbf{r},t)={c}_{0 \sigma  }{u^*_{\sigma}}e^{i{\varepsilon}_0t}e^{-i({\mathbf{k}_{0}{\mathbf{r}}}-\alpha )}
\end{eqnarray}
is given by 
\begin{eqnarray} \label{eq100}
{\hat {\bar \psi }}_{0}= {\hat c}_{\sigma }^{\dagger }{c}_{0 \sigma  }{{\bar u}_{\sigma}}e^{-i({\mathbf{k}_{0}{\mathbf{r}}}-\alpha )},
\end{eqnarray}
where the particle spin $\sigma =\pm 1/2$. The operators of the finite fields 
\begin{eqnarray} \label{eq101}
{\psi} (\mathbf{r},t) =\sum_{n=1}^N{\psi _{0 n}(\mathbf{r},t)}
\end{eqnarray}
and 
\begin{eqnarray} \label{eq102}
{\psi}^* (\mathbf{r},t) =\sum_{n=1}^N{\psi _{0 n}^*(\mathbf{r},t)}, 
\end{eqnarray}
which are the superpositions of the respective unit-fields $ \psi_{0 n} (\mathbf{r},t)$ and $ \psi_{0 n}^* (\mathbf{r},t)$, could be presented respectively as
\begin{eqnarray} \label{eq103}
{\hat \psi}= \sum_{n=1}^N{\hat \psi}_{0 n}
\end{eqnarray}
and
\begin{eqnarray} \label{eq104}
{\hat {\bar \psi }} =\sum_{n=1}^N{\hat {\bar \psi }}_{0n}.
\end{eqnarray}
It is not necessary to repeat all the mathematical procedures of Secs. (2.1.2.) and (2.1.3.) for the unit-fields (spinor particles). One can easily demonstrate that the physical parameters [for instance, energy, momentum and number of the unit-waves (spinors)] are described by the respective equations of Secs. (2.1.2.) and (2.1.3.). In the equations, however, the summations of the kind 
\begin{eqnarray} \label{eq105}
\sum_{n=1}^N(...)+\sum_{n\neq{m}}^{N^2-N}(...)
\end{eqnarray}
should be replaced by 
\begin{eqnarray} \label{eq106}
\sum_{n=1; {\sigma }_n} (...) + \sum_{n \neq m; {{\sigma}_n , {\sigma}_m}} (...)  {\delta _{{\sigma}_n {\sigma}_m}},
\end{eqnarray}
where the Pauli exclusion principle for the fermions (unit-waves) should be taken into account in specific (concrete) calculations of the field parameters. For details, see Sec. II and comments to Eqs. (\ref{eq120}) and (\ref{eq121}), and Figs. (5)-(8)]. The Pauli principle states that no two identical fermions may occupy the same quantum state simultaneously. It could be noted again that the dynamics of the above-considered spinor finite-fields is determined by the Euler-Lagrange (Dirac) equation of motion with the initial and boundary conditions imposed. The physical parameters of the cross-correlating relativistic fields are described by the respective equations of Secs. (2.1.2.) and (2.1.3.), where the unit-fields (spinor particles or electrons) have the relativistic energy ${\varepsilon}_0=( \mathbf{k}_0^2 + m_0^2 )^{1/2}$. The physical parameters of the non-relativistic ($ \mathbf{k}_0 \rightarrow 0$) cross-correlating fields are described by these equations, where the unit-field energy is given by $ \varepsilon_0 \approx ( \mathbf{k}_0^2 / 2m_0) + m_0$. In the present model, the transition from the relativistic cross-correlating spinor finite fields to non-relativistic ones {\it{is similar}} to the transition from the relativistic infinite field describing by the Dirac equation of motion with the resonator-like boundary conditions to the finite waves of probabilities (finite wave-functions), whose dynamics is determined by the Schr{\"o}dinger equation with the quantum-mechanics boundary conditions imposed. It should be mentioned that the positive energy-mass of a spinor antiparticle described in the Part II of the present study is different from the negative energy-mass of antiparticle (positron) in the Dirac field theory, which is based on the Copenhagen-Dirac postulate of "interference-less", self-interfering particles and the particle (antiparticle) energy-mass ${\varepsilon}_0=\pm ({\mathbf{k}}_0^2+m_0^2)^{1/2}$. Notice, in contrast to the traditional interpretation of the Dirac infinite field by the quantum field theory, the Dirac equation was originally formulated and interpreted by the author as a {\it{single-particle equation}} analogous to the Schr{\"o}dinger single-particle equation.

\section{3. Forces mediated by gradients of the cross-correlation energy}

This section considers the general properties of the forces (interactions) between the interfering particles (bodies) causing by gradients of the cross-correlation energy. In modern physics, generally speaking, there are two kinds of physical mechanisms that describe the forces (interactions) between particles~\cite{Dir1,Dir2,Einst2,Land,Jack,Bere,Itz,Ryde,Pes,Grei,Wein}. According to quantum field theories, the forces acting upon a particle are seen as the action of the respective gauge-boson field that is present at the particle location. In the perturbative approximation, the forces are attributed to the exchange of the field virtual particles (gauge bosons). In the Einstein theory of general relativity, the gravitational interaction between two objects (particles) is not viewed as a force, but rather, objects moving freely in gravitational fields travel under their own inertia in straight lines through "curve" spacetime. I should show that the physical mechanism behind the forces inducing by gradients of the cross-correlation energy is similar to the two aforementioned traditional mechanisms. 

Let me describe the forces mediating by gradients of the cross-correlation energy of the interfering particles (unit-fields) in the form, which does not depend on the kind of the particles. For the sake of simplicity, we first consider the cross-correlation (interaction) of the two unit-fields (particles). According to the present model [for the scalar bosons that interfere with each other, see Eqs. (\ref{eq6}), (\ref{eq7}) and (\ref{eq9})], the total energy ${\cal H}$ of the two arbitrary unit-fields ${\psi _{01}(\mathbf{r},t)}$ and ${\psi _{02}(\mathbf{r},t)}$ at the time moment $ t=t_{12}$ is calculated by integrating the Hamiltonian density $h$ in the space. In the general form, the Hamiltonian contains the canonical energies of the single fields $\psi _{01}$ and $\psi _{02}$, which occupy respectively the volumes $V_1\equiv V_{11}$ and $V_2 \equiv V_{22}$, and the cross-correlation energies of the fields overlapped in the common volume ${ V_{12}}={ V_{21}}$ at the time moment $ t_{12}$:
\begin{eqnarray}  \label{eq107}
{\cal H}= \sum_{n=1}^2{\cal H}_{nn}+\sum_{n\neq m}^{2}{\cal H}_{nm}=\sum_{n=1}^2{\int_{V_{nn}}}{h}_{nn}d^3x+\sum_{n\neq{m}}^{2}{\int_{V_{nm}}}{h}_{nm}d^3x,
\end{eqnarray}
where $h_{nm}$ is the Hamiltonian density, which takes into account the interference between particles, namely the interference of the unit-fields ${\psi _{01}(\mathbf{r}_1,t_{12})} \equiv {\psi _1 (\mathbf{r}_1)}$ with ${\psi _{02}(\mathbf{r}_2,t_{12})} \equiv {\psi _2 (\mathbf{r}_2)}$. In the case of the two identical particles, the unit-fields are described by ${\psi _1 (\mathbf{r}_1)} \equiv {\psi(\mathbf{r})}$, ${\psi _2 (\mathbf{r}_2)} \equiv {\psi}(\mathbf{r+R})$, ${\psi^* _1 (\mathbf{r}_1)} \equiv {\psi^* (\mathbf{r})}$ and ${\psi^* _2 (\mathbf{r}_2)} \equiv {\psi^* }(\mathbf{r+R})$, where $R$ is the distance between the unit-fields (particles) [see, the illustration for the non-complex unit-fields in Fig. 2]. 
\begin{figure}
\begin{center}
\includegraphics[keepaspectratio, width=0.5\columnwidth]{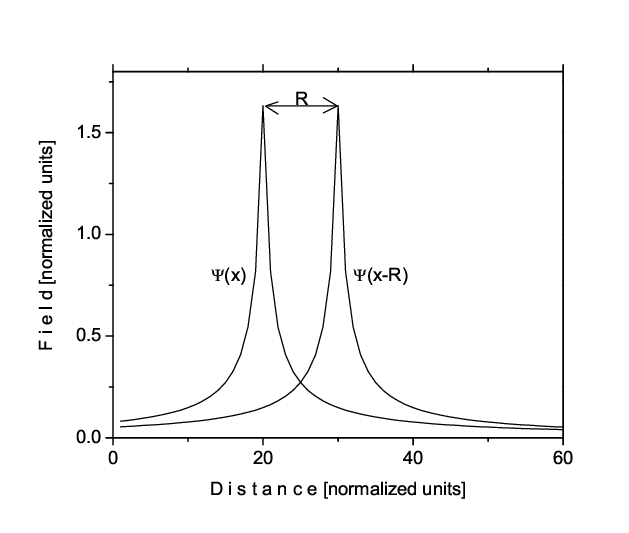}
\end{center}
\caption{The two identical unit-fields (particles) with $\psi_1(x) \equiv \psi (x)$ and $\psi_2(x) \equiv \psi (x-R)$, where $R$ is the distance between the unit-fields (particles).}
\label{fig:Fig2}
\end{figure}
The particle energies ${\cal H}_{11}$ and ${\cal H}_{22}$, which depend on neither the coordinate $\mathbf{r}$ nor the time $t$ after the integration over the volumes ${V_{11}}$ and ${V_{22}}$, are permanent (constant). The cross-correlation energy ${\cal H}_{12,21}\equiv {\cal H}_{12}+{\cal H}_{21}$, however, depends on the overlapping (interference) of the fields ${\psi _{1}(\mathbf{r})}$ with ${\psi _{2}(\mathbf{r+R})}$. For more details see Secs. (1) and (2). The cross-correlation energy ${\cal H}_{12,21}$, which depends on the distance $\mathbf{R}$ between the particles (unit-fields), can be presented as the potential energy ${\cal H}_{12,21}({\mathbf{R}})={\cal H}_{12}({\mathbf{R}})+{\cal H}_{21}({\mathbf{R}})$. Then the force ${\mathbf{F}}$ causing by the gradient of the cross-correlation energy is given by
\begin{eqnarray} \label{eq108}
{\mathbf{F}}=- \frac {\partial {{\cal H}}}{\partial {\mathbf{R}}}=-\frac {\partial {{\cal H}}}{\partial {R^i}}\mathbf{e}_i,
\end{eqnarray}
with
\begin{eqnarray} \label{eq109}
\frac {\partial {{\cal H}}}{\partial {R^i}}={\int_{ V_{12}}}\frac {\partial }{\partial {R^i}}[{h}_{12}({\mathbf{r}},{\mathbf{r}}+{\mathbf{R}})+{h}_{21}({\mathbf{r}},{\mathbf{r}}+{\mathbf{R}})]d^3x,
\end{eqnarray}
where ${h}_{12}({\mathbf{r}},{\mathbf{r}}+{\mathbf{R}})+{h}_{21}({\mathbf{r}},{\mathbf{r}}+{\mathbf{R}})={h}_{12,21}({\mathbf{r}},{\mathbf{r}})$ denotes the density of the cross-correlation energy, and $\mathbf{r}={x}^i \mathbf{e}_i$ and $\mathbf{R}={R}^i \mathbf{e}_i$. In an ensemble of the $N$ unit-fields (particles), the force ${\mathbf{F}}_q$ acting upon a particle having the index $q$ is seen as the action of the $N-1$ unit-fields, which are present at the particle location:  
\begin{eqnarray} \label{eq110}
{\mathbf{F}}_q=- \frac {\partial {{\cal H}}}{\partial {\mathbf{R}}_{qn}}=-\frac {\partial {{\cal H}}}{\partial {R^i_{qn}}}\mathbf{e}_i,
\end{eqnarray}
with
\begin{eqnarray} \label{eq111}
\frac {\partial {{\cal H}}}{\partial {R^i_{qn}}}=\sum_{n\neq q}^{N-1}{\int_{ V_{qn}}}\frac {\partial }{\partial {R^i_{qn}}}[{h}_{qn}({\mathbf{r}}_{q},{\mathbf{r}}_{q}+{\mathbf{R}}_{qn})+{h}_{nq}({\mathbf{r}}_{q},{\mathbf{r}}_{q}+{\mathbf{R}}_{nq})]d^3x_{q},
\end{eqnarray}
where ${\mathbf{r}}_{n}={\mathbf{r}}_{q}+{\mathbf{R}}_{qn}$ is the spatial coordinate in the field $\psi _n({\mathbf{r}}_n)$. It should be stressed that Eqs. (\ref{eq107})-(\ref{eq111}) do not depend on the kind of the unit-fields (particles). One can easily apply these equations for the boson and fermion fields of any kind. The quantization (second quantization) of the non-quantum fields by replacing the fields by the respective field operators does not change Eqs. (\ref{eq107})-(\ref{eq111}) [see, Sec. 2]. Thus the quantum cross-correlation energies and respective forces of the fields of operators are equal to the non-quantum values of the "non-quantum" unit-fields of the matter (mass-energy) described in Sec. (2). Although the properties of the force (\ref{eq108}) depend on the concrete forms of the unit-fields ${\psi _{01}(\mathbf{r}_1)}$ and ${\psi _{02}(\mathbf{r}_2)}$, the several general properties can be derived without knowledge of the exact forms of the fields. For instance, if the Hamiltonian (\ref{eq107}) is not invariant under the $U(1)$ local gauge transformation ($\psi _{0n} \rightarrow \psi ' _{0n}=e^{i\alpha _n} \psi _{0n}$), then the gauge symmetry breaking gives rise to the {\it{phase-dependent interference}} between the unit-fields ${\psi _{01}(\mathbf{r},t)} \equiv \phi _{01} (\mathbf{r},t) e^{i\alpha _{1}} $ and ${\psi _{02}(\mathbf{r},t)} \equiv \phi _{2}(\mathbf{r},t) e^{i\alpha _{2}}$ resulting into the phase-dependent interaction between particles by the gradient of the cross-correlation energy. The force (\ref{eq108}) is attractive if the gradient of the cross-correlation energy of the two-unit fields (particles) is positive. In such a case, the maximum and minimum of the total energy (\ref{eq107}) are achieved in the limits $ \mathbf{R}{\rightarrow}\infty$ and $ \mathbf{R}{\rightarrow}$ 0, respectively. In the case of the negative gradient of the cross-correlation energy, the force (\ref{eq108}) is repulsive. The maximum and minimum of the total energy of the repulsive unit-fields (particles) correspond respectively to the limits $ \mathbf{R}{\rightarrow}$ 0 and $ \mathbf{R}{\rightarrow}\infty$. 

In the terms of virtual processes (transitions) of the traditional particle field theory, the quantum cross-correlation energy and the respective force associated with the gradient of this energy can be considered as the product of the virtual exchange of the indistinguishable unit-fields (particles) [see, Eq. (\ref{eq64})]. That is to say that the cross-correlation energy and the force could be attributed to the simultaneous creation/destruction of the unit-field ${\psi _{01}(\mathbf{r})}$ and the simultaneous destruction/creation of the unit-field ${\psi _{02}(\mathbf{r})}$. The cross-correlation in such a virtual process is associated with the indistinguishableness of identical unit-fields ${\psi _{01}(\mathbf{r})}$ and ${\psi _{02}(\mathbf{r})}$ under the interference (interaction). The above-described picture of the virtual quantum exchange of the indistinguishable unit-fields (particles) under the quantum interference, which induces the interaction force associated with the gradient of the cross-correlation energy, is quite similar to the exchange of virtual particles for a short time ($\Delta t \leq 1/\Delta \varepsilon $) in the perturbative particle field theory, where the forces are due to the exchange of the virtual particles (gauge bosons). It should be stressed, however, that the cross-correlation energy associated with the interference between particles (unit-fields) does not exist in quantum mechanics and particle field theory based on the traditional interpretation of quantum interference~\cite{Dir1,Dir2,Land,Jack,Bere,Itz,Ryde,Pes,Grei,Wein}. The picture of interaction of the particles by the cross-correlation of the unit-fields is also {\it{in agreement}} with the Einstein theory of general relativity. According to this theory, the massive object motion, which in classical mechanics is ascribed to the action of the attractive force between two objects, corresponds to inertial motion of the objects within a "curve" geometry of spacetime~\cite{Einst2}. In general relativity there is no force deflecting objects from their natural, straight paths. Instead, the motion corresponds to changes in the properties of spacetime (spacetime field), which in turn changes the straight paths that objects will naturally follow. The field curvature is caused by the energy-momentum of matter. That is to say that the massive object determines the curvature of the spacetime field, in turn, the spacetime field governs the matter motion. The physical mechanism of the forces associated with the cross-correlation of massive unit-fields (particles) under the classical or quantum interference (interaction), which is described by Eqs. (\ref{eq107})-(\ref{eq111}), is very {\it{similar}} to the Einstein mechanism. Indeed, the motion of the first unit-field (particle) $\psi _{1}$ in the presence of the second unit-field (particle) $\psi _{2}$ can be described by the following equation 
\begin{eqnarray}  \label{eq112}
\frac {d {\mathbf{p}_1}}{d {t}}=-\frac {\partial {{\cal H}}}{\partial {R^i}}\mathbf{e}_i,
\end{eqnarray}
where ${\mathbf{p}_1} \equiv {\mathbf{k}_1}$ is the momentum of the first unit-field (particle) at the time moment $ t=t_{12}$, and $-[{\partial {{\cal H}}}/{\partial {R^i}}]\mathbf{e}_i={\mathbf{F}}$ denotes the gravitational force. The motion is determined by the gradient ("curvature") of the density cross-correlation energy (spacetime-like field term)
\begin{eqnarray}  \label{eq113}
{h}_{12,21}({\mathbf{r}},{\mathbf{r}}+{\mathbf{R}})={h}_{12}({\mathbf{r}},{\mathbf{r}}+{\mathbf{R}})+{h}_{21}({\mathbf{r}},{\mathbf{r}}+{\mathbf{R}}),
\end{eqnarray} 
namely by the gradient ("curvature") 
\begin{eqnarray} \label{eq114}
\frac {\partial }{\partial {R^i}}{h}_{12,21}({\mathbf{r}},{\mathbf{r}}+{\mathbf{R}})=\frac {\partial }{\partial {R^i}}[{h}_{12}({\mathbf{r}},{\mathbf{r}}+{\mathbf{R}})+{h}_{21}({\mathbf{r}},{\mathbf{r}}+{\mathbf{R}})] 
\end{eqnarray}
of the spacetime-like field term in Eqs. (\ref{eq108}) and (\ref{eq112}). The change of the particle motion corresponds to changes in the properties of the spacetime-like field term, which, in turn, changes the straightest-possible paths that the first particle will follow without presence of the non-quantum or quantum unit-field of the second particle. The gradient "curvature" (\ref{eq114}) of the spacetime-like field term (\ref{eq113}) is caused by the presence of  the energy-momentum of matter of the second particle (unit-field). Without presence of the second particle, the gradient "curvature" (\ref{eq114}) of the spacetime-like field term (\ref{eq113}) is zero, and the first particle moves freely (${d {\mathbf{p}_1}}/{d {t}}=0$). The change of the properties of the spacetime-like field term (\ref{eq113}) can be considered as the change of the gradient ("curvature") (\ref{eq114}) of the spacetime-like field term. Some physicists say that the massive bodies "tell" the spacetime how to "curve"; the spacetime term "tells" the bodies how to move. In the present model, the particles "tell" the spacetime-like field term (\ref{eq113}) how to "curve", and the spacetime-like field term "tells" the particles how to move. Mathematically, the general relativity replaces the gravitational potential of classical physics by a symmetric rank-two field tensor. In the present model, the gradient ("curvature") of the cross-correlation energy is also viewed rather as a symmetric rank-two field tensor than the gravitational potential of Newton's mechanics. The force (\ref{eq108}) corresponds to the attractive gravitational force if the gradient ("curvature") of the cross-correlation energy of the two unit-fields (particles) is positive. The force (\ref{eq108}) corresponds to the repulsive gravitational force, if such a force does exist, in the case of the negative gradient of the cross-correlation energy. Although the cross-correlation mechanism {\it{compare well}} with the Newton and Einstein mechanisms of gravitation forces, the cross-correlation energy-mass is the {\it{nonexistent energy-mass}} in both the classical physics and Einstein general relativity. It should be stressed again that the above-presented analysis does not depend on the kind of the cross-correlating fields. 

In order to make the above-presented general analysis more transparent let me consider several particular examples. The following examples describe the static forces inducing by gradients of the cross-correlation energy in the ensemble of static (time-independent) boson scalar unit-fields ${\psi _{0n}(\mathbf{r},t)}={\psi _{0n}(\mathbf{r})}$. The static unit-fields ${\psi _{0n}(\mathbf{r})} \equiv {\psi _n (\mathbf{r})}$ can be treated formally as the fields with ${\partial}{\psi _n (\mathbf{r},t)}/{\partial {t}}=0$. The total energy of the two boson unit-fields is found by using Eq. (\ref{eq6}). For the two unit-fields (particles) with ${\psi_1 (\mathbf{r})}\equiv \psi  _1 (\mathbf{r}) e^{i\alpha _1}$, ${\psi _2 (\mathbf{r})}\equiv \psi  _2 (\mathbf{r}) e^{i\alpha _2}$, ${\psi ^* _1 (\mathbf{r})}\equiv \psi ^*  _1 (\mathbf{r}) e^{-i\alpha _1}$ and ${\psi ^* _2 (\mathbf{r})}\equiv \psi ^* _2 (\mathbf{r}) e^{-i\alpha _2}$, which can be presented respectively as ${\psi _1 (\mathbf{r})}=\phi  (\mathbf{r}) e^{i\alpha _1}$,  ${\psi _2 (\mathbf{r})}=\phi  (\mathbf{r+R}) e^{i\alpha _2}$, ${\psi ^* _1 (\mathbf{r})}=\phi ^* (\mathbf{r}) e^{-i\alpha _1}$,  ${\psi ^* _2 (\mathbf{r})}=\phi ^* (\mathbf{r+R}) e^{-i\alpha _2}$   [an illustration for the non-complex unit-fields (${\psi _1 (\mathbf{r})}= {\psi ^* _1 (\mathbf{r})}$,  ${\psi _2 (\mathbf{r})}= {\psi ^* _2 (\mathbf{r})}$) is shown in Fig. 2],  Eqs. (\ref{eq6}) and (\ref{eq107}) yielded the total energy
\begin{eqnarray}  \label{eq115}
{\cal H} = {\cal H}_1 + {\cal H}_2 + {\cal H}_{12,21}(R),
\end{eqnarray}
where 
\begin{eqnarray}  \label{eq116}
{\cal H}_1 + {\cal H}_2 = \int _{0}^{\infty}[m_1^2{\phi ^*}({\mathbf {r}}){\phi }({\mathbf {r}})+\nabla {\phi ^*}({\mathbf {r}}) \nabla {\phi }({\mathbf {r}})]d^3x +\nonumber \\ + \int_{0}^{\infty}[m_2^2{\phi ^*}({\mathbf {r+R}}){\phi}({\mathbf {r+R}})+\nabla {\phi ^*}({\mathbf {r+R}}) \nabla {\phi}({\mathbf {r+R}})]d^3x.
\end{eqnarray}
is the part of the total energy, which does not depend on the distance $R$. Here, the term ${\cal H}_{12,21}(R)$ denotes the cross-correlation energy ${\cal H}_{12}(R) + {\cal H}_{21}(R)$, and the self-correlation energies are given by ${\cal H}_{11} \equiv {\cal H}_1$ and ${\cal H}_{22} \equiv {\cal H}_2$. In order to satisfy the Einstein theory of relativity, the energy (\ref{eq116}) should be equal to the Planck-Einstein energy ${\varepsilon}=({\mathbf{k}}_1^2+m_1^2)^{1/2}+({\mathbf{k}}_2^2+m_2^2)^{1/2}$ of the two non-correlating unit-fields (particles):
\begin{eqnarray}  \label{eq117}
{\cal H}_1 + {\cal H}_2 = ({\mathbf{k}}_1^2+m_1^2)^{1/2}+({\mathbf{k}}_2^2+m_2^2)^{1/2}.  
\end{eqnarray} 
The cross-correlation energy ${\cal H}_{12,21}(R)$, which depends on the distance $R$, is given by
\begin{eqnarray} \label{eq118}
{\cal H}_{12,21}(R)= \int_{0}^{\infty}(m_1m_2 [e^{-i(\alpha _1-\alpha _2)}{\phi}^*({\mathbf {r}}) \phi ({\mathbf {r}}+ {\mathbf {R}}) +  e^{i(\alpha _1-\alpha _2)}{\phi}({\mathbf {r}}) \phi ^*({\mathbf {r}} + {\mathbf {R}})]+\nonumber \\  + [e^{-i(\alpha _1-\alpha _2)} \nabla {\phi ^*}({\mathbf {r}}) \nabla {\phi}({\mathbf {r}}+{\mathbf {R}}) + e^{i(\alpha _1-\alpha _2)}\nabla {\phi}({\mathbf {r}}) \nabla {\phi ^*}({\mathbf {r}}+{\mathbf {R}})])d^3x. 
\end{eqnarray}
The force corresponding to the gradient of the cross-correlation energy is given by Eq. (\ref{eq108}), where
\begin{eqnarray}  \label{eq119}
{\frac {\partial {\cal H}}{\partial {R^i}}} = \int_{0}^{\infty}{\frac {\partial }{\partial {R^i}}}(m_1m_2 [e^{-i(\alpha _1-\alpha _2)}{\phi}^*({\mathbf {r}}) \phi ({\mathbf {r}}+ {\mathbf {R}}) +  e^{i(\alpha _1-\alpha _2)}{\phi}({\mathbf {r}}) \phi ^*({\mathbf {r}} + {\mathbf {R}})]+\nonumber \\  + [e^{-i(\alpha _1-\alpha _2)} \nabla {\phi ^*}({\mathbf {r}}) \nabla {\phi}({\mathbf {r}}+{\mathbf {R}}) + e^{i(\alpha _1-\alpha _2)}\nabla {\phi}({\mathbf {r}}) \nabla {\phi ^*}({\mathbf {r}}+{\mathbf {R}})])d^3x.
\end{eqnarray}
Notice, the integral in Eq. (\ref{eq119}), which corresponds to the term (\ref{eq118}), depends on the coordinates, phases and the pure geometry of the boson fields. Remember, that the operator representation [see, Sec. (2)] of the unit-fields ${\psi(\mathbf{r})}$ and ${\psi({\mathbf{r}}+{\mathbf{R}})}$ shows that the force (\ref{eq108}) can be viewed as the product of the virtual exchange of the unit-fields (particles). In other words, the force could be interpreted as the product of the virtual process of the creation/destruction of the particle in the point $(\mathbf{r},t)$ of the unit-field ${\psi(\mathbf{r},t)}$ and the simultaneous destruction/creation of the unit-field (particle) in the same point $({\mathbf{r}},t)$ of the unit-field ${\psi({\mathbf{r}}+{\mathbf{R}},t)}$ at the every time moment $t$. The pure geometrical nature of the force can be emphasized by using the unit masses $m_1=1$ and $m_2=1$. In such a representation, the gradient ("curvature") (\ref{eq119}) of the cross-correlation energy (\ref{eq118}) is viewed rather as a symmetric rank-two field tensor of the Einstein general relativity than the gravitational potential of classical physics. From the point of view of the present model of the particle interactions, which is based on the gradient of the cross-correlation energy, the Einstein and quantum descriptions (interpretations) of the forces {\it{are equivalent}}. The difference between the models is {\it{in absence}} of the cross-correlation energy mediated by the interference between particles (bodies) in both the Einstein relativity and the particle field theory based on the canonical interpretation of quantum interference. The values of the energy and force could be calculated by using the concrete field functions ${{\psi}}({\mathbf{r}})$ and ${{\psi}}({\mathbf{r}}+{\mathbf{R}})$. The attractive or repulsive character of the force (\ref{eq108}) is determined by the sign ${\cal S} =\pm 1$ of the cross-correlation energy ${\cal H}_{12}(R)$. For an example, in the case of ${{\phi}}({\mathbf{r}}) = {{\phi ^*}}({\mathbf{r}})$ and ${{\phi}}({\mathbf{r+R}}) = {{\phi ^*}}({\mathbf{r+R}})$, the expressions (\ref{eq118}) and (\ref{eq119}) can be presented in the very simple forms
\begin{eqnarray} \label{eq120}
{\cal H}_{12,21}(R)=2{\cal S}G \int_{0}^{\infty}[m_1{\phi}({\mathbf {r}})m_2 \phi ({\mathbf {r}}+ {\mathbf {R}})+\nabla {\phi}({\mathbf {r}}) \nabla {\phi}({\mathbf {r}}+{\mathbf {R}})]d^3x 
\end{eqnarray}
and 
\begin{eqnarray} \label{eq121}
{\frac {\partial {\cal H}}{\partial {R^i}}} = 2{\cal S}G\int_{0}^{\infty}{\frac {\partial }{\partial {R^i}}}[m_1{\phi}({\mathbf {r}})m_2 \phi ({\mathbf {r}}+ {\mathbf {R}})+\nabla {\phi}({\mathbf {r}}) \nabla {\phi}({\mathbf {r}}+{\mathbf {R}})]d^3x,
\end{eqnarray}
where ${\cal S}=\cos (\alpha  _1-\alpha  _2)/|\cos (\alpha  _1-\alpha  _2)|$ and $G=|\cos (\alpha  _1-\alpha  _2)|$. The sing ${\cal S}$ and the coupling parameter $G$ are dimensionless constants, which are determined by the experimental data for the energy and force of the concrete boson unit-fields. If Eqs. (\ref{eq120}) and (\ref{eq121}) are applied formally to the fermion particles (electrons) with the spins $s_1=\pm 1/2$ and $s_2=\pm 1/2$, then the sign ${\cal S}=\cos (\alpha  _1-\alpha  _2)/|\cos (\alpha  _1-\alpha  _2)|$ can be considered as the sign of the product $SG\sim s_1 s_1$, which does not depend on the intrinsic or non-intrinsic angular momentums of the unit-fields (particles). Notice, the Planck-Einstein energy (\ref{eq117}) of the two unit-fields (particles) does not depend on the $SG$ product associated with the spins $s_1$ and $s_2$. The cross-correlation energy (\ref{eq120}) may be {\it{reinterpreted}} as the Heisenberg-Dirac energy of the exchange interaction of the spins.  The cross-correlation force mediating by the gradient of the energy (\ref{eq120}) would be attractive if the electrons have the different spins, $s_1=\pm 1/2$ and $s_2=\mp 1/2$. This case corresponds to the phase difference $\pi /2 <\alpha  _1-\alpha _2<{3\pi}/2 $. The unit-waves with the phase difference $\alpha  _1-\alpha _2=\pi /2$ or $\alpha _1-\alpha _2=3\pi /2$ do not interfere (interact) one with another. The force would be repulsive if the electrons are identical, $s_1=s_2=\pm 1/2$. The repulsive force corresponds to the phase difference  $0 <\alpha _1-\alpha _2<{\pi}/2 $ or $3{\pi }/2 <\alpha _1-\alpha _2<2{\pi}$. That is to say that the two identical fermions (unit-fields) cannot occupy the same single-particle states because the overlapped unit-fields cannot share the common volume due to the repulsive force. In the present model this could be interpreted as the Pauli exclusion principle. That means that the physical mechanism (interpretation) behind the Pauli principle in the present model is different from the quantum mechanics and particle field theory based on the waves of probabilities or the waves of operators. For instance, in quantum mechanics, the two-particle system is described by a symmetric (boson) or antisymmetric (fermion) state, which has been mathematically constructed by using the probability amplitudes (wave-functions) of the particles. If the electrons are the same, the mathematically engineered antisymmetric expression of the state gives zero. Therefore, in an antisymmetric state, two identical particles cannot occupy the same single-particle states. This is known as the Pauli exclusion principle, which in quantum mechanics has the pure {\it{mathematical interpretation}} based on the mathematical properties of the antisymmetric state (function). In particle field theory, the Pauli exclusion principle is attributed to the mathematical properties of the canonical commutation relation (pure mathematical object) of fermions. The mathematical construction of the antisymmetric state (function) in quantum mechanics or the canonical commutation relation of fermions in particle field theory yields the Fermi-Dirac statistics, whose interpretation is also pure {\it{mathematical}}. Similarly, the modelling of the boson gas by a symmetric state (function) or the use of the canonical commutation relation of bosons leads mathematically to the Bose-Einstein statistics. In the present model, the physical mechanism behind the Bose-Einstein statistics, Fermi-Dirac statistics and Pauli exclusion principle is attributed to the attractive and repulsive forces associating with the subtractive and additive interference (cross-correlation) of the real unit-waves (bosons or fermions) of the matter (mass-energy). It should be noted (see, Eq. (\ref{eq118})) that the sign ${\cal S}_1$ of the product $m_1m_2 [e^{-i(\alpha _1-\alpha _2)}{\phi}^*({\mathbf {r}}) \phi ({\mathbf {r}}+ {\mathbf {R}}) +  e^{i(\alpha _1-\alpha _2)}{\phi}({\mathbf {r}}) \phi ^*({\mathbf {r}} + {\mathbf {R}})]$ can be different from the sign ${\cal S}_2$ of the product  $[e^{-i(\alpha _1-\alpha _2)} \nabla {\phi ^*}({\mathbf {r}}) \nabla {\phi}({\mathbf {r}}+{\mathbf {R}}) + e^{i(\alpha _1-\alpha _2)}\nabla {\phi}({\mathbf {r}}) \nabla {\phi ^*}({\mathbf {r}}+{\mathbf {R}})]$. Therefore the expressions (\ref{eq120}) and (\ref{eq121}) can be presented in the forms containing the signs $S_1$ and $S_2$ as
\begin{eqnarray}
{\cal H}_{12,21}(R)= \int_{0}^{\infty}(S_1m_1m_2 |e^{-i(\alpha _1-\alpha _2)}{\phi}^*({\mathbf {r}}) \phi ({\mathbf {r}}+ {\mathbf {R}}) +  e^{i(\alpha _1-\alpha _2)}{\phi}({\mathbf {r}}) \phi ^*({\mathbf {r}} + {\mathbf {R}})|+\nonumber \\  + S_2|e^{-i(\alpha _1-\alpha _2)} \nabla {\phi ^*}({\mathbf {r}}) \nabla {\phi}({\mathbf {r}}+{\mathbf {R}}) + e^{i(\alpha _1-\alpha _2)}\nabla {\phi}({\mathbf {r}}) \nabla {\phi ^*}({\mathbf {r}}+{\mathbf {R}})|)d^3x. 
\end{eqnarray}
and 
\begin{eqnarray}
{\frac {\partial {\cal H}}{\partial {R^i}}} = \int_{0}^{\infty}{\frac {\partial }{\partial {R^i}}}(S_1m_1m_2 |e^{-i(\alpha _1-\alpha _2)}{\phi}^*({\mathbf {r}}) \phi ({\mathbf {r}}+ {\mathbf {R}}) +  e^{i(\alpha _1-\alpha _2)}{\phi}({\mathbf {r}}) \phi ^*({\mathbf {r}} + {\mathbf {R}})|+\nonumber \\  + S_2|e^{-i(\alpha _1-\alpha _2)} \nabla {\phi ^*}({\mathbf {r}}) \nabla {\phi}({\mathbf {r}}+{\mathbf {R}}) + e^{i(\alpha _1-\alpha _2)}\nabla {\phi}({\mathbf {r}}) \nabla {\phi ^*}({\mathbf {r}}+{\mathbf {R}})|)d^3x.
\end{eqnarray}
In the case of ${\cal S}_1 \neq {\cal S}_2$, the interference of unit-fields could be neither purely constructive nor purely destructive interference. Respectively, the interaction of unit-fields would not be purely repulsive or purely attractive. Thus the statistics would be different from the Fermi-Dirac or Bose-Einstein statistics. Some general properties of the energy and force determining by Eqs.~(123)-(126) can be derived without knowledge of the exact functions ${{\phi}}({\mathbf{r}})$ and ${{\phi}}({\mathbf{r}}+{\mathbf{R}})$. In the limit $ \mathbf{R}{\rightarrow}\infty$, the cross-correlation energy and force are equal to zero due to non-overlapping of the fields ${{\phi}}({\mathbf{r}})$ and ${{\phi}}({\mathbf{r}}+{\mathbf{R}})$. The total energy (\ref{eq115}) of the two unit-fields (particles), which can have the positive or negative gradient of the cross-correlation energy for the different distances ${\mathbf{R}}$, is equal to the Planck-Einstein energy ${\varepsilon}=({\mathbf{k}}_1^2+m_1^2)^{1/2}+({\mathbf{k}}_2^2+m_2^2)^{1/2}$ at $ \mathbf{R}{\rightarrow}\infty$. In the limit $ \mathbf{R}{\rightarrow}0$, the energy (\ref{eq115}) of the two particles depends on the additive or subtractive character of the cross-correlation and the nature of the particles. Also note that the annihilation (pure destructive interference) of the particle and antiparticle yields the energy $\varepsilon =0$. The energy of the particles with repulsive forces tends to infinity.
\begin{figure}
\begin{center}
\includegraphics[keepaspectratio, width=0.5\columnwidth]{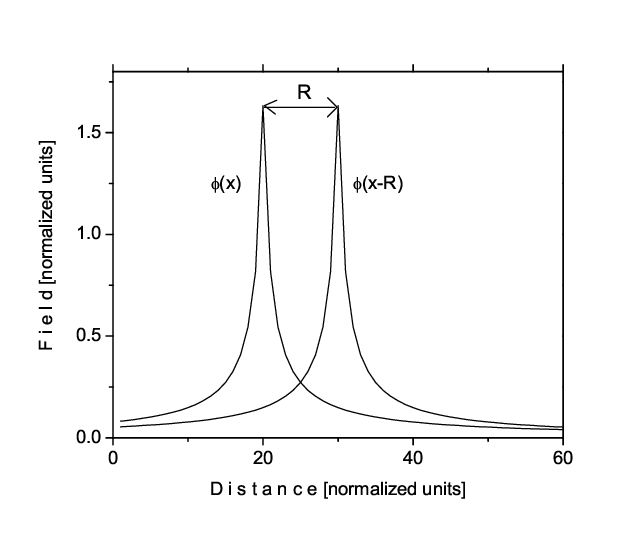}
\end{center}
\caption{The two identical unit-fields (particles) with ${{\phi}}({\mathbf{r}})\equiv {{\phi}}(x)=(3m/8)^{-1/2}(1+|x|)^{-1}$ and ${{\phi}}({\mathbf{r}}+{\mathbf{R}})\equiv {{\phi}}(x-R)=(3m/8)^{-1/2}(1+|x-R|)^{-1}$, where $R$ is the distance between the unit-fields (particles).}
\end{figure}
\begin{figure}
\begin{center}
\includegraphics[keepaspectratio, width=0.5\columnwidth]{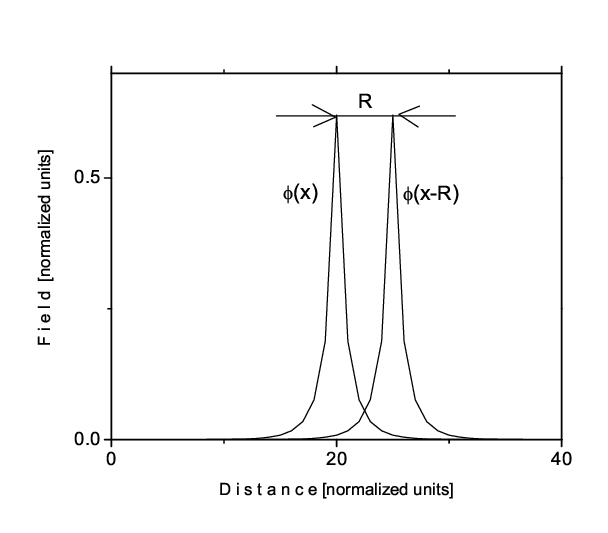}
\end{center}
\caption{The two identical unit-fields (particles) with $ {{\phi}}({\mathbf{r}})\equiv {{\phi}}(x)=(m/2.6)^{-1/2}(1+|x|)^{-1}e^{-(1/2)|x|}$ and ${{\phi}}({\mathbf{r}}+{\mathbf{R}})\equiv {{\phi}}(x-R)=(m/2.6)^{-1/2}(1+|x-R|)^{-1}e^{-(1/2)|x-R|}$, where $R$ is the distance between the unit-fields (particles).}
\label{fig:Fig4}
\end{figure}

The above-considered general properties and behaviors of the energy and force could be understand better by considering the one-dimensional (1-D in space) scalar unit-fields (bosons) with the two simplest forms of the function ${\phi (\mathbf{r})}$, namely the static unit-fields with the functions [see, the illustrations in Figs. (3) and (4)]
\begin{eqnarray}  \label{eq122}
{{\phi}}({\mathbf{r}})\equiv {{\phi}}(x)=(3m/8)^{-1/2}(1+|x|)^{-1} 
\end{eqnarray} 
and
\begin{eqnarray} \label{eq123}
{{\phi}}({\mathbf{r}})\equiv {{\phi}}(x)=(m/2.6)^{-1/2}(1+|x|)^{-1}e^{-(1/2)|x|},
\end{eqnarray} 
which could be considered as the quasi-solutions of the 1-D Euler-Lagrange equation of motion having the form of the Klein-Gordon-Fock equation (see, Part II). In order to emphasize the pure geometrical character of Eqs. (121)-(126) for these fields, I have performed the all computations with the unit masses $m_1=1$ and $m_2=1$. Notice, the introduction of the unity into the fields ${{\phi}}({\mathbf{r}})\sim (|x|)^{-1}$ and ${{\phi}}({\mathbf{r}})\sim (|x|)^{-1}e^{-(1/2)|x|}$ has been performed for the mathematical simplicity, only. The fields (\ref{eq122}) and (\ref{eq123}) do satisfy the Planck-Einstein energy ${\varepsilon}=({\mathbf{k}}_1^2+m_1^2)^{1/2}+({\mathbf{k}}_2^2+m_2^2)^{1/2}$ of the two interference-less unit-fields (particles) without the divergences in the point $x=0$. Figures (5) - (8) show the calculated energy and force for the fields (\ref{eq122}) and (\ref{eq123}). The pair of the unit-fields ${{\phi}}({\mathbf{r}})\equiv {{\phi}}(x)=(3m/8)^{-1/2}(1+|x|)^{-1}$ and ${{\phi}}({\mathbf{r}}+{\mathbf{R}})\equiv {{\phi}}(x-R)=(3m/8)^{-1/2}(1+|x-R|)^{-1}$ could be considered as the particle pair, which obeys the inverse square law, with the "relatively weak" (do not confuse with electroweek) long-range interaction associated with the Coulomb or gravitational force [Figs. (5) and (6)].
\begin{figure}
\begin{center}
\includegraphics[keepaspectratio, width=0.5\columnwidth]{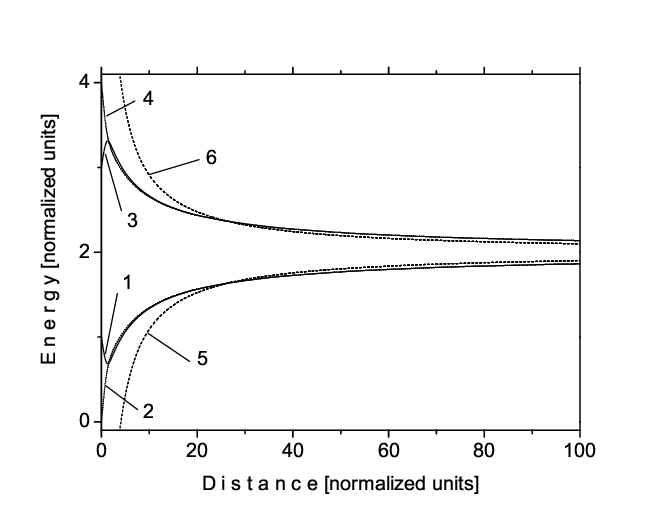}
\end{center}
\caption{The calculated energies (1,2) and (3,4) are compared with the energies $\varepsilon (R)=(m_1+m_2)+{\cal S}G10(1+|R|)^{-1}$, where the signs ${\cal S}=-1$ and ${\cal S}=+1$ correspond to the curves (5) and (6), respectively. The signs (${\cal S}_1=1$, ${\cal S}_2=-1$) and  (${\cal S}_1=-1$, ${\cal S}_2=1$) correspond to the curves (1) and (3), respectively. The signs (${\cal S}_1={\cal S}_2=-1$) and  (${\cal S}_1={\cal S}_2=1$) correspond to the curves (2) and (4), respectively.}
\label{fig:Fig5}
\end{figure}
\begin{figure}
\begin{center}
\includegraphics[keepaspectratio, width=0.5\columnwidth]{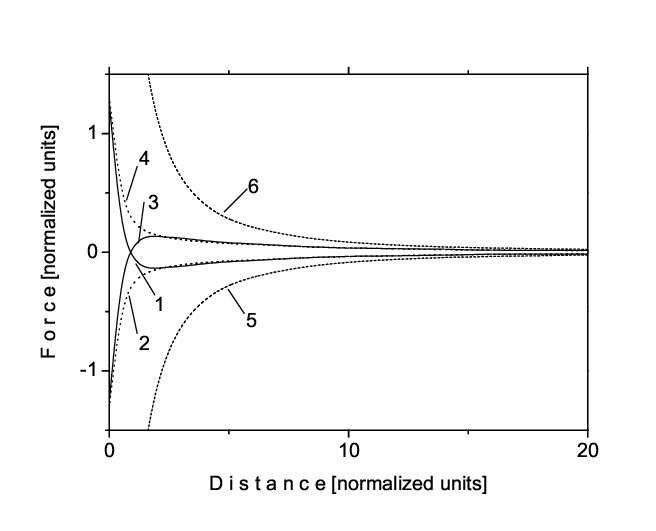}
\end{center}
\caption{The calculated forces (1,2) and (3,4) correspond to the calculated energies (1,2) and (3,4) of Fig. 5. The forces (1,2) and (3,4) are compared with the forces $-{\partial \varepsilon (R)}/ {\partial {R}}$, where the energies $\varepsilon (R)=(m_1+m_2)+{\cal S}G10(1+|R|)^{-1}$ and the signs ${\cal S}=-1$ and ${\cal S}=+1$ correspond to the curves (5) and (6), respectively. The signs (${\cal S}_1=1$, ${\cal S}_2=-1$) and  (${\cal S}_1=-1$, ${\cal S}_2=1$) correspond to the curves (1) and (3), respectively. The signs (${\cal S}_1={\cal S}_2=-1$) and  (${\cal S}_1={\cal S}_2=1$) correspond to the curves (2) and (4), respectively.}
\label{fig:Fig6}
\end{figure}
Figures (5) and (6) show the calculated energy and force, respectively. The calculated energy and force are compared with the energy $\varepsilon (R)=(m_1+m_2)+{\cal S}G10(1+|R|)^{-1}$ and force $-{\partial \varepsilon (R)}/ {\partial {R}}$, which are similar to the Newton potential energy and force. The signs ${\cal S}=+1$ and ${\cal S}=-1$ correspond to the repulsive and attractive forces, respectively. In the calculations,  the cross-correlation (interaction, coupling) parameter is given by $G$=1. The force (\ref{eq108}) slowly increases with decreasing the distance $R$, $|F(R)| \sim |R|^{-2}$. At the distance less than approximately 1, the attractive and repulsive forces exchange the roles in the case of the curves (1) and (3) of Figs. (5) and (6). The calculated energy (${\cal H}=1.9$) approaches the total energy (${\cal H}=2$) of the free particles at the critical distance $R_c\approx 10^2$. In the case of the curves (1) and (3) of Figs. (5) and (6), the energies have extreme values at the distance $R\approx 1$. In the case of the curves (2) and (4) of Figs. (5) and (6), the energies have maximum or minimum values at the distance $R=0$. The bounded unit-field pare with the minimum energy at the distance $R\neq \infty $ could be considered as a composite particle. Note that the considerable energy-mass defect $\Delta {\cal H}={\cal H}-({\cal H}_{1}+{\cal H}_{2})={\cal H}_{12}$ exists in such a case [see, the comments to Eq. (\ref{eq35})]. The calculations have been performed for the fields with the coupling parameter $G$=1. The field (\ref{eq122}) could be attributed to the one-dimensional gravitational field if the sign-coupling dimensionless constant $SG$ is equal to the respective Newton constant of gravitation. If the field ${\phi}(\mathbf{r})$ is associated with both the unit-mass $m$ and unit-charge $q$, then the fields ${\phi}_1$ and ${\phi}_2$ are presented as ${\phi}_1(m,x) = {\phi}_1(m,q,x)$ and ${\phi}_2(m,x) ={\phi}_2(m,q,x)$. The energy (mass) of the composite one-dimensional field is described by Eq. (\ref{eq115}), where the parameters should correspond to the Planck-Einstein energy of two charged particles. That means the energy (\ref{eq116}) is equal to the Planck-Einstein energy ${\cal H}_1 + {\cal H}_2 = m_1+m_2$ of the two charged, massive unit-fields (with the spin $s=0$ or $s\neq 0$) at the rest in the limit $R\rightarrow \infty$. The charges of the particles (as well as the particle spins) are attributed to the sign-coupling product $SG\equiv S \gamma  q_1  q_2$, which characterizes the sign and value of the gradient of the cross-correlation (interaction, coupling) energy under interference of the two charged unit-waves [see, also comments to Eqs. (\ref{eq120}) and (\ref{eq121})]. The unit-fields could be attributed to the one-dimensional Coulomb fields if the sign-coupling dimensionless constant is equal to the respective Coulomb constant. The pair of the unit-fields $ {{\phi}}({\mathbf{r}})\equiv {{\phi}}(x)=(m/2.6)^{-1/2}(1+|x|)^{-1}e^{-(1/2)|x|}$ and ${{\phi}}({\mathbf{r}}+{\mathbf{R}})\equiv {{\phi}}(x-R)=(m/2.6)^{-1/2}(1+|x-R|)^{-1}e^{-(1/2)|x-R|}$ could be considered as the particle pairs, which exhibits the exponential law, with the strong or weak short-range interaction associated with the respective strong or weak force. The calculated energy and force of the composite field  ${{\psi}}({\mathbf{r}})+{{\psi}}({\mathbf{r}}+{\mathbf{R}})$ is shown in Figs. (7) and (8). \begin{figure}
\begin{center}
\includegraphics[keepaspectratio, width=0.5\columnwidth]{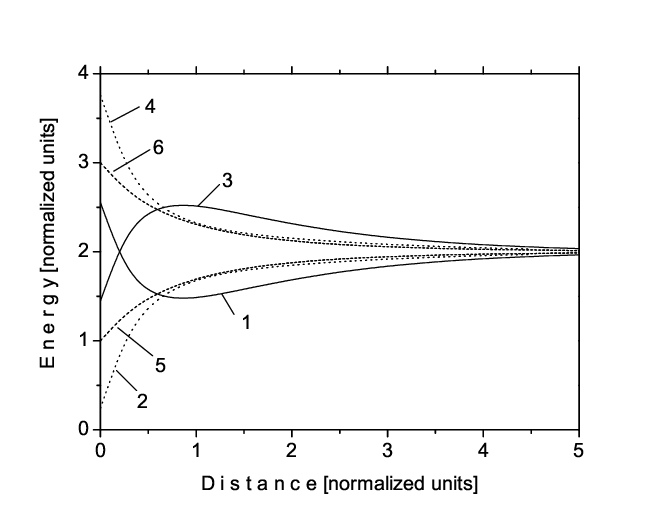}
\end{center}
\caption{The calculated energies (1,2) and (3,4) are compared with the energies $\varepsilon (R)=(m_1+m_2)+{\cal S}G(1+|R|)^{-1}e^{-(1/2)|R|}$, where the signs ${\cal S}=-1$ and ${\cal S}=+1$ correspond to the curves (5) and (6), respectively. The signs (${\cal S}_1=1$, ${\cal S}_2=-1$) and  (${\cal S}_1=-1$, ${\cal S}_2=1$) correspond to the curves (1) and (3), respectively. The signs (${\cal S}_1={\cal S}_2=-1$) and  (${\cal S}_1={\cal S}_2=1$) correspond to the curves (2) and (4), respectively.}
\label{fig:Fig7}
\end{figure}
\begin{figure}
\begin{center}
\includegraphics[keepaspectratio, width=0.5\columnwidth]{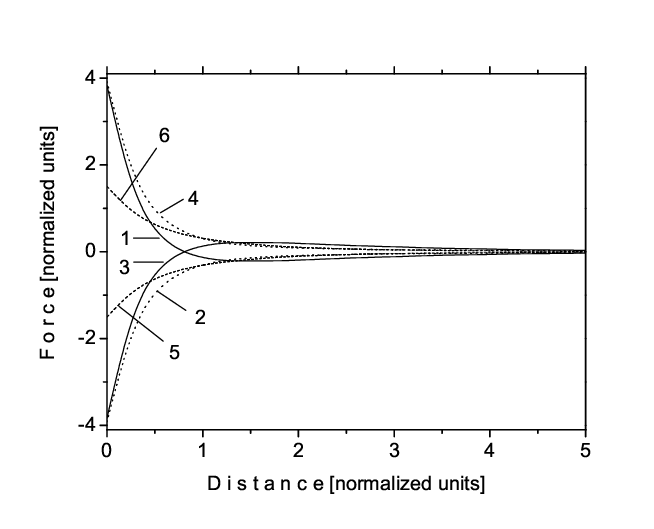}
\end{center}
\caption{The calculated forces (1,2) and (3,4) correspond to the calculated energies (1,2) and (3,4) of Fig. 7. The forces (1,2) and (3,4) are compared with the forces $-{\partial \varepsilon (R)}/ {\partial {R}}$, where the energies $\varepsilon (R)=(m_1+m_2)+{\cal S}G(1+|R|)^{-1}e^{-(1/2)|R|}$ and the signs ${\cal S}=-1$ and ${\cal S}=+1$ correspond to the curves (5) and (6), respectively. The signs (${\cal S}_1=1$, ${\cal S}_2=-1$) and  (${\cal S}_1=-1$, ${\cal S}_2=1$) correspond to the curves (1) and (3), respectively. The signs (${\cal S}_1={\cal S}_2=-1$) and  (${\cal S}_1={\cal S}_2=1$) correspond to the curves (2) and (4), respectively.}
\label{fig:Fig8}
\end{figure}
The calculated energy and force are compared with the energy $\varepsilon (R)=(m_1+m_2)+{\cal S}G(1+|R|)^{-1}e^{-(1/2)|R|}$ and force $-{\partial \varepsilon (R)}/ {\partial {R}}$, which correspond to the Yukawa-like potential energy and force. The calculations have been performed for the coupling parameter $G$=1. The repulsive and attractive forces correspond to the signs ${\cal S}=+1$ and ${\cal S}=-1$, respectively. The force (\ref{eq108}) rapidly decreases with increasing the distance $R$, $|F(R)| \sim |R|^{-1}e^{-|R|}$. At the distance less than approximately 1, the attractive and repulsive forces exchange the roles in the case of the curves (1) and (3) of Figs. (7) and (8). The calculated energy (${\cal H}=1.9$) approaches the total energy (${\cal H}=2$) of the free particles at the critical distance $R_c\approx 5$, which is short in comparison to the critical distance $R_c\approx 10^2$ of the weakly interacting (mass-induced or charge-induced) fields. In the case of the curves (1) and (3) of Figs. (7) and (8), the energies have extreme values at the distance $R\approx 1$. In the case of the curves (2) and (4) of Figs. (7) and (8), the energies have maximum or minimum values at the distance $R=0$. The bounded unit-field pare with the minimum energy at the distance $R\neq \infty $ could be considered as a composite particle. Note that the considerable energy-mass defect $\Delta {\cal H}={\cal H}_{12}$ exists in such a case [see, also the comments to Eq. (\ref{eq35})]. The energy associated with the energy-mass defect is usually called the binding energy. The field (\ref{eq123}) could be attributed to the one-dimensional Yukawa field if the sign-coupling product is equal to the respective Yukawa constant. Respectively, the field (\ref{eq123}) could be attributed to the strong or weak short-range interaction if the sign-coupling product $SG$ is equal to the respective strong or weak force. I have presented the computer results for the "weak" (\ref{eq122}) and "strong" (\ref{eq123}) boson scalar unit-fields. The results for the vector boson fields and fermion fields are similar to the boson scalar unit-fields. One can easily compute the energy and force by using the presented model for the vector boson fields and the spinor fields, as well as for the composite fields composed from both the boson and spinor fields. It should be mentioned again that the above-described bounded unit-field pair, the pair that has the minimum energy at the distance $R\neq \infty $, is considered as a stable composite particle. To provide the stable state at the distance $R=0 $, the sign ${\cal S}_1=-1$ of the product ${\psi}({\mathbf {r}})\psi ({\mathbf {r}}+ {\mathbf {R}}) $ should not be different from the sign ${\cal S}_2=-1$ of the product  $\nabla {\psi}({\mathbf {r}}) \nabla {\psi}({\mathbf {r}}+{\mathbf {R}}) $. In the case of ${\cal S}_1=-1$ and ${\cal S}_2=-1$, the two-particle state (particle) is the degenerate state because the two unit-fields (particles) are indistinguishable, in any spacetime point. In other words, such a particle does not have any internal structure (substructure). The stable non-degenerate state (composite particle with measurable internal structure) is provided by the signs ${\cal S}_1=1$ and ${\cal S}_2=-1$ at the distance $0<R\neq \infty $. 

The Einstein and traditional quantum-physics descriptions of the forces are {\it{equivalent}} from the point of view of the present model of the particle interactions, which is based on the gradient of the cross-correlation energy connected with the interference between unit-fields (particles). The difference between the models is in {\it{absence}} of the cross-correlation energy mediated by the interference between particles in both the Einstein relativity and traditional quantum-physics theories. In other words, the cross-correlation energy-mass associated with the cross-correlation (interference) of bodies (particles) {\it{does not exist}} in the Newton (classical) mechanics, Einstein relativity, quantum-mechanics and quantum-field theories. The interference-less behaviour of particles in these theories is provided by the spatial separation of the massive point-like particles, which can not occupy the same spacetime point. Indeed, in the Newton mechanics and Einstein special relativity of the interference-less particles, the material point-like particles are always separated by the "straight" empty space. The material interference-free bodies are separated by the "curve" empty space in the Einstein general relativity. The physical picture is more complicated in quantum mechanics and quantum field theories, where the "straight" or "curve" empty space (vacuum) between the interference-less point particles (bosons or fermions) is full-filled by the virtual particles (gauge bosons). Although the bosons can occupy the same spacetime point, the cross-correlation energy of the bosons in this point is equal to zero due to the Copenhagen-Dirac postulate. The identical fermions can not share the same spacetime point. Therefore the cross-correlation energy of the identical fermions in this point is equal to zero. The cross-correlation energy of non-identical fermions occupying the same spacetime point is equal to zero due to the Copenhagen-Dirac postulate. The non-zero cross-correlation energy induced by the interference between particles could give natural explanation of the invisible energy-mass, which in astrophysics and cosmology is usually called the "dark" energy-mass. Indeed, in order to account for the well-known discrepancies between measurements based on the mass of the visible matter in astronomy and cosmology and definitions of the energy (mass) made through dynamical or general relativistic means, the present model does not need in hypothesizing the existence of "dark" energy-mass. In the present model, the "dark" cosmological energy-mass as well as the well-known spiral cosmological structures are attributed simply to the coherent cross-correlation energy-mass of the moving cosmological objects. For instance, the unit-fields (particles) of cosmological objects moving in the same direction with the same velocity have the same moments $\mathbf{k_n}=\mathbf{k}$. According to the present model [for more details, see Part II and Sec. (2) of Part I], the interference between the macroscopic bodies (fields) composed from the unit-fields does induce the cross-correlation energy-mass, which would be undetectable by emitted or scattered electromagnetic radiation in astronomical and cosmological observations. In this regard, note that an infinite unit-field (for instance, ${{\psi}}({\mathbf{r}})\sim |\mathbf{r}|^{-1}$) occupying the finite volume ($V\neq \infty$) of the {\it{visible}} Universe is overlapped with the superposition of all other infinite unit-fields. The energy of such an infinite unit-field includes both the canonical energy of the single unit-field (particle) and the cross-correlation energy ($\varepsilon _{cc}$) associated with the interference (cross-correlation) with the superposition of all other infinite unit-fields, namely with the physical vacuum of the visible Universe. The cross-correlation energy in such a case may be considered as the energy of interaction of the unit-field (particle) with the {\it{virtual particles}} of the physical vacuum of the visible Universe. Notice, in the case of homogeneous distribution of unit-fields (particles), the cross-correlation energy $\varepsilon _{cc}\rightarrow 0$ if the volume $V\rightarrow  \infty$. Also note that the {\it{vacuum-particle}}  cross-correlation energy, which is equal to zero for the constant velocity ($\mathbf{v}=const.$) in inertial coordinate systems having the shift symmetry of time, {\it{increases}} per the time unit $\delta t$ with the increase $\delta \mathbf{v}$ or $\delta \mathbf{k}$ of the velocity $\mathbf{v}$ or the momentum $\mathbf{k}$ of the unit-field (particle) due to the interference  (interaction) of the accelerating unit-field with the physical vacuum. The increase $\delta \varepsilon _{cc}$ of the cross-correlation (interaction) energy caused by the interference between the accelerating unit-field and the physical vacuum could be interpreted as the mass increase, in agreement with the Einstein special relativity relation $m=m_0/(1-v^2/c^2)^{1/2}$. That give, probably for the first time, a microscopic explanation (physical  interpretation) of this phenomenon. One should not confuse here the {\it{absolute (true) vacuum}} associated with the {\it{"straight"}} geometry of the {\it{non-material}} spacetime field of the Einstein special relativity with the {\it{physical (non-absolute) vacuum}} connected with the "{\it{curve}}" geometry of the {\it{non-material}} spacetime field of the Einstein general relativity. The energies of the "{\it{straight}}" spacetime (non-material absolute vacuum) of the Einstein spacial relativity and Newton mechanics are equal to {\it{zero}}, while the "{\it{curve}}", {\it{non-material}}, geometrical spacetime (mathematical object) of the Einstein general relativity has the {\it{non-zero values}} depending on the spatial distribution of masses. In the present model, the spacetime energy is equal to zero only if the non-material spacetime is not occupied by the material unit-fields (particles). In other words, the energy of an empty (non-material) spacetime of the present model is equal to zero. Notice, the spacetime energy of the Universe occupied by the {\it{infinite}} unit-fields contains both the energies of material unit-fields (particles) and the cross-correlation energy of interfering unit-fields. Therefore, the Universe occupied by the {\it{infinite}} unit-fields always has the non-zero value of the spacetime energy associated with the cross-correlation of unit-fields. The spacetime not occupied by the {\it{finite}} unit-fields obeys the zero value of energy. 

The above-presented analysis of the interactions of unit-fields has showed that the physical pictures of the {\it{non-zero}} energies of the physical vacuum  (spacetime) of the Einstein general relativity, the traditional quantum-field theory and the present model are {\it{formally identical}}. Nevertheless, from a point of view of the {\it{materialistic philosophy}}, the physical ({\it{materialistic}}) interpretation of the non-zero spacetime energy of the physical vacuum in a volume occupied by the finite or infinite {\it{material unit-fields}} is more natural and transparent physically than the interpretations based on the use of the Einstein field of {\it{non-material}}, geometrical, "curve" spacetime or the field of {\it{virtual}} (simultaneously created and destroyed) unobservable particles mediated by the perturbation mathematical approximations of the traditional theory of quantum fields. For the {\it{non-materialistic}} explanation (interpretation) of the non-zero spacetime energy of the physical vacuum in a finite volume one may use the {\it{non-material}}, "curve" spacetime field or the field of {\it{virtual (unobservable)}} particles assuming that such fields do exist in the volume without  presence of the material particles inside or outside this volume. Another example of the {\it{non-materialistic}} explanation (interpretation) of the non-zero spacetime energy of the physical vacuum is the so-called String Theory, where the {\it{non-material spacetime coordinates}} (covariant or non-covariant ones) mediate {\it{material}} properties of the {\it{non-material}} spacetime and give to the spacetime the form of a non-virtual or virtual material particle (sring). In the same manner, the {\it{non-material wave-function}} of canonical quantum mechanics and the {\it{non-material wave of operators}} of traditional quantum field theories, which are connected with a point-like material particle (a non-virtual or virtual one), give the {\it{material properties}} to this particle. Remember, the  non-material wave-function and the non-material wave of operators are conventionally considered to be inseparable from the material point-like particle.    

The general properties of the forces induced by the cross-correlation energy of the two interfering particles have been {\it{illustrated}} numerically by using the quasi-solutions of the 1-D Euler-Lagrange equation of motion having the form of the Klein-Gordon-Fock equation for the scalar-boson unit-field. One can easily follow the model for an arbitrary number of the vector-boson and fermion unit-fields having the arbitrary values of the rest masses ($m_{n}\equiv m_{0n}$), charges ($q_{n}\equiv q_{0n}$) and spins ($s_{n}\equiv s_{0n}$) of the unit-fields (particles). The fields of different kinds, generally speaking, may cross-correlate or not cross-correlate with each other. For instance, the electron and photon fields do cross-correlate (interact) with each other. The gravitational (mass-induced) field also does cross-correlate (interact) with the charge-induced EM field of photons. In other words, the gravitational field deflects photons. The cross-correlation of the different fields will be considered in details in Part II of the present study. It should be also mentioned that a tiny attractive force between closely placed metal plates in the Casimir effect, which {\it{according}} to Hendrik B. G. Casimir and Dirk Polder is due to the van der Waals force between polarizable molecules of the metallic plates~\cite{Casi}, is {\it{attributed}} in the present model to the cross-correlation energy of the plates. The cross-correlation energy could be {\it{reinterpreted}} as the zero-point energy (vacuum energy) associated with the virtual exchange of particles of the quantum fields associated with the molecules in the traditional quantum physics. The gradient of the cross-correlation energy falls off rapidly with the distance [see, Figs. (5)-(8)]. Therefore the force has a measurable value only when the distance between plates is extremely small.  

\section{4. The non-zero cross-correlation energy mediated by interference}

Physical phenomena involving the interference (cross-correlation) of classical or quantum finite-fields would be affected by the interference between particles if the cross-correlation energy of the fields does not equal to zero. Before considering the conditions of the non-zero cross-correlation energy let me briefly show how the cross-correlation model presented in Secs. (2) and (3) addresses the principal questions of Sec. (1), which have not been explained by the quantum mechanics and particle field theory based on the Copenhagen-Dirac interpretation of quantum interference.  

\subsection{4.1. The present model versus the canonical interpretation of quantum interference}

In the canonical quantum mechanics and quantum field theory based on the Copenhagen-Dirac postulate of "interference-less", self-interfering particles, the wave-functions ${\psi _n(\mathbf{r},t)}$ of particles are not additive [$\psi (\mathbf{r},t) {\neq}\sum_{n=1}^N{\psi _n(\mathbf{r},t)}$] due to the non-additivity of the probability amplitudes (the probabilities of quantum alternatives are not additive). In other words, the particles do not interfere with each other. The interference (cross-correlation) between particles or antiparticles in these theories never occurs in agreement with the Copenhagen-Dirac postulate because of the gauge symmetry of the mathematically constructed (engineered) Hamiltonians under the $U(1)$ local  gauge transformation ($\psi _{0n} \rightarrow \psi ' _{0n}=e^{i\alpha _n} \psi _{0n}$). The absence of both the interference between particles and the respective cross-correlation energy is provided by the $U(1)$ gauge symmetry of the Hamiltonians constructed to be independent from the phases $\alpha _n$. In the present model, both the non-quantum and quantum (operator) fields are modelled as superpositions of the interfering unified-fields. The Hamiltonians of such non-quantum and quantum fields are not invariant under the $U(1)$ local gauge transformation given rise to the interference between the non-quantum or quantum fields and the dependence of the Hamiltonians (energies) on the field phases. Naturally, the dependence of energy on the unit-field phases could determine the uncertainty of energy, for instance, in the tunnelling process of any kind involving the phase uncertainty ("hidden" parameter). The {\it{general}} correspondence principle says that the quantum and classical treatments must be in agreement not only for a very large number of particles. In the present model, the non-quantum and quantum (operator) treatments are in agreement for an arbitrary number of particles (unit-fields). The Copenhagen interpretation (philosophy) of the de Broglie wave associated with a particle as the wave of probability presents a more or less intuitively transparent background for the physical interpretation of quantum mechanics. However, the Copenhagen interpretation of the de Broglie wave using the pure mathematical object (probability) did not solve really the problem of physical interpretation of quantum mechanics. The probability or the wave of probability is not a real physical matter. In particle field theory, up to now, the physical interpretation of the wave (field) of operators does not exist. An operator is considered rather as a pure mathematical object than a real material substance. Note, in this connection, that the problem of physical (materialistic) interpretation associated with the concept of probability does not exist in the Newton and Einstein mechanics. For an example, the real propagation of material particles in a gas is associated with the real mass, speed and kinetic energy. Each particle of the gas has the real (concrete, particular) speed and kinetic energy that can be equal or not to the pure mathematical counterparts, the average or expected values. In other words, the every particle of the gas propagates rather with a concrete (specific) speed than with the average or expected velocity. The average (expected) speed and the respective kinetic energy are the pure mathematical objects calculated by using the probability (pure mathematical object) associated with other pure mathematical objects, the Bose-Einstein and Fermi-Dirac statistics. In the present model, the unit-wave associated with a boson or fermion particle, unlike the non-material wave of probability or operators connected with a material point-like particle in quantum mechanics or particle field theory, is a real, finite unit-wave (unit-field) of the matter (mass-energy) with the Bose-Einstein or Fermi-Dirac cross-correlation properties. The Copenhagen-Dirac interpretation of quantum interference does strictly exclude existence of the particle interference and cross-correlation energy. The interference between bodies (particles) and the respective cross-correlation energy do not exist also in the Einstein general relativity. The classical wave physics and the present model do use the interference and cross-correlation energy in description of the pure additive or subtractive interference. For both the classical and quantum (operator) fields, in the present model, the two-times increase of the wave amplitude does increase the wave energy in four times, and the wave with zero amplitude has zero energy. The quantum field energy calculated by using the present model is equal to the classical value. In order to account for the well-known discrepancies between measurements based on the mass of the visible matter in astronomy and cosmology and definitions of the mass made through dynamical or general relativistic means, the present model does not need in a hypothesis of the existence of "dark" energy-mass. In the present model, the "dark" cosmological energy-mass as well as the well-known spiral cosmological structures are  attributed to the cross-correlation energy-mass of the moving and interfering cosmological objects (particles). In particle field theory, the particle energy ${\varepsilon}_0=\omega _0+ (1/2)\omega _0$ is different from the Planck-Einstein particle energy  ${\varepsilon}_0 =\omega _0=(  {\mathbf  k}_0^2   + m_0^2)^{1/2}$~\cite{Planck,Einst1}. The quantum energy ${\varepsilon}_0=(1/2)\omega _0$, which is identified as the vacuum energy associated with the particle, does contradict both the Einstein theory and classical physics of the empty space. In the present model, both the non-quantum and quantum energies of a particle are equal to the Planck-Einstein particle energy ${\varepsilon}_0 =\omega _0=(  {\mathbf  k}_0^2   + m_0^2)^{1/2}$.  The positive energy-mass ${\varepsilon}_0=({\mathbf{k}}_0^2+m_0^2)^{1/2}$ of antiparticles in the present model is different from the negative energy-mass of the Dirac antiparticles satisfying the energy-mass relation ${\varepsilon}_0=-({\mathbf{k}}_0^2+m_0^2)^{1/2}$. In the present model, the non-quantum and quantum energies of the empty spacetime (true vacuum) are identified as the Newton-Planck-Einstein vacuum energy ${\varepsilon}_{vac}=0$ of the empty straight-spacetime (absolute vacuum). In quantum theories of particle physics, the forces acting upon a particle are seen as the action of the respective gauge-boson (material) field that is present at the particle location. In perturbative particle field theory, the forces are attributed to the exchange of the field virtual particles (material gauge-bosons). Such a model is inconsistent with the Einstein theory of general relativity, where the gravitational interaction between two massive objects (particles) is not viewed as a force, but rather, objects moving freely in non-material gravitational fields travel under their own inertia in straight lines through non-material "curve" spacetime~\cite{Einst2}. From the point of view of the present model of the particle interactions, which is based on the gradient ("spatial curvature") of the cross-correlation energy of the interfering particles (material unit-fields), the Einstein and quantum descriptions (interpretations) of the forces are formally equivalent. One could mention also the problem associated with the nonconservation of the number of particles and energy in perturbative particle field theory based on the Copenhagen-Dirac postulate of "interference-free", self-interfering particles. The renormalization procedures do solve the problem, but not for a general case~\cite{Beth,Tomo,Schw,Feyn}. In the present model, such a kind of the problems can be solved in a general form without the use of the particular renormalization procedures of the traditional quantum field theories (see, the calculations of energies and forces in Sec. (3)  and Part II). To this end, the superluminal signaling in quantum mechanics (the Einstein-Podolsky-Rosen paradox), which is associated with the J.S. Bell inequalities~\cite{Bell}, does contradict the Einstein special relativity. The discussion of the problem associated with the superluminal signalling in quantum mechanics and particle field theory will be presented in Sec. (5).

\subsection{4.2. Conditions of the non-zero cross-correlation energy}

Any physical phenomenon involving interference (cross-correlation) of the classical or quantum fields would be affected by the interference between particles if the cross-correlation energy of the interfering particles (unit-fields) does not equal to zero. In the present model, there is an uncertainty of the energy and moment of the interfering unit-waves (particles) due to the phase-dependent energy (momentum) of the superposition of unit-fields. This uncertainty, however, should not be confused with the {\it{mathematical uncertainty}} of the energy and momentum attributing to the {\it{uncertainty principle}}~\cite{Heis1,Heis2} for the single particle. In quantum mechanics, lack of commutation of the time derivative operator with the time operator itself {\it{formally}} (mathematically) results into an uncertainty principle for time and energy: the longer the period of time, the more precisely energy can be defined. In addition, the non-commutation of the momentum operator with the coordinate operator mathematically results into the Heisenberg position-momentum uncertainty 
\begin{eqnarray} \label{eq124}
\Delta x \Delta k \geq  1/2,
\end{eqnarray} 
which means it is impossible mathematically to determine simultaneously both the position and momentum of a particle with any great degree of accuracy. In quantum mechanics, the physical mechanism behind the position-momentum uncertainty is attributed to the "compression" of the de Broglie wave $\psi_0 (\mathbf{r},t)$ of a particle. According to the Copenhagen (canonical) interpretation of quantum mechanics, the probability amplitude attributed to the non-material wave-function $\psi_0 (\mathbf{r},t)$ of a material particle determines the probability to find the single particle at the time $t$ and position $\mathbf{r}$. The value $\psi _0^*(\mathbf{r},t)\psi_0 (\mathbf{r},t)$ is interpreted as the probability density. To obtain an accurate reading of the position of a particle, the mathematical non-material objects (wave-function and probability) associated with the particle must be "compressed" as much as possible by the material boundaries of other, usually macroscopic, material objects. That means that the non-material wave of probability must be made up of increasing numbers of mathematical sine waves added together in the Fourier decomposition of the de Broglie wave. The momentum of the particle is proportional to the wavelength of one of the non-material sine waves, but it could be any of them. Thus a more precise position measurement by adding together more non-material sine waves means the momentum measurement becomes less precise (and vice versa). The considered mathematical interpretation sometimes is mentioned incorrectly as the physical one. Indeed, the non-material wave-function of a material particle could not be {\it{physically}} compressed by the material boundaries. Moreover, the particle (wave-function) could not be {\it{physically}} compressed by the particles (wave-functions) of material boundaries, because a particle interferes (interacts) in canonical quantum mechanics only with itself. 

In the present model, there are uncertainties of the position-momentum of both the single unit-field (particle) and the superposition of $N$ unit-fields (particles). The interpretation of the position-momentum uncertainty of the single unit-field (particle) is different from the above-presented mathematical interpretation of quantum mechanics.  For the sake of simplicity, let me consider a scalar boson. The energy of a scalar boson ({\it{material}} unit-field) depends on the unit-field ${\psi}_0$ and the spatial "curvature" $\nabla {\psi}_0$ of the unit-field in the Hamiltonian density (2). The unit-field function is determined mathematically by the Euler-Lagrange equation of motion with the initial and boundary conditions imposed. The energy of the free unit-field (\ref{eq16}) is equal to the Planck-Einstein particle energy ${\varepsilon}=({\mathbf{k}}^2+m^2)^{1/2}$ [see, (\ref{eq15})]. The momentum ${\mathbf{k}}$ of the free particle (material unit-field) located infinitely far from the material boundaries can have an arbitrary value. The value depends solely on the unit-field gradient ("curvature") $\nabla {\psi}_0$. The localization of the material unit-field ${\psi}_0(x)$ by the material boundaries in the region $\Delta x= x_2-x_1$ means physically and mathematically that the real parts of the fields must vanish on the boundaries, $Re[{\psi}(x_1)]=Re[{\psi}(x_2)]=0$. The physical interaction of the unit-field with the boundaries results into selection of the unit-field with the spatial "curvature" determining by the condition $\nabla {\psi_0 (x)}^*\nabla {\psi_0 (x)}=k^2\nabla {\psi_0 (x)}^*\nabla {\psi_0 (x)}$, where $k=n\pi /\Delta x$ (n=0,1, 2, ...) and $\Delta k = \Delta k_{min} = \pi /\Delta x$. If the value $n$ of the unit-field is unknown, then these relations can be presented as the position-momentum uncertainty  
\begin{eqnarray} \label{eq125} 
\Delta x \Delta k\geq \pi, 
\end{eqnarray} 
which is similar to the Heisenberg uncertainty (\ref{eq124}). The relation (\ref{eq125}) can be rewritten for the energy by using the Planck-Einstein particle energy as
\begin{eqnarray} \label{eq126}
\Delta x \Delta [ {\varepsilon}^2 - m^2 ]^{1/2} \geq {\pi}. 
\end{eqnarray}      
The physical mechanism behind the uncertainties (\ref{eq125}) and (\ref{eq126}) could be attributed to the increase of the spatial "curvature" (gradient) $\nabla {\psi_0 (x)}$ of the real material unit-field by interference (interaction) of the  material unit-field ${\psi_0 (x)}$ with the particles of material boundaries of macroscopic objects. That is to say that the uncertainties are due to the physical interaction  (interference) of the particle (material unit-field) with the particles (material unit-fields) of the external material boundaries. Such a {\it{physical}} interpretation of the uncertainties, the increase of the "curvature" (gradient) of the material unit-field by the particles of the external boundaries under the spatial localization of the unit-field (particle), is different from the {\it{mathematical}} interpretation based on the "physical compression" of the pure mathematical objects  (wave-function and probability) connected with with a material particle. Remember, a particle {\it{physically}} interferes (interacts) in canonical quantum mechanics only with itself. It should be stressed that Eqs. (\ref{eq125}) and (\ref{eq126}) should be used for the particle (unit-field) and not for the empty "straight" space (Newton-Einstein vacuum) with $k={\varepsilon}=m=0$. Equations (\ref{eq125}) and (\ref{eq126}) could be generalized for the $N$ unit-waves (particles) localized in the region $\Delta x$ by using Eqs. (\ref{eq20}) and (\ref{eq34}). If the unit-waves have the same phases, then Eqs. (\ref{eq20}), (\ref{eq34}), (\ref{eq125}), and (\ref{eq126}) yielded
\begin{eqnarray} \label{eq127}
\Delta x \Delta k \geq \pi N^2 
\end{eqnarray}
and 
\begin{eqnarray} \label{eq128}
\Delta x \Delta [ {\varepsilon}^2 - m^2 ]^{1/2} \geq {\pi} N^2. 
\end{eqnarray}
If the phases are unknown, then Eqs. (\ref{eq127}) and (\ref{eq128}) could be rewritten as 
\begin{eqnarray} \label{eq129}
\Delta x \Delta k  \geq 0 
\end{eqnarray}
and 
\begin{eqnarray} \label{eq130}
\Delta x \Delta [ {\varepsilon}^2 - m^2 ]^{1/2}  \geq 0. 
\end{eqnarray}
In such a case, the boundaries do determine the "curvatures" $\nabla {\psi}_{0n}$ of the unit-fields ${\psi}_{0n}$, but not select the energy and momentum of the unit-field superposition due to the phase uncertainty. The energy and momentum are determined rather by the cross-correlation of the unit-fields than by the interaction of the fields with the boundaries. In order to derive the energy-time uncertainty, one should consider the unit-field ${\psi_0 (x,t)}={\psi_0 (x)}{\psi_0 (t)}$ of the matter and the localization of the field ${\psi_0 (t)}$ in the time interval $\Delta t= t_2-t_1$. The energy ${\varepsilon}$ of the field ${\psi_0 (x,t)}$ depends on the field temporal "curvature" ${\partial {\psi _0}}/{\partial t}$ [see, Eq. (\ref{eq2})]. The localization of the unit-field ${\psi _0}(t)$ in the time interval $\Delta t= t_2-t_1$ means mathematically that the real parts of the fields must vanish on the time boundaries, $Re[{\psi _0}(t_1)]=Re[{\psi _0}(t_2)]=0$. The time-boundary condition enforces the unit-field to have the temporal "curvature" determining by the condition $[{\partial {\psi _0}^*}/{\partial t}][{\partial {\psi _0}}/{\partial t}]={\varepsilon}^2[{\partial {\psi _0}^*}/{\partial t}][{\partial {\psi _0}}/{\partial t}]$, where ${\varepsilon}=n\pi /\Delta t$ (n=1, 2, ...) and $\Delta {\varepsilon} = \Delta {\varepsilon}_{min} = \pi /\Delta t$. If the value $n$ is unknown, then these relations can be presented as the energy-time uncertainty  
\begin{eqnarray} \label{eq131}
{\Delta \varepsilon} \Delta t \geq \pi. 
\end{eqnarray} 
The energy-time uncertainty (\ref{eq131}) is caused by the increase of the temporal "curvature" ${\partial {\psi_0 (t)}}/{\partial t}$ of the real unit-field ${\psi_0 (t)}$ of the matter (mass-energy) with decreasing the localization interval $\Delta t$. Such a physical interpretation of the time-energy uncertainty is different from the traditional interpretations of quantum mechanics. Note that the above-presented analysis is applicable for the particles of another kind. For instance, one can easily rewrite Eqs. (\ref{eq125})-(\ref{eq131}) for other boson or fermion unit-fields (particles). In the context of the above-presented interpretations of the uncertainties, one could mention again that the Copenhagen interpretation of the de Broglie wave associated with a particle as the wave of probability presents a more or less intuitively transparent background for the physical interpretation of the quantum mechanics. The Copenhagen interpretation of the de Broglie wave using the pure mathematical object (probability), however, does not solve really the problem of physical interpretation of the Heisenberg position-momentum and time-energy uncertainties. The wave of probability is not a real physical matter (mass-energy), which can be "compressed" by the physical (material)  boundaries in some incorrect interpretations of the canonical quantum mechanics. The particle field theory, up to now, does not give any physical interpretation of the wave (field) of operators. Indeed, an operator should be considered rather as a mathematical object than a real physical matter. Naturally, the spatial and/or temporal "compression" of the mathematical object (probability or operator) would not have any physical meaning. In the present model, the unit-wave associated with a boson or fermion particle, unlike the wave of probability or operators in quantum mechanics or particle field theory, is a real finite unit-wave (unit-field) of the matter (mass-energy), whose "curvature" (gradient) can be changed spatially and/or temporally. The physical mechanism behind the position-momentum and time-energy uncertainties is attributed to the increase of the spatial ($\nabla {\psi _0}$) and/or temporal (${\partial {\psi_0 }}/{\partial t}$) "curvatures" (gradients) of a real unit-field of the matter under the spatial or temporal localization of the unit-field by interaction with other microscopical or macroscopic objects. It can be mentioned that the position-time (spacetime) uncertainty is discussed in Ref.~\cite{Kukh1}(e).

It should be stressed again that any physical phenomenon involving the interference (cross-correlation) of the boson and/or fermion fields (particles) would be affected by the field interference if the respective cross-correlation energy is not equal to zero. The static unit-fields (${\partial}{\psi _{0n}(\mathbf{r},t)}/{\partial {t}}=0$) have been considered in Sec. (3). It has been shown as an example that the cross-correlation energy of the static boson scalar fields does not equal to zero at the distance $R\leq R_c$, where $R_c\approx 10^2$ and $R_c\approx 5$ for the weak and strong unit-fields, respectively. The cross-correlation energies of the transient boson fields (waves), such as the non-quantum finite waves composed from the non-quantum boson unit-waves, are described by the cross-correlation terms in Eqs. (\ref{eq12}) and (\ref{eq19}). The second quantization of the non-quantum finite-waves by replacing the waves (fields) by the field operators has yielded the quantum cross-correlation energy (\ref{eq64}). The non-quantum and quantum cross-correlation energies in Eqs. (\ref{eq12}), (\ref{eq19}), and (\ref{eq64}) are equal to each other. The time-dependent cross-correlation energy of the wave-like fields composed from the boson unit-waves and/or the fermion unit-waves [see, Sec. (2)] is proportional to the terms 
\begin{eqnarray} \label{eq132}
\frac {1}{V_{nm}} {\int_{V_{nm}}}[e^{{-i( \Delta \mathbf{k}_{0nm} {\mathbf{r}}-\Delta \varepsilon_{0nm} {t}-\Delta \alpha _{nm}})}+e^{{i( \Delta \mathbf{k}_{0nm} {\mathbf{r}}-\Delta \varepsilon_{0nm} {t}-\Delta \alpha _{nm}})}]d^3x,
\end{eqnarray}
which have the non-negligible values in the case of  
\begin{eqnarray} \label{eq133}
\mathbf {k}_{0n} \rightarrow  \mathbf {k}_{0m}
\end{eqnarray}
(remember, $\varepsilon_{0n} \rightarrow  \varepsilon_{0m}$ if $\mathbf {k}_{0n} \rightarrow  \mathbf {k}_{0m}$ and $m_{0n}=m_{0m}$) and/or in the case of
\begin{eqnarray} \label{eq134}
V_{nm}(k_{xn}-k_{xm})(k_{yn}-k_{ym})(k_{zn}-k_{zm}) \rightarrow 0,
\end{eqnarray}
where $\mathbf{k}_n \equiv \mathbf{k}_{0n}$, $\mathbf{k}_m \equiv \mathbf{k}_{0m}$ and $\mathbf{k}_n \neq \mathbf{k}_m$. Notice, in the case of $\mathbf{k}_n \neq \mathbf{k}_m$ and $V_{nm} >(k_{xn}-k_{xm})(k_{yn}-k_{ym})(k_{zn}-k_{zm})$, the terms (\ref{eq132}) are equal to zero if the unit-waves are confined by the resonator-like boundaries. The experimental realization of the condition (\ref{eq133}) is extremely difficult (see, Ref. \cite{Kukh1}(a-c)). {\it{The creation of the unit-fields with the momentums $\mathbf {k}_n \rightarrow  \mathbf {k}_m$ and $\mathbf {k}_n =  \mathbf {k}_m$will be considered in details in the following section devoted to the Bose-Einstein condensation, super-radiation, Bosenova effect, superfluidity, superconductivity, supermagnetism, and quantum anomalous and fractional Hall effects.}} In the case of the ordinary physical systems, which are not associated with physical super-properties, the cross-correlation energy is equal to zero due to the conditions $\mathbf {k}_n \neq \mathbf {k}_m$ and/or $V_{nm} >> 1 / (k_{xn}-k_{xm})(k_{yn}-k_{ym})(k_{zn}-k_{zm})$. While, the sub-wavelength [$V_{nm} << 1 / (k_{xn}-k_{xm})(k_{yn}-k_{ym})(k_{zn}-k_{zm})$] systems obey the non-zero cross-correlation energy. The case (\ref{eq134}) corresponds to the {\it{pure constructive or destructive interference in the sub-wavelength volume}}. For instance, the physical super-properties of the sub-wavelength systems could be attributed to the cross-correlation energy in the {\it{near-field diffraction (scattering)}} of waves, as well as in the {\it{tunnelling effects and virtual-particle phenomena}} of any kind (for instance, in quantum-mechanics tunnelling, wave-mechanical tunnelling, evanescent wave coupling, forces or effects by virtual particles, and Casimir effect). Note, in this regard, that the classical-mechanics particles that do not have enough energy to classically surmount a barrier will not be able to reach the other side. In canonical quantum mechanics, the particles with an arbitrary small energy always have the  non-zero probability to overcame the energy barrier of any value. Although the probability of tunnelling decreases for taller and wider barriers, the particles would surmount the all potential barriers after a some time leading to decay of the all-known composite particles (nucleus, atoms and molecules). {\it{In contrast to the tunnelling and virtual processes of the canonical quantum mechanics and particle field theory, which may involve the infinite energy connected with the Heisenberg uncertainty principle, the realization (reality) of such physical phenomena in the present model is strictly limited by the finite value of the cross-correlation energy}}. If the cross-correlation energy is smaller than the potential barrier, then the unit-fields will never surmount the energy barrier. Although the unit-field may occupy a small volume outside the barrier, the main part of the unit-field should be always inside the barrier. Naturally, the evanescent part of the unit-field may overlap (interact) with the unit-fields that occupy the external space of the potential barrier. It could be also noted again that the spatially infinite fields are overlapped in the infinite volume. Unlike the fields occupying the infinite volume, the interference of the finite fields requires the creation of conditions of the field overlapping. In order to interfere, the finite fields or their parts should occupy (share) the same volume $V_{nm}$ at the same time moment  $ t_{nm}=t_n= t_m$. The cross-correlation energy decreases if the volume ${V_{nm}}$ decreases, for instance, due to the movement of unit-fields (particles) away from each other. One should not confuse vanishing the cross-correlation energy by the non-overlapping of the finite fields with the case of the overlapped "orthogonal fields" [see, Sec. (2)]. For instance, the cross-correlation energies of the two unit-fields (particles) before and after a collision (interference) are equal to zero due to the non-overlapping of the unit-fields. During the elastic collision (interaction) of the unit-fields, the vanishing of the cross-correlation energy is caused by the orthogonality of the unit-fields ($\mathbf {k}_{0n} \neq \mathbf {k}_{0m}$) in the Hilbert space. Such a process could be considered as an elastic collision of the particles. In the case of the non-orthogonal unit-fields [see, Secs. (2) and (3)], the cross-correlation (interference) of the unit-waves can lead to the non-elastic collision. For instance, the Raman or another nonlinear scattering of a photon by an atom (composite boson-like particle) should be considered as the non-elastic collision. So far the conditions of the non-zero cross-correlation energy have been analyzed for the unit-waves that have the permanent (constant) phases. In the general case, the phases may have the time-dependent values $\alpha _n=\alpha _n(t)$ and $\alpha _m=\alpha _m(t)$. The cross-correlation energy averaged over the time interval $\Delta t$ could be considered as the time-independent energy $\langle {\cal H} \rangle_{\Delta t}$ of the interfering unit-waves: 
\begin{eqnarray} \label{eq135}
\langle{\cal H}\rangle_{\Delta t}=\frac {1}{\Delta t}{\int_{{\Delta t}}}{\int_{0}^{\infty}}\langle h \rangle d^3xdt=\sum_{n=1}^N{\cal H}_{nn}+\sum_{n\neq m}^{N^2-N}\langle{\cal H}_{nm}\rangle_{\Delta t},
\end{eqnarray}
where the self-correlation energy $\langle{\cal H}_{nn}\rangle_{\Delta t}={\cal H}_{nn}$ is a time-independent value, and the time-averaged ("time-independent") cross-correlation energy is given by
\begin{eqnarray} \label{eq136}
\sum_{n\neq m}^{N^2-N}\langle{\cal H}_{nm} \rangle_{\Delta t}=\sum_{n\neq m}^{N^2-N}\frac {1}{\Delta t}{\int_{{\Delta t}}}{\int_{0}^{\infty}}\langle h_{nm} \rangle d^3xdt.
\end{eqnarray}
In the case of the incoherent unit-fields having absolutely random phases, both the term (\ref{eq132}) averaged over any time $\Delta t$ and the respective energy (\ref{eq136}) are equal to zero due to the non-correlation of such fields. Naturally, the increase of the degree of coherence (phase correlation) increases the time-averaged cross-correlation energy of the partially coherent unit-fields. Notice, the cross-correlation energy (\ref{eq136}) averaged over the time $\Delta t >> 1/\Delta \varepsilon_{0nm}$ does vanish if the {\it{coherent}} unit-fields (particles) with the constant phases $\alpha _n$ and $\alpha _m$ are {\it{distinguishable}} ($\mathbf {k}_{0n} \neq \mathbf {k}_{0m}$ and/or $m_{0n} \neq m_{0m}$, respectively $\varepsilon_{0n} \neq \varepsilon_{0m}$). {\it{The property explains physically which particles are identical or non-identical. The coherency of the identical ($\mathbf {k}_{0n} = \mathbf {k}_{0m}$ and $m_{0n} = m_{0m}$, respectively $\varepsilon_{0n} = \varepsilon_{0m}$) unit-fields (particles) should be considered as the necessary condition of existence of the cross-correlation energy}}. One should not confuse here the coherency and cross-correlation of the two fields in the same [$(P,t)=(P_1,t_1)=(P_2,t_2)$] spacetime points, which is related to the cross-correlation energy, with the coherency associated with the cross-correlation between the fields in the different [$(P_1,t_1)\neq (P_2,t_2)$] spacetime points of the amplitude or intensity interferometry.   

\subsection{4.3. Quantum coherence and correlation}

The basic properties of both the classical and quantum coherence are usually clarified in the context of the Young double-slit experiment [Sec. (1)]. Let me briefly describe the coherence properties in the frame of quantum mechanics and compare them with the present model. In quantum mechanics, the dynamics of quantum particles in the double-slit experiment describes the relationship between the classical waves and quantum particles. The quantum particle is described by a wave-function associated with the wave of probability, which contains all the information about the state of the quantum system. According to the Copenhagen-Dirac interpretation of {\it{quantum interference}} each particle interferes only with itself. In the Young experiment, each particle can go through either slit and hence has two paths that it can take to a particular final position ($P,t$). In quantum mechanics, these two paths interfere. If there is pure destructive interference, the particle never arrives at that particular position.  In quantum mechanics the ability of a particle to interfere with itself is called quantum coherence. The quantum description of perfectly coherent paths is called a pure state, in which the two paths (wave-functions) are combined in a superposition. The correlation between the two wave-functions exceeds what would be predicted for classical correlation between the two wave-functions (see, the Bell inequalities~\cite{Bell}). It should be noted in this regard that the generally accepted wave-function of a particle (for instance, a photon) associated with the wave of probability does not exist in particle field theory. Therefore, strictly speaking, the particle field theory could not consider the quantum (probabilistic) interference phenomena in the Young double-slit experiment. In addition, in contrast with the Copenhagen-Dirac postulate of the "interference-less" particles, the particle of quantum field theories interferes (interacts) with the particles of the slits due to the boundary conditions. If the quantum mechanical system of the above-mentioned two wave-functions (or any system of wave-functions of entangled particles in quantum mechanics) is decohered, which would occur, for instance, in a measurement via eigenselection, then there is no longer any phase relationship (correlation) between the two states. The quantum description of imperfectly coherent (partially coherent) paths is called a mixed state, described by a density matrix and is analogous to a classical system of mixed probabilities, where the correlations are classical. One should not confuse here the self-interference of the wave of operators associated with a particle in particle field theory with the self-interference of the wave of probability associated with the above-mentioned wave-functions of a particle in the Copenhagen interpretation of quantum mechanics. The wave of operators associated with a particle is not attributed to the probability amplitude of the wave of probability associated with the wave-function of the particle. Unlike in quantum mechanics, in particle field theory the generally accepted wave-function of a particle (antiparticle) does not exist up to now. Therefore the interference mechanism behind the self-interference properties (Dirac's one-particle self-interference) of a particle in particle field theory should be attributed rather to the mathematical properties of the wave of operators than the wave of probability of quantum mechanics. It should be stressed again that the particle should interfere only with itself in the Young experiment according to the Copenhagen-Dirac interpretation of {\it{quantum interference}}. However, that {\it{contradict}} the fact that the {\it{particle interferes (interacts) also with the particles of the slits due to the boundary conditions imposed by the slits given rise to the "hidden" interference between particles}}.

In the present model, the coherence properties of the unit-waves (particles) in the Young experiment are similar to the aforementioned quantum mechanical properties. The difference between the two models is in the interpretation of the interference phenomenon and in the absence of the interference between particles and the cross-correlation energy in the quantum mechanical model based on the canonical interpretation of quantum interference. The present model considers a quantum particle as an indivisible unit-field (unit-wave) of the matter (mass-energy). The dynamics of the unit-wave is described by the Euler-Lagrange equation of motion with the initial and boundary conditions imposed. The energy of the field, which depends on the spatial ($\nabla {\psi}_0$) and temporal (${\partial {\psi}}_0/{\partial t}$) "curvatures" of the field, is given by the Hamiltonian of the system. The spatial and temporal "curvatures" of the unit-wave can be changed by the interference (interaction) of the unit-field with the particles (unit-fields) of boundaries of other microscopic or macroscopic objects, but the unit-field cannot be divided into the different independent parts. In the Young experiment, the indivisibility of the unit-field (unit-wave) means the unit-field passing though the double slit is not divided by the slits into two different independent fields (waves). The spatial and temporal "curvatures" and the shape of the unit-wave are changed by interaction (interference) of the unit-field with the unit-fields of the boundaries (slits), but the unit-field is not divided into the independent parts (unit-waves). The collision (interaction) of the unit-field (for instance, photon) with the slits is elastic, linear process. Therefore the energy of the unit-field is conserved. Notice, the particle (photon) and the respective energy can be divided into the independent parts in the case of the nonlinear processes (inelastic collisions), only. In the Young experiment, the interaction of the unit-field with the slits does result into reshaping of the unit-field in the transverse and longitudinal directions. The interference pattern on the observation screen is produced by the interference of the two parts of the same unit-field (photon), which has been reshaped by the slits before the interference on the observation screen. That is to say that the interference pattern is formed on the slits by the change of the shape and spatial "curvature" of the unit-wave by the interaction of the unit-field with the slits. This mechanism can be represented as interference of the unit-field with itself on the observation screen. The interaction of the unit-field with any additional object after the interaction with the slits would change the spatial "curvature", shape and phase of the unit-wave and destroy the interference pattern. In such a case, the unit-field is decohered by the additional interaction process. The described mechanism of the self-interference is similar to that in the classical description of the Young experiment. The presented mechanism, however, does not require the formation of two independent waves. The two parts of the reshaped unit-wave (photon) could interfere on the observation screen like the two independent coherent waves in the classical model of the Young experiment. One should not confuse here the one-particle self-interference of the wave of probability, which is associated with the single particle (photon) in the canonical interpretation of quantum mechanics, with the transverse and longitudinal reshaping of the real indivisible unit-wave (photon) by the slits. In the present model, the {\it{reshaped}} unit-wave (particle) can interfere with itself, as well with other coherent unit-waves (photons). That provides a simple, non-probabilistic explanation of the famous paradox of canonical quantum mechanics associated with the non-classical properties of the quantum macroscopic system, namely the Schr{\"o}dinger's cat thought experiment. It should be also noted that a wave-function of the entangled (strongly correlated) particles is considered in the present model rather as an indivisible unit-field than a wave-function of a few-body quantum system of independent unit-waves (particles). In other words, the entangled particles are considered as the parts of the same indivisible unit-wave reshaped in the transverse and/or longitudinal direction into the entangled parts. Respectively, the parts of the same indivisible unit-wave reshaped in the transverse and/or longitudinal direction are considered as the entangled particles. As a consequence of the indivisibility of the reshaped unit-wave, which is composed from the entangled parts ("entangled particles"), the interference properties of the quantum wave of strongly correlated (entangled) particles are similar to that of an ordinary (non-composite) unit-wave. If the correlation (cross-correlation) between the different parts of the field reshaped in a liner or nonlinear process is very weak then the parts of the composite wave could be considered as independent unit-waves (particles). 

The unit-wave (particle) could interfere not only with itself, but also with another coherent unit-wave. The interference (cross-correlation) of the two unit-waves can result into creation of the positive or negative cross-correlation energy, which does increase or decrease the energy of the unit-field pair [see, Secs. (2)-(4)]. It is interesting that the cross-correlation of two unit-waves in the conventional Young setup does not affect the energy of the unit-field pair. Indeed, in the conventional Young double-slit experiment the two coherent unit-waves produced by the two slits separated by the distance ${\Lambda}\gg 2\pi /k_0 $ are orthogonal in the far-field diffraction zone because of the different momentums (${\mathbf k}_{01} /k_{01} \neq {\mathbf k}_{02}/k_{02}, k_{01}=k_{02}=k_0$). According to the expression (\ref{eq132}), the cross correlation term (energy) of the orthogonal unit-waves vanishes. Respectively, the field energy is given by ${\varepsilon} = 2k_0$. The situation is completely different in the case of Young's subwavelength system, where ${\Lambda}\ll 2\pi /k_0$ and correspondingly ${\mathbf k}_{01}/k_{01}={\mathbf k}_{02}/k_{02}$ and $k_{01}=k_{02}=k_0$. In the case of $\alpha _1-\alpha _2=0$, the interference creates the extra energy, ${\varepsilon} = 4k$. In Refs.~\cite{Kukh1}(a) and (c), this phenomenon has been described and interpreted as a {\it{classical analog}} of the Dicke {\it{quantum}} superradiance~\cite{Dick} of a subwavelength ensemble of excited atoms. The interference completely destroys the energy (${\varepsilon}$ = 0) at the phase condition $\alpha _1-\alpha _2=\pi$. A simple analysis of the Hamiltonians (\ref{eq12}), (\ref{eq19}) and (\ref{eq62}) shows that the addition of waves is not so efficient at larger spacing of the slits (${\Lambda}\sim 2\pi /k_0 $), but still leads to the interference-induced enhancements and resonances (versus wavelength) in the total energy emitted by the slits. In such a case the field energy [$0\leq {\varepsilon}_0 \leq 4k_0 $] depends on the values $k_{01}$, $k_{02}$, $\alpha _{1}$, $\alpha _{2}$ and ${\Lambda}$ (for more details, see the study~\cite{Kukh1}(a-c) and references therein). In the case of the Young's two-source system that consists of the two sub-wavelength (${\Lambda}\sim 2\pi /k_0 $) dipole radiators, the dependence is especially simple. One should simply consider the radiation of the two almost identical dipoles in the near-field zone or far-field zone. For instance the total time-averaged radiated power $W$ of the two classical dipoles separated by the distance ${\Lambda} \ll 2\pi /k_0 $ is given by $W=(4{\omega}^4/3  )|{\mathbf d}_{01}+{\mathbf d}_{02}|^2\approx 4W_{01}$, where $W_{01}\approx  (4{\omega _{0}}^4/3  )|{\mathbf d}_{01}|^2$ is the power of the first radiator with the dipole ${\mathbf d}_{01}\approx {\mathbf d}_{02}$. Here, the value $W_{12,21}=(4{\omega}^4/3  )|2{\mathbf d}_{01}{\mathbf d}_{02}|$ could be considered as the cross-correlation power (energy) of the two dipols. The total radiated power $W_Q$ of the two quantum dipoles separated by the distance ${\Lambda} \ll 2\pi /k_0 $ is also given by $W_Q \approx 4W_{Q01}$. The quantum expression for the power of the two-dipole radiation differs from the classical one only in that it contains the matrix element of the dipole momentum instead of the dipole momentum itself. The superradiance of the two classical or quantum dipoles is very similar to the Dicke  superradiance~\cite{Dick}. This simple analysis of two classical and quantum dipoles also could be considered as a proof of existence of a {\it{classical analogue}} [Refs.~\cite{Kukh1}(a) and (c)] of the Dicke {\it{quantum}} superradiance. Notice, in the case of the number $N>1$ of quantum dipoles, the Pauli exclusion principle for the electrons of the dipoles should be taken into account. The Pauli principle states that no two identical electrons may occupy the same quantum state simultaneously. It should be stressed again that the cross-correlation energy of the interfering classical waves or particles (unit-waves), which does not exist in quantum mechanics and particle field theory, plays a key role in description of the aforementioned interference phenomena. The cross correlation energy could be important also for understanding the enhancement or suppression of the energy and momentum [Sec. (2.1.2.)] in other coherent processes involving classical waves or particles. Here, one could mention the enhancement or suppression of the energy and momentum of a light pulse propagating in a dispersive medium. The phases of the different Fourier $\mathbf k$-components of the wavepacket propagating in the same direction ($\mathbf k_{0n}/k_{0n}=\mathbf k_0/k_0$) can be changed under the propagation. According to the present model the phase modification could result into the interference-induced enhancement or suppression of the pulse energy and momentum. The interference phenomenon could affect also the energy and momentum of a wave-packet scattering by the subwavelength scatters, such as nano-apertures, nano-particles and other nano-objects. For more examples and details, see the study \cite{Kukh1} and refrerences therein. 

Nowadays, the problem associated with the superluminal signalling in quantum mechanics (the Einstein-Podolsky-Rosen paradox ~\cite{Ein}), which follows from the J.S. Bell inequalities~\cite{Bell}, but contradicts the Einstein special relativity, is considered as a most serious incompatibility between the two fundamental theories. Therefore, it is important to consider this problem in more details. The superluminal cross-correlation (signalling) between the two entangled particles exceeds what would be predicted for classical (non-superluminal) correlation between the two wave-functions (see, the Bell inequalities~\cite{Bell}). In the present model, the entangled particles are considered as the parts of the same indivisible unit-wave of the physical matter, which has been reshaped in the transverse and/or longitudinal direction into the entangled parts [strongly correlated (entangled) particles] by interaction (interference) of the unit-field with the fields of other microscopic or macroscopic objects. As a consequence of the indivisibility of the reshaped unit-wave, which is composed from the entangled parts ("entangled particles"), the interference properties of the quantum wave of strongly correlated (entangled) particles are similar to that of an ordinary (non-composite) unit-wave. The shape, temporal and spatial "curvatures" of the ordinary or composite unit-field can be changed, but the field cannot be divided into different independent parts. In order to satisfy the experimentally observed superluminal signalling one have to assume that any local change of the shape and/or "curvature" of the ordinary or composite unit-field of the physical matter does propagate with superluminal (linear and/or rotational) velocity within the field. For instance, the superluminal rotational velocity within the spinor unit-field (electron) could be formally associated with the quantum spin of the particle. More precisely, in the present model, the quantum spin of the particle is attributed to the sign-coupling product $SG$ (for details, see Part II), which characterizes the sign and value of the the cross-correlation (interaction, coupling) energy under the interference of two unit-waves [also, see comments to Eqs. (\ref{eq120}) and (\ref{eq121}), and Figs. (5)-(8)]. The superluminal velocity (signalling) within the ordinary or composite unit-field of the matter is attributed in the present model rather to the physical properties of the field medium than the empty space (vacuum) of the Einstein relativity. Therefore the superluminal signalling within the ordinary or composite unit-field of the physical matter does not contradict the Einstein non-superluninal velocity of the signalling between the point particles separated by the "straight" or "curve" empty space (Einstein's vacuum). It is not easy to find a classical analogue of the superluminal signalling within the ordinary unit-field or the unit-field like material medium of the composite unit-field (entangled particles). The superluminal signalling within the medium of an ordinary unit-field could be considered (interpreted), for instance, as a non-quantum or quantum analog of the wave propagation in the material body, which has the "modulus of bulk elasticity" bigger than the "coefficient of stiffness" of any classical material medium. If the incompressible medium of the ordinary unit-field has an infinite elasticity, the signal could propagate within the ordinary unit-field with infinite speed. The composite unit-field [entangled (strongly correlated) particles] could be considered as a non-quantum or quantum analogue of Newton's cradle, a series of swinging spheres. In the Newton cradle, the identical bodies (particles) exchange velocities (momentums). If the first particle has nonzero initial velocity and the second particle is at rest, then after collision the first particle will be at rest and the second particle will travel with the initial velocity of the first particle. In the case of perfectly elastic collision of the totally rigid particles (the parts of the composite unit-field), the exchange of velocities (momentums) provides a so-called "quantum jump" taking zero time. Thus the signalling ("quantum leap") in the Newton cradle of unit-fields [strongly correlated (entangled) particles] is performed with infinite speed. Although the Newton cradle is one dimensional, the two or three dimensional cradle also could be considered as an analogue of the strongly correlated (entangled) quantum particles. Notice, according to the present model, the Bell superluminal signals~\cite{Bell} (the Einstein-Podolsky-Rosen paradox) and the well-known objects in astronomy, which propagate with velocities greater than the velocity of light, involve no physics incompatible with the theory of special relativity. In the present model, the {\it{superluminar velocities of signals, the superluminar "quantum leaps" and the superluminar collapses of wavefunctions}} of any kind are attributed rather to the physical properties of the unit-field like material mediums than the "straight" or "curve" empty space of the Einstein special or general relativity (for details, see Part II). 

In the context of the above-presented consideration of the quantum coherence and correlation, one should not confuse a real finite-wave or real finite unit-wave of the real physical matter (mass-energy) with the wave of probability in quantum mechanics or with the wave of operators in particle field theory. One should not confuse also the first-order coherency (cross-correlation) of the two fields in the same [$(P,t)=(P_1,t_1)=(P_2,t_2)$] spacetime points, which are related to the cross-correlation energy, with the second or higher order coherency (cross-correlation) between the fields in the same [$(P,t)=(P_1,t_1)=(P_2,t_2)$]  or different [$(P_1,t_1)\neq (P_2,t_2)$] spacetime points of the amplitude or intensity interferometry. The degree of coherency of the fields in the first or the second order, in the same or different spacetime points can be measured by using the conventional amplitude or intensity interferometry. It could be noted in this regard that any coherent or incoherent field is self-coherent. In the present model, the self-coherency is associated with the conventional energy of a field, which is called the self-correlation energy or self-energy. Two finite waves (beams) do interfere coherently in a space point if the phase difference between the waves in this point is constant in the time. The coherency of the two fields in the same point is attributed in the present model to the cross-correlation and cross-correlation energy in this point. If the phase difference is random or changing the fields are incoherent or partially coherent in the point. In the points where the fields are incoherent, the interference of the fields yields zero value of the cross-correlation energy. That is to say that the interference between two (or more) finite fields (waves) in a space point establishes a correlation (cross-correlation) between these waves in this point. The coherent superposition of the wave amplitudes is attributed to the ordinary (first-order) interference, which corresponds to the first-order cross-correlation and the respective (first-order) cross-correlation energy. In analogy to the first order coherence, the second order interference generalizes the interference between amplitudes to that between squares of amplitudes. The high-order cross-correlations have quite different properties in comparison to the first-order correlation. For instance, in the case of the complex, scalar, electric waves $E_1=|E_1|e^{i\alpha_1 }$ and $E_2=|E_2|e^{i\alpha_1 }$, the first-order interference (cross-correlation) attributed to the amplitude interferometry does associate with the {\it{phase-dependent intensity (energy)}} $I=(E_1+E_2)(E_1+E_2)^*=I_1+I_2+2\sqrt{I_1}\sqrt{I_2}cos(\alpha _1-\alpha _1)$, where $0\leq  I \leq I_{max}$, $I_1=E_1E^*_1$ and $I_2=E_2E^*_2$. The second-order cross-correlation (interference) associated with the intensity interferometry corresponds to the {\it{phase-independent intensity (energy) squared}} $I^2=(I_1+I_2)^2=I^2_1+I^2_2+2I_1I_2$, where $I^2_1+ I^2_2 \leq  I^2  \leq I^2_{max}$. Although the first-order cross-correlation term $2\sqrt{I_1}\sqrt{I_2}cos(\alpha _1-\alpha _1)$ associated with the {\it{first-order}} cross-correlation of the waves $E_1$ and $E_2$ depends on the phases $\alpha _1$ and $\alpha _2$, the second-order cross-correlation term $2I_1I_2$ attributed to the {\it{second-order}} cross-correlation of the wave intensities $I_1$ and $I_2$ does not depend on the wave phases. One should not confuse here definitions of the first-order and second-order cross-correlation {\it{terms}} with the {\it{degrees}} of first-order and second-order coherences of the amplitude and intensity interferometries. Notice, in the Newton mechanics, the first order cross-correlation does reflect formally a fact that the two-times increase of the particle velocity $v$ increases the particle energy ${\varepsilon}$ in four times, ${\varepsilon}=m(2v)^2/2$. The free particle with zero velocity ($v=0$) has zero energy (${\varepsilon}=0$). In canonical quantum mechanics and particle field theory, where a particle is associated with a wave of probability or a wave (field) of operators, the interference and correlations between two or more particles (waves) in the same [$(P_1,t_1)=(P_2,t_2)$] or different [$(P_1,t_1)\neq (P_2,t_2)$] spacetime points are described mathematically by the second or higher order correlation functions. Note that the third-order correlation between the waves $E_1$ and $E_2$ corresponds the value $I^3=(I_1+I_2)^3$, which does not depend on the wave phases. It should be stressed that the second-order correlation, which is also independent from the wave phases, does associate with the wave intensity (energy) squared. One could distinguish the second-order correlations attributed to the intensity interferometry associated with the Bose-Einstein correlations of bosons and the Fermi-Dirac second-order correlations of fermions. While in the Fermi-Dirac second-order correlations the particles are antibunched, in the Bose-Einstein correlations they are bunched. Another distinction between the Bose-Einstein and Fermi-Dirac correlations is that only the Bose-Einstein correlations can present quantum coherence. The Copenhagen-Dirac postulate of the "interference-less", self-interfering particles strictly forbids existence of both the interference (cross-correlation) between particles and the respective cross-correlation energy. According to the canonical interpretation, the interference between two different particles never occurs and each particle interferes (cross-correlates) only with itself. Nevertheless, in the canonical quantum physics and particle field theory based on the Copenhagen-Dirac postulate, the interference between particles and the respective cross-correlation energy are permitted, due to the unclear reasons, in the second or higher order of interference (cross-correlation) in the same and/or different spacetime points. It is not completely clear up to now how the particles, which are free from the interference and cross-correlation with each other, provide the Bose-Einstein and Fermi-Dirac cross-correlations. Although the second-order, cross-correlation terms of the Bose-Einstein and Fermi-Dirac correlations in the quantum physics and particle field theory are attributed to the {\it{energy}} of bosons and fermions, the second-order terms should be associated rather with the {\it{energy squared}} than with the {\it{energy}}. The solution of the problem is analyzed and interpreted in Part II.

\section{5. The role of cross-correlation energy in several basic physical phenomena}

The interference and coherence of the unit-waves (particles) is the necessary condition of existence of the cross-correlation and cross-correlation energy in an ensemble of the unit-waves (Secs. 4.2. and 4.3.). As an example, the present section shows a key role of the interference-induced positive and negative cross-correlation energies, which are the nonexistent energies in quantum physics and particle field theory, in several basic coherent phenomena, such as the Bose-Einstein condensation, super-radiation, Bosenova effect, superfluidity, superconductivity, supermagnetism, and quantum anomalous and fractional Hall effects. The coherent phenomena are first considered by using the traditional models of quantum mechanics and quantum field theory. Then they are compared with the predictions of the present model and the experimental observations. 

The conventional interpretation of the quantum interference {\it{strictly forbids}} existence of the interference between particles. Nevertheless, in canonical quantum mechanics and field theory based on the Copenhagen-Dirac postulate of "interference-less", self-interfering particles, the large-scale (macroscopic) {\it{quantum coherence and interference between particles}} are permitted. That leads to the several basic coherent physical phenomena in the coherent quantum systems. For instance, the superposition of particles obeying the properties of Bose-Einstein condensation, super-radiation, Bosenova effect, superfluidity, superconductivity, supermagnetism, and quantum anomalous and fractional Hall effects are examples of the coherent phenomena and coherent quantum systems. The Bose-Einstein condensation, which is associated with the second-order cross-correlation of bosons, is a consequence of the Bose-Einstein statistics and thus applicable to any kind of bosons. The Bose-Einstein condensation is at the origin of the most important condensed matter phenomena, superconductivity and superfluidity. The Bose-Einstein correlations manifest themselves also in hadron interferometry and the Bose-Einstein correlation between particles and anti-particles. Another example, which emphasizes the non-classical properties of quantum coherence in macroscopic systems, is the Schr{\"o}dinger's cat thought experiment. In the canonical quantum mechanics and particle field theory based on the Copenhagen-Dirac postulate, the aforementioned phenomena could be attributed only to the {\it{self-coherence and self-interference (self-correlation)}} of particles, which {\it{do not require the interference between particles (wave-functions)}} and the particle coherency. In other words, the postulate strictly forbids existence of both the interference (cross-correlation) between particles and the respective cross-correlation energy. The interference between two different particles never occurs and each particle interferes (cross-correlates) only with itself. Nevertheless, the interference and cross-correlation in the traditional models of the aforementioned coherent phenomena are permitted, due to the unclear reasons, in the second-order (or higher) of {\it{interference and cross-correlation between particles}}. Although the traditional models of the coherent quantum phenomena work perfectly, the permission of interference between particles has not been explained up to now. According to the present model, the coherent phenomena are simply the direct consequences of the cross-correlation energy mediated by the interference between particles (unit-fields). The coherence of the unit-waves (particles) is the necessary condition of existence of the interference, cross-correlation and cross-correlation energy in an ensemble of the unit-waves (particles). The coherency of the unit-waves is destroyed by collisions of the coherent unit-fields (particles) with the incoherent or partially coherent particles of the external microscopic or macroscopic objects, for instance with the material boundaries and/or detectors. Any superposition of the unit-waves (particles) obeying the thermodynamical equilibrium with the incoherent unit-fields (particles) of the high-temperature ($T >> 0$) boundaries has zero cross-correlation energy. 

Let me begin consideration of the aforementioned coherent phenomena with a brief general analysis of the super-radiation properties of several coherent quantum systems by using the present model and then compare them with the canonical quantum mechanics and experimental observations. The active medium of lasers is an example of the coherent quantum system, which is out of the high-temperature thermodynamical equilibrium. The process of induced emission of a photon by an excited atom of the laser medium, which is described by the Einstein stimulated emission coefficient, results into creation of the two photons. The two photons (electromagnetic unit-waves) of the same frequency are emitted at the same time, therefore they would be coherent. The unit-waves are identical, $\mathbf {k}_{01} = \mathbf {k}_{02} \equiv \mathbf {k}_{0}$ and $\alpha _{1}=\alpha _{2}$. The cross-correlation energy  ${\varepsilon}_{12,21}$ of the coherent photon pair that satisfies the condition (\ref{eq133}) is then given by ${\varepsilon}_{12,21}=2k_0$. Thus the total energy ${\varepsilon}=2k_0+2k_0=4k_0$ of the coherent photon pair is different from the Dirac energy ${\varepsilon}=2k_0$ (or more precisely ${\varepsilon}=2[k_0+(1/2)k_0]$) of the interference-less (correlation-free) photons. The energy ${\varepsilon}=4k_0$, however, is in agreement with the well-known experimental fact that the energy ${\varepsilon}$ of lasers operating in the super fluorescent mode is given by ${\varepsilon}\sim N^2$, where $N$ is the number of the excited atoms. Another example of the super-radiation is the Dicke superradiance of the subwavelength ensemble of excited atoms~\cite{Dick}. Although the canonical quantum mechanics and quantum field theory strictly forbid existence of the interference, cross-correlation and cross-correlation energy of particles [according to the Copenhagen-Dirac postulate, a particle can interfere (correlate) only with itself],  the interference and cross-correlation energy of particles appear in the Dicke quantum model as the result of the particular mathematical approach (approximation). The interference and cross-correlation energy of particles do appear in the  quantum-mechanical model due to the pure mathematical reason (approximation), namely due to the description of the radiating gas by a single quantum wave-function, which is composed from {\it{the cross-correlating mathematical wave-functions}} of the electrons of the individual atoms. The "mathematical cross-correlation (interference)" induced by the cross-correlating mathematical wave-functions and the calculated cross-correlation energy then are incorrectly {\it{interpreted}} as the physical interference and energy of the correlated (coherent) motion of electrons in the atoms. Thus the canonical quantum mechanics, which is based on the Copenhagen-Dirac postulate of "interference-free" particles, correctly describes the super-radiance if the mathematical structure of quantum mechanics is modified by the mathematical approximation. In other words, any {\it{physical interpretation}} of the Dicke model, in fact, would be the interpretation of the mathematical approximation that does not compare well with the basic principle (Copenhagen-Dirac postulate) of the canonical quantum mechanics. In the Dicke model, the momentums ${\mathbf k}_{0n}$ and phases $\alpha _n$ of the light unit-waves (photons) produced by the excited atoms in the far-field zone are the same, ${\mathbf k}_{0n}={\mathbf k}_{0}$ and $\alpha _n=\alpha $. According to both the Dicke model and present model, the radiated energy scales as the number of atoms squared, ${\varepsilon}\sim N^2$. In addition to the Dicke superradiance, the present model predicts the total destruction of the photons by the pure subtractive interference at  the phase condition $\alpha _n - \alpha _m = \pi$. Note that the energy $\Delta \varepsilon$ spent on the excitation of an atom is given by $\Delta \varepsilon =k_0$. Thus the superradiance of two atoms produces the extra energy $\varepsilon = 4k_0-2k_0$, which in the present model is attributed to the cross-correlation energy ${\varepsilon}_{12}=2k_0$. Notice, the physical mechanism behind the interference-induced extra energy is simply the four-times increase of the wave energy by the two-times increase of the wave amplitude. The energy conservation of the total macroscopic system under superradiance or superfluorescence is obtained if one takes into consideration a fact that the excitation of an atom has the probabilistic character. The averaged energy spent on the excitation of one atom, for instance, by an electron beam, usually exceeds the energy $\Delta \varepsilon =k_0$. The creation of the electron beam also requires the energy $\varepsilon >k_0$. Although the radiated energy increases under the additive interference, the total energy of the physical system is conserved. The creation of conditions of the pure additive interference requires the "additional" energy, which is added to the total physical system before the interference. The physical system does not obey the shift symmetry of time because of the probabilistic character [irreversible character associated with scattering and/or diffraction of unit-fields (particles)] of the excitation of an atom. Therefore, the energy nonconservation associated with the cross-correlation energy does not contradict the Noether theorem. The energy conservation of the total physical system is provided rather by the probabilistic exchange of the energy with the environment than by the shift symmetry of time. In contrast to the pure constructive or destructive interference, the creation of conditions of the normal (ordinary), for instance, in classical interferometers, does not require the additional energy. The energy conservation of the total physical system is provided by the shift symmetry of time of such a system. Such systems do not include the irreversible processes associated with the irreversible (probabilistic) scattering and/or diffraction of the unit-fields (particles).

In the canonical quantum mechanics and particle field theory, the physical mechanism behind the superfluidity, superconductivity, and supermagnetism is usually clarified by considering a Bose-Einstein condensate~\cite{Bose,Einst3,Lond}. It is generally accepted that the mechanism of a Bose-Einstein condensate is the large-scale quantum coherence. In the present model, the large-scale classical and quantum coherence (interference) of the boson unit-fields also leads to the aforementioned coherent phenomena. The difference between the present model and the traditional quantum mechanical models, is in the {\it{interpretation of quantum interference}} [see, the analysis of Young's experiment in Sec. (4.3.)] and in the {\it{absence}} of both the interference between particles and the respective cross-correlation energy in the canonical quantum mechanics and particle field theory based on the Copenhagen-Dirac  postulate of "interference-less", self-interfering particles. Let me briefly describe the Bose-Einstein condensation in the frame of canonical quantum mechanics and compare that with the present model. It is generally accepted that all the atoms (boson-like particles) that make up the Bose-Einstein condensate are {\it{coherent}}. The atoms are thus necessarily all described by a single quantum wave-function, which according to quantum mechanics is responsible for the condensate super-properties. Einstein has demonstrated, by using rather the {\it{statistical}} approach than the quantum mechanical arguments, that cooling boson atoms to a very low temperature would cause them to condense into the lowest accessible quantum state, resulting in a Bose-Einstein condensate (BEC)~\cite{Bose,Einst3}. This transition occurs below a critical temperature
\begin{eqnarray} \label{eq137}
T_c\approx 3.31(mk_B)^{-1}(N/V)^{2/3},
\end{eqnarray}
which is calculated by integrating over all momentum states the expression for maximum number of excited particles~\cite{Einst3}:  
\begin{eqnarray} \label{eq138}
N=V{\int_{0}^{\infty}}(2\pi )^{-3}(e^{-k^2/2mk_BT_c}-1)^{-1} d^3k,
\end{eqnarray}
where $N/V$ is the particle density, and $m$ is the particle mass. In the traditional description of BEC, all the boson-like particles that make up the condensate are modeled to be {\it{in-phase (coherent)}}. Respectively, the all particles are described by a single quantum wave-function $\Psi (\mathbf{r},t)$, which determines the super-properties of BEC. As long as the number of particles of BEC is fixed, the values $\Psi^*(\mathbf{r})\Psi (\mathbf{r})$ and $N=\int \Psi^*(\mathbf{r})\Psi (\mathbf{r}) d^3x$ are interpreted as the particle density and the total number of atoms, respectively. The dynamics of wave-function $\Psi (\mathbf{r})$ of the ground state of the BEC system of identical bosons is conventionally described by using the Hartree-Fock approximation (extension of the mean field theory). In the Hartree-Fock approximation, the total wave-function of the system of $N$ bosons is taken as a product of single-particle functions, $\Psi (\mathbf{r}_1, \mathbf{r}_2, ..., \mathbf{r}_N)=\psi_1 (\mathbf{r}_1)\psi_2 (\mathbf{r}_2)...\psi_N (\mathbf{r}_N)\equiv \psi (\mathbf{r}_1)\psi (\mathbf{r}_2)...\psi (\mathbf{r}_N)$. Note that the wave-functions (particles) are indistinguishable in such a case. In other words, the boson wave-functions are the same, $\psi_n (\mathbf{r}_n)\equiv \psi(\mathbf{r}_n)$ and  $\psi^*_n (\mathbf{r}_n)\equiv \psi^*(\mathbf{r}_n)$, where $n=1, ..., N$. Provided essentially all bosons have condensed to the ground state, and treating the bosons by the mean field theory, the energy $\varepsilon = {\cal H}$ associated with the stat $\Psi$ is given by the model Hamiltonian of the BEC system based on the Hartree-Fock approximation as
\begin{eqnarray} \label{eq139}
{\cal H}={\int_{0}^{\infty}} h  d^3x=\sum_{n=1}^N {\cal H}_{nn}+\sum_{n\neq m}^{N^2-N}{\cal H}_{nm},
\end{eqnarray}
with 
\begin{eqnarray} \label{eq140}
\sum_{n=1}^{N}{\cal H}_{nn} =\sum_{n=1}^{N}{\int_{0}^{\infty}}[(1/2m_0) \psi^*_n (\mathbf{r}_n) \nabla ^2{\psi (\mathbf{r}_n)} + U_{ext}(\mathbf{r}_n){\psi^*_n (\mathbf{r}_n)}{\psi_n (\mathbf{r}_n)}]d^3x_n
\end{eqnarray}
and 
\begin{eqnarray} \label{eq141}
\sum_{n\neq m}^{N^2-N}{\cal H}_{nm} =\sum_{n\neq m}^{N^2-N}{\int_{0}^{\infty}}{\int_{0}^{\infty}}{\psi^* (\mathbf{r}_n)}{\psi^* (\mathbf{r}_m)}{U(|\mathbf{r}_n -\mathbf{r}_m|)}{\psi (\mathbf{r}_n)}{\psi (\mathbf{r}_m)}d^3x_nd^3x_m,
\end{eqnarray}
where $m_0$ is the mass of the boson, $U_{ext}(\mathbf{r})$ is the external potential, ${U(|\mathbf{r}_n -\mathbf{r}_m|)}=(g/2){\delta (|\mathbf{r}_n -\mathbf{r}_m|)}$ is the interaction (potential) energy of the interaction of the n-th particle with the m-th particle, $g$ is the coupling parameter that represents the value of the inter-particle interaction, and ${\delta (|\mathbf{r}_n -\mathbf{r}_m|)}$ denotes the Dirac delta-function.  Minimizing this energy with respect to infinitesimal variations in $\psi (\mathbf{r})$ and holding the number of bosons constant yields the well-known Gross-Pitaevski model equation (GPE) of motion for the single-particle wave-function~\cite{Gros,Pit}:
\begin{eqnarray} \label{eq142}
i\frac {\partial {\psi(\mathbf{r},t)}}{\partial t}=[(-1/2m_0){\nabla}^2 + U_{ext}(\mathbf{r})+g{\psi^* (\mathbf{r})}{\psi (\mathbf{r})}]\psi (\mathbf{r}). 
\end{eqnarray}
If the single-particle wave-function satisfies the Gross-Pitaevski equation, the total wave-function $\Psi$ minimizes the expectation value of the Hamiltonian (\ref{eq139}) under normalization condition $N=\int \Psi^*(\mathbf{r})\Psi (\mathbf{r}) d^3x$. The Hamiltonian (\ref{eq139}) and GPE (\ref{eq142}) provide a good description of the behavior of the Bose-Einstein condensates and are thus conventionally used for clarification of the basic properties of BEC. The super-properties of BEC describing by Eqs. (\ref{eq139}) - (\ref{eq142}) have their origins in the interaction between the particles describing by the interaction energy $\sum_{n\neq m}^{N^2-N}{\cal H}_{nm}$ and the respective potential energy $g{\psi^* (\mathbf{r})}{\psi (\mathbf{r})}$. This becomes evident by equating the coupling constant $g$ of interaction in Eqs. (\ref{eq141}) and (\ref{eq142}) with zero, on which the ordinary Hamiltonian ${\cal H}=\sum_{n=1}^N {\cal H}_{nn}$ and Schr\"{o}dinger equation describing a particle inside a trapping potential are recovered. It should be stressed that the same result is obtained if one excludes the cross-correlation (interference) between the wave-functions of the individual bosons from the term (\ref{eq141}), ${\int_{0}^{\infty}}{\int_{0}^{\infty}}{\psi^* (\mathbf{r}_n)}{\psi^* (\mathbf{r}_m)}(g/2){\delta (|\mathbf{r}_n-\mathbf{r}_m|)}{\psi (\mathbf{r}_n)}{\psi (\mathbf{r}_m)}d^3x_nd^3x_m = (g/2){\int_{0}^{\infty}}[{\psi^*_n (\mathbf{r})}{\psi_m (\mathbf{r})}][{\psi^*_m (\mathbf{r})}{\psi_n (\mathbf{r})}]d^3x= 0$. Here, $\psi (\mathbf{r}_n)\equiv \psi_n(\mathbf{r}_n)$, $\psi (\mathbf{r}_m)\equiv \psi_m(\mathbf{r}_m)$, $\psi^* (\mathbf{r}_n)\equiv \psi^*_n(\mathbf{r}_n)$ and  $\psi^* (\mathbf{r}_m)\equiv \psi^*_m(\mathbf{r}_m)$ due to the indistinguishableness of the bosons (wave-functions). With the help of the present model, one can easily recognize the trial mathematically constructed  second-order cross-correlation term (\ref{eq141}) as the counterpart [see, Eq. (\ref{eq6})] of the first-order cross-correlation energy of the Bose-Einstein condensate. Although the canonical quantum mechanics strictly forbids the existence of the interference, cross-correlation and cross-correlation energy [a particle can interfere (correlate) only with itself], in the quantum mechanical description of BEC, the interference between particles and the respective cross-correlation energy do appear in the second order of the cross-correlation as the result of using the Hartree-Fock mathematical approximation. In other words, the interference between particles and cross-correlation energy do appear in the BEC quantum-mechanical model due to the Hartree-Fock  approximation of the quantum system by using a trial {\it{mathematical}} wave-function $\Psi (\mathbf{r}_1, \mathbf{r}_2, ..., \mathbf{r}_N)$, which is the product of {\it{the mathematically cross-correlating single functions}} associated with the single particles, $\Psi (\mathbf{r}_1, \mathbf{r}_2, ..., \mathbf{r}_N)=\psi_1 (\mathbf{r}_1)\psi_2 (\mathbf{r}_2)...\psi_N (\mathbf{r}_N)\equiv \psi (\mathbf{r}_1)\psi (\mathbf{r}_2)...\psi (\mathbf{r}_N)$. Then the calculated cross-correlation term (\ref{eq141}), which has the form of the {\it{second-order}} cross-correlation function $\sum_{n\neq m}^{N^2-N}{\cal H}_{nm} =(g/2)\sum_{n\neq m}^{N^2-N}{\int_{0}^{\infty}}{\psi^*_n (\mathbf{r})}{\psi_m (\mathbf{r})}{\psi ^*_m (\mathbf{r})}{\psi_n (\mathbf{r})}d^3x \neq 0$, is interpreted as the interaction energy. Remember, the second-order cross-correlation is attributed rather to the {\it{square of energy}} in intensity interferometry than the {\it{energy}} in amplitude interferometry (see, Sec. 4.3.). Thus the {\it{"modified" quantum mechanics}} [the quantum mechanics based on the interference and cross-correlation energy of particles automatically inserted into the model by using the mathematical approximations, which are inconsistent with the Copenhagen-Dirac postulate of "interference-less", self-interfering particles] correctly describes BEC. In the present model, the super-properties of BEC are similar to the quantum mechanical properties describing by Eqs. (\ref{eq139}) - (\ref{eq142}), which use the "modified" quantum mechanics. The difference between the "modified" quantum mechanics of BEC and the present model is in the {\it{interpretation}} of quantum interference and in the {\it{absence}} of the interference, cross-correlation and cross-correlation energy of particles in the canonical quantum mechanics based on the Copenhagen-Dirac postulate. 

It should be stressed again in the context of the above-presented consideration that a particle of canonical quantum mechanics associated with the wave of probability is free from the interference (cross-correlation) with other particles. The Hartree-Fock mathematical approximation {\it{automatically inserts}} the "hidden" interference between particles and the respective cross-correlation energy into BEC. Such a kind of the {\it{"hidden"  interference}} (mathematical cross-correlation), however, is different from the ordinary interference of the present model. Indeed, the cross-correlation energy associated with the approximation-induced cross-correlation ("hidden" interference) between two or more particles in the same [$(\mathbf{r}_n)=(\mathbf{r}_m)=(\mathbf{r})$] points of BEC is described in the Hartree-Fock approximation by the second order cross-correlation function ${\int_{0}^{\infty}}{\int_{0}^{\infty}}{\psi^* (\mathbf{r}_n)}{\psi^* (\mathbf{r}_m)}(g/2){\delta (|\mathbf{r}_n-\mathbf{r}_m|)}{\psi (\mathbf{r}_n)}{\psi (\mathbf{r}_m)}d^3x_nd^3x_m = (g/2){\int_{0}^{\infty}}[{\psi^*_n (\mathbf{r})}{\psi_m (\mathbf{r})}][{\psi^*_m (\mathbf{r})}{\psi_n (\mathbf{r})}]d^3x\neq 0$, which {\it{does not depend}} on the values of the particle phases. In other words, the  quantum Hamiltonian (\ref{eq139}) of BEC constructed by using the Hartree-Fock  approximation is invariant under the $U(1)$ local gauge transformation ($\psi _{n} \rightarrow \psi ' _{n}=e^{i\alpha _n} \psi _{n}$). The $U(1)$ gauge symmetry of the trial Hamiltonian (\ref{eq139}) provides the very particular (in comparison to the ordinary interference) mathematical cross-correlation ("hidden" interference) and the respective cross-correlation energy, which {\it{do not depend}} on the phases of the wave-functions of particles. {\it{In such a case, the boson-like particles that make up the Bose-Einstein condensate may have absolutely different phases providing the BEC and its super-properties without the particle coherence}}. However, it is generally accepted that all the wave-functions of the boson-like particles that make up the BEC {\it{must be in-phase}}. That indicates that the cross-correlation ("hidden" interference) in the traditional models of BEC is provided rather by the mathematical approximation than the large-scale classical or quantum coherence. The "hidden" interference is usually {\it{interpreted}} as the inter-particle interaction describing by the interaction (potential) energy. The present model describes the cross-correlation energy associated with the interference and correlation between particles (unit-fields) by the first-order cross-correlation function, {\it{which does depend on the unit-field phases}}. The cross-correlation energy in the present model does not depend on the unit-field phases only if all the particles (unit-fields) are incoherent. Indeed, the time-averaged cross-correlation energy (\ref{eq136}) of the unit-fields with random phases is equal to zero due to the non-correlation of such fields. One could distinguish the Bose-Einstein second-order bunched correlation of bosons attributed to the intensity interferometry from the first-order cross-correlation and cross-correlation energy of the unit-fields (bosons) associated with the amplitude interferometry. It should be mentioned again that the Copenhagen-Dirac postulate strictly excludes existence of both the interference (cross-correlation) between different particles and the respective cross-correlation energy. According to the postulate, the interference between different particles never occurs, each particle interferes (cross-correlates) only with itself. Therefore the interference and cross-correlation of coherent particles in the traditional models of the Bose-Einstein condensate should be attributed rather to the {\it{self-interference}} of particles than the {\it{interference (cross-correlation) between particles}}. In these models, the interference, cross-correlation and cross-correlation energy of particles do appear in the second-order of interference (cross-correlation) due to the pure mathematical reason (the Hartree-Fock approximation), which does not compare well with the canonical interpretation of quantum interference. Moreover, the second-order interference corresponds to the "intensity interferometry", which is associated rather with the energy (intensity) squared $\varepsilon ^2$ than the energy $\varepsilon = {\cal H}$. The problem is analyzed in details in Part II of the present study.

The above-described {\it{differences}} between the traditional quantum theory of BEC and the present model are {\it{summarized and interpreted}} as follows. The interference-induced cross-correlation energy, which does not exist in the canonical quantum mechanics based on the Copenhagen-Dirac postulate of {\it{quantum self-interference}}, is permitted in the BEC quantum model in the form of the interaction energy (\ref{eq141}). According to the postulate, the BEC and its extraordinary properties are provided by the self-coherence of particles. Although the particles (wave-functions) of BEC are modelled {\it{to be in-phase}}, the {\it{quantum self-interference}} of a particle does not require coherency with other particles. The self-interfering particles (wave-functions) may have absolutely different phases providing the BEC and its extraordinary properties. The second-order, phase-independent cross-correlation term (\ref{eq141}) is responsible for the interaction ("hidden" interference) between the different particles. The cross-correlation integral in the expression (\ref{eq141}) is usually {\it{interpreted}} as exchange one. According to quantum mechanics, the integral describes the quantum exchange of particles associated with the indistinguishableness of the identical bosons. The quantum exchange of bosons is somewhat similar to the exchange of virtual particles in a short time ($\Delta t \leq 1/\Delta \varepsilon $) in the perturbation quantum theory. The BEC quantum system is described mathematically by a trial mathematical wave-function composed from {\it{the cross-correlating, single mathematical functions associated with the single particles}} [$\Psi (\mathbf{r}_1, \mathbf{r}_2, ..., \mathbf{r}_N)=\psi_1 (\mathbf{r}_1)\psi_2 (\mathbf{r}_2)...\psi_N (\mathbf{r}_N)\equiv \psi (\mathbf{r}_1)\psi (\mathbf{r}_2)...\psi (\mathbf{r}_N)$], which yields the mathematical phase-independent cross-correlation ("hidden" interference) between particles and the respective phase-independent cross-correlation energy. In the model, the "hidden" interference is {\it{interpreted}} as the inter-particle interaction having the interaction energy mediated by the quantum exchange of particles. Remember, the same mathematical approach yielding the "hidden" interference between particles induced by the pure mathematical reason (approximation), which contradicts the Copenhagen-Dirac postulate, has been originally used by R. H. Dicke for the mathematical construction of the quantum-mechanical model of the superradiance~\cite{Dick}. The model Hamiltonian (\ref{eq139}) of BEC is also engineered mathematically by using the same (Hartree-Fock) approximation, which yields the interference and cross-correlation energy of particles, in the second order of the mathematical cross-correlation. The dynamics of the trial single-particle wave-function of BEC is described by the Gross-Pitaevski model equation derived by using the trial Hamiltonian (\ref{eq139}). As an example of the mathematical ("hidden") interference, one could mention also the well-known Bogoliubov treatment of the Gross-Pitaevskii equation, which help to find the elementary excitations of a Bose-Einstein condensate and to demonstrate the BEC super-fluidity. To that purpose, the condensate wave-function is approximated by a sum of the equilibrium wave-function and a small perturbation. In the model, the Bose gas does exhibit an energy (velocity) gap. The energy (velocity) gap, according to Landau's criterion, shows that the condensate is a super-fluid, meaning that if an object is moved in the condensate at an energy (velocity) smaller than the energy (velocity) gap, it will not be energetically favourable to produce excitations. The object will move without dissipation, which is a characteristic of a superfluid. The same result was found also by using the {\it{second quantization}} of the Bogoliubov wave-function of BEC. For the {\it{comparison}}, in the present model, the physical mechanism of the BEC superfluidity is not mediated by mathematical approximations. In the model, a quantum particle is considered as an indivisible unit-field of the physical matter. The dynamics of the unit-field is described by the Euler-Lagrange equation of motion~\cite{Land,Jack,Bere,Itz,Ryde,Pes,Grei,Wein} with the initial and boundary conditions imposed. The energy of the superposition of material unit-fields is given by the respective Hamiltonian of the system, which contains the cross-correlation energy $\sum_{n\neq m}^{N^2-N}{\cal H}_{nm}$ [in the first order of the cross-correlation (interference)] without mathematical approximations [see, Secs. (2)-(4)]. The interaction (cross-correlation) energy is given by $\sum_{n\neq m}^{N^2-N}{\cal H}_{nm}$, where ${\cal H}_{nm}={\int_{0}^{\infty}}[\frac {\partial {\psi ^*_n}}{\partial t} \frac {\partial {\psi_m}}{\partial t} +\nabla {\psi ^*_n}\cdot \nabla {\psi_m} + m_n{\psi^*_n}m_m{\psi_m}]d^3x$. The field of  BEC is considered as the superposition of individual coherent or partially coherent unit-fields (bosons), $\Psi (\mathbf{r}_1, \mathbf{r}_2, ..., \mathbf{r}_N)=\sum_{n=1}^{N}\psi_n (\mathbf{r}_n)$. The interaction (cross-correlation) energy $\sum_{n\neq m}^{N^2-N}{\cal H}_{nm}$ of the bosons (harmonic wave-like unit-fields) has been calculated in Sec. (2). The interaction of the boson unit-waves of BEC with any perturbing wave is possible only in the case $\mathbf{k'}={\mathbf{k}}_0$ [see, the condition (\ref{eq132})], where $\mathbf{k'}$ and ${\mathbf{k}}_0={\mathbf{k}}_n={\mathbf{k}}_m$ are the momentums of the perturbing wave and the boson unit-waves of BEC, respectively. If the absolute values and/or directions of the momentums are different, then the time-averaged  energy of the interaction (cross-correlation) of the BEC with the perturbing wave is zero. That is to say that the perturbation could not resist the BEC current providing the phenomena of superfluidity. The interpretation is somewhat different from the traditional mechanism, which does use the energy (velocity) gap in the explanation of superfluidity.   

Although the quantum mechanical model and the present models are very {\it{similar}} in many aspects (see, the above-presented discussion), the following basic properties of BEC do exist in the present model, only. In the present model [see, Secs. (2)-(4)], the total energy ${\cal H}={\cal N}{\varepsilon}_0$ of the cross-correlating unit-waves (bosons) in BEC is given by 
\begin{eqnarray} \label{eq143}
0\leq{ \cal H }\leq N^2{{\varepsilon}_0},
\end{eqnarray}
where ${\varepsilon}_0=({\mathbf{k}}_0^2+m_0^2)^{1/2}$ is the particle energy in the ground state, and 
\begin{eqnarray} \label{eq144}
0\leq{{\cal N}}\leq N^2
\end{eqnarray}
is the effective number of the unit-waves (particles) $\psi (\mathbf{r}_n) \equiv  {\psi _{0n}}(\mathbf{r})$. The effective number of particles ${\cal N}$, which is determined by the phases $\alpha _{n}$ of the unit-waves, can be a non-integer:
\begin{eqnarray} \label{eq145}
{\cal N}=N+\sum_{n\neq{m}}^{N^2-N}e^{i\Delta \alpha _{nm}}.
\end{eqnarray}
The momentum $\mathbf{P}$ of BEC is given by 
\begin{eqnarray} \label{eq146}
\mathbf{P}={\cal N}{\mathbf{k}_{0} }, 
\end{eqnarray}
where ${\mathbf{k}_0 }=\mathbf{k}_{0g}$ is the momentum of the unit-wave (particle) in the ground state. Notice, the value $k_{0g}^2 =3\pi^2V^{-2/3}$ if BEC is closed into the resonator-like box of the volume $V$, and the external potential $U(\mathbf{r}){\rightarrow}0$. If the electric unit-charge $q_0$ is associated with the mass $m_0$ of the unit-wave ${\psi _{0n}}(\mathbf{r})\equiv \psi (\mathbf{r}_n)$ [see comments to Eqs. (\ref{eq120}) and (\ref{eq121}), and Figs. (5)-(8)], then the charge ${{\cal Q}}$ ($0\leq{{\cal Q}}\leq N^2{q}_0$) of the cross-correlating unit-fields is given by 
\begin{eqnarray} \label{eq147}
{\cal Q}={\cal N}{q}_0,
\end{eqnarray}
where the value ${\cal N}$ can be a non-integer. Here, the Pauli exclusion principle for the fermions (unit-waves) should be taken into account in concrete calculations of the field parameters. The Pauli principle states that no two identical fermions  may occupy the same quantum state simultaneously. Therefore, the charged unit-field should be a boson-like composite unit-field (particle). Similarly, if the magnetic moment $\vec {\mu }_0$ is associated with the mass $m_0$ of the unit-field ${\psi _{0n}}(\mathbf{r})\equiv \psi (\mathbf{r}_n)$ [see comments to Eqs. (\ref{eq120}) and (\ref{eq121}), and Figs. (5)-(8)], then the effective magnetic moment ${\vec {M}}$ ($0\leq{{ \vec {M}}}\leq N^2{\vec {\mu }}_0$) of the cross-correlating unit-fields is given by 
\begin{eqnarray} \label{eq148}
{\vec {M}}={\cal N}{\vec {\mu }}_0,
\end{eqnarray}
where the value ${\cal N}$ can be a non-integer. 

It is generally accepted that all boson-like particles that make up BEC at the temperature $T<<T_c$ are in-phase (see, the above-considered conventional quantum mechanical description of BEC). In such a case, the effective number of particles (\ref{eq145}) in BEC is given by 
\begin{eqnarray} \label{eq149}
{\cal N}=N^2.
\end{eqnarray}
Respectively, Eqs. (\ref{eq143})-(\ref{eq148}) yielded the extraordinary values 
\begin{eqnarray} \label{eq150}
{\cal H }= N^2{{\varepsilon}_0},
\end{eqnarray}
\begin{eqnarray} \label{eq151}
\mathbf{P}=N^2{\mathbf{k}_{0} }, 
\end{eqnarray}
\begin{eqnarray} \label{eq152}
{\cal Q}=N^2{q}_0,
\end{eqnarray}
and 
\begin{eqnarray} \label{eq153}
{\vec {M}}=N^2{\vec {\mu }}_0
\end{eqnarray}
of the energy, momentum, charge and magnetic moment of BEC, which are responsible for the BEC super-properties. 
Notice, although the boson parameters associated with the charge have been formally introduced [see, Eqs. (\ref{eq147}), (\ref{eq148}), (\ref{eq152}) and (\ref{eq153})] into the BEC model, the {\it{nonexistence}} of the boson/antiboson charges is determined rather by the experimental data than the model (see, Part II). The anomalous values of the physical parameters of BEC at the low temperatures ($T<<T_c$) are obtained also in the case of annihilation of the unit-waves ${\psi _{0n}}(\mathbf{r})$ and ${\psi _{0m}}(\mathbf{r})$, which is provided by the pure subtractive interference of the unit-waves at the phase condition $\Delta \alpha _{nm}=\alpha _{n}-\alpha _{m}=\pm \pi$. The physical parameters of the annihilated BEC do vanish, ${\cal N}={\cal H }=\mathbf{P}={\cal Q}={\vec {M}}=0$. If the boson gas is in the thermodynamical equilibrium at the temperatures $T>>T_c$, the cross-correlation energy is equal to zero and the BEC energy $\langle {\cal H}\rangle_T = N \langle {{\varepsilon}_0}\rangle_T$ is equal to the ordinary mean thermal energy:  
\begin{eqnarray} \label{eq154}
\langle {\cal H}\rangle_T = N(3/2)k_BT,
\end{eqnarray}
where $N$ is the total number of the particles. Notice, Eq. (\ref{eq154}) uses the non-relativistic particle energy $ \varepsilon_0 \approx ( \mathbf{k}_0^2 / 2m_0)$, where the irrelevant relativistic part (energy $ \varepsilon = m_0$) has been discarded. At the intermediate temperatures ($T\sim T_c$) the boson gas contains the particles with the extraordinary and ordinary physical properties. In other words, the gas contains the $N_{ic}<N$ incoherent (ordinary) unit-waves and the $N_{c}=N-N_{ic}$ coherent (extraordinary) unit-waves. If the coherent particles (unit-waves) are in-phase, the effective number of such particles is given by ${\cal N}_c=N_c^2=(N-N_{ic})^2$. The phase transition from the incoherent (ordinary) state of the gas to the BEC state takes place below a critical temperature $T_c$, which is calculated by assuming the equality of the effective number of coherent particles to the number of incoherent particles (${\cal N}_c=N_{ic}$):
\begin{eqnarray} \label{eq155}
N^2_c=N-N_c.
\end{eqnarray}
According to Eq. (\ref{eq149}), even the small fraction ($N_c<<N_{ic}$) of the coherent unit-waves (particles) would dominate (${\cal N}_c>>N_{ic}$) the gas physical properties if   
\begin{eqnarray} \label{eq156}
N_c > [N+(1/4)]^{1/2}-(1/2).
\end{eqnarray}
For instance, below the critical temperature, the coherent physical properties of the one hundred and one ($N_c=101$) coherent particles would dominate the ten thousand ($N_{ic}=10000$) incoherent particles of BEC [see, Eqs. (\ref{eq149})-(\ref{eq153})]. If the dependence of the number of coherent particles on the temperature is known, then the value $T_c$ is calculated by using Eq. (\ref{eq155}). For instance, at the exponential decrease of the number of coherent particles with increasing the gas temperature, the value $N_c$ is given by
\begin{eqnarray} \label{eq157}
N_c=N^2{\exp}(-(3/2)k_BT(k^2_{0g}/2m)^{-1}+1),
\end{eqnarray}
where $N_c=N^2$ at the condition $(3/2)k_BT=k^2_{0g}/2m$. In the case of BEC closed into the resonator-like box of the volume $V$ and the external potential $U(\mathbf{r}){\rightarrow}0$, the expression (\ref{eq155}) yields the following equation for the calculation of the critical temperature  $T_c$:
\begin{eqnarray} \label{eq158}
e^2N^3{\exp}(-2\pi ^{-2}mk_BV^{2/3}T_c)+eN{\exp}(-\pi ^{-2}mk_BV^{2/3}T_c)-1=0,
\end{eqnarray}
where $e\equiv \exp(1)$. In the limits of the low and high temperatures, Eq. (\ref{eq158}) yields the analytical solutions
\begin{eqnarray} \label{eq159}
T_c\approx (1/2)\pi ^2(mk_B)^{-1}V^{-2/3}(e^2N^3+eN-1)(2e^2N^3+eN)^{-1}
\end{eqnarray}
and  
\begin{eqnarray} \label{eq160}
T_c\approx \pi ^2(mk_B)^{-1}V^{-2/3}{\ln}N,
\end{eqnarray}
respectively. The critical temperature (\ref{eq159}) is somewhat different from the Einstein critical temperature (\ref{eq137}). However, the value (\ref{eq160}) compares well with the temperature (\ref{eq137}). The exact dependence of the critical temperature $T_c$ on the particle density, which is described by Eq. (\ref{eq158}), has more complicated character in comparison to the Einstein model. One can easily calculate the critical temperatures by using the present model also for more sophisticated experimental conditions in the Bode-Einstein condensate. The physical mechanism behind the critical temperature $T_c$ of the phase transition in the Bose-Einstein condensate could be clarified better by considering the simplest BEC. The bounded unit-field boson pair, the pair that has the minimum energy at the distance $R\neq \infty $, has been considered in Sec. (4) as a stable composite particle. The composite particle can occupy the stable state even when the first and second particles have zero momentums, $\mathbf{k}_1=\mathbf{k}_2=0$. In the frame of the present model, a composite particle could be considered as the simplest (two-particle, $N=2$) Bose-Einstein condensate with $T_c\approx (1/2)\pi ^2(mk_B)^{-1}V^{-2/3}(e^22^3+2e-1)(2e^22^3+2e)^{-1}$. The two-particle condensate (composite particle) dissociates into the two independent (free) particles if the mean thermal kinetic energy $\langle {\cal H}\rangle_T = 2(3/2)k_BT$ of the two particles is bigger than the energy-mass defect $\Delta {\cal H}={\cal H}-({\cal H}_{11}+{\cal H}_{22})={\cal H}_{12} +{\cal H}_{21}$ [see comments to Eqs. (\ref{eq9}), (\ref{eq35}), (\ref{eq118})-(\ref{eq123}), and Figs. (5)-(8)] of the composite particle, $\langle {\cal H}\rangle_T > {\cal H}_{12} +{\cal H}_{21}$. The dissociation at the temperature $T > T_c$, where $T_c= [{\cal H}_{12} +{\cal H}_{21}]/3k_B$, is the physical mechanism behind the phase transition in the simplest (two-particle, $N=2$) Bose-Einstein condensate describing by the Einstein critical temperature $T_c\approx 3.31(mk_B)^{-1}(2/V)^{2/3}$ [see, Eq. (\ref{eq137})]. The physical mechanism attributed to the difference between the mean thermal kinetic energy of particles and the energy-mass defect (potential energy) of the Bose-Einstein condensate is {\it{somewhat different}} from the thermodynamical interpretations of the critical temperature of the phase transition in the phenomenological and microscopic models of BEC based on the Copenhagen-Dirac postulate and the Hartree-Fock mathematical approximation in the canonical quantum mechanics.

Let me now briefly consider the {\it{other coherent quantum phenomena}} associated with the macroscopic quantum coherence (interference), namely the superfluidity, superconductivity, supermagnetism, Bosenova effect, and quantum anomalous and fractional Hall effects. I should first demonstrate the relationships of the aforementioned extraordinary physical phenomena with the Bose-Einstein condensation, and then compare them with the present model. It is convenient to begin such a kind of analysis with consideration of the superconductivity. The phenomenological semi-microscopic Ginzburg-Landau theory (GLT) of superconductivity~\cite{Ginz}, which combines Landau's phenomenological macroscopic theory of second-order phase transitions with a Schr{\"o}dinger-like wave equation, had a great success in explanation of the macroscopic properties of superconductors. Although the Landau theory and GLT are constructed phenomenologically, they are usually {\it{interpreted}} as the semi-microscopic mean field theories (MFT). Any true microscopical MFT replaces all microscopic interactions to any one particle with an average or effective interaction. That reduce the multi-particle problem into an effective one-particle problem. The approximation mathematical procedure is quit similar to the Hartree-Fock mathematical approach, which yields the "hidden" interference, cross-correlation and cross-correlation energy of particles in the BEC quantum mechanical model. Probably, the MFT nature of the Landau theory and GLT is a mathematical reason of a great success of semi-microscopic Ginzburg-Landau model in explanation of the macroscopic properties of superconductors. The true microscopic fundamental quantum models (the Bardeen-Cooper-Schrieffer (BCS) and Bogoliubov quantum models) of superconductivity have been proposed by J. Bardeen, L. N. Cooper and J. R. Schrieffer~\cite{Bard} and independently N. N. Bogoliubov~\cite{Bogo}. The "hidden" interference between particles and the respective cross-correlation energy in these two fundamental models also may be {\it{interpreted}} as the result of the Copenhagen-Dirac postulate of "interference-less", self interfering particles and the Hartree-Fock like mathematical approximation. Indeed, in the normal state of a metal, the electrons move independently, whereas in the BCS state, they are bound into "Cooper pairs" by the attractive interaction through the exchange of phonons. The BCS theory explains the superconducting current as a superfluid of Cooper pairs. Thus the superconductivity is {\it{interpreted}} as a macroscopic effect, which results from "condensation" of Cooper pairs. The pairs have boson properties, while bosons, at sufficiently low temperature, can form the above-described Bose-Einstein condensate. The BCS method~\cite{Bard} of the mathematical approximation, which yields the "hidden" interference, cross-correlation and cross-correlation energy in the quantum model of superconductivity based on the Copenhagen-Dirac postulate of "interference-less", self interfering particles is quite similar to the Hartree-Fock approximation (MFT extension) of the above-described model of BEC. In the BEC model, a trial mathematical wave-function $\Psi (\mathbf{r}_1, \mathbf{r}_2, ..., \mathbf{r}_N)$ of the quantum many-particle system is the product $\Psi (\mathbf{r}_1, \mathbf{r}_2, ..., \mathbf{r}_N)=\psi_1 (\mathbf{r}_1)\psi_2 (\mathbf{r}_2)...\psi_N (\mathbf{r}_N)\equiv \psi (\mathbf{r}_1)\psi (\mathbf{r}_2)...\psi (\mathbf{r}_N)$ of the cross-correlating, mathematical, single functions $\psi_n (\mathbf{r}_N)$ associated with the $N$ single bosons yielding the "hidden" interference (mathematical cross-correlation) between the bosons. The BCS model may be interpreted as a model, where a trial mathematical wave-function $\Psi$ of the BCS quantum system of the $N$ electrons is the product
\begin{eqnarray} \label{eq161}
\Psi (\mathbf{R}_1, \mathbf{R}_2, ..., \mathbf{R}_{N/2})=\psi_1 (\mathbf{R}_1)\psi_2 (\mathbf{R}_2)...\psi_{N/2} (\mathbf{R}_{N/2})\equiv \psi (\mathbf{R}_1)\psi (\mathbf{R}_2)...\psi (\mathbf{R}_{N/2})
\end{eqnarray}
of the cross-correlating functions $\psi_n (\mathbf{R}_n)$ associated with the $N/2$ electron pairs (Cooper's bosons) having the coordinates $\mathbf{R}_n$. In the Bogoliubov model of superconductivity, the approximation mathematical method, which yields the "hidden" interference, cross-correlation and cross-correlation energy in the quantum model of superconductivity based on the Copenhagen-Dirac postulate of "interference-less", self interfering particles, could be associated with the Bogoliubov transformations~\cite{Bogo}. This extremely elegant and sophisticated transformation method is very similar to the Hartree-Fock approximation (MFT extension). The Bogoliubov mathematical approach (approximation), however, uses rather the mathematically cross-correlating operators than the cross-correlating, single-particle mathematical functions $\psi (\mathbf{r}_n)$. In the Bogoliubov model, the "hidden" interference, cross-correlation and cross-correlation energy of particles are mediated by the "cross-correlating" commutators with the commutator relation 
\begin{eqnarray} \label{eq162}
[\hat b, \hat b^{\dagger }]= (|u|^2+|v|^2)[\hat a, \hat a^{\dagger }],
\end{eqnarray}
where
\begin{eqnarray} \label{eq163}
\hat b = u\hat a +v \hat a^{\dagger }, 
\end{eqnarray}
and
\begin{eqnarray} \label{eq164}
\hat b^{\dagger }= u^*\hat a^{\dagger } +v^* \hat a.  
\end{eqnarray}
Here, $\hat a$ and $\hat a^{\dagger }$ are canonical annihilation and creation operators of the single boson particle. The Bogoliubov transformation is a canonical transformation of the operators $\hat b$ and $\hat b^{\dagger }$. A superfluid is the Bose-Einstein condensate, which does possess some hydrodynamical properties that do not appear in ordinary fluid. Although the quantum mechanics of the superfluidity based on the Copenhagen-Dirac postulate of "interference-less", self interfering particles and the Hartree-Fock like or operator-based mathematical approximations is very similar to the superconductivity, the microscopic details of the phenomena are different. If superfluid particles are bosons, their superfluidity is regarded as a consequence of Bose-Einstein condensation in the system of interacting boson particles. If the particles are fermions, then the superfluidity is described by the generalization of BCS quantum theory of superconductivity. In the generalized theory, Cooper pairing takes place between atoms rather than electrons, and the attractive interaction between them is mediated by spin fluctuations rather than phonons. A unified description of superconductivity and superfluidity is currently possible only by the quantum field theory in terms of the Hamiltonian-Lagrangian gauge symmetry and gauge symmetry breaking. The present model provides a unified description of the two phenomena in terms of the cross-correlation energy associated with the ordinary interference of the boson or boson-like unit-waves (particles). The superfluid boson current density $\mathbf{j}$ is simply given by $\mathbf{j}={\mathbf{P}}/mV = {\cal N}_c{\mathbf{k}_{0} }/mV$, where ${\cal N}_c$ ($0\leq{{\cal N}_c}\leq N^2$) is the effective number of the $N_c$ coherent unit-waves (bosons), and $N$ is the total number of bosons. The density ${\mathbf{j}}_e$ of the electric super-current of the Cooper electron pairs (boson unit-waves) having the charge $2e$ and mass $2m_e$ is given by ${\mathbf{j}}_e=2e{\mathbf{P}}/2m_eV=e{\cal N}_c{\mathbf{k}}_{0} /m_eV$, where $0\leq{{\cal N}_c}\leq N^2$ is the effective number of the coherent Cooper bosons. Thus even small fraction ($N_c << N$) of the coherent unit-waves (particles) would produce the coherent current, whose physical properties would dominate the current associated with the incoherent unit-waves (particles). If all the BEC particles are in-phase, the effective number ${\cal N}_c=N^2$ of the coherent unit-waves (bosons) would provide the maximum value of the BEC current. It should be stressed again that interaction of the boson unit-waves of BEC with any perturbing wave is possible only in the case $\mathbf{k'}={\mathbf{k}}_0$, where $\mathbf{k'}$ and ${\mathbf{k}}_0$ are the momentums of the perturbing wave and the unit-waves of BEC, respectively [see, the condition (\ref{eq132})]. If the directions or absolute values of these momentums are different, then the energy of the interaction (cross-correlation) of the Bose-Einstein condensate with the perturbing wave is zero. In other words, the perturbation does not resist the BEC current providing the phenomena of superfluidity and superconductivity.   

Superdiamagnetism (perfect diamagnetism) is a phenomenon occurring in certain materials at low temperatures, characterized by the complete absence of magnetic permeability and the exclusion of the interior magnetic field. According to the "modified" quantum mechanics, superdiamagnetism is a feature of superconductivity, which is described by the quantum mechanics by a single wave-function~\cite{Lond} of the BEC state. The Bose-Einstein condensation also applies to quasiparticles in solids. For instance, a magnon in an antiferromagnet carries spin 1 and thus obeys the Bose-Einstein statistics. A magnetic ordering at the temperatures lower than the point of condensation is the analog of superfluidity. The anomalous magnetic properties of the material, in the present model, are attributed to the extraordinary magnetic moment [see, Eqs. (\ref{eq148}) and (\ref{eq153})] of the Bose-Einstein condensate. The quantum anomalous and fractional states of particles in quantum anomalous and fractional Hall effects are not connect directly with BEC. Indeed, the quantum anomalous Hall effect or most commonly the anomalous Hall effect is usually attributed   either to a disorder-related effect due to spin-dependent scattering of the charges or an effect which can be described in terms of the so-called Pancharatnam-Berry phase effect. Naturally, such effects can be easily reinterpreted as the "hidden" cross-correlation (interference) between particles. The fractional quantum Hall effect is a physical phenomenon in which the quantum system behaves as if it was composed of particles with charge smaller than the elementary charge. Fractionally-charged quasiparticles, for instance, in the very elegant theory proposed by R.B. Laughlin ~\cite{Laug}, is based on the trial mathematical wave-function 
\begin{eqnarray} \label{eq165}
\Psi (\mathbf{r}_1, \mathbf{r}_2, ..., \mathbf{r}_N)=[\Pi _{N\geq i \geq j \geq 1}(r_i-r_j)^n] \Pi ^N _k \psi (r_k) 
\end{eqnarray}
for the ground state, as well as its quasiparticle and quasihole excitations, which is quite similar to the trial  mathematical wave-function (\ref{eq161}), which yields the "hidden" interference (mathematical cross-correlation) between particles and the respective cross-correlation energy. According to the theory of composite fermions, which was proposed by J. K. Jain ~\cite{Jain}, in the repulsive interactions, two (or, in general, an even number) flux quanta are captured by each electron, forming integer-charged quasiparticles called composite fermions. In this very effective theory, the fractional states of electrons are understood as the integer quantum Hall effect of composite fermions, which may be {\it{reinterpreted}} as products of the "hidden" interference of particles. The present model {\it{interprets}} these effects simply in terms of the interfering unit-waves (particles) with the effective number ($0\leq{{\cal N}}\leq N^2$) of particles, which can be fractional number bigger or smaller than one [see, Eq. (\ref{eq147})].

It is impossible in the present study to consider all the super-properties and coherent effects of the Bose-Einstein condensate. However, one should mention the so-called Bosenova or Bose-supernova effect~\cite{Ande} recently observed in the Bose-Einstein condensate. The effect is a small explosion, which can be induced by changing the magnetic field in which the Bose-Einstein condensate is located, so that the BEC quantum wave-function's self-attraction becomes repulsive. Under the explosion, a considerable part of the particles in the condensate disappears. The "missing" particles (atoms) are undetectable in the experiments probably because they form into molecules or they get enough energy from somewhere to fly away fast enough. Several mean-field quantum theories have been proposed to explain the phenomenon. It seems that in the canonical quantum theories this characteristic of the Bose-Einstein condensate remains unexplained, because the energy of a particle near absolute zero appears to be insufficient to cause the implosion. The present model describes the Bosenova explosion in terms of the repulsive force associated with the gradient of the cross-correlation energy ($0\leq{ \cal H }\leq N^2{{\varepsilon}_0}$) of the interfering boson or boson-like particles (unit-waves). The number of detectable particles, which is given by the effective number ($0\leq{{\cal N}}\leq N^2$) of particles, varies with changing the interference conditions in the Bose-Einstein condensate. For the details of the interaction mechanism, see Part II.

The all above-considered traditional models give the absolutely {\it{correct description and explanation}} of the extraordinary coherent physical phenomena, such as Bose-Einstein condensation, super-radiation, Bosenova effect, superfluidity, superconductivity, supermagnetism, and quantum anomalous and fractional Hall effects. The models are based on the canonical quantum mechanics and quantum field theory based on the Copenhagen-Dirac postulate of "interference-less", self interfering particles, which are modified by the different, generally-accepted mathematical approximations. The trial Hamiltonians of the models are invariant under the $U(1)$ local gauge transformation. Therefore {\it{the interference}} between particles and the respective cross-correlation energy in the models, which do not depend on the values of the particle phases, are provided mathematically by the second order cross-correlation functions. In other words, the $U(1)$ gauge symmetry of the trial Hamiltonians provides the very particular (in comparison to ordinary interference) mathematical cross-correlation ("hidden" interference) and the respective cross-correlation energy, which do not depend on the phases of the probability waves and the waves of operators associated with the particles. The {\it{interpretations}} of such models are usually incorrectly based on the interpretations of the mathematical approximations used in the models. One could also stress that the Hamiltonians of both the modified and canonical quantum theories are constructed to be invariant under the $U(1)$ local gauge transformation ($\psi _{n} \rightarrow \psi ' _{n}=e^{i\alpha _n} \psi _{n}$) that provides the "interference-less" behaviour of particles due to the independence of the Hamiltonians on the phases $\alpha _n$ of the waves of probabilities and operators. 

In quantum mechanics, particle field theory and the present model, the Bose-Einstein condensation of bosons is intrinsically related to the Bose-Einstein statistics, Fermi-Dirac statistics and Pauli exclusion principle. The physical mechanism behind these phenomena in the present model, however, is different from the traditional mechanisms of quantum mechanics and particle field theory based on the waves (fields) of probabilities or the waves (fields) of operators. Unfortunately, the traditional mechanisms (interpretations) are extremely complicated and non-transparent. Therefore the traditional mechanisms should be discussed again, in the context of the present model, by emphasizing their pure mathematical nature associated with the different (powerful) mathematical approximations, which are based on the use of mathematical objects (mathematically  correlating wave-functions and commutators). For the sake of simplicity, the traditional mechanisms will be clarified by considering the most simple systems, namely the two-particle objects. In the canonical quantum mechanics, the two-particle system is described by an antisymmetric (fermion) or symmetric (boson) state, which is   mathematically constructed by using the mathematical trial wave-functions associated with the probability amplitudes of particles. According to the {\it{spin-statistics theorem}} of quantum mechanics the integer spin particles are bosons, while the half-integer spin particles are fermions. If the fermions (for instance, electrons) are the same, the mathematically constructed (trial) antisymmetric expression of the state gives zero. Therefore, in an antisymmetric state, two identical fermions cannot occupy the same single-particle states. This is usually {\it{interpreted}} as the Pauli exclusion principle, which in such an interpretation form has a mathematical character based on the mathematical properties of the antisymmetric state (antisymmetric function). In particle field theory, the Pauli exclusion principle is sometimes attributed to the mathematical properties of the canonical commutation relations (mathematical objects) associated with the fermion operators. The mathematical construction of the antisymmetric state (mathematical function associated with the wave of probability) in the canonical quantum mechanics or the commutation relation (trial commutator) of the fermion operators in particle field theory mathematically yields the Fermi-Dirac statistics (correlations), whose {\it{interpretations}} also have pure mathematical character. The Pauli exclusion principle and Fermi-Dirac statistics mathematically forbid the Bose-Einstein condensation of particles (fermions). The modeling of the two-particle boson gas by the use of the mathematically constructed symmetric state (mathematical function) of the canonical quantum mechanics or the use of the mathematically constructed canonical and Bogoliubov's commutation relations for the mathematical operators in the quantum field theory does result into the Bose-Einstein statistics (correlations), which mathematically permit the Bose-Einstein condensation. According to the quantum mechanics and particle field theory based on the Copenhagen-Dirac postulate of "interference-less", self-interfering particles and the different, generally-accepted mathematical approximations, the Bose-Einstein condensation, Bose-Einstein statistics, Fermi-Dirac statistics and Pauli exclusion principle have their physical origins in the interaction ("hidden" interference) between the particles describing by the interaction (cross-correlation) energy $\sum_{n\neq m}^{N^2-N}{\cal H}_{nm}$. The interaction energy is connected with the "hidden" interference mediated by the different mathematical approximations, which are based on the  cross-correlating non-material wave-functions or the non-material operator commutators. In the present model, the physical mechanism behind the Bose-Einstein statistics, Fermi-Dirac statistics and Pauli exclusion principle is attributed simply to the repulsive or attractive forces associating with the additive or subtractive interference (cross-correlation) of the real unit-waves (bosons or fermions) of the matter (mass-energy).

It should be stressed again that the {\it{unified description}} of the interfering elementary particles (unit-fields) providing the coherent (extraordinary) physical phenomena is given in Part II of the present study, where the Hamiltonian ${\cal H}_{11}$ of a unit-field (particle) is associated with the unit-field energy squared (${\cal H}_{11}={\varepsilon}_0^{2} ={\mathbf{k}}_0^2+m_0^2$). Respectively, the Hamiltonian ${\cal H}$ of the field composed from the $N$ unit-fields (particles) is attributed to the total field-energy squared (${\cal H}={\varepsilon}^{2}=\sum_{n=1}^N{\cal H}_{nn}+\sum_{{n\neq m}}^{N^2-N}{\cal H}_{nm}$). In such a case, the physical parameters of the interfering unit-fields are described by the respective equations of Secs. (3)-(5) with $0\leq{ \varepsilon  }\leq N{{\varepsilon}_0}$ and $0\leq{{\cal N}}\leq N$. The interference and cross-correlation energy of these unit-fields is attributed to the second-order cross-correlation of the intensity interferometry. In Part I of the present study, the Hamiltonian ${\cal H}_{11}$ of a particle is associated with the unit-field energy ${\cal H}_{11}={\varepsilon}_0 =[{\mathbf{k}}_0^2+m_0^2]^{1/2}$, and the Hamiltonian ${\cal H}$ of the field composed from the $N$ particles is attributed to the total-energy ${\cal H}={\varepsilon}=\sum_{n=1}^N{\cal H}_{nn}+\sum_{{n\neq m}}^{N^2-N}{\cal H}_{nm}$. The energy of the cross-correlating unit-fields is then given by $0\leq{ \cal H }\leq N^2{{\varepsilon}_0}$ with the respective effective number $0\leq{{\cal N}}\leq N^2$ of unit-fields. The interference and cross-correlation energy of such unit-fields is associated with the first-order cross-correlation of the amplitude interferometry. 

\section{6. The present model versus traditional analysis of a many-particle system}

The following {\it{general analysis}} could make more transparent {\it{the methodology and philosophy of the present model}} based on the non-quantum and quantum interference between particles and the respective cross-correlation energies. The canonical quantum mechanics strictly forbids the existence of the interference between particles and the respective cross-correlation energy [a quantum particle interferes (correlates) only with itself]. The Hamiltonians of the canonical quantum mechanics are invariant under the $U(1)$ local gauge transformation ($\psi _{n} \rightarrow \psi ' _{n}=e^{i\alpha _n} \psi _{n}$). The $U(1)$ gauge symmetry provides the interference-less behavior of particles due to the independence of the canonical Hamiltonians on the phases $\alpha _n$. In the "modified" quantum mechanics based on the Copenhagen-Dirac postulate of "interference-less", self interfering particles, the "hidden" interference  between particles and the respective cross-correlation energy are mediated by using the different powerful mathematical approximations, which do not compare well with the basic physical principle (the Copenhagen-Dirac postulate of interference-less particles) of the traditional quantum mechanics and particle field theory. In such an approach, a mathematical wave-function of the quantum system is usually constructed mathematically as a product of the single-particle mathematical functions that cross-correlate in the second or higher order yielding the mathematical cross-correlation ("hidden" interference) between particles and the respective cross-correlation energy. The mathematical approach mediates the "hidden" interference and cross-correlation energy of particles [in the second-order or higher of the interference (cross-correlation)] in the quantum mechanical {\it{multi-particle}} systems. Although the mathematically constructed trial Hamiltonians of the "modified" quantum mechanics are invariant under the $U(1)$ local gauge transformation, the interference between particles and the respective cross-correlation energy are provided mathematically by the second or higher order cross-correlation functions, which do not depend on the values of the particle phases. That is to say that the $U(1)$ gauge symmetry of the trial Hamiltonians provides the very particular (in comparison to the ordinary interference) mathematical cross-correlation ("hidden" interference) and the respective cross-correlation energy, which do not depend on the phases of the probability waves of particles. {\it{In such a case, the particles (wave-functions) can have absolutely different phases providing the coherent quantum phenomena without the coherence of particles (wave-functions)}}. That indicates that the cross-correlation ("hidden" interference) is provided rather by the mathematical approximation than the large-scale classical or quantum coherence. For the sake of simplicity, the mathematical approach in the above-considered models of the coherent quantum phenomena has been attributed to the Hartree-Fock approximation and mean-field theory (MFT). Nevertheless, one can easily demonstrate that the "hidden" interference between particles and cross-correlation energy are mediated by other powerful mathematical approximations (approaches), such as the MFT extensions (e.g. Hartree-Fock and random phase approximations), many-body perturbation theory, Green's function-based methods, configuration interaction, coupled cluster, various Monte-Carlo approaches, density functional theory, and lattice gauge theory. In the {\it{quantum field theory}}, the "hidden" interference between particles and the respective cross-correlation energy are mathematically induced by the mathematical approximations associated with  trial cross-correlating operators (the Bogoliubov-like operators). The mathematical construction of the antisymmetric state (function) in quantum mechanics or the canonical commutation relation for fermions in particle field theory automatically yields the Fermi-Dirac statistics and Pauli exclusion principle, whose {\it{interpretations}} are usually mathematical ones. Modelling the boson gas by the symmetric state (function) or the use of the canonical commutation relation of bosons leads mathematically to the Bose-Einstein statistics. Naturally, the  {\it{interpretations}} of the mathematical approximations could have the mathematical character. In the quantum mechanics and quantum field theory based the Copenhagen-Dirac postulate of "interference-less" particles, the "hidden" interference, cross-correlation and cross-correlation energy of particles appears as a result of interaction ("hidden" interference) of the particles describing by the different approximate mathematical methods in the non-operator or operator form, which are inconsistent with the basic principle [a particle interferes (correlates) only with itself] of the canonical quantum mechanics and quantum field theory. The {\it{interpretations}} of the mathematical approximations are sometimes incorrectly considered as the physical interpretations of the "modified" quantum mechanics and particle field theory of the Bose-Einstein condensation, superfluidity, superconductivity, supermagnetism, super-radiation, Bosenova effect, and quantum anomalous and fractional Hall effects. In the frame of such an {\it{approach}}, the coherent quantum phenomena are described by the {\it{phase-independent Hamiltonians}}. The {\it{physical mechanism}} behind the "hidden" interference and cross-correlation energy of particles {\it{is mediated by the different mathematical approximations}}, which do not compare well with the Copenhagen-Dirac postulate of "interference-less", self interfering particles. Thus the coherent quantum phenomena may be attributed formally even to the mathematical errors mediated by the mathematical approximations. 

A particle of the canonical quantum mechanics associated with the wave of probability is free from the interference (cross-correlation) with other particles. The above-discussed mathematical approach, which does not compare well with  the basic principle of canonical quantum mechanics and quantum field theory (the Copenhagen-Dirac postulate of "interference-less" particles"), has been extremely useful for the construction ("mathematical engineering") of the models that {\it{successfully}} describe and explain the {\it{coherent quantum phenomena}}, such as the Bose-Einstein condensation, superfluidity, superconductivity, supermagnetism, super-radiation, Bosenova effect, and quantum anomalous and fractional Hall effects. Therefore this approximate mathematical approach was {\it{generally accepted}} in the past. It is very {\it{effectively}} used in the quantum mechanics and quantum field theory up to now. Although the mathematical cross-correlation ("hidden" interference) between particles and the respective cross-correlation energy are closely connected with the mathematical approximations, these theories rely on a set of approximations specific to the particular physical problem. The "hidden" interference between particles and the cross-correlation energy have appeared in the quantum mechanics and particle field theory already at the very beginning of consideration of the physical problems pertaining to the properties of quantum systems made of a large number of particles (many-body problem). The "hidden" interference and cross-correlation between particles do appear in the "modified" quantum mechanics and quantum field theory in the form of the interaction (cross-correlation) energy}}. It is impossible in the present study to analyze a myriad of models of such quantum theories. The conventional approach of the {\it{mathematical insertion}} of the "hidden" interference, cross-correlation energy, Bose-Einstein statistics, Fermi-Dirac statistics and Pauli exclusion principle into the canonical quantum mechanics and quantum field theory, which strictly forbid existence of the interference between particles and the respective cross-correlation energy, is illustrated by using the following traditional (typical) analysis of a many-particle quantum system. 

As long as the number of particles of a non-relativistic quantum system is fixed the system can be described by a wave-function, which contains all the information about the state of that system. This is the so-called first quantization approach in the traditional analysis of a many-particle quantum system. The interaction-free (interference-less) particles are described by the Schr{\"o}dinger equation of the canonical quantum mechanics of a single particle. The wave-functions $\psi_n (\mathbf{r})$ and $\psi_m (\mathbf{r})$ of the interaction-free particles must not correlate with each other in the points $\mathbf{r}$: 
\begin{eqnarray}  \label{eq166}
\int \psi^*_n(\mathbf{r}) \psi _m (\mathbf{r}) d^3x \equiv  \langle \psi _n (\mathbf{r})| \psi _m (\mathbf{r})\rangle=\delta _{nm},
\end{eqnarray}
where $\delta _{nm}$ is the Kronecker delta. Notice, the value (\ref{eq166}) providing the interference-less behavior of particles is equal to zero if the functions do not overlap spatially with each other or if they are orthogonal in the Hilbert space. The de Broglie wave [wave-function $\psi_n (\mathbf{r})$] associated with the $n$-th particle has a probabilistic interpretation in position space. The values $\psi^*_n(\mathbf{r})\psi_n (\mathbf{r})$ and $\int \psi^*_n(\mathbf{r})\psi_n (\mathbf{r}) d^3r=N=1$ are {\it{interpreted}} respectively as the probability (particle) density and the total number ($N$=1) of particles for the $n$-th particle. For the sake of simplicity, consider a system composed from two identical particles. In the case of the multi-particle ($N>2$) quantum system, the consideration is the same as the analysis of two particles. The non-relativistic energy ${\cal H}$ of the interaction-free particles, which is a superposition of the non-relativistic canonical Hamiltonians of the single particles, is given by the Hamiltonian 
\begin{eqnarray}  \label{eq167}
{\cal H}= {\cal H}_1+{\cal H}_2=\langle \psi _1 (\mathbf{r}_1)| ({\hat p}_1^2/2m_1) |\psi _1 (\mathbf{r}_1)\rangle + \langle \psi _2 (\mathbf{r}_2)| ({\hat p}_2^2/2m_2) |\psi _2 (\mathbf{r}_2) \rangle .
\end{eqnarray}
Notice, the Hamiltonian (\ref{eq167}) of the interaction-less particles is invariant under the $U(1)$ local gauge transformation ($\psi _{n} \rightarrow \psi ' _{n}=e^{i\alpha _n} \psi _{n}$) giving rise to the interference-less behavior of particles. According to quantum theory, the particles do not possess definite positions during the periods between measurements. Instead, they are governed by wave-functions that give the probability of finding a particle at each position. As time passes, the wave-functions tend to spread out and overlap. Once this happens, it becomes impossible to determine (distinguish), in a subsequent measurement, which of the particle positions correspond to those measured earlier. Intuitively, the quantum state [trial wave-function $ \Psi (\mathbf{r}_1, \mathbf{r}_2)$] of the system should be 
\begin{eqnarray}  \label{eq168}
\Psi (\mathbf{r}_1, \mathbf{r}_2)=\psi _1 (\mathbf{r}_1)\psi _2 (\mathbf{r}_2)
\end{eqnarray}
or 
\begin{eqnarray}  \label{eq169}
\Psi (\mathbf{r}_1, \mathbf{r}_2)=\psi _1 (\mathbf{r}_2)\psi _2 (\mathbf{r}_1),
\end{eqnarray}
for instance, see the pioneer studies \cite{Dick} and \cite{Heit}. The engineering of the trial wave-function is simply the canonical way of constructing a basis for a tensor product space from the individual spaces. We can use the functions (\ref{eq168}) and (\ref{eq169}) to form symmetric and antisymmetric functions of the system. The need for symmetric (bosonic) or antisymmetric (fermionic) states and functions is sometimes {\it{incorrectly}} regarded as an empirical fact. It turns out, for the mathematical reasons ultimately based on quantum field theory, that we must have the exchange symmetric states
\begin{eqnarray}  \label{eq170}
\Psi (\mathbf{r}_1, \mathbf{r}_2)=\psi _1 (\mathbf{r}_1)\psi _2 (\mathbf{r}_2) \pm   \psi _1 (\mathbf{r}_2)\psi _2 (\mathbf{r}_1), 
\end{eqnarray}
which includes the exchange of coordinates $\mathbf{r}_1$ and $\mathbf{r}_2$ (not virtual particles). The states, where this is a sum, are known as symmetric; the states involving the difference are called antisymmetric. If $\psi_1 (\mathbf{r})$ and $\psi _2(\mathbf{r})$ are the same [$\psi_1 (\mathbf{r}_1)=\psi_2 (\mathbf{r}_1)$ and $\psi_1 (\mathbf{r}_2)=\psi_2 (\mathbf{r}_2)$], the antisymmetric expression gives zero. The same result is obtained if $\psi_1 (\mathbf{r}_1)=\psi_1 (\mathbf{r}_2)$ and $\psi_2 (\mathbf{r}_2)=\psi_2 (\mathbf{r}_1)$. In other words, in an antisymmetric state, two identical particles cannot occupy the same single-particle states. That is sometimes {\it{interpreted}} as the Pauli exclusion principle, which describes disappearance (annihilation) of both the wave-function [$\Psi (\mathbf{r}_1, \mathbf{r}_2)=0$] and the object composed from the two identical particles. Note, in this regard, that Wolfgang Pauli has never considered the exclusion principle as the above-described annihilation (disappearance) of the identical particles. The question "Why are the particles identical?" is usually {\it{interpreted}} as a wrong one, because the question arises from mistakenly regarding individual particles as fundamental objects, when in fact it is only the particle (quantum) field is fundamental. The trial Hamiltonian of the system describing by the mathematically constructed functions (\ref{eq168}), (\ref{eq169}) or (\ref{eq170}) can be written by using {\it{the trial wave-function $\Psi (\mathbf{r}_1, \mathbf{r}_2)$}}, for instance, in the form 
\begin{eqnarray}  \label{eq171}
{\cal H}= {\cal H}_1+{\cal H}_2 + [{\cal H}_{12}+{\cal H}_{21}] =\langle \Psi (\mathbf{r}_1, \mathbf{r}_2)|( {\hat p}_1^2/2m_1 +  {\hat p}_1^2/2m_1) |\Psi (\mathbf{r}_1, \mathbf{r}_2)\rangle + \langle \Psi (\mathbf{r}_1, \mathbf{r}_2)|U(|\mathbf{r}_1-\mathbf{r}_2|)|\Psi (\mathbf{r}_1, \mathbf{r}_2)\rangle,
\end{eqnarray}
which is different from the Hamiltonian (\ref{eq167}) of the interaction-less particles. Therefore, the new term ${\cal H}_{12}+{\cal H}_{21} = \langle \Psi (\mathbf{r}_1, \mathbf{r}_2)|U(|\mathbf{r}_1-\mathbf{r}_2|)|\Psi (\mathbf{r}_1, \mathbf{r}_2)\rangle$ is usually {\it{interpreted}} as the potential energy of the interaction of particles with each other. Notice, the interaction term in the form $\langle \Psi (\mathbf{r}_1, \mathbf{r}_2)|U(|\mathbf{r}_1-\mathbf{r}_2|)|\Psi (\mathbf{r}_1, \mathbf{r}_2)\rangle$ can include also the potential energy associated with the interaction of particles with the external macroscopic field [for instance, see Eqs. (\ref{eq139})-(\ref{eq141})]. Although the wave-functions $\psi_1 (\mathbf{r})$ and $\psi _2(\mathbf{r})$ do not cross-correlate (interfere) with each other, in agreement with the canonical quantum mechanics [see, Eqs. (\ref{eq166}) and (\ref{eq167})], the "hidden" interference, cross-correlation and the cross-correlation energy ${\cal H}_{12}+{\cal H}_{21}$ do appear in the "modified" quantum mechanics in the form of the interaction integrals  
\begin{eqnarray}  \label{eq172}
{\cal H}_{12}+{\cal H}_{21}  \sim \int \int \psi^*_1(\mathbf{r}_1)\psi^*_2(\mathbf{r}_2)U(|\mathbf{r}_1-\mathbf{r}_2|)\psi_1(\mathbf{r}_1)\psi_2(\mathbf{r}_2) d^3x_1 d^3x_2,
\end{eqnarray}
or
\begin{eqnarray}  \label{eq173}
{\cal H}_{12}+{\cal H}_{21}  \sim \int \int \psi^*_1(\mathbf{r}_2)\psi^*_2(\mathbf{r}_1)U(|\mathbf{r}_1-\mathbf{r}_2|)\psi_1(\mathbf{r}_2)\psi_2(\mathbf{r}_1) d^3x_1 d^3x_2,
\end{eqnarray}
in the case of the trial wave-functions (\ref{eq168}) or (\ref{eq169}), respectively. The more complicated trial wave-function (\ref{eq170}) yields the interaction energy that includes the terms  (\ref{eq172}) and (\ref{eq173}) and the Heisenberg-Dirac exchange integrals
\begin{eqnarray}  \label{eq174}
 \sim \int \int \psi^*_1(\mathbf{r}_1)\psi^*_2(\mathbf{r}_2)U(|\mathbf{r}_1-\mathbf{r}_2|)\psi_1(\mathbf{r}_2)\psi_2(\mathbf{r}_1) d^3x_1 d^3x_2
\end{eqnarray} 
and 
\begin{eqnarray}  \label{eq175}
\sim \int \int \psi^*_1(\mathbf{r}_2)\psi^*_2(\mathbf{r}_1)U(|\mathbf{r}_1-\mathbf{r}_2|)\psi_1(\mathbf{r}_1)\psi_2(\mathbf{r}_2) d^3x_1 d^3x_2.
\end{eqnarray} 
Notice, the object describing by the antisymmetric function (state) (\ref{eq170}) composed from the identical ($\psi_1 =\psi _2$) particles does annihilate, because the wave-function $\Psi$ and the total energy (\ref{eq171}) do vanish. Although this phenomenon is usually interpreted as the Pauly exclusion principle, the Pauly principle was originally formulated without involving annihilation of the identical particles. According to the usual {\it{interpretations}} of the "modified" quantum mechanics, the coordinate-exchange integrals describe the quantum exchange of particles associated with the indistinguishableness of the identical particles. Although the trial Hamiltonian (\ref{eq171}) is invariant under the $U(1)$ local gauge transformation ($\psi _{n} \rightarrow \psi ' _{n}=e^{i\alpha _n} \psi _{n}$), the mathematical cross-correlation ("hidden" interference) between particles and the respective cross-correlation energy, which do not depend on the particle phases $\alpha _n$, are provided by the interaction integrals of this Hamiltonian. The interference and cross-correlation of the fields ${\psi}_1 (\mathbf{r})$ and ${\psi}_2  (\mathbf{r})$ in the trial Hamiltonian  (\ref{eq171}) are especially transparent in the case of $U(|\mathbf{r}_1-\mathbf{r}_2|)=(g/2){\delta (|\mathbf{r}_1-\mathbf{r}_2|})$, where $g$ is the coupling parameter that represents the inter-particle interaction (interference), for instance, in the Gross-Pitaevski model of the Bose-Einstein condensate [see, Eqs. (\ref{eq141}) and (\ref{eq142})]. In such a case, the interference and cross-correlation between particles do appear in the "modified" quantum mechanics in the form of the cross-correlation (interaction) energy ${\cal H}_{12}+{\cal H}_{21}  \sim \int [\psi^*_1(\mathbf{r})\psi_2(\mathbf{r})][\psi^*_2(\mathbf{r})\psi_1(\mathbf{r})] d^3x \neq 0$ with the second-order cross-correlation in the same ($\mathbf{r}_1 = \mathbf{r}_2 = \mathbf{r}$)  points $\mathbf{r}_1$ and $\mathbf{r}_2$. One could compare this expression with Eq. (\ref{eq166}), which describes the interference-free and correlation-free particles. The value (\ref{eq166}) providing the interference-less behaviour of particles does vanish if the functions do not overlap spatially with each other or if they are orthogonal in the Hilbert space. If the interaction potential is constant (for instance, $U(|\mathbf{r}_1-\mathbf{r}_2|)=1$), then the interference and cross-correlation energy do appear in the "modified" quantum mechanics in the form of the cross-correlation integrals (\ref{eq172}), (\ref{eq173}), (\ref{eq174}) and (\ref{eq175}) with $U(|\mathbf{r}_1-\mathbf{r}_2|)=1$, which contain the second-order cross-correlation (interference) in the different ($\mathbf{r}_1 \neq \mathbf{r}_2$) points $\mathbf{r}_1$ and $\mathbf{r}_2$, but do not contain directly the potential value. Thus in the above-described "modified" quantum mechanics, which does not compare well with the basic quantum mechanical principle [a particle interferes (correlates) only with itself], the interference and cross-correlation of particles do appear in the second-order of interference (cross-correlation) in the same and/or different space points. The $U(1)$ gauge symmetry of the trial Hamiltonian (\ref{eq171}) provides the very particular mathematical cross-correlation ("hidden" interference) and the respective cross-correlation energy, which do not depend on the phases of the probability waves of particles. The mathematical "engineering" of the symmetric or antisymmetric state [the trial mathematical wave-function associated with the waves of probability of the identical (indistinguishable) particles] yields automatically  the Bose-Einstein statistics or the Fermi-Dirac correlations (statistics) and the Pauli exclusion principle, whose {\it{interpretations}} are mathematical ones. The trial mathematical wave-function of two identical particles is symmetric or antisymmetric with respect to the permutation of the two particles, depending whether one considers bosons or fermions. In a particular case of the non-identical particles there is no permutation symmetry. Consequently there should be no Bose-Einstein or Fermi-Dirac correlation between these particles. In the trial wave-function formalism, the difference between particles could be attributed only to the difference in mass and/or charge. The difference in the spin of bosons or fermions results respectively into the Bose-Einstein or Fermi-Dirac cross-correlations (statistics). Note again that the {\it{interpretations}} of the model are also based on interpretations of the mathematical approximations, which are inconsistent with the basic principle [a particle interferes (correlates) only with itself] of canonical quantum mechanics. It should be also noted that in condensed matter physics, the quantum states with ill-defined particle numbers are particularly important for describing the various superfluids. Many of the defining characteristics of a superfluid arise from the notion that its quantum state is a superposition of states (wave-functions) with different particle numbers $N$. For instance, the concept of a coherent quantum state used to model the laser and the BCS ground state refers to a state with an ill-defined particle number but a well-defined phase. In the present model, the cross-correlation (interaction) integral $\langle \Psi (\mathbf{r}_1, \mathbf{r}_2)|U(|\mathbf{r}_1-\mathbf{r}_2|)|\Psi (\mathbf{r}_1, \mathbf{r}_2)\rangle$ is associated with the cross-correlation energy ${\cal H}_{12}+{\cal H}_{21}$ induced by the interference (cross-correlation) of the unit-fields $\psi_1 (\mathbf{r})$ and $\psi _2(\mathbf{r})$ of the real physical matter [see, Secs. (2)-(5)]. The physical mechanism behind the Bose-Einstein statistics, Fermi-Dirac statistics and Pauli exclusion principle is attributed to the attractive and repulsive forces associating with the subtractive and additive interference (cross-correlation) of the real unit-waves (bosons or fermions) of the physical matter. The additive and subtractive interference (the repulsive and attractive forces) correspond to the fermion and boson unit-waves, respectively. The time-averaged cross-correlation energy (\ref{eq136}) of the incoherent unit-fields having random phases is equal to zero due to the non-correlation of such fields. The increase of the degree of coherence (phase correlation) increases the time-averaged cross-correlation energy of the partially coherent unit-fields. The cross-correlation energy (\ref{eq136}) averaged over the time $\Delta t >> 1/ \Delta \varepsilon_{0nm}$ does vanish if the {\it{coherent}} unit-fields (particles) with the constant phases $\alpha _n$ and $\alpha _m$ are {\it{non-identical}} ($\mathbf {k}_{0n} \neq \mathbf {k}_{0m}$ and/or $m_{0n} \neq m_{0m}$, respectively $\varepsilon_{0n} \neq \varepsilon_{0m}$). The property gives the physical explanation of the identity or non-identity of the unit-fields (particles). The coherency of the identical unit-fields (particles) is the necessary condition of existence of the cross-correlation energy. 

Consider now the above-presented mathematical construction ("engineering") of the multi-particle quantum mechanical system also from a point of view of quantum field theory. Particle field theory deals with quantum systems where particles and antiparticles are created and destroyed. The description of the systems with the non fixed number of the particles demands a more general theoretical approach called second quantization. In quantum field theory, unlike in quantum mechanics, position is not an observable, and thus, one does not need the concept of a position-space probability density. The wave-function $\psi_n (\mathbf{r})$ does not have a probabilistic interpretation in position space. The quantum mechanical system is considered as a quantum field. The field elementary degrees of freedom are the occupation numbers, and each occupation number is indexed by a number $n$, indicating which of the single-particle states $\psi _1, \psi _2 , ...,\psi _n ,...,\psi _N $ it refers to. In the above-described "modified" quantum mechanics, which does not compare well with the Copenhagen-Dirac quantummechanical principles, the trial Hamiltonian describing the multi-particle system in space representation is given by 
\begin{eqnarray}  \label{eq176}
{\cal H}=(-1/2m)\sum_{n=1}^N \int \psi^*_n (\mathbf{r}_n){\nabla}_n^2\psi_n (\mathbf{r}_n)d^3x_n + \sum_{n \neq m}^{N^2-N}\int \int \psi^*_n (\mathbf{r}_n)\psi^*_m (\mathbf{r}_m) U(|\mathbf{r}_n-\mathbf{r}_m|)\psi_n (\mathbf{r}_n) \psi_m (\mathbf{r}_m) d^3x_n d^3x_m,
\end{eqnarray}
where the indexes $n$ and $m$ run over all particles [also, see Eqs. (\ref{eq139}) - (\ref{eq141}) and (\ref{eq171})]. In the quantum field theory, the respective trial Hamiltonian is given by 
\begin{eqnarray}  \label{eq177}
{\cal H}=(-1/2m) \int \Psi^* (\mathbf{r}) {\nabla}^2 \Psi (\mathbf{r})d^3x  + \int \int \Psi^* (\mathbf{r}) \Psi^* ({\mathbf{r}}') U (|{\mathbf{r}}-{\mathbf{r}}'|) \Psi (\mathbf{r})\Psi ({\mathbf{r}}') d^3xd^3x',
\end{eqnarray}
which looks like an expression for the expectation value of the energy, with $ \Psi $ playing the role of the trial wave-function. The properties of this field are explored {\it{mathematically}} by defining creation ($ {\hat a}^{\dagger}$) and annihilation ($\hat a$) quantum operators with the commutation relations imposed, which add and subtract particles. This is the second quantization approach. The field is considered as a set of degrees of freedom indexed by position, where the second quantization indexes the field by enumerating the single-particle quantum states. The second quantization procedure relies on the particles being identical. It is impossible to construct mathematically a quantum field theory from a distinguishable many-particle system, because there would be no way of separating and indexing the degrees of freedom. From the point of view of the above-described quantum field model, the particles are identical if and only if they are excitations of the same quantum field. Such an interpretation is the mathematical interpretation of the trial mathematical object, which contradicts the basic principles of canonical quantum mechanics. Remember, the question "Why are the particles identical in quantum mechanics?" is usually incorrectly {\it{interpreted}} as a wrong one, because the question arises from mistakenly regarding individual particles as fundamental objects, when in fact it is only the field (trial mathematical object constructed by using the different mathematical approximations) that is fundamental. Mathematically, the second quantization of the field is performed by placing the trial fields $\Psi$ and $\Psi ^*$ into the infinite resonator and subsequent replacement of these fields by the respective multimode field operators $\hat \Psi$ and $\hat \Psi ^{\dagger }$. Thus the trial Hamiltonian (\ref{eq177}) of the trial quantum field can be rewritten in terms of the trial field of operators by using the creation and annihilation operators. In such a case, the "hidden" inter-particle interference and the respective interaction energy are provided rather by the trial cross-correlating operators (mathematical objects) with the trial (Bogoliubov-like) commutator relations for the particles and antiparticles [the observable bosons and fermions or the virtual gauge bosons associated with the gauge field $U (|{\mathbf{r}}-{\mathbf{r}}'|)$ and the gauge symmetry breaking] than by the interference of the waves of probabilities associated with the single-particle functions $\psi_n (\mathbf{r}_n)$ and $\psi _m(\mathbf{r}_m)$. That is to say that the mathematical cross-correlation between particles and the respective cross-correlation energy are mediated by the mathematical cross-correlation between the operators in the interaction integrals of the trial Hamiltonian operator. The trial Hamiltonian (\ref{eq177}) and the respective Hamiltonian operator are invariant under the $U(1)$ local gauge transformation given rise to the very particular mathematical cross-correlation ("hidden" interference) and the respective energy, which do not depend on the particle phases. The mathematical construction of Hamiltonians for bosons and fermions by using the respective trial commutator relations automatically yields the particle interference, cross-correlation (interaction) energy, particle number, Bose-Einstein statistics, Fermi-Dirac correlations (statistics) and Pauli exclusion principle. If one considers the possibility that non-identical particles are virtually related in the sense that they can annihilate and transform into identical particles, there should be a new kind of the Bose-Einstein or Fermi-Dirac correlations (statistics). Note again that the {\it{interpretations}} of the above-presented quantum-field model are based on interpretations of the mathematical approximations, which do not compare well with the basic principle [a particle interferes (correlates) only with itself] of canonical quantum mechanics and quantum field theory. Unfortunately, in the frame of such an approach, some "physical" {\it{interpretations}} may be attributed formally even to the interpretations of the pure mathematical errors mediated by the mathematical approximations. In the present model, the cross-correlation terms in Eqs. (\ref{eq176}) and (\ref{eq177}) are considered as the cross-correlation energies mediated by the interference and cross-correlation between the particles (unit-fields) of the physical matter [see, Secs. (2)-(5)]. The physical mechanism behind the Bose-Einstein statistics, Fermi-Dirac statistics and Pauli exclusion principle is attributed to the attractive and repulsive forces associating with the subtractive and additive interference (cross-correlation) of the unit-waves (bosons or fermions) of the real physical substance (mass-energy). The additive and subtractive interference (the repulsive and attractive forces) correspond to the identical fermion and boson unit-waves, respectively. The interference of non-identical particles or antiparticles (unit-waves) should result into a new kind of the Bose-Einstein or Fermi-Dirac correlations (statistics), namely the correlations between non-identical particles (unit-waves). Notice, the cross-correlation energy (\ref{eq136}) averaged over the time $\Delta t >> 1/\Delta \varepsilon_{0nm}$ does vanish if the {\it{coherent}} unit-fields (particles) with the constant phases $\alpha _n$ and $\alpha _m$ are {\it{non-identical}} ($\mathbf {k}_{0n} \neq \mathbf {k}_{0m}$ and/or $m_{0n} \neq m_{0m}$, respectively $\varepsilon_{0n} \neq \varepsilon_{0m}$).

It should be stressed that the "modified" quantum mechanics and the particle field theory based on the "hidden" interference between particles and the respective cross-correlation energy inserting into the models by the different powerful mathematical approximations have been constructed to give absolutely {\it{correct description and explanation}} of the basic coherent quantum phenomena, such as the particle interference (cross-correlation), cross-correlation energy, interaction, particle number, Bose-Einstein statistics, Fermi-Dirac statistics and Pauli exclusion principle. Naturally, these theories give also the correct description and interpretation of the extraordinary coherent physical phenomena, such as the Bose-Einstein condensation, super-radiation, Bosenova effect, superfluidity, superconductivity, supermagnetism, and quantum anomalous and fractional Hall effects. The Hamiltonians of the models are invariant under the $U(1)$ local gauge transformation providing the very particular, in comparison to the ordinary interference, mathematical cross-correlation ("{\it{hidden" interference}}) and the respective cross-correlation energy, which do not depend on the particle phases. Such a mathematical approach ("mathematical engineering"), which has been generally accepted in the very past, is effectively used in the quantum mechanics and quantum field theory up to now. The present model could be considered as a generalization of the "modified" quantum mechanics and particle field theory by taking into account rather the {\it{real, direct, physical interference}} between {\it{material}} unit-fields (particles) and the interference-induced energy than the {\it{"hidden", approximation-induced mathematical interference}} between the {\it{non-material}} waves of probabilities or operators associated with material point-like particles and the respective cross-correlation (interaction) energy. In the frame of the present model, the hypothetical "hidden" variable of Einstein, Podolsky and Rosen \cite{Ein}, which according to the authors should be added to the canonical quantum mechanics to avoid its indeterminism and superluminal signalling, cold be attributed to the "{\it{non-hidden}}", {\it{physical}} interference between {\it{material unit-fields}}.

\subsection{7. SUMMARY AND CONCLUSIONS}

In Part I of the present study, the interference between material unit-fields (particles) and the interference-induced positive and negative cross-correlation energies, which do not exist from the point of view of both the Newton mechanics, Einstein relativity, canonical quantum mechanics and particle field theory, have been investigated. In order to make the results understandable not only to specialists in the research field, the model concepts where analyzed and reanalyzed in the context of each section of Secs. (1)-(7). In the model, for both the classical and quantum fields, the two-times increase of the wave amplitude does increase the wave energy in four times, and the wave with zero amplitude has zero energy. The problem of nonconservation of the energy and number of particles by the cross-correlation was overcame by taking into consideration the fact that creation of the conditions of the pure additive or subtractive interference of the waves (fields) requires "additional" energy that must be added or subtracted from the physical system. Then the "additional" energy does provide conservation of the total energy of the system. The Hamiltonians that describe the energy inducing by the cross-correlation (interference) in the basic classical and quantum fields have been derived. The conditions of pure constructive or destructive interference were found by using these Hamiltonians. The influence of the cross-correlation energy on the basic physical properties of boson and fermion fields was demonstrated. The energy, mass, charge, and momentum of the interfering fields were calculated. The calculated energy ${\varepsilon}_0=\omega _0=(  {\mathbf  k}_0^2   + m_0^2)^{1/2}$ of a unit-field (particle) is equal to the Planck-Einstein energy of the particle, but is different from the value ${\varepsilon}_0=\omega _0+ (1/2)\omega _0$ of traditional quantum field theory. The positive energy-mass ${\varepsilon}_0=({\mathbf{k}}_0^2+m_0^2)^{1/2}$ of antiparticles in the present model is different from the negative energy-mass of the Dirac antiparticles in the Dirac model, which is based on the Dirac interpretation of quantum interference and the particle (antiparticle) energy-mass relation ${\varepsilon}_0=\pm ({\mathbf{k}}_0^2+m_0^2)^{1/2}$. The calculated vacuum energy, in agreement with the Einstein and classical physics of the empty space, is equal to zero. Here, one should not confuse the true vacuum of the Einstein or Newton model with the physical vacuum of the present model occupied by the interfering {\it{finite}} and {\it{infinite}} unite-fields.  The Copenhagen interpretation (philosophy) of the de Broglie wave associated with a particle as the wave of probability presents a more or less intuitively transparent background for the physical interpretation of quantum mechanics. In particle field theory, up to now, it is not completely clear how to interpret physically the wave (field) of operators. In the present model, the unit-wave associated with a boson or fermion particle, unlike the wave of probability or operators in quantum mechanics and particle field theory, is a real, finite unit-wave (unit-field) of the boson or fermion matter (mass-energy), whose curvature (gradient) can be changed spatially and/or temporally. The physical mechanism behind the position-momentum and time-energy uncertainties is attributed to the increase of the spatial ($\nabla {\psi}_0$) and/or temporal (${\partial {\psi }_0}/{\partial t}$) curvatures (gradients) of a real, material  unit-field under the spatial or temporal localization of the unit-field by the interaction (interference) with the unit-fields (particles ) of other microscopical or macroscopical objects. In order to account for the well-known discrepancies between measurements based on the mass of the visible matter in astronomy and cosmology and definitions of the energy (mass) made through dynamical or general relativistic means, the present model does not need in hypothesizing the existence of "dark" energy-mass. In the present model, the "dark" cosmological energy as well as the well-known spiral cosmological structures are associated with the cross-correlation energy of the moving cosmological objects. According to the present model, the Bell superluminal signals~\cite{Bell} (the Einstein-Podolsky-Rosen paradox) and the well-known superluminal objects in astronomy, if they really propagate with velocities greater than the velocity of light, involve no physics incompatible with the theory of special relativity. The superluminar velocities and the superluminar "quantum leaps" of any kind are attributed rather to the physical properties of the unit-field like material mediums than the empty space (vacuum) of the Einstein special or general relativity. It has been also shown that the positive or negative gradient of the cross-correlation energy does mediate the attractive or repulsive forces, respectively. These forces could be attributed to all known classical and quantum fields (interactions), for instance, to the gravitational and Coulomb fields. In the present model, the physical mechanism behind the Bose-Einstein statistics, Fermi-Dirac statistics and Pauli exclusion principle is attributed to the attractive and repulsive forces associating with the subtractive and additive interference (cross-correlation) of the unit-waves (bosons or fermions) of the real physical matter. If the interference of unit-fields is neither pure constructive nor pure destructive (the interaction is not repulsive or attractive), then the respective statistics are different from the Fermi-Dirac or Bose-Einstein statistics. Such parastatistics could be associated, for instance, with the so-called fractional and braid statistics. To this end, the model showed a key role of the cross-correlation energy in several basic physical phenomena, such as Bose-Einstein condensation, super-radiation, Bosenova effect, superfluidity, superconductivity, supermagnetism, and quantum anomalous and fractional Hall effects. 

{\it{In the "modified" quantum mechanics and particle field theories, the mathematical cross-correlation ("hidden" interference) is associated with the pure mathematical approximations, which do not compare well with the basic physical principle (the Copenhagen-Dirac postulate of interference-less particles) of the canonical versions of  these theories. Physically, the "hidden" interference is interpreted as the inter-particle interaction associated with the interaction energy. The present model could be considered as a generalization of the "modified" quantum mechanics and particle field theory. The generalization is performed by taking into account rather the real, direct, physical interference between material unit.fields and the interference-induced energy than the "hidden", mathematical interference between non-material waves of probabilities or operators associated with the material point-like particles and the respective cross-correlation energy.}} The general properties of the forces mediated by the gradients of cross-correlation energy under the interference between particles (unit-fields) have been illustrated by the numerical solutions for the pairs of scalar boson unit-fields. One can easily follow the model for an arbitrary number of the boson or fermion unit-fields (for an example, for the particles or antiparticles of the Standard Model of Particle Physics) with the arbitrary values of the rest masses $m_{0n}$, charges $q_{0n}$ and spins $s_{0n}$ of the unit-fields, as well as for the composite fields composed from both the boson and fermion unit-fields. The concept of the interference between particles and the interference-induced energy may be easily extended to other quantum field theories. For instance, if one assumes that quantum particles are not zero-dimensional objects of the traditional quantum mechanics or particle field theories, but rather one-dimensional strings of the String Theory that can move and vibrate, giving the observed particles the physical properties, the cross-correlation and the cross-correlation energy could be attributed to the interference of the vibrations in the overlapped strings. It should be stressed in this regard that the unit-fields of the present model are not the point particles or strings. The unit-fields have simultaneously the properties of the point particles, strings and 3-dimensional (in space) waves. A quasi-unified description of the classical and quantum fields and interactions, which does not include gravity, is currently possible only in terms of the gauge symmetry breaking in the Standard Model of Particle Physics (SM). The present study (Part I) has developed the theoretical background for unified description of the all-known classical and quantum fields and interactions. The background is provided in the extremely trivial (in comparison to SM) manner, namely in terms of the interference, cross-correlation and cross-correlation energy of the 3-dimensional unit-fields. The next study (Part II) unifies the all-known classical and quantum fields, particles and interactions (including gravitation) by using this theoretical background. The unification is performed by the direct generalization of the Einstein relativistic energy-mass relation  ${\varepsilon}^{2}={\mathbf{k}}^2+m^2$ for the interfering particles (bodies). The unit-fields have simultaneously properties of the point particles and waves of quantum mechanics, the point particles of SM, the strings of theory of strings and the 3-dimensional waves of theory of classical fields. Therefore the unification of fields and interactions is considered as a generalization of the classical and quantum field theories, SM and string theory. 

\section*{Acknowledgments}
The study was supported in part by the Framework for European Cooperation in the field of Scientific and Technical Research (COST, Contract No MP0601) and the Hungarian Research and Development Program (KPI, Contract GVOP 0066-3.2.1.-2004-04-0166/3.0). 




{\begin{center} \bf \large ENERGY MEDIATED BY INTERFERENCE OF PARTICLES (Parts I-IV): The Way to Unified Classical and Quantum Fields and Interactions \\ 
\vspace{0.2cm}
{\it{Part II. Unification of Classical and Quantum Fields and Interactions}} \end{center}}


{\begin{center} S. V. Kukhlevsky \end{center}} 
{\begin{center}{\it Department of Physics, Faculty of Natural Sciences,\\ University of Pecs, Ifjusag u. 6, H-7624 Pecs, Hungary} \end{center}}


\begin{quote}
\small A quasi-unified description of the classical and quantum fields, which defines the electromagnetic, weak and strong interactions, but does not include gravity, is currently possible only in the frame of the Standard Model of Particle Physics (SM) in terms of the gauge symmetry breaking. Part I of the present study has developed the theoretical background for unified description of the all-known classical and quantum fields in terms of the interference between particles and the respective cross-correlation energy, which do not exist from the point of view of quantum mechanics and SM. Part II uses this background for unification of the electromagnetic, weak, strong and gravitational fields and interactions. The unification is performed by generalization of the Einstein energy-mass relation ${\varepsilon}^{2} ={\mathbf{k}}^2+m^2$ for the interfering unit-fields associated with the interacting particles. The unit-fields obey properties of the point particles and waves of quantum mechanics, the point particles of SM, the strings of theories of strings and the 3-D waves of theories of classical fields. 
\end{quote}



%
%
\section{1. Introduction}
A generally accepted quasi-unified description of the classical and quantum fields, which defines the electromagnetic, weak, and strong interactions, but does not include gravity, is currently provided by the Standard Model of Particle Physics (SM) in terms of the gauge symmetry relations and the gauge symmetry breaking. SM is based on the three independent gauge interactions, symmetries and coupling constants, which do associate with the electromagnetic, weak and strong interactions of particles. There are several similar candidate models (so-called Grand Unified Theories) in which at high energy, the three gauge interactions of SM are merged into one single interaction characterized by one larger gauge symmetry and thus one unified coupling constant. Some models, which exhibit similar properties, do not unify all interactions using one simple Lie group as the gauge symmetry, but do so using semi-simple groups or other super-symmetries. Grand Unified Theories are considered as an intermediate step towards the so-called a Theory of Everything that would unify gravity with the electromagnetic, weak, and strong interactions. For an example, the String Theory is currently considered as a candidate for a Theory of Everything. For the detailed description of the aforementioned generally-accepted models and candidate theories see the canonical studies and traditional textbooks, as well as any modern paper, review paper or textbook published in huge amount in the literature devoted to the quantum fields and interactions (for instance, see Refs. \cite{1Planck,1Einst1,1Bohr,1Einst2,1Noeth,1Brog,1Born,1Schr,1Ferm,1Heis1,1Dir1,1Wign,1Weyl,1Heis2,1Amb,1Paul,1Dir2,1Tomo,1Beth,1Schw,1Feyn,1Dyso,1Yang,1Namb,1Gold,1Glash,1Engl,1Wein1,1Sal,1Higg,1Land,1Jack,1Bere,1Itz,1Ryde,1Pes,1Grei,1Wein} and references therein). 

Part I of the present study has developed the theoretical background for a unified description of the classical and quantum fields and interactions in terms of the interference between elementary particles (indivisible unit-fields) and the respective cross-correlation energy, which do not exist from the point of view of the canonical quantum mechanics, quantum field theories, SM and string theories. The Hamiltonians that describe the cross-correlation energy mediating by interference of the basic classical and quantum fields composed from the 3-dimensional (in space) unit-fields of real physical matter (mass-energy) have been derived by the generalization of the traditional Hamiltonians of the classical and quantum field theories for the superposition of interfering unit-fields associated with the interacting particles. It has been shown that the gradient of the cross-correlation energy induced by the interference between particles (unit-fields) mediates the attractive or repulsive forces, which could be attributed to all known classical and quantum fields (for details, see Part I). Part II of the present study uses this theoretical background for unified description of the fundamental (electromagnetic, weak, strong and gravitational) fields and interactions.  However, unlike in Part I, the unification is performed rather by the generalization of the basic (energy-mass) relation of the Einstein special relativity than the traditional Hamiltonians of the classical and quantum field theories. The model unifies the all-known fields, particles and interactions by the straightforward generalization of the Einstein relativistic energy-mass relation ${\varepsilon}^{2} ={\mathbf{k}}^2+m^2$ for the interacting particles and bodies, which are composed from the interfering, indivisible unit-fields associated with the elementary particles. The unit-fields have simultaneously properties of the point particles and waves of quantum mechanics (particle-wave duality), the point particles of SM, the strings of theories of strings and the 3-D (in space) waves of theories of classical fields. Therefore the unification of the fundamental fields and interactions is considered as the further development and generalization of the canonical quantum mechanics, classical and quantum field theories, SM and string theories. 

Part II is organized as follows. Section 1 provides a brief introduction to the problem associated with the unification of classical and quantum fields and interactions. In order to make the unification of the fields, particles and interactions understandable to non-experts in the research field, the unification in Part II is performed in many philosophical, mathematical and physical details. The model concepts are  reanalyzed in the mathematical context of each section. Section 2 summarizes and briefly interprets the basic physical concepts of SM emphasizing the methodological and philosophical aspects of this model. Then these concepts are compared with the physical principles of the present model. Section 3 presents a single elementary particle as a single unit-field. Unification of a single elementary particle and a single unit-field is performed by the generalization of the Einstein energy-mass relation ${\varepsilon}^{2} ={\mathbf{k}}^2+m^2$ for the single unit-field. The interacting elementary particles of a multi-particle system and the interfering (cross-correlating) unit-fields are unified in Sec. 4. The unification is performed by generalization of the Einstein energy-mass relation for the interfering unit-fields associated with the interacting particles. Section 5 describes the internal structures [generators and associate components (ACs)] of a unit-field associated with fundamental (gravitational, electromagnetic, weak and strong) fields. The interference (interaction) of the structured unit-fields (elementary particles) having the arbitrary generators and the associate components is considered in Sec. 6.  In order to make the general equations of Sec. 6 more transparent, Sec. 7 analysis the interference (interaction) of the unit-fields (particles) having the concrete generators and associate components of the unit-fields, namely the de Broglie generators and the total associate components containing the spherically symmetric Laplace ACs and  Helmholtz ACs. Section 7 unifies the unit-fields corresponding to the experimentally observed particles obeying the different combinations of the gravitational, electromagnetic, weak and strong interactions. The results of unification of the all-known elementary particles and interactions are summarized in Sec. 8.

\section{2. Basic concepts of the present model versus standard model} 

The basic physical concepts (principles) of the present model of the {\it{unified fields, elementary particles and interactions}} are different from the Standard Model of Particle Physics. The standard model is usually interpreted as one of the most complicated physical models, which is based on the sophisticated gauge symmetry relations and the gauge symmetry breaking. In the present model, the unification of the all-known classical and quantum fields, particles and interactions is performed in the extremely simple and transparent manner, namely in terms of the interference between particles (unit-fields) and the respective cross-correlation energy, which do not exist from the point of view of SM. Before the unification of fields, particles and interactions let me summarize and briefly interpret the basic physical concepts of SM emphasizing the methodological and philosophical aspects of this model [Sec. (2.1.)]. Then these concepts will be compared with the physical concepts of the present model [Sec. (2.2.)]. 

\subsection{2.1. Basic concepts of Standard Model of Particle Physics}

The four known fundamental interactions of nature, all of which are non-contact forces, are electromagnetism, weak interaction, strong interaction and gravitation. The Standard Model provides a quasi-unified description of the classical and quantum fields, which define the electromagnetic, weak and strong interactions, but does not include gravity. SM is usually considered as the final outcome of quantum field theory combining the basic physical conceptions of canonical quantum mechanics with the special relativity~\cite{1Planck,1Einst1,1Bohr,1Einst2,1Noeth,1Brog,1Born,1Schr,1Ferm,1Heis1,1Dir1,1Wign,1Weyl,1Heis2,1Amb,1Paul,1Dir2,1Tomo,1Beth,1Schw,1Feyn,1Dyso}. In the generally-accepted form, SM has been developed around 1968 by Sheldon Glashow, Abdus Salam and Steven Weinberg (for instance, see the studies \cite{1Yang,1Namb,1Gold,1Glash,1Engl,1Wein1,1Sal,1Higg} and references therein). SM describes nature except gravity in terms of the fundamental material fields, e.g., the fields for the respective elementary particles. The elementary particle of the respective fundamental (global) field, which is infinite in space, is associated with the field "quanta". An elementary particle [a quanta of matter (energy-mass)] of the fundamental material field is assumed not to be made up of smaller particles. In other words, an elementary particle does not has a substructure. The quarks, anti-quarks, leptons, anti-leptons and gauge bosons are elementary particles, the building blocks of the gravity-less matter in SM. All other particles are made from these elementary particles. Quarks (up, down, charm, strange, top, bottom) and Leptons (electron neutrino, electron, muon neutrino, muon, tau neutrino, tau) are fermions. The respective anti-quarks and anti-leptons are also fermions. Gauge bosons (photon, $W$-boson, $Z$-boson, gluon, Higgs boson) naturally are bosons. The elementary particles are modelled in SM as {\it{point (0-D) particles}}, although some other theories posit a physical dimension. For instance, in the String Theory, particles are one-dimensional (1-D) strings that can move and vibrate, giving the physical properties to the elementary particles. 

In the conceptual picture of the SM fundamental interactions (interactive forces), the gravitation-less physical matter consists of the aforementioned (0-D) point-like fermions, which carry properties of charges and intrinsic angular momentums (half-integer spins $s = \pm 1/2$). The interactive forces are the ways that the fermions interact with one another by means of the electromagnetic, weak-nuclear and strong-nuclear forces having the non-contact nature.  Phenomenologically (macroscopically), the forces acting upon a fermion particle are seen as action of the respective boson field that is present at the particle location. The electromagnetic, weak-nuclear and strong-nuclear forces are associated with electromagnetic, weak-nuclear and strong-nuclear fields. Microscopically, the electromagnetic, strong and weak interactions between fermions are described in terms of the mathematical approximation method known as perturbation theory. The {\it{point fermions}} separated by the true vacuum  ({\it{"straight" empty spacetime}} of Einstein special relativity) attract or repel each other by the force mediating by the virtual exchange of the respective gauge {\it{point bosons}} through the vacuum. The bosons are the virtual particles because they are created and exist only in the exchange process. The exchange of bosons does transport momentum and energy between the fermions, thereby changing their momentum and energy. The interaction results into attracting or repelling fermions characterizing by the interactive force that has the absolute value and direction. The exchange can also carry a charge between the fermions, changing the charges of the fermions and turning them from one type of fermion to another. Finally, since the {\it{point boson}} transports one unit of intrinsic angular momentum (spin $s = \pm 1$), the spin of the {\it{point fermion}} can vary, for instance, from +1/2 to -1/2 under the boson exchange. It should be stressed that the microscopic interpretation of the interactions using the mathematical approximations associated with the perturbation theory is rather a pure mathematical interpretation than physical one. Moreover, such an interpretation does not adequately describe some physical phenomena. For an example, the perturbation-based model does not compare well with the Einstein theory of general relativity. Indeed, one can assume that the non-contact gravitational interaction between particles is provided by the Newton force, which is associated with the gravitational field mediated by the mass of interacting particles. Then there should be hypothetical gravitons, the supersymmetric partners of the gauge bosons of SM, that would carry the gravitational force in the perturbation approximation of gravitational interaction. In the Einstein general relativity, however, the gravitational interaction between two particles is not viewed as a force, but rather particles moving freely in gravitational fields travel under their own inertia in straight lines through "curve" spacetime. Thus the force of gravity is considered as the result of the geometry of "curve" spacetime. 

One can present also the slightly more detailed picture of the SM concepts associated with the fields, particles and interactions. In SM the electromagnetism and weak interaction are considered as two different aspects of the same force. The electromagnetism is the force (interaction) that acts between electrically charged fermions (quarks and leptons). The interaction acting between the fermions at rest is known as the electrostatic (Coulomb) force mediating by the electric field. The combined effect of electric forces acting between the fermions moving relative to each other is known as Lorentz's force inducing by the electric and magnetic fields. In the perturbation approximation of the electromagnetic interaction, the carriers of the electromagnetic force are the virtual photons (mass-less gauge bosons) associated with the electric charge of quarks and leptons. In the frame of the Heisenberg energy uncertainty relation and the perturbation approximation, the long range of the electromagnetic interaction is attributed to absence of the rest mass of photons. The field strength of the electromagnetic interaction is adjusted in SM by the electromagnetic coupling constant. The electromagnetism is called the quantum electrodynamics. The weak interaction (weak nuclear force), which affects all known fermions (quarks and leptons), is considered as being caused by the simultaneous emission and absorption (virtual exchange) of the $W$ and $Z$ virtual gauge bosons (massive particles). The typical field strength of the week interaction, which is adjusted by the respective (electro-week) coupling constant, is several orders of magnitude less than that of both electromagnetism and the strong nuclear force. The short range of the weak interaction is attributed to the heaviness of $W$ and $Z$ bosons. In contrast to the other interactions, the weak interaction may change not only the energy, momentum and spin of fermions, but also the quark flavor from one of six to another. The weak interaction, which is left-right asymmetric, does conserve CPT, but violates CP symmetry. Although the electromagnetic and weak interactions (forces) appear very different at low energies, SM unifies the electromagnetism and weak interaction into a single electroweak force above the unification energy ($\sim 100$ GeV). The quantum chromodynamics (QCD) of SM describes the strong force (color force) between the fractionally charged quarks interacting by means of eight gluons (mass-less gauge bosons) associated with the color charges of quarks. The color force is about one hundred times stronger than electromagnetic force, which in turn is orders of magnitude stronger than the weak interaction and gravitation. The strong force is assumed to be mediated by gluons, acting upon quarks, antiquarks, and the gluons themselves. On the short distance, the strong force holds quarks and gluons together to form the proton, neutron and other particles. On the longer distance, the color force binds protons and neutrons together to form the nucleus. In such a case, the force is considered as the residuum of the strong interaction between the quarks that make up the protons and neutrons. The lines of force of the gluons interacting with each other at long distances collimate into strings. In this way, the mathematical formalism of QCD not only explains how quarks interact over short distances, but also the string-like behavior, which they manifest over longer distances. Finally, the still undiscovered Higgs boson of the hypothetical Higgs quantum field, which has a non-zero expectation value in empty space, gives rest mass to the all observed elementary particles, except the photon and gluon, without breaking the so-called gauge invariance of the mathematical formalism of SM.   

Mathematically, the electromagnetic, weak-nuclear and strong-nuclear fields and interactions (forces) are modeled in the frame of the Lagrangian or Hamiltonian formalism, which controls the dynamics of the three fundamental fields. Each kind of point-like particles is described in terms of the respective dynamical field which exists and evolutes according to the Euler-Lagrange or Hamiltonian-Jacobi equation of motion with the initial and boundary conditions imposed in the 4-D spacetime. The most general renormalizable Lagrangian (Hamiltonian) of the fields is constructed by postulating a set of symmetries of the field-spacetime system that satisfies the experimentally observed particle (field) content. Naturally, the fundamental fields must obey the experimentally observed global Poincar{\'e} symmetry [$Translations \times SO^+(1,3)$], which consists of the ordinary translational symmetry, rotational symmetry and the inertial coordinate system invariance of Einstein's theory of special relativity. The local $SU(3)\times SU(2)\times U(1)$ gauge symmetry is an internal symmetry constructed and adjusted to provide the three experimentally observed fundamental interactions and the number and kind of elementary particles and anti-particles. The hypercharge $U(1)$, weak isospin $SU(2)$ and colour $SU(3)$ symmetries correspond to the electromagnetic, weak and strong interactions, respectively. These gauge symmetries obey the twelve flavours of elementary fermions, plus their corresponding antiparticles, as well as the four elementary gauge bosons that mediate the electromagnetic, strong and weak forces. SM contains the nineteen free (adjustable) parameters, including coupling constants of the fundamental interactions, whose numerical values are established experimentally. The observable quantities associated with a fundamental field of SM are not changed under the respective transformation attributed to the symmetry of the Lagrangian, even though the transformed field configuration vary in spacetime. All the changes of the field configuration mediated by the gauge transformation do cancel each other when written in terms of the observable quantities (energies, momentums, spins, charges, etc.). In other words, the different configurations of a fundamental field associated with the respective gauge symmetry (gauge invariance) of the field have identical observable quantities. The global Poincar{\'e} symmetry [$Translations \times SO^+(1,3)$] conserves energy, momentum and angular momentum. The local gauge symmetry [$SU(3)\times SU(2)\times U(1)$] conserves color charge, weak isospin, electric charge and weak hypercharge. The accidental symmetry (continuous $U(1)$ global symmetry) conserves the barion, electron, muon and tau numbers. From the point of view of the symmetry formalism of SM, the mathematical nature of a gauge transformation, which is a transformation from one field configuration to another, determines the mathematical nature of the gauge boson. Physically, the existence of the electromagnetic, strong and weak interactions arises in SM from a type of gauge symmetry relating to the fact that all elementary particles of a given type are experimentally indistinguishable from each other. The quantum properties of the classical (non-quantum) fundamental fields of SM are explored mathematically by defining creation and annihilation quantum operators with the commutation relations imposed, which add and subtract particles. This is known as the second quantization approach. The second quantization is performed by replacing the classical (non-quantum) spatially-infinite fields by the respective multimode field operators.

To this end, one could mention some candidate models closely connecting with with the basic physical and mathematical concepts of SM. The current best models attempting to unify all fundamental particles and interactions of SM and gravitation call for the symmetries, known as super-symmetries, which are different from the $U(1)$, $SU(2)$ and $SU(3)$ gauge symmetries. For instance, the Grand Unified Theories (GUTs) are proposals to show that the three fundamental interactions, other than gravity, arise from a single interaction characterized by one larger gauge symmetry with one unified coupling constant that breaks down at low energy levels to the three fundamental interactions with the respective coupling constants. GUTs predict relationships among the constants of nature, such as the electromagnetic, week and strong interaction constants, which are unrelated in SM. Some theories beyond SM look for a graviton to complete SM, while others, emphasize the possibility that the spacetime itself may have quantum nature (in form of the so-called "atoms of space and time") and more than four spacetime dimensions. Theories of Everything try to integrate GUTs with quantum gravity theories, which include the string theory, loop quantum gravity and twistor theory.  Some of these theories include a hypothetical fifth fundamental force to explain the recently-discovered accelerating expansion of the universe, giving rise to a need of the possible modifications of the Einstein general relativity. The fifth fundamental force has also been suggested to explain phenomena such as CP violations, cosmological "dark" matter (energy-mass), and "dark" flow. In some supersymmetric theories, the new particles (moduli) acquire their masses through supersymmetry breaking effects mediating even more fundamental forces, which are different from the forces of SM and Einstein's gravitation. For the detailed description of the physical concepts and mathematical formulations of SM and candidate models see the canonical studies and traditional textbooks, as well as any paper, review paper or textbook published in the literature on the quantum fields and interactions (for instance, see Refs. \cite{1Planck,1Einst1,1Bohr,1Einst2,1Noeth,1Brog,1Born,1Schr,1Ferm,1Heis1,1Dir1,1Wign,1Weyl,1Heis2,1Amb,1Paul,1Dir2,1Tomo,1Beth,1Schw,1Feyn,1Dyso,1Yang,1Namb,1Gold,1Glash,1Engl,1Wein1,1Sal,1Higg,1Land,1Jack,1Bere,1Itz,1Ryde,1Pes,1Grei,1Wein} and references therein).

\subsection{2.2. Critical analysis of basic concepts of Standard Model of Particle Physics}

The classical (non-quantum) and quantum fields of SM are modelled by using simultaneously the energy conservation law and the Dirac interpretation of quantum interference associated with the Copenhagen interpretation of quantum mechanics. According to Paul Dirac, under the quantum interference each particle interferes only with itself. In other words, the fundamental (global) spatially-infinite fields of the particles in SM do satisfy both the energy conservation law and the interference-less properties of particles. Part I of the present study pointed out that the phenomenon of redistribution of the field energy (intensity) by the interaction of quantum fields of SM is quit similar to the redistribution of the field energy (intensity) under the ordinary interference of classical fields. Therefore, the interaction of fields can be considered, at least formally, as the interference of fields. Under the Dirac quantum interference, however, the interference (cross-correlation) between two different particles never occurs. Thus the interaction of particles in SM {\it{is not attributed to the interference and the respective cross-correlation energy}} of classical (non-quantum) or quantum fields. This conceptual aspect of SM did not attract any critical attention of researchers {\it{due to the huge progress of the quantum physics, quantum field theory and SM based on the standard interpretation of quantum mechanics}}. It is clear now that the {\it{postulate of "interference-less", self interfering particles}} has played a key role in the method and direction of development of the all modern theories of fields, particles and interactions. The {\it{Copenhagen-Dirac postulate}} is the {\it{common background}} of the canonical quantum mechanics, traditional quantum field theories and SM, which do provide adequate description almost of the all-known physical phenomena.

Let me describe in more details how the {\it{postulate of interference-less particles}} relates to quantum mechanics and SM. In agreement with the Dirac postulate based on the Copenhagen interpretation of quantum mechanics, the particles of quantum mechanics and SM have been modeled to be free from the interference (cross-correlation) with other particles. Therefore it was extremely difficult for physicists to find a mathematical method of the "legal" insertion of the interference between particles and the respective cross-correlation energy into canonical quantum physics and SM. The mathematical cross-correlation ("hidden" interference) between particles and the respective cross-correlation energy have been automatically inserted into the quantum mechanics and SM by using the different powerful mathematical approximations, which do not compare well with the Copenhagen-Dirac postulate (see, Part I). The "hidden" interference {\it{contradicts, in an extremely non-trivial manner,}} the basic principles of canonical quantum mechanics and quantum field theory based on the Copenhagen ("probabilistic") interpretation of quantum mechanics and the Dirac postulate of interference-less particles. The "hidden" interference has been originally used in the construction ("mathematical engineering") of models for successful explanations of the coherent quantum phenomena, such as the Bose-Einstein condensation, superfluidity, superconductivity, supermagnetism, super-radiation, and quantum anomalous and fractional Hall effects. The "hidden" insertion of the interference between particles into the models yielded the correct description of the particle interference (interaction), cross-correlation (interaction) energy, particle number, Bose-Einstein statistics, Fermi-Dirac correlations (statistics) and Pauli exclusion principle. Then the mathematical approach in the form of {\it{"mathematical engineering" that does not require any critical analysis of the physical concepts}}  has been successfully extended also to quantum field theories and SM. The "mathematical engineering" based on the {\it{"hidden" interference}} mediated by the pure mathematical reasons (approximations), which contradict the Copenhagen-Dirac postulate, was generally accepted by the physics community in the very past. It is effectively used in quantum mechanics, quantum field theories and SM up to now without any critical analysis.

Although the "hidden" interference between particles and the respective cross-correlation energy are closely connected with the pure mathematical approximations, SM relies on a set of approximations specific to the particular physical problem. The "hidden" mathematical interference between particles and the respective cross-correlation energy have appeared in quantum mechanics, quantum field theories and SM already at the very beginning of consideration of the physical problems pertaining to the properties of quantum systems made of a large number of particles (many-body problem). The "hidden" interference (cross-correlation) between particles did appear in these theories in the form of the interaction (cross-correlation) energy describing by the interaction integrals in the mathematically constructed trial Hamiltonians. In SM the "hidden" interference (interaction) of the {\it{non-quantum}} (classical) fields is associated with the interaction energy of the fields describing by the trial {\it{non-quantum}} field Hamiltonian. Then the quantum properties of classical fundamental fields of SM are explored mathematically by defining creation (${\hat a}^{\dagger}$) and annihilation ($\hat a$) quantum operators with the commutation relations imposed, which add and subtract particles. That is known as the second quantization approach. Mathematically, the second quantization is performed by inserting the spatially-infinite classical fields $\Psi$ and $\Psi ^*$ into an infinite resonator and subsequent replacement of these fields by the respective multimode field operators $\hat \Psi$ and $\hat \Psi ^{\dagger }$. The "hidden" inter-particle interference and the respective interaction energy are provided in SM by the trial cross-correlating operators (pure mathematical objects) with the trial (Bogoliubov-like) commutator relations for the particles and antiparticles, namely the experimentally observable bosons and fermions and the virtual (unobservable) gauge bosons associated with the gauge fields, the gauge symmetry and the gauge symmetry breaking (spontaneous or not). In other words, the {\it{"hidden" interference}} between particles and the respective cross-correlation energy associated with the quantum fields are mediated automatically by the {\it{cross-correlation}} between the operators in the interaction integrals of the trial Hamiltonian field-operator of SM. The trial, artificially-constructed field Hamiltonians and the respective Hamiltonian operators of SM are invariant under the $U(1)$ local gauge transformation given rise to the very particular mathematical cross-correlation ("hidden" interference) and the respective energy, which do not depend on the particle phases. {\it{In such a case, the all-known coherent quantum phenomena are provided in the modified quantum mechanics and SM not only by the coherent particles (wave-functions), but also by the incoherent particles that have absolutely different phases}}. That means that the cross-correlation ("hidden" interference) is provided rather by the pure mathematical approximations than the large-scale classical or quantum coherence. The {\it{physical mechanism}} behind the "hidden" interference and cross-correlation energy of particles {\it{is mediated by the different mathematical approximations}}, which do not compare well with the Copenhagen-Dirac postulate of "interference-less", self interfering particles. In the frame of such an approach, the coherent quantum phenomena may be attributed even to the mathematical errors mediated by the mathematical approximations. Although SM is usually considered to be the unique and correct outcome of combining the rules of quantum mechanics with special relativity, it seems that SM contradicts at least formally to the basic (Copenhagen-Dirac) physical principle [a particle interferes (correlates) only with itself] of the canonical quantum mechanics and particle field theory. That is to say that the all-known physical interpretations of SM are based on interpretations of the pure mathematical approximations, which do contradict the basic postulate of the modern quantum physics, namely the Copenhagen-Dirac postulate of interference-less particles associated with the Copenhagen interpretation of quantum mechanics. 

It could be also mentioned that the Copenhagen interpretation (philosophy) of the de Broglie wave using the pure mathematical object (probability) did not solve really the problem of physical interpretation of quantum mechanics. In any physical interpretation, the probability or the wave of probability would be rather a pure mathematical (non-material) object than a real material substance. Unfortunately, up to now, both the {\it{Schr{\"o}dinger interpretation}} of the de Broglie wave using arguments of classical theories of real material waves and the {\it{Einstein determinism}} of classical physics over the Copenhagen probabilistic quantum physics are rejected in principle by the scientific community. {\it{It could be also noted that the Klein-Gordon-Fock and Dirac relativistic equations in the modern generally-accepted interpretation describe the global infinite fields of particles, while they have been originally formulated and interpreted by the authors as single-particle equations analogous to the Schr{\"o}dinger equation}}. In SM, up to now, it is not completely clear how to interpret physically the field (wave) of operators. In any interpretation, an operator should be considered rather as a pure mathematical object than a real physical matter. For more details of the above-mentioned methodological and philosophical problems, see Part I.

\subsection{2.3. Basic concepts of the present model of the unified fields, particles and interactions}

The basic physical concepts (principles) of the present model, which do associate with the {\it{unified fields, particles and interactions}}, are different from the Standard Model of Particle Physics. However, the present model {\it{does not use}} the extra dimensions of spacetime, the fifth ("dark") fundamental energy-mass and force or other {\it{hypothetical objects}} of the modern candidate models, such as the Grand Unified Theories and Theories of Everything. Part II of the present study deals with the Einstein special relativity and quantum mechanics unified by using the concepts that have been introduced and developed in Part I. Let me now {\it{summarize and briefly interpret}} these concepts emphasizing their {\it{methodological and philosophical}} aspects in the context of the above-described conceptual picture of SM. 

The {\it{most basic concept}} of the present model is extremely simple and transparent. The concept is based on the well-known law of physics, namely the {\it{interference phenomenon}}. In contrast to the non-material waves associated with the wave functions of canonical quantum mechanics and the non-material operator fields of SM, which deal with the {\it{interference-less}} (each particle interferes only with itself) wave functions and {\it{spatially-infinite}} fundamental fields of the {\it{operators}}, the present model suggests existence of an arbitrary number of the {\it{interfering}}, {\it{spatially-finite}} fundamental fields (beams) of the {\it{real physical matter (mass-energy)}}. The spatially-finite material fields ${\psi _{n}(\mathbf{r},t)}$ of the fundamental kind {\it{may interfere (cross-correlate)}} with each other and/or with the finite fields of another fundamental kind. The fundamental spatially-finite fields ${\psi _{n}(\mathbf{r},t)}$ are composed from the respective indivisible unit-fields ${\psi _{0n}(\mathbf{r},t)}$ of the mass-energy,  $\psi_{n} (\mathbf{r},t) =\sum_{n=1}^{N}{\psi _{0n}(\mathbf{r},t)}$. The unit-field (unit-wave) ${\psi _{0n}(\mathbf{r},t)}$ could interfere with itself as well with the other ($n \neq m$) unit-fields ${\psi _{0m}(\mathbf{r},t)}$. The {\it{real, material, spatially-finite unit-field (unit-wave) ${\psi _{0n}(\mathbf{r},t)}$, which does associate with the respective elementary particle and its physical matter (mass-energy)}}, is not the wave of probability of quantum mechanics or the field (wave) of operators of SM. In other words, the unit-field is not explained as a point particle or a wave of quantum mechanics (particle-wave dualism) or a point particle of SM. The unit-field has simultaneously {\it{properties of the point particle and wave of quantum mechanics, the point particle of SM and the 3-D (in space) wave of theory of classical fields.}} An indivisible unit-field may exist in the infinite number of configurations. The {\it{exact form and dynamics}} of the indivisible unit-field ${\psi _{0n}(\mathbf{r},t)}$, which depends on the experimental conditions, {\it{is determined by the field Lagrangian (Hamiltonian) and the Euler-Lagrange (Hamiltonian-Jacobi) equation of motion with the initial and boundary conditions imposed}}. In SM, up to now, it is not completely clear how to interpret physically the field (wave) of operators. An operator is rather a pure mathematical object than a real physical substance. In the present model, the division of the "non-quantum" field $\psi_n$ of the real physical matter (mass-energy) into the indivisible "non-quantum" unit-fields $ \psi_{0n} $ of the mass-energy, in fact, is {\it{the second quantization of the field without the use of the non-material fields (waves) of operators or probabilities}}. 

The interference between unit-fields (particles) could affect many basic physical properties of the particles. Under the quantum interference each particle of the canonical quantum mechanics can interfere only with itself. In the present model, unlike in the Dirac model of quantum interference, the {\it{particles (unit-fields) may  interfere (interact) also with each other}}. Thus {\it{the attractive or repulsive forces}} associated with interaction of the particles {\it{are attributed to the interference (cross-correlation)}} of the interfering (interacting) unit-fields. That is to say that the positive or negative gradient of the field cross-correlation energy does induce the attractive or repulsive forces that redistribute the field energy (intensity). In the frame of such an approach, the quantum and classical descriptions are in agreement for an arbitrary number of particles (unit-fields). It should be stressed that the physical system of the interfering (interacting) fields includes not only the interfering fields, but also the environment (material boundaries or other fields) that could provide the pure constructive or destructive interference. In other words, the unit-fields (particles) do interfere and interact not only with each other, but also with the environment. Although the field energy is changed under the pure constructive or destructive interference of fields, the total energy of the physical system is conserved. The pure additive or subtractive interference is provided by the "additional" energy, which should be added or subtracted from the total physical system before the interference of fields (particles) with each other. In such a case, the physical system of the interfering (interacting) fields does not obey the shift symmetry of time. The {\it{energy conservation}} of the total physical system {\it{is provided rather by the exchange of the field energy}} with the environment than by the shift symmetry of time and the Dirac quantum interference associated with "interference-less" self-interfering  particles. The interaction (cross-correlation) energy is attributed in the present model to the gradient of cross-correlation energy mediating by the interference between the unit-fields (particles) of the physical matter (mass-energy). Consequently, the {\it{physical mechanism}} behind the number of particles, cross-correlation (interference), cross-correlation energy, interference (interaction), Bose-Einstein statistics, Fermi-Dirac statistics and Pauli exclusion principle {\it{is associated with the repulsive and attractive forces mediating by the additive and subtractive interference (cross-correlation)}} of the unit-waves (bosons or fermions) of the physical matter (mass-energy). This mechanism plays a key role in the basic coherent quantum phenomena, such as Bose-Einstein condensation, super-radiation, Bosenova effect, superfluidity, superconductivity, supermagnetism, and quantum anomalous and fractional Hall effects. For more details and explanations, see Part I. 

In the {\it{standard model}} of fields, particles and interactions, the {\it{mathematical cross-correlation ("hidden" interference)}} is associated with the pure mathematical approximations, which do not compare well with the basic physical principle (the Copenhagen-Dirac postulate of "interference-less" self-interfering particles) of canonical quantum physics. Physically, the "hidden" interference is interpreted as the inter-particle interaction and the interaction energy. In the present model of the {\it{unified fields, particles and interactions}}, which could be considered as a generalization of SM, the unification is performed by taking into account rather the interference-induced energy mediated by the {\it{real, direct, physical}} interference between {\it{material}} unit-fields than the "hidden", {\it{mathematical}} interference between {\it{non-material}} waves of probabilities or operators associated with the point-like particles and the respective cross-correlation (interaction) energy. {\it{The "non-hidden" interference between material unit-fields may be considered as the EPR hypothetical "hidden" variable, which according to Einstein, Podolsky and Rosen should be added to quantum mechanics to avoid its indeterminism and superluminal signalling (see, also discussions related to the EPR paradox associated with superluminal signalling in Part I)}}. In Part I, the Hamiltonians that describe the cross-correlation energy mediating by interference of the basic classical and quantum material fields composed from the 3-dimensional (in space) unit-fields of physical matter have been derived by the generalization of the traditional Hamiltonians of the classical and quantum field theories for the superposition of the interfering unit-fields (interacting particles). Part II uses this background for unification of the electromagnetic, weak, strong and gravitational fields, particles and interactions. However, unlike in Part I, the unification is performed rather by the generalization of the basic (mass-energy) relation of the Einstein relativity than the traditional Hamiltonians of quantum field theory. The {\it{model unifies}} the all-known fields, particles and interactions by the {\it{straightforward generalization of the Einstein relativistic energy-mass relation ${\varepsilon}^{2} ={\mathbf{k}}^2+m^2$}} for the {\it{interfering (interacting) particles and bodies}} composed from the {\it{interfering unit-fields}}. The unit-fields {\it{exhibit simultaneously}} properties of the point particles and waves of quantum mechanics (particle-wave duality), the point particles of SM, the strings of theories of strings and the 3-D (in space) waves of theories of classical fields. Therefore the {\it{unification}} of the electromagnetic, weak, strong and gravitational fields and interactions could be considered not only as the {\it{generalization}} of SM, but also as {\it{the further development and unification}} of the quantum mechanics, classical and quantum field theories, SM and string theories. In order to make the unification understandable to the non-specialists in classical and quantum fields, the following analysis is performed in many philosophical, physical and mathematical details.  

\section{3. Unification of a single elementary particle and a single unit-field: A single particle ($N=1$)}

According to the basic physical concept of the present model, the fundamental fields $\psi_n (\mathbf{r},t) =\sum_{n=1}^N{\psi _{0n}(\mathbf{r},t)}$ are composed from the interfering, indivisible unit-fields ${\psi _{0n}(\mathbf{r},t)}$ of the mass-energy associated with the elementary particles. The unification of the fundamental (electromagnetic, weak, strong and gravitational) fields, particles and interactions in the model is based on the generalization of the famous energy-mass relation ${\varepsilon}^{2} ={\mathbf{k}}^2+m^2$ of the Einstein special relativity for the case of the interacting particles and bodies composed from the interfering (cross-correlating) unit-fields. Section 3 begins the unification with the generalization of the Einstein energy-mass relation for the unit-field $\psi _{0}$ associated with a {\it{single (N=1)}} elementary particle, which is {\it{free from interactions}} with other particles. The generalized mass-energy relation then yields the equation of motion for the unit-field. The exact form (configuration) and dynamics of the single unit-field (elementary particle) is determined by the equation of motion with the initial and boundary conditions imposed. The generalization, in fact, is simply the unification of the Einstein relativity of a point-like particle and the quantum mechanics of a wave-like (de Broglie) particle. Mathematically, the generalization of the Einstein energy-mass relation is performed by using the second or first derivatives of a unit-field, since these two approaches are equivalent in the case of the de Broglie wave  (unit-field) associated with a free particle. For the sake of generality, the model is presented also in the alternative form by using the unit-field Lagrangian that corresponds to the generalized energy-mass relation for the unit-field $\psi _{0}$. In such a case, the model is formulated in the frame of the Lagrangian formalism, where the configuration and dynamics of the single unit-field  is determined by the field Lagrangian and the Euler-Lagrange equation of motion with the initial and boundary conditions imposed.  

\subsection{3.1. The energy-mass relation and equation of motion for a free unit-field associated with a free particle}

\subsubsection{3.1.1. The model based on the {\bf{2nd}} derivatives of a free unit-field}

\vspace{0.4cm}

{\it{1. The model 1st-version based on the straightforward generalization of the Einstein energy-mass relation for a free unit-field by using the 2nd derivatives}}

\vspace{0.4cm}

The basic relation of Einstein's relativity for a single, point-like particle, which is free from interactions with other particles, is given by 
\begin{eqnarray} \label{eq1abc}
{\varepsilon}_0^{2} = \mathbf{k}_{0}^2+m_{0}^2, 
\end{eqnarray} 
where ${\varepsilon}_{0}$, $\mathbf{k}_{0}$ and $m_{0}$ are respectively the energy, momentum and rest mass of the particle. The straightforward generalization of the Einstein relativistic energy-mass relation (\ref{eq1abc}), which is associated with a free {\it{point-particle}} located in the spacetime point $(\mathbf{r},t)$, to the case of the free {\it{spatially-extended unit-field}} ${\psi _{0}(\mathbf{r},t)}$ occupying the finite or infinite volume $V$ is based on the use of the canonical approach of quantum mechanics. The replacement of the classical energy, momentum and mass of the point particle in Eq. (\ref{eq1abc}) by the respective quantum mechanical values associated with the operators of energy (${\hat \varepsilon}={\frac {\partial } {\partial t}}$), momentum (${\hat \mathbf{k}}=\nabla$) and mass (${\hat m}=m$) of the unit-field is performed by using the well-known, particular unit-field configuration ${\psi _{0}(\mathbf{r},t)}$, namely the de Broglie wave 
\begin{eqnarray} \label{eq2abc}
{\psi _{0}(\mathbf{r},t)} = V^{-1/2}e^{i(\mathbf{k}_0\mathbf{r}-{\varepsilon}_0t - \alpha_0)}
\end{eqnarray}
associated with the free particle. This simple procedure yields the generalized relativistic energy-mass relation 
\begin{eqnarray} \label{eq3abc}
{\varepsilon}_0^{2} =  {{{\frac 1 2}\int_{V}}  \psi_{0} ^* \left( -\ddot{\psi}_0 - \nabla ^2{\psi_{0}} + m_0^2\psi_{0}\right)d^3x}
\end{eqnarray}
for the free {\it{unit-field}}, where the quantities
\begin{eqnarray} \label{eq4abc}
m_0^{2} = {{\int_{V}}  \psi_{0} ^* \left(m_0^2\psi_{0}\right)d^3x},
\end{eqnarray}
\begin{eqnarray} \label{eq5abc}
\mathbf{k}_0^2={{\int_{V}}  \psi_{0} ^* \left(-\nabla ^2{\psi_{0}} \right)d^3x},
\end{eqnarray}
\begin{eqnarray} \label{eq6abc}
{\varepsilon}_0^{2} = {{\int_{V}}  \psi_{0} ^* \left(-\ddot{\psi_0} \right)d^3x} 
\end{eqnarray}
and 
\begin{eqnarray} \label{eq7abc}
\mathbf{k}_0^2 + m_0^{2}  = {{\int_{V}}  \psi_{0} ^* \left(-\nabla ^2{\psi_{0}} + m_0^2\psi_{0}\right)d^3x}
\end{eqnarray}
provide the Einstein relativistic energy-mass relation (\ref{eq1abc}). Notice, the values (\ref{eq3abc}) - (\ref{eq7abc}) do not depend on the wave phase $\alpha_0$. Also note that  
the unit-field (\ref{eq2abc}) is normalized by ${{\int_{V}}\psi_{0} ^*\psi_{0}d^3x}=1$. Comparison of the right-hand sides of Eq. (\ref{eq6abc}) and Eq. (\ref{eq7abc}) yields the relativistic equation of motion 
\begin{eqnarray} \label{eq8abc}
\square {\psi}_0 + m_0^2 {\psi}_0=0
\end{eqnarray}
for the free unit-field. I  use the conventional notations $ {\frac {\partial^2 {\psi_0}} {\partial t^2}} \equiv \ddot{\psi}_0$ and $\square \equiv {\frac {{\partial}^2 }{\partial {t^2}}} - {\nabla}^2$. Notice, the generalized relativistic energy-mass relation (\ref{eq3abc}) and the respective equation of motion (\ref{eq8abc}) do use the second derivatives of the free unit-field. The equivalent relativistic energy-mass relation and equation of motion, which are based on the first derivatives of the free unit-field, are presented in Sec. (3.1.2.). Although Eqs. (\ref{eq3abc}) and (\ref{eq8abc}) have been derived by using the particular unit-field configuration [de Broglie wave (\ref{eq2abc})], the present model assumes that Eq. (\ref{eq8abc}) does determine the all possible unit-field configurations associated with any relativistic or non-relativistic free elementary particle that satisfies the Einstein energy-mass relation (\ref{eq1abc}). Moreover, the generalized energy-mass relation (\ref{eq3abc}) for the unit-field configurations determining by the equation of motion (\ref{eq8abc}) with the initial and boundary conditions imposed is valid for the all-known free elementary particles. The value of the Einstein energy-mass given by Eq. (\ref{eq1abc}) for a free {\it{point-particle}} is equal to the generalized energy-mass determining by Eq. (\ref{eq3abc}) for the free {\it{unit-field}} (unit-wave) associated with this particle. Notice, Eq. (\ref{eq8abc}) is indistinguishable from the Klein-Gordon-Fock equation of the relativistic quantum field theory. The Klein-Gordon-Fock equation in the generally accepted interpretation (formulation) of the quantum field theory, however, describes the global infinite field of bosons and anti-bosons, while Eq. (\ref{eq8abc}) is formulated as a single-particle relativistic equation similar to the Schr{\"o}dinger equation for the wave-function of a free non-relativistic elementary particle of any kind. Equation (\ref{eq8abc}) with the initial and boundary conditions imposed describes both the unit-field configuration and its dynamics.

In the case of a non-relativistic particle ($\mathbf{k}_0^2 << m_0^{2}$), Eqs. (\ref{eq1abc}) - (\ref{eq8abc}) may be simplified. The Einstein energy-mass relation (\ref{eq1abc}) is simplified to the form  
\begin{eqnarray} \label{eq9abc}
{\varepsilon}_0 \approx  {\frac {\mathbf{k}_{0}^2} {2m_0} }+ m_{0},
\end{eqnarray}
which is similar to the energy-mass relation ${\varepsilon} = \frac {\mathbf{k}_{0}^2} {2m_0}$ of the Newton mechanics if the particle energy ${\varepsilon}$ is defined as ${\varepsilon} = {\varepsilon}_0 - m_{0}$. The replacement of the classical energy, momentum and mass of the point particle in Eq. (\ref{eq9abc}) by the respective quantum mechanical values of the de Broglie wave (\ref{eq2abc}) yields the non-relativistic energy-mass relation
\begin{eqnarray} \label{eq10abc}
{\varepsilon}_0 =  {{{\frac 1 2}\int_{V}}  \psi_{0} ^* \left( i\dot{\psi}_0-{\frac 1 {2m_0} } {\nabla ^2{\psi_{0}}} + m_0\psi_{0}\right)d^3x},
\end{eqnarray}
where
\begin{eqnarray} \label{eq11abc}
{\varepsilon}_0 = {{\int_{V}}  \psi_{0} ^* \left(i\dot{\psi_0} \right)d^3x} 
\end{eqnarray}
and 
\begin{eqnarray} \label{eq12abc}
{\frac {\mathbf{k}_{0}^2} {2m_0} }+ m_{0} = {{\int_{V}}  \psi_{0} ^* \left( -{\frac 1 {2m_0} } {\nabla ^2{\psi_{0}}} + m_0\psi_{0}\right)d^3x}.
\end{eqnarray}
The equation of non-relativistic motion
\begin{eqnarray} \label{eq13abc}
i\dot{\psi}_0 = -{\frac 1 {2m_0} } {\nabla ^2{\psi_{0}}} + m_0\psi_{0}
\end{eqnarray}
is obtained by comparison of the right-hand sides of Eq. (\ref{eq11abc}) and Eq. (\ref{eq12abc}). In the case of the stationary unit-field ${\psi _{0}(\mathbf{r},t)}={\psi _{0}(\mathbf{r})}e^{-i{\varepsilon}_0t}$, Eq. (\ref{eq13abc}) may be written as the Schr{\"o}dinger stationary equation 
\begin{eqnarray} \label{eq14abc}
\varepsilon \psi _{0}(\mathbf{r}) = - {\frac 1 {2m_0} } {\nabla ^2{\psi _{0}(\mathbf{r})}}
\end{eqnarray}
of the non-relativistic quantum mechanics of a free particle by using the aforementioned definition ${\varepsilon} = {\varepsilon}_0 - m_{0}$. It should be stressed that Eqs. (\ref{eq1abc}) - (\ref{eq8abc}) are valid for both the relativistic and non-relativistic particles, while the approximations (\ref{eq9abc}) - (\ref{eq14abc}) describe only the non-relativistic particle. Also note that replacement of the unit-field (\ref{eq2abc}) by the unit-field ${\psi _{0}(\mathbf{r},t)}=V^{-1/2}e^{-i(\mathbf{k}_0\mathbf{r}-{\varepsilon}_0t)}$ does not modify the relativistic equations (\ref{eq3abc}) - (\ref{eq8abc}), while the  replacement does change the sign of the non-relativistic energy $(\ref{eq11abc})$ from $(+)$ to $(-)$.  

The equation of motion (\ref{eq8abc}) describes the configuration and dynamics of the free unit-field ${\psi}_{0}$ associated with a single elementary particle that is free from interactions with other particles. In other words, the unit-field configuration ${\psi}_{0}$ and the energy ${\varepsilon}_0$ of a free particle determining by Eqs. (\ref{eq1abc}) - (\ref{eq8abc}) do not depend on the particle spin and charge. {\it{The different configurations ${\psi}_{0}$ of a single unit-field corresponding to the different solutions of Eq. (\ref{eq8abc}) with the initial and boundary conditions imposed could be attributed to the different momentums $\mathbf{k}_{0}$ of a single elementary particle having  the rest mass $m_0 $}}. The dynamical configuration $\psi_{0}$ corresponding to a solution of Eq. (\ref{eq8abc}) for a single unit-field with the rest mass $m_0 \neq 0$ could be considered as the configuration describing the dynamics of a massive particle. The unit-field configuration $\psi_{0}$ determining by the equation 
\begin{eqnarray} \label{eq15abc}
\square {\psi}_{0} =0
\end{eqnarray}
describes dynamics of a mass-less ($m_0 = 0$) particle, which has the energy squared given by 
\begin{eqnarray} \label{eq16abc}
{\varepsilon}_0^{2} = {{{\frac 1 2}\int_{V}}  \psi_{0} ^* \left(-\ddot{\psi} - \nabla ^2{\psi_{0}} \right)d^3x}. 
\end{eqnarray}
Notice, Eqs. (\ref{eq1abc}) - (\ref{eq8abc}) are valid for both the massive and mass-less particles, while the non-relativistic approximations (\ref{eq9abc}) - (\ref{eq14abc}) do not have any physical sense in the case of $m_0 = 0$. 

\vspace{0.4cm}

{\it{2. The model 2nd-version based on the generalization of the Einstein energy-mass relation for a free unit-field by using the Euler-Lagrange formalism and the 2nd derivatives}}

\vspace{0.4cm}

Mathematically, the standard model of particle physics is formulated in the frame of the Lagrangian or Hamiltonian formalism. Each kind of point-like particles is described in terms of the respective dynamical field which exists and evolutes according to the Euler-Lagrange or Hamiltonian-Jacobi equation of motion with the initial and boundary conditions imposed. The Lagrangian (Hamiltonian) of the field is constructed by postulating a set of symmetries of the field-spacetime system. For instance, the fundamental fields must obey the inertial coordinate system invariance of the Einstein theory of special relativity. The above-presented model is based on the straightforward generalization of the relativistic energy-mass relation of the Einstein theory. The model can be reformulated by using the Lagrangian formalism, where the configuration of a free unit-field (elementary particle) and its dynamics are determined by the unit-field Lagrangian and the Euler-Lagrange equation of motion with the initial and boundary conditions imposed. One can easily demonstrate that the unit-field Lagrangian corresponding to the generalized relativistic energy-mass relation (\ref{eq3abc}) is given by
\begin{eqnarray} \label{eq17abc}
L = {{{\frac 1 2}\int_{V}}  \psi_{0} ^* \left( \ddot{\psi}_0 - \nabla ^2{\psi_{0}}+ m_0^2\psi_{0}\right)d^3x},
\end{eqnarray}
where the unit-field is normalized by 
\begin{eqnarray} \label{eq18abc}
{{\int_{V}}\psi_{0} ^*\psi_{0}d^3x}=1. 
\end{eqnarray}
Equation (\ref{eq17abc}) can be written as 
\begin{eqnarray} \label{eq19abc}
L = {\int_{V}}{\cal L}d^3x, 
\end{eqnarray}
where
\begin{eqnarray} \label{eq20abc}
{\cal L} = {\frac 1 2}\psi_{0} ^* ( \ddot{\psi}_0 - \nabla ^2{\psi_{0}}+ m_0^2\psi_{0})
\end{eqnarray}
denotes the Lagrangian density, which may be presented as
\begin{eqnarray} \label{eq21abc}
{\cal L} = {\frac 1 2}\psi_{0} ^* ( {\partial}_{\mu}{\partial}^{\mu}\psi_{0} + m_0^2\psi_{0})
\end{eqnarray}
using the conventional notations ${\partial}_{\mu}{\partial}^{\mu} \equiv {\frac {{\partial}^2 }{\partial {t^2}}} - {\nabla}^2$. It could be noted that the unit-field $\psi_0$ associated with the particle may be considered also in the frame of the Hamiltonian-Jacobi formalism. In such a case, the Hamiltonian corresponding to the Lagrangian (\ref{eq17abc}) has the form
\begin{eqnarray} \label{eq22abc}
H = {\int_{V}}{\cal H}d^3x, 
\end{eqnarray}
where
\begin{eqnarray} \label{eq23abc}
{\cal H}={\frac 1 2}\psi_{0} ^* ( -\ddot{\psi}_0 - \nabla ^2{\psi_{0}}+ m_0^2\psi_{0}).
\end{eqnarray}
is the Hamiltonian density. In {\it{contrast}} to Part I, whose theoretical background uses the energy definition 
\begin{eqnarray} \label{eq24abc}
{\varepsilon}=H
\end{eqnarray}
of the canonical quantum field theory, the square of energy (\ref{eq3abc}) is now given by another definition: 
\begin{eqnarray} \label{eq25abc}
{\varepsilon}_0^{2}=H={{{\frac 1 2}\int_{V}}  \psi_{0} ^* \left( -\ddot{\psi}_0 - \nabla ^2{\psi_{0}}+ m_0^2\psi_{0}\right)d^3x}.
\end{eqnarray}
Notice, for the unit-field normalized by Eq. (\ref{eq18abc}), the energy squared (\ref{eq25abc}) is indistinguishable from the value (\ref{eq3abc}). In the frame of the Lagrange formalism, the configuration and dynamics of the unit-field $\psi_0$ is determined by the Euler-Lagrange equation 
\begin{eqnarray} \label{eq26abc}
\left[  \frac {{\delta }{\cal L}} {\delta  {\psi}^*}-{\partial}_{\mu}\frac {{\delta }{\cal L} } {\delta {\partial}_{\mu}{\psi}^*} \right]=0
\end{eqnarray}
with the initial and boundary conditions imposed. Notice, the volume $V$ of the unit-field determining by the unit-field boundaries, in contrast to the infinite fields of SM, may be finite or infinite. For the Lagrangian density (\ref{eq21abc}), Eq. (\ref{eq26abc}) yields the Euler-Lagrange relativistic equation  
\begin{eqnarray} \label{eq27abc}
\square {\psi}_0 + m_0^2 {\psi}_0=0,
\end{eqnarray}
which is indistinguishable from Eq. (\ref{eq8abc}). 

In the case of a non-relativistic unit-field, the Lagrangian that corresponds to Eq. (\ref{eq10abc}) is given by
\begin{eqnarray} \label{eq28abc}
L = {{{\frac 1 2}\int_{V}}  \psi_{0} ^* \left( -i\dot{\psi}_0-{\frac 1 {2m_0} } {\nabla ^2{\psi_{0}}} + m_0\psi_{0}\right)d^3x},
\end{eqnarray}
where
\begin{eqnarray} \label{eq29abc}
{\cal L} = {\frac 1 2}\psi_{0} ^* \left( -i\dot{\psi}_0 - {\frac 1 {2m_0} } {\nabla ^2{\psi_{0}}} + m_0\psi_{0}\right)
\end{eqnarray}
denotes the Lagrangian density. The unit-field Hamiltonian $H$ corresponding to the Lagrangian (\ref{eq26abc}) has the form (\ref{eq22abc}), where the Hamiltonian density is given by
\begin{eqnarray} \label{eq30abc}
{\cal H}={\frac 1 2}\psi_{0} ^* \left( i\dot{\psi}_0 - {\frac 1 {2m_0} } {\nabla ^2{\psi_{0}}} + m_0\psi_{0}\right).
\end{eqnarray}
Correspondingly, the non-relativistic energy ${\varepsilon}$ of the unit-field is determined by the expression  
\begin{eqnarray} \label{eq31abc}
{\varepsilon}_0=H={{{\frac 1 2}\int_{V}}  \psi_{0} ^* \left( i\dot{\psi}_0 - {\frac 1 {2m_0} } {\nabla ^2{\psi_{0}}} + m_0\psi_{0}\right)d^3x},
\end{eqnarray}
which is indistinguishable from the energy (\ref{eq10abc}) for the unit-field normalized by Eq. (\ref{eq19abc}). For the Lagrangian density (\ref{eq29abc}), the Euler-Lagrange equation (\ref{eq26abc}) yields the equation
\begin{eqnarray} \label{eq32abc}
i\dot{\psi}_0 = -{\frac 1 {2m_0} } {\nabla ^2{\psi_{0}}} + m_0\psi_{0},
\end{eqnarray}
which is indistinguishable from Eq. (\ref{eq13abc}). Notice, Eq. (\ref{eq17abc}) - Eq. (\ref{eq32abc}) do use the normalization (\ref{eq18abc}). The normalization gives rise to the probabilistic (canonical) interpretation of the unit-field  ${\psi_0}$ in Eq. (\ref{eq32abc}), where the unit-field $\psi_0 (\mathbf{r},t)$ can be reinterpreted according to the Copenhagen philosophy of quantum mechanics of point particles. The Copenhagen interpretation of the value $\rho (\mathbf{r},t)\equiv \psi^*(\mathbf{r},t)\psi (\mathbf{r},t)$ as the probability density, however, can lead to the well-known philosophical and physical problems associated, for instance, with the wave-particle duality and the negative probability density. In the case of the stationary unit-field, Eq. (\ref{eq32abc}) is equivalent to the Schr{\"o}dinger stationary equation (\ref{eq14abc}) for the Copenhagen wave of probability determining by the wave-function of a free elementary particle.

\subsubsection{3.1.2. The model based on the {\bf{1st}} derivatives of a free unit-field}

\vspace{0.4cm}

{\it{1. The model 3-rd version based on the straightforward generalization of the Einstein energy-mass relation for a free unit-field by using the 1st derivatives}}

\vspace{0.4cm}

The unit-field model presented in Sec. (3.1.1.) is based on the use of the second derivatives of a unit-field. Equations (\ref{eq3abc}) - (\ref{eq8abc}) and (\ref{eq10abc}) - (\ref{eq32abc}),  have been derived by using Eqs. (\ref{eq4abc}), (\ref{eq5abc}) and (\ref{eq6abc}) for the squares of mass, momentum and energy of the de Broglie wave (unit-field). Here, I present the alternative, equivalent formulation of the model, where Eqs. (\ref{eq4abc}), (\ref{eq5abc}) and (\ref{eq6abc}) are replaced respectively by 
\begin{eqnarray} \label{eq33abc}
m_0^{2} = {{\int_{V}}  m_0^2 \psi_{0} ^* \psi_{0}d^3x},
\end{eqnarray}
\begin{eqnarray} \label{eq34abc}
\mathbf{k}_0^2= {{\int_{V}}(\nabla{\psi_{0}^*}) (\nabla{\psi_{0}} )d^3x}
\end{eqnarray}
and
\begin{eqnarray} \label{eq35abc}
{\varepsilon}_0^{2} = {{\int_{V}} \left( \dot{\psi_0 ^*} \right)  \left(\dot{\psi_0} \right)d^3x},
\end{eqnarray} 
where the unit-field is normalized by ${{\int_{V}}\psi_{0} ^*\psi_{0}d^3x}=1$. The use of Eqs. (\ref{eq34abc}) and (\ref{eq35abc}), which are based on the first derivatives of a unit-field, yields the generalized relativistic energy-mass relation 
\begin{eqnarray} \label{eq36abc}
{\varepsilon}_0^{2} = {{{\frac 1 2}\int_{V}}  \left(  \dot{\psi_0 ^*} \dot{\psi_0} +  \nabla \psi_{0}^*  \nabla \psi_{0} + m_0^2 \psi_{0} ^* \psi_{0}   \right)d^3x}.
\end{eqnarray}
Notice, the energy squared (\ref{eq36abc}) is different from the values (\ref{eq3abc}) and (\ref{eq25abc}). The value (\ref{eq36abc}) is equal to the values (\ref{eq3abc}) and (\ref{eq25abc})  only in the very particular case, when the unit-field $\psi_0$ has the de Broglie configuration (\ref{eq2abc}). The equation of motion corresponding to Eq. (\ref{eq36abc}) is given by the expression 
\begin{eqnarray} \label{eq37abc}
\dot{\psi_0 ^*} \dot{\psi_0} = \nabla \psi_{0}^*  \nabla \psi_{0} + m_0^2 \psi_{0} ^* \psi_{0},
\end{eqnarray}
which is more complicate than Eqs. (\ref{eq8abc}) and (\ref{eq27abc}).

In the case of a non-relativistic unit-field (particle), the non-relativistic energy ${\varepsilon}_0$ corresponding to Eq. (\ref{eq9abc}) is determined by the relation 
\begin{eqnarray} \label{eq38abc}
{\varepsilon}_0 = {{\frac 1 2} {\int_{V}}  \left( i\psi_{0} ^*  \dot{\psi}_0+{\frac 1 {2m_0} } {\nabla {\psi_{0}^*}\nabla {\psi_{0}}} + m_0\psi_{0} ^*  \psi_{0}\right)d^3x}, 
\end{eqnarray}
which is different from the values (\ref{eq10abc}) and (\ref{eq31abc}). The energy (\ref{eq38abc}) is equal to the values (\ref{eq10abc}) and (\ref{eq31abc})  only when the unit-field $\psi_0$ has the de Broglie configuration (\ref{eq2abc}). The equation of motion corresponding to Eq. (\ref{eq38abc}) is given by the formula 
\begin{eqnarray} \label{eq39abc}
i\psi_{0} ^*  \dot{\psi}_0={\frac 1 {2m_0} } {\nabla {\psi_{0}^*}\nabla {\psi_{0}}} + m_0\psi_{0} ^*  \psi_{0},
\end{eqnarray}
which is different from Eq. (\ref{eq13abc}) and (\ref{eq32abc}). 

\vspace{0.4cm}

{\it{2. The model 4th-version based on the generalization of the Einstein energy-mass relation for a free unit-field by using the Euler-Lagrange formalism and the 1st derivatives}}

\vspace{0.4cm}

The unit-field Lagrangian that corresponds to the energy-mass relation (\ref{eq36abc}) with the unit-field normalization (\ref{eq18abc}) is given by
\begin{eqnarray} \label{eq40abc}
L = {{{\frac 1 2}\int_{V}} \left(  \dot{\psi_0 ^*} \dot{\psi_0} -  \nabla \psi_{0}^*  \nabla \psi_{0} - m_0^2 \psi_{0} ^* \psi_{0}   \right)d^3x},
\end{eqnarray}
where
\begin{eqnarray} \label{eq41abc}
{\cal L} ={\frac 1 2} \left(  \dot{\psi_0 ^*} \dot{\psi_0} -  \nabla \psi_{0}^*  \nabla \psi_{0} - m_0^2 \psi_{0} ^* \psi_{0} \right)
\end{eqnarray}
denotes the Lagrangian density, which may be presented in the covariant form as
\begin{eqnarray} \label{eq42abc}
{\cal L} = {\frac 1 2} ( {\partial}_{\mu} \psi_{0} ^* {\partial}^{\mu}\psi_{0} - m_0^2\psi_{0}).
\end{eqnarray}
The unit-field Hamiltonian $H$ corresponding to the Lagrangian (\ref{eq40abc}) has the form (\ref{eq22abc}), where the Hamiltonian density is given by
\begin{eqnarray} \label{eq43abc}
{\cal H}={\frac 1 2} \left(  \dot{\psi_0 ^*} \dot{\psi_0} +  \nabla \psi_{0}^*  \nabla \psi_{0} + m_0^2 \psi_{0} ^* \psi_{0} \right).
\end{eqnarray}
Thus the square of energy is given by the expression
\begin{eqnarray} \label{eq44abc}
{\varepsilon}_0^{2}=H={{{\frac 1 2}\int_{V}} \left(  \dot{\psi_0 ^*} \dot{\psi_0} +  \nabla \psi_{0}^*  \nabla \psi_{0} + m_0^2 \psi_{0} ^* \psi_{0}   \right)d^3x},
\end{eqnarray}
which is different from the relation ${\varepsilon}_0=H$ given by Eq. (15) of Part I of the present study. If the unit-field is normalized by Eq. (\ref{eq18abc}), then the energy squared (\ref{eq44abc}) is indistinguishable from the value (\ref{eq36abc}). For the Lagrangian density (\ref{eq42abc}) the Euler-Lagrange equation (\ref{eq26abc}) yields the equation 
\begin{eqnarray} \label{eq45abc}
\square {\psi}_0 + m_0^2 {\psi}_0=0,
\end{eqnarray}
which is indistinguishable from Eqs. (\ref{eq8abc}) and (\ref{eq27abc}), but is different from the more complicated equation (\ref{eq37abc}). Notice, the equations of motions (\ref{eq8abc}), (\ref{eq27abc}) and (\ref{eq37abc}) describe the configuration and dynamics of the unit-field ${\psi}_{0}$ associated with a massive elementary particle having the rest mass $m_0 \neq 0$.  The unit-field configuration $\psi_{0}$ determining by the equation 
\begin{eqnarray} \label{eq46abc}
\square {\psi}_{0} =0
\end{eqnarray}
describes dynamics of a mass-less ($m_0 = 0$) particle, with the energy squared given by  
\begin{eqnarray} \label{eq47abc}
{\varepsilon}_0^{2} = H= {{{\frac 1 2}\int_{V}} \left( \dot{\psi_0 ^*} \dot{\psi_0} +  \nabla \psi_{0}^*  \nabla \psi_{0} \right)d^3x} .
\end{eqnarray}

In the case of a non-relativistic unit-field, the Lagrangian that corresponds to Eq. (\ref{eq9abc}) is given by
\begin{eqnarray} \label{eq48abc}
L = {{{\frac 1 2}\int_{V}}  \left( i\psi_{0} ^*  \dot{\psi}_0-{\frac 1 {2m_0} } {\nabla {\psi_{0}^*}\nabla {\psi_{0}}} - m_0\psi_{0} ^*  \psi_{0}\right)d^3x},
\end{eqnarray}
where
\begin{eqnarray} \label{eq49abc}
{\cal L} = {\frac 1 2}\left( i\psi_{0} ^*  \dot{\psi}_0-{\frac 1 {2m_0} } {\nabla {\psi_{0}^*}\nabla {\psi_{0}}} - m_0\psi_{0} ^*  \psi_{0}\right)
\end{eqnarray}
denotes the Lagrangian density, which was derived by using the first derivatives of the unit-field. The unit-field Hamiltonian $H$ corresponding to the Lagrangian (\ref{eq48abc}) has the form (\ref{eq22abc}), where the Hamiltonian density is given by
\begin{eqnarray} \label{eq50abc}
{\cal H}={\frac 1 2}\left( i\psi_{0} ^*  \dot{\psi}_0+{\frac 1 {2m_0} } {\nabla {\psi_{0}^*}\nabla {\psi_{0}}} + m_0\psi_{0} ^*  \psi_{0}\right).
\end{eqnarray}
Correspondingly, the non-relativistic energy ${\varepsilon}$ of the unit-field is determined as  
\begin{eqnarray} \label{eq51abc}
{\varepsilon}_0=H={{{\frac 1 2}\int_{V}}  \left( i\psi_{0} ^*  \dot{\psi}_0+{\frac 1 {2m_0} } {\nabla {\psi_{0}^*}\nabla {\psi_{0}}} + m_0\psi_{0} ^*  \psi_{0}\right)d^3x}.
\end{eqnarray}
In the case of the unit-field normalized by Eq. (\ref{eq18abc}), the energy squared (\ref{eq51abc}) is indistinguishable from the value (\ref{eq36abc}). For the Lagrangian density (\ref{eq49abc}), the Euler-Lagrange equation (\ref{eq26abc}) yields the expression 
\begin{eqnarray} \label{eq52abc}
i\dot{\psi}_0 = -{\frac 1 {2m_0} } {\nabla ^2{\psi_{0}}} + m_0\psi_{0},
\end{eqnarray}
which is indistinguishable from Eqs. (\ref{eq13abc}) and (\ref{eq32abc}). Notice, Eqs. (\ref{eq33abc}) - (\ref{eq44abc}) are valid for both the massive and mass-less particles, while the non-relativistic approximations (\ref{eq48abc}) - (\ref{eq52abc}) do not have any physical meaning in the case of $m_0 = 0$. It should be stressed that  Eqs. (\ref{eq40abc}) - (\ref{eq52abc}) do use the normalization (\ref{eq18abc}) that gives rise to the {\it{probabilistic (canonical) interpretation}} of the unit-field ${\psi_0}$ in Eqs. (\ref{eq45abc}) and (\ref{eq52abc}) as the wave-function of a free elementary particle associated with the Copenhagen wave of probability. That means that the non-relativistic unit-field could be interpreted as the Schr{\"o}dinger wave-function associated with the  non-relativistic wave of probability.

\subsection{3.2. Physical properties of the single particle based on the generalized energy-mass relation for the single unit-field}

Physical properties of the above-described free unit-field (particle) are summarized and interpreted as follows. The generalized Einstein energy-mass relations (\ref{eq3abc}) and (\ref{eq36abc}) and the respective equations of motion (\ref{eq8abc}) and (\ref{eq45abc}) have been derived by using the de Broglie wave (\ref{eq2abc}), which is the particular configuration of a free unit-field associated with an elementary particle. The present model assumes that the energy-mass relations (\ref{eq3abc}) and (\ref{eq36abc}) with the respective equations of motion (\ref{eq8abc}) and (\ref{eq45abc}) do determine the unit-field configuration of any relativistic or non-relativistic free elementary particle that satisfies the Einstein energy-mass relation (\ref{eq1abc}). That is to say that the generalized energy-mass relations (\ref{eq3abc}) and (\ref{eq36abc}) for the free unit-field determining by the equations of motion (\ref{eq8abc}) and (\ref{eq45abc}) with the initial and boundary conditions imposed are {\it{valid}} for the {\it{all-known free elementary particles}}. There is no difference between the values of the Einstein energy-mass determining by Eq. (\ref{eq1abc}) of a free point particle and the generalized energy-mass given by Eqs. (\ref{eq3abc}) and (\ref{eq36abc}) for the free unit-field associated with this particle. Equations (\ref{eq8abc}) and (\ref{eq45abc}) look like the Klein-Gordon-Fock equation of the relativistic quantum field theory. However, in contrast to the Klein-Gordon-Fock equation of quantum field theory describing the global infinite field of particles (scalar bosons), Eqs. (\ref{eq8abc}) and (\ref{eq45abc}) are formulated and interpreted as a single-particle relativistic equation similar to the Schr{\"o}dinger equation for the de Broglie wave of the matter associated with a non-relativistic particle.  {\it{It could be mentioned again that the Klein-Gordon-Fock equation has been originally formulated and interpreted by the authors as a single-particle equation analogous to the Schr{\"o}dinger equation}}.

The equations of motion (\ref{eq8abc}) and (\ref{eq45abc}) describe the configuration and dynamics of the free unit-field ${\psi}_{0}$ that determine the physical properties of a single elementary particle that is {\it{free}} from interactions with other particles. The unit-field configuration ${\psi}_{0}$ and the self-energy ${\varepsilon_0}$ of a {\it{free particle}} determining by Eqs. (\ref{eq1abc}) - (\ref{eq32abc}) or Eqs. (\ref{eq33abc}) - (\ref{eq52abc}) do not depend on the spin and charge of the particle. The {\it{kind}} (type) of a free particle and the respective unit-field configuration are determined in Eqs. (\ref{eq1abc}) - (\ref{eq52abc}) solely by the particle rest-mass $m_0$. The free particles (unit-fields) could be {\it{distinguished}} from each other by the particle {\it{rest-mass}} $m_0$. In other words, the free particles (unit-fields) having the {\it{different rest masses}} could be considered as {\it{different particles (unit-fields)}}. For instance, the dynamical configuration $\psi_{0}$ corresponding to a solution of Eq. (\ref{eq8abc}) or (\ref{eq45abc}) for a single unit-field with the rest mass $m_0 \neq 0$ could be considered as the configuration describing the dynamics of a {\it{massive}} particle. The unit-field configuration $\psi_{0}$ determining by the equation (\ref{eq15abc}) or (\ref{eq46abc}) describes dynamics of the {\it{mass-less}} ($m_0 = 0$) particle. For the given rest-mass $m_0$, Eqs. (\ref{eq8abc}) and (\ref{eq45abc}) with the initial and boundary conditions imposed describe not only the unit-field {\it{configuration}} and the {\it{kind}} of a free particle, but also its {\it{dynamics}}. The different configurations ${\psi}_{0}$ of a single unit-field corresponding to the different solutions of Eq. (\ref{eq8abc}) or (\ref{eq45abc}) could be attributed to the different momentums $\mathbf{k}_{0}$ of a single elementary particle having  the rest mass $m_0 $. Although the unit-field $\psi_{0}$ describing by Eqs. (\ref{eq1abc}) - (\ref{eq32abc}) or Eqs. (\ref{eq33abc}) - (\ref{eq52abc}) is material one, it is not measurable quantity. The experimentally observable quantities are the rest mass $m_0$, momentum $\mathbf{k}_{0}$ and energy ${\varepsilon_0}$ of the unit-field (particle). These quantities control the unit-filed configuration and its dynamics and {\it{vice versa}}. For an example, the "annihilation" of the rest mass $m_0\neq 0$ in the energy-mass relations (\ref{eq3abc}) and (\ref{eq36abc}) and the respective equations of motion  (\ref{eq8abc}) and (\ref{eq45abc}), yields the change of the unit-field configuration and the kind of a particle. The annihilation yields the new (mass-less) unit-field configuration corresponding to the new (mass-less) particle, which is determined by the energy-mass relation (\ref{eq16abc}) or (\ref{eq47abc}) and the respective equation of motion  (\ref{eq15abc}) or (\ref{eq46abc}). According to{\it{ the energy conservation law}} the unit-field energy can be converted from one form to another, but it cannot be created or destroyed. Therefore the energy of the new (mass-less) unit-field determining by Eq. (\ref{eq16abc}) or (\ref{eq47abc}) must be equal respectively to the energy of the annihilated massive unit-field given by Eq. (\ref{eq3abc}) or (\ref{eq36abc}). It should be mentioned that the relativistic equations (\ref{eq1abc}) - (\ref{eq8abc}), (\ref{eq15abc}) - (\ref{eq27abc}), (\ref{eq33abc}) - (\ref{eq37abc}) and  (\ref{eq40abc}) - (\ref{eq47abc}) are valid for both the massive and mass-less particles, while the non-relativistic approximations (\ref{eq9abc}) - (\ref{eq14abc}), (\ref{eq28abc}) - (\ref{eq32abc}), (\ref{eq38abc}) , (\ref{eq39abc}), (\ref{eq48abc}) -(\ref{eq52abc}) do not have any physical sense in the case of $m_0 = 0$. In other words, the non-relativistic equations could not describe the creation of a mass-less ($m_0 = 0$) particle under the annihilation of a massive ($m_0 \neq 0$) particle (unit-field).
 
From a point of view of the energy, a {\it{point particle}} and the {\it{unit-wave}} associated with this particle are equivalent (indistinguishable) objects. Indeed, the relativistic particle satisfies simultaneously the relativistic energy-mass relation (\ref{eq1abc}) for a point particle of Einstein's special relativity and the generalized (equivalent) relativistic energy-mass relation (\ref{eq3abc}) or (\ref{eq36abc}) for the unit-wave associated with this particle. In addition, the non-relativistic energy (\ref{eq9abc}) of a point particle of the Einstein non-relativistic theory (Newton mechanics) is equal to the non-relativistic energy (\ref{eq10abc}) or (\ref{eq51abc}) of the non-relativistic unit-wave. The equality of energies of the point particle and the unit-wave could be considered as the {\it{particle-wave duality}} similar to the particle-wave duality in the Schr{\"o}dinger (non-relativistic) quantum mechanics of  the de Broglie wave associated with a free particle. The formulation of the present model in the frame of the Lagrangian-Hamiltonian formalism with the normalization (\ref{eq18abc}) gives rise to {\it{the probabilistic (quantum mechanical) interpretation}} of the unit-field (wave) ${\psi_0}$. However, in contrast to the quantum-mechanical wave-function $\psi _0 (\mathbf{r},t)$ of a point particle located in the spatiotemporal point $(\mathbf{r},t)$ with the probability density $\psi^*_0 (\mathbf{r},t)\psi_0 (\mathbf{r},t)$, the {\it{material}} unit-field $\psi_0 (\mathbf{r},t)$ associated with the particle mass-energy {\it{does really exist}} in the each spatiotemporal point $(\mathbf{r},t)$ of the unit-field. {\it{That is the fundamental (principal) difference between the canonical quantum mechanics and the present model}}. It could be also mentioned that the two versions of the unit-field model, which are based on the 1st or 2nd derivatives of the unit-field, have the different forms (\ref{eq3abc}) and (\ref{eq36abc}) of the generalized Einstein energy-mass relation, but the indistinguishable equations of motion (\ref{eq8abc}) and (\ref{eq45abc}). The energy-mass relations (\ref{eq3abc}) and (\ref{eq36abc}) are equivalent for the de Broglie wave (\ref{eq2abc}) associated with a single particle that is free from interactions with other particles. In other words, for the unit-field configuration (\ref{eq2abc}), one can use the 1st or 2nd derivatives of the unit-field in Eqs. (\ref{eq3abc}), (\ref{eq36abc}), (\ref{eq8abc}) and (\ref{eq45abc}). Although these two versions of the model are equivalent for the de Broglie wave (\ref{eq2abc}), Sec. (4) shows that they may give different results for more complicated unit-field configurations.  

\subsection{3.3. The role of initial and boundary conditions in dynamics of a single unit-field} 

The {\it{kind}} of a free elementary particle in Eqs. (\ref{eq1abc}) - (\ref{eq52abc}) of the above-presented conceptual model of a free unit-field (particle) is determined by the particle {\it{rest-mass}} $m_0$, only. The {\it{rest mass}} $m_0$ could be considered as a particular kind of the {\it{initial conditions}}. For a given rest-mass $m_0$, the equation of motion (\ref{eq8abc}) or (\ref{eq45abc}) with the initial and boundary conditions imposed describes {\it{dynamics}} of the unit-field configuration. The dynamical configurations ${\psi}_{0}$ corresponding to the different solutions of the equation of motion could be attributed to the different momentums $\mathbf{k}_{0}$ of the same particle. That is to say that the unit-field associated with an elementary particle having the rest mass $m_0$ does exist in different configurations corresponding to the different solutions of the equation of motion with the actual initial and boundary conditions. A solution to the differential  equation of motion is unambiguously determined if the value of the unit-field function (Dirichlet boundary conditions) or the normal derivative (Neumann boundary conditions) of the unit-field function is specified on the boundaries. {\it{In the present study, a unit-field is considered as the free unit-field if it does interact only with the boundaries imposed by the empty, "straight" spacetime (true vacuum)}}. For instance, a massive, free unit-field describing by Eqs. (\ref{eq8abc}) or (\ref{eq45abc}) has a boundary between the massive unit-field and the mass-less true vacuum. That means that the massive, free unit-field does interact only with the mass-less boundary of the true vacuum (empty, "straight" spacetime). A mass-less, free unit-field, which is described by Eqs. (\ref{eq15abc}) or (\ref{eq46abc}), has a boundary between the mass-less ($m_0=0$) unit-field and the mass-less true vacuum. In such a case, a mass-less, free unit-field interacts with the mass-less boundary of the empty, "straight" spacetime. A free unit-field before interaction with the material boundaries is considered as a free unit-field. The material boundaries can be modified by using the external energy, namely the energy that does not associate with the unit-field energy. Interaction of a unit-field with the actual material boundaries may change or not change the absolute value $|\mathbf{k}_{0}|$ of the unit-field momentum $\mathbf{k}_{0}$. The interaction that changes the value $|\mathbf{k}_{0}|$ is called a non-elastic process. The process could be interpreted as the exchange of energy between the unit-field and the boundary. Analysis of such a process requires the detailed model of interaction of the unit-field (particle) with the unit-fields (particles) of the material boundary. If the interaction does not affect the momentum absolute value $|\mathbf{k}_{0}|$, then the interaction has an elastic character. {\it{The elastic interaction of a single unit-field with the material boundaries can be treated phenomenologically by using the Dirichlet or Neumann boundary conditions for the equation of motion (\ref{eq8abc}) or (\ref{eq45abc})}}. The evolution of the unit-field configuration under the free-space propagation and elastic interaction of the single unit-field with a material boundary is illustrated by the following three examples. 

Let me first consider the simplest, relativistic unit-field configuration that corresponds to the free plane-wave, namely the time-harmonic solutions of the relativistic equations (\ref{eq8abc}), (\ref{eq27abc}), (\ref{eq39abc}) and (\ref{eq45abc}) for the empty, "straight" spacetime. At the initial time moment $t = 0$, such a unit-field is given by the de Broglie wave 
\begin{eqnarray} \label{eq53abc}
\psi_0 (\mathbf{r},t)=a_0 e^{i({\mathbf{k}_0{\mathbf{r}}}-\omega_0 t - \alpha_0  )}
\end{eqnarray}
with the unit-wave amplitude $a_0 = V^{-1/2}$ and the phase $\alpha_0=0$, where $V\neq \infty$ is the volume occupying by the spatially finite unit-field $\psi_0 (\mathbf{r},t)=a_0 e^{i{\mathbf{k}_0{\mathbf{r}}}}$. Notice, the unit-wave amplitude $a_0$ decreases with increasing the volume $V$. It could be also mentioned that the aforementioned amplitude $a_0 = V^{-1/2}$ of the unit-wave (\ref{eq53abc}) is different from the value $a_0=(2V_0 {\varepsilon}_0)^{-1/2}$ of the unit-wave (16) in Part I of the present study as well as from the amplitude  $a_{\bf k}=(2V_0 {\varepsilon}_{\bf k})^{-1/2}$ of the scalar bosons and anti-bosons in the global quantum fields (3) and (4) of Part I. In Eq. (\ref{eq53abc}), $\omega_0 ={\varepsilon}_0$ and $\mathbf{k}_0$ denote respectively the unit-wave frequency and wave-number. The unit-field (particle) has the momentum $\mathbf{k}_{0}$ given by
\begin{eqnarray} \label{eq54abc}
\mathbf{k}_0={{\int_{V}}  \psi_{0} ^* \left(-i\nabla {\psi_{0}} \right)d^3x}.
\end{eqnarray}
The unit-field frequency (energy) satisfies the Einstein energy-mass relation (\ref{eq1abc}), which has the form of the relativistic dispersion relation ${\omega_0}^2=k_0^2+m_0^2$. The unit-field propagates in the free-space along the direction $\mathbf{k}_0/k_0$ with the so-called group velocity $v_g = | \mathbf{v}_g | = d \omega_0 /dk_0$. The use of the Einstein dispersion relation yields the group velocity
\begin{eqnarray} \label{eq55abc}
v_g = 1/(1+m_0^2/\mathbf{k}_0^2)^{1/2}.
\end{eqnarray} 
Each point of the unit-field propagates within the unit-field with the phase velocity $\mathbf{v} = (\omega_0 /k_0)(\mathbf{k}_0 /k_0) $. The value $v=| \mathbf{v} |= \omega_0 /k_0$ is given by 
\begin{eqnarray} \label{eq56abc}
v = (1+m_0^2/\mathbf{k}_0^2)^{1/2}.
\end{eqnarray} 
The group velocity satisfies the inequality $v_g \leq c$ for any value of $m_0$ or $k_0$ (in the natural units, $v_g = (1+m_0^2/\mathbf{k}_0 ^2)^{1/2} \leq 1$). That means that a signal associated with the massive unit-field (particle) propagates through the true vacuum with the speed $v_g \leq c$. A massive particle is considered to be at the rest ($v_g =0$) if the particle momentum $k_0=0$. Each point of the massive unit-field propagates within the unit-field with the phase velocity (\ref{eq56abc}), where $v \rightarrow 0$ if the unit-field is associated with the relativistic particle (${k}_0 \rightarrow \infty$). In the case of the mass-less ($m_0=0$) unit-field, which is described by Eqs. (\ref{eq8abc}), (\ref{eq27abc}), (\ref{eq39abc}) and (\ref{eq45abc}) with $m_0=0$ or simply by Eqs. (\ref{eq15abc}) and (\ref{eq46abc}), the dispersion relation is given by ${\omega_0}^2=k_0^2$. The respective group velocity $v_g$ is equal to the {\it{speed of light}} ($v_g=c=1$). The phase velocity $v$ of the mass-less unit-field also is equal to the {\it{speed of light}} ($v=c=1$). Thus a signal associated with the mass-less unit-field (particle) propagates through both the true vacuum and the unit-field medium with the luminal velocity ($v_g=v=c=1$). In the non-relativistic case ($\mathbf{k}_0^2<< m_0^2$), the group velocity (\ref{eq55abc}) of the unit-field (\ref{eq53abc}) is equal to the {\it{non-relativistic (Newton) velocity}} $v_{g}=k_0/m_0$ of the non-relativistic particle associated with the unit-field. Notice, the similar dynamics of the the non-relativistic unit-field (\ref{eq53abc}) is predicted by the non-relativistic equations (\ref{eq13abc}), (\ref{eq14abc}), (\ref{eq32abc}), (\ref{eq39abc}) and (\ref{eq52abc}) with the respective non-relativistic dispersion relation ${\omega_0}=(k_0^2+m_0^2)^{1/2}\approx (k_0^2/2m_0) +m_0$. In such a case, the group and phase velocities are given by $v_{g}\approx k_0/m_0$ and $v\approx m_0/k_0$, respectively. The phase velocity of the non-relativistic unit-field determining by Eq. (\ref{eq56abc}) is also given by the value $v\approx m_0/k_0$. A signal associated with the phase velocity propagates through the unit-field having the momentum $k_0 \rightarrow 0$ with the {\it{infinite speed}} ($v \rightarrow \infty$). The unit-field configuration (\ref{eq53abc}) and momentum (\ref{eq54abc}) do not change under propagation of the unit-field in the true vacuum. Indeed, at any time moment $t = t'$, the unit-field has the configuration 
\begin{eqnarray} \label{eq57abc}
\psi_0 (\mathbf{r}',t')=a_0 e^{i({\mathbf{k}_0{\mathbf{r}}'}-\omega_0 t' )}
\end{eqnarray} 
with the unit-field momentum $\mathbf{k}_0$ and frequency $\omega_0$. Here, $a_0 = V^{-1/2}$ is the unit-field amplitude, and $V$ is the volume occupying by the unit-field at the moment $t = t'$. 

Let me now suppose that at the time moment $t=t''$ the unit-field $\psi_0 (\mathbf {r}'', t'')=a_0 e^{i({\mathbf{k}_0{\mathbf{r}}''}-\omega_0 t'' )}$ begins propagation through the "transparent" aperture in the "non-transparent" infinitely thin material screen placed in the true vacuum. Under the propagation, the unit-field interacts with the material boundaries of the screen. The evolution of the unit-field configuration behind the screen at the time $t>t''$ can be described using the well-known Fresnel-Kirchhoff integral theorem as   
\begin{eqnarray} \label{eq58abc}
{\psi_0 (P,t)} = {\frac {1} {4\pi}}{\int} {{\int}_S}\left[ {\psi_0}{\frac {\partial} {\partial n}} \left (  {\frac {e^{i(\mathbf{k}_0 \mathbf{R}-\omega_0 t)}} {R}} \right) - {\frac {e^{i(\mathbf{k}_0 \mathbf{R}-\omega_0 t)}} {R}}{\frac {\partial {\psi_0}} {\partial n}} \right] dS,
\end{eqnarray} 
where $R$ is the distance from the point $P$ to the point $(x'',y'',z'')$ of the unit-field $\psi_0 \equiv \psi_0 (\mathbf{r}'',t'')$ on the surface $S$ of the aperture, and $\partial / \partial n$ denotes differentiation along the normal to the surface of integration. The Fresnel-Kirchhoff integral theorem (\ref{eq58abc}) for the mass-less ($m_0=0$) wave is usually regarded as an integral form of the wave equation $\square {\psi_0} = 0$ with the dispersion relation ${\omega_0 }^2=\mathbf{k}_0 ^2$ and the conventional boundary conditions imposed by the "non-transparent" boundaries. The diffraction integral (\ref{eq58abc}) for an optical wave is considered as one of the possible mathematical formulations of the Huygens-Fresnel diffraction principle in optics. In the case of a massive ($m_0 \neq 0) $ unit-wave, the integral theorem (\ref{eq58abc}) could be regarded as an integral form of the equation $\square {\psi_0} + m_0^2 {\psi_0} = 0$ with the energy-mass (dispersion) relation ${\omega _0}^2=\mathbf{k}_0 ^2+m_0^2$. Respectively, Eq. (\ref{eq58abc}) could be considered as one of the possible mathematical formulations of the Huygens-Fresnel principle for diffraction of the massive unit-wave on the aperture. The expression (\ref{eq58abc}) describes the unit-field $\psi_0 (P,t)$ in any point $P$ behind the screen by using the unit-field distribution $\psi_0 (\mathbf{r}'', t'')$ with the amplitude $a_0$ on the surface $S$ of the aperture at the time moment $t=t''$. The interaction of the unit-field with the material boundaries is described {\it{phenomenologically}}, namely in terms of the boundary conditions imposed on the unit-field on the screen boundaries. The interaction is considered as an {\it{elastic}} process that does not change the absolute value $|\mathbf{k}_0|$ of the unit-field momentum. Thus the unit-fields (\ref{eq53abc}) and (\ref{eq57abc}) have the same value $\mathbf{k}_0$ of momentums, while the diffracted unit-field (\ref{eq58abc}) does have the definite absolute value $|\mathbf{k}_0|$ with the indefinite direction $\mathbf{k}_0 / |\mathbf{k}_0|$ of propagation. In other words, the diffracted unit-field ${\psi_0 (P,t)}$ is considered as the free unit-field $\psi_0 (\mathbf {r}'', t'')=a_0 e^{i({\mathbf{k}_0{\mathbf{r}}''}-\omega_0 t'')}$ reshaped by the aperture. The diffracted unit-field (\ref{eq58abc}) having the definite absolute value $|\mathbf{k}_0|$ propagates in the uncertain direction within the interval determining by the Heisenberg uncertainty relation $\Delta {\mathbf{k}_0} \Delta {\mathbf{x}} \sim 1$, where $|\Delta {\mathbf{x}}|$ is the aperture dimension. That gives the new physics inside the Heisenberg uncertainty principle. The reshaping can be interpreted as the signalling (exchange of the information) between the points $(x'',y'',z'')$ and $P$. The signal from the point $(x'',y'',z'')$ to the point $P$ of the unit-field (\ref{eq58abc}) propagates in the form of the spherical wave $R^{-1}e^{i(\mathbf{k}_0 \mathbf{R}-\omega_0 t)}$ with the phase velocity $v=\omega_0 /k_0=(1+m_0^2/ \mathbf{k}_0^2)^{1/2}$. In the case of a non-relativistic particle ($k_0\rightarrow 0$), the signal propagates {\it{within}} the unit-field (\ref{eq58abc}) with the phase velocity $v \rightarrow \infty$. It should be stressed that the unit-field configuration (\ref{eq58abc}) describes the {\it{indivisible}} unit-field. In other words, the unit-field (\ref{eq58abc}) consists of the Huygens-Fresnel superposition of the spherical elementary waves, which are {\it{inseparable}} from each other. A simple analysis shows that the above-considered dynamics of the unit-field configuration under the elastic collision of a unit-field with the material boundaries could be observed in many other experimental conditions, for instance, in the Young two-slit experiment. 

Consider now the third typical example of the evolution of the unit-field configuration under the elastic interaction of the single unit-field with a material boundary. Let me assume that the unit-field (\ref{eq53abc}) with the amplitude $a_0 = V^{-1/2}$ is placed into the center of an empty box, which has the volume $V_b=L_b^3>>V$ and the "non-transparent" material boundaries. After the time $ t \sim  L_b/v_g$, the unit-field would interact with the box boundary. The elastic interaction could be described {\it{phenomenologically}} by using the Neumann boundary condition imposed on the normal derivative of the unit-field function on the "non-transparent" boundaries. The elastic interaction of the unit-field with the box boundary results into the elastic reflection of the unit-field from the boundary that changes the momentum of the unit-field (\ref{eq53abc}) from $\mathbf{k}_0$ to $-\mathbf{k}_0$. After the reflection along the normal to the surface, the unit-field propagates in the direction $-{\mathbf{k}_0}/k_0$ as the unit-wave  
\begin{eqnarray} \label{eq59abc}
\psi_{0} (\mathbf{r},t)=a_0 e^{-i({\mathbf{k}_{0}{\mathbf{r}}}-\omega_{0} t )}.
\end{eqnarray}
The reflection may be attributed to the phase change of the unit-field (\ref{eq53abc}): $a_0 e^{-i({\mathbf{k}_{0}{\mathbf{r}}}-\omega_{0} t )}= a_0 e^{i({\mathbf{k}_{0}{\mathbf{r}}}-\omega_{0} t \pm \pi )}$. The total time of the reflection is given by $ t \sim V^{1/3} /v_g$. The unit-fields (\ref{eq53abc}) and (\ref{eq59abc}) have the group and phase velocities determined respectively by Eqs. (\ref{eq55abc}) and (\ref{eq56abc}). The localization (confinement) of the unit-field (\ref{eq53abc}) by the "non-transparent" material box with the volume $V_b=V=L^3$ results into the discretization of the unit-field momentum ${k}_{0}$ and frequency (energy) $\omega _0$ under the Dirichlet boundary conditions imposed on the unit-field on the "non-transparent" boundaries. In such a case, the unit-field could have one of the possible configurations
\begin{eqnarray} \label{eq60abc}
\psi_{0nml} (\mathbf{r},t)=L^{-3/2} e^{i({\mathbf{k}_{0nml}{\mathbf{r}}}-\omega_{0nml} t )}
\end{eqnarray}
with the squares of the wavenumber (momentum) and frequency (energy) given respectively by 
\begin{eqnarray} \label{eq61abc}
{\mathbf{k}}_{0nml}^2=(\pi /L)^2[n^{-2}+m^{-2}+l^{-2}]
\end{eqnarray}
and 
\begin{eqnarray} \label{eq62abc}
{\omega _{0mnl}}^2= \mathbf{k}_{0mnl} ^2+m_0^2,
\end{eqnarray}
where $n$, $m$ and $l$ are the non-zero integer numbers. The group and phase velocities of the unit-field (\ref{eq60abc}) are determined by Eqs. (\ref{eq55abc}) and (\ref{eq56abc}) with the discrete momentum (\ref{eq61abc}) and frequency (\ref{eq62abc}). The localization (compression) of the unit-field (\ref{eq53abc}) results also into the Heisenberg uncertainty relation (see, the relation (130) of Part I of the present study). The unit-field (\ref{eq60abc}) does not include the reflected wave (\ref{eq59abc}). That means that the velocity of the box is equal to the group velocity of the unit-field. The well-known ({\it{standing-wave}}) configuration $\psi_{0nml} (\mathbf{r},t)=(L^{-3/2}/2) [e^{i({\mathbf{k}_{0nml}{\mathbf{r}}}-\omega_{0nml} t )} + e^{-i({\mathbf{k}_{0nml}{\mathbf{r}}}-\omega_{0nml} t )}]$, which includes the reflected wave, has the group velocity $v_{g}=0$ corresponding to the zero momentum (\ref{eq54abc}). The elastic reflection of a unit-field and the spectral discretization of the wavenumber (momentum) and energy (frequency) could be associated with many other experimental conditions, where a unit-field interacts with the resonator-like boundaries. Such a behaviour of the de Broglie (quantum) wave is well-known in the canonical quantum mechanics.

The most unexpected behavior of a unit-field does associate with the superluminal values of the phase velocity (\ref{eq56abc}) at the particular experimental conditions. In the case of ${k}_0 \rightarrow 0$, a signal associated with the phase velocity can propagate {\it{within}} the massive ($m_0\neq 0$) unit-field with the {\it{infinite speed}} [$v = (1+m_0^2/ \mathbf{k}_0^2)^{1/2} \rightarrow \infty$]. In the classical (Young-type) diffraction experiment, any instant change of the unit-field boundaries by instant insertion of any additional object into the experiment should also result into superluminal modification of the diffracted non-relativistic massive unit-field. The modification of the mass-less ($m_0=0$) unit-field in such an experiment would take place with the speed of light $v=\omega_0 /k_0=c=1$. It should be stressed that the superluminal reshaping of a massive unit-field, which is provided with the superluminal phase velocity {\it{within}} the unit-field, does not contradict the Einstein special relativity. In the Einstein theory, the signaling between massive bodies (particles) separated by the empty, "straight" spacetime is provided by a mass-less electromagnetic wave propagating between the particles with the {\it{group speed of light}}. The superluminal reshaping of a massive unit-field with the superluminal phase velocity, which is somewhat similar to the instantaneous collapse of a wave function and the superluminar "quantum leap" in the Copenhagen interpretation of quantum mechanics, does not affect the signaling between massive particles (unit-fields) separated by the true vacuum. The Einstein signaling by a massive ($m_0\neq 0$) particle is associated with the group velocity (\ref{eq55abc}) of the massive unit-fields (\ref{eq53abc}) that is not superluminal [$v_g =1/(1+m_0^2/ \mathbf{k}_0^2)^{1/2}\leq 1$]. In agreement with the Einstein theory of relativity, any mass-less ($m_0=0$) unit-field of the present model propagates through the true vacuum with the {\it{group velocity of light}} ($v_g=v=c=1$). Notice, any change of the boundaries placed infinitely far from the spatially finite unit-field would not affect the configuration and momentum of the unit-field. The modification of a unit-field occurs only after interaction of the unit-field with the boundaries. 

The signaling by the massive unit-field (\ref{eq53abc}) associated with a massive ($m_0\neq 0$) particle is provided with the velocity (\ref{eq55abc}) that could not be superluminal [$v_g =1/(1+m_0^2/ \mathbf{k}_0^2)^{1/2}\leq 1$]. In contrast to the Einstein theory of point particles, the present model at the particular experimental conditions predicts the signalling by the material field with the {\it{superluminal group velocity}}. For an example, the spatially evanescent unit-field 
\begin{eqnarray} \label{eq63abc}
\psi_0 (\mathbf{r},t)=a_0 e^{i({\mathbf{k}_0{\mathbf{r}}}-\omega_0 t )}=a_0 e^{-{\tilde{\mathbf{k}}_0}{\mathbf{r}}} e^{-i\omega_0 t }
\end{eqnarray}
satisfies the equations of motion (\ref{eq8abc}) and (\ref{eq45abc}) under the dispersion relation 
\begin{eqnarray} \label{eq64abc}
{\omega_0 }^2 = - \tilde{\mathbf{k}}_0 ^2+m_0^2,
\end{eqnarray}
which is {\it{formally}} indistinguishable from the Einstein mas-energy relation ${\omega_0 }^2 = \mathbf{k}_0^2+m_0^2$ in the case of the imaginary momentum $\mathbf{k}_0=i \tilde{\mathbf{k}}_0$. The group velocity (\ref{eq55abc}) of the unit-field (\ref{eq63abc}) is given by 
\begin{eqnarray} \label{eq65abc}
v_g =1/(1-m_0^2/\tilde{\mathbf{k}}_0^2)^{1/2}.
\end{eqnarray}
The relativistic ($\tilde{\mathbf{k}}_0^2 \rightarrow m_0^2$) unit-wave (\ref{eq63abc}) propagates with the superluminal group velocity $v_g \rightarrow \infty$. The relativistic unit-field 
\begin{eqnarray} \label{eq66abc}
\psi_0 (\mathbf{r},t)=a_0 e^{i({\mathbf{k}_0{\mathbf{r}}}-\omega_0 t )}=a_0 e^{-{\tilde{\mathbf{k}}_0}{\mathbf{r}}} 
e^{-{{\tilde\omega}_0 t}}, 
\end{eqnarray}
which is evanescent in both the space and time, also propagates with the {\it{superluminal group velocity}} (\ref{eq65abc}) under the dispersion relation 
\begin{eqnarray} \label{eq67abc}
{\tilde\omega}_0^2=\tilde{\mathbf{k}}_0 ^2-m_0^2.
\end{eqnarray}
The dispersion (\ref{eq67abc}) {\it{formally}} satisfies the Einstein mas-energy relation ${\omega_0 }^2 = \mathbf{k}_0^2+m_0^2$ for the imaginary momentum $\mathbf{k}_0=i \tilde{\mathbf{k}}_0$ and frequency $\omega_0=-i {\tilde\omega}_0$. The evanescent unit-field (unit-wave), which has the imaginary values of the frequency (energy) and/or wave-number (momentum), does associate with the particle that may propagate with the {\it{superluminal group velocity}}. It could be mention that superluminal particles are usually called {\it{tachions}}. The dispersion relations (\ref{eq64abc}) and (\ref{eq67abc}), strictly speaking, are different from the Einstein mass-energy relation (\ref{eq1abc}) associated with the real frequency (energy), wave-number (momentum) and mass. Moreover, the temporal evanescence (virtual existence) of the unit-field (\ref{eq66abc}) associated with the disappearing particle does contradict the law of energy conservation. Therefore the superluminal unit-fields (\ref{eq63abc}) and (\ref{eq66abc}) and the respective superluminal particles (tachions) probably do not really exist. Nevertheless, {\it{the superluminal field $\psi(\mathbf{r},t)$, which is similar to the unit-field (\ref{eq63abc}) or (\ref{eq66abc}), may be created experimentally by using the method of Fourier's decomposition}}. In such a case, the composite-field  $\psi(\mathbf{r},t)= \Sigma _i a_i\psi_{0i} (\mathbf{r},t)$ having the superluminal field configuration is composed from the Fourier components (non-evanescent unit-fields $\psi_{0i} (\mathbf{r},t)$ associated with the temporally non-evanescent particles) that satisfy the Einstein mas-energy relation ${\omega_{0i} }^2 = \mathbf{k}_{0i}^2+m_{0i}^2$ with the real values ${\omega_{0i} }$, $\mathbf{k}_{0i}$ and $m_{0i}$. The temporal evanescence of the superluminal configuration associated with the  field superposition $\Sigma _i a_i\psi_{0i} (\mathbf{r},t)$ is provided rather by redistribution of the non-evanescent unit-fields $a_i\psi_{0i} (\mathbf{r},t)$ than the disappearance of these unit-fields (particles). The signalling by the composite, evanescent field (wave) having the {\it{superluminal group velocity}} could play important role in explanation of the many superluminal physical processes, such as the near-field diffraction (scattering) of waves, the tunnelling effects and the virtual-particle phenomena of any kind. It could be also mentioned that the superluminal signaling by the relativistic neutrinos has been recently observed in the CERN-LNGS experiment. The superlunimal signaling in this experiment, {\it{if such a behavior really exists}}, may be attributed to the superluminal group velocity ($v_g>1$) of the composite evanescent field associated with the neutrinos. 

\section{4. Unification of interacting elementary particles and interfering (cross-correlating) unit-fields: A multi-particle system ($N\geq 2$)}

In the above-presented conceptual model, a single elementary particle that is free from interaction with other particles has been presented as the free, indivisible unit-field associated with this particle. Unification of a free particle and a free unit-field was performed by the generalization of the energy-mass relation ${\varepsilon}^{2} ={\mathbf{k}}^2+m^2$ of the Einstein theory of a point particle to the case of the unit-field associated with the particle. Then the generalized mass-energy relation yielded the equation of motion for the unit-field. A unit-field was considered to be free if it interacts with the boundaries imposed by the free space (vacuum), only. It was also shown that the elastic interaction of a unit-field with the material boundaries can be treated phenomenologically by using the different boundary conditions for the equation of motion of a free unit-field (particle). In contrast to a free particle, the analysis of non-elastic interaction of a unit-field (particle) with material boundaries or another unit-field (particle) requires a detailed (microscopic) model of interaction of the unit-fields with each other. Section 4 unifies the fundamental (electromagnetic, weak, strong and gravitational) fields, particles and interactions by the further generalization of the Einstein energy-mass relation for the {\it{interacting particles and bodies}} composed from the {\it{interfering unit-fields}}. It is assumed that any fundamental field $\psi (\mathbf{r},t) =\sum_{n=1}^N{\psi _{0n}(\mathbf{r},t)}$ is composed from the {\it{interfering, indivisible unit-fields}} ${\psi _{0n}(\mathbf{r},t)}$ associated with the {\it{interacting elementary particles}}. Section 4 begins the unification with the generalization of the Einstein energy-mass relation for the composite field $\psi (\mathbf{r},t) ={\psi _{01}(\mathbf{r},t)}+ {\psi _{02}(\mathbf{r},t)}$ associated with the composite particle composed from two ($N=2$) indistinguishable elementary particles. Then the generalization of the Einstein energy-mass relation is performed for an arbitrary number $N$ of the unit-fields (elementary particles). Mathematically, the generalization of the Einstein energy-mass relation is performed by using the second or first derivatives of a unit-field. Although these two approaches are equivalent in the case of the de Broglie wave associated with a free particle, the use of the second or first derivatives may give the different results for the interfering unit-fields. For the sake of generality, the unification of interacting elementary particles and interfering (cross-correlating) unit-fields is presented also in the alternative form by using the composite-field Lagrangian (Hamiltonian) that corresponds to the generalized energy-mass relation for the interfering unit-fields. In such a case, the model is formulated in the frame of the Lagrangian formalism, where the configuration and dynamics of the composite field  is determined by the composite-field Lagrangian and the Euler-Lagrange equation of motion with the initial and boundary conditions imposed. 

\subsection{4.1. The energy-mass relation and equation of motion for a composite particle composed from interacting {\it{point-like particles}}: Interaction as cross-correlation (interference) of {\it{point-like particles}}}

For the sake of simplicity, let me first consider the simplest composite particle, namely the {\it{point-like particle}} composed from {\it{two point-particles}} of the Einstein theory of special relativity. It is assumed that the first point-like elementary particle has the mass $m_{01}$, momentum $\mathbf{k}_{01}$ and energy ${\varepsilon}_{01}$ in the spacetime point $(\mathbf{r}_{1},t_{1})$. The second elementary particle is characterized by the respective parameters $m_{02}$, $\mathbf{k}_{02}$, ${\varepsilon}_{02}$ and $(\mathbf{r}_{2},t_{2})$. In the case of $(\mathbf{r}_{1},t_{1})=(\mathbf{r}_{2},t_{2})=(\mathbf{r},t)$, the two identical [$m_{01}=m_{02}=m_0$ and $\mathbf{k}_{01}=\mathbf{k}_{02}=\mathbf{k}_0$] point-particles could be considered as the {\it{degenerate}}, composite point-particle with the common energy ${\varepsilon}$ determining by the energy superposition principle: 
\begin{eqnarray} \label{eq68abc}
{\varepsilon} = {\varepsilon}_{01}+{\varepsilon}_{02}. 
\end{eqnarray} 
The Einstein energy-mass relation (\ref{eq1abc}) for the degenerate, composite particle is given by 
\begin{eqnarray} \label{eq69abc}
{\varepsilon}^{2} = {\mathbf{k}}^2+m^2, 
\end{eqnarray} 
where the values ${\varepsilon}={\varepsilon}_{01}+{\varepsilon}_{02}=2{\varepsilon}_{0}$, $\mathbf{k}=\mathbf{k}_{01}+\mathbf{k}_{02}=2\mathbf{k}_{0}$ and $m = m_{01} + m_{02} = 2m_{0}$ are respectively the sums (superpositions) of the energies, momentums and rest-masses of the first and second particles. That means that the particles satisfy the superposition principle also for the momentums and masses. In the degenerate particle, the energies, momentums and rest-masses of the indistinguishable particles are additive. The basic relation (\ref{eq69abc}) of Einstein's relativity for the degenerate, composite point-particle located in the spatiotemporal point $(\mathbf{r},t)$ can be rewritten as 
\begin{eqnarray} \label{eq70abc}
{\varepsilon}^{2} = {\varepsilon}_{01}^{2} + {\varepsilon}_{02}^{2} + {\cal E}_{12}+{\cal E}_{21}={\varepsilon}_{01}^{2} + {\varepsilon}_{02}^{2} + {\cal E}_{12,21}. 
\end{eqnarray} 
Here, ${\varepsilon}\equiv {\varepsilon}(\mathbf{r},t)$ is the common energy of the composite particle, and $ {\varepsilon}_{01} \equiv {\varepsilon}_{01} (\mathbf{r},t) = ( \mathbf{k}_{01}^2 + m_{01}^2)^{1/2}$ and $ {\varepsilon}_{02} \equiv {\varepsilon}_{02} (\mathbf{r},t) = ( \mathbf{k}_{02}^2 + m_{02}^2)^{1/2}$ denote respectively the Einstein relativistic energies of the first and second particles. The value ${\cal E}_{12,21}(\mathbf{r},t)= {\cal E}_{12}(\mathbf{r},t)+{\cal E}_{21}(\mathbf{r},t)$ logically to call the {\it{total cross-correlation term}} or simply the {\it{cross-correlation term}}. The term associates with the cross-correlation ("interference") of the two point-particles in the spatiotemporal point $(\mathbf{r},t)$. The first cross-correlation term ${\cal E}_{12}(\mathbf{r},t)={\varepsilon}_{01} (\mathbf{r},t){\varepsilon}_{02} (\mathbf{r},t) $ and the second one ${\cal E}_{21}(\mathbf{r},t)={\varepsilon}_{02} (\mathbf{r},t){\varepsilon}_{01} (\mathbf{r},t)$ associated respectively with the first and second point-particles satisfy the relation 
\begin{eqnarray} \label{eq71abc}
{\cal E}_{12}={\cal E}_{21}=(1/2){\cal E}_{12,21}.
\end{eqnarray} 
The relation (\ref{eq70abc}) may be represented {\it{formally}} even in the more general form by using the notations ${\varepsilon}\equiv {\varepsilon}(\mathbf{r}_1, \mathbf{r}_2,t_1,t_2)$, $ {\varepsilon}_1 \equiv {\varepsilon}_{01} (\mathbf{r}_1 ,t_1) = ( \mathbf{k}_{01}^2 + m_{01}^2)^{1/2}$, $ {\varepsilon}_2 \equiv {\varepsilon}_{02} (\mathbf{r}_2 ,t_2) = ( \mathbf{k}_{02}^2 + m_{02}^2)^{1/2}$ and ${\cal E}_{12,21}\equiv {\cal E}_{12,21}(\mathbf{r}_1, \mathbf{r}_2,t_1,t_2)= {\varepsilon}_{01} (\mathbf{r}_1 ,t_1){\varepsilon}_{02} (\mathbf{r}_2 ,t_2) + {\varepsilon}_{02} (\mathbf{r}_2 ,t_2){\varepsilon}_{01} (\mathbf{r}_1 ,t_1) $, where the cross-correlation term ${\cal E}_{12,21}(\mathbf{r}_1, \mathbf{r}_2,t_1,t_2)$ does associate with the cross-correlation of the two particles in the points $\mathbf{r}_1$, $\mathbf{r}_2$, $t_1$ and $t_2$. It should be stressed, however, that Eq. (\ref{eq70abc}) could be used only in the case of the indistinguishable ($m_{01}=m_{02}=m_0$ and $\mathbf{k}_{01}=\mathbf{k}_{02}=\mathbf{k}_0$) point-like particles located in the same spacetime point $(\mathbf{r}_{1},t_{1})=(\mathbf{r}_{2},t_{2})=(\mathbf{r},t)$. Otherwise Eq. (\ref{eq70abc}) does not have any  physical sense. Although the cross-correlation term ${\cal E}_{12,21}={\varepsilon}_{01}{\varepsilon}_{02}+{\varepsilon}_{02}{\varepsilon}_{01}$ can be introduced formally into the Einstein energy-mass relation (see, Eq. \ref{eq70abc}), the cross-correlation term does not play any important role in the Einstein theory of special relativity. The cross-correlation term ${\cal E}_{12,21}={\varepsilon}_{01}{\varepsilon}_{02}+{\varepsilon}_{02}{\varepsilon}_{01}$ associated with the {\it{square of energies}} in Eq. (\ref{eq70abc}) does not result (${\varepsilon} \neq {\varepsilon}_{01}+{\varepsilon}_{02}+ {\varepsilon}_{12,21}$, where ${\varepsilon}_{12,21}\neq 0$) into the cross-correlation of {\it{energies}} [see, Eq. (\ref{eq68abc})]. That is to say that the Einstein special relativity describes the particles, which are free (${\varepsilon}_{12,21}= 0$) from the interaction (cross-correlation) energies. {\it{The cross-correlation ("interference") between the energies of point-particles does appear in the relativistic and non-relativistic theories of the interacting point-particles in the form of the interaction (potential) energy ${\varepsilon}_{12,21}\neq 0$.}} Indeed, in the all relativistic theories of {\it{interacting}} particles based on the Einstein theory of special relativity, the common (total) energy ${\varepsilon}$ of the composite point-particle composed at the time moment $t$ from the interacting point-particles is presented as 
\begin{eqnarray} \label{eq72abc}
{\varepsilon}(\mathbf{r}_1, \mathbf{r}_2,t) = {\varepsilon}_{01} (\mathbf{r}_1 ,t) + {\varepsilon}_{02} (\mathbf{r}_2 ,t) + {\varepsilon}_{12}(|\mathbf{r}_1-\mathbf{r}_2| ,t) +{\varepsilon}_{21}(|\mathbf{r}_2-\mathbf{r}_1| ,t) 
\end{eqnarray} 
or
\begin{eqnarray} \label{eq73abc}
{\varepsilon}(\mathbf{r}_1, \mathbf{r}_2,t) = {\varepsilon}_{01} (\mathbf{r}_1 ,t) + {\varepsilon}_{02} (\mathbf{r}_2 ,t) + {\varepsilon}_{12,21}(|\mathbf{r}_1-\mathbf{r}_2| ,t), 
\end{eqnarray} 
where the values $ {\varepsilon}_{01} (\mathbf{r}_1,t) = ( \mathbf{k}_{01}^2 + m_{01}^2)^{1/2}$ and $ {\varepsilon}_{02} (\mathbf{r}_2,t) = ( \mathbf{k}_{02}^2 + m_{02}^2)^{1/2}$ denote respectively the relativistic energies of the first and second point-particles;  the cross-correlation energy ${\varepsilon}_{12,21}(|\mathbf{r}_1-\mathbf{r}_2| ,t)={\varepsilon}_{12}(|\mathbf{r}_1-\mathbf{r}_2| ,t) +  {\varepsilon}_{21}(|\mathbf{r}_2-\mathbf{r}_1| ,t)$ denotes the interaction energy associated with the four fundamental interactions of nature, namely the gravitation, electromagnetism, weak interaction and strong interaction. One should not confuse here the cross-correlation energy ${\varepsilon}_{12,21}(\mathbf{r}_1, \mathbf{r}_2 ,t)$, which associates with the cross-correlation in the energies of two particles, with the cross-correlation term ${\cal E}_{12,21}(\mathbf{r}_1, \mathbf{r}_2 ,t)$ attributed to the cross-correlation in the square of energies. The {\it{interaction}} energy ${\varepsilon}_{12,21}(|\mathbf{r}_1-\mathbf{r}_2| ,t)$ is considered as the {\it{cross-correlation ("interference")}} energy ${\varepsilon}_{12,21}(\mathbf{r}_1, \mathbf{r}_2 ,t) ={\varepsilon}_{12,21}(|\mathbf{r}_1-\mathbf{r}_2| ,t)$ that does associate with the cross-correlation ("interference") of the two point-particles in the spatiotemporal points $(\mathbf{r}_1,t)$ and $(\mathbf{r}_2,t)$. The energies ${\varepsilon}_{12}$ and ${\varepsilon}_{21}$, which are attributed respectively to the cross-correlation of the first particle with the second one and {\it{vice versa}}, satisfy the relation 
\begin{eqnarray} \label{eq74abc}
{\varepsilon}_{12}={\varepsilon}_{21}=(1/2){\varepsilon}_{12,21},
\end{eqnarray} 
where ${\varepsilon}_{12,21}\neq 0$. In the non-relativistic ($\mathbf{k}_{01}^2 << m_{01}^2$) theories, the common energy ${\varepsilon}$ of the composite point-particle composed from the interacting point-particles is given by Eq. (\ref{eq73abc}), where the non-relativistic energies of the first ($ {\varepsilon}_{01} \approx  {\frac {\mathbf{k}_{01}^2} {2m_{01}} }+ m_{01}$) and second ($ {\varepsilon}_{02} \approx  {\frac {\mathbf{k}_{02}^2} {2m_{02}} }+ m_{02}$) point-particles are replaced by the energies $ {\varepsilon}_{1} = {\varepsilon}_{01} -m_{01}=  {\frac {\mathbf{k}_{01}^2} {2m_{01}} }$ and $ {\varepsilon}_{2} = {\varepsilon}_{02} - m_{02}= {\frac {\mathbf{k}_{02}^2} {2m_{02}} }$, respectively. In the non-relativistic and relativistic cases, the cross-correlation (interaction) energy ${\varepsilon}_{12,21}(|\mathbf{r}_1-\mathbf{r}_2| ,t)$ may have the same or different forms. The gradients of the {\it{interaction}} ({\it{cross-correlation}}) energy ${\varepsilon}_{12,21}(|\mathbf{r}_1-\mathbf{r}_2| ,t)$, which are called the {\it{interaction}} forces, are given by 
\begin{eqnarray} \label{eq75abc}
\mathbf{F}_{12}(\mathbf{r}_{1} ,t) =-{\frac {\partial}  {\partial \mathbf{R}_{12}}} {\varepsilon}_{12,21}(|\mathbf{R}_{12}| ,t)
\end{eqnarray} 
and 
\begin{eqnarray} \label{eq76abc}
\mathbf{F}_{21}(\mathbf{r}_{2} ,t) =-{\frac {\partial}  {\partial \mathbf{R}_{21}}} {\varepsilon}_{12,21}(|\mathbf{R}_{21}| ,t), 
\end{eqnarray} 
where $\mathbf{R}_{12}=\mathbf{r}_1-\mathbf{r}_2$ and $\mathbf{R}_{21}=\mathbf{r}_2-\mathbf{r}_1$. The {\it{interaction}} forces $\mathbf{F}_{12}$ and $\mathbf{F}_{21}$,
which could be considered as the {\it{cross-correlation ("interference")}} forces, act respectively upon the first and second point-particles. The {\it{interaction}} ({\it{cross-correlation}}) forces satisfy the relation
\begin{eqnarray} \label{eq77abc}
\mathbf{F}_{12}= - \mathbf{F}_{21}.
\end{eqnarray} 
due to the relation $\mathbf{R}_{12}=-\mathbf{R}_{21}$. The physical action of the first particle onto the second particle is indivisible from the action the second particle onto the first one. That is to say that the energy ${\varepsilon}_{12}$ is indivisible and indistinguishable from the energy ${\varepsilon}_{21}$. Therefore the interactive force $\mathbf{F}_{12}$ can not exist without existence of the interactive force $\mathbf{F}_{21}$. The kind of the interaction (cross-correlation) force acting between the {\it{two}} point-particles is determined by the kind of interaction (cross-correlation) energy and {\it{vice versa. }} Equations (\ref{eq75abc}) and (\ref{eq76abc}) do determine the four fundamental (g{\it{gravitational, electromagnetic, weak and strong}}) forces by the two components 
\begin{eqnarray} \label{eq78abc}
 \varepsilon_{12} = (1/2) \varepsilon_{12,21}
\end{eqnarray} 
and 
\begin{eqnarray} \label{eq79abc}
 \varepsilon_{21} = (1/2) \varepsilon _{12,21}
\end{eqnarray} 
of the interaction (cross-correlation) energy $\varepsilon_{12,21}$, where the energy $\varepsilon_{12,21}$ is associated with the {\it{gravitational, electromagnetic, weak and strong}} interactions. The energy of cross-correlation (interaction) and the respective force depend on the {\it{masses, charges, isospins, momentums and spins}} of the particles. Correspondingly, the kind of the point-like elementary particle is characterized by the {\it{mass, charge (electric charge, weak hyper-charge or color charge), weak isospin and intrinsic angular momentum (spin) }}of the particle. The equation of motion for the composite particle is derived by assuming that the variation of the energy [${\varepsilon}_{01} (\mathbf{r}_1 ,t) + {\varepsilon}_{02} (\mathbf{r}_2 ,t)$] is related to the variation of the total interaction (cross-correlation) energy ${\varepsilon}_{12,21}(|\mathbf{r}_1-\mathbf{r}_2| ,t)$ as    
\begin{eqnarray} \label{eq80abc}
\delta [{\varepsilon}_{01} (\mathbf{r}_1 ,t) + {\varepsilon}_{02} (\mathbf{r}_2 ,t)] = - \delta {\varepsilon}_{12,21}(|\mathbf{r}_1-\mathbf{r}_2| ,t)
\end{eqnarray} 
The relation (\ref{eq80abc}) is associated with the conservation of energy at the time moment $t$, where the variation of interaction (cross-correlation) energy $\delta {\varepsilon}_{12,21}(|\mathbf{r}_1-\mathbf{r}_2| ,t)$ is converted into the variation of the Einstein energy  $\delta [{\varepsilon}_{01} (\mathbf{r}_1 ,t) + {\varepsilon}_{02} (\mathbf{r}_2 ,t)]$  and {\it{vice versa}}. Notice, the two point-particles could be considered as the free particles that does not associate with the compose particle if ${\varepsilon}_{12,21}(|\mathbf{r}_1-\mathbf{r}_2| ,t) << [{\varepsilon}_{01} (\mathbf{r}_1 ,t) + {\varepsilon}_{02} (\mathbf{r}_2 ,t)] $. Equation (\ref{eq80abc}) yields the equation of motion, which can be presented in the general form as
\begin{eqnarray} \label{eq81abc}
{\frac {\partial {\varepsilon}_{01} (\mathbf{r}_1 ,t)}  {\partial \mathbf{r}_{1}}} {d \mathbf{r}_{1}} + {\frac {\partial {\varepsilon}_{02} (\mathbf{r}_2 ,t)}  {\partial \mathbf{r}_{2}}} {d \mathbf{r}_{2}} + {\frac {\partial {\varepsilon}_{01} (\mathbf{r}_1 ,t)}  {\partial t}} {d t} + {\frac {\partial {\varepsilon}_{02} (\mathbf{r}_2 ,t)}  {\partial t}} {d t} = \nonumber \\   = -{\frac {\partial {\varepsilon}_{12,21}(|\mathbf{r}_1-\mathbf{r}_2| ,t)}  {\partial \mathbf{r}_{1}}} {d \mathbf{r}_{1}} - {\frac {\partial {\varepsilon}_{12,21}(|\mathbf{r}_1-\mathbf{r}_2| ,t)}  {\partial \mathbf{r}_{2}}} {d \mathbf{r}_{2}} - {\frac {\partial {\varepsilon}_{12,21}(|\mathbf{r}_1-\mathbf{r}_2| ,t)}  {\partial t}} {d t} - {\frac {\partial {\varepsilon}_{12,21}(|\mathbf{r}_1-\mathbf{r}_2| ,t)}  {\partial t}} {d t}.            
\end{eqnarray} 
 The equation of motion (\ref{eq81abc}) may be simplified in some particular cases. For an example, in the case of 
\begin{eqnarray} \label{eq82abc}
\delta [{\varepsilon}_{01} (\mathbf{r}_1 ,t) + {\varepsilon}_{02} (\mathbf{r}_2 ,t)] = - \delta {\varepsilon}_{12,21}(|\mathbf{r}_1-\mathbf{r}_2| ,t)=0,
\end{eqnarray} 
Eq. (\ref{eq81abc}) describes the {\it{stationary condition}} of the composite particle in which the total energy ${\varepsilon}(\mathbf{r}_1, \mathbf{r}_2,t)$, the total interaction (cross-correlation) energy ${\varepsilon}_{12,21}(|\mathbf{r}_1-\mathbf{r}_2| ,t)$ and the Einstein energy  $[{\varepsilon}_{01} (\mathbf{r}_1 ,t) + {\varepsilon}_{02} (\mathbf{r}_2 ,t)]$ of point-particles do not depend on the time moment $t$. In other words, the stationary composite particle has the constant energies ${\varepsilon}(\mathbf{r}_1, \mathbf{r}_2)$, ${\varepsilon}_{12,21}(|\mathbf{r}_1-\mathbf{r}_2| )$, ${\varepsilon}_{01} (\mathbf{r}_1)$ and ${\varepsilon}_{02} (\mathbf{r}_2)$. Another typical example that associates with the composite particle is the movement of the first particle in the coordinate system, where the second particle is at the rest. In such a case, Eq. (\ref{eq81abc}) yields the equation of motion for the first particle:
\begin{eqnarray} \label{eq83abc}
{\frac {\partial {\mathbf{k}_{01}} (\mathbf{r}_1 ,t)}  {\partial t}} = \mathbf{F}_{12}(\mathbf{r}_{1} ,t), 
\end{eqnarray} 
where the interaction (cross-correlation) force acting upon the particle is given by 
\begin{eqnarray} \label{eq84abc}
 \mathbf{F}_{12}(\mathbf{r}_{1} ,t) =-{\frac {\partial}  {\partial \mathbf{r}_{1}}} {\varepsilon}_{12,21}(|\mathbf{r}_{1}| ,t). 
\end{eqnarray} 
It should be stressed that Eqs.  (\ref{eq72abc}) - (\ref{eq84abc}) in the case of gravitational interaction are valid for the Newton gravitation, but {\it{do not compare well}} with the Einstein general relativity. Indeed, the gravitational interaction between two particles in the Einstein relativity is not viewed as an interaction force mediated by the gradient of the interaction (cross-correlation) energy, but rather particles moving freely in gravitational fields travel under their own inertia in straight lines through "curve" spacetime. In other words, the force of gravity and the potential energy of gravitational interaction are explained as the pure geometrical result attributed to the geometry of spacetime.

The above-presented analysis, which considers the energies and forces associated with the {\it{two}} elementary point-particles, is valid also in the case of a composite particle composed from the $N>2$ point-particles. I will not present the multi-particle ($N>2$) analysis, because the equations that correspond to Eqs.  (\ref{eq68abc}) - (\ref{eq84abc}) for $N>2$ are well-known. Let me only mention several facts associated with the {\it{degenerate}} point-particle composed from the arbitrary number $N$ of the indistinguishable, interaction-less point-particles of the Einstein special relativity. The squares of energy, momentum and mass of the {\it{degenerate}} particle are given respectively by ${\varepsilon}^2=N^2{\varepsilon}_0^2$, ${\mathbf{k}}^2=N^2{\mathbf{k}}_0^2$ and $m^2=N^2m_0^2$, where ${\varepsilon}_0^2=\mathbf{k}_{0}^2+m_{0}^2$. Notice, the energy ${\varepsilon}=N{\varepsilon}_0$, momentum ${\mathbf{k}}=N{\mathbf{k}}_0$ and mass $m=Nm_0$ of the {\it{degenerate}} particle is similar to the energy and momentum of the $N$ bosons of the global infinite field describing by the Klein-Gordon-Fock equation of quantum field theory.

\subsection{4.2. The energy-mass relation and equation of motion for the interfering (cross-correlating) unit-fields associated with elementary particles}

\subsubsection{4.2.1. The model based on the {\bf{2nd}} derivatives of the composite field}

\vspace{0.4cm}

{\it{1. The model 1st-version based on the straightforward generalization of the Einstein energy-mass relation for the composite field by using the 2nd derivatives}}

\vspace{0.4cm}

The spatially-finite or infinite unit-fields $\psi_{01}(\mathbf{r}_1,t_1)$ and $\psi_{02}(\mathbf{r}_2,t_2)$ associated with the 1st and the 2nd elementary particles are not the Einstein point-like particles with the spateotemporal coordinates $(\mathbf{r}_1,t_1)$ and $(\mathbf{r}_2,t_2)$. {\it{Therefore Eqs. (\ref{eq68abc}) - (\ref{eq70abc}) should be generalized for the arbitrary spacetime coordinates $(\mathbf{r}_{1},t_{1})$ and $(\mathbf{r}_{2},t_{2})$ of the composite field $\psi = \psi (\mathbf{r}_1, \mathbf{r}_2,t_1,t_2)$ composed from the unit-fields $\psi _{01}(\mathbf{r}_1,t_1)$ and $\psi _{02}(\mathbf{r}_2,t_2)$.}} The common energy ${\varepsilon}$ of the composite particle may be attributed to the composite field $ \psi (\mathbf{r}_1, \mathbf{r}_2,t_1,t_2)={\psi _{01}(\mathbf{r}_1,t_1)}+{\psi _{02}(\mathbf{r}_2,t_2)}$ only in the case of $t_1=t_2=t$. Indeed, the cross-correlation of the unit-fields $ \psi _{01}(\mathbf{r}_1,t_1)$ and $\psi _{02}(\mathbf{r}_2,t_2)$ at the different ($t_1 \neq t_2$) time moments, in the common or different inertial coordinate systems, does associate rather with the amplitude or intensity interferometry (see, Part I) than the cross-correlation term ${\varepsilon}_{12,21}$ of the common energy ${\varepsilon}$ of the composite particle. At the time moment $t$ of the inertial coordinate system, which is the common system for the first and second unit-fields, the composite field $ \psi (\mathbf{r}_1, \mathbf{r}_2,t)$ may be presented as the field ${\psi (\mathbf{r}, t)}= \psi_{01} (\mathbf{r}_1,t) + \psi_{02} (\mathbf{r}_2,t) $ with the different ($\mathbf{r}_1 \neq \mathbf{r}_2$) space coordinates. The common field that has the common temporal coordinate may be presented also as the field ${\psi (\mathbf{r}, t)}= \psi_{01} (\mathbf{r},t) + \psi_{02} (\mathbf{r},t) $ with the common ($\mathbf{r}_1=\mathbf{r}_2=\mathbf{r}$) space coordinate. The common field $\psi (\mathbf{r},t)$ composed from the identical unit-fields $\psi _{01}(\mathbf{r},t)\equiv \psi (\mathbf{r}_1,t)$ and $\psi _{02} (\mathbf{r},t)\equiv \psi (\mathbf{r}_2,t)$ associated with the particles has the form 
\begin{eqnarray} \label{eq85abc}
{\psi (\mathbf{r}, t)}= \psi (\mathbf{r}_1,t) + \psi (\mathbf{r}_2,t) = {\psi _{01}(\mathbf{r},t)}+{\psi _{02}(\mathbf{r},t)}. 
\end{eqnarray}
Notice, the composite field (\ref{eq85abc}) is in agreement with the {\it{field superposition principle}} (see, Part I). The field superposition principle may be interpreted as a particular form of the {\it{energy superposition principle}} ${\varepsilon} ={\varepsilon}_{01}+{\varepsilon}_{02}$ for two point particles, which are considered as the composite particle with the common energy ${\varepsilon}$ determining by the energy superposition principle ${\varepsilon} = {\varepsilon}_{01}+{\varepsilon}_{02}$. 

The generalization of the Einstein energy-mass relation (\ref{eq1abc}) for the composite field (\ref{eq85abc}) of the physical matter, which is associated with the energies and masses of the first and second particles in the common volume $V$, is perform by the replacement  
\begin{eqnarray} \label{eq86abc}
\psi_0 (\mathbf{r},t) \rightarrow {\psi (\mathbf{r}, t)},
\end{eqnarray} 
where the unit-field $\psi_0 (\mathbf{r},t)$ is replaced by the composite field (\ref{eq85abc}) in the all equations of Sec. 3 that include the field $\psi_0 (\mathbf{r},t)$. 
This procedure first yields the replacements  
\begin{eqnarray} \label{eq87abc}
{\dot {\psi_0}} (\mathbf{r},t) \rightarrow [{\dot \psi_{01}} (\mathbf{r},t) + {\dot \psi_{02}} (\mathbf{r},t)]
\end{eqnarray} 
and
\begin{eqnarray} \label{eq88abc}
\nabla {\psi_0 (\mathbf{r},t)} \rightarrow [\nabla {\psi_{01} (\mathbf{r},t)} + \nabla {\psi_{02} (\mathbf{r},t)}]
\end{eqnarray} 
and then leads to the energy-mass relation with the respective equation of motion for the composite field associated with the composite particle. 

For the superposition (\ref{eq85abc}), the replacements (\ref{eq86abc}) - (\ref{eq88abc}) in Eqs. (\ref{eq3abc}) - (\ref{eq7abc}) yielded the energy-mass relation 
\begin{eqnarray} \label{eq89abc}
{\varepsilon}^{2} = {\varepsilon}_{01}^{2} + {\varepsilon}_{02}^{2} + {\cal E}_{12}+{\cal E}_{21}={\varepsilon}_{01}^{2} + {\varepsilon}_{02}^{2} + {\cal E}_{12,21}, 
\end{eqnarray} 
where
\begin{eqnarray} \label{eq90abc}
{\varepsilon}_{01}^{2} = {{{\frac 1 2}\int_{V}}  \psi_{01} ^* \left( -\ddot{\psi}_{01} - \nabla ^2{\psi_{01}} + m_{0}^2\psi_{01}\right)d^3x},
\end{eqnarray}
\begin{eqnarray} \label{eq91abc}
{\varepsilon}_{02}^{2} = {{{\frac 1 2}\int_{V}}  \psi_{02} ^* \left( -\ddot{\psi}_{02} - \nabla ^2{\psi_{02}} + m_{0}^2\psi_{02}\right)d^3x},
\end{eqnarray}
\begin{eqnarray} \label{eq92abc}
{\cal E}_{12}= {{{\frac 1 2}\int_{V}}  \psi_{01} ^* \left( -\ddot{\psi}_{02} - \nabla ^2{\psi_{02}} + m_{0}^2\psi_{02}\right)d^3x},
\end{eqnarray}
\begin{eqnarray} \label{eq93abc}
{\cal E}_{21}= {{{\frac 1 2}\int_{V}}  \psi_{02} ^* \left( -\ddot{\psi}_{01} - \nabla ^2{\psi_{01}} + m_{0}^2\psi_{01}\right)d^3x},
\end{eqnarray}
and ${\cal E}_{12,21}\equiv {\cal E}_{12}+{\cal E}_{21}$. One can easily demonstrate the {\it{very important inequality  ${\cal E}_{12,21}\leq {\varepsilon}_{01}^{2} + {\varepsilon}_{02}^{2}$}} associated with the interaction of two unit-fields (particles). Notice, Eq. (\ref{eq89abc}) is indistinguishable from Eq. (\ref{eq70abc}). For the composite field (\ref{eq85abc}), the relativistic equation of motion 
\begin{eqnarray} \label{eq94abc}
\square \cdot [{\psi}_{01}+{\psi}_{02}] + m_0^2 \cdot [{\psi}_{01}+{\psi}_{02}]=0
\end{eqnarray}
 is obtained by the replacements (\ref{eq86abc}) - (\ref{eq88abc}) in Eq. (\ref{eq8abc}). For the composite field 
\begin{eqnarray} \label{eq95abc}
\psi (\mathbf{r},t)=\sum_{n=1}^N{\psi _{0n}(\mathbf{r},t)},
\end{eqnarray}
which is composed from the $N$ unit-fields $\psi _{0n}(\mathbf{r},t)$ associated with the $N$ particles having the masses $m_{0n}=m_{0}$, the above-described procedure yields the energy-mass relation
\begin{eqnarray} \label{eq96abc}
{\varepsilon}^2 = \sum_{n=1}^N{{\varepsilon}}_{0n}^2+\sum_{n\neq m}^{N^2-N}{{\cal E}}_{nm},
\end{eqnarray}
where
\begin{eqnarray} \label{eq97abc}
{\varepsilon}_{0n}^{2} = {{{\frac 1 2}\int_{V}}  \psi_{0n} ^* \left( -\ddot{\psi}_{0n} - \nabla ^2{\psi_{0n}} + m_{0}^2\psi_{0n}\right)d^3x}
\end{eqnarray}
and
\begin{eqnarray} \label{eq98abc}
{\cal E}_{nm}= {{{\frac 1 2}\int_{V}}  \psi_{0n} ^* \left( -\ddot{\psi}_{0m} - \nabla ^2{\psi_{0m}} + m_{0}^2\psi_{0m}\right)d^3x}.
\end{eqnarray}
The respective relativistic equation of motion is given by
\begin{eqnarray} \label{eq99abc}
\square \cdot \left[ \sum_{n=1}^N{\psi _{0n}(\mathbf{r},t)} \right]+ m_0^2 \cdot \left[ \sum_{n=1}^N{\psi _{0n}(\mathbf{r},t)}\right]=0.
\end{eqnarray}
 
In the case of the composite particle composed from the non-relativistic particles, the replacements (\ref{eq86abc}) - (\ref{eq88abc}) in Eqs. (\ref{eq10abc}) - (\ref{eq14abc}) yielded the non-relativistic energy-mass relation
\begin{eqnarray} \label{eq100abc}
{\varepsilon} = {\varepsilon}_{01}  + {\varepsilon}_{02}  + {\varepsilon}_{12} +{\varepsilon}_{21}= {\varepsilon}_{01}  + {\varepsilon}_{02}  + {\varepsilon}_{12,21}
\end{eqnarray} 
with the respective non-relativistic equation of motion
\begin{eqnarray} \label{eq101abc}
i[\dot{\psi}_{01}+\dot{\psi}_{02}]= -{\frac 1 {2m_0} } {\nabla ^2 [{\psi}_{01}+{\psi}_{02}]} + m_0[{\psi}_{01}+{\psi}_{02}],
\end{eqnarray}
where
\begin{eqnarray} \label{eq102abc}
{\varepsilon}_{01} =  {{{\frac 1 2}\int_{V}}  \psi_{01} ^* \left( i\dot{\psi}_{01}-{\frac 1 {2m_0} } {\nabla ^2{\psi_{01}}} + m_0\psi_{01}\right)d^3x},
\end{eqnarray}
\begin{eqnarray} \label{eq103abc}
{\varepsilon}_{02} =  {{{\frac 1 2}\int_{V}}  \psi_{02} ^* \left( i\dot{\psi}_{02}-{\frac 1 {2m_0} } {\nabla ^2{\psi_{02}}} + m_0\psi_{02}\right)d^3x},
\end{eqnarray}
\begin{eqnarray} \label{eq104abc}
{\varepsilon}_{12} =  {{{\frac 1 2}\int_{V}}  \psi_{01} ^* \left( i\dot{\psi}_{02}-{\frac 1 {2m_0} } {\nabla ^2{\psi_{02}}} + m_0\psi_{02}\right)d^3x},
\end{eqnarray}
\begin{eqnarray} \label{eq105abc}
{\varepsilon}_{21} = {{{\frac 1 2}\int_{V}}  \psi_{02} ^* \left( i\dot{\psi}_{01}-{\frac 1 {2m_0} } {\nabla ^2{\psi_{01}}} + m_0\psi_{01}\right)d^3x},
\end{eqnarray}
and ${\varepsilon}_{12,21}\equiv {\varepsilon}_{12} +{\varepsilon}_{21}$. Notice, Eq. (\ref{eq100abc}) is indistinguishable from Eqs. (\ref{eq72abc}) and (\ref{eq73abc}). For the composite field (\ref{eq95abc}), the procedure yields 
the non-relativistic energy-mass relation
\begin{eqnarray} \label{eq106abc}
{\varepsilon} = \sum_{n=1}^N{{\varepsilon}}_{0n}+\sum_{n\neq m}^{N^2-N}{{\varepsilon}}_{nm}, 
\end{eqnarray} 
with the respective non-relativistic equation of motion
\begin{eqnarray} \label{eq107abc}
i \left[ \sum_{n=1}^N{\dot{\psi} _{0n}(\mathbf{r},t)} \right] = -{\frac 1 {2m_0} } {\nabla ^2 \left[ \sum_{n=1}^N{\psi _{0n}(\mathbf{r},t)}\right]} + m_0\left[ \sum_{n=1}^N{\psi _{0n}(\mathbf{r},t)}\right],
\end{eqnarray}
where
\begin{eqnarray} \label{eq108abc}
{\varepsilon}_{0n} =  {{{\frac 1 2}\int_{V}}  \psi_{0n} ^* \left( i\dot{\psi}_{0n}-{\frac 1 {2m_0} } {\nabla ^2{\psi_{0n}}} + m_0\psi_{0n}\right)d^3x}
\end{eqnarray}
and
\begin{eqnarray} \label{eq109abc}
{\varepsilon}_{nm} = {{{\frac 1 2}\int_{V}}  \psi_{0n} ^* \left( i\dot{\psi}_{0m}-{\frac 1 {2m_0} } {\nabla ^2{\psi_{0m}}} + m_0\psi_{0m}\right)d^3x}.
\end{eqnarray}

\vspace{0.4cm}

2. {\it{The model 2nd-version based on the generalization of the Einstein energy-mass relation for the composite field by using the Euler-Lagrange formalism and the 2nd derivatives}}

\vspace{0.4cm}

For the composite field (\ref{eq86abc}) composed from the two unit-fields, the replacements (\ref{eq86abc}) - (\ref{eq88abc}) in Eqs. (\ref{eq17abc}) - (\ref{eq25abc}) yielded the relativistic energy-mass relation (\ref{eq89abc}), where
\begin{eqnarray} \label{eq110abc}
{\varepsilon}_{01}^{2} =  {{{\frac 1 2}\int_{V}}  \psi_{01} ^* \left( -\ddot{\psi}_{01} - \nabla ^2{\psi_{01}} + m_{0}^2\psi_{01}\right)d^3x},
\end{eqnarray}
\begin{eqnarray} \label{eq111abc}
{\varepsilon}_{02}^{2} = {{{\frac 1 2}\int_{V}}  \psi_{02} ^* \left( -\ddot{\psi}_{02} - \nabla ^2{\psi_{02}} + m_{0}^2\psi_{02}\right)d^3x},
\end{eqnarray}
\begin{eqnarray} \label{eq112abc}
{\cal E}_{12}= {{{\frac 1 2}\int_{V}}  \psi_{01} ^* \left( -\ddot{\psi}_{02} - \nabla ^2{\psi_{02}} + m_{0}^2\psi_{02}\right)d^3x}
\end{eqnarray}
and 
\begin{eqnarray} \label{eq113abc}
{\cal E}_{21}=  {{{\frac 1 2}\int_{V}}  \psi_{02} ^* \left( -\ddot{\psi}_{01} - \nabla ^2{\psi_{01}} + m_{0}^2\psi_{01}\right)d^3x},
\end{eqnarray}
where ${\cal E}_{12,21}\equiv {\cal E}_{12}+{\cal E}_{21} \leq {\varepsilon}_{01}^{2} + {\varepsilon}_{02}^{2}$. The respective relativistic equation of motion  (\ref{eq94abc}) is obtained by the replacements (\ref{eq86abc}) - (\ref{eq88abc}) in Eqs. (\ref{eq26abc}) and (\ref{eq27abc}). For the composite field (\ref{eq95abc}), which is composed from the $N$ unit-fields $\psi _{0n}(\mathbf{r},t)$ associated with the $N$ particles, the procedure yields the relativistic energy-mass relation (\ref{eq96abc}), 
where
\begin{eqnarray} \label{eq114abc}
{\varepsilon}_{0n}^{2} = {{{\frac 1 2}\int_{V}}  \psi_{0n} ^* \left( -\ddot{\psi}_{0n} - \nabla ^2{\psi_{0n}} + m_{0}^2\psi_{0n}\right)d^3x} 
\end{eqnarray}
and
\begin{eqnarray} \label{eq115abc}
{\cal E}_{nm}= {{{\frac 1 2}\int_{V}}  \psi_{0n} ^* \left( -\ddot{\psi}_{0m} - \nabla ^2{\psi_{0m}} + m_{0}^2\psi_{0m}\right)d^3x}.
\end{eqnarray}
The respective relativistic equation of motion is given by  Eq. (\ref{eq99abc}).
 
In the case of the composite particle composed from the two non-relativistic particles, the replacements (\ref{eq86abc}) - (\ref{eq88abc}) in Eqs. (\ref{eq28abc}) - (\ref{eq32abc}) yielded the non-relativistic energy-mass relation (\ref{eq100abc}) with the respective non-relativistic equation of motion (\ref{eq101abc}), where
\begin{eqnarray} \label{eq116abc}
{\varepsilon}_{01} = {{{\frac 1 2}\int_{V}}  \psi_{01} ^* \left( i\dot{\psi}_{01}-{\frac 1 {2m_0} } {\nabla ^2{\psi_{01}}} + m_0\psi_{01}\right)d^3x},
\end{eqnarray}
\begin{eqnarray} \label{eq117abc}
{\varepsilon}_{02} =  {{{\frac 1 2}\int_{V}}  \psi_{02} ^* \left( i\dot{\psi}_{02}-{\frac 1 {2m_0} } {\nabla ^2{\psi_{02}}} + m_0\psi_{02}\right)d^3x},
\end{eqnarray}
\begin{eqnarray} \label{eq118abc}
{\varepsilon}_{12} =  {{{\frac 1 2}\int_{V}}  \psi_{01} ^* \left( i\dot{\psi}_{02}-{\frac 1 {2m_0} } {\nabla ^2{\psi_{02}}} + m_0\psi_{02}\right)d^3x}
\end{eqnarray}
and 
\begin{eqnarray} \label{eq119abc}
{\varepsilon}_{21} =  {{{\frac 1 2}\int_{V}}  \psi_{02} ^* \left( i\dot{\psi}_{01}-{\frac 1 {2m_0} } {\nabla ^2{\psi_{01}}} + m_0\psi_{01}\right)d^3x}.
\end{eqnarray}
For the composite field (\ref{eq95abc}), which is composed from the $N$ unit-fields $\psi _{0n}(\mathbf{r},t)$, the replacements (\ref{eq86abc}) - (\ref{eq88abc}) in Eqs. (\ref{eq28abc}) - (\ref{eq32abc}) yielded the non-relativistic energy-mass relation (\ref{eq106abc}) with the respective non-relativistic equation of motion
(\ref{eq107abc}), where
\begin{eqnarray} \label{eq120abc}
{\varepsilon}_{0n} =  {{{\frac 1 2}\int_{V}}  \psi_{0n} ^* \left( i\dot{\psi}_{0n}-{\frac 1 {2m_0} } {\nabla ^2{\psi_{0n}}} + m_0\psi_{0n}\right)d^3x}
\end{eqnarray}
and
\begin{eqnarray} \label{eq121abc}
{\varepsilon}_{nm} =  {{{\frac 1 2}\int_{V}}  \psi_{0n} ^* \left( i\dot{\psi}_{0m}-{\frac 1 {2m_0} } {\nabla ^2{\psi_{0m}}} + m_0\psi_{0m}\right)d^3x}.
\end{eqnarray}

\subsubsection{4.2.2. The model based on the {\bf{1st}} derivatives of the composite field}

\vspace{0.4cm}

{\it{1. The model 3rd-version based on the straightforward generalization of the Einstein energy-mass relation for the composite field by using the 1st derivatives}}

\vspace{0.4cm}

For the superposition (\ref{eq85abc}), the replacements (\ref{eq86abc}) - (\ref{eq88abc}) in Eqs. (\ref{eq34abc}) - (\ref{eq36abc}) yielded the energy-mass relation 
\begin{eqnarray} \label{eq122abc}
{\varepsilon}^{2} = {\varepsilon}_{01}^{2} + {\varepsilon}_{02}^{2} + {\cal E}_{12}+{\cal E}_{21}={\varepsilon}_{01}^{2} + {\varepsilon}_{02}^{2} + {\cal E}_{12,21}, 
\end{eqnarray} 
where
\begin{eqnarray} \label{eq123abc}
{\varepsilon}_{01}^{2} = {{{\frac 1 2}\int_{V}} \left(  \dot{\psi_{01} ^*} \dot{\psi_{01}} +  \nabla \psi_{01}^*  \nabla \psi_{01} + m_0^2 \psi_{01} ^* \psi_{01}   \right)d^3x},
\end{eqnarray}
\begin{eqnarray} \label{eq124abc}
{\varepsilon}_{02}^{2} = {{{\frac 1 2}\int_{V}} \left(  \dot{\psi_{02} ^*} \dot{\psi_{02}} +  \nabla \psi_{02}^*  \nabla \psi_{02} + m_0^2 \psi_{02} ^* \psi_{02}   \right)d^3x},
\end{eqnarray}
\begin{eqnarray} \label{eq125abc}
{\cal E}_{12}= {{{\frac 1 2}\int_{V}}  \left(  \dot{\psi_{01} ^*} \dot{\psi_{02}} +  \nabla \psi_{01}^*  \nabla \psi_{02} + m_0^2 \psi_{01} ^* \psi_{02}   \right)d^3x}
\end{eqnarray}
and 
\begin{eqnarray} \label{eq126abc}
{\cal E}_{21}= {{{\frac 1 2}\int_{V}}  \left(  \dot{\psi_{02} ^*} \dot{\psi_{01}} +  \nabla \psi_{02}^*  \nabla \psi_{01} + m_0^2 \psi_{02} ^* \psi_{01}   \right)d^3x},
\end{eqnarray}
where ${\cal E}_{12,21}\equiv {\cal E}_{12}+{\cal E}_{21} \leq {\varepsilon}_{01}^{2} + {\varepsilon}_{02}^{2}$.  Notice, Eq. (\ref{eq122abc}) is indistinguishable from Eqs. (\ref{eq89abc}) and (\ref{eq70abc}). The respective relativistic equation of motion 
\begin{eqnarray} \label{eq127abc}
[\dot{\psi_{01} ^*}+\dot{\psi_{02} ^*}] [\dot{\psi_{01}}+\dot{\psi_{02}}] = [\nabla \psi_{01}^* +\nabla  \psi_{02}^*]  [\nabla \psi_{01} +\nabla  \psi_{02}] + m_0^2 [ \psi_{01}^* + \psi_{02}^*]  [ \psi_{01} +  \psi_{02}],
\end{eqnarray}
is obtained by the replacements (\ref{eq86abc}) - (\ref{eq88abc}) in Eq. (\ref{eq37abc}). For the composite field 
\begin{eqnarray} \label{eq128abc}
\psi (\mathbf{r},t)=\sum_{n=1}^N{\psi _{0n}(\mathbf{r},t)},
\end{eqnarray}
which is composed from the $N$ unit-fields $\psi _{0n}(\mathbf{r},t)$ associated with the $N$ particles having the masses $m_{0n}=m_{0}$, the above-described procedure yields the energy-mass relation
\begin{eqnarray} \label{eq129abc}
{\varepsilon}^2 = \sum_{n=1}^N{{\varepsilon}}_{0n}^2+\sum_{n\neq m}^{N^2-N}{{\cal E}}_{nm},
\end{eqnarray}
where
\begin{eqnarray} \label{eq130abc}
{\varepsilon}_{0n}^{2} = {{{\frac 1 2}\int_{V}}  \left(  \dot{\psi_{0n} ^*} \dot{\psi_{0n}} +  \nabla \psi_{0n}^*  \nabla \psi_{0n} + m_0^2 \psi_{0n} ^* \psi_{0n}   \right)d^3x}
\end{eqnarray}
and
\begin{eqnarray} \label{eq131abc}
{\cal E}_{nm}= {{{\frac 1 2}\int_{V}}  \left(  \dot{\psi_{0n} ^*} \dot{\psi_{0m}} +  \nabla \psi_{0n}^*  \nabla \psi_{0m} + m_0^2 \psi_{0n} ^* \psi_{0m}   \right)d^3x}.
\end{eqnarray}
The respective relativistic equation of motion is given by
\begin{eqnarray} \label{eq132abc}
\left[ \sum_{n=1}^N{\dot{\psi_{0n} ^*}(\mathbf{r},t)} \right] \left[ \sum_{n=1}^N{\dot{\psi_{0n} }(\mathbf{r},t)} \right] = \left[ \sum_{n=1}^N{\nabla{\psi_{0n} ^*}(\mathbf{r},t)} \right] \left[ \sum_{n=1}^N{\nabla{\psi_{0n} }(\mathbf{r},t)} \right]+ m_0^2 \left[ \sum_{n=1}^N{{\psi_{0n} ^*}(\mathbf{r},t)} \right] \left[ \sum_{n=1}^N{{\psi_{0n} }(\mathbf{r},t)} \right]=0.
\end{eqnarray}
Notice, he relativistic energy-mass relation (\ref{eq129abc}) is indistinguishable from Eq. (\ref{eq96abc}). 
 
In the case of the composite particle composed from the non-relativistic particles, the replacements (\ref{eq86abc}) - (\ref{eq88abc}) in Eqs. (\ref{eq38abc}) and (\ref{eq39abc}) yielded the non-relativistic energy-mass relation
\begin{eqnarray} \label{eq133abc}
{\varepsilon} = {\varepsilon}_{01}  + {\varepsilon}_{02}  + {\varepsilon}_{12} +{\varepsilon}_{21}= {\varepsilon}_{01}  + {\varepsilon}_{02}  + {\varepsilon}_{12,21} 
\end{eqnarray} 
with the respective non-relativistic equation of motion
\begin{eqnarray} \label{eq134abc}
i[\psi_{01} ^* +\psi_{02} ^* ] [ \dot{\psi}_{01} +  \dot{\psi}_{01}]={\frac 1 {2m_0} } [\nabla \psi_{01}^* +\nabla  \psi_{02}^*]  [\nabla \psi_{01} +\nabla  \psi_{02}] + m_0[ \psi_{01}^* + \psi_{02}^*]  [ \psi_{01} +  \psi_{02}],
\end{eqnarray}
where
\begin{eqnarray} \label{eq135abc}
{\varepsilon}_{01}={{{\frac 1 2}\int_{V}}  \left( i\psi_{01} ^*  \dot{\psi}_{01}+{\frac 1 {2m_0} } {\nabla {\psi_{01}^*}\nabla {\psi_{01}}} + m_0\psi_{01} ^*  \psi_{01}\right)d^3x}.
\end{eqnarray}
\begin{eqnarray} \label{eq136abc}
{\varepsilon}_{02}={{{\frac 1 2}\int_{V}}  \left( i\psi_{02} ^*  \dot{\psi}_{02}+{\frac 1 {2m_0} } {\nabla {\psi_{02}^*}\nabla {\psi_{02}}} + m_0\psi_{02} ^*  \psi_{02}\right)d^3x}.
\end{eqnarray}
\begin{eqnarray} \label{eq137abc}
{\varepsilon}_{12}={{{\frac 1 2}\int_{V}}  \left( i\psi_{01} ^*  \dot{\psi}_{02}+{\frac 1 {2m_0} } {\nabla {\psi_{01}^*}\nabla {\psi_{02}}} + m_0\psi_{01} ^*  \psi_{02}\right)d^3x}.
\end{eqnarray}
and 
\begin{eqnarray} \label{eq138abc}
{\varepsilon}_{21}={{{\frac 1 2}\int_{V}}  \left( i\psi_{02} ^*  \dot{\psi}_{01}+{\frac 1 {2m_0} } {\nabla {\psi_{02}^*}\nabla {\psi_{01}}} + m_0\psi_{02} ^*  \psi_{01}\right)d^3x}.
\end{eqnarray}
Notice, Eq. (\ref{eq134abc}) is indistinguishable from Eqs. (\ref{eq100abc}), (\ref{eq72abc}) and (\ref{eq73abc}). For the composite field (\ref{eq95abc}), the procedure yields 
the non-relativistic energy-mass relation
\begin{eqnarray} \label{eq139abc}
{\varepsilon} = \sum_{n=1}^N{{\varepsilon}}_{0n}+\sum_{n\neq m}^{N^2-N}{{\varepsilon}}_{nm}, 
\end{eqnarray} 
with the respective non-relativistic equation of motion
\begin{eqnarray} \label{eq140abc}
i\left[ \sum_{n=1}^N{{\psi_{0n} ^*}(\mathbf{r},t)} \right] \left[ \sum_{n=1}^N{\dot{\psi_{0n} }(\mathbf{r},t)} \right]={\frac 1 {2m_0} } \left[ \sum_{n=1}^N{\nabla{\psi_{0n} ^*}(\mathbf{r},t)} \right] \left[ \sum_{n=1}^N{\nabla{\psi_{0n} }(\mathbf{r},t)} \right]+ m_0 \left[ \sum_{n=1}^N{{\psi_{0n} ^*}(\mathbf{r},t)} \right] \left[ \sum_{n=1}^N{{\psi_{0n} }(\mathbf{r},t)} \right],
\end{eqnarray}
where
\begin{eqnarray} \label{eq141abc}
{\varepsilon}_{0n}={{{\frac 1 2}\int_{V}}  \left( i\psi_{0n} ^*  \dot{\psi}_{0n}+{\frac 1 {2m_0} } {\nabla {\psi_{0n}^*}\nabla {\psi_{0n}}} + m_0\psi_{0n} ^*  \psi_{0n}\right)d^3x}.
\end{eqnarray}
and
\begin{eqnarray} \label{eq142abc}
{\varepsilon}_{nm}={{{\frac 1 2}\int_{V}}  \left( i\psi_{0n} ^*  \dot{\psi}_{0m}+{\frac 1 {2m_0} } {\nabla {\psi_{0n}^*}\nabla {\psi_{0m}}} + m_0\psi_{0n} ^*  \psi_{0m}\right)d^3x}.
\end{eqnarray}

\vspace{0.4cm}

2. {\it{The model 4th-version based on the generalization of the Einstein energy-mass relation for the composite field by using the Euler-Lagrange formalism and the 1st derivatives}}

\vspace{0.4cm}

For the composite field (\ref{eq86abc}) composed from the two unit-fields, the replacements (\ref{eq86abc}) - (\ref{eq88abc}) in Eqs. (\ref{eq40abc}) - (\ref{eq44abc}) yielded the {\it{relativistic}} energy-mass relation
\begin{eqnarray} \label{eq143abc}
{\varepsilon}^{2} = {\varepsilon}_{01}^{2} + {\varepsilon}_{02}^{2} + {\cal E}_{12}+{\cal E}_{21}={\varepsilon}_{01}^{2} + {\varepsilon}_{02}^{2} + {\cal E}_{12,21}, 
\end{eqnarray} 
where
\begin{eqnarray} \label{eq144abc}
{\varepsilon}_{01}^{2} = {{{\frac 1 2}\int_{V}} \left(  \dot{\psi_{01} ^*} \dot{\psi_{01}} +  \nabla \psi_{01}^*  \nabla \psi_{01} + m_0^2 \psi_{01} ^* \psi_{01}   \right)d^3x},
\end{eqnarray}
\begin{eqnarray} \label{eq145abc}
{\varepsilon}_{02}^{2} = {{{\frac 1 2}\int_{V}} \left(  \dot{\psi_{02} ^*} \dot{\psi_{02}} +  \nabla \psi_{02}^*  \nabla \psi_{02} + m_0^2 \psi_{02} ^* \psi_{02}   \right)d^3x},
\end{eqnarray}
\begin{eqnarray} \label{eq146abc}
{\cal E}_{12}= {{{\frac 1 2}\int_{V}}  \left(  \dot{\psi_{01} ^*} \dot{\psi_{02}} +  \nabla \psi_{01}^*  \nabla \psi_{02} + m_0^2 \psi_{01} ^* \psi_{02}   \right)d^3x}
\end{eqnarray}
and 
\begin{eqnarray} \label{eq147abc}
{\cal E}_{21}= {{{\frac 1 2}\int_{V}}  \left(  \dot{\psi_{02} ^*} \dot{\psi_{01}} +  \nabla \psi_{02}^*  \nabla \psi_{01} + m_0^2 \psi_{02} ^* \psi_{01}   \right)d^3x},
\end{eqnarray}
where ${\cal E}_{12,21}\equiv {\cal E}_{12}+{\cal E}_{21} \leq {\varepsilon}_{01}^{2} + {\varepsilon}_{02}^{2}$. It could be mentioned that Eq. (\ref{eq143abc}) is indistinguishable from Eqs. (\ref{eq122abc}), (\ref{eq89abc}) and (\ref{eq70abc}). The respective relativistic equation of motion  
\begin{eqnarray} \label{eq148abc}
\square \cdot [{\psi}_{01}+{\psi}_{02}] + m_0^2 \cdot [{\psi}_{01}+{\psi}_{02}]=0,
\end{eqnarray}
which is indistinguishable from Eq. (\ref{eq94abc}), is obtained by the replacements (\ref{eq86abc}) - (\ref{eq88abc}) in Eq. (\ref{eq45abc}). For the composite field (\ref{eq95abc}), which is composed from the $N$ unit-fields $\psi _{0n}(\mathbf{r},t)$ associated with the $N$ particles, the procedure yields the relativistic energy-mass relation 
\begin{eqnarray} \label{eq149abc}
{\varepsilon}^2 = \sum_{n=1}^N{{\varepsilon}}_{0n}^2+\sum_{n\neq m}^{N^2-N}{{\cal E}}_{nm},
\end{eqnarray}
where
\begin{eqnarray} \label{eq150abc}
{\varepsilon}_{0n}^{2} = {{{\frac 1 2}\int_{V}} \left(  \dot{\psi_{0n} ^*} \dot{\psi_{0n}} +  \nabla \psi_{0n}^*  \nabla \psi_{0n} + m_0^2 \psi_{0n} ^* \psi_{0n}   \right)d^3x},
\end{eqnarray}
and
\begin{eqnarray} \label{eq151abc}
{\cal E}_{nm} = {{{\frac 1 2}\int_{V}} \left(  \dot{\psi_{0n} ^*} \dot{\psi_{0m}} +  \nabla \psi_{0n}^*  \nabla \psi_{0m} + m_0^2 \psi_{0n} ^* \psi_{0m}   \right)d^3x}.
\end{eqnarray}
Notice, Eq. (\ref{eq149abc}) is indistinguishable from Eq. (\ref{eq129abc}). The respective relativistic equation of motion is given by  
\begin{eqnarray} \label{eq152abc}
\square \cdot \left[ \sum_{n=1}^N{\psi _{0n}(\mathbf{r},t)} \right]+ m_0^2 \cdot \left[ \sum_{n=1}^N{\psi _{0n}(\mathbf{r},t)}\right]=0,
\end{eqnarray}
which is indistinguishable from Eq. (\ref{eq99abc}).
 
In the case of the composite particle composed from the two non-relativistic particles, the replacements (\ref{eq86abc}) - (\ref{eq88abc}) in Eqs. (\ref{eq48abc}) - (\ref{eq52abc}) yielded the {\it{non-relativistic}} energy-mass relation
\begin{eqnarray} \label{eq153abc}
{\varepsilon} = {\varepsilon}_{01}  + {\varepsilon}_{02}  + {\varepsilon}_{12} +{\varepsilon}_{21}= {\varepsilon}_{01}  + {\varepsilon}_{02}  + {\varepsilon}_{12,21} 
\end{eqnarray} 
and the respective non-relativistic equation of motion
\begin{eqnarray} \label{eq154abc}
i[\dot{\psi}_{01}+\dot{\psi}_{02}]= -{\frac 1 {2m_0} } {\nabla ^2 [{\psi}_{01}+{\psi}_{02}]} + m_0[{\psi}_{01}+{\psi}_{02}],
\end{eqnarray}
where
\begin{eqnarray} \label{eq155abc}
{\varepsilon}_{01}={{{\frac 1 2}\int_{V}}  \left( i\psi_{01} ^*  \dot{\psi}_{01}-{\frac 1 {2m_0} } {\nabla {\psi_{01}^*}\nabla {\psi_{01}}} + m_0\psi_{01} ^*  \psi_{01}\right)d^3x},
\end{eqnarray}
\begin{eqnarray} \label{eq156abc}
{\varepsilon}_{02}={{{\frac 1 2}\int_{V}}  \left( i\psi_{02} ^*  \dot{\psi}_{02}-{\frac 1 {2m_0} } {\nabla {\psi_{02}^*}\nabla {\psi_{02}}} + m_0\psi_{02} ^*  \psi_{02}\right)d^3x},
\end{eqnarray}
\begin{eqnarray} \label{eq157abc}
{\varepsilon}_{12}={{{\frac 1 2}\int_{V}}  \left( i\psi_{01} ^*  \dot{\psi}_{02}-{\frac 1 {2m_0} } {\nabla {\psi_{01}^*}\nabla {\psi_{02}}} + m_0\psi_{01} ^*  \psi_{02}\right)d^3x}
\end{eqnarray}
and 
\begin{eqnarray} \label{eq158abc}
{\varepsilon}_{21}={{{\frac 1 2}\int_{V}}  \left( i\psi_{02} ^*  \dot{\psi}_{01}-{\frac 1 {2m_0} } {\nabla {\psi_{02}^*}\nabla {\psi_{01}}} + m_0\psi_{02} ^*  \psi_{01}\right)d^3x}
\end{eqnarray}
Notice, Eq. (\ref{eq153abc}) is indistinguishable from Eqs. (\ref{eq72abc}), (\ref{eq73abc}) and (\ref{eq100abc}); Eq. (\ref{eq154abc}) is indistinguishable from Eq. (\ref{eq101abc}). For the composite field (\ref{eq95abc}), which is composed from the $N$ unit-fields $\psi _{0n}(\mathbf{r},t)$, the replacements (\ref{eq86abc}) - (\ref{eq88abc}) in Eqs. (\ref{eq48abc}) - (\ref{eq52abc}) yielded the non-relativistic energy-mass relation 
\begin{eqnarray} \label{eq159abc}
{\varepsilon} = \sum_{n=1}^N{{\varepsilon}}_{0n}+\sum_{n\neq m}^{N^2-N}{{\varepsilon}}_{nm}, 
\end{eqnarray} 
with the respective non-relativistic equation of motion
\begin{eqnarray} \label{eq160abc}
i \left[ \sum_{n=1}^N{\dot{\psi} _{0n}(\mathbf{r},t)} \right] = -{\frac 1 {2m_0} } {\nabla ^2 \left[ \sum_{n=1}^N{\psi _{0n}(\mathbf{r},t)}\right]} + m_0\left[ \sum_{n=1}^N{\psi _{0n}(\mathbf{r},t)}\right],
\end{eqnarray}
where
\begin{eqnarray} \label{eq161abc}
{\varepsilon}_{0n}={{{\frac 1 2}\int_{V}}  \left( i\psi_{0n} ^*  \dot{\psi}_{0n}-{\frac 1 {2m_0} } {\nabla {\psi_{0n}^*}\nabla {\psi_{0n}}} + m_0\psi_{0n} ^*  \psi_{0n}\right)d^3x},
\end{eqnarray}
and
\begin{eqnarray} \label{eq162abc}
{\varepsilon}_{nm}={{{\frac 1 2}\int_{V}}  \left( i\psi_{0n} ^*  \dot{\psi}_{0m}-{\frac 1 {2m_0} } {\nabla {\psi_{0n}^*}\nabla {\psi_{0m}}} + m_0\psi_{0n} ^*  \psi_{0m}\right)d^3x}.
\end{eqnarray}
Notice, Eqs. (\ref{eq159abc}) and (\ref{eq160abc}) are indistinguishable from Eqs. (\ref{eq106abc}) and (\ref{eq107abc}), respectively.

\subsection{4.3. Physical properties of a composite particle based on the generalized energy-mass relation for the composite field }

\vspace{0.4cm}

1. {\it{The interaction energies and forces associated with the relativistic and non-relativistic interfering (cross-correlating) unit-fields}}

\vspace{0.4cm}

The above-presented models (Secs. 4.1. and 4.2.) are based on the generalization of the energy-mass relation of the Einstein special relativity, where the {\it{kind of an elementary particle}} is characterized only by its {\it{mass}}. Section 4.2. has presented a composite particle of the Einstein special relativity as the {\it{composite field}}, where the unit-fields of the composite field do associate with the massive ($m_{0n}=m_0 \neq 0$) or mass-less ($m_{0n}=m_0 = 0$) {\it{elementary}} particles of the Einstein relativity. In the case of the {\it{relativistic}} composite field (\ref{eq85abc}) satisfying the condition
\begin{eqnarray} \label{eq163abc}
{\cal E}_{12,21}= {\varepsilon}_{01} {\varepsilon}_{02} + {\varepsilon}_{02} {\varepsilon}_{01},  
\end{eqnarray}
the relativistic mass-energy relations (\ref{eq89abc}), (\ref{eq122abc}) and (\ref{eq143abc}) correspond to the relativistic energy-mass relation (\ref{eq70abc}) attributed to the two particles. If the the total cross-correlation term is given by    
\begin{eqnarray} \label{eq164abc}
{\cal E}_{12,21} \neq {\varepsilon}_{01} {\varepsilon}_{02} + {\varepsilon}_{02} {\varepsilon}_{01},  
\end{eqnarray}
then the comparison of the mass-energy relations (\ref{eq89abc}), (\ref{eq122abc}) and (\ref{eq143abc}) with the energy-mass relation (\ref{eq70abc}) {\it{indicates unambiguously existence of the interaction (non-zero cross-correlation)}} between the two unit-fields (particles). The further comparison of the total energy determining by the mass-energy relations (\ref{eq89abc}), (\ref{eq122abc}) and (\ref{eq143abc}) with the total energy (\ref{eq72abc}) or (\ref{eq73abc}) attributed to the interacting particles yields the total interaction (cross-correlation) energy 
\begin{eqnarray} \label{eq165abc}
{\varepsilon}_{12,21}=({\varepsilon}_{01}^{2} + {\varepsilon}_{02}^{2} + {\cal E}_{12,21})^{1/2}-({\varepsilon}_{01} + {\varepsilon}_{02}),
\end{eqnarray} 
where the total cross-correlation term 
\begin{eqnarray} \label{eq166abc}
{\cal E}_{12,21}=2{\cal E}_{12}=2{\cal E}_{21}\leq {\varepsilon}_{01}^{2} + {\varepsilon}_{02}^{2}
\end{eqnarray} 
is determined by Eqs. (\ref{eq92abc}), (\ref{eq93abc}), (\ref{eq112abc}), (\ref{eq113abc}), (\ref{eq125abc}), (\ref{eq126abc}), (\ref{eq146abc}) and (\ref{eq147abc}). The gradient of the interaction energy ${\varepsilon}_{12,21}$ attributed to both the first and second point-particles satisfy the relation 
\begin{eqnarray} \label{eq167abc}
{\varepsilon}_{12,21}=2{\varepsilon}_{12}=2{\varepsilon}_{21},
\end{eqnarray} 
where ${\varepsilon}_{12,21}\neq 0$ is given by Eq. (\ref{eq165abc}). The gradients of the interaction (cross-correlation) energies ${\varepsilon}_{12}$ and ${\varepsilon}_{21}$ determining by Eqs. (\ref{eq165abc}) - (\ref{eq167abc}) automatically yielded the relativistic interaction (cross-correlation) forces
\begin{eqnarray} \label{eq168abc}
\mathbf{F}_{12}(\mathbf{r}_1 ,t) =-{\frac {\partial}  {\partial \mathbf{R}_{12}}} {\varepsilon}_{12,21}(|\mathbf{R}_{12}| ,t)
\end{eqnarray} 
and 
\begin{eqnarray} \label{eq169abc}
\mathbf{F}_{21}(\mathbf{r}_2,t) = {\frac {\partial}  {\partial \mathbf{R}_{12}}} {\varepsilon}_{12,21}(|\mathbf{R}_{21}|,t), 
\end{eqnarray} 
acting respectively upon the first and second unit-fields (particles). The forces satisfy the relation 
\begin{eqnarray} \label{eq170abc}
\mathbf{F}_{12}(\mathbf{r}_{1},t)=-\mathbf{F}_{21}(\mathbf{r}_{2},t). 
\end{eqnarray}
It should be stressed that Eqs. (\ref{eq165abc}) - (\ref{eq170abc}) are valid for both the {\it{strong-relativistic }} (${\varepsilon}_{01}^{2} + {\varepsilon}_{02}^{2} \sim  {\cal E}_{12,21}$) and {\it{weak-relativistic}}  (${\cal E}_{12,21}<<{\varepsilon}_{01}^{2} + {\varepsilon}_{02}^{2} $) interactions. In the case of the {\it{weak-relativistic}} interaction, which is the typical situation in the most of physical experiments, Eq. (\ref{eq165abc}) simplifies to the equation   
\begin{eqnarray} \label{eq171abc}
{\varepsilon}_{12,21}\approx ({\varepsilon}_{01}^{2}-{\varepsilon}_{01}) + ({\varepsilon}_{02}^{2}-{\varepsilon}_{02}) + (1/2){\cal E}_{12,21},
\end{eqnarray} 
which yields the {\it{weak-relativistic}} interaction (cross-correlation) forces
\begin{eqnarray} \label{eq172abc}
\mathbf{F}_{12}(\mathbf{r}_1 ,t)  \approx  - {\frac 1  {2}}{\frac {\partial}  {\partial \mathbf{R}_{12}}} {\cal E}_{12,21}(|\mathbf{R}_{12}| ,t)
\end{eqnarray} 
and 
\begin{eqnarray} \label{eq173abc}
\mathbf{F}_{21}(\mathbf{r}_2,t) \approx {\frac 1  {2}}{\frac {\partial}  {\partial \mathbf{R}_{12}}} {\cal E}_{12,21}(|\mathbf{R}_{12}|,t), 
\end{eqnarray} 
The interaction between the 1st and 2nd interacting elementary unit-fields (particles) results into the attraction or repelling of the unit-fields (particles) characterizing by the interactive forces (\ref{eq168abc}), (\ref{eq169abc}), (\ref{eq172abc}) and (\ref{eq173abc}) that have the absolute values and directions. In the case of the negative values ${\varepsilon}_{12}(|\mathbf{r}| ,t)$, ${\varepsilon}_{21}(|\mathbf{r}| ,t)$, ${\cal E}_{12}(|\mathbf{r}| ,t)$ and ${\cal E}_{21}(|\mathbf{r}| ,t)$, the interactive forces (\ref{eq168abc}), (\ref{eq169abc}), (\ref{eq172abc}) and (\ref{eq173abc}) are {\it{attractive}}. The interactive forces (\ref{eq168abc}), (\ref{eq169abc}), (\ref{eq172abc}) and (\ref{eq173abc}) are {\it{repulsive}} if the cross-correlation parameters ${\varepsilon}_{12}(|\mathbf{r}| ,t)$, ${\varepsilon}_{21}(|\mathbf{r}| ,t)$, ${\cal E}_{12}(|\mathbf{r}| ,t)$ and ${\cal E}_{21}(|\mathbf{r}| ,t)$ have the positive values. In Eqs. (\ref{eq165abc}) - (\ref{eq173abc}), one should not confuse the {\it{strong-relativistic }} and {\it{weak-relativistic}} interactions with the {\it{weak}} and {\it{strong}} interactions of SM.

In the case of the {\it{non-relativistic}} composite field (\ref{eq85abc}) composed from the two interfering (cross-correlating) non-relativistic unit-fields, the comparison of the non-relativistic mass-energy relations (\ref{eq100abc}), (\ref{eq133abc}) and (\ref{eq153abc}) with the non-relativistic energy-mass relation (\ref{eq72abc}) or (\ref{eq73abc}) attributed to the two {\it{interacting}}, non-relativistic particles yielded the total non-relativistic cross-correlation (interaction) energy
\begin{eqnarray} \label{eq174abc}
{\varepsilon}_{12,21}={\varepsilon}_{12}+{\varepsilon}_{12}, 
\end{eqnarray} 
where the non-relativistic cross-correlation (interaction) energies ${\varepsilon}_{12}=(1/2){\varepsilon}_{12,21}$ and ${\varepsilon}_{21}=(1/2){\varepsilon}_{12,21}$ associated with the first and the second non-relativistic unit-fields (particles) are determined by the non-relativistic equations (\ref{eq104abc}), (\ref{eq105abc}), (\ref{eq118abc}), (\ref{eq119abc}), (\ref{eq137abc}), (\ref{eq138abc}), (\ref{eq157abc}) and (\ref{eq158abc}). The non-relativistic interaction (cross-correlation) forces $\mathbf{F}_{12}(\mathbf{r} ,t)$ and $\mathbf{F}_{21}(\mathbf{r} ,t)$ are determined by Eqs. (\ref{eq168abc}) - (\ref{eq170abc}), where the gradients of the interaction (cross-correlation) energies ${\varepsilon}_{12}$ and ${\varepsilon}_{21}$ are calculated by using the relations (\ref{eq104abc}), (\ref{eq105abc}), (\ref{eq118abc}) (\ref{eq119abc}), (\ref{eq137abc}), (\ref{eq138abc}), (\ref{eq157abc}) and (\ref{eq158abc}). 

The above presented analysis of the two interfering unit-fields (particles) is valid for any number $N$ of the interfering unit-fields (particles) of the composite field (particle). In the case of $N>2$, one can easily demonstrate that Eqs. (\ref{eq163abc}) - (\ref{eq174abc}) describe the interaction (cross-correlation) of unit-fields (particles) in the each pair of the unit-fields (particles). 

\vspace{0.4cm}

2. {\it{The probability density, energy and effective number of the relativistic and non-relativistic interfering (cross-correlating) unit-fields and their connection with the Euler-Lagrange and Hamilton-Jacoby  formalisms}}

\vspace{0.4cm}

The interaction (cross-correlation) relations (\ref{eq163abc}) - (\ref{eq174abc}) have been derived by using the two equivalent versions of the model of a composite field (particle) $\psi (\mathbf{r},t)=\sum_{n=1}^N{\psi _{0n}(\mathbf{r},t)}$. The 1st and 2nd model versions are based on the straightforward generalization of the Einstein energy-mass relation for the interfering unit-fields by using respectively the 1st and 2nd derivatives of the unit-fields. The model versions have been presented also in the more general form by using the Euler-Lagrange formalism. In addition, it was demonstrated that the model versions can be easily rewritten in the frame of the Hamilton-Jacoby formalism by using the Hamiltonians ${\cal H}_{0n}$ of the unit-field $\psi_{0n}$. It should be stressed that the Euler-Lagrange and  Hamilton-Jacoby  formalisms for the the {\it{single interference-free}} unit-field ${\psi _{0n}(\mathbf{r},t)}$ is based on the normalization 
\begin{eqnarray} \label{eq175abc}
{{\int_{V}}\psi _{0n} ^* \psi _{0n} d^3x}=1, 
\end{eqnarray}
which could give rise to {\it{the probabilistic ("Copenhagen quantum-mechanical") interpretation}} of the unit-field (unit-wave) $\psi _{0n}$ associated with the $n$-th single particle, where the value $\psi _{0n}^* \psi _{0n}$ is interpreted as the probability density to find the {\it{single}} particle in the spacetime point ${(\mathbf{r},t)}$. However, unlike in the canonical quantum mechanics based on the quantum-mechanical (non-material)  wave-function $\psi _{0n} (\mathbf{r},t)$ describing a material point-particle located in the spatiotemporal point $(\mathbf{r},t)$ with the probability density $\psi^*_{0n} (\mathbf{r},t)\psi_{0n} (\mathbf{r},t)$, the {\it{material}} unit-field (particle) $\psi_{0n} (\mathbf{r},t)$ associated with the particle mass-energy {\it{does really exist}} in the each spatiotemporal point $(\mathbf{r},t)$ of the unit-field. That is the principal difference between the canonical (probabilistic) quantum mechanics and the present model (See, Sec. 3.2.). The {\it{material}} composite field $\psi (\mathbf{r},t)=\sum_{n=1}^N{\psi _{0n}(\mathbf{r},t)}$  associated with the composite-particle mass and energy  also {\it{really exists}} in the each spatiotemporal point $(\mathbf{r},t)$ of the composite field. The composite field $\psi (\mathbf{r},t)$ obeys the inequality
\begin{eqnarray} \label{eq176abc}
{{\int_{V}}\psi ^*\psi d^3x} \neq 1,
\end{eqnarray}
which is not consistent with the normalization (\ref{eq18abc}) of the {\it{single interference-free}} unit-field $\psi_{0n}$. Therefore the composite field ${\psi}$ (composite particle) {\it{could not be associated with the fields (waves) of probabilities of the "Copenhagen" (canonical) quantum mechanics}}. The non-probabilistic interpretation of the composite field does not contradict the traditional quantum field theories, which also do not use the concept of the fields (waves) of probabilities. 

Consider now the energy and number of the relativistic and non-relativistic interfering (cross-correlating) unit-fields and their connection with the Hamilton formalism of Part I of the present study. In the {\it{relativistic}} case, the {\it{relativistic}} Hamiltonian ${\cal H}_{01}$ of a unit-field (particle) is associated with the {\it{unit-field energy squared}}: 
\begin{eqnarray} \label{eq177abc}
{\cal H}_{01}={\varepsilon}_{01}^{2},
\end{eqnarray} 
where ${\varepsilon}_{01}^{2} ={\mathbf{k}}_{01}^2+m_{01}^2$. The relativistic Hamiltonian ${\cal H}$ of the composite field (particle) composed from the $N$ unit-fields (particles) is attributed to the {\it{total field-energy squared}}:
\begin{eqnarray} \label{eq178abc}
{\cal H}={\varepsilon}^{2},
\end{eqnarray} 
where 
\begin{eqnarray} \label{eq179abc}
{\cal H}=\sum_{n=1}^N{\cal H}_{nn}+\sum_{{n\neq m}}^{N^2-N}{\cal H}_{nm},
\end{eqnarray} 
\begin{eqnarray} \label{eq180abc}
{\varepsilon}^{2}= \sum_{n=1}^N{\varepsilon}_{0n}^2+\sum_{{n\neq m}}^{N^2-N}{\cal E}_{nm},
\end{eqnarray} 
\begin{eqnarray} \label{eq181abc}
{\cal H}_{nn}={\varepsilon}_{0n}^2
\end{eqnarray} 
and 
\begin{eqnarray} \label{eq182abc}
{\cal H}_{nm}={\cal E}_{nm}
\end{eqnarray} 
Here, the values ${\varepsilon}_{0n}^2$ are calculated by using Eqs. (\ref{eq97abc}), (\ref{eq114abc}), (\ref{eq130abc}) and (\ref{eq150abc}), where the cross-correlation term ${\cal E}_{nm}$ is determined by Eqs. (\ref{eq98abc}), (\ref{eq115abc}), (\ref{eq131abc})and (\ref{eq151abc}). One can easily demonstrate that the physical parameters of the interfering {\it{relativistic}} unit-fields (particles) are described by the energy ${ \varepsilon  }$ in the energy interval 
\begin{eqnarray} \label{eq183abc}
0\leq{ \varepsilon  }\leq N{{\varepsilon}_{01}}
\end{eqnarray} 
with the particle relativistic energy ${\varepsilon}_{01} =[{\mathbf{k}}_{01}^2+m_{01}^2]^{1/2}$ and the effective number ${\cal N}={ \varepsilon  }/{\varepsilon}_{01}$ of particles in the number interval 
\begin{eqnarray} \label{eq184abc}
0\leq {{\cal N}}\leq N.
\end{eqnarray} 
If the relativistic {\it{identical unit-fields}} add coherently, then the composite-field energy and effective number of the unit-fields do scale as the number of unit-fields (particles): 
\begin{eqnarray} \label{eq185abc}
{ \varepsilon  }= N{{\varepsilon}_{01}}
\end{eqnarray} 
and
\begin{eqnarray} \label{eq186abc}
{\cal N}= N.
\end{eqnarray} 
In the {\it{non-relativistic}} case, the {\it{non-relativistic}} Hamiltonian ${\cal H}_{01}$ of a unit-field (particle)  is attributed to the {\it{unit-field energy}}: 
\begin{eqnarray} \label{eq187abc}
{\cal H}_{01}={\varepsilon}_{01},
\end{eqnarray} 
where the the non-relativistic particle energy ${\varepsilon}_{01}$ is given by Eq. (\ref{eq9abc}) as ${\varepsilon}_{01} =  {\frac {\mathbf{k}_{01}^2} {2m_{01}} }+ m_{01}$. The non-relativistic Hamiltonian ${\cal H}$ of the composite field (particle) composed from the $N$ unit-fields (particles) is attributed to the {\it{total field-energy}}
\begin{eqnarray} \label{eq188abc}
{\cal H}={\varepsilon}
\end{eqnarray} 
with  
\begin{eqnarray} \label{eq189abc}
{\cal H}=\sum_{n=1}^N{\cal H}_{nn}+\sum_{{n\neq m}}^{N^2-N}{\cal H}_{nm},
\end{eqnarray} 
\begin{eqnarray} \label{eq190abc}
{\varepsilon}= \sum_{n=1}^N{\varepsilon}_{0n}+\sum_{{n\neq m}}^{N^2-N}{\varepsilon}_{nm},
\end{eqnarray} 
\begin{eqnarray} \label{eq191abc}
{\cal H}_{nn}={\varepsilon}_{0n}
\end{eqnarray} 
and 
\begin{eqnarray} \label{eq192abc}
{\cal H}_{nm}={\varepsilon}_{nm},
\end{eqnarray} 
where the unit-field energy ${\varepsilon}_{nm}$ is determined by Eqs. (\ref{eq108abc}), (\ref{eq120abc}), (\ref{eq141abc}) and (\ref{eq161abc}), where the cross-correlation (interaction) energy  ${\varepsilon}_{nm}$ is calculated by using Eqs. (\ref{eq109abc}), (\ref{eq121abc}), (\ref{eq142abc}) and (\ref{eq162abc}). The energy of the cross-correlating {\it{non-relativistic}} unit-fields is then given by 
\begin{eqnarray} \label{eq193abc}
0\leq{ \cal H }\leq N^2{{\varepsilon}_{01}}
\end{eqnarray} 
with the particle energy ${\varepsilon}_{01} =  {\frac {\mathbf{k}_{01}^2} {2m_{01}} }+ m_{01}$ and the  respective effective number 
\begin{eqnarray} \label{eq194abc}
0\leq{{\cal N}}\leq N^2
\end{eqnarray} 
of the unit-fields (particles). If the identical unit-fields {\it{add coherently}}, then the composite-field energy and effective number of the unit-fields do scale as the number of unit-fields (particles) squared: 
\begin{eqnarray} \label{eq195abc}
{ \varepsilon  }= N^2{{\varepsilon}_{01}}
\end{eqnarray} 
and
\begin{eqnarray} \label{eq196abc}
{\cal N}= N^2.
\end{eqnarray} 
Notice, the squares of energy, momentum and mass of the {\it{degenerate}}, composite field $\psi (\mathbf{r},t) = N {\psi _{01}(\mathbf{r},t)}$ associated with the {\it{degenerate}}, composite particle is given respectively by ${\varepsilon}^2=N^2{\varepsilon}_{01}^2$, ${\mathbf{k}}^2=N^2{\mathbf{k}}_0^2$ and $m^2=N^2m_{01}^2$, where ${\varepsilon}_{01}^2=\mathbf{k}_{01}^2+m_{01}^2$. The energy ${\varepsilon}=N{\varepsilon}_{01}$, momentum ${\mathbf{k}}=N{\mathbf{k}}_{01}$ and mass $m=Nm_{01}$ of the {\it{degenerate}} field (particle) is similar to the energy and momentum of the $N$ bosons of the global infinite field of bosons describing by the Klein-Gordon-Fock equation of quantum field theory.

Part I of the present study has used the Hamiltonian-energy relation ${\cal H}={\varepsilon}(k)$ of the {\it{traditional quantum field theory}}, where the relativistic Hamiltonian ${\cal H}_{01}$ is associated with the {\it{relativistic particle energy}} as ${\cal H}_{01}={\varepsilon}_{01} =({\mathbf{k}}_{01}^2+m_{01}^2)^{1/2}$. In such a case, the {\it{relativistic}} Hamiltonian ${\cal H}$ of the field composed from the $N$ unit-fields (particles) is associated with the field total energy, ${\varepsilon}= {\cal H}=\sum_{n=1}^N{\cal H}_{nn}+\sum_{{n\neq m}}^{N^2-N}{\cal H}_{nm}$. The energy of the cross-correlating unit-fields is then satisfy the inequality $0\leq{ \cal H }\leq N^2{{\varepsilon}_{01}}$ with the  respective effective number $0\leq{{\cal N}}\leq N^2$ of the unit-fields (particles), where ${\varepsilon}_{01} =({\mathbf{k}}_{01}^2+m_{01}^2)^{1/2}$. {\it{It is clear now that the relations (\ref{eq191abc}) and (\ref{eq192abc}), which have been derived in Part I for the relativistic [${\varepsilon}_{0n} =({\mathbf{k}}_{0n}^2+m_{0n}^2)^{1/2}$] unit-fields by using the traditional quantum field theory, are valid only for the non-relativistic [${\varepsilon}_{0n} \approx ({\mathbf{k}_{0n}^2} /2m_{0n}) + m_{0n}$] unit-fields}}. The relativistic unit-fields (particles) should obey the relativistic relations (\ref{eq181abc}) and (\ref{eq182abc}), which are completely different from the non-relativistic relations (\ref{eq191abc}) and (\ref{eq192abc}). That also means that the total energy ${\varepsilon}$ of the composite field (particle) composed from the $N$ relativistic or non-relativistic unit-fields (elementary particles) does not overcome the relativistic limit: 
\begin{eqnarray} \label{eq197abc}
{\varepsilon} \leq N({\mathbf{k}}_{01}^2+m_{01}^2)^{1/2}.
\end{eqnarray} 
It can be mentioned again that the above-presented relativistic equations based on the relativistic energy ${\varepsilon}_{0n} =({\mathbf{k}}_{0n}^2+m_{0n}^2)^{1/2}$ are considerably simplified in the case of the mass-less ($m_{0n}=0$) unit-fields associated with the mass-less ($m_{0n}=0$) particles. While the non-relativistic equations based on the non-relativistic approximation ${\varepsilon}_{0n} \approx ({\mathbf{k}_{0n}^2} /2m_{0n}) + m_{0n}$ do not have any physical meaning for the mass-less unit-fields (particles). 

\vspace{0.4cm}

3. {\it{The physical interpretation of the effective number of the unit-fields (particles): The normal and virtual unit-fields (elementary particles)}}

\vspace{0.4cm}

The physical interpretation of the effective number ${\cal N}={ \varepsilon  }/{\varepsilon}_{01}$ of the $N$ interference-less unit-fields (particles) is very simple: 
\begin{eqnarray} \label{eq198abc}
{\cal N}=N.
\end{eqnarray} 
The physical meaning of the effective number ${\cal N}$ of the $N$ interfering (cross-correlating) unit-fields is more complicated. For the sake of simplicity let me first interpret the number of particles in the case of the composite field (particle) composed from the {\it{two}} interfering (interacting) unit-fields. The energy-mass relation for the composite relativistic field is given by Eqs. (\ref{eq89abc}), (\ref{eq122abc}) and (\ref{eq143abc}) as
\begin{eqnarray} \label{eq199abc}
{\varepsilon}^{2} = {\varepsilon}_{01}^{2} + {\varepsilon}_{02}^{2} + {\cal E}_{12}+{\cal E}_{21}, 
\end{eqnarray} 
where the cross-correlation terms ${\cal E}_{12}$ and ${\cal E}_{21}$ are determined by the relativistic equations (\ref{eq92abc}), (\ref{eq93abc}), (\ref{eq112abc}), (\ref{eq113abc}), (\ref{eq125abc}), (\ref{eq126abc}), (\ref{eq146abc}) and (\ref{eq147abc}). For the composite non-relativistic field, the energy-mass relation is given by Eqs. (\ref{eq100abc}), (\ref{eq133abc}) and (\ref{eq153abc}) as
\begin{eqnarray} \label{eq200abc}
{\varepsilon} = {\varepsilon}_{01}  + {\varepsilon}_{02}  + {\varepsilon}_{12} +{\varepsilon}_{21},
\end{eqnarray} 
where the non-relativistic cross-correlation (interaction) energies ${\varepsilon}_{12}$ and ${\varepsilon}_{21}$ are determined by the non-relativistic equations (\ref{eq104abc}), (\ref{eq105abc}), (\ref{eq118abc}), (\ref{eq119abc}), (\ref{eq137abc}), (\ref{eq138abc}), (\ref{eq157abc}) and (\ref{eq158abc}). According to the Einstein special relativity the energies ${\varepsilon}_{01}$ and ${\varepsilon}_{02}$ in the non-relativistic relation (\ref{eq200abc}) could be attributed to the physical substance of the 1st and 2nd non-relativistic unit-fields (elementary particles). Correspondingly, the energies ${\varepsilon}_{21}$ and ${\varepsilon}_{21}$ could be attributed, at least formally, to the physical matter (mass-energy) of the 3rd and 4th non-relativistic unit-fields (particles) of interaction. The 3rd and 4th {\it{non-relativistic unit-fields (particles)} of interaction}, which are attributed to the interaction of the 1st and 2nd non-relativistic unit-fields (elementary particles), could be interpreted as the carriers of the {\it{non-relativistic interaction (force)}}. The 3rd and 4th {\it{unit-fields (particles)}} of interaction are the virtual unit-fields (particles) because they are created and exist only in the exchange (interaction) process. In other words, the interaction between the 1st and 2nd interacting (interfering) normal unit-fields (particles) could be considered as being caused (mediated) by the simultaneous emission and absorption (virtual exchange) of the 3rd and 4th {\it{virtual unit-fields (particles)}} of interaction. The exchange of 3rd and 4th {\it{virtual unit-fields (particles)}} of interaction does transport momentum and energy between the 1st and 2nd normal unit-fields (elementary particles), thereby changing their momentum and energy. The interaction results into the attraction or repelling of the 1st and 2nd non-relativistic interacting elementary unit-fields (particles) characterizing by the interactive force that has the absolute value and direction. The interaction may be interpreted also in the frame of the perturbation approximation of the traditional quantum field theory as follows. {\it{From a point of view of the energy, a {\it{point-like elementary  particle}} and the {\it{unit-wave}} associated with this particle are equivalent (indistinguishable) objects (Sec. 3)}}. Therefore the 1st and 2nd elementary point-particles separated by the vacuum attract or repel each other by the force mediating by the virtual exchange of the 3rd and 4th {\it{virtual point-particles}} of interaction through the vacuum. The interpretation of the relativistic relation (\ref{eq199abc}) is quite similar to the above-considered non-relativistic equation (\ref{eq200abc}). The energies squared ${\varepsilon}_{01}^2$ and ${\varepsilon}_{02}^2$ could be attributed to the physical substance (mass-energy) of the 1st and 2nd {\it{relativistic normal unit-fields (elementary particles)}}. The cross-correlation terms ${\cal E}_{12}$ and ${\cal E}_{21}$ could be attributed to the physical substance of the 3st and 4th {\it{relativistic virtual unit-fields (particles)}} of the relativistic interaction. Thus the relativistic interaction between the two normal, relativistic unit-fields (elementary particles) could be considered (interpreted) as the interplay of the {\it{four}} unit-fields (particles), where the composite field (particle) is composed from the 1st and 2nd {\it{normal, relativistic unit-fields (elementary particles)}} and the 3rd and 4th {\it{virtual, relativistic unit-fields (elementary particles)}}. The effective number ${\cal N}$ of the interacting unit-fields (particles) in the interval (\ref{eq184abc}) or (\ref{eq194abc}) has been obtained by comparison of the total energy ${\varepsilon}$ of the composite field (particle) with the energies ${\varepsilon}_{01}$ and ${\varepsilon}_{02}$ of the 1st and 2nd identical (${\varepsilon}_{01} = {\varepsilon}_{02}$) normal unit-fields (elementary particles): ${\cal N}={ \varepsilon  }/{\varepsilon}_{01}$ in the interval $0\leq {{\cal N}}\leq N$ or $0\leq{{\cal N}}\leq N^2$ for $N=2$. Although the effective number ${\cal N}$ of the interfering unit-fields (interacting particles) is different from the number ({\it{four}}) of the interplaying unit-fields (particles), the two interpretations describe the same physical process, namely the interference (interaction) of the two ($N=2$) elementary particles. One can easily extend the above-presented analysis to the multi-particle ($N >2$) system. In such a case, the each pair of the {\it{normal unit-fields (elementary particles)}} is associated with the pair of {\it{virtual unit-fields (elementary particles)}}, while the effective number ${\cal N}$ of the interacting unit-fields (particles) is given by ${\cal N}={ \varepsilon  }/{\varepsilon}_{01}$ in the interval $0\leq {{\cal N}}\leq N$ or $0\leq{{\cal N}}\leq N^2$ for $N>2$.

\vspace{0.4cm}

4. {\it{The energy conservation and the non-conservation of mass and number of the normal and virtual unit-fields (elementary particles): Annihilation of unit-fields (elementary particles)}}

\vspace{0.4cm}

For the two normal interfering unit-fields (interacting elementary particles), the energy conservation law applied to Eqs. (\ref{eq199abc}) and (\ref{eq200abc}) is given respectively by the relations 
\begin{eqnarray} \label{eq201abc}
{\varepsilon}_{01}^{2}(t) + {\varepsilon}_{02}^{2}(t) + {\cal E}_{12}(t)+{\cal E}_{21}(t)={\varepsilon}_{01}^{2}(t') + {\varepsilon}_{02}^{2}(t') + {\cal E}_{12}(t')+{\cal E}_{21}(t')
\end{eqnarray} 
and 
\begin{eqnarray} \label{eq202abc}
{\varepsilon}_{01}(t)  + {\varepsilon}_{02}(t)  + {\varepsilon}_{12}(t) +{\varepsilon}_{21}(t)={\varepsilon}_{01}(t')  + {\varepsilon}_{02}(t')  + {\varepsilon}_{12}(t') +{\varepsilon}_{21}(t')
\end{eqnarray} 
for the relativistic and non-relativistic unit-fields (particles). Here, the left- and right-hand sides of Eq. (\ref{eq201abc}) are the squares of total energies of the interfering {\it{relativistic}} unit-fields  (particles) at the time moments $t$ and $t'$, respectively. While, the left- and right-hand sides of Eq. (\ref{eq202abc}) describe the total energies of the interfering {\it{non-relativistic}} unit-fields at the time moments $t$ and $t'$, respectively. From the point of view of the energy, Eqs. (\ref{eq201abc}) and (\ref{eq202abc}) describe the interplay of the four unit-fields (particles), namely the 1st and 2nd normal, massive unit-fields (elementary particles) and 3rd and 4th virtual, mass-less unit-fields (elementary particles). The relativistic equation (\ref{eq201abc}) describes the energy conservation of the four unit-fields (particles) associated with the {\it{massive}} and/or {\it{mass-less}} unit-fields (particles). For an example, the left-hand side of Eq. (\ref{eq201abc}) may be associated with the massive and mass-less unit-fields (particles). The right-hand side of Eq. (\ref{eq201abc}) may contain the  mass-less unit-fields (particles), only. In such a case, the rest-mass of the 1st and 2nd normal unit-fields (elementary particles) does {\it{annihilate}} and then convert into the energy of the mass-less particles. If the left-side of Eq. (\ref{eq201abc}) does include only the mass-less unit-fields (particles), then the energy of the mass-less particles converts into the mass-energy of the massive and mass-less unit-fields (particles). It should be stressed that the non-relativistic equation (\ref{eq202abc}) based on the non-relativistic approximation ${\varepsilon}_{0n} \approx ({\mathbf{k}_{0n}^2} /2m_{0n}) + m_{0n}$ does not have any physical sense for the {\it{mass-less}} unit-fields (particles). In the case of ${\cal E}_{12}(t')={\cal E}_{21}(t')=0$ or ${\varepsilon}_{12}(t') ={\varepsilon}_{21}(t')=0$, the 3rd and 4th virtual unit-fields (elementary particles) disappear:
\begin{eqnarray} \label{eq203abc}
{\varepsilon}_{01}^{2}(t) + {\varepsilon}_{02}^{2}(t) + {\cal E}_{12}(t)+{\cal E}_{21}(t)={\varepsilon}_{01}^{2}(t') + {\varepsilon}_{02}^{2}(t')
\end{eqnarray} 
and 
\begin{eqnarray} \label{eq204abc}
{\varepsilon}_{01}(t)  + {\varepsilon}_{02}(t)  + {\varepsilon}_{12}(t) +{\varepsilon}_{21}(t)={\varepsilon}_{01}(t')  + {\varepsilon}_{02}(t'). 
\end{eqnarray} 
That is to say that the 3rd and 4th virtual unit-fields (particles) do {\it{annihilate}} and the respective virtual energies and masses convert into the normal energies and masses of the 1st and 2nd normal unit-fields (particles). The cross-correlation terms [${\cal E}_{12}(t')$ and ${\cal E}_{21}(t')$] and the cross-correlation energies [${\varepsilon}_{12}(t')$ and ${\varepsilon}_{21}(t')$] do vanish in the two typical composite systems. The first system corresponds to the 1st and 2nd unit-fields, which are orthogonal in the Hilbert space. Such unit-fields correspond to the eigensolutions of the equations of motion of the composite system. In the second system, the 1st and 2nd unit-fields (particles) are separated by the infinite distance. The above-presented analysis can be easily extended to the multi-particle ($N >2$) system. In such a case, the each pair of the normal unit-fields (elementary particles) is associated with the pair of virtual unit-fields (elementary particles).  

\vspace{0.4cm}

5. {\it{Global}} ($N \rightarrow \infty$) {\it{Composite Field associated with the Universe of Einstein's elementary particles}}

\vspace{0.4cm}

In the above-presented conceptual picture of the unit-fields (elementary particles) combining the basic physical conceptions of canonical quantum mechanics with the special relativity, the Global Composite Field composed from the unit-fields associated with the all ($N \rightarrow \infty$) elementary particles satisfying the Einstein special relativity could be considered as the Global Composite Field of the Einstein Universe. In such a picture, the Global Composite Field  
\begin{eqnarray} \label{eq205abc}
\Psi (\mathbf{r},t)=\sum_{n=1}^{N \rightarrow \infty}{\psi _{0n}(\mathbf{r},t)}
\end{eqnarray} 
attributed to the Global Composite Particle is composed from the interfering or non-interfering material unit-fields associated with the material elementary particles of the Einstein special relativity. The elementary particle of the Global Composite Field is associated with the "field quanta" (unit-field). The unit-field (quanta of energy-mass) of the Global Composite Field is assumed not to be made up of smaller unit-fields (elementary particles). 

Let me indicate the formal and conceptual differences between the formulation of the Global Quantum (Composite) Field (\ref{eq205abc}) and the modeling of the Fundamental Non-Quantum Fields and the Fundamental Quantum Fields of Operators associated with the fundamental (electromagnetic, weak and strong) interactions of the traditional quantum field theories and SM. The present model follows the scheme [{\it{a unit-field (particle) $\rightarrow$ Global Quantum (Composite) Field }}], while the traditional quantum field theories and SM follow the inverse approach [{\it{Global Non-Quantum Field $\rightarrow$ a particle}}]. That means the present model regards individual particles (unit-fields) as fundamental objects, while the field theories and SM assume that only the Global field is fundamental. In other words, the present model of the Global Quantum Field $\Psi (\mathbf{r},t)$ of the Einstein Universe {\it{ does begin}} the field description with determination of a "field quanta", namely formulation of a single unit-field ${\psi _{0n}(\mathbf{r},t)}$ and its equation of motion {\it{by using the generalization of the Einstein energy-mass relation}}. That is different from the canonical approach (formulation) of traditional quantum field theories and SM, which {\it{first look for}} the Global, Fundamental, Non-Quantum Field and its Lagrangian (Hamiltonian) {\it{by postulating a set of symmetries}} of the Field Lagrangian (Hamiltonian) and {\it{then}} construct the Fundamental Quantum Field of Operators determining the fundamental elementary particle ("field quanta"). Indeed, in the traditional quantum field theories and SM, the fundamental elementary particles of the Fundamental Non-Quantum  Fields $\Psi_i (\mathbf{r},t)$ associated with the three fundamental [electromagnetic ($i=1$), weak ($i=2$) and strong ($i=3$)] interactions are found by using the Dirac {\it{second}} quantization procedure based on the formal replacement of the Fundamental Non-Quantum Field $\Psi_i (\mathbf{r},t)$ by the Operator of Fundamental Field $\hat \Psi_i (\mathbf{r},t)$ composed from the creation (${\hat a^{\dagger}}_{i, 0n}$) and destruction (${\hat a}_{i, 0n}$) operators of the respective ($i$-th) fundamental elementary particles. The present model deals with the {\it{one unified field}} [Global Quantum  Field $\Psi (\mathbf{r},t)$] composed from the unit-fields (particles), while the traditional quantum field theories and SM consider the {\it{six}} fundamental fields, namely the {\it{three}} Fundamental {\it{Non-quantum}} Fields $\Psi_i (\mathbf{r},t)$ and the {\it{three}} Fundamental {\it{Quantum}} Fields of Operators $\hat \Psi_i (\mathbf{r},t)$. Here, one should not confuse the Global {\it{Quantum}} Field (\ref{eq205abc}) of the present model, which is not a field of operators, with the Fundamental {\it{Quantum}} Field of Operators of the traditional quantum field theories and SM.   

The conceptual difference between the approach (formulation) of the present model and the traditional quantum field theories and SM could be illustrated by the following concrete example. Formally, Eqs. (\ref{eq8abc}) and (\ref{eq45abc}) look like the Klein-Gordon-Fock equation of the relativistic quantum field theory and SM. However, in contrast to the Klein-Gordon-Fock equation describing the Infinite Non-quantum Field $\Psi_{SB} (\mathbf{r},t)$ of Particles (Scalar Bosons), Eqs. (\ref{eq8abc}) and (\ref{eq45abc}) are formulated and interpreted in the present model as the single-particle relativistic equation for the unit-fields ${\psi _{0n}(\mathbf{r},t)}$ of the  Global Composite Field (\ref{eq205abc}), which is similar to the Schr{\"o}dinger equation for the de Broglie wave of the physical matter associated with a non-relativistic particle. In the traditional quantum field theories and SM, the Infinite Non-quantum Fields $\Psi_{SB} (\mathbf{r},t)$ and $\Psi^*_{SB} (\mathbf{r},t)$ placed into the Infinite ($V\rightarrow \infty$) Resonator of Universe, are given by the Fourier discrete (modal) decompositions that satisfies the Klein-Gordon-Fock equation of motion with the boundary conditions of the Universe Resonator: 
\begin{eqnarray} \label{eq206abc}
\Psi_{SB}=\sum_{n=1}^{N \rightarrow \infty}{(2V{\varepsilon}_{\mathbf{k}_n})^{-1/2}}({a}_{\mathbf{k}_n}e^{-i{\mathbf{k}_n{\mathbf{r}}}}+{b}_{\mathbf{k}_n}e^{i{\mathbf{k}_n{\mathbf{r}}}})
\end{eqnarray}
and
\begin{eqnarray} \label{eq207abc}
\Psi_{SB}^*=\sum_{n=1}^{N \rightarrow \infty}{(2V{\varepsilon}_{\mathbf{k}_n})^{-1/2}}({a}_{\mathbf{k}_n} e^{i{\mathbf{k}_n{\mathbf{r}}}}+{b}_{\mathbf{k}_n}e^{-i{\mathbf{k}_n{\mathbf{r}}}}),
\end{eqnarray}
where ${ \varepsilon}_{ \mathbf {k}_n } =\omega _{ \mathbf {k}_n}= ({ \mathbf {k}_n }^2 + m^2)^{1/2}$ is the Planck-Einstein energy of the $n$-th boson, which has the discrete (quantum) wave-number (\ref{eq61abc}) and frequency (\ref{eq62abc}) with $n=m=l$ due to the quantization imposed by the boundary conditions. Notice, the different resonator modes associated with the different particles are orthogonal in the Hilbert space, providing the interference-less and interaction-less behavior of the different modes and particles. In other words, the Infinite {\it{Non-quantum}} Fields $\Psi_{SB} (\mathbf{r},t)$ and $\Psi^*_{SB} (\mathbf{r},t)$, in fact, are the Infinite {\it{Quantum}} Fields, with the "hidden" quantization provided by the use of  the Universe Resonator. The second quantization of the fields (\ref{eq206abc}) and (\ref{eq207abc}) yields the Infinite Quantum Fields of Operators (Infinite Quantum Field-Operators):
\begin{eqnarray} \label{eq208abc}
\hat \Psi_{SB}=\sum_{n=1}^{N \rightarrow \infty}{(2V{\varepsilon}_{\mathbf{k}_n})^{-1/2}}({\hat a}_{\mathbf{k}_n}e^{-i{\mathbf{k}_n{\mathbf{r}}}}+{\hat b}_{\mathbf{k}_n}^{\dagger}e^{i{\mathbf{k}_n{\mathbf{r}}}})
\end{eqnarray}
and 
\begin{eqnarray} \label{eq209abc}
\hat \Psi_{SB} ^{\dagger }
=\sum_{n=1}^{N \rightarrow \infty}{(2V{\varepsilon}_{\mathbf{k}_n})^{-1/2}}({\hat a}_{\mathbf{k}_n}^{\dagger} e^{i{\mathbf{k}_n{\mathbf{r}}}}+{\hat b}_{\mathbf{k}_n}e^{-i{\mathbf{k}_n{\mathbf{r}}}}).
\end{eqnarray}
The operators ${ \hat a^{\dagger}}_{ \mathbf {k}_n }$, $ { \hat a }_{ \mathbf {k}_n}$, ${ \hat b^{ \dagger }}_{ \mathbf {k}_n }$, and ${ \hat b}_{\mathbf {k}_n }$ are
respectively the creation and destruction operators for the $n$-th particle (boson) and the $n$-th antiparticle (antiboson) inside the Resonator of Universe. The interference-less and interaction-less behavior of the resonator-mode operators associated with the particles is provided by the canonical commutation relations $[ \hat a _{k_n},\hat a^{ \dagger }_{k_m}]$ = $[ \hat b_{k_n}, \hat b^{\dagger }_{k_m}]$ = $\delta _{nm}$ for the bosons and antibosons (the other operator pairs commute). Here, $\delta _{nm}$ is the Kronecker symbol. The details and interpretations of the Infinite Non-quantum Fields (\ref{eq206abc}) and (\ref{eq207abc}) and the Infinite Quantum Fields of Operators  (\ref{eq208abc}) and (\ref{eq209abc}) have been presented in Sec. 2.1.2. of Part I. Although the Global {\it{Quantum}} Field (\ref{eq205abc}) of the present model is somewhat similar to the Infinite Non-quantum Fields (\ref{eq206abc}) and (\ref{eq207abc}) and the Infinite Quantum Fields of Operators  (\ref{eq208abc}) and (\ref{eq209abc}) of the traditional quantum field theories and SM, the Global {\it{Quantum}} Field is not a Field of Operators and its quantization does not require the {\it{resonator-like boundaries of Universe}}. Furthermore, the Global {\it{Quantum}} Field (\ref{eq205abc}) describes both the interacting (interfering) and non-interacting (interference-free) unit-fields associated with the elementary particles of {\it{the Einstein special relativity}}, while the Infinite Non-quantum Fields (\ref{eq206abc}) and (\ref{eq207abc}) and the Infinite Quantum Fields of Operators  (\ref{eq208abc}) and (\ref{eq209abc}) deal with the interaction-free particles of the very {\it{particular kind}} (scalar bosons).

\section{5. Substructures of a unit-field associated with fundamental (gravitational, electromagnetic, weak and strong) fields: The generator and associate components of the unit-field}

Equations (\ref{eq1abc}) - (\ref{eq205abc}) do not indicate the dependence of the {\it{single}} unit-field ${\psi}_{0n}$ and its energy ${\varepsilon_{0n}}$ on the intrinsic angular-momentum (spin) and charge of the unit-filed (particle). Indeed, the unit-field ${\psi}_{0n}$ is determined as the {\it{solution}} of the equation of motion (\ref{eq8abc}) or (\ref{eq45abc}) with the initial and boundary conditions imposed, which do not contain the spin and charge of the particle. That is to say that the unit-field ${\psi}_{0n}$ depends solely on the particle rest-mass $m_0$, which could be considered as the particular initial conditions of Eqs. (\ref{eq8abc}) and (\ref{eq45abc}). In Eqs. (\ref{eq1abc}) - (\ref{eq205abc}), one could distinguish the massive ($m_{0n} \neq 0$) and mass-less ($m_{0n} =0$) unit-fields ${\psi}_{0n}$. The unit-fields $\psi_{0n}$ corresponding to the solutions of Eq. (\ref{eq8abc}) or (\ref{eq45abc}) with the rest masses $m_{0n} \neq 0$ could be considered as the massive unit-fields associated with the massive particles. The unit-fields $\psi_{0n}$ determining by the equations (\ref{eq15abc}) and (\ref{eq46abc}) correspond to the mass-less unit-fields ( particles). The different configurations of a unit-field ${\psi}_{0n}$  corresponding to the different solutions of Eq. (\ref{eq8abc}) or (\ref{eq45abc}) for the given rest-mass $m_{0n}$ of the $n$-th particle are attributed to the different momentums $\mathbf{k}_{0n}$ of the particle. In such a case, the model describes the interfering (interacting) and non-interfering (non-interacting) unit-fields ${\psi}_{0n}$ of Global Composite Field (\ref{eq205abc}) attributed to the elementary particles satisfying Einstein's special relativity {\it{in the general form}}, which does not indicate the {\it{kind (type)}} of the unit-field (elementary particle). The description could be considered as the {\it{unified model of the unified unit-fields (particles)}}. In the {\it{unified model}} [Eqs. (\ref{eq1abc}) - (\ref{eq205abc})], a {\it{unified unit-field}} ${\psi}_{0n}$ does associate with the $n$-th unified {\it{point-particle}} of Einstein's special relativity that does not has a substructure. In other words, the unified unit-field ${\psi}_{0n}$ is considered as the 4-dimensional structure-less object in the spacetime, namely as a structure-less quanta of energy-mass of the Global Unified Field (\ref{eq205abc}). Naturally, {\it{the appropriate  substructure (internal structure)}} of the unit-field $\psi_{0n}$ associated with the respective {\it{kind}} of an elementary particle of the Universe could determine the four fundamental (gravitational, electromagnetic, weak and strong) interactions, forces and energies of the Nature. 

The appropriate structuring of the unit-field ${\psi}_{0n}$ is not a very simple problem. The internal substructure of the unit-field ${\psi} _{0n}$ should determine the {\it{kind}} of the elementary particle and its energy in agreement with the four fundamental ({\it{gravitational, electromagnetic, weak and strong}}) interactions, forces and energies associated with the rest masses, intrinsic angular momentums (spins) and charges of the respective kinds of elementary particles. An analysis shows that the unit-field ${\psi} _{0n}$ should be separated (divided) into the substructures as follows. The structuring of the {\it{unified}} unit-field ${\psi} _{0n}$ could be presented {\it{in general form}} as the {\it{indivisible connection}} of the {\it{unit-field generator}} ${\tilde \psi} _{0n}$ with the gravitational ($\phi_{Gn}$), electric ($\phi_{En}$), weak ($\phi_{Wn}$) and strong ($\phi_{Sn}$) {\it{associate-components}} ($ACs$) of the unit-field:
\begin{eqnarray} \label{eq210abc}
{\psi _{0n}} =  {\tilde \psi} _{0n} (\phi _{Gn}+\phi _{En}+\phi _{Wn}+\phi_{Sn})= {\tilde \psi} _{0n} \sum_{I=1}^{4 } \phi_{In}.
\end{eqnarray} 
The four unit-field components ${\tilde \psi} _{0n}\phi_{In}$ are attributed to the four fundamental [gravitational ($I$=1), electromagnetic ($I$=2), weak ($I$=3) and strong ($I$=4)] fields characterized by the respective masses, charges and spins. In the case of $\sum_{I=1}^{4 }\phi_{In}=1$, the structure-less unified unit-field ${\psi _{0n}}$ attributed to the $n$-th unified elementary particle {\it{is indistinguishable from the unit-field generator ${\tilde \psi} _{0n}$}} (for comparison, see the relevant examples in Part I of the present study). If the {\it{generator}} ${\tilde \psi} _{0n}$ of the unit-field component ${\tilde \psi} _{0n}\phi_{In}$ {\it{vanishes}}, then the {\it{unit-field component disappear}} (${\tilde \psi} _{0n}\phi_{In}=0$). That means that the associate-component $\phi_{In}$ of the unit-field structure ${\tilde \psi} _{0n} \sum_{I=1}^{4 } \phi_{In}$ is the secondary object, which is indivisibly connected with the unit-field generator ${\tilde \psi} _{0n}$. The use of the {\it{structured}} unit-field (\ref{eq210abc}) instead of the {\it{structure-less}} unified unit-field $\psi _{0n}$ yields the {\it{four-component version}} of the unified model [Eqs. (\ref{eq1abc}) - (\ref{eq205abc})]. If the unit-field associating with an elementary particle does not obey some fundamental interaction properties due to the absence of some associate-components attributed to the rest mass, charge or spin, then the unit-field structure simplifies to the one-, two- or three-component unit-field. The use of the one-, two-,  three- or four-component unit-field instead of the structure-less unified unit-field naturally yields the {\it{one-, two-,  three or four-component version}} (sub-model) of the unified model [Eqs. (\ref{eq1abc}) - (\ref{eq205abc})]. The {\it{associate-components}} $\phi_{In}$ may be interpreted as the four "dressings" of the {\it{unit-field}} generator ${\tilde \psi} _{0n}$. In such a picture, the {\it{unit-field generator}} ${\tilde \psi} _{0n}$ is a "carrier" of these "dressings" characterizing by the respective masses, charges and spins, which are responsible for the four fundamental (gravitational, electromagnetic, weak and strong) fields and interactions.

Although the unit-field representation (\ref{eq210abc}) could be somewhat strange for the non-experts in the quantum fields and SM, the representation reflects just a fact that the gravitational, electromagnetic, weak and strong fields satisfy the {\it{superposition principle}}. Notice, the alternative representation ${\psi _{0n}} =  {\tilde \psi} _{0n} (\phi _{Gn}\phi _{En}\phi _{Wn}\phi_{Sn})$ of the unit-field ${\psi _{0n}} $ results into incorrect vanishing of the unit-field in the region of the spacetime, where one of the associate-components does vanish, while other {\it{ACs}} do not. The gravitational, electromagnetic, weak and strong fields have different physical natures. In order to separate the physical natures of the unit-field associate-components ({\it{ACs}}) from each other, any mathematical product of the cross-correlation between the associate components of the different natures is assumed to be equal to zero. In other words, one should suppose that the associate components $\phi_{In}$ and $\phi_{Jn}$ should satisfy the orthogonality-like gauge (restriction):
\begin{eqnarray} \label{eq211abc}
\phi_{In}\phi_{Jn}=  \phi_{In}\phi_{Jn} \delta _{IJ},
\end{eqnarray} 
where $\delta _{IJ}$ is the Kronecker delta. Respectively, the values like $\nabla \phi_{In}\nabla\phi_{Jn}$ and $\dot {\phi} _{In}\dot {\phi} _{Jn}$ should be also vanished ($\nabla \phi_{In}\nabla \phi_{Jn}=\nabla \phi_{In}\nabla \phi_{Jn} \delta _{IJ}$ and $\dot {\phi} _{In}\dot {\phi} _{Jn}=\dot {\phi} _{In}\dot {\phi} _{Jn} \delta _{IJ}$) for $I \neq J$. The orthogonality-like gauge (\ref{eq211abc}), which could be mathematically interpreted as the orthogonality of the associate-components in the Hilbert (mathematical) space, will be used in the following sections to exclude the cross-correlation (interference) between the associate components of different physical natures in the physical spacetime. Mathematically, the orthogonality-like gauge (\ref{eq211abc}) means that the four {\it{ACs}} may be formally represented as the components of the four-vector or the four-component one-column (one-row) matrix in the Hilbert space. Such a mathematical representation is relevant, for instance, to the matrix mechanics of Heisenberg, the Pauly and Dirac matrices in quantum physics and the tensor formalism in Einstein's relativity, which exclude the cross-correlation (interference) of the different components of the $n$-vector or the $n$-component one-column (one-row) matrix. Nevertheless, for the sake of {\it{mathematical simplicity}}, instead of the aforementioned mathematical formalisms I will use the representation (\ref{eq210abc}) together with the orthogonality-like gauges (restrictions) imposed on the unit-field generator ${\tilde \psi} _{0n}$ and the associate-components ${\phi} _{In}$. In order to satisfy the physical properties of the experimentally observed elementary particles, the unit-field generator and associate-components should be calibrated (fixed) by the additional restriction (gauge) conditions, which will be presented in the following sections.  

\subsection{5.1. The one-component unit-field associated with the gravitational, electromagnetic, weak or strong ACs}

The one-component unit-field (particle) with the {\it{pure gravitational, electric, weak or strong}} associate-component ({\it{AC}}) is described by the respective one-component sub-model, which is based on the use of the {\it{one-component}} unit-field ${\psi _{0n}} = {\tilde \psi} _{0n} \phi_{In}$ determined by the equation
\begin{eqnarray} \label{eq212abc}
{\psi _{0n}} =  {\tilde \psi} _{0n} \phi_{Gn}={\tilde \psi} _{0n} \phi_{1n},
\end{eqnarray} 
\begin{eqnarray} \label{eq213abc}
{\psi _{0n}} =  {\tilde \psi} _{0n} \phi_{En}={\tilde \psi} _{0n} \phi_{2n},
\end{eqnarray} 
\begin{eqnarray} \label{eq214abc}
{\psi _{0n}} =  {\tilde \psi} _{0n} \phi_{Wn}={\tilde \psi} _{0n} \phi_{3n},
\end{eqnarray} 
or
\begin{eqnarray} \label{eq215abc}
{\psi _{0n}} =  {\tilde \psi} _{0n} \phi_{Sn}={\tilde \psi} _{0n} \phi_{4n},
\end{eqnarray} 
which is associated respectively with the pure gravitation, electromagnetism, weak interaction or strong interaction. Naturally, the unit-fields (\ref{eq212abc}) - (\ref{eq215abc}) are described by the unified model [Eqs. (\ref{eq1abc}) - (\ref{eq205abc})], in which the structure-less unified unit-field ${\psi _{0n}}$ is replaced by the one-component structured unit-field ${\psi _{0n}} = {\tilde \psi} _{0n} \phi_{In}$, where $I$ = 1, 2 3 or 4. The {\it{new equations}} for the unit-fields (\ref{eq212abc}) - (\ref{eq215abc}) would be distinguished from each other only by the index $I$. That is to say that the mathematical background of description of the one-component unit-fields is completely unified. The different physical properties of the pure gravitational ($I$=1), electromagnetic ($I$=2), weak ($I$=3) and strong ($I$=4) unit-fields (particles) are attributed to the different properties of the gravitational ($\phi_{1n}$), electromagnetic ($\phi_{2n}$), weak ($\phi_{3n}$) and strong ($\phi_{4n}$) associate-components of the unit-field. Therefore {\it{the gravitational, electric, weak and strong associate-components ($ACs$) of the unit-field should be considered as the mediators of the respective interactions and forces, while the unit-field generator ${\tilde \psi} _{0n}$ should be interpreted as a "carrier" of these $ACs$}}. It is not necessary to present the all new equations corresponding to Eqs. (\ref{eq1abc}) - (\ref{eq205abc}). Let me show only the equations, which determine the most basic physical properties of the pure gravitational, electromagnetic, weak or strong fields, particles and interactions.   

\subsubsection{5.1.1. The energy-mass relation and equation of motion for the single one-component unit-field with the pure gravitational, electromagnetic, weak or strong AC: The  one-component unit-field gauges}

\vspace{0.4cm}

{\it{1. The model 1st-version based on the straightforward generalization of the Einstein energy-mass relation for the single one-component unit-field by using the 2nd derivatives}}

\vspace{0.4cm}

The generalized energy-mass relation for the $n$-th one-component {\it{relativistic}} unit-field [(\ref{eq212abc}) - (\ref{eq215abc})] is given by Eq. (\ref{eq3abc}) as
\begin{eqnarray} \label{eq216abc}
{\varepsilon}_{0n}^{2} =  {{{\frac 1 2}\int_{V}}  \psi_{0n} ^* \left( -\ddot{\psi}_{0n} - \nabla ^2{\psi_{0n}} + m_{0n}^2\psi_{0n}\right)d^3x},  
\end{eqnarray}
where ${\psi _{0n}} = {\tilde \psi} _{0n} \phi_{In}$. The respective relativistic equation of motion, which is given by 
Eq. (\ref{eq8abc}) as
\begin{eqnarray} \label{eq217abc}
\square {\psi}_{0n} + m_{0}^2 {\psi}_{0n}=0,
\end{eqnarray}
yields the one {\it{general equation of motion}}
\begin{eqnarray} \label{eq218abc}
\phi_{In}(\square {\tilde \psi} _{0n} + m_{0}^2 {\tilde \psi} _{0n})  + {\tilde \psi} _{0n}\square \phi_{In}+2\dot {\tilde \psi} _{0n} \dot {\phi} _{In}-2\nabla {\tilde \psi} _{0n} \nabla { \phi} _{In}= 0
\end{eqnarray}
for the unit-field generator ${\tilde \psi} _{0n}$ and associate-component ${\phi} _{In}$.

Some solutions of the general equation (\ref{eq218abc}) correspond to the unit-fields (elementary particles) with the physical properties, which have not been observed experimentally. In order to select the solutions satisfying only the experimentally observed properties of the elementary particles, the field generator ${\tilde \psi} _{0n}$ and the associate-component ${\phi} _{In}$ should be calibrated (fixed) by the additional restriction (gauge) conditions. Such a procedure is similar to the procedure of fixing (choosing) a field gauge in the traditional field theories. An analysis of the consequences that follow from the use of different calibrations ({\it{gauges}}) of the unit-field generator ${\tilde \psi} _{0n}$ and the associate-component ${\phi} _{In}$ suggests the following two gauges. The first calibration (static gauge)
\begin{eqnarray} \label{eq219abc}
\dot {\phi} _{In}= 0
\end{eqnarray}
means that the associate-component ${\phi} _{In}$ is static one. The second restriction (gauge) 
\begin{eqnarray} \label{eq220abc}
\dot {\tilde \psi} _{0n} \dot {\phi} _{In} - \nabla {\tilde \psi} _{0n} \nabla { \phi} _{In}= 0
\end{eqnarray}
should be attributed to the orthogonality of the gradients $\nabla {\tilde \psi} _{0n}$ and $\nabla { \phi} _{In}$ in the Hilbert (mathematical) space, which means that the gradients $\nabla {\tilde \psi} _{0n}$ and $\nabla { \phi} _{In}$ do not cross-correlate (interfere) in the physical spacetime. The procedure of fixing the gauges (\ref{eq219abc}) and (\ref{eq220abc}) in Eq. (\ref{eq218abc}) yields the simpler equation of motion
\begin{eqnarray} \label{eq221abc}
\phi_{In}(\square {\tilde \psi} _{0n} + m_{0}^2 {\tilde \psi} _{0n})  + {\tilde \psi} _{0n}\nabla ^2 \phi_{In} = 0.
\end{eqnarray}
In addition to the gauges (\ref{eq219abc}) and (\ref{eq220abc}), one may use the Laplace gauge 
\begin{eqnarray} \label{eq222abc}
\nabla ^2 { \phi} _{In}= 0
\end{eqnarray}
or the Helmholtz (eigen) gauge 
\begin{eqnarray} \label{eq223abc}
\nabla ^2 { \phi} _{In}+\Gamma_{In}^2  { \phi} _{In}=0
\end{eqnarray}
imposed on the associate-component ${ \phi} _{In}$ by the Laplace (\ref{eq222abc}) or Helmholtz (\ref{eq223abc}) equation. The equation (\ref{eq221abc}) with the gauge  (\ref{eq222abc}) or  (\ref{eq223abc}) imposed on the unit-field {\it{AC}} yields the respective equation for the unit-field generator ${\tilde \psi} _{0n}$:  
\begin{eqnarray} \label{eq224abc}
\square {\tilde \psi} _{0n} + m_{0}^2 {\tilde \psi} _{0n} = 0
\end{eqnarray}
\begin{eqnarray} \label{eq225abc}
\square {\tilde \psi} _{0n} + (m_{0}^2 + \Gamma _{In}^2){\tilde \psi} _{0n} = 0.
\end{eqnarray}
In Eqs. (\ref{eq223abc}) and (\ref{eq225abc}), which depend on the {\it{eigen parameter}} $\Gamma _{In}^2$, the value $\Gamma _{In}$ can be the real ($\Gamma _{In} = |\Gamma _{In}|$) or imaginary ($\Gamma _{In} =i |\Gamma _{In}|$) value. The equation (\ref{eq224abc}) describes a unit-field (particle) with the mass $m_0$, while Eq. (\ref{eq225abc}) corresponds to a unit-field (particle) with the mass $(m_{0}^2 + \Gamma _{In}^2)^{1/2} $. Naturally, Eqs. (\ref{eq222abc}) and (\ref{eq224abc}) should describe the unit-fields (elementary particles) whose physical properties are different from the unit-fields (elementary particles) describing by Eqs. (\ref{eq223abc}) and (\ref{eq225abc}). The unit-fields (particles) with the different physical properties could be considered as the unit-fields (particles) of {\it{different kinds}}. Notice, Eqs. (\ref{eq224abc}) and (\ref{eq225abc}) are coupled to the associate-component ${\phi}_{0I}$, whose configuration is determining respectively by Eq. (\ref{eq222abc}) or (\ref{eq223abc}), only through the orthogonality condition (\ref{eq220abc}). If the eigen-parameter ${\Gamma ^2_{In}} =  0$, then Eqs. (\ref{eq222abc}) and (\ref{eq223abc}) are indistinguishable from each other. In such a case, the equations (\ref{eq224abc}) and (\ref{eq225abc}) are equivalent. In other words, they describe the same unit-field with the particle mass $m_0$. That means Eqs. (\ref{eq222abc}) and (\ref{eq224abc}) are the particular cases of the general equations (\ref{eq223abc}) and (\ref{eq225abc}). In addition, one should describe properties of the {\it{mass-less}} unit-fields (particles), which have the zero rest-mass ($m_0=0$) with the respective energy-mass relation ${\varepsilon}_{0n}^{2} =\mathbf{k}_{0n}^2$. The {\it{mass-less}}, one-component unit-fields (particles) are described simply by Eq. (\ref{eq218abc}) with the rest mass $m_0=0$, which has the form $\phi_{In}\square {\tilde \psi} _{0n}  + {\tilde \psi} _{0n}\square \phi_{In} = 0$. The unit-fields that satisfy this equation have the {\it{structure-less}} form (\ref{eq53abc}) of the the time-harmonic plane waves $\psi_{0n} (\mathbf{r},t)= \tilde \psi_{0n} (\mathbf{r},t) { \phi} _{In} (\mathbf{r},t) = a_{0n} e^{i({\mathbf{k}_{0n}{\mathbf{r}}}-{\varepsilon}_{0n} t )}$ with the generator $\tilde \psi_{0n} (\mathbf{r},t)={\sqrt  {a_{0n}}} e^{i([{\mathbf{k}_{0n}/2]{\mathbf{r}}}-[{\varepsilon}_{0n}/2] t )}$ and associate-component $\phi_{In} (\mathbf{r},t)={\sqrt  {a_{0n}}} e^{i([{\mathbf{k}_{0n}/2]{\mathbf{r}}}-[{\varepsilon}_{0n}/2] t )}$ that are {\it{indistinguishable from each other}}. In contrast to the massive unit-fields (particles), which obey the {\it{time-independent}} {\it{ACs}} ($\dot {\phi} _{In}= 0$), the mass-less unit-fields particles have the {\it{time-dependent}} {\it{ACs}} ($\dot {\phi} _{In}\neq 0$). In such a case the gauge (\ref{eq220abc}) provides the balance between the temporal and spatial variations of the unit-field. Notice, in the case of $\sum_{I=1}^{4 }\phi_{In}=1$, the structure-less unified unit-field ${\psi _{0n}}$ attributed to the $n$-th unified elementary particle {\it{is indistinguishable from the unit-field generator ${\tilde \psi} _{0n}$}} (for comparison, see Eq. (\ref{eq210abc}) and the relevant examples in Sec. I of the present study). 

The use of the gauges (\ref{eq219abc}) and (\ref{eq220abc}) in the general equation (\ref{eq216abc}) yields the unit-field energy squared as  
\begin{eqnarray} \label{eq226abc}
{\varepsilon}_{0n}^{2} =  {{{\frac {1} {2 }}\int_{V}}  {\tilde \psi} _{0n} ^* {\phi_{In}}^*  \left[ -\ddot {\tilde \psi} _{0n} - \nabla ^2{\tilde \psi} _{0n} +  m_{0}^2 {\tilde \psi} _{0n} + {\tilde \psi} _{0n} (\phi_{In})^{-1} \nabla ^2 \phi_{In} \right] \phi_{In} d^3x},
\end{eqnarray}
where the unit-field generator ${\tilde \psi} _{0n}$ is a solution of Eq. (\ref{eq224abc}) or (\ref{eq225abc}), while the unit-field associate-component $\phi_{In}$ is a solution of Eq. (\ref{eq222abc}) or (\ref{eq223abc}), respectively. The well-known stationary solutions of Eqs. (\ref{eq224abc}) and (\ref{eq225abc}) are given by the time-harmonic plane waves 
\begin{eqnarray} \label{eq227abc}
\tilde \psi_{0n} (\mathbf{r},t)=a_{0n} e^{i({\mathbf{k}_{0n}{\mathbf{r}}}-{\varepsilon}_{0n} t + \alpha _n )} 
\end{eqnarray}
of the form (\ref{eq53abc}) with the phase $\alpha_n \neq 0 $. Such solutions are also called the plane waves or the de Broglie waves.  In the case of the unit-field with the generator (\ref{eq227abc}) and the associate-component $\phi_{In}$ calibrated by the Laplace equation (\ref{eq222abc}), the unit-field energy squared is given by the (\ref{eq226abc}) as
\begin{eqnarray} \label{eq228abc}
{\varepsilon}_{0n}^{2} =  {{{\frac 1 2} \left[ {\varepsilon}_{0n}^{2} + (\mathbf{k}_{0n}^2+m_{0}^2) \right]\int_{V}}   {\tilde \psi} _{0n} ^* {\tilde \psi} _{0n}{\phi} _{In} ^* {\phi} _{In} d^3x}  =   \nonumber  \\  =   (\mathbf{k}_{0n}^2+m_{0}^2)  \int_{V} {\tilde \psi} _{0n} ^* {\tilde \psi} _{0n}{\phi} _{In} ^* {\phi} _{In} d^3x,
\end{eqnarray}
where the value ${\varepsilon}_{0n}^{2}$ {\it{does not depend on the unit-field generator phase}} $\alpha_n$. The square of energy (\ref{eq228abc}) of the one-component unit-field is consistent with the Einstein energy-mass relations ${\varepsilon}_{0}^{2} =\mathbf{k}_{0}^2+m_{0}^2$ and ${\varepsilon}_{0}^{2} =\mathbf{k}_{0}^2$ for the massive and mass-less particles of the Einstein special relativity under the normalization (\ref{eq18abc}), which here is given by the relation $\int_{V} {\tilde \psi} _{0n} ^* {\tilde \psi} _{0n}{\phi} _{In} ^* {\phi} _{In} d^3x= a_{0n}^2 \int_{V} {\phi} _{In} ^* {\phi} _{In} d^3x = 1$. For the unit-field with the generator (\ref{eq227abc}) and the associate-component $\phi_{In}$ calibrated by the Helmholtz equation (\ref{eq223abc}), the unit-field energy squared is given by the (\ref{eq226abc}) as
\begin{eqnarray} \label{eq229abc}
{\varepsilon}_{0n}^{2} = {{{\frac 1 2} \left[ {\varepsilon}_{0n}^{2} + (\mathbf{k}_{0n}^2+(m_{0}^2+\Gamma_{In}^2 ))  \right] \int_{V}}   {\tilde \psi} _{0n} ^* {\tilde \psi} _{0n}{\phi} _{In} ^* {\phi} _{In} d^3x} =   \nonumber  \\  = \left[ \mathbf{k}_{0n}^2+(m_{0}^2  + \Gamma_{In}^2 )  \right] \int_{V} {\tilde \psi} _{0n} ^* {\tilde \psi} _{0n}{\phi} _{In} ^* {\phi} _{In} d^3x,  
\end{eqnarray}
which is in agreement with the Einstein energy-mass relation ${\varepsilon}_{0}^{2} = \mathbf{k}_{0n}^2+(m_{0}^2+\Gamma_{In}^2) $ for a particle of the mass $(m_{0}^2+\Gamma_{In}^2)^{1/2}$ under the above-mentioned normalization (\ref{eq18abc}). The additional (gauge) mass ${\Gamma _{In}}$ could be attributed to the mass of {\it{AC}} of the unit-field (particle). In the case of the {\it{AC}} calibrated by the gauge (\ref{eq223abc}) with ${\Gamma _{In}}=0$, the gauge mass ${\Gamma _{In}}$ of the {\it{AC}} is equal to zero. Since the value ${\varepsilon}_{0n}^{2}\geq 0$, the unit-fields (particles) with the parameters ${\Gamma _{In}}= i | {\Gamma _{In}}|$ under the condition $ | {\Gamma _{In}}|\geq m_0$ could exist only as the moving ($\left[ \mathbf{k}_{0n}^2+(m_{0}^2  - |\Gamma_{In}|^2 )  \right] > 0$) unit-fields (particles).  

In the case of a {\it{non-relativistic}} unit-field [(\ref{eq212abc}) - (\ref{eq215abc})] calibrated by Eqs. (\ref{eq219abc}) and (\ref{eq220abc}) with the generator (\ref{eq227abc}) and the associate-component $\phi_{In}$ calibrated by the Laplace equation (\ref{eq222abc}), the non-relativistic energy-mass relation (\ref{eq10abc}) yields the equation
\begin{eqnarray} \label{eq230abc}
{\varepsilon}_{0n} = {{{\frac 1 2} \left[ \varepsilon _{0n} + \left( {\frac {\mathbf{k}_{0n}^2} {2m_0} }+ m_{0} \right) \right] \int_{V}}   {\tilde \psi} _{0n} ^* {\tilde \psi} _{0n}{\phi} _{In} ^* {\phi} _{In} d^3x}=   \nonumber  \\  =  {  \left( {\frac {\mathbf{k}_{0n}^2} {2m_0} }+ m_{0} \right)  \int_{V}}   {\tilde \psi} _{0n} ^* {\tilde \psi} _{0n}{\phi} _{In} ^* {\phi} _{In} d^3x,
\end{eqnarray}
which is consistent with the Einstein energy-mass relation ${\varepsilon}_{0} ={\frac {\mathbf{k}_{0n}^2} {2m_0} }+ m_{0}$ for a non-relativistic (Newton) particle under the normalization (\ref{eq18abc}). The equation of non-relativistic motion (\ref{eq13abc}) 
\begin{eqnarray} \label{eq231abc}
i\dot {\psi}_{0n} = -{\frac 1 {2m_0} } {\nabla ^2{ \psi_{0n}}} + m_0  \psi_{0n}
\end{eqnarray}
with the aforementioned calibrations yields the non-relativistic equation
\begin{eqnarray} \label{eq232abc}
i\dot {\tilde \psi}_{0n} = -{\frac 1 {2m_0} } {\nabla ^2{\tilde \psi_{0n}}} + m_0 \tilde \psi_{0n}
\end{eqnarray}
for the unit-field generator${\tilde \psi}_{0n}$ coupled to the associate-component ${\phi} _{In}$ only through the orthogonality gauge (\ref{eq220abc}). For the stationary unit-field generator ${\tilde \psi}_{0n}$, Eq. (\ref{eq232abc}) yields the Schr{\"o}dinger stationary equation 
\begin{eqnarray} \label{eq233abc}
\varepsilon \tilde \psi_{0n} = - {\frac 1 {2m_0} } {\nabla ^2{\tilde \psi_{0n}}},
\end{eqnarray}
where $\varepsilon = \varepsilon _{0n} -m_0$. The use of the gauge (\ref{eq223abc}) instead (\ref{eq222abc}) yields the non-relativistic equation of motion 
\begin{eqnarray} \label{eq234abc}
i\dot {\tilde \psi}_{0n} = -{\frac 1 {2m_0} } {\nabla ^2{\tilde \psi_{0n}}} + \left(  m_0  + {\frac {\Gamma_{In}^2 } {2m_0} }  \right) \tilde \psi_{0n}
\end{eqnarray}
with the non-relativistic energy
\begin{eqnarray} \label{eq235abc}
{\varepsilon}_{0n} = {{{\frac 1 2} \left[ \varepsilon _{0n} +  \left( {\frac {\mathbf{k}_{0n}^2} {2m_0} }+ \left(  m_0  + {\frac {\Gamma_{In}^2 } {2m_0} }  \right) \right)  \right] \int_{V}}   {\tilde \psi} _{0n} ^* {\tilde \psi} _{0n}{\phi} _{In} ^* {\phi} _{In} d^3x} =   \nonumber  \\  = {   \left( {\frac {\mathbf{k}_{0n}^2} {2m_0} }+   m_0  + {\frac {\Gamma_{In}^2 } {2m_0} }  \right)  \int_{V}}   {\tilde \psi} _{0n} ^* {\tilde \psi} _{0n}{\phi} _{In} ^* {\phi} _{In} d^3x ,
\end{eqnarray}
which is consistent with the Einstein energy-mass relation $ {\frac {\mathbf{k}_{0n}^2} {2m_0} }+ \left(  m_0  + {\frac {\Gamma_{In}^2 } {2m_0} }  \right)$ for a non-relativistic (Newton) particle of the mass $\left(  m_0  + {\frac {\Gamma_{In}^2 } {2m_0} }  \right)^{1/2}$ under the normalization (\ref{eq18abc}). For the stationary conditions, Eq. (\ref{eq234abc}) is presented as the Schr{\"o}dinger stationary equation 
\begin{eqnarray} \label{eq236abc}
\varepsilon \tilde \psi_{0n} = - {\frac 1 {2m_0} } {\nabla ^2{\tilde \psi_{0n}}},
\end{eqnarray}
where $\varepsilon = \varepsilon _{0n} -\left( m_0 + {\frac {\Gamma_{In}^2 } {2m_0} }  \right)$. The physical properties of the non-relativistic unit-field (elementary particle) describing by Eqs. (\ref{eq230abc}) - (\ref{eq233abc}) are different from the respective Eqs. (\ref{eq234abc}) -  (\ref{eq236abc}). That means that the equations describe the non-relativistic unit-fields (particles) of {\it{different kind}}.

\vspace{0.4cm}

{\it{2. The model 2-nd-version based on the generalization of the Einstein energy-mass relation for the one-component unit-field by using the Euler-Lagrange formalism and the 2nd derivatives}}

\vspace{0.4cm}

The Euler-Lagrange formalism (\ref{eq17abc}) - (\ref{eq32abc}), which is based on the generalized energy-mass relation 
\begin{eqnarray} \label{eq237abc}
{\varepsilon}_{0n}^{2} =  {{{\frac 1 2}\int_{V}}  \psi_{0n} ^* \left( -\ddot{\psi}_{0n} - \nabla ^2{\psi_{0n}} + m_{0n}^2\psi_{0n}\right)d^3x} 
\end{eqnarray}
and the respective relativistic equation of motion
\begin{eqnarray} \label{eq238abc}
\square {\psi}_{0n} + m_{0}^2 {\psi}_{0n}=0,
\end{eqnarray}
can be easily rewritten for the one-component unit-field ${\psi _{0n}} = {\tilde \psi} _{0n} \phi_{In}$ given by Eqs. (\ref{eq212abc}) - (\ref{eq215abc}). This simple procedure yields the {\it{equations of motions}}, which are {\it{indistinguishable}} from the respective equations (\ref{eq217abc}) - (\ref{eq225abc}), (\ref{eq231abc}) - (\ref{eq234abc}) and (\ref{eq236abc}). The energy-mass relations in the Euler-Lagrange formalism have the forms 
\begin{eqnarray} \label{eq239abc}
{\varepsilon}_{0n}^{2} =  {{{\frac 1 2}\int_{V}}  \psi_{0n} ^* \left( -\ddot{\psi}_{0n} - \nabla ^2{\psi_{0n}} + m_{0n}^2\psi_{0n}\right)d^3x},  
\end{eqnarray}
\begin{eqnarray} \label{eq240abc}
{\varepsilon}_{0n}^{2} =  {{{\frac {1} {2 }}\int_{V}}  {\tilde \psi} _{0n} ^* {\phi_{In}}^*  \left[ -\ddot {\tilde \psi} _{0n} - \nabla ^2{\tilde \psi} _{0n} +  m_{0}^2 {\tilde \psi} _{0n} + {\tilde \psi} _{0n} (\phi_{In})^{-1} \nabla ^2 \phi_{In} \right] \phi_{In} d^3x}, 
\end{eqnarray}
\begin{eqnarray} \label{eq241abc}
{\varepsilon}_{0n}^{2} =  { (\mathbf{k}_{0n}^2+m_{0}^2) \int_{V}}   {\tilde \psi} _{0n} ^* {\tilde \psi} _{0n}{\phi} _{In} ^* {\phi} _{In} d^3x, 
\end{eqnarray}
\begin{eqnarray} \label{eq242abc}
{\varepsilon}_{0n}^{2} =  {(\mathbf{k}_{0n}^2+(m_{0}^2+\Gamma_{In}^2 )) \int_{V}}   {\tilde \psi} _{0n} ^* {\tilde \psi} _{0n}{\phi} _{In} ^* {\phi} _{In} d^3x,  
\end{eqnarray}
\begin{eqnarray} \label{eq243abc}
{\varepsilon}_{0n} =  {   \left( {\frac {\mathbf{k}_{0n}^2} {2m_0} }+ m_{0} \right)  \int_{V}}   {\tilde \psi} _{0n} ^* {\tilde \psi} _{0n}{\phi} _{In} ^* {\phi} _{In} d^3x
\end{eqnarray}
and
\begin{eqnarray} \label{eq244abc}
{\varepsilon}_{0n} =  {  \left[ {\frac {\mathbf{k}_{0n}^2} {2m_0} }+ \left(  m_0  + {\frac {\Gamma _{In}^2} {2m_0} }  \right) \right]  \int_{V}}   {\tilde \psi} _{0n} ^* {\tilde \psi} _{0n}{\phi} _{In} ^* {\phi} _{In} d^3x,
\end{eqnarray}
which are indistinguishable respectively from the energy-mass relations (\ref{eq216abc}), (\ref{eq226abc}), (\ref{eq228abc}), (\ref{eq229abc}),  (\ref{eq230abc}) and (\ref{eq235abc}). Notice, Eqs. (\ref{eq239abc}) - (\ref{eq244abc}) are consistent with the Einstein energy-mass relation under the normalization (\ref{eq18abc}) given by the equation
 \begin{eqnarray} \label{eq245abc}
\int_{V} {\tilde \psi} _{0n} ^* {\tilde \psi} _{0n}{\phi} _{In} ^* {\phi} _{In} d^3x=1.
\end{eqnarray}

\vspace{0.4cm}
 
{\it{3. The model 3rd-version based on the straightforward generalization of the Einstein energy-mass relation for the one-component unit-field by using the 1st derivatives}}

\vspace{0.4cm}

The straightforward generalization of the Einstein energy-mass relation by using the 1st derivatives of the one-component {\it{relativistic}} unit-field  ${\psi _{0n}} = {\tilde \psi} _{0n} \phi_{In}$ is given by Eq. (\ref{eq36abc}) as
\begin{eqnarray} \label{eq246abc}
{\varepsilon}_{0n}^{2} = {{{\frac 1 2}\int_{V}}  \left(  \dot{\psi_{0n} ^*} \dot{\psi_{0n}} +  \nabla \psi_{0n}^*  \nabla \psi_{0n} + m_0^2 \psi_{0n} ^* \psi_{0n}   \right)d^3x},
\end{eqnarray}
The respective equation of motion is given by Eq. (\ref{eq37abc}) in the form
\begin{eqnarray} \label{eq247abc}
\dot{\psi_{0n} ^*} \dot{\psi_{0n}} = \nabla \psi_{0n}^*  \nabla \psi_{0n} + m_0^2 \psi_{0n} ^* \psi_{0n}, 
\end{eqnarray}
where ${\psi _{0n}} = {\tilde \psi} _{0n} \phi_{In}$. For the unit-field calibrated by the gauges 
\begin{eqnarray} \label{eq248abc}
\dot {\phi} _{In}= 0
\end{eqnarray}
\begin{eqnarray} \label{eq249abc}
\nabla {\tilde \psi} _{0n} \nabla { \phi} _{In}= 0
\end{eqnarray}
 the equation of motion (\ref{eq247abc}) can be presented as a system of two equations
\begin{eqnarray} \label{eq250abc}
\dot{\tilde \psi_{0n} ^*} \dot{\tilde \psi_{0n}} = \nabla \tilde \psi_{0n}^*  \nabla \tilde \psi_{0n} + \left(m_0^2 +{\frac {\nabla {\phi_{In}^*}\nabla {\phi_{In}}}   {\phi_{In} ^*  \phi_{In}} } \right)\tilde \psi_{0n} ^* \tilde \psi_{0n}. 
\end{eqnarray}
for the unit-field generator $ \psi_{0n}$ and addition-component $ \phi_{In}$. The use of the additional gauge
\begin{eqnarray} \label{eq251abc}
\nabla {\phi}^* _{In} \nabla { \phi} _{In}= 0
\end{eqnarray}
yields the equations
\begin{eqnarray} \label{eq252abc}
{\varepsilon}_{0n}^{2} = {{{\frac 1 2}\int_{V}}  \left[  \dot{\tilde \psi_{0n} ^*} \dot{\tilde \psi_{0n}} + \nabla \tilde \psi_{0n}^*  \nabla \tilde \psi_{0n} + m_0^2 \tilde \psi_{0n} ^* \tilde \psi_{0n}  \right]  \phi_{In} ^*  \phi_{In}d^3x},
\end{eqnarray}
and
\begin{eqnarray} \label{eq253abc}
\dot{\tilde \psi_{0n} ^*} \dot{\tilde \psi_{0n}} = \nabla \tilde \psi_{0n}^*  \nabla \tilde \psi_{0n} + m_0^2 \tilde \psi_{0n} ^* \tilde \psi_{0n}. 
\end{eqnarray}
which are consistent with the Einstein energy-mass relation ${\varepsilon}_{0}^{2} = \mathbf{k}_{0n}^2+m_{0}^2 $ for a particle of the mass $m_{0}$. The use of the gauge 
\begin{eqnarray} \label{eq254abc}
\nabla {\phi}^* _{In} \nabla { \phi} _{In}= \Gamma _{In}^2 {\phi}^* _{In} { \phi} _{In}
\end{eqnarray}
instead of the calibration (\ref{eq251abc}) yields
\begin{eqnarray} \label{eq255abc}
{\varepsilon}_{0n}^{2} = {{{\frac 1 2}\int_{V}}  \left[  \dot{\tilde \psi_{0n} ^*} \dot{\tilde \psi_{0n}} + \nabla \tilde \psi_{0n}^*  \nabla \tilde \psi_{0n} + (m_0^2 + \Gamma _{In}^2 )\tilde \psi_{0n} ^* \tilde \psi_{0n} \right]  \phi_{In} ^*  \phi_{In}d^3x},
\end{eqnarray}
and
\begin{eqnarray} \label{eq256abc}
\dot{\tilde \psi_{0n} ^*} \dot{\tilde \psi_{0n}} = \nabla \tilde \psi_{0n}^*  \nabla \tilde \psi_{0n} + (m_0^2 + \Gamma _{In}^2)\tilde \psi_{0n} ^* \tilde \psi_{0n}. 
\end{eqnarray}
which are consistent with the Einstein energy-mass relation ${\varepsilon}_{0}^{2} = \mathbf{k}_{0n}^2+ (m_{0}^2 + \Gamma _{In}^2  )$ for a particle of the mass $(m_{0}^2 + \Gamma _{In} ^2)^{1/2}$ under the unit-field normalization (\ref{eq18abc}).

In the case of the one-component {\it{non-relativistic}} unit-field ${\psi _{0n}} = {\tilde \psi} _{0n} \phi_{In}$, the non-relativistic energy ${\varepsilon}_{0n}$ that corresponds to Eq. (\ref{eq38abc}) is determined as 
\begin{eqnarray} \label{eq257abc}
{\varepsilon}_{0n}^{2} = {{\frac 1 2} {\int_{V}}  \left( i\psi_{0n} ^*  \dot{\psi}_{0n}+{\frac 1 {2m_0} } {\nabla {\psi_{0n}^*}\nabla {\psi_{0n}}} + m_0\psi_{0n} ^*  \psi_{0n}\right)d^3x}.
\end{eqnarray}
The respective equation of motion is given by the formula (\ref{eq39abc}) in the form 
\begin{eqnarray} \label{eq258abc}
i\psi_{0n} ^*  \dot{\psi}_{0n}={\frac 1 {2m_0} } {\nabla {\psi_{0n}^*}\nabla {\psi_{0n}}} + m_0\psi_{0n} ^*  \psi_{0n},
\end{eqnarray}
where ${\psi _{0n}} = {\tilde \psi} _{0n} \phi_{In}$. For the unit-field calibrated by the gauges (\ref{eq248abc}) and (\ref{eq249abc}), the equation of motion (\ref{eq258abc}) can be presented as the equation
\begin{eqnarray} \label{eq259abc}
i\tilde \psi_{0n} ^*  \dot{\tilde \psi}_{0n}={\frac 1 {2m_0} } {\nabla {\tilde \psi_{0n}^*}\nabla {\tilde \psi_{0n}}} + \left(m_0 +{\frac {\nabla {\phi_{In}^*}\nabla {\phi_{In}}}   {\phi_{In} ^*  \phi_{In}} } \right)\tilde \psi_{0n} ^*  \tilde \psi_{0n},
\end{eqnarray}
for the unit-field generator $\tilde \psi_{0n}$. The use of the additional gauge (\ref{eq251abc}) or (\ref{eq254abc}) respectively yields the equations
\begin{eqnarray} \label{eq260abc}
{\varepsilon}_{0n}^{2} = {{\frac 1 2} {\int_{V}}  \left( i {\tilde \psi_{0n} ^*}  \dot{\tilde \psi}_{0n}+{\frac 1 {2m_0} } {\nabla {\tilde \psi_{0n}^*}\nabla {\tilde \psi_{0n}}} + m_0{\tilde \psi_{0n} ^* } {\tilde \psi_{0n}}\right)  \phi_{In} ^*  \phi_{In}d^3x}
\end{eqnarray}
and
\begin{eqnarray} \label{eq261abc}
i{\tilde \psi_{0n} ^* } \dot{\tilde \psi}_{0n}={\frac 1 {2m_0} } {\nabla {\tilde \psi_{0n}^*}\nabla {\tilde \psi_{0n}}} + m_0 {\tilde \psi_{0n} ^*}  {\tilde \psi_{0n}}
\end{eqnarray}
or the equations 
\begin{eqnarray} \label{eq262abc}
{\varepsilon}_{0n}^{2} = {{\frac 1 2} {\int_{V}}  \left[ i {\tilde \psi_{0n} ^*}  \dot{\tilde \psi}_{0n}+  {\frac 1 {2m_0} } {\nabla {\tilde \psi_{0n}^*}\nabla {\tilde \psi_{0n}}} +  \left(  m_0 + {\frac {\Gamma  _{In}^2} {2m_0} } \right) {\tilde \psi_{0n} ^* } {\tilde \psi_{0n}}\right]  \phi_{In} ^*  \phi_{In} d^3x} 
\end{eqnarray}
and
\begin{eqnarray} \label{eq263abc}
i{\tilde \psi_{0n} ^* } \dot{\tilde \psi}_{0n}={\frac 1 {2m_0} } {\nabla {\tilde \psi_{0n}^*}\nabla {\tilde \psi_{0n}}} +  \left(  m_0 + {\frac {\Gamma _{In}^2} {2m_0} } \right) {\tilde \psi_{0n} ^*}  {\tilde \psi_{0n}},
\end{eqnarray}
which are consistent with the Einstein energy-mass relations ${\varepsilon}_{0} =  {\frac {\mathbf{k}_{0n}^2} {2m_0} }  + m_{0}$ or ${\varepsilon}_{0} =  {\frac {\mathbf{k}_{0n}^2} {2m_0} }  + \left(  m_0 + {\frac {\Gamma _{In}^2} {2m_0} } \right)$ for a non-relativistic particle of the mass $m_{0}$ or $\left(  m_0 + {\frac {\Gamma _{In}^2} {2m_0} } \right)^{1/2}$ under the unit-field normalization (\ref{eq18abc}).

\vspace{0.4cm}

{\it{4. The model 4th-version based on the generalization of the Einstein energy-mass relation for the one-component unit-field by using the Euler-Lagrange formalism and the 1st derivatives}}

\vspace{0.4cm}

In the model version based on on the generalization of the Einstein energy-mass relation by using the Euler-Lagrange formalism and the 1st derivatives of the {\it{relativistic}} unit-field (\ref{eq215abc}), the square of relativistic energy and the equation of relativistic motion are given respectively by Eqs. (\ref{eq44abc}) and (\ref{eq45abc}) as  
\begin{eqnarray} \label{eq264abc}
{\varepsilon}_{0n}^{2}={{{\frac 1 2}\int_{V}} \left(  \dot{\psi_{0n} ^*} \dot{\psi_{0n}} +  \nabla \psi_{0n}^*  \nabla \psi_{0n} + m_0^2 \psi_{0n} ^* \psi_{0n}   \right)d^3x}
\end{eqnarray}
and
\begin{eqnarray} \label{eq265abc}
\square {\psi}_{0n} + m_0^2 {\psi}_{0n}=0,
\end{eqnarray}
where ${\psi _{0n}} = {\tilde \psi} _{0n} \phi_{In}$ with the calibrations (\ref{eq248abc}), (\ref{eq249abc}) and (\ref{eq251abc}) or the gauges (\ref{eq248abc}), (\ref{eq249abc}) and (\ref{eq254abc}). In the case of the {\it{non-relativistic}} unit-field (\ref{eq215abc}), the non-relativistic energy ${\varepsilon}_{0n}$ and the equation of non-relativistic motion are given respectively by Eqs. (\ref{eq51abc}) and (\ref{eq52abc}) as 
\begin{eqnarray} \label{eq266abc}
{\varepsilon}_{0n}={{{\frac 1 2}\int_{V}}  \left( i\psi_{0n} ^*  \dot{\psi}_{0n}+{\frac 1 {2m_0} } {\nabla {\psi_{0n}^*}\nabla {\psi_{0n}}} + m_0\psi_{0n} ^*  \psi_{0n}\right)d^3x}.
\end{eqnarray}
and 
\begin{eqnarray} \label{eq267abc}
i\dot{\psi}_{0n} = -{\frac 1 {2m_0} } {\nabla ^2{\psi_{0n}}} + m_0\psi_{0n},
\end{eqnarray}
where ${\psi _{0n}} = {\tilde \psi} _{0n} \phi_{In}$. The relativistic (\ref{eq265abc}) and non-relativistic (\ref{eq267abc}) equations of motion for the unit-field ${\psi _{0n}} = {\tilde \psi} _{0n} \phi_{In}$ are indistinguishable from Eqs. (\ref{eq217abc}) and (\ref{eq231abc}). That means that the relativistic and non-relativistic equations of motions in the 4th version of the unit-field model are equivalent to the respective equations of the 1st version. The relativistic (\ref{eq264abc}) and non-relativistic (\ref{eq266abc}) energy-mass relations are indistinguishable from the relativistic equations (\ref{eq252abc}) and (\ref{eq255abc}) and the non-relativistic equations (\ref{eq260abc}) and (\ref{eq262abc}).

\subsection{5.2. The one-, two-, three- and four-component unit-fields (particles)}

The {\it{one-component}} unit-field with the pure gravitational ($G$), electric ($E$), weak ($W$) or strong ($S$) associate-component, which has been considered in Sec. (5.1.), is given in the general form as 
\begin{eqnarray} \label{eq268abc}
{\psi _{0n}} =  {\tilde \psi} _{0n} \phi_{1n}
\end{eqnarray} 
\begin{eqnarray} \label{eq269abc}
{\psi _{0n}} =  {\tilde \psi} _{0n} \phi_{2n}
\end{eqnarray} 
\begin{eqnarray} \label{eq270abc}
{\psi _{0n}} =  {\tilde \psi} _{0n} \phi_{3n}
\end{eqnarray} 
\begin{eqnarray} \label{eq271abc}
{\psi _{0n}} =  {\tilde \psi} _{0n} \phi_{4n}.
\end{eqnarray} 
The {\it{two-component}} unit-field combines the $G$, $E$, $W$ and $S$ associate-components as
\begin{eqnarray} \label{eq272abc}
{\psi _{0n}} =  {\tilde \psi} _{0n} (\phi_{1n}+ \phi_{2n})
\end{eqnarray} 
\begin{eqnarray} \label{eq273abc}
{\psi _{0n}} =  {\tilde \psi} _{0n} (\phi_{1n} + \phi_{3n})
\end{eqnarray} 
\begin{eqnarray} \label{eq274abc}
{\psi _{0n}} =  {\tilde \psi} _{0n} (\phi_{1n}+ \phi_{4n})
\end{eqnarray} 
\begin{eqnarray} \label{eq275abc}
{\psi _{0n}} =  {\tilde \psi} _{0n}( \phi_{2n}+ \phi_{3n})
\end{eqnarray} 
\begin{eqnarray} \label{eq276abc}
{\psi _{0n}} =  {\tilde \psi} _{0n} (\phi_{2n}+ \phi_{4n})
\end{eqnarray}
\begin{eqnarray} \label{eq277abc}
{\psi _{0n}} =  {\tilde \psi} _{0n} (\phi_{3n}+ \phi_{4n}).
\end{eqnarray}
The {\it{three-component}} unit-field combines the $G$, $E$, $W$ and $S$ associate-components as 
\begin{eqnarray} \label{eq278abc}
{\psi _{0n}} =  {\tilde \psi} _{0n} (\phi_{1n}+ \phi_{2n}+ \phi_{3n})
\end{eqnarray}
\begin{eqnarray} \label{eq279abc}
{\psi _{0n}} =  {\tilde \psi} _{0n} (\phi_{2n}+ \phi_{3n}+ \phi_{4n})
\end{eqnarray}
\begin{eqnarray} \label{eq280abc}
{\psi _{0n}} =  {\tilde \psi} _{0n} (\phi_{1n} + \phi_{2n}+ \phi_{4n})
\end{eqnarray} 
\begin{eqnarray} \label{eq281abc}
{\psi _{0n}} =  {\tilde \psi} _{0n} (\phi_{1n} + \phi_{3n}+ \phi_{4n}).
\end{eqnarray}
The {\it{four-component}} unit-field has the form
\begin{eqnarray} \label{eq282abc}
{\psi _{0n}} =  {\tilde \psi} _{0n} ( \phi_{1n}+ \phi_{2n}+ \phi_{3n}+ \phi_{4n}).
\end{eqnarray} 
The orthogonality-like condition  (\ref{eq211abc}) means that the interference (cross-correlation) $G-G$, $E-E$, $W-W$ and $S-S$ of the unit-field components of the same physical nature is permitted in the present model, while the interactions (cross-correlations) $G-E$, $G-W$, $G-S$, $E-W$, $E-S$ and $W-S$ of associate-components with the different physical natures is excluded. For an example, the condition  (\ref{eq211abc}) permits the Newton gravitational force $F_N\sim m_{01}m_{02}/R^2$ and Coulomb electrical force $F_N\sim q_{01}q_{02}/R^2$, but does not support the Newton-Coulomb like forces $F_{C-N}\sim q_{01}m_{02}/R^2 $ and $F_{N-C}\sim m_{01}q_{02}/R^2 $, where $q$ denotes the electric charge of the unit-field (particle). One can easily extend the present model to a such kind of forces, if they do exist somewhere in the Universe.

In the {\it{general (unified) form}}, the one-, two-, three- or four-component unit-field can be presented as  
\begin{eqnarray} \label{eq283abc}
{\psi _{0n}} =   {\tilde \psi} _{0n} \Phi^{a}_{n},
\end{eqnarray} 
where $a$ = 1, 2, 3 or 4, and the {\it{total associate-components (TACs)}} $\Phi^{1}_{n}$,  $\Phi^{2}_{n}$, $\Phi^{3}_{n}$ and $\Phi^{4}_{n}$ correspond the one-, two-, three- and four-component unit-fields describing respectively by Eqs. (\ref{eq268abc}) - (\ref{eq271abc}), Eqs. (\ref{eq272abc}) - (\ref{eq277abc}), Eqs. (\ref{eq278abc}) - (\ref{eq281abc}) and Eq. (\ref{eq282abc}). The  gravitational ($I=1$), electric ($I=2$), weak ($I=3$) and strong ($I=4$) {\it{associate-components (ACs)}} $\phi_{In}$ have the common unit-field generator ${\tilde \psi} _{0n}$. If the unit-field generator ${\tilde \psi} _{0n}$ vanishes, then the all components ${\tilde \psi} _{0n}\phi_{In}$ of the unit-field ${\psi _{0n}} $ disappear. That is to say that the unit-field ${\psi _{0n}} $ is annihilated by vanishing the unit-field generator ${\tilde \psi} _{0n}$. The zero total associate-component ($\Phi^{a}_{n}=0$) of the unit-field also annihilates the unit-field (particle).

\subsubsection{5.2.1. The energy-mass relations and equations of motions for the single unit-fields with one-, two-, three- and four-component ACs: The multi-component unit-field gauges}

Section (5.1) has presented the 1st, 2nd, 3rd and 4th versions of the energy-mass relations and equations of motions for the single, one-component, {\it{structured}} unit-field. Section (5.2) generalizes the results of Sec. (5.1) for the {\it{two-, three- and four-component}} unit-fields. Notice, the 1st, 2nd, 3rd and 4th versions of Sec. 3 describing the structure-less (unified) unit-field and the all four versions 1st, 2nd, 3rd and 4th versions of Sec. (5.1) dealing with the {\it{structured}} unit-field are considered to be equivalent ones, Sec. (5.1) yielded the energy-mass relations that require the different gauges [(\ref{eq248abc}), (\ref{eq249abc}) and (\ref{eq251abc}) or (\ref{eq248abc}), (\ref{eq249abc}) and (\ref{eq254abc})] in order to {\it{satisfy}} the Einstein energy-mass relation. Since the four versions are considered to be equivalent, I present the generalization for the {\it{two-, three- and four-component}} unit-fields by using the 1-st version of the model, only. One can easily rewrite the formulas for the 2nd, 3rd and 4th versions of the energy-mass relations and equations of motions.

Let me now derive the energy-mass relations and equations of motions for the single, multi-component unit-field (\ref{eq283abc}) with the total associate-component $\Phi^{a}_{n}$ by using the {\it{1-st version}} of the unit-field model, which is based on the straightforward generalization of the Einstein energy-mass relation for the multi-component unit-field by using the {\it{2nd derivatives}} (for the one-component unit-field  ${\psi _{0n}} =   {\tilde \psi} _{0n} \phi _{In}$ the procedure is given in Sec. 5.1.1). The generalized energy-mass relation for the $n$-th one-component {\it{relativistic}} unit-field (\ref{eq283abc}) is given by Eq. (\ref{eq3abc}) as
\begin{eqnarray} \label{eq284abc}
{\varepsilon}_{0n}^{2} =  {{{\frac 1 2}\int_{V}}  \psi_{0n} ^* \left( -\ddot{\psi}_{0n} - \nabla ^2{\psi_{0n}} + m_{0n}^2\psi_{0n}\right)d^3x},  
\end{eqnarray}
where the unit-field ${\psi _{0n}} =   {\tilde \psi} _{0n} \Phi^{a}_{n}$ has the {\it{generator}} ${\tilde \psi} _{0n}$ and the {\it{total associate-component}} $\Phi^{a}_{n}$. The respective relativistic equation of motion, which is given by Eq. (\ref{eq8abc}) as
\begin{eqnarray} \label{eq285abc}
\square {\psi}_{0n} + m_{0}^2 {\psi}_{0n}=0,
\end{eqnarray}
yields the one general equation of motion
\begin{eqnarray} \label{eq286abc}
\Phi^{a}_{n} (\square {\tilde \psi} _{0n} + m_{0}^2 {\tilde \psi} _{0n})  + {\tilde \psi} _{0n}\square \Phi^{a}_{n} +2\dot {\tilde \psi} _{0n}\dot  \Phi^{a}_{n}+2\nabla {\tilde \psi} _{0n} \nabla \Phi^{a}_{n} = 0
\end{eqnarray}
for the unit-field generator ${\tilde \psi} _{0n}$ and the total  associate-component $\Phi^{a}_{n}$ of the unit-field ${\psi _{0n}} =   {\tilde \psi} _{0n} \Phi^{a}_{n}$. Notice, the masses $m_{0n}$ of the same particles are given by $m_{0n}= m_{0}$, while the different particles have the different masses ($m_{0n}\neq m_{0}$). In order to select the solutions to Eq. (\ref{eq286abc}) that satisfy only the experimentally observed properties of the elementary particles, the field generator ${\tilde \psi} _{0n}$ and the total  associate-component $\Phi^{a}_{n}$ should be calibrated (fixed) by the following gauges. The first calibration (static gauge)
\begin{eqnarray} \label{eq287abc}
\dot \Phi^{a}_{n} = 0
\end{eqnarray}
means that the {\it{total associate-component}} ($TAC$) is static one. The second restriction (gauge) 
\begin{eqnarray} \label{eq288abc}
\dot {\tilde \psi} \dot \Phi^{a}_{n} -\nabla {\tilde \psi} _{0n} \nabla \Phi^{a}_{n} = 0
\end{eqnarray}
should be attributed to the orthogonality of the gradients $\nabla {\tilde \psi} _{0n}$ and $\nabla \Phi^{a}_{n}$ in the Hilbert (mathematical) space, which means that the gradients $\nabla {\tilde \psi} _{0n}$ and $\nabla \Phi^{a}_{n}$ do not cross-correlate (interfere) in the physical spacetime. Equation (\ref{eq286abc}) for the unit-field calibrated by the gauges (\ref{eq287abc}) and (\ref{eq288abc}) yields the equation of motion
\begin{eqnarray} \label{eq289abc}
\square {\tilde \psi} _{0n} + \left[ m_{0}^2 +  {\Gamma ^a_{n}}^2 \right] {\tilde \psi} _{0n}  = 0
\end{eqnarray}
for the unit-field generator ${\tilde \psi} _{0n}$. The common "eigen-like function" ${\Gamma ^a_{n}}^2$, which is given by the expression
\begin{eqnarray} \label{eq290abc}
{\Gamma ^{a} _{n}}^2 = \left( \nabla ^2 \Phi^{a}_{n} \right) \left( {\Phi^{a}_{n}}\right)^{-1}, 
\end{eqnarray}
depends on the total associate-component $\Phi^{a}_{n}$. For the associate-components $ \phi_{In}$ calibrated by the gauges (\ref{eq287abc}) and (\ref{eq288abc}), the use of the additional [Helmholt (${\Gamma ^2_{In}} \neq 0$) or Laplace (${\Gamma ^2_{In}} = 0$)]  gauge
\begin{eqnarray} \label{eq291abc}
\nabla ^2 \phi_{In} + {\Gamma ^2_{In}} \phi_{In} = 0
\end{eqnarray}
in the {\it{general relativistic equation of motion}} (\ref{eq286abc}) yields the {\it{relativistic}} equation 
\begin{eqnarray} \label{eq292abc}
\square {\tilde \psi} _{0n} + \left[ m_{0}^2 + {\Gamma ^a_{n}}^2 \right] {\tilde \psi} _{0n}  = 0,
\end{eqnarray}
which determines the relativistic motion of the unit-field generator ${\tilde \psi} _{0n}$. For the unit-field with the one-component $TAC$, the equation (\ref{eq292abc}) has the form (\ref{eq225abc}) with the total relativistic eigen-parameter 
\begin{eqnarray} \label{eq293abc}
{\Gamma ^a_{n}}^2={\Gamma ^2_{In}}\equiv {\Gamma ^2_{in}}.
\end{eqnarray}
In the case of the two-, three- and four-component unit-fields the total relativistic einen-parameters have respectively the forms
\begin{eqnarray} \label{eq294abc}
{\Gamma ^a_{n}}^2 = \frac {{\Gamma ^2_{in}} \phi_{in} + {\Gamma ^2_{jn}} \phi_{jn}} { \phi_{in} +  \phi_{jn}},
\end{eqnarray}
\begin{eqnarray} \label{eq295abc}
{\Gamma ^a_{n}}^2 = \frac {{\Gamma ^2_{in}} \phi_{in} + {\Gamma ^2_{jn} \phi_{jn}} + {\Gamma ^2_{kn}} \phi_{kn}} { \phi_{in} +  \phi_{jn} + \phi_{kn}}
\end{eqnarray}
and  
\begin{eqnarray} \label{eq296abc}
{\Gamma ^a_{n}}^2 = \frac {{\Gamma ^2_{Gn}} \phi_{Gn} + {\Gamma ^2_{Cn}}\phi_{Cn} + {\Gamma ^2_{Wn}}\phi_{Wn} + {\Gamma ^2_{Sn}}\phi_{Sn}   } { \phi_{Gn} +  \phi_{Cn} +  \phi_{Wn} +  \phi_{Sn}}.
\end{eqnarray}
Here, the indexes $i$, $j$ and $k$ are given by $i = G,C,W,S$, $j = G,C,W,S$ and $k = G,C,W,S$ with $i \neq j \neq k$. Notice, Eqs. (\ref{eq293abc}) - (\ref{eq295abc}) have been derived by using the orthogonality-like condition (\ref{eq211abc}). The multi-component unit-field (\ref{eq283abc}) describing by Eq. (\ref{eq292abc}) and the calibration (\ref{eq291abc}) obeys the particle mass $\left[ m_{0}^2 + {\Gamma ^a_{n}}^2\right]^{1/2}$, where the additional (gauge) mass ${\Gamma ^a_{n}}$ could be attributed to the mass of the total associate component ({\it{TAC}}) of the unit-field (particle). 

The solution to Eq. (\ref{eq292abc}) for {\it{a single unit-field (particle)}} placed in the free space is a time-harmonic stationary plane-wave 
\begin{eqnarray} \label{eq297abc}
\tilde \psi_{0n} (\mathbf{r},t)=a_0 e^{i({\mathbf{k}_{0n}{\mathbf{r}}}-{\varepsilon}_{0n} t )}. 
\end{eqnarray}
of the form (\ref{eq53abc}), which could be associated with the de Broglie wave. For the unit-field with the one-component {\emph{TAC}}, the use of the gauges (\ref{eq287abc}), (\ref{eq288abc}) and (\ref{eq291abc}) and the replacement of the unit-field generator ${\tilde \psi} _{0n}$ by the de Broglie plane wave (\ref{eq297abc}) in the relativistic equation (\ref{eq284abc}) yields the relativistic unit-field energy (\ref{eq229abc}) in the form
\begin{eqnarray} \label{eq298abc}
{\varepsilon}_{0n}^{2} = (\mathbf{k}_{0n}^2+m_{0}^2  + \Gamma_{in}^2 ) \int_{V} {\phi} _{in} ^* {\phi} _{in} d^3x. 
\end{eqnarray}
In the case of the two-, three- and four-component unit-fields, the procedure yields the relativistic unit-field energy given respectively by
\begin{eqnarray} \label{eq299abc}
{\varepsilon}_{0n}^{2} =  \left[ \mathbf{k}_{0n}^2+m_{0}^2  + \frac {\Gamma_{in}^2  \int_{V} {\phi} _{in} ^* {\phi} _{in} d^3x +\Gamma_{jn}^2  \int_{V} {\phi} _{jn} ^* {\phi} _{jn} d^3x } {\int_{V} ({\phi} _{in} ^* {\phi} _{in} + {\phi} _{jn} ^* {\phi} _{jn} )d^3x   } \right]  {\int_{V} ({\phi} _{in} ^* {\phi} _{in} + {\phi} _{jn} ^* {\phi} _{jn} )d^3x }, 
\end{eqnarray}
\begin{eqnarray} \label{eq300abc}
{\varepsilon}_{0n}^{2} =  \left[ \mathbf{k}_{0n}^2+m_{0}^2  + \frac {\Gamma_{in}^2  \int_{V} {\phi} _{in} ^* {\phi} _{in} d^3x +\Gamma_{jn}^2  \int_{V} {\phi} _{jn} ^* {\phi} _{jn} d^3x +\Gamma_{kn}^2  \int_{V} {\phi} _{kn} ^* {\phi} _{kn} d^3x } {\int_{V} ({\phi} _{in} ^* {\phi} _{in} + {\phi} _{jn} ^* {\phi} _{jn}+ {\phi} _{kn} ^* {\phi} _{kn} )d^3x   } \right] 
\cdot    \nonumber  \\ \cdot  {\int_{V} ({\phi} _{in} ^* {\phi} _{in} + {\phi} _{jn} ^* {\phi} _{jn}+ {\phi} _{kn} ^* {\phi} _{kn} )d^3x  } 
\end{eqnarray}
and 
\begin{eqnarray} \label{eq301abc}
{\varepsilon}_{0n}^{2} =  \left[ \mathbf{k}_{0n}^2+m_{0}^2  + \frac {\Gamma_{Gn}^2  \int_{V} {\phi} _{Gn} ^* {\phi} _{Gn} d^3x +\Gamma_{Cn}^2  \int_{V} {\phi} _{Cn} ^* {\phi} _{Cn} d^3x +\Gamma_{Wn}^2  \int_{V} {\phi} _{Wn} ^* {\phi} _{Wn} d^3x +\Gamma_{Sn}^2  \int_{V} {\phi} _{Sn} ^* {\phi} _{Sn} d^3x } {\int_{V} ({\phi} _{Gn} ^* {\phi} _{Gn} + {\phi} _{Cn} ^* {\phi} _{Cn}+ {\phi} _{Wn} ^* {\phi} _{Wn} + {\phi} _{Sn} ^* {\phi} _{Sn} )d^3x   } \right] 
\cdot    \nonumber  \\ \cdot  {\int_{V} ({\phi} _{Gn} ^* {\phi} _{Gn} + {\phi} _{Cn} ^* {\phi} _{Cn}+ {\phi} _{Wn} ^* {\phi} _{Wn} + {\phi} _{Sn} ^* {\phi} _{Sn} )d^3x }  
\end{eqnarray}
for the two-, three- and four-component {\emph{TACs}}. Notice, in the case of $\Gamma_{Gn}^2= \Gamma_{Cn}^2 = \Gamma_{Wn}^2 = \Gamma_{Sn}^2 =0$, the square of total mass $M^2 = m_{0}^2 {\int_{V} ({\phi} _{Gn} ^* {\phi} _{Gn} + {\phi} _{Cn} ^* {\phi} _{Cn}+ {\phi} _{Wn} ^* {\phi} _{Wn} + {\phi} _{Sn} ^* {\phi} _{Sn} )d^3x }$ could be associated with the masses contained in the gravitational, electric, weak and strong components of the unit-field. The terms that include the parameters $\Gamma_{Gn}^2 \neq 0$, $\Gamma_{Cn}^2 \neq 0$, $\Gamma_{Wn}^2 \neq 0$ and $\Gamma_{Sn}^2 \neq 0$ also could be attributed to the masses of the gravitational, electric, weak and strong components of the unit-field (particle). It will be shown in the subsequent subsections that $\Gamma_{Gn}^2 = 0$, $\Gamma_{Cn}^2 = 0$, $\Gamma_{Wn}^2 \neq 0$ and $\Gamma_{Sn}^2 \neq 0$. That means that the total mass $M$ of the particle is associated with the rest-mass $m_{0}$ attributed to the gravitational and electric components of the unit-field and the gauge masses of the weak and strong components of the unit-field, which are proportional to the eigen parameters $\Gamma_{Wn}$ and $\Gamma_{Sn}$. The mechanism provides the non-zero gauge-mass of the unit-field (particle) independently from the value of the rest-mass $m_{0}^2=0$ attributed to the gravitational and electric components of the unit-field.  

In the case of a {\it{non-relativistic unit-field}} (\ref{eq283abc}), the above-described calibrations in the {\it{non-relativistic equation of motion}} (\ref{eq13abc}) yielded the equation
\begin{eqnarray} \label{eq302abc}
i\dot {\tilde \psi}_{0n} = -{\frac 1 {2m_0} } {\nabla ^2{\tilde \psi_{0n}}} + \left[ m_0+ {\frac {{\Gamma ^a_{In}}^2} {2m_0} }  \right] \tilde \psi_{0n},
\end{eqnarray}
which determines the non-relativistic motion of the unit-field generator ${\tilde \psi}_{0n}$. For the unit-field with the one-component {\it{TAC}}, the total non-relativistic eigen-parameter ${\Gamma ^a_{n}}^2=(1/2m_0){\Gamma ^2_{In}}\equiv (1/2m_0){\Gamma ^2_{in}}$. In the case of the two-, three- and four-component unit-fields the total non-relativistic einen-parameters have respectively the forms (\ref{eq293abc}) - (\ref{eq295abc}) multiplied by the factor $(1/2m_0)$. The solution to Eq. (\ref{eq302abc}) for a single unit-field (particle) placed in the free space,  is the de Broglie wave $\tilde \psi_{0n} (\mathbf{r},t)=a_0 e^{i({\mathbf{k}_{0n}{\mathbf{r}}}-{\varepsilon}_{0n} t )}$ of the form (\ref{eq53abc}). For the unit-field with the one-component {\emph{TAC}}, the use of the gauges (\ref{eq287abc}), (\ref{eq288abc}) and (\ref{eq291abc}) and the replacement of the unit-field generator ${\tilde \psi} _{0n}$ by the de Broglie wave in the non-relativistic energy-mass relation (\ref{eq10abc}) yields the relativistic unit-field energy in the form
\begin{eqnarray} \label{eq303abc}
{\varepsilon}_{0n} = \left[ {\frac { \mathbf{k}_{0n}^2} {2m_0} }+ m_{0}  + {\frac {    \Gamma_{in}^2 } {2m_0} } \right]  \int_{V} {\phi} _{in} ^* {\phi} _{in} d^3x. 
\end{eqnarray}
In the case of the two-, three- and four-component unit-fields, the procedure yields the relativistic unit-field energy given respectively by
\begin{eqnarray} \label{eq304abc}
{\varepsilon}_{0n} =  \left[ {\frac { \mathbf{k}_{0n}^2} {2m_0} }+m_{0}  + {\frac { 1} {2m_0} } \frac {\Gamma_{in}^2  \int_{V} {\phi} _{in} ^* {\phi} _{in} d^3x +\Gamma_{jn}^2  \int_{V} {\phi} _{jn} ^* {\phi} _{jn} d^3x } {\int_{V} ({\phi} _{in} ^* {\phi} _{in} + {\phi} _{jn} ^* {\phi} _{jn} )d^3x   } \right]  {\int_{V} ({\phi} _{in} ^* {\phi} _{in} + {\phi} _{jn} ^* {\phi} _{jn} )d^3x }, 
\end{eqnarray}
\begin{eqnarray} \label{eq305abc}
{\varepsilon}_{0n} =  \left[ {\frac { \mathbf{k}_{0n}^2} {2m_0} }+m_{0}  + {\frac { 1} {2m_0} } \frac {\Gamma_{in}^2  \int_{V} {\phi} _{in} ^* {\phi} _{in} d^3x +\Gamma_{jn}^2  \int_{V} {\phi} _{jn} ^* {\phi} _{jn} d^3x +\Gamma_{kn}^2  \int_{V} {\phi} _{kn} ^* {\phi} _{kn} d^3x } {\int_{V} ({\phi} _{in} ^* {\phi} _{in} + {\phi} _{jn} ^* {\phi} _{jn}+ {\phi} _{kn} ^* {\phi} _{kn} )d^3x   } \right] \cdot    \nonumber  \\ \cdot  {\int_{V} ({\phi} _{in} ^* {\phi} _{in} + {\phi} _{jn} ^* {\phi} _{jn}+ {\phi} _{kn} ^* {\phi} _{kn} )d^3x  } 
\end{eqnarray}
and 
\begin{eqnarray} \label{eq306abc}
{\varepsilon}_{0n} ={\int_{V} ({\phi} _{Gn} ^* {\phi} _{Gn} + {\phi} _{Cn} ^* {\phi} _{Cn}+ {\phi} _{Wn} ^* {\phi} _{Wn} + {\phi} _{Sn} ^* {\phi} _{Sn} )d^3x } \cdot    \nonumber  \\ \cdot  \left[ {\frac { \mathbf{k}_{0n}^2} {2m_0} }+m_{0} +   {\frac { 1} {2m_0} } \frac {\Gamma_{Gn}^2  \int_{V} {\phi} _{Gn} ^* {\phi} _{Gn} d^3x +\Gamma_{Cn}^2  \int_{V} {\phi} _{Cn} ^* {\phi} _{Cn} d^3x +\Gamma_{Wn}^2  \int_{V} {\phi} _{Wn} ^* {\phi} _{Wn} d^3x +\Gamma_{Sn}^2  \int_{V} {\phi} _{Sn} ^* {\phi} _{Sn} d^3x } {\int_{V} ({\phi} _{Gn} ^* {\phi} _{Gn} + {\phi} _{Cn} ^* {\phi} _{Cn}+ {\phi} _{Wn} ^* {\phi} _{Wn} + {\phi} _{Sn} ^* {\phi} _{Sn} )d^3x   } \right] 
\end{eqnarray}
respectively for the two-, three- and four-component {\emph{TACs}}.

\subsection{5.3. Categorization of the single one-, two-, three- and four-component unit-fields (particles) by using symmetry of the unit-field associate-components (ACs): Kinds of the unit-field ACs}

\subsubsection{5.3.1. Categorization according to the symmetries of the coordinate systems of ACs}

In the model of a structure-less unit-field (elementary particle) describing by Eqs. (\ref{eq1abc}) - (\ref{eq205abc}), the {\it{kind}} of the unit-field (particle)  has been determined solely by the particle rest-mass $m_0$. The structure-less unit-field ${\psi}_{0n}$ depends only on the particle rest-mass, which could be considered as the particular initial conditions of the equation of motion. In such a case, the different configurations of the structure-less unit-field ${\psi}_{0n}$  corresponding to the different solutions of the equation of motion for a given rest-mass $m_{0n}$ are attributed to the different momentums $\mathbf{k}_{0n}$ of the same particle. In other words, the structure-less unit-field ${\psi}_{0n}$ with the rest mass $m_0= 0$ or $m_0\neq 0$ is considered as the 4-dimensional (in spacetime) structure-less object, namely as the structure-less quanta of energy-mass. The unit-field (particle) with the internal structure (substructure) corresponding to the gravitational,  electromagnetic, weak and strong fields and interactions has been described by Eqs. (\ref{eq210abc}) - (\ref{eq306abc}). In the model of the structured unit-fields (\ref{eq283abc}), the unit-fields ${\psi _{0n}} =   {\tilde \psi} _{0n} \Phi^{a}_{n}$ are distinguished from each other not only by the rest mass $m_0$, but also by the associate-components $\phi_{in}$ of the total associate component $\Phi^{a}_{n}$. That means that the structured unit-fields with the different {\it{ACs}} correspond to the unit-fields (particles) of the {\it{different kinds}}. It was indicated that the two calibrations (gauges) of the associate-components $\phi_{in}$ could be consistent with the Einstein special relativity and the experimentally observed properties of the elementary particles. In order to satisfy these conditions the unit-field {\it{ACs}} should be calibrated by the gauge (\ref{eq291abc}) in the form of the Laplace (${\Gamma ^2_{In}} = 0$) or Helmholt (${\Gamma ^2_{In}} \neq 0$) equation. The Laplace and Helmholtz equations (gauges) describe the {\it{ACs}} of the {\it{two different kinds}}, namely the {\it{Laplace and Helmholtz associate-components}} with the respective gauge masses ${\Gamma _{In}} = 0$ and ${\Gamma _{In}} \neq 0$. 

The Laplace and Helmholtz equations have the infinite number of possible solutions (configurations). The number of the symmetric {\it{ACs}}, which are symmetric in space, however, could be finite due to the finite number of possible symmetries. {\it{One of the ways of categorization of the all possible configurations of the ACs, which corresponds to the different solutions of the Laplace or Helmholtz equation is the spatial symmetry of the AC}}. A solution to the Laplace or Helmholtz equation is uniquely determined if the value of the function is specified on all boundaries (Dirichlet boundary conditions) or if the normal derivative of the function is known on all boundaries (Neumann boundary conditions). The Helmholtz equation has solutions in all 11 well-known coordinate systems, namely in {\it{Cartesian, circular cylindrical, conical, confocal ellipsoidal, elliptic cylindrical, oblate spheroidal, parabolic, parabolic cylindrical, paraboloidal, prolate spheroidal and spherical systems}}. Naturally, the Helmholtz associate-component $\phi_{in}$ can have any configuration that corresponds to a solution of the Helmholtz equation in any coordinate system of the 11 systems. The spatial symmetry of the Helmholtz {\it{AC}} could be determined by the symmetry of the coordinate system. Thus the 11 {\it{kinds}} of the Helmholtz {\it{ACs}} could be categorized by the 11 spatial symmetries of the unit-field associate-components according to the spatial symmetries of the 11 coordinate systems. In addition to the aforementioned 11 coordinate systems and symmetries, the Laplace equation has solutions in the two other coordinate systems, namely in the bispherical and toroidal ones. In such a case, the Laplace {\it{AC}} of the unit-field (particle) can have any configuration that corresponds to a solution of the Laplace equation in one of the 13 coordinate systems. The 13 {\it{kinds}} of the Laplace {\it{AC}} could be categorized by the spatial symmetry of the Laplace associate-component in the 13 coordinate systems. The all 13 coordinate systems and the respective Laplace and Helmholtz {\it{ACs}} are summarized (categorized) in Table \ref{tab:t1}. 

\begin{table}[t]
\centering
\begin{tabular}{lll}
\hline\noalign{\smallskip}
{\it{Coordinate system}} & {\it{Solutions (configurations) of the unit-field ACs }}  \\
\noalign{\smallskip}\hline\noalign{\smallskip}
Cartesian & Exponential, circular and hyperbolic functions  \\
\noalign{\smallskip}\hline
Spherical & Legendre polynomial, power and circular functions  \\
\noalign{\smallskip}\hline
Cylindrical & Bessel, exponential and circular functions  \\
\noalign{\smallskip}\hline
Conical & Ellipsoidal harmonics and power functions   \\
\noalign{\smallskip}\hline
Oblate spheroidal & Legendre polynomial and circular functions  \\
\noalign{\smallskip}\hline
Prolate spheroidal & Legendre polynomial and circular functions  \\
\noalign{\smallskip}\hline
Elliptic cylindrical & Mathieu and circular functions \\
\noalign{\smallskip}\hline
Parabolic cylindrical & Parabolic cylinder, Bessel and circular functions  \\
\noalign{\smallskip}\hline
Parabolic & Bessel and circular functions \\
\noalign{\smallskip}\hline
Paraboloidal & Circular functions \\
\noalign{\smallskip}\hline
Confocal ellipsoidal & Ellipsoidal harmonics  \\
\noalign{\smallskip}\hline
Bispherical & Functions based on a multiplicative factor \\
\noalign{\smallskip}\hline
Toroidal & Functions based on a multiplicative factor \\
\noalign{\smallskip}\hline
\end{tabular}
\caption{The all possible coordinate systems and the respective unit-field ACs. Notice, the Laplace equation does have solutions in the bispherical and toroidal coordinate systems, while the Helmholtz equation does not.}
\label{tab:t1}
\end{table}

Strictly speaking, the configurations of the Laplace and Helmholtz {\it{ACs}} summarized in Table \ref{tab:t1} are 3-dimensional objects. Nevertheless, in principle, the more exotic (low-dimensional) {\it{ACs}} satisfying the Laplace or Helmholtz equation also can be considered. For instance, the Laplace and Helmholtz {\it{ACs}} having the low-dimensional configurations could be presented as the 0-D (point), 1-D (straight string), 2-D (open curved string, closed string, open membrane or closed membrane) {\it{ACs}}. {\it{It should be stressed that the creation of such exotic configurations (points, strings and membranes) from the 3-dimensional Laplace and Helmholtz {\it{ACs}} is practically impossible}}. Indeed, the low-dimensional {\it{ACs}} have the Dirac $\delta$-like distributions at least in one of the three spatial dimensions. Therefore the unit-fields with the low-dimensional {\it{ACs}} obey infinite energies [see, Eq. (\ref{eq284abc})]. Although the low-dimensional {\it{ACs}} obey the infinite energies, in contrast to the 3-dimensional {\it{ACs}} having the finite-energies, they are very important for the simplified (0-D, 1-D or 2-D) theoretical analysis of the unit-fields (particles) and for comparison of the present model with the modern string theories.

In the modern conceptual picture of the Universe, the all elementary particles have been created in the early stage of the cosmological Big-Bang. The non-elementary particles of the Universe are the products of fusion of the elementary particles in the late stages of the Big-Bang. One may suppose that after creation of the elementary particles they satisfy to the Einstein special relativity. In such a case, the all 13 {\it{kinds}} of the Laplace unit-field {\it{ACs}} and the all 11 {\it{kinds}} of the Helmholtz unit-field {\it{ACs}}, which do satisfy the {\it{generalized Einstein energy-mass relation and respective equations of motions, may exist somewhere in the Universe}}. Although the all 24 {\it{kinds}} of the unit-field {\it{ACs}} may exist in principle, the only unit-field {\it{ACs}} that obey physical properties of the {\it{experimentally observed elementary particles}} will be considered in the following sections. One can easily follow the present model for any hypothetical unit-field {\it{AC}} of the 24 {\it{kinds}} of the unit-field {\it{ACs}} of the elementary particles that have not been yet observed experimentally.

\subsubsection{5.3.2. The single unit-fields (particles) with the spherical and cylindrical symmetries of the unit-field ACs: Physical explanation of a particle spin}

As an example, consider here the unit-fields (particles) with the {\it{spherical}} and {\it{cylindrical}} symmetries of the Laplace and Helmholtz {\it{ACs}}. For the well-known solutions of the Laplace and Helmholtz differential equations in other 11 coordinate systems see mathematical textbooks. 

\vspace{0.2cm}
1. {\it{The single unit-fields (particles) with the spherical symmetries of the Laplace ACs}}
\vspace{0.2cm}

Let me first consider the unit-field (particle) ${\psi _{0n}} =   {\tilde \psi} _{0n} \Phi^{a}_{n}$ whose {\it{AC}} is described by the Laplace equation (calibration). In spherical polar coordinates [$\mathbf{r} \equiv (r, \theta , \varphi )$], the Laplace (${\Gamma ^2_{In}} = 0$) equation (\ref{eq291abc}) for the unit-field associate-component $\Phi^{a}_{n}$ has the form
\begin{eqnarray} \label{eq307abc}
\nabla^2 {\Phi^{a}_{n}} (\mathbf{r}) =    \left[     \frac {1} {r}    \frac {\partial ^2} {\partial r^2} r - \frac {\hat{\mathbf{L}}^2} { r^2} \right] {\Phi^{a}_{n}}(\mathbf{r}) =0,
\end{eqnarray}
where ${\hat{\mathbf{L}}}^2$ is the square of the angular momentum operator ${\hat{\mathbf{L}}}=i(\mathbf{r}\times \nabla)$. The two linearly independent well-known solutions to the Laplace equation (\ref{eq307abc}) in a ball centered at the origin, which are non-singular and singular at the origin, are called respectively the regular and irregular solid harmonics. The simplest regular and irregular {\it{ACs}} are given by the regular 
\begin{eqnarray} \label{eq308abc}
{\Phi^{a}_{n}}(\mathbf{r}) =  r^l Y_l^m (\theta, \varphi  ),
\end{eqnarray}
and irregular  
\begin{eqnarray} \label{eq309abc}
{\Phi^{a}_{n}}(\mathbf{r}) = r^{-l-1} Y_l^m (\theta, \varphi  )
\end{eqnarray}
solid harmonics, where the spherical harmonics $Y_l^m (\theta, \varphi  )$ of degree $l$ and order $m$ are the joint eigenfunctions of the square of the orbital angular momentum operator ${\hat{\mathbf{L}}}^2$ and the generator of rotations around the azimuthal axis ${\hat{L_z}}$:
\begin{eqnarray} \label{eq310abc}
{\hat{\mathbf{L}}}^2  Y_l^m (\theta, \varphi  ) = l(l+1) Y_l^m (\theta, \varphi  )
\end{eqnarray}
and 
\begin{eqnarray} \label{eq311abc}
{\hat{L_z}}  Y_l^m (\theta, \varphi  ) = -i \frac {\partial } {\partial {\varphi }} Y_l^m (\theta, \varphi  ) = m Y_l^m (\theta, \varphi  ). 
\end{eqnarray}
The ball occupied by the solid harmonic may have the infinite radius ($R = \infty$). For a given non-negative integer $l$, there are $2l+1$ independent solutions of the form (\ref{eq308abc}), one for each integer $m$ with $-l \leq  m \leq  l$. Under the standard probabilistic normalization (in the Dirac notations)
\begin{eqnarray} \label{eq312abc}
\langle Y_{l_1}^{m_1} | Y_{l_2}^{m_2}  \rangle \equiv  \int_{\theta =0}^\pi   \int_{\varphi  =0}^{2\pi} {Y_{l_1}^{m_1}}^*  Y_{l_2}^{m_2} d\Omega  =\delta _{l_1,l_2} \delta _{m_1,m_2},
\end{eqnarray}
the Laplace spherical harmonics are given by
\begin{eqnarray} \label{eq313abc}
Y_l^m (\theta, \varphi  ) =  \left[     \frac {(2l+1)} {4\pi }    \frac {(l-m)!} {(l+m)!} \right]^{1/2} P_l^m (cos \theta )e^{i m \varphi}, 
\end{eqnarray}
where $d\Omega = sin \theta d \varphi d \theta$, $P_l^m (cos \theta) $ is an associated Legendre polynomial, and $\delta _{l_1,l_3}$ and $\delta _{m_1,m_2}$ denote the Kronecker symbols. The harmonic degree $l$ logically to call the {\it{orbital quantum number}}. While the harmonic order $m$ could be denoted as the {\it{"magnetic" quantum number}}. 

The 3-D unit-field (particle) ${\psi _{0n}} =   {\tilde \psi} _{0n} \Phi^{a}_{n}$ is not a point particle of classical mechanics, quantum mechanics or SM. Therefore, it obeys {\it{naturally}} the intrinsic (internal) angular momentum (spin) $\langle {\hat \mathbf{L}} \rangle \equiv \langle {\hat \mathbf{s}} \rangle$ associated with the unit-field {\it{AC}}. In the case of the unit-field with the unit-field associate-component $\Phi^{a}_{n}$ satisfying Eq. (\ref{eq308abc}) or (\ref{eq309abc}), {\it{the square of a unit-field spin}} is given by 
\begin{eqnarray} \label{eq314abc}
\langle {\hat{\mathbf{s}}}^2 \rangle = \langle Y_l^m | {\hat{\mathbf{s}}}^2  |Y_l^m  \rangle = l(l+1),
\end{eqnarray}
while {\it{the z-component of the intrinsic orbital angular momentum (spin)}} is equal to  
\begin{eqnarray} \label{eq315abc}
\langle \hat{s_z} \rangle = \langle Y_l^m | \hat{s_z}  |Y_l^m  \rangle  = m.
\end{eqnarray}
The numbers $l$ and $m$ denote respectively the {\it{intrinsic orbital quantum number}} and the {\it{intrinsic "magnetic" quantum number}}. The unit-fields (particles) with the Laplace {\it{ACs}} and the {\it{integer}} spins describing by Eqs. (\ref{eq310abc}) - (\ref{eq315abc}) could correspond to the elementary {\it{bosons}}. The present model (see, Sec. 5.2.) does not need the probabilistic normalization (\ref{eq312abc}). 
The regular  
\begin{eqnarray} \label{eq316abc}
{\Phi^{a}_{n}}(\mathbf{r}) =  r^l A_{lm}Y_l^m (\theta, \varphi  ),
\end{eqnarray}
and irregular   
\begin{eqnarray} \label{eq317abc}
{\Phi^{a}_{n}}(\mathbf{r}) = r^{-l-1} A_{lm}Y_l^m (\theta, \varphi  )
\end{eqnarray}
{\it{ACs}}, where $A_{lm}$ is an arbitrary constant, are also the legal solutions of Eq. (\ref{eq307abc}). In the case of  Eqs. (\ref{eq316abc}) and (\ref{eq317abc}), the square of a unit-field spin is given by 
\begin{eqnarray} \label{eq318abc}
\langle {\hat{\mathbf{s}}}^2 \rangle = \langle A_{lm} Y_l^m | {\hat{\mathbf{s}}}^2  | A_{lm} Y_l^m  \rangle (\langle A_{lm} Y_l^m | A_{lm} Y_l^m  \rangle) ^{-1}=  l(l+1) ,
\end{eqnarray}
while the z-component of the spin is equal to  
\begin{eqnarray} \label{eq319abc}
\langle \hat{s_z} \rangle = \langle A_{lm} Y_l^m | \hat{s_z}  | A_{lm} Y_l^m  \rangle (\langle A_{lm} Y_l^m | A_{lm} Y_l^m  \rangle) ^{-1} = m.
\end{eqnarray}
The unit-fields (particles) with the Laplace {\it{ACs}} and the spins describing by Eqs. (\ref{eq318abc}) and (\ref{eq319abc}) could correspond to the elementary particles describing by {\it{integer}} spins. It should be noted in this regard that the classical (Newton) intrinsic orbital angular momentum $\mathbf{L}=\mathbf{r}\times  \mathbf{p}$ and relativistic (Einstein)  intrinsic orbital angular momentum tensor $L^{nm}= \sum (x^np^m-x^mp^n)$ of a point particle are always equal to zero if $p\neq \infty$. In canonical quantum mechanics and SM, the spins of the point (structure-less) particles have been engineered formally by using the operator or matrix formalism. In the present model, the unit-fields (particles) obey {\it{naturally}} the spins attributed to the physical properties of the respective {\it{ACs}} [see, (\ref{eq307abc}) - (\ref{eq319abc})], which do not require the physical interpretation of any abstract mathematical object, such as an operator or matrix. The angular distribution of the {\it{AC}} function does not change in the conservative physical processes due to the conservation law for the spin and energy. Therefore the knowledge of the exact angular distribution of the {\it{AC}} function could not be important for description of the physical processes. For such a description, the model of the unit-field spin can be represented as the formal matrix algebra or operator formalism, like the matrix algebra and operator formalism of the spin used in the canonical quantum mechanics and SM. The transition from the {\it{AC}} function associated with the material associate-component of the real (material) unit-field to the matrix representation or operator formalism is quit similar to the transition from the Schr{\"o}dinger wave model of quantum mechanics to the Heisenberg or Pauli matrix (operator) models based on the matrix (operator) commutation relations. {\it{The presented model, probably for the first time, presents a physical explanation of a particle spin without using the physical interpretations of the abstract matrix algebra or operator formalism}}.

The general regular solution to Laplace's equation (\ref{eq307abc}) is a linear combination of the regular solid harmonics multiplied by the appropriate scale factors $A_{lm}$: 
\begin{eqnarray} \label{eq320abc}
{\Phi^{a}_{n}}(\mathbf{r}) = \sum_{l=0}^{\infty}\sum_{m=-l}^{l}A_{lm} r^l Y_l^m (\theta , \varphi),
\end{eqnarray}
where $A_{lm}$ are the scaling constants. Correspondingly, the general irregular solution to Laplace's equation (\ref{eq307abc}) is a linear combination of the irregular solid harmonics multiplied by the scale factors $B_{lm}$: 
\begin{eqnarray} \label{eq321abc}
{\Phi^{a}_{n}}(\mathbf{r}) = \sum_{l=0}^{\infty}\sum_{m=-l}^{l}B_{lm} r^{-l-1} Y_l^m (\theta , \varphi),
\end{eqnarray}
where $B_{lm}$ denotes the appropriate scaling constants. Then the general solution to Eq. (\ref{eq307abc}), which includes both the regular and irregular solid harmonics is given by
\begin{eqnarray} \label{eq322abc}
{\Phi^{a}_{n}}(\mathbf{r}) = \sum_{l=0}^{\infty}\sum_{m=-l}^{l}[A_{lm} r^l  + B_{lm} r^{-l-1}]Y_l^m (\theta , \varphi).
\end{eqnarray}
The linear combinations of the regular and irregular solid harmonics can describe the complicated configurations, such as the asymmetric {\it{ACs}}. The transformations associated, for instance, with the group of rotations or translations, as well as the conformal symmetry properties of the solutions of the types (\ref{eq308abc}), (\ref{eq309abc}), (\ref{eq316abc}), (\ref{eq317abc}), (\ref{eq320abc}) - (\ref{eq322abc}), have been investigated in the past by the well-known "potential theory" and the theory of Laplace's equation (for the details, see the mathematical textbooks). 

The relativistic and non-relativistic energies of the unit-field (particle) ${\psi _{0n}} =   {\tilde \psi} _{0n} \Phi^{a}_{n}$ with the Laplace {\it{ACs}} are determined respectively by Eq. (\ref{eq298abc}) - Eq. (\ref{eq301abc}) and Eq. (\ref{eq303abc}) - Eq. (\ref{eq306abc}). For an example, for the two-component unit-fields (particles) with the Laplace gravitational and electrical {\it{ACs}} ($\Gamma ^2_{Gn} = 0$ and $\Gamma ^2_{En} = 0$, for details see Sec. 7.2), the relativistic and non-relativistic energies are given by Eqs. (\ref{eq299abc}) and  (\ref{eq304abc}) as 
\begin{eqnarray} \label{eq323abc}
{\varepsilon}_{0n}^{2} =  \left[ \mathbf{k}_{0n}^2+m_{0}^2   \right]  {\int_{V} ({\phi} _{Gn} ^* {\phi} _{Gn} + {\phi} _{Cn} ^* {\phi} _{Cn} )d^3x }, 
\end{eqnarray}
and 
\begin{eqnarray} \label{eq324abc}
{\varepsilon}_{0n} =  \left[ {\frac { \mathbf{k}_{0n}^2} {2m_0} }+m_{0}  \right]  {\int_{V} ({\phi} _{Gn} ^* {\phi} _{Gn} + {\phi} _{Cn} ^* {\phi} _{Cn} )d^3x }, 
\end{eqnarray}
respectively, where the normalization ${\int_{V} ({\phi} _{Gn} ^* {\phi} _{Gn} + {\phi} _{Cn} ^* {\phi} _{Cn} )d^3x }=1$ provides the Einstein relativistic and non-relativistic energy-mass relations. 

\vspace{0.2cm}
2. {\it{The single unit-fields (particles) with the spherical symmetries of the Helmholtz ACs}}
\vspace{0.2cm}

The Helmholtz (${\Gamma ^2_{In}} \neq 0$) equation (\ref{eq291abc}) describing the Helmholtz {\it{AC}} of the unit-field (particle) ${\psi _{0n}} =   {\tilde \psi} _{0n} \Phi^{a}_{n}$ in spherical coordinates has the form
\begin{eqnarray} \label{eq325abc}
[\nabla^2 + {\Gamma ^{a} _{In}}^2] {\Phi^{a}_{n}}(\mathbf{r}) =    \left[     \frac {1} {r}    \frac {\partial ^2} {\partial r^2} r - \frac {\hat{\mathbf{L}}^2} { r^2}+ {\Gamma ^{a} _{In}}^2 \right] {\Phi^{a}_{n}}(\mathbf{r}) =0,
\end{eqnarray}
where the eigen parameter ${\Gamma ^{a} _{In}}^2 $ denotes a positive or negative constant corresponding to the real (${\Gamma ^{a} _{In}} =  |{\Gamma ^{a} _{In}} | $) or imaginary (${\Gamma ^{a} _{In}} = i |{\Gamma ^{a} _{In}} | $) value of ${\Gamma ^{a} _{In}}$.

The two linearly independent well-known solutions to Eq. (\ref{eq325abc}) are related to the spherical Bessel functions of the first and second kind. The first solution in a ball centered at the origin, which is associated with the spherical Bessel functions of the first kind, is given by
\begin{eqnarray} \label{eq326abc}
{\Phi^{a}_{n}}(\mathbf{r}) = j_l({\Gamma ^{a} _{In}} r)  A_{lm} Y_l^m (\theta, \varphi),
\end{eqnarray}
where $A_{lm}$ is the scaling constant that determines the value of the unit-field (particle) spin, $j_l({\Gamma ^{a} _{In}} r) $ is the {\it{spherical}} Bessel function of the first kind with the real (${\Gamma ^{a} _{In}} r = |{\Gamma ^{a} _{In}} | r$) or imaginary (${\Gamma ^{a} _{In}} r = i |{\Gamma ^{a} _{In}} | r$) argument, and $Y_l^m (\theta, \varphi)$ denotes the spherical harmonic determining by Eqs. (\ref{eq310abc}) - (\ref{eq312abc}). In the terms of the {\it{modified}} Bessel functions of the first kind, which have the real arguments $|{\Gamma ^{a} _{In}} | r$, the {\it{spherical}} Bessel function of the first kind with the imaginary argument is given by
\begin{eqnarray} \label{eq327abc}
j_l (i |{\Gamma ^{a} _{In}} | r) = i^{l+1/2} \frac {I_{l+1/2}(|{\Gamma ^{a} _{In}} | r)} {(\pi ^{-1}2i|{\Gamma ^{a} _{In}} | r)^{1/2}},
\end{eqnarray}
where $I_{l+1/2}(|{\Gamma ^{a} _{In}} | r) $ is the {\it{modified}} Bessel function of the first kind with the real argument. The second solution in a ball centered at the origin, which is associated with the {\it{spherical}} Bessel functions of the second kind, is given by
\begin{eqnarray} \label{eq328abc}
{\Phi^{a}_{n}}(\mathbf{r}) = y_l({\Gamma ^{a} _{In}} r)  A_{lm} Y_l^m (\theta, \varphi),
\end{eqnarray}
where $A_{lm}$ is the scaling constant that determines the value of the unit-field (particle) spin, $y_l({\Gamma ^{a} _{In}} r) $ is the {\it{spherical}} Bessel function of the second kind with the real (${\Gamma ^{a} _{In}} r = |{\Gamma ^{a} _{In}} | r$) or imaginary (${\Gamma ^{a} _{In}} r = i |{\Gamma ^{a} _{In}} | r$) argument, and $Y_l^m (\theta, \varphi)$ denotes the spherical harmonic determining by Eqs. (\ref{eq310abc}) - (\ref{eq312abc}). In the terms of the {\it{modified}} Bessel functions of the first and second kind, which have the real arguments $|{\Gamma ^{a} _{In}} | r$, the {\it{spherical}} Bessel function of the first kind with the imaginary argument is given by
\begin{eqnarray} \label{eq329abc}
y_l (i |{\Gamma ^{a} _{In}} | r) = i^{l+3/2} \frac {I_{l+1/2}( |{\Gamma ^{a} _{In}} | r)} {(\pi ^{-1}2i |{\Gamma ^{a} _{In}} | r)^{1/2}} - \frac {2} {\pi } i^{-(l+1/2)} \frac {K_{l+1/2}(|{\Gamma ^{a} _{In}} | r)} {(\pi ^{-1}2i |{\Gamma ^{a} _{In}} | r)^{1/2}},
\end{eqnarray}
where $I_{l+1/2}(|{\Gamma ^{a} _{In}} | r) $ and $K_{l+1/2}(|{\Gamma ^{a} _{In}} | r)$ are respectively the {\it{modified}} Bessel function of the first and second kinds with the real argument. Notice, the following useful relations hold among the functions: $K_{l+1/2}(|{\Gamma ^{a} _{In}} | r)= [\pi /2sin(\pi [l+1/2])][I_{-l-1/2}(|{\Gamma ^{a} _{In}} | r) - I_{l+1/2}(|{\Gamma ^{a} _{In}} | r)] $. The linear combinations 
\begin{eqnarray} \label{eq330abc}
{\Phi^{a}_{n}}(\mathbf{r}) = \sum_{l=0}^{l_{max}}\sum_{m=-l}^{l}A_{lm}  j_l({\Gamma ^{a} _{In}} r) Y_l^m (\theta, \varphi)
\end{eqnarray}
and
\begin{eqnarray} \label{eq331abc}
{\Phi^{a}_{n}}(\mathbf{r}) = \sum_{l=0}^{l_{max}}\sum_{m=-l}^{l}B_{lm}  y_l({\Gamma ^{a} _{In}} r) Y_l^m (\theta, \varphi),
\end{eqnarray}
are also solutions of Eq. (\ref{eq320abc}). The superposition of the linear combinations (\ref{eq323abc}) and (\ref{eq326abc}) gives the general solution 
\begin{eqnarray} \label{eq332abc}
{\Phi^{a}_{n}}(\mathbf{r}) = \sum_{l=0}^{\infty}\sum_{m=-l}^{l}[A_{lm} j_l({\Gamma ^{a} _{In}} r) + B_{lm} y_l({\Gamma ^{a} _{In}} r)]Y_l^m (\theta, \varphi).
\end{eqnarray}
in a ball centered at the origin. The spin of the unit-field (particle) with the Helmholtz {\it{AC}} (\ref{eq326abc}) or (\ref{eq328abc}) are described by Eqs. (\ref{eq318abc}) and (\ref{eq319abc}). The unit-fields with the {\it{ACs}} (\ref{eq330abc}) - (\ref{eq332abc}) could describe the more complicated spins of the unit-fields (particles).

The relativistic and non-relativistic energies of the unit-field (particle) ${\psi _{0n}} =   {\tilde \psi} _{0n} \Phi^{a}_{n}$ with the Helmholtz {\it{ACs}} are determined respectively by Eq. (\ref{eq298abc}) - Eq. (\ref{eq301abc}) and Eq. (\ref{eq303abc}) - Eq. (\ref{eq306abc}). For an example, for the four-component unit-fields (particles) with the Laplace gravitational and electrical {\it{ACs}} ($\Gamma ^2_{Gn} = 0$ and $\Gamma ^2_{En} = 0$) and the Helmholtz weak-interaction and strong-interaction {\it{ACs}} ($\Gamma ^2_{Gn} \neq 0$ and $\Gamma ^2_{En} \neq 0$) , for details see Sec. 7.2) , the relativistic and non-relativistic energies are given by Eqs. (\ref{eq301abc}) and  (\ref{eq306abc}) as 
\begin{eqnarray} \label{eq333abc}
{\varepsilon}_{0n}^{2} =  \left[ \mathbf{k}_{0n}^2+m_{0}^2  + \frac {\Gamma_{Wn}^2  \int_{V} {\phi} _{Wn} ^* {\phi} _{Wn} d^3x +\Gamma_{Sn}^2  \int_{V} {\phi} _{Sn} ^* {\phi} _{Sn} d^3x } {\int_{V} ({\phi} _{Gn} ^* {\phi} _{Gn} + {\phi} _{Cn} ^* {\phi} _{Cn}+ {\phi} _{Wn} ^* {\phi} _{Wn} + {\phi} _{Sn} ^* {\phi} _{Sn} )d^3x   } \right] 
\cdot    \nonumber  \\ \cdot  {\int_{V} ({\phi} _{Gn} ^* {\phi} _{Gn} + {\phi} _{Cn} ^* {\phi} _{Cn}+ {\phi} _{Wn} ^* {\phi} _{Wn} + {\phi} _{Sn} ^* {\phi} _{Sn} )d^3x },  
\end{eqnarray}
and 
\begin{eqnarray} \label{eq334abc}
{\varepsilon}_{0n} ={\int_{V} ({\phi} _{Gn} ^* {\phi} _{Gn} + {\phi} _{Cn} ^* {\phi} _{Cn}+ {\phi} _{Wn} ^* {\phi} _{Wn} + {\phi} _{Sn} ^* {\phi} _{Sn} )d^3x } \cdot    \nonumber  \\ \cdot  \left[ {\frac { \mathbf{k}_{0n}^2} {2m_0} }+m_{0} +   {\frac { 1} {2m_0} } \frac {\Gamma_{Wn}^2  \int_{V} {\phi} _{Wn} ^* {\phi} _{Wn} d^3x +\Gamma_{Sn}^2  \int_{V} {\phi} _{Sn} ^* {\phi} _{Sn} d^3x } {\int_{V} ({\phi} _{Gn} ^* {\phi} _{Gn} + {\phi} _{Cn} ^* {\phi} _{Cn}+ {\phi} _{Wn} ^* {\phi} _{Wn} + {\phi} _{Sn} ^* {\phi} _{Sn} )d^3x   } \right] ,
\end{eqnarray}
respectively. The relativistic (\ref{eq333abc}) and non-relativistic (\ref{eq334abc}) energies of the unit-field do not depend on the intrinsic orbital and azimuthal angular momentums (spins) of the unit-field. That is to say that the unit-field spin does not depend on the momentum $\mathbf{k}_{0n}$ and energy ${\varepsilon}_{0n}$ of the unit-field (particle) and {\it{vise versa}}. The momentum $\mathbf{k}_{0n}$ and energy ${\varepsilon}_{0n}$ can be changed, however, that would not affect the the spin of the unit-field (particle). In Eqs. (\ref{eq333abc}) and (\ref{eq334abc}), the {\it{gauge (TACs) masses}} 
\begin{eqnarray} \label{eq335abc}
\left[ {\Gamma_{Wn}^2  \int_{V} {\phi} _{Wn} ^* {\phi} _{Wn} d^3x +\Gamma_{Sn}^2  \int_{V} {\phi} _{Sn} ^* {\phi} _{Sn} d^3x } \right]^{1/2}   
\end{eqnarray}
and 
\begin{eqnarray} \label{eq336abc}
{\frac { 1} {2m_0} } \left[ {\Gamma_{Wn}^2  \int_{V} {\phi} _{Wn} ^* {\phi} _{Wn} d^3x +\Gamma_{Sn}^2 \int_{V} {\phi} _{Sn} ^* {\phi} _{Sn} d^3x } \right]
\end{eqnarray}
could be associated with the effective relativistic rest-masses $ \left( m_{0}^2 + \left[ {\Gamma_{Wn}^2  \int_{V} {\phi} _{Wn} ^* {\phi} _{Wn} d^3x +\Gamma_{Sn}^2 \int_{V} {\phi} _{Sn} ^* {\phi} _{Sn} d^3x } \right] \right) ^{1/2}$ and the effective non-relativistic rest-masses $\left(  m_0  + {\frac { 1} {2m_0} } \left[ {\Gamma_{Wn}^2  \int_{V} {\phi} _{Wn} ^* {\phi} _{Wn} d^3x +\Gamma_{Sn}^2  \int_{V} {\phi} _{Sn} ^* {\phi} _{Sn} d^3x } \right]  \right)$ of the unit-fields (particles). Alternatively, the parameters (\ref{eq335abc}) and (\ref{eq336abc}) can be attributed to the effective relativistic $ \left( \mathbf{k}_{0n}^2 + \left[ {\Gamma_{Wn}^2  \int_{V} {\phi} _{Wn} ^* {\phi} _{Wn} d^3x + \Gamma_{Sn}^2 \int_{V} {\phi} _{Sn} ^* {\phi} _{Sn} d^3x } \right] \right) ^{1/2}$ and non-relativistic $ \left( \mathbf{k}_{0n}^2 + \left[ {\Gamma_{Wn}^2  \int_{V} {\phi} _{Wn} ^* {\phi} _{Wn} d^3x + \Gamma_{Sn}^2 \int_{V} {\phi} _{Sn} ^* {\phi} _{Sn} d^3x } \right] \right) ^{1/2}$ momentums of the the relativistic and non-relativistic unit-fields (particles).
 
\vspace{0.2cm}
3. {\it{The single unit-fields (particles) with the cylindrical symmetries of the Laplace and Helmholtz ACs}}
\vspace{0.2cm}

The Laplace (${\Gamma ^2_{In}} = 0$) or Helmholtz (${\Gamma ^2_{In}} \neq 0$) gauge (\ref{eq291abc}) describing the unit-fields ${\psi _{0n}} =   {\tilde \psi} _{0n} \Phi^{a}_{n}$ with Laplace or Helmholtz {\it{AC}} in cylindrical coordinates [$\mathbf{r} \equiv (r, \varphi , z)$] has the form
\begin{eqnarray} \label{eq337abc}
\nabla^2 {\Phi^{a}_{n}}(\mathbf{r}) =    \left[     \frac {1} {r}    \frac {\partial } {\partial r}  \left( r  \frac {\partial } {\partial r}   \right)  +  \frac {\partial ^2} {\partial z^2}    +   \frac {1} {r^2}   \frac {\partial ^2} {\partial {\varphi }^2}        \right] {\Phi^{a}_{n}}(\mathbf{r}) =0
\end{eqnarray}
or 
\begin{eqnarray} \label{eq338abc}
[\nabla^2 + {\Gamma ^a_{In}}^2] {\Phi^{a}_{n}}(\mathbf{r}) =   \left[     \frac {1} {r}    \frac {\partial } {\partial r}  \left( r  \frac {\partial } {\partial r}   \right)  +  \frac {\partial ^2} {\partial z^2}    +   \frac {1} {r^2}   \frac {\partial ^2} {\partial {\varphi }^2}  + {\Gamma ^a_{In}}^2 \right] {\Phi^{a}_{n}}(\mathbf{r}) =0,
\end{eqnarray}
respectively. The well-known general solution of the Helmholtz equation (\ref{eq338abc}) in a cylinder of radius $R$ is given by
\begin{eqnarray} \label{eq339abc}
{\Phi^{a}_{n}}(\mathbf{r}) = \sum_{n=0}^{\infty}\sum_{m=0}^{\infty} \left[ A_{nm} J_m  \left( r\sqrt {n^2 +{\Gamma ^a_{In}}^2 } \right) + B_{nm} Y_m \left( r\sqrt {n^2 + {\Gamma ^a_{In}}^2 } \right) \right]  \cdot  \nonumber \\  \cdot    \left[ C_n e^{-nz}+D_n e^{nz} \right]    \left[ E_m e^{im \varphi }+F_m e^{-im \varphi }\right],
\end{eqnarray}
where $A_{nm}$,  $B_{nm}$, $C_{n}$,  $D_{n}$, $E_{m}$ and $F_{m}$ are the scaling constants, $J_m \left( r\sqrt {n^2 +\ {\Gamma ^a_{In}}^2} \right) $ and $Y_m \left( r\sqrt {n^2 + {\Gamma ^a_{In}}^2 } \right) $ are respectively the {\it{ordinary}} Bessel functions of the first and second kinds with the real ($ r\sqrt {n^2 +\gamma ^2 } =r | \sqrt {n^2 +\gamma ^2 }|$) or imaginary ($ r\sqrt {n^2 +\gamma ^2 } = r i | \sqrt {n^2 +\gamma ^2 }|$) arguments, and $n$ and $m$ denote positive integers. Here, the functions $Y_m \left( r\sqrt {n^2 + {\Gamma ^a_{In}}^2 } \right) $ should not be confused with the Laplace spherical harmonics (\ref{eq313abc}). The simplest solutions of Eq. (\ref{eq338abc}) have the forms
\begin{eqnarray} \label{eq340abc}
{\Phi^{a}_{n}}(\mathbf{r}) = J_m \left( r\sqrt {n^2 +\ {\Gamma ^a_{In}}^2} \right) e^{-nz} A_{m} e^{im \varphi }
\end{eqnarray}
\begin{eqnarray} \label{eq341abc}
{\Phi^{a}_{n}}(\mathbf{r}) = J_m \left( r\sqrt {n^2 +\ {\Gamma ^a_{In}}^2} \right) e^{-nz} A_{m} e^{-im \varphi }
\end{eqnarray}
\begin{eqnarray} \label{eq342abc}
{\Phi^{a}_{n}}(\mathbf{r}) = J_m \left( r\sqrt {n^2 +\ {\Gamma ^a_{In}}^2} \right) e^{nz} A_{m} e^{im \varphi }
\end{eqnarray}
\begin{eqnarray} \label{eq343abc}
{\Phi^{a}_{n}}(\mathbf{r}) = J_m \left( r\sqrt {n^2 +\ {\Gamma ^a_{In}}^2} \right) e^{nz} A_{m} e^{-im \varphi }
\end{eqnarray}
\begin{eqnarray} \label{eq344abc}
{\Phi^{a}_{n}}(\mathbf{r}) = Y_m \left( r\sqrt {n^2 +\ {\Gamma ^a_{In}}^2} \right) e^{-nz} A_{m} e^{im \varphi }
\end{eqnarray}
\begin{eqnarray} \label{eq345abc}
{\Phi^{a}_{n}}(\mathbf{r}) = Y_m \left( r\sqrt {n^2 +\ {\Gamma ^a_{In}}^2} \right) e^{-nz} A_{m} e^{-im \varphi }
\end{eqnarray}
\begin{eqnarray} \label{eq346abc}
{\Phi^{a}_{n}}(\mathbf{r}) = Y_m \left( r\sqrt {n^2 +\ {\Gamma ^a_{In}}^2} \right) e^{nz} A_{m} e^{im \varphi }
\end{eqnarray}
\begin{eqnarray} \label{eq347abc}
{\Phi^{a}_{n}}(\mathbf{r}) = Y_m \left( r\sqrt {n^2 +\ {\Gamma ^a_{In}}^2} \right) e^{nz} A_{m} e^{-im \varphi }
\end{eqnarray}
where $n$ and $m$ are positive integers {\it{independent}} from each other, and $A_{m}$ denotes the scaling factor. The {\it{ACs}} (\ref{eq339abc}) - (\ref{eq347abc}) with the eigen parameter ${\Gamma ^a_{In}}^2 =0$ are the solutions to the Laplace equation (\ref{eq337abc}). That means that the unit-field (particle) with the Helmholtz {\it{AC}} calibrated by the gauge (\ref{eq291abc}) with the eigen parameter ${{\Gamma ^a_{In}}^2}=0$ is indistinguishable from the unit-field (particle) describing by the Laplace gauge.

The angular components $A_{m}e^{im \varphi }$ and $A_{m}e^{-im \varphi }$ of the unit-fields (particles) with the Helmholtz (${\Gamma ^a_{In}}^2 \neq 0$) or Laplace (${\Gamma ^a_{In}}^2 = 0$)  {\it{ACs}} (\ref{eq340abc}) - (\ref{eq347abc}) are the eigenfunctions of the generator ${\hat{L_z}} \equiv {\hat{s_z}}$ of rotations around the azimuthal axis:
\begin{eqnarray} \label{eq348abc}
{\hat{s_z}}  A_{m}e^{im \varphi } \equiv  -i \frac {\partial } {\partial {\varphi }} A_{m}e^{im \varphi } = m A_{m}e^{im \varphi } 
\end{eqnarray}
\begin{eqnarray} \label{eq349abc}
{\hat{s_z}}  A_{m}e^{-im \varphi } \equiv  -i \frac {\partial } {\partial {\varphi }} A_{m}e^{-im \varphi } = - m A_{m}e^{-im \varphi }. 
\end{eqnarray}
The respective z-components of the intrinsic angular orbital momentums (spins) are given by 
\begin{eqnarray} \label{eq350abc}
\langle \hat{s_z} \rangle = \langle A_{m}e^{im \varphi } | \hat{s_z}  | A_{m}e^{im \varphi }  \rangle (\langle A_{m}e^{im \varphi } |A_{m}e^{im \varphi }  \rangle)^{-1} = m
\end{eqnarray}
\begin{eqnarray} \label{eq351abc}
\langle \hat{s_z} \rangle = \langle A_{m}e^{-im \varphi } | \hat{s_z}  | A_{m}e^{-im \varphi } \rangle (\langle A_{m}e^{im \varphi } |A_{m}e^{im \varphi }  \rangle)^{-1} = -  m.
\end{eqnarray}
That means that Eqs. (\ref{eq350abc}) and (\ref{eq351abc}) for the probabilistic ($A_{m}=1$) and non-probabilistic ($A_{m}\neq 1$) normalizations give the same result:
\begin{eqnarray} \label{eq352abc}
\langle \hat{s_z} \rangle = \langle e^{im \varphi } | \hat{s_z}  | e^{im \varphi }  \rangle (\langle e^{im \varphi } |e^{im \varphi }  \rangle)^{-1} = m
\end{eqnarray}
\begin{eqnarray} \label{eq353abc}
\langle \hat{s_z} \rangle = \langle e^{-im \varphi } | \hat{s_z}  | e^{-im \varphi } \rangle (\langle e^{im \varphi } |e^{im \varphi }  \rangle)^{-1} = -  m.
\end{eqnarray}
\begin{eqnarray} \label{eq354abc}
\langle \hat{s_z} \rangle = \langle A_{m}e^{im \varphi } | \hat{s_z}  | A_{m}e^{im \varphi }  \rangle (\langle A_{m}e^{im \varphi } |A_{m}e^{im \varphi }  \rangle)^{-1} = m
\end{eqnarray}
\begin{eqnarray} \label{eq355abc}
\langle \hat{s_z} \rangle = \langle A_{m}e^{-im \varphi } | \hat{s_z}  | A_{m}e^{-im \varphi } \rangle (\langle A_{m}e^{im \varphi } |A_{m}e^{im \varphi }  \rangle)^{-1} = -  m.
\end{eqnarray}
In contrast to the the unit-fields (particles) with the Laplace and Helmholtz spherical {\it{ACs}}, the unit-fields (particles) with the Laplace and Helmholtz cylindrical {\it{ACs}} have the azimuthal spins, only.

It should be stressed again that the all 24 {\it{kinds}} of the unit-fields (elementary particles) can exist in principle. As an example, Sec. 5 has showed the unit-fields (particles) with the Laplace and Helmholtz {\it{ACs}} having {\it{spherical}} and {\it{cylindrical}} symmetries. {\it{The model, probably for the first time, provided a physical explanation a particle spin}}. For the well-known solutions of the Laplace and Helmholtz differential equations corresponding to the unit-field {\it{ACs}} that are symmetric in the {\it{Cartesian, conical, confocal ellipsoidal, elliptic cylindrical, oblate spheroidal, parabolic, parabolic cylindrical, paraboloidal, prolate spheroidal, bispherical or toroidal system}} see mathematical textbooks.

\section{6. The interference (interaction) of the structured unit-fields (elementary particles) with the arbitrary generators ${\tilde \psi _{0n}}$ and the multi-component TACs $\Phi^{a}_{1}$} 

Before considering the concrete (experimentally observed) unit-fields (particles), let me present in the {\it{general forms}} the relativistic and non-relativistic energy-mass relations and equations of motions of a composite field (particle), which is composed from the {\it{arbitrary}} unit-fields (particles) whose generators and associate-components ({\it{ACs)}} do satisfy the gauges (\ref{eq287abc}),  (\ref{eq288abc}) and (\ref{eq291abc}). The simplest composite field (particle) is composed from the two interfering (cross-correlating) unit-fields associated with the two interacting elementary particles. The common field $\psi (\mathbf{r},t)$ composed from the interfering (cross-correlating) unit-fields $\psi _{01}(\mathbf{r},t)$  and $\psi _{02}(\mathbf{r},t)$ associated with the respective interacting particles has the form (\ref{eq85abc}), which for the structured unit-fields 
\begin{eqnarray} \label{eq356abc}
\psi _{01}={\tilde \psi _{01}}\Phi^{a}_{1}
\end{eqnarray}
and
\begin{eqnarray} \label{eq357abc}
\psi _{02}={\tilde \psi _{02}}\Phi^{a}_{2},
\end{eqnarray}
with the one-, two-, three- or four-component {\it{TACs}} is presented as
\begin{eqnarray} \label{eq358abc}
{\psi }={\tilde \psi _{01}}\Phi^{a}_{1}+{\tilde \psi _{02}}\Phi^{a}_{2}. 
\end{eqnarray}
The generalized basic relation of Einstein's special relativity for the composition (\ref{eq358abc}) of the {\it{relativistic}} unit-fields is given by Eq. (\ref{eq89abc}) as  
\begin{eqnarray} \label{eq359abc}
{\varepsilon}^{2} = {\varepsilon}_{01}^{2} + {\varepsilon}_{02}^{2} + {\cal E}_{12}+{\cal E}_{21}={\varepsilon}_{01}^{2} + {\varepsilon}_{02}^{2} + {\cal E}_{12,21}. 
\end{eqnarray} 
For  the composition (\ref{eq358abc}) of the {\it{non-relativistic}} unit-fields, the non-relativistic energy-mass relation (\ref{eq100abc}) has the form
\begin{eqnarray} \label{eq360abc}
{\varepsilon} = {\varepsilon}_{01}  + {\varepsilon}_{02}  + {\varepsilon}_{12} +{\varepsilon}_{21}= {\varepsilon}_{01}  + {\varepsilon}_{02}  + {\varepsilon}_{12,21}. 
\end{eqnarray} 

For the composite field (particle)
\begin{eqnarray} \label{eq361abc}
\psi =\sum_{n=1}^N \tilde  \psi _{0n}\Phi^{a}_{n},
\end{eqnarray}
which is composed from the $N$ relativistic unit-fields ${\tilde \psi _{01}}\Phi^{a}_{n}$ associated with the $N\geq 1$ relativistic particles, the {\it{relativistic}} energy-mass relation (\ref{eq96abc}) has the form  
\begin{eqnarray} \label{eq362abc}
{\varepsilon}^2 = \sum_{n=1}^N{{\varepsilon}}_{0n}^2+\sum_{n\neq m}^{N^2-N}{{\cal E}}_{nm}.
\end{eqnarray}
The non-relativistic energy-mass relation for the composition (\ref{eq381abc}) of the {\it{non-relativistic}} unit-fields is given by Eq. (\ref{eq106abc}) as
\begin{eqnarray} \label{eq363abc}
{\varepsilon} = \sum_{n=1}^N{{\varepsilon}}_{0n}+\sum_{n\neq m}^{N^2-N}{{\varepsilon}}_{nm}.
\end{eqnarray}

\subsection{6.1. The {\it{relativistic}} interference (interaction) of the structured unit-fields (particles) with the {\it{arbitrary generators ${\tilde \psi _{0n}}$ and the multi-component TACs $\Phi^{a}_{1}$}}: The relativistic energy-mass relations and equations of motions}

In the case of the {\it{relativistic}} composite field (\ref{eq358abc}), which is composed from the two structured unit-fields (particles), the relativistic energy squared ${\varepsilon}_{01}^{2}$ of the first unit-field (\ref{eq356abc}) in the relativistic energy-mass relation (\ref{eq359abc}) is given by Eq. (\ref{eq90abc}), which {\it{under action}} of the gauges (\ref{eq287abc}),  (\ref{eq288abc}) and (\ref{eq291abc}) has the form (\ref{eq301abc}):
\begin{eqnarray} \label{eq364abc}
{\varepsilon}_{01}^{2} =  {{{\frac {1} {2 }}\int_{V}}  {\tilde \psi} _{01} ^* {\Phi^{a}_{1}}^*  \left[ -\ddot {\tilde \psi} _{01} - \nabla ^2{\tilde \psi} _{01} +  m_{0}^2 {\tilde \psi} _{01} + {\tilde \psi} _{01} (\Phi^{a}_{1})^{-1} \nabla ^2 \Phi^{a}_{1}  \right] \Phi^{a}_{1}d^3x}.
\end{eqnarray}
Similarly, the relativistic energy squared of the second unit-field (\ref{eq357abc}) has the general form
\begin{eqnarray} \label{eq365abc}
{\varepsilon}_{02}^{2} =  {{{\frac {1} {2 }}\int_{V}}  {\tilde \psi} _{02} ^* {\Phi^{a}_{2}}^*  \left[ -\ddot {\tilde \psi} _{02} - \nabla ^2{\tilde \psi} _{02} +  m_{0}^2 {\tilde \psi} _{02} + {\tilde \psi} _{02} (\Phi^{a}_{2})^{-1} \nabla ^2 \Phi^{a}_{2} \right] \Phi^{a}_{2} d^3x}.
\end{eqnarray}
The relativistic cross-correlation terms associated with the first and second unit-fields are given by Eqs. (\ref{eq92abc}) and (\ref{eq93abc}) with the aforementioned gauges in the general form respectively as
\begin{eqnarray} \label{eq366abc}
{\cal E}_{12}=  {{{\frac {1} {2 }}\int_{V}}  {\tilde \psi} _{01} ^* {\Phi^{a}_{1}}^*  \left[ -\ddot {\tilde \psi} _{02} - \nabla ^2{\tilde \psi} _{02} +  m_{0}^2 {\tilde \psi} _{02} + {\tilde \psi} _{02} (\Phi^{a}_{2})^{-1} \nabla ^2 \Phi^{a}_{2} \right] \Phi^{a}_{2} d^3x}
\end{eqnarray}
and
\begin{eqnarray} \label{eq367abc}
{\cal E}_{21}=  {{{\frac {1} {2 }}\int_{V}}  {\tilde \psi} _{02} ^* {\Phi^{a}_{2}}^*  \left[ -\ddot {\tilde \psi} _{01} - \nabla ^2{\tilde \psi} _{01} +  m_{0}^2 {\tilde \psi} _{01} + {\tilde \psi} _{01} (\Phi^{a}_{1})^{-1} \nabla ^2 \Phi^{a}_{1}  \right] \Phi^{a}_{1}d^3x},
\end{eqnarray}
respectively. The total relativistic cross-correlation term ${\cal E}_{12,21} = {\cal E}_{12} + {\cal E}_{21}$ associated with both the first and second unit-fields is given by  
\begin{eqnarray} \label{eq368abc}
{\cal E}_{12,21} = {{{\frac {1} {2 }}\int_{V}}  {\tilde \psi} _{01} ^* {\Phi^{a}_{1}}^*  \left[ -\ddot {\tilde \psi} _{02} - \nabla ^2{\tilde \psi} _{02} +  m_{0}^2 {\tilde \psi} _{02} + {\tilde \psi} _{02} (\Phi^{a}_{2})^{-1} \nabla ^2 \Phi^{a}_{2} \right] \Phi^{a}_{2} d^3x} +  \nonumber \\  + {{{\frac {1} {2 }}\int_{V}}  {\tilde \psi} _{02} ^* {\Phi^{a}_{2}}^*  \left[ -\ddot {\tilde \psi} _{01} - \nabla ^2{\tilde \psi} _{01} +  m_{0}^2 {\tilde \psi} _{01} + {\tilde \psi} _{01} (\Phi^{a}_{1})^{-1} \nabla ^2 \Phi^{a}_{1}  \right] \Phi^{a}_{1}d^3x}.
\end{eqnarray}
The relativistic equation of motion of the composite field (\ref{eq358abc}) is given by Eq. (\ref{eq94abc}) or (\ref{eq286abc}) with the gauges (\ref{eq287abc}),  (\ref{eq288abc}) and (\ref{eq291abc}) or the calibrations (\ref{eq287abc}),  (\ref{eq288abc}) and (\ref{eq293abc}) in the general form
\begin{eqnarray} \label{eq369abc}
{\Phi^{a}_{1}} \left( \square {\tilde \psi} _{01} + m_{0}^2 {\tilde \psi} _{01} + {\tilde \psi} _{01} (\Phi^{a}_{1})^{-1} \nabla ^2 \Phi^{a}_{1} \right) + {\Phi^{a}_{2}} \left( \square {\tilde \psi} _{02} +  m_{0}^2 {\tilde \psi} _{02} + {\tilde \psi} _{02} (\Phi^{a}_{2})^{-1} \nabla ^2 \Phi^{a}_{2}\right)  = 0.
\end{eqnarray}
Notice, the unit-fields ${\tilde \psi} _{01} {\Phi^{a}_{1}}$ and ${\tilde \psi} _{02}  {\Phi^{a}_{2}}$ of the composite field do not obey the normalizations $(\ref{eq18abc})$ and $(\ref{eq245abc})$ of the single unit-fields associated with the single particles. In the case of ${\Phi^{a}_{1}}={\Phi^{a}_{2}}=const.$, Eq. (\ref{eq369abc}) describes the correlation-free (interaction-free) unit-fields that could be attributed to the intersection-free particles (bosons). 

In the case of the {\it{relativistic}} composite field (\ref{eq361abc}), which is composed from the $N$ structured relativistic unit-fields, the relativistic energy squared ${\varepsilon}_{0n}^{2}$ of the $n$-th unit-field $\tilde  \psi _{0n}\Phi^{a}_{n}$ in the relativistic energy-mass relation (\ref{eq362abc}) is given by Eq. (\ref{eq97abc}) in the general form as
\begin{eqnarray} \label{eq370abc}
{\varepsilon}_{0n}^{2} =  {{{\frac {1} {2 }}\int_{V}}  {\tilde \psi} _{0n} ^* {\Phi^{a}_{n}}^*  \left[ -\ddot {\tilde \psi} _{0n} - \nabla ^2{\tilde \psi} _{0n} +  m_{0}^2 {\tilde \psi} _{0n} + {\tilde \psi} _{0n} (\Phi^{a}_{n})^{-1} \nabla ^2 \Phi^{a}_{n}  \right] \Phi^{a}_{n}d^3x}.
\end{eqnarray}
The relativistic cross-correlation terms associated with the $n$-th unit-fields is given by Eq. (\ref{eq98abc}) with the aforementioned gauges in the form 
\begin{eqnarray} \label{eq371abc}
{\cal E}_{nm}=  {{{\frac {1} {2 }}\int_{V}}  {\tilde \psi} _{0n} ^* {\Phi^{a}_{n}}^*  \left[ -\ddot {\tilde \psi} _{0m} - \nabla ^2{\tilde \psi} _{0m} +  m_{0}^2 {\tilde \psi} _{0m} + {\tilde \psi} _{0m} (\Phi^{a}_{m})^{-1} \nabla ^2 \Phi^{a}_{m} \right] \Phi^{a}_{m} d^3x}
\end{eqnarray}
The relativistic cross-correlation terms associated with the $m$-th unit-fields is given by Eq. (\ref{eq98abc}) with the aforementioned gauges in the form 
\begin{eqnarray} \label{eq372abc}
{\cal E}_{mn}=  {{{\frac {1} {2 }}\int_{V}}  {\tilde \psi} _{0m} ^* {\Phi^{a}_{m}}^*  \left[ -\ddot {\tilde \psi} _{0n} - \nabla ^2{\tilde \psi} _{0n} +  m_{0}^2 {\tilde \psi} _{0n} + {\tilde \psi} _{0n} (\Phi^{a}_{n})^{-1} \nabla ^2 \Phi^{a}_{n} \right] \Phi^{a}_{n} d^3x}
\end{eqnarray}
The total relativistic cross-correlation term associated with both the $n$-th and $m$-th unit-fields is given by the equation ${\cal E}_{nm,mn} = {\cal E}_{nm} + {\cal E}_{mn}$ in the form
\begin{eqnarray} \label{eq373abc}
{\cal E}_{nm,mn} = {{{\frac {1} {2 }}\int_{V}}  {\tilde \psi} _{0n} ^* {\Phi^{a}_{n}}^*  \left[ -\ddot {\tilde \psi} _{0m} - \nabla ^2{\tilde \psi} _{0m} +  m_{0}^2 {\tilde \psi} _{0m} + {\tilde \psi} _{0m} (\Phi^{a}_{m})^{-1} \nabla ^2 \Phi^{a}_{m} \right] \Phi^{a}_{m} d^3x}   +  \nonumber \\  + {{{\frac {1} {2 }}\int_{V}}  {\tilde \psi} _{0m} ^* {\Phi^{a}_{m}}^*  \left[ -\ddot {\tilde \psi} _{0n} - \nabla ^2{\tilde \psi} _{0n} +  m_{0}^2 {\tilde \psi} _{0n} + {\tilde \psi} _{0n} (\Phi^{a}_{n})^{-1} \nabla ^2 \Phi^{a}_{n} \right] \Phi^{a}_{n} d^3x}.
\end{eqnarray}
respectively. The relativistic equation of motion of the composite field (\ref{eq361abc}) is given by Eq. (\ref{eq99abc}) in the general form as
\begin{eqnarray} \label{eq374abc}
\sum_{n=1}^{N } {\Phi^{a}_{n}} \left( \square {\tilde \psi} _{0n} + m_{0}^2 {\tilde \psi} _{0n} + {\tilde \psi} _{0n} (\Phi^{a}_{n})^{-1} \nabla ^2 \Phi^{a}_{n} \right)   = 0.
\end{eqnarray}
The unit-fields ${\tilde \psi} _{0n} {\Phi^{a}_{n}}$ and  ${\tilde \psi} _{0m} {\Phi^{a}_{m}}$ of the composite field do not obey the normalizations $(\ref{eq18abc})$ and $(\ref{eq245abc})$ of the single unit-fields associated with the single particles. It should be stressed that the relativistic energy of the $n$-th unit-field, which is determined by Eq. (\ref{eq370abc}), {\it{does not depend on the unit-field phase}} $\alpha_n$. While the total relativistic cross-correlation term (\ref{eq373abc}) associated with the $n$-th and $m$-th unit-fields {\it{depends on the phases}} $\alpha_n$ and $\alpha_m$ of the interfering unit-fields (particles).

\subsection{6.2. The {\it{non-relativistic}} interference (interaction) of the structured unit-fields (particles) with the {\it{arbitrary generators ${\tilde \psi _{0n}}$ and the multi-component TACs $\Phi^{a}_{1}$}}: The non-relativistic energy-mass relations and equations of motions}

In the case of the {\it{non-relativistic}} composite field (\ref{eq358abc}), which is composed from the two structured non-relativistic unit-fields, the non-relativistic energy  ${\varepsilon}_{01}$ of the first unit-field (\ref{eq356abc}) in the relativistic energy-mass relation (\ref{eq360abc}) is given by Eq. (\ref{eq102abc}) or Eq. (\ref{eq302abc}), which {\it{under action}} of the calibrations (gauges) (\ref{eq287abc}),  (\ref{eq288abc}) and (\ref{eq291abc}) has the form
\begin{eqnarray} \label{eq375abc}
{\varepsilon}_{01} =  {{{\frac {1} {2 }}\int_{V}}  {\tilde \psi} _{01} ^* {\Phi^{a}_{1}}^* \left[ i\dot {\tilde \psi} _{01} - {\frac {1} {2m_0} }  \nabla ^2{\tilde \psi} _{01} +  m_{0}{\tilde \psi} _{01} + {\frac {1} {2m_0} } {\tilde \psi} _{01} (\Phi^{a}_{1})^{-1} \nabla ^2 \Phi^{a}_{1}   \right] \Phi^{a}_{1}d^3x}. 
\end{eqnarray}
Similarly, the relativistic energy of the second unit-field (\ref{eq357abc}) has the general form
\begin{eqnarray} \label{eq376abc}
{\varepsilon}_{02} =  {{{\frac {1} {2 }}\int_{V}}  {\tilde \psi} _{02} ^* {\Phi^{a}_{2}}^* \left[ i\dot {\tilde \psi} _{02} - {\frac {1} {2m_0} }  \nabla ^2{\tilde \psi} _{02} +  m_{0}{\tilde \psi} _{02} + {\frac {1} {2m_0} } {\tilde \psi} _{02} (\Phi^{a}_{2})^{-1} \nabla ^2 \Phi^{a}_{2}   \right]\Phi^{a}_{2}d^3x}.
\end{eqnarray}
The non-relativistic cross-correlation energies associated with the first and second unit-fields are given by Eqs. (\ref{eq104abc}) and (\ref{eq105abc}) with the aforementioned gauges in the general form respectively as
\begin{eqnarray} \label{eq377abc}
{\varepsilon}_{12} =  {{{\frac {1} {2 }}\int_{V}}  {\tilde \psi} _{01} ^* {\Phi^{a}_{1}}^* \left[ i\dot {\tilde \psi} _{02} - {\frac {1} {2m_0} }  \nabla ^2{\tilde \psi} _{02} +  m_{0}{\tilde \psi} _{02} + {\frac {1} {2m_0} } {\tilde \psi} _{02} (\Phi^{a}_{2})^{-1} \nabla ^2 \Phi^{a}_{2}   \right]\Phi^{a}_{2}d^3x} 
\end{eqnarray}
and
\begin{eqnarray} \label{eq378abc}
{\varepsilon}_{21} =  {{{\frac {1} {2 }}\int_{V}}  {\tilde \psi} _{02} ^* {\Phi^{a}_{2}}^* \left[ i\dot {\tilde \psi} _{01} - {\frac {1} {2m_0} }  \nabla ^2{\tilde \psi} _{01} +  m_{0}{\tilde \psi} _{01} + {\frac {1} {2m_0} } {\tilde \psi} _{01} (\Phi^{a}_{1})^{-1} \nabla ^2 \Phi^{a}_{1}    \right]\Phi^{a}_{1}d^3x}.
\end{eqnarray}
The total cross-correlation energy is given by ${\varepsilon}_{12,21}= {\varepsilon}_{12}+{\varepsilon}_{21}$. The non-relativistic equation of motion of the composite field (\ref{eq358abc}) is given by Eq. (\ref{eq101abc}) with the gauges (\ref{eq287abc}),  (\ref{eq288abc}) and (\ref{eq291abc}) in the form
\begin{eqnarray} \label{eq379abc}
{\Phi^{a}_{1}} \left( - i\dot {\tilde \psi} _{01} - {\frac {1} {2m_0} }  \nabla ^2{\tilde \psi} _{01} +m_{0}{\tilde \psi} _{01} + {\frac {1} {2m_0} } {\tilde \psi} _{01} (\Phi^{a}_{1})^{-1} \nabla ^2 \Phi^{a}_{1}\right) +  \nonumber \\  + {\Phi^{a}_{2}} \left(- i\dot {\tilde \psi} _{02} - {\frac {1} {2m_0} }  \nabla ^2{\tilde \psi} _{02} + m_{0}{\tilde \psi} _{02} + {\frac {1} {2m_0} } {\tilde \psi} _{02} (\Phi^{a}_{2})^{-1} \nabla ^2 \Phi^{a}_{2} \right)  = 0.
\end{eqnarray}
Notice, the unit-fields ${\tilde \psi} _{01} {\Phi^{a}_{1}}$ and ${\tilde \psi} _{02}  {\Phi^{a}_{2}}$ of the composite field do not obey the normalizations $(\ref{eq18abc})$ and $(\ref{eq245abc})$ of the single unit-fields associated with the single particles.

In the case of the {\it{non-relativistic}} composite unit-field (\ref{eq361abc}), which is composed from the $N$ Laplace or Helmholtz unit-fields, the relativistic energy ${\varepsilon}_{01}$ of the $n$-th unit-field $\tilde  \psi _{0n}\Phi^{a}_{n}$ in the relativistic energy-mass relation (\ref{eq363abc}) is given by Eq. (\ref{eq108abc}) in the general form as
\begin{eqnarray} \label{eq380abc}
{\varepsilon}_{0n} =  {{{\frac {1} {2 }}\int_{V}}  {\tilde \psi} _{0n} ^* {\Phi^{a}_{n}}^* \left[ i\dot {\tilde \psi} _{0n} - {\frac {1} {2m_0} }  \nabla ^2{\tilde \psi} _{0n} +  m_{0}{\tilde \psi} _{0n} + {\frac {1} {2m_0} } {\tilde \psi} _{0n} (\Phi^{a}_{n})^{-1} \nabla ^2 \Phi^{a}_{n}   \right]\Phi^{a}_{n}d^3x}.
\end{eqnarray}
The non-relativistic cross-correlation energies associated with the $n$-th and $m$-th unit-fields are given by Eq. (\ref{eq109abc}) in the general form as
\begin{eqnarray} \label{eq381abc}
{\varepsilon}_{nm} =  {{{\frac {1} {2 }}\int_{V}}  {\tilde \psi} _{0n} ^* {\Phi^{a}_{n}}^* \left[ i\dot {\tilde \psi} _{0m} - {\frac {1} {2m_0} }  \nabla ^2{\tilde \psi} _{0m} +  m_{0}{\tilde \psi} _{0m} + {\frac {1} {2m_0} } {\tilde \psi} _{0m} (\Phi^{a}_{m})^{-1} \nabla ^2 \Phi^{a}_{m}   \right]\Phi^{a}_{m}d^3x}. 
\end{eqnarray}
and
\begin{eqnarray} \label{eq382abc}
{\varepsilon}_{mn} =  {{{\frac {1} {2 }}\int_{V}}  {\tilde \psi} _{0m} ^* {\Phi^{a}_{m}}^* \left[ i\dot {\tilde \psi} _{0n} - {\frac {1} {2m_0} }  \nabla ^2{\tilde \psi} _{0n} +  m_{0}{\tilde \psi} _{0n} + {\frac {1} {2m_0} } {\tilde \psi} _{0n} (\Phi^{a}_{n})^{-1} \nabla ^2 \Phi^{a}_{n}   \right]\Phi^{a}_{n}d^3x}.
\end{eqnarray}
respectively. The total cross-correlation energy is given by ${\varepsilon}_{12,21}= {\varepsilon}_{12}+{\varepsilon}_{21}$. The relativistic equation of motion of the Laplace or Helmholtz composite field (\ref{eq361abc}) is given by Eq. (\ref{eq107abc}) in the general form as
\begin{eqnarray} \label{eq383abc}
\sum_{n=1}^{N } {\Phi^{a}_{n}} \left(- i\dot {\tilde \psi} _{0n} - {\frac {1} {2m_0} }  \nabla ^2{\tilde \psi} _{0n} +  m_{0}{\tilde \psi} _{0n} + {\frac {1} {2m_0} } {\tilde \psi} _{0n} (\Phi^{a}_{n})^{-1} \nabla ^2 \Phi^{a}_{n}  \right)   = 0.
\end{eqnarray}
The unit-fields ${\tilde \psi} _{0n} {\Phi^{a}_{n}}$ of the composite field do not obey the normalizations $(\ref{eq18abc})$ and $(\ref{eq245abc})$ of the single unit-fields associated with the single particles. Notice, the non-relativistic energy (\ref{eq380abc}) of the $n$-th unit-field {\it{does not depend on the unit-field phase}} $\alpha_n$. While the total non-relativistic cross-correlation energy (\ref{eq382abc}) associated with the $n$-th and $m$-th unit-fields {\it{depends on the phases}} $\alpha_n$ and $\alpha_m$ of the interfering unit-fields (particles).

\section{7. The interference (interaction) of the unit-fields (particles) having the de Broglie generators ${\tilde \psi} _{0n}$ and the Laplace or Helmholtz one-component ACs ${\phi_{in}}$}

\subsection{7.1. The {\it{general forms}} of the interaction terms and energies of the unit-fields (particles) having the {\it{de Broglie generators}} ${\tilde \psi} _{0n}$ and the Laplace or Helmholtz {\it{arbitrary one-component ACs}} ${\phi_{in}}$}

The above-presented Eqs. (\ref{eq364abc}) - (\ref{eq374abc}) and Eqs. (\ref{eq375abc}) - (\ref{eq383abc}) describe respectively the relativistic and non-relativistic unit-fields (particles) in the {\it{general form}} associated with the {\it{arbitrary}} configurations of the unit-field generators and associate-components ({\it{ACs)}} satisfying to the gauges (\ref{eq287abc}),  (\ref{eq288abc}) and (\ref{eq291abc}). Naturally, the equations that describe the unit-fields (particles) with the {\it{concrete}} ({\it{ACs)}} and ({\it{TACs)}} {\it{corresponding to the concrete, experimentally-observed elementary particles}} should be also presented. 

For the sake of simplicity, I present here the equations for the {\it{two}} unit-fields, namely for the unit-fields $\psi _{01}={\tilde \psi _{01}}\Phi^{a}_{1}$ (\ref{eq356abc}) and $\psi _{02}={\tilde \psi _{02}}\Phi^{a}_{2}$ (\ref{eq357abc}) of the composite field ${\psi }={\tilde \psi _{01}}\Phi^{a}_{1}+{\tilde \psi _{02}}\Phi^{a}_{2}$ (\ref{eq358abc}), where the total associate-components ({\it{TACs}}) have the {\it{one-component forms}} $\Phi^{a}_{1} = {\phi_{i1}}$ and  $\Phi^{a}_{2} = {\phi_{i2}}$. The simplest solutions to the relativistic Eqs. (\ref{eq369abc}) and non-relativistic (\ref{eq379abc}) equations of motions, which couple the first unit-field to the second one  and {\it{vice versa}}, are the de Broglie unit-waves 
\begin{eqnarray} \label{eq384abc}
\tilde \psi_{01} (\mathbf{r},t)=a_0 e^{i({\mathbf{k}_{01}{\mathbf{r}}}-{\varepsilon}_{01} t +\alpha_1)} 
\end{eqnarray}
\begin{eqnarray} \label{eq385abc}
\tilde \psi_{02} (\mathbf{r},t)=a_0 e^{i({\mathbf{k}_{02}{\mathbf{r}}}-{\varepsilon}_{02} t + \alpha_2)} 
\end{eqnarray}
of the form (\ref{eq297abc}) with the momentums $\mathbf{k}_{01} \neq \mathbf{k}_{02}$, energies $\varepsilon_{01} \neq \varepsilon_{02}$ and phases $\alpha_1 \neq \alpha_2$. 

In the case of the {\it{relativistic}} composition of the unit-fields (\ref{eq384abc}) and (\ref{eq385abc}), the {\it{relativistic}} interaction (cross-correlation) terms are given by  Eqs. (\ref{eq366abc}) and (\ref{eq367abc}) as
\begin{eqnarray} \label{eq386abc}
{\cal E}_{12}= {{\int_{V}}  \left[ \mathbf{k}_{02}^2+ \left( m_{01}m_{02} + {\Gamma _{i1}}{\Gamma _{i2}} \right) \right] a_0^2 {\phi_{i1}}^*\phi_{i2}e^{-i({\mathbf{k}_{01}{\mathbf{r}}}-{\varepsilon}_{01} t )} e^{i({\mathbf{k}_{02}{\mathbf{r}}}-{\varepsilon}_{02} t +\alpha_1 )}d^3x}
\end{eqnarray}
and
\begin{eqnarray} \label{eq387abc}
{\cal E}_{21}= {{\int_{V}} \left[ \mathbf{k}_{01}^2+ \left( m_{01}m_{02} + {\Gamma _{i1}}{\Gamma _{i2}} \right) \right] a_0^2 {\phi_{i2}}^*\phi_{i1} e^{i({\mathbf{k}_{01}{\mathbf{r}}}-{\varepsilon}_{01} t )} e^{-i({\mathbf{k}_{02}{\mathbf{r}}}-{\varepsilon}_{02} t + \alpha_2)}d^3x},
\end{eqnarray}
where the eigen parameter ${\Gamma ^2_{in}}$ is determined by Eq. (\ref{eq291abc}) in the form of the Helmholt (${\Gamma ^2_{in}} \neq 0$) or Laplace (${\Gamma ^2_{in}} = 0$) gauge
\begin{eqnarray} \label{eq388abc}
\nabla ^2 \phi_{in} + {\Gamma ^2_{in}} \phi_{in} = 0.
\end{eqnarray}
Then the total relativistic interaction (cross-correlation) term is given by
\begin{eqnarray} \label{eq389abc}
{\cal E}_{12,21}={\cal E}_{12}+{\cal E}_{21}    
\end{eqnarray}
where $\mathbf{k}_{01} \neq \mathbf{k}_{02}$, $\varepsilon_{01} \neq \varepsilon_{02} $, $m_{01}\neq m_{02} $ and ${\Gamma_ {i1}} \neq {\Gamma_ {i2}}$, and ${\Delta  \mathbf{k}_{12}}=\mathbf{k}_{01} - \mathbf{k}_{02}$, ${\Delta {\varepsilon}_{12}  }= \varepsilon_{01} - \varepsilon_{02}$ and $\Delta \alpha_{12}=\alpha_{1}-\alpha_{2}$. It should be mentioned that the {\it{relativistic}} interaction (cross-correlation) terms in the 4th version of the model (Sec. 5.2.1.) are described by Eqs. (\ref{eq386abc}) and (\ref{eq387abc}), where the parameters $\mathbf{k}_{02}^2$ and $\mathbf{k}_{01}^2$ should be replaced respectively by the values $\mathbf{k}_{01}\mathbf{k}_{02}$ and $\mathbf{k}_{02}\mathbf{k}_{01}$. In the case of $\mathbf{k}_{01} = \mathbf{k}_{02}$, the terms of the 1-st and 4-th versions are indistinguishable from each other. 

The {\it{non-relativistic}} energies associated with the first and second {\it{relativistic}} unit-fields are given by Eqs. (\ref{eq377abc}) and (\ref{eq378abc}) with the aforementioned gauges in the general form respectively as
\begin{eqnarray} \label{eq390abc}
{\varepsilon}_{12}=  {{\int_{V}}  \left( {\frac {1} {2m_0} } \mathbf{k}_{02}^2+ \left[ m_{0} +{\frac {1} {2m_0} } {\Gamma _{i1}}{\Gamma _{i2}} \right] \right) a_0^2 {\phi_{i1}}^*\phi_{i2}e^{-i({\mathbf{k}_{01}{\mathbf{r}}}-{\varepsilon}_{01} t + \alpha_1)} e^{i({\mathbf{k}_{02}{\mathbf{r}}}-{\varepsilon}_{02} t +\alpha_2 )} d^3x}
\end{eqnarray}
and
\begin{eqnarray} \label{eq391abc}
{\varepsilon}_{21}= {{\int_{V}} \left( {\frac {1} {2m_0} } \mathbf{k}_{01}^2 +\left[ m_{0} + {\frac {1} {2m_0} } {\Gamma _{i1}}{\Gamma _{i2}} \right] \right) a_0^2 {\phi_{i2}}^*\phi_{i1} e^{i({\mathbf{k}_{01}{\mathbf{r}}}-{\varepsilon}_{01} t + \alpha_1)} e^{-i({\mathbf{k}_{02}{\mathbf{r}}}-{\varepsilon}_{02} t +\alpha_2 )}d^3x},
\end{eqnarray}
where the value $m_0={\sqrt{m_{01}}}{\sqrt{m_{02}}}$. Then the total non-relativistic interaction (cross-correlation) energy is given by
\begin{eqnarray} \label{eq392abc}
{\varepsilon}_{12,21}={\varepsilon}_{12}+{\varepsilon}_{21}.    
\end{eqnarray}
Notice, the {\it{non-relativistic}} interaction (cross-correlation) energies in the 4th version of the model (Sec. 5.2.1.) are described by Eqs. (\ref{eq390abc}) and (\ref{eq391abc}), where the parameters $\mathbf{k}_{02}^2$ and $\mathbf{k}_{01}^2$ should be replaced by the values $\mathbf{k}_{01}\mathbf{k}_{02}$ and  $\mathbf{k}_{02}\mathbf{k}_{01}$. Naturally, in the case of $\mathbf{k}_{01} = \mathbf{k}_{02}$, the interaction energies of the 1-st and 4-th versions are indistinguishable from each other. 

Equations (\ref{eq384abc}) - (\ref{eq392abc}) describe the interaction energies of the unit-fields (particles) having the de Broglie generators ${\tilde \psi} _{0n}$ and the {\it{arbitrary}} Laplace (${\Gamma ^2_{in}}=0$) or Helmholtz (${\Gamma ^2_{in}} \neq 0$) {\it{ACs}} ${\phi_{in}}$ calibrated by the gauge (\ref{eq291abc}). It should be stressed that the total relativistic cross-correlation term (\ref{eq389abc}) and the total non-relativistic cross-correlation energy (\ref{eq392abc}) associated with the 1-st and 2-nd unit-fields {\it{do depend on the phases}} $\alpha_1$ and $\alpha_2$ of the interfering unit-fields (particles). In the case of unit-fields (particles) having the probabilistic (for instance, thermal) distributions of the values $\mathbf{k}_n$, $\varepsilon_{n}$ and $\alpha_{12}$ in macroscopic bodies (macroscopic composite "particles"), the values (\ref{eq389abc}) and (\ref{eq392abc}) {\it{should be averaged}} using these probabilistic distributions.

\subsection{7.2. The interference (interaction) of the unit-fields having the {\it{de Broglie generators}} ${\tilde \psi} _{0n}$ and the {\it{spherically symmetric, one-component Laplace-ACs}} ${\phi_{in}}$: The non-quantum and quantum pure {\it{gravitations}} and pure {\it{electromagnetism}}}

\vspace{0.2cm}
1. {\it{The Newton-like and Lorentz-like non-quantum and quantum pure gravitations }}
\vspace{0.2cm}

Let me consider, for the sake of simplicity, the equations for the {\it{two}} unit-fields, namely for the unit-fields $\psi _{01}={\tilde \psi _{01}}\Phi^{a}_{1}$ (\ref{eq356abc}) and $\psi _{02}={\tilde \psi _{02}}\Phi^{a}_{2}$ (\ref{eq357abc}) of the composite field ${\psi }={\tilde \psi _{01}}\Phi^{a}_{1}+{\tilde \psi _{02}}\Phi^{a}_{2}$ (\ref{eq358abc}), where the total associate-components ({\it{TACs}}) have the {\it{one-component forms}} $\Phi^{a}_{1} = {\phi_{i1}}$ and  $\Phi^{a}_{2} = {\phi_{i2}}$, and the relevant values are given by $\mathbf{k}_{01} = \mathbf{k}_{02}= \mathbf{k}_{0}$, $\varepsilon_{01} = \varepsilon_{02} =\varepsilon_{0} $, $m_{01}=m_{02}=m_{0}$, $\alpha_{1}=\alpha_{2}=\alpha _0$ and ${\Gamma_ {i1}}={\Gamma_ {i2}}={\Gamma_ {i}}$. An analysis of the unit-fields (particles) (\ref{eq356abc}) and (\ref{eq357abc}) having the de Broglie generators $\tilde \psi_{01} (\mathbf{r},t)=a_0 e^{i({\mathbf{k}_{01}{\mathbf{r}}}-{\varepsilon}_{01} t +\alpha _1 )}$ (\ref{eq384abc}) and $\tilde \psi_{02} (\mathbf{r},t)=a_0 e^{i({\mathbf{k}_{02}{\mathbf{r}}}-{\varepsilon}_{02} t +\alpha _2)}$ (\ref{eq385abc}) and the {\it{spherically symmetric}} regular (\ref{eq316abc}) or irregular (\ref{eq317abc}) {\it{Laplace-ACs}} shows that only the {\it{one-component unit-fields}} (\ref{eq356abc}) and (\ref{eq357abc}) with the {\it{irregular}} {\it{Laplace-ACs}} and the intrinsic orbital quantum number $l_n=1$ and the respective intrinsic "magnetic" quantum numbers $s_n \equiv m_n = -1, 0,1$ of the unit-field spin may be considered as candidates for the pure-gravitational interaction with the interaction energy $\sim 1/R$. Among the candidates, the only unit-fields with the quantum numbers $l_n = 1$ and $s_n=0$ satisfy the experimentally observed physical properties of the gravitation. Indeed, in the case of the first {\it{irregular}} {\it{Laplace-AC}} [${\phi_{i1}}={\phi_{G1}}(\bf{r})$] centered at the origin and the second {\it{irregular}} {\it{Laplace-AC}} [${\phi_{i2}}={\phi_{G2}}(\bf{r}-\bf{R})$] located at the distance $R=|\bf{R}|\equiv |\bf{R}_{12}|$ in the azimuthal direction, the relativistic gravitational interaction (cross-correlation) total term (\ref{eq389abc}) has the form
\begin{eqnarray} \label{eq393abc}
{\cal E}_{12,21}(R,l_1,l_2,s_1,s_2) = \mathbf{k}_{01} \mathbf{k}_{02} \left[  {{ {{A_{1s_1}} {A_{1s_2}}} { a_0^2}} \int_{\theta =0}^\pi d\theta sin\theta  \int_{\varphi  =0}^{2\pi} d \varphi  \int_{0}}^{\infty}r^2dr {\frac {1} {r^2}}{\frac {f(r,R,\theta , \varphi , s_1, s_2 ) } { |  {\bf{r}}-{\bf{R}}|^{2}} }   \right]  +   \nonumber  \\  +   m_{01} m_{02}\left[  {{{{A_{1s_1}} {A_{1s_2}}} { a_0^2}} \int_{\theta =0}^\pi d\theta sin\theta  \int_{\varphi  =0}^{2\pi} d \varphi  \int_{0}}^{\infty}r^2dr {\frac {1} {{r}^2}}{\frac {f(r,R,\theta , \varphi ,s_1, s_2 ) } { |  {\bf{r}}-{\bf{R}}|^{2}} } \right] , 
\end{eqnarray}
where
\begin{eqnarray} \label{eq394abc}
{\Gamma ^2_{i1}}= {\Gamma ^2_{i2}}={\Gamma ^2_{G1}}= {\Gamma ^2_{G2}}=0 \nonumber  \\
l_1=l_2=1  
\end{eqnarray}
and the {\it{Laplace-AC}} amplitude ${A_{1_ns_n}}$ can be an arbitrary constant [see, Eq. (\ref{eq317abc})]. Here, the angle factor $f(r,R,\theta , \varphi , s_1, s_2 )$ associated with the spherical harmonics is given by 
\begin{eqnarray} \label{eq395abc}
f(r,R,\theta , \varphi , s_1, s_2 )= {Y_{l_1}^{s_1}}^* (\theta , \varphi) Y_{l_2}^{s_2} (\theta , \varphi _2 ) + {Y_{1_1}^{s_1}} (\theta , \varphi) {Y_{1_2}^{s_2}}^* (\theta , \varphi _2), 
\end{eqnarray}
with 
\begin{eqnarray} \label{eq396abc}
{Y_{1}^{-1}}(\theta , \varphi )=  \sqrt{\frac {3}{8\pi}}  sin\theta e^{-i\varphi},
\end{eqnarray}
\begin{eqnarray} \label{eq397abc}
{Y_{1}^{-1}}(\theta , \varphi_2 )= \sqrt{\frac {3}{8\pi}} sin\theta e^{-i\varphi_2},
\end{eqnarray}
\begin{eqnarray} \label{eq398abc}
{Y_{1}^{0}}(\theta , \varphi )={Y_{1}^{0}}(\theta , \varphi_2 )= \sqrt{\frac {3}{4\pi}} cos\theta ,
\end{eqnarray}
\begin{eqnarray} \label{eq399abc}
{Y_{1}^{1}}(\theta , \varphi )= -\sqrt{\frac {3}{8\pi}} sin\theta e^{i\varphi}
\end{eqnarray}
and
\begin{eqnarray} \label{eq400abc}
{Y_{1}^{1}}(\theta , \varphi_2 )= -\sqrt{\frac {3}{8\pi}} sin\theta e^{i\varphi_2},
\end{eqnarray}
where the intrinsic orbital quantum numbers have the values $l_1=l_2=1 $, the intrinsic "magnetic" quantum numbers have the values $s_1 = -1, 0, 1$ and $s_2 = -1, 0, 1$, and the angle $ \varphi _2 =\varphi _2 (r, R, \varphi)$ denotes the azimuthal angle associated with the vector $\bf{r}-\bf{R}$ in the spherical coordinate system of the second {\it{AC}}. Notice, a simple analysis shows that Eqs. (\ref{eq390abc}) - (\ref{eq392abc}) derived by using the {\it{non-relativistic}} (Newton) approximation of the unit-field (particle) energy {\it{do not describe correctly the gravitational interactions}}. That means that the gravitationally interacting unit-fields (elementary particles) are {\it{relativistic objects}} describing rather by the Einstein particle energy than by the Newton particle energy. In order to exclude the singularities in the particle self-energy (\ref{eq228abc}) and the cross-correlation energy (\ref{eq393abc}) associated with the points $r=0$ and ${\bf{r}}-{\bf{R}}=0$, one should suppose that the gravitational {\it{ACs}}  ${\phi_{G1}}(\bf{r})$ and ${\phi_{G2}}(\bf{r}-\bf{R})$ are hollow around these points in the bolls of the gravitational radius $r_{0}^G$. The assumption yields the relativistic gravitational interaction (cross-correlation) term with the modified integration boundaries:
\begin{eqnarray} \label{eq401abc}
{\cal E}_{12,21}(R,s_1,s_2 ) = \mathbf{k}_{01} \mathbf{k}_{02}f_G +   m_{01} m_{02}f_G , 
\end{eqnarray}
where the gravitational geometrical factor 
\begin{eqnarray} \label{eq402abc}
f_G=f_G(R,s_1,s_2 ) 
\end{eqnarray}
is given by
\begin{eqnarray} \label{eq403abc}
f_G(R,s_1,s_2 ) =  {{{A_{1s_1}} {A_{1s_2}}} { a_0^2}} \int_{\theta =0}^\pi d\theta sin\theta  \int_{\varphi  =0}^{2\pi} d \varphi  \int_{r_{0}^G}^{R-r_{0,low}^G/ 2} \int_{R+r_{0,up}^G / 2}^{\infty} dr {\frac {f(r,R,\theta , \varphi , s_1, s_2 ) } {| {\bf{r}}-{\bf{R}}|^{2}}} .
\end{eqnarray}
The value $|{\bf{r}}-{\bf{R}}|$, which is calculated by using a simple geometrical analysis of the two vectors, is a function of the parameters $r$, $R$, $\theta$ and $\varphi$. Correspondingly, the low ($r_{0,low}^G$) and upper ($r_{0,up}^G$) integration limits are calculated by using this geometrical analysis as functions of the parameters $r$, $R$, $\theta$, $\varphi$ and $r_0^G$. The radius 
\begin{eqnarray} \label{eq404abc}
r = r_0^G
\end{eqnarray}
could be considered as the classical radius of a gravitationally interacting particle. Among the unit-fields (particles) with the intrinsic "magnetic" quantum numbers $s_n = -1, 0, 1$ only the unit-field (particles) with $s_n =0$ demonstrate the "{\it{spin-less}}" behavior [${\cal E}_{12,21}(R,s_1,s_2 ) = {\cal E}_{12,21}(R)$] that correspond to the {\it{non-quantum  (Newton and Einstein) gravitation}}. The infinite upper integration boundary in Eqs. (\ref{eq393abc}) and (\ref{eq403abc}) corresponding to the infinite {\it{ACs}} is in agreement with a fact that the Newton and Einstein interactions are described by the infinite gravitational fields. Although the interaction energy of unit-fields (particle) with the spins $s_n = -1,1$ obey the gravitation-like dependence $\sim 1/R$, {\it{the only unit-fields with the spins $s_n=0$ do satisfy the spin-independent interaction of the Newton and Einstein gravitations}}. In the case of the spins $s_1=s_2=0$, the relativistic gravitational interaction (cross-correlation) total term is given by Eqs. (\ref{eq401abc}) - (\ref{eq403abc}) as
\begin{eqnarray} \label{eq405abc}
{\cal E}_{12,21}(R) = {\frac {\mathbf{k}_{01}} {m_{01}} } {\frac {\mathbf{k}_{02}} {m_{02}} } \left[ 2\gamma _G{\frac {m_{01} m_{02}} {R} } \right]  +     \left[ 2\gamma _G{\frac {m_{01} m_{02}} {R} } \right]  
\end{eqnarray}
with the gravitational parameter 
\begin{eqnarray} \label{eq406abc}
\gamma _G= ({1}  / 2)  R {{{A_{1s_1}} {A_{1s_2}}} { a_0^2}} \int_{\theta =0}^\pi d\theta sin\theta  \int_{\varphi  =0}^{2\pi} d \varphi  \int_{r_{0}^G}^{R-r_{0,low}^G/ 2} \int_{R+r_{0,up}^G / 2}^{\infty} dr {\frac {f(r,R,\theta , \varphi , 0, 0 ) } {| {\bf{r}}-{\bf{R}}|^{2}}}.   
\end{eqnarray}
The strength of gravitational interaction is determined by the gravitational parameter $\gamma _G$, while the range of the gravitational interaction is attributed to the long-range geometrical factor $f_G(R)\sim 1/R$ of the unit-fields with the {\it{irregular}} {\it{Laplace-ACs}}. In order to be in agreement with the Newton and Einstein gravitations, the gravitational parameter (\ref{eq406abc}), which does not depend on the value $R$ at the distances $R>>r_0^G$, should be adjusted to the value of the {\it{Newton gravitational constant}}. Notice, the dependence $\gamma _G=\gamma _G(R, r_0^G)$ at the extremely small ($R \sim  r_0^G$) distances between the two unit-fields (particles) gives the new inside the physics of the non-quantum and quantum gravitations.  

The {\it{weak-relativistic}} (${\cal E}_{12,21}<<{\varepsilon}_{01}^{2} + {\varepsilon}_{02}^{2}$) interaction forces (\ref{eq172abc}) and (\ref{eq173abc}) acting upon the first and second unit-fields (particles) have respectively the forms
\begin{eqnarray} \label{eq407abc}
\mathbf{F}_{12}(\mathbf{R} ) = -{\frac 1  {2}}{\frac {\partial}  {\partial \mathbf{R}}} {\cal E}_{12,21}(R) = {\frac {\mathbf{k}_{01}} {m_{01}} }\times {\frac {\mathbf{k}_{02}} {m_{02}} }\times \left[ \gamma _G{\frac {m_{01} m_{02}} {R^2}  {\frac {\mathbf{R}} {R}} } \right] +  \left[ \gamma _G {\frac {m_{01} m_{02}} {R^2} {\frac {\mathbf{R}} {R} } } \right] 
\end{eqnarray} 
and  
\begin{eqnarray} \label{eq408abc}
\mathbf{F}_{21}(\mathbf{R} ) = {\frac 1  {2}}{\frac {\partial}  {\partial \mathbf{R}}} {\cal E}_{12,21}(R) = -{\frac {\mathbf{k}_{01}} {m_{01}} }\times {\frac {\mathbf{k}_{02}} {m_{02}} }\times \left[ \gamma _G{\frac {m_{01} m_{02}} {R^2} }{\frac {\mathbf{R}} {R} } \right] -  \left[ \gamma _G {\frac {m_{01} m_{02}} {R^2} } {\frac {\mathbf{R}} {R} } \right] ,
\end{eqnarray} 
The forces (\ref{eq407abc}) and (\ref{eq408abc}) acting between the "moving" (${\mathbf{k}_{01}}={\mathbf{k}_{02}}={\mathbf{k}_{0}}\neq 0$) unit-fields could be considered as the {\it{Lorentz-like gravitational forces}}. The present model, which modifies the Einstein general relativity, predicts the relativistic gravitational interaction (cross-correlation) energy 
\begin{eqnarray} \label{eq409abc}
{\cal E}^f_{12,21}(R) = {\frac {{k}_{01}} {m_{01}} } {\frac {{k}_{02}} {m_{02}} } \left[ 2\gamma _G{\frac {m_{01} m_{02}} {R} } \right]  
\end{eqnarray}
and the respective forces 
\begin{eqnarray} \label{eq410abc}
\mathbf{F}^f_{12}(\mathbf{R} ) = {\frac {\mathbf{k}_{01}} {m_{01}} }\times {\frac {\mathbf{k}_{02}} {m_{02}} }\times \left[ \gamma _G{\frac {m_{01} m_{02}} {R^2}  {\frac {\mathbf{R}} {R}} } \right] 
\end{eqnarray} 
\begin{eqnarray} \label{eq411abc}
\mathbf{F}^f_{21}(\mathbf{R} ) = -{\frac {\mathbf{k}_{01}} {m_{01}} }\times {\frac {\mathbf{k}_{02}} {m_{02}} }\times \left[ \gamma _G{\frac {m_{01} m_{02}} {R^2} }{\frac {\mathbf{R}} {R} } \right] ,
\end{eqnarray} 
that do not exist according to the modern field theories. That means that the present model, in addition to the {\it{Newton gravitational force and interaction-energy}} associated with the unit-fields at "rest" (${\mathbf{k}_{01}}={\mathbf{k}_{02}}=0$), predicts the {\it{Lorentz-like gravitational force and interaction-energy}} attributed to the "moving" (${\mathbf{k}_{01}}={\mathbf{k}_{02}}\neq 0$) unit-fields. The "new"  gravitational interaction (cross-correlation) energy ${\cal E}^f_{12,21}(R)$ and the "new" force $\mathbf{F}^f_{12}(\mathbf{R} )= - \mathbf{F}^f_{21}(\mathbf{R} )$ describing by Eqs.(\ref{eq409abc}) - (\ref{eq411abc}) logically to consider as the {\it{fifth}} fundamental energy and force, which {\it{could explain}} the recently-discovered accelerating expansion of  Universe, as well as the CP violations, the cosmological "dark" matter (energy-mass) and the "dark" flow. The gravitational interaction acting between the unit-fields at "rest" in the present model is given by the {\it{Newton gravitational interaction-energy}}:
\begin{eqnarray} \label{eq412abc}
{\cal E}^N_{12,21}(R) =  \left[ 2\gamma _G{\frac {m_{01} m_{02}} {R} } \right] .  
\end{eqnarray}
and the {\it{Newton gravitational forces}}
\begin{eqnarray} \label{eq413abc}
\mathbf{F}_{12}^N(\mathbf{R} ) =    \left[ \gamma _G {\frac {m_{01} m_{02}} {R^2} {\frac {\mathbf{R}} {R} } } \right] 
\end{eqnarray} 
\begin{eqnarray} \label{eq414abc}
\mathbf{F}_{21}^N(\mathbf{R} ) =  - \left[ \gamma _G {\frac {m_{01} m_{02}} {R^2} {\frac {\mathbf{R}} {R} } } \right]. 
\end{eqnarray} 
One can easily recognize the formal similarity between the above-presented gravitational model and the classical (non-quantum) electrostatics and magnetostatics. Equations (\ref{eq405abc}) - (\ref{eq414abc}) may be presented in the terms of the "electrostatic-like" and "magnetostatic-like" gravitational fields and "potentials" as
\begin{eqnarray} \label{eq415abc}
{\cal E}_{12,21}(R) = {\cal E}^f_{12,21}(R) + {\cal E}^N_{12,21}(R),
\end{eqnarray}
\begin{eqnarray} \label{eq416abc}
{\cal E}^f_{12,21}(R) = {k}_{01}U^f_2(R)  + {k}_{02}U^f_1(R)
\end{eqnarray}
\begin{eqnarray} \label{eq417abc}
\mathbf{F}^f_{12}(\mathbf{R} ) = \mathbf{k}_{01}\times {\bf{B}}^f_2(\bf{R}) 
\end{eqnarray} 
\begin{eqnarray} \label{eq418abc}
\mathbf{F}^f_{21}(\mathbf{R} ) = \mathbf{k}_{02}\times {\bf{B}}^f_1(\bf{R})
\end{eqnarray} 
\begin{eqnarray} \label{eq419abc}
{\cal E}^N_{12,21}(R) = {m}_{01}U^N_2(R)  + {m}_{02}U^N_1(R)
\end{eqnarray}
\begin{eqnarray} \label{eq420abc}
\mathbf{F}^N_{12}(\mathbf{R} ) =   m_{01} {\bf{E}}^N_2(\bf{R}) 
\end{eqnarray} 
\begin{eqnarray} \label{eq421abc}
\mathbf{F}^N_{21}(\mathbf{R} ) =  m_{02} {\bf{E}}^N_2(\bf{R}),
\end{eqnarray} 
where $U^f_2(R)={k}_{02} \gamma _G{\frac {1} {R}}$ and $U^f_1(R)={k}_{01} \gamma _G{\frac {1} {R}}$ are respectively the magnetic-like gravitational potentials {\it{mediating}} by the "{\it{moving masses}}" of the second and first unit-fields (particles); ${ \bf{B} }^f_2 ({\bf{R}}) = {\mathbf{k}_{02}}\times \gamma _G  {\frac {1} {R^2}  {\frac {\mathbf{R}} {R}} } $ and ${ \bf{B} }^f_1 ({\bf{R}}) = -{\mathbf{k}_{01}}\times \gamma _G  {\frac {1} {R^2}  {\frac {\mathbf{R}} {R}} } $ are respectively the magnetic-like gravitational fields inducing by the "moving masses" of these unit-fields (particles);  $U^N_2(R) = {m}_{02} \gamma _G{\frac {1} {R}}$ and $U^N_1(R) = {m}_{01} \gamma _G{\frac {1} {R}}$ are respectively the Newton (electric-like) gravitational potentials {\it{mediating}} by the "{{non-moving masses}}" of the second and first unit-fields (particles) at "rest"; ${ \bf{E} }^N_2 ({\bf{R}}) = {m}_{02} \gamma _G  {\frac {1} {R^2}  {\frac {\mathbf{R}} {R}} } $ and ${ \bf{E} }^N_1 ({\bf{R}}) = -{m}_{01} \gamma _G  {\frac {1} {R^2}  {\frac {\mathbf{R}} {R}} } $ are respectively the strengths of the Newton (electric-like) gravitational fields mediating by the "non-moving" masses of these unit-fields (particles). Thus, in the terms of electrostatics, the {\it{gravitational force}} acting between the unit-fields (particles) at rest could be interpreted as the Newton ("electrostatic") force inducing by the electric-like gravitational field of the "{{non-moving masses}}".  The {\it{fifth gravitational force (interaction energy)}} could be considered as the {\it{"magnetostatic" gravitational force (interaction energy)}} mediating by the magnetic-like gravitational field of the "{\it{moving masses}}". In the terms of electrostatics and magnetostatics, the combined action of the electric-like and magnetic-like gravitational fields could be interpreted as the Lorentz-like gravitational force mediating by the electric-like and magnetic-like gravitational fields. 

Equations (\ref{eq407abc}) - (\ref{eq421abc}) are valid for the {\it{weak relativistic}} (${\cal E}_{12,21} << {\varepsilon}_{01}^{2} + {\varepsilon}_{02}^{2}$) interaction, which is the case of the most of experimental conditions. The other two interaction conditions, namely the {\it{strong relativistic}} and {\it{non-relativistic}} gravitational interactions, are also possible. Under the {\it{strong relativistic}} (${\cal E}_{12,21} \sim {\varepsilon}_{01}^{2} + {\varepsilon}_{02}^{2}$) gravitational interaction, which corresponds to the extremely small distance $R\sim r_0$, one should use the above-derived equations with the relations (\ref{eq165abc}), (\ref{eq168abc}) and (\ref{eq169abc}) instead of the relations (\ref{eq172abc}) and (\ref{eq173abc}). The {\it{strong-relativistic}} gravitational interaction forces (\ref{eq168abc}) and (\ref{eq169abc}) acting upon the first and second unit-fields (particles) have respectively the forms
\begin{eqnarray} \label{eq422abc}
\mathbf{F}_{12}(\mathbf{R} ) = -{\frac 1  {2}}{\frac {\partial}  {\partial \mathbf{R}}} {\varepsilon }_{12,21}(R) 
\end{eqnarray} 
and  
\begin{eqnarray} \label{eq423abc}
\mathbf{F}_{21}(\mathbf{R} ) = {\frac 1  {2}}{\frac {\partial}  {\partial \mathbf{R}}} {\varepsilon }_{12,21}(R) ,
\end{eqnarray} 
where the gravitational interaction energy ${\varepsilon }_{12}(R)$ is given by Eq. (\ref{eq165abc}) as 
\begin{eqnarray} \label{eq424abc}
{\varepsilon}_{12,21}=({\varepsilon}_{01}^{2} + {\varepsilon}_{02}^{2} + {\cal E}_{12,21})^{1/2}-({\varepsilon}_{01} + {\varepsilon}_{02}),
\end{eqnarray} 
with the relativistic cross-correlation (interaction) total term ${\cal E}_{12,21}$ given by Eqs. (\ref{eq393abc}) - (\ref{eq404abc}). It should be stressed that Eqs. (\ref{eq393abc}) - (\ref{eq421abc}) are valid in the particular case of the de Broglie generators of the unit-fields having the same moments. Although the distance $R\sim r_0$ probably can be realized in some very particular experimental conditions, for instance inside a cosmological black hole, the unit-field generators of such unit-fields (particles) are not the de Broglie waves. In other words, the {\it{strong relativistic}} gravitational interaction of the real unit-fields (particles) should be described rather by the equations (\ref{eq422abc}) - (\ref{eq424abc}) than Eqs. (\ref{eq405abc}) - (\ref{eq421abc}).  

The gravitational interaction between the two unit-fields determining by Eqs. (\ref{eq393abc}) - (\ref{eq414abc}) may be interpreted not only by using the "electric-like" and "magnetic-like" gravitational fields and potentials [Eqs. (\ref{eq415abc}) - (\ref{eq421abc})], but also in the terms of the virtual particles. The relativistic energy of the composite field (particle) composed from the gravitationally  interacting unit-fields (particles) is given by Eq. (\ref{eq199abc}) as 
\begin{eqnarray} \label{eq425abc}
{\varepsilon}^{2} = {\varepsilon}_{01}^{2} + {\varepsilon}_{02}^{2} + {\cal E}_{12}+{\cal E}_{21} ={\varepsilon}_{01}^{2} + {\varepsilon}_{02}^{2} + {\cal E}_{12,21}. 
\end{eqnarray} 
The energies squared ${\varepsilon}_{01}^2$ and ${\varepsilon}_{02}^2$ could be attributed to the "moving" and "non-moving" {\it{masses}} of the 1st and 2nd relativistic {\it{normal}} unit-fields (elementary particles). The gravitational cross-correlation terms ${\cal E}_{12}={\cal E}_{12}(m_{01},m_{02})$ and ${\cal E}_{21}={\cal E}_{21}(m_{02},m_{01})$ could be attributed to the "moving" and "non-moving" {\it{masses}} of the 1st and 2nd relativistic {\it{normal}} unit-fields. The gravitational cross-correlation terms ${\cal E}_{12}$ and ${\cal E}_{21}$ associated with the masses and gravitational interactions may be considered as the 3rd and 4th {\it{virtual}} relativistic unit-fields (elementary particles) attributed to the relativistic gravitational interaction. Thus the relativistic gravitational interaction between the two {\it{normal}} relativistic unit-fields (elementary particles) could be considered (interpreted) as the interplay of the {\it{four}} unit-fields (particles), where the composite field (particle) is composed from the 1st and 2nd {\it{normal}} unit-fields (particles) and the 3rd and 4th {\it{virtual}} unit-fields (particles). In the case of the {\it{weak relativistic}} (${\cal E}_{12,21} << {\varepsilon}_{01}^{2} + {\varepsilon}_{02}^{2}$) interaction, the energy ${\varepsilon}_{12,21}={\varepsilon}_{12,21}({\cal E}_{12,21})$ of the 3rd and 4th {\it{virtual}} unit-fields (particles) is given by Eq. (\ref{eq171abc}). For more details of the virtual-particle interpretation see the comments to Eq. (\ref{eq199abc}). The {\it{virtual relativistic unit-fields (elementary particles)}}, which {\it{carry}} the gravitational force and energy, logically to call the {\it{virtual gravitons}}. Since the physical properties of the interaction energy attributed to the energy of the 3rd and 4th {\it{virtual}} relativistic unit-fields (elementary particles) are determined by the gauge (\ref{eq291abc}) with the eigen parameter ${\Gamma ^2 _{G}}={\Gamma _{G1}}{\Gamma _{G2}}=0$, which could be interpreted in the total (effective) mass $(m_{01}m_{02} + {\Gamma _{G1}}{\Gamma _{G2}})^{1/2}=(m_{01}m_{02} )^{1/2}$ as the {\it{gauge-mass squared}} of the gravitational AC, the virtual particles may be considered as the {\it{mass-less gauge bosons}} ({\it{virtual mass-less gauge gravitons}}). The name {\it{virtual graviton}} corresponds to the hypothetical carrier of the quantum gravitation in the modern literature devoted to the quantum gravitational fields and interactions. In the frame of the Heisenberg energy uncertainty relation and the perturbation approximation, the long range of the gravitational interaction may be formally (phenomenologically) attributed to absence of the rest mass of gravitons. However, the present model gives the microscopic explanation of the phenomenon. The long range of the gravitational interaction is attributed to the long-range geometrical factor $f_G(R)\sim 1/R$ (\ref{eq403abc}) of the unit-fields with the {\it{irregular}} {\it{Laplace-ACs}}, where the field strength of the gravitational interaction is determined by the gravitation constant $\gamma _G$ (\ref{eq406abc}). One should not confuse the {\it{non-virtual}}, mass-less, one-component gravitational unit-fields ({\it{non-virtual gravitons}}) describing by Eq. (\ref{eq218abc}) with the rest mass $m_0=0$ and the index $I=G$,  which has the form $\phi_{Gn}\square {\tilde \psi} _{0n}  + {\tilde \psi} _{0n}\square \phi_{Gn} = 0$. The {\it{non-virtual}} gravitational unit-fields (gravitons) that satisfy this equation have the {\it{structure-less}} form (\ref{eq53abc}) of the the time-harmonic plane waves $\psi_{0n} (\mathbf{r},t)= \tilde \psi_{0n} (\mathbf{r},t) { \phi} _{Gn} (\mathbf{r},t) = a_{0n} e^{i({\mathbf{k}_{0n}{\mathbf{r}}}-{\varepsilon}_{0n} t )}$ with the generator $\tilde \psi_{0n} (\mathbf{r},t)={\sqrt  {a_{0n}}} e^{i([{\mathbf{k}_{0n}/2]{\mathbf{r}}}-[{\varepsilon}_{0n}/2] t )}$ and associate-component $\phi_{Gn} (\mathbf{r},t)={\sqrt  {a_{0n}}} e^{i([{\mathbf{k}_{0n}/2]{\mathbf{r}}}-[{\varepsilon}_{0n}/2] t )}$ that are {\it{indistinguishable from each other}}. The {\it{non-virtual gravitons}} have the zero rest-mass ($m_0=0$) with the respective energy-mass relation ${\varepsilon}_{0n}^{2} =\mathbf{k}_{0n}^2$. In contrast to the massive unit-fields (particles), which obey the {\it{time-independent}} {\it{ACs}} ($\dot \phi_{Gn}=0$), the {\it{non-virtual}}, mass-less, one-component gravitational unit-fields ({\it{non-virtual gravitons}}) have the {\it{time-dependent}} {\it{ACs}} ($\dot \phi_{Gn}\neq 0$). In such a case the gauge (\ref{eq220abc}) provides the balance between the temporal and spatial variations of the non-virtual gravitational unit-field (non-virtual graviton). Notice, in the case of the unit-field $\psi_{0n} (\mathbf{r},t)= \tilde \psi_{0n}{ \Phi}^a _{n}$ with the {\it{TAC}} $\Phi^a_{n}=1$, the structure-less unified unit-field ${\psi _{0n}}$ attributed to the $n$-th unified elementary particle {\it{is indistinguishable from the unit-field generator ${\tilde \psi} _{0n}$}} (for comparison, see Eq. (\ref{eq210abc}) and the relevant examples in Sec. I of the present study).

For the sake of simplicity, I have presented the model [Eqs. (\ref{eq393abc}) - (\ref{eq425abc})] for the two ($N=2$) unit-fields (particles) with the moments $\mathbf{k}_{01}=\mathbf{k}_{02}=\mathbf{k}_{0}$. One can easily follow the model for an arbitrary number $N$ of the unit-fields (particles) having the de Broglie generators $\tilde \psi_{0n} (\mathbf{r},t)=a_0 e^{i({\mathbf{k}_{0n}{\mathbf{r}}}-{\varepsilon}_{0n} t )}$ with the moments $\mathbf{k}_{0n}=\pm \mathbf{k}_{0}$. The model with $N>>1$ describes the {\it{classical (non-quantum)}} gravitational fields and interactions that correspond to the case of Newton's and Einstein's gravitations ({\it{non-quantum gravitations}}). The model based on Eqs. (\ref{eq422abc}) - (\ref{eq424abc}) corresponds to the case of {\it{quantum gravitation}}. It should be stressed that among the unit-fields (particles) with the intrinsic "magnetic" quantum numbers $s_n = -1, 0, 1$ of the unit-field spin only the unit-fields (particles) with $s_n =0$ demonstrate the {\it{spin-less (non-quantum) behavior}} [${\cal E}_{12,21}(R,s_1,s_2 ) = {\cal E}_{12,21}(R)$] that satisfies the non-quantum  (Newton and Einstein) gravitation [Eqs. (\ref{eq405abc}) - (\ref{eq421abc})]. The {\it{spin-less (non-quantum) gravitational interaction}} of the unit-field (particles) with the gravitational {\it{ACs}} having $s_n =0$ could be considered as the {\it{pure-attractive}} interaction of boson-like unit-fields ({\it{bosons}}) that {\it{explains}} the {\it{Bose-Einstein condensation and statistics}} of such unit-fields (particles). The {\it{spin-dependent  (quantum) gravitational interaction}} between the two unit-fields is described by the interaction energy ${\cal E}_{12,21}(R,s_1,s_2 ) $ determining by Eqs. (\ref{eq393abc}) - (\ref{eq404abc}) with the intrinsic "magnetic" quantum numbers ($s_1=1$, $s_2=1$), ($s_1=-1$, $s_2=-1$), ($s_1=1$, $s_2=-1$) or ($s_1=-1$, $s_2=1$). If the unit-field (particles) with $s_n=1$ and $s_n=-1$ do exist somewhere in Universe, then Eqs. (\ref{eq172abc}) and (\ref{eq173abc}) and  Eqs. (\ref{eq168abc}) and (\ref{eq169abc}) should determine the {repulsive ($s_1=s_2$) and attractive ($s_1\neq s_2$) {\it{quantum-gravitational forces}} between the unit-fields (particles). The {\it{spin-dependent (quantum) gravitational interaction}} of the unit-field (particles) with the gravitational {\it{ACs}} having $s_n \neq0$ could be considered as the gravitational interaction of the fermion-like unit-fields ({\it{fermions}}). The {\it{spin-dependent (quantum) gravitational interaction}} that {\it{attract or repel}} the fermion-like unit-fields (particles) could be considered as a {\it{physical background}} of the {\it{Pauli exclusion principle and the Fermi-Dirac statistic}} of the gravitationally interacting unit-fields (particles). It should be also noted that in the case of the two {\it{normal}} unit-fields (elementary particles) having the same "magnetic" numbers $s_1=s_2$, the unit-fields (particles) can {\it{annihilate}}. Indeed, the unit-fields are the solutions of the equations of motions of the present model for both the $A_{1s_n}$ and $-A_{1s_n}$ amplitudes. The {\it{Laplace-AC}} amplitude ${A_{1_ns_n}}$ can be an arbitrary constant [see, Eq. (\ref{eq317abc})]. The two unit-fields with $s_1=s_2$ can satisfy the {\it{annihilation condition}} in the case ${A_{1s_1}} = - {A_{1s_2}}$. The unit-field (particle) with the positive value of  ${A_{1s_1}}$ could be considered as a {\it{normal unit-field (particle)}}, while the unit-field (particle) with the negative value of  ${A_{1s_2}}$ could be considered as an {\it{anti unit-field}} ({\it{anti-particle}}). 

The annihilation of the gravitationally interacting unit-fields with the spins $s_1=s_2=0$ does not require adjustments of the unit-field spins in contrast to the gravitationally interacting unit-fields with the spins $s_1=s_2=\pm 1$ . The more easy annihilation of the gravitationally interacting unit-fields with the spins $s_n=0$ in comparison to the unit-fields having the spins $s_1=s_2= \pm 1$ should result into domination of the non-zero spin unit-fields. The absence of the gravitationally interacting unit-fields with the spins $s_n\neq 0$, if they really have been created under the cosmological Big Bang, could be considered as a fact that such unit-fields (particles) have  been produced in the negligible amount in comparison to the gravitationally interacting unit-fields with the spins $s_1=s_2=0$. It should be noted again that Eqs. (\ref{eq393abc}) - (\ref{eq421abc}) are valid in the particular case of the {\it{de Broglie generators}} $\tilde \psi_{01} (\mathbf{r},t)=a_0 e^{i({\mathbf{k}_{01}{\mathbf{r}}}-{\varepsilon}_{01} t )}$ (\ref{eq384abc}) and $\tilde \psi_{02} (\mathbf{r},t)=a_0 e^{i({\mathbf{k}_{02}{\mathbf{r}}}-{\varepsilon}_{02} t )}$ (\ref{eq385abc}). If the unit-field generators are not the de Broglie time-harmonic plane-waves, then the pure gravitational interaction of the unit-fields (particles) are determined rather by the general Eqs. (\ref{eq364abc}) - (\ref{eq374abc}) than the particular Eqs. (\ref{eq393abc}) - (\ref{eq421abc}). One can easily follow the present model for any hypothetical unit-field (particle) with the spherically symmetric {\it{regular}} {\it{Laplace-AC}} (\ref{eq316abc}) if such a kind of elementary particles exists somewhere in the Universe. Notice, the unit-fields with the spherically symmetric {\it{regular}} {\it{Laplace-ACs}} (\ref{eq316abc}) should have the finite external dimensions.

\vspace{0.2cm}
2. {\it{The Coulomb and Lorentz interactions of the electrically charged unit-fields (particles): The non-quantum and quantum pure electromagnetisms}}
\vspace{0.2cm}

Among the unit-fields (particles) (\ref{eq356abc}) and (\ref{eq357abc}) having the de Broglie generators (\ref{eq384abc}) and (\ref{eq385abc}) and the spherically symmetric regular (\ref{eq316abc}) or irregular (\ref{eq317abc}) {\it{Laplace-ACs}} only the {\it{one-component unit-fields}} (\ref{eq356abc}) with the {\it{irregular Laplace-ACs}} and the intrinsic orbital quantum number $l_n=1$ and the intrinsic "magnetic" quantum numbers $s_n = -1, 0, 1$ may be considered as candidates for the pure electromagnetism describing by the Coulomb and Lorentz interactions of the electrically charged unit-fields (particles) with the interaction energy $\sim 1/R$. Among the candidates, the only unit-fields with the quantum numbers $l_n = 1$ and $s_n=\pm 1$ obey the physical properties of the quantum electromagnetism. Since the {\it{ACs}} of the electrically-charged unit-fields (particles) are indistinguishable from the {\it{ACs}} of the gravitationally interacting unit-fields (particles), the electrically and gravitationally interacting unit-fields are described by the similar interaction (cross-correlation) total terms. Although the descriptions are similar in many aspects, the physical nature of the electrical {\it{ACs}} associated with the electric charges of the unit-fields is different from the gravitational {\it{ACs}} associated with the unit-field masses. For the comparison, see the previous section devoted to the gravitation. In the case of the first {\it{irregular Laplace-AC}} ${\phi_{i1}}(\bf{r})={\phi_{C1}}(\bf{r})$ centered at the origin and the second {\it{irregular Laplace-AC}} ${\phi_{i2}}(\bf{r}-\bf{R})={\phi_{C2}}(\bf{r}-\bf{R})$ located at the distance $R=|\bf{R}|\equiv |\bf{R}_{12}|$ in the azimuthal direction, the relativistic electric (Coulomb) interaction (cross-correlation) total term (\ref{eq389abc}) has the form (\ref{eq393abc}):
\begin{eqnarray} \label{eq426abc}
{\cal E}_{12,21}(R,l_1,l_2,s_1,s_2) = \mathbf{k}_{01} \mathbf{k}_{02} \left[  {{ {{A_{1s_1}} {A_{1s_2}}} { a_0^2}} \int_{\theta =0}^\pi d\theta sin\theta  \int_{\varphi  =0}^{2\pi} d \varphi  \int_{0}}^{\infty}r^2dr {\frac {1} {r^2}}{\frac {f(r,R,\theta , \varphi , s_1, s_2 ) } { |  {\bf{r}}-{\bf{R}}|^{2}} }   \right]  +   \nonumber  \\  +   m_{01} m_{02}\left[  {{{{A_{1s_1}} {A_{1s_2}}} { a_0^2}} \int_{\theta =0}^\pi d\theta sin\theta  \int_{\varphi  =0}^{2\pi} d \varphi  \int_{0}}^{\infty}r^2dr {\frac {1} {{r}^2}}{\frac {f(r,R,\theta , \varphi ,s_1, s_2 ) } { |  {\bf{r}}-{\bf{R}}|^{2}} } \right] , 
\end{eqnarray}
where
\begin{eqnarray} \label{eq427abc}
{\Gamma ^2_{i1}}= {\Gamma ^2_{i2}}=0 ={\Gamma ^2_{C1}}= {\Gamma ^2_{C2}}=0 \nonumber  \\
l_1=l_2=1 .  
\end{eqnarray}
Here, the angle factor $f(r,R,\theta , \varphi , s_1, s_2 )$ associated with the spherical harmonics is given by 
\begin{eqnarray} \label{eq428abc}
f(r,R,\theta , \varphi , s_1, s_2 )= {Y_{l_1}^{s_1}}^* (\theta , \varphi) Y_{l_2}^{s_2} (\theta , \varphi _2 ) + {Y_{1_1}^{s_1}} (\theta , \varphi) {Y_{1_2}^{s_2}}^* (\theta , \varphi _2), 
\end{eqnarray}
where the intrinsic orbital quantum numbers have the values $l_1=l_2=1 $, the intrinsic "magnetic" quantum numbers have the values $s_1 = -1, 0, 1$ and $s_2 = -1, 0, 1$. The other parameters have the meanings of the previous section devoted to the gravitation. Notice, a simple analysis also shows that Eqs. (\ref{eq390abc}) - (\ref{eq392abc}) derived by using the {\it{non-relativistic}} (Newton) approximation of the unit-field (particle) energy {\it{do not describe correctly the interaction of electrically charged unit-fields (particles)}}. That means that the electrically charged unit-fields (elementary particles) are {\it{relativistic objects}} describing rather by the Einstein particle energy than by the Newton particle energy. In order to exclude the singularities in the particle self-interaction energy (\ref{eq228abc}) and the cross-interaction energy (\ref{eq426abc}) associated with the points $r=0$ and ${\bf{r}}-{\bf{R}}=0$, one should suppose that the electric (Coulomb) {\it{ACs}} ${\phi_{C1}}(\bf{r})$ and ${\phi_{C2}}(\bf{r}-\bf{R})$ are hollow around these points in the bolls of the Coulomb radius $r_0^C$, which could be different in the general case from the gravitational radius ($r_0^C \neq r_0^G$). The calibration of unit-fields by the cut-off of the AC fields around the singular points yields the relativistic electric (Coulomb) interaction (cross-correlation) term with the modified integration boundaries:
\begin{eqnarray} \label{eq429abc}
{\cal E}_{12,21}(R,s_1,s_2 ) = \mathbf{k}_{01} \mathbf{k}_{02}f_C +   m_{01} m_{02}f_C , 
\end{eqnarray}
where the electrical (Coulomb) geometrical factor 
\begin{eqnarray} \label{eq430abc}
f_C=f_C(R,s_1,s_2 ) 
\end{eqnarray}
is given by
\begin{eqnarray} \label{eq431abc}
f_C(R,s_1,s_2 ) =  {{{A_{1s_1}} {A_{1s_2}}} { a_0^2}} \int_{\theta =0}^\pi d\theta sin\theta  \int_{\varphi  =0}^{2\pi} d \varphi  \int_{r_{0}^C}^{R-r_{0,low}^C/ 2} \int_{R+r_{0,up}^C / 2}^{\infty} dr {\frac {f(r,R,\theta , \varphi , s_1, s_2 ) } {| {\bf{r}}-{\bf{R}}|^{2}}} .
\end{eqnarray}
A simple analysis shows that the radius 
\begin{eqnarray} \label{eq432abc}
r = r_0^C
\end{eqnarray}
could be considered as the classical radius of an electrically interacting particle. The introduction of the integration boundary $r_0^C$ of an electrically interacting unit-field (particle) into the present model {\it{explains in the microscopical details}} why the canonical quantum mechanics, quantum field theories and SM of the dimension-less point-particles do require {\it{different renormalization procedures}} to avoid the field singularities and divergences. In fact, the {\it{renormalization procedure}} is a "hidden" introduction of the dimension $r_0^C$ of the dimension-less  point-particle. Among the unit-fields (particles) with the intrinsic magnetic quantum numbers $s_n = -1, 0, 1$ only the unit-field (particles) with $s_n =0$ demonstrate the "{\it{spin-less}}" behavior [${\cal E}_{12,21}(R,s_1,s_2 ) = {\cal E}_{12,21}(R)$] that satisfy the non-quantum  (Coulomb and Lorentz) interactions. The infinite upper integration limit in Eqs. (\ref{eq426abc}) and (\ref{eq431abc}) corresponding to the infinite {\it{ACs}} of the unit-fields is in agreement with the infinite fields of the Coulomb and Lorentz interactions. If the spin numbers $s_1=s_2=0$, then the relativistic electric  interaction (cross-correlation) total term is given by Eqs. (\ref{eq426abc}) - (\ref{eq431abc}) as
\begin{eqnarray} \label{eq433abc}
{\cal E}_{12,21}(R) = {\frac {\mathbf{k}_{01}} {m_{01}} } {\frac {\mathbf{k}_{02}} {m_{02}} } \left[ 2\gamma _C{\frac {q_{01} q_{02}} {R} } \right]  +     \left[ 2\gamma _C{\frac {q_{01} q_{02}} {R} } \right] ,  
\end{eqnarray}
where the strength of interaction is determined by the interaction parameter 
\begin{eqnarray} \label{eq434abc}
\gamma _C = ({1}  / 2)  R {{  [{A_{1s_1}}(q_{01}/m_{01})]  [{A_{1s_2}}(q_{02}/m_{02}) }] { a_0^2}} \int_{\theta =0}^\pi d\theta sin\theta  \int_{\varphi  =0}^{2\pi} d \varphi  \int_{r_{0}^G}^{R-r_{0,low}^G/ 2} \int_{R+r_{0,up}^G / 2}^{\infty} dr {\frac {f(r,R,\theta , \varphi , 0, 0 ) } {| {\bf{r}}-{\bf{R}}|^{2}}},
\end{eqnarray}
while the long range of the interaction is attributed to the long-range geometrical factor $f_C(R)\sim 1/R$ of the unit-fields with the {\it{irregular}} Laplace-ACs. Notice, the {\it{Laplace-AC}} amplitude ${A_{1_ns_n}}$ {\it{can be an arbitrary constant}} [see, Eq. (\ref{eq317abc})]. In Eq. (\ref{eq434abc}), the amplitude ${A_{1_ns_n}}$ is presented in the form ${A_{1_ns_n}}(q_{0n}/m_{0n})$ that corresponds to the electric (electromagnetic) interaction mediating by the electrical (Coulomb) charges. To satisfy the Coulomb and Lorentz interactions, the electrical parameter (\ref{eq434abc}) that does not depend on the value $R$ at the distances $R>>r_0^C$, should be adjusted to the {\it{Coulomb constant}}. The value $q_{01}=\pm |q_{01}|$ and $q_{02}=\pm |q_{02}|$ could be interpreted as the {\it{electric charges}} of the first and second unit-fields (particles). Unlike in the traditional quantum field theories, the {\it{Laplace-AC}} amplitude associated with the {\it{electric charge}} of the electromagnetically interacting unit-field (particle) gives the {\it{physical (non-phenomenological) explanation}} of the {\it{electric charge}}. Notice, the fact that the electrical interaction parameter $\gamma _C=\gamma _C(R, r_0^G)$ depends on the length scale at the extremely small ($R \sim  r_0^C$) distances between the two unit-fields (particles) gives the new inside the physics of the non-quantum and quantum electromagnetism.

The {\it{weak-relativistic}} (${\cal E}_{12,21}<<{\varepsilon}_{01}^{2} + {\varepsilon}_{02}^{2}$) interaction forces (\ref{eq172abc}) and (\ref{eq173abc}) acting upon the first and second {\it{electrically charged}} unit-fields (particles) have respectively the forms
\begin{eqnarray} \label{eq435abc}
\mathbf{F}_{12}(\mathbf{R} ) = -{\frac 1  {2}}{\frac {\partial}  {\partial \mathbf{R}}} {\cal E}_{12,21}(R) = {\frac {\mathbf{k}_{01}} {m_{01}} }\times {\frac {\mathbf{k}_{02}} {m_{02}} }\times \left[ \gamma _C{\frac {q_{01} q_{02}} {R^2}  {\frac {\mathbf{R}} {R}} } \right] +  \left[ \gamma _C {\frac {q_{01} q_{02}} {R^2} {\frac {\mathbf{R}} {R} } } \right] 
\end{eqnarray} 
and  
\begin{eqnarray} \label{eq436abc}
\mathbf{F}_{21}(\mathbf{R} ) = {\frac 1  {2}}{\frac {\partial}  {\partial \mathbf{R}}} {\cal E}_{12,21}(R) = -{\frac {\mathbf{k}_{01}} {m_{01}} }\times {\frac {\mathbf{k}_{02}} {m_{02}} }\times \left[ \gamma _C{\frac {q_{01} q_{02}} {R^2} }{\frac {\mathbf{R}} {R} } \right] -  \left[ \gamma _C {\frac {q_{01} q_{02}} {R^2} } {\frac {\mathbf{R}} {R} } \right] ,
\end{eqnarray} 
The forces (\ref{eq435abc}) and (\ref{eq436abc}) acting between the "moving" (${\mathbf{k}_{01}}={\mathbf{k}_{02}}={\mathbf{k}_{0}}\neq 0$) unit-fields could be considered as the {\it{Lorentz forces}}. Naturally, Eqs (\ref{eq433abc}) - (\ref{eq436abc}) may be interpreted using the conventional terms 
\begin{eqnarray} \label{eq437abc}
{\cal E}_{12,21}(R) = {\cal E}^M_{12,21}(R) + {\cal E}^C_{12,21}(R),
\end{eqnarray}
\begin{eqnarray} \label{eq438abc}
{\cal E}^M_{12,21}(R) = q_{01}{\frac {{k}_{01}} {m_{01}}}U^M_2(R)  + q_{02}{\frac {{k}_{02}} {m_{02}}}U^M_1(R)
\end{eqnarray}
\begin{eqnarray} \label{eq439abc}
\mathbf{F}^M_{12}(\mathbf{R} ) =   q_{01} {\frac {\mathbf{k}_{01}} {m_{01}}} \times{\bf{B}}^M_2(\bf{R}) 
\end{eqnarray} 
\begin{eqnarray} \label{eq440abc}
\mathbf{F}^M_{21}(\mathbf{R} ) = q_{02} {\frac {\mathbf{k}_{02}} {m_{02}}} \times {\bf{B}}^M_1(\bf{R})
\end{eqnarray} 
\begin{eqnarray} \label{eq441abc}
{\cal E}^C_{12,21}(R) = {q}_{01}U^C_2(R)  + {q}_{02}U^C_1(R)
\end{eqnarray}
\begin{eqnarray} \label{eq442abc}
\mathbf{F}^C_{12}(\mathbf{R} ) =   q_{01} {\bf{E}}^C_2(\bf{R}) 
\end{eqnarray} 
\begin{eqnarray} \label{eq443abc}
\mathbf{F}^C_{21}(\mathbf{R} ) =  q_{02} {\bf{E}}^C_2(\bf{R}),
\end{eqnarray} 
of the electrostatics and magnetostatics, where $U^M_2(R)=q_{02}{\frac {{k}_{02}} {m_{02}}} \gamma _C{\frac {1} {R}}$ and $U^M_1(R)= q_{01}{\frac {{k}_{01}} {m_{01}}} \gamma _C{\frac {1} {R}}$ denote respectively the magnetic potentials {\it{mediating}} by the "{\it{moving charges}}" of the second and first unit-fields (particles); ${ \bf{B} }^M_2 ({\bf{R}}) = q_{02}{\frac {\mathbf{k}_{02}} {m_{02}}} \times\gamma _C  {\frac {1} {R^2}  {\frac {\mathbf{R}} {R}} } $ and ${ \bf{B} }^M_1 ({\bf{R}}) = -q_{01}{\frac {\mathbf{k}_{01}} {m_{01}}} \times \gamma _C  {\frac {1} {R^2}  {\frac {\mathbf{R}} {R}} } $ are respectively the magnetic fields inducing by the "moving charges" of these unit-fields;  $U^C_2(R) = {q}_{02} \gamma _C{\frac {1} {R}}$ and $U^C_1(R) = {q}_{01} \gamma _C{\frac {1} {R}}$ denote respectively the Coulomb potentials {\it{mediating}} by the "{{non-moving charges}}" of the second and first unit-fields (particles) at "rest"; ${ \bf{E} }^C_2 ({\bf{R}}) = {q}_{02} \gamma _C {\frac {1} {R^2}  {\frac {\mathbf{R}} {R}} } $ and ${ \bf{E} }^C_1 ({\bf{R}}) =- {q}_{01} \gamma _C  {\frac {1} {R^2}  {\frac {\mathbf{R}} {R}} } $ are respectively the strengths of the Coulomb fields mediating by the "non-moving" charges of these unit-fields. Thus, in the terms of electrostatics, the electrostatic force acting between the charged unit-fields (particles) at rest could be interpreted as the Coulomb force inducing by the electric field of the "{{non-moving charges}}". The magnetostatic force (interaction energy) could be considered as the magnetostatic force (interaction energy) mediating by the magnetic field of the "{\it{moving charges}}". The combined action of the electric and magnetic fields could be interpreted as the Lorentz force mediating by the electric and magnetic fields. One could mention that the above-presented "elementary"  magnetic and electric fields do satisfy the well-known electrodynamic relation ${\bf B}={\bf v}\times {\bf E}=m^{-1}{\bf k}\times {\bf E}$ for the macroscopic magnetic and electric fields.

Equations (\ref{eq433abc}) - (\ref{eq443abc}) describe the {\it{weak relativistic}} (${\cal E}_{12,21} << {\varepsilon}_{01}^{2} + {\varepsilon}_{02}^{2}$) interactions between {\it{charged unit-fields}}. The other two interaction conditions, namely the {\it{strong relativistic}} and {\it{non-relativistic}} interactions between {\it{electrically charged unit-fields}}, are also possible. Under the {\it{strong relativistic}} (${\cal E}_{12,21} \sim {\varepsilon}_{01}^{2} + {\varepsilon}_{02}^{2}$) interaction, which corresponds the extremely small distance $R\sim r_0^C$, one should use the above-derived equations with the relations (\ref{eq165abc}), (\ref{eq168abc}) and (\ref{eq169abc}) instead of the relations (\ref{eq172abc}) and (\ref{eq173abc}). The {\it{strong-relativistic}} interaction forces (\ref{eq168abc}) and (\ref{eq169abc}) acting upon the first and second unit-fields (particles) have respectively the forms
\begin{eqnarray} \label{eq444abc}
\mathbf{F}_{12}(\mathbf{R} ) = -{\frac 1  {2}}{\frac {\partial}  {\partial \mathbf{R}}} {\varepsilon }_{12,21}(R) 
\end{eqnarray} 
and  
\begin{eqnarray} \label{eq445abc}
\mathbf{F}_{21}(\mathbf{R} ) = {\frac 1  {2}}{\frac {\partial}  {\partial \mathbf{R}}} {\varepsilon }_{12,21}(R) ,
\end{eqnarray} 
where the interaction energy ${\varepsilon }_{12}(R)$ is given by Eq. (\ref{eq165abc}) as 
\begin{eqnarray} \label{eq446abc}
{\varepsilon}_{12,21}=({\varepsilon}_{01}^{2} + {\varepsilon}_{02}^{2} + {\cal E}_{12,21})^{1/2}-({\varepsilon}_{01} + {\varepsilon}_{02}),
\end{eqnarray} 
with the relativistic cross-correlation (interaction) total term ${\cal E}_{12,21}$ given by Eqs. (\ref{eq419abc}) - (\ref{eq425abc}). It should be stressed that Eqs. (\ref{eq433abc}) - (\ref{eq443abc}) are valid in the particular case of the de Broglie electromagnetic generators ${\tilde \psi _{0n}}$ of the unit-fields having the same moments. Although the distance $R\sim r_0$ can be easily realized experimentally, the unit-field generators of such unit-fields (particles) are not the de Broglie waves because of the strong interaction. That is to say that the {\it{strong relativistic}} interaction of the charged unit-fields (particles) should be described rather by the equations (\ref{eq435abc}) - (\ref{eq437abc}) than Eqs. (\ref{eq433abc}) - (\ref{eq443abc}). 

The Coulomb and Lorentz interaction between the two electrically-charged unit-fields determining by Eqs. (\ref{eq426abc}) - (\ref{eq446abc}) may be interpreted not only by using the electric and magnetic fields and potentials [Eqs. (\ref{eq437abc}) - (\ref{eq443abc})], but also in the terms of the virtual particles. The relativistic energy of the composite charged-field (particle) composed from the electromagnetically interacting unit-fields (particles) is given by Eq. (\ref{eq199abc}) as 
\begin{eqnarray} \label{eq447abc}
{\varepsilon}^{2} = {\varepsilon}_{01}^{2} + {\varepsilon}_{02}^{2} + {\cal E}_{12}+{\cal E}_{21} ={\varepsilon}_{01}^{2} + {\varepsilon}_{02}^{2} + {\cal E}_{12,21}. 
\end{eqnarray} 
The energies squared ${\varepsilon}_{01}^2$ and ${\varepsilon}_{02}^2$, which do not depend on the electric charges of the 1st and the 2nd unit-fields (elementary particles), could be attributed to the "moving" and "non-moving" {\it{masses}} of the 1st and 2nd relativistic {\it{normal}} unit-fields (particles). The electrical cross-correlation terms ${\cal E}_{12}={\cal E}_{12}(q_{01},q_{02})$ and ${\cal E}_{21}={\cal E}_{21}(q_{02},q_{01})$ could be attributed to the "moving" and "non-moving" {\it{electric charges}} of the 1st and 2nd relativistic {\it{normal}} unit-fields. The electrical cross-correlation terms ${\cal E}_{12}$ and ${\cal E}_{21}$ associated with the electric charges and electrical interactions could be considered as the 3rd and 4th {\it{virtual}} relativistic unit-fields (elementary particles) attributed to the relativistic electrical interaction of the 1st and 2nd {\it{normal}} electrically charged unit-fields. Thus the relativistic electromagnetic interaction between the two {\it{normal}} electrically charged unit-fields (elementary particles) could be considered (interpreted) as the interplay of the {\it{four}} unit-fields (particles), where the composite field (particle) is composed from the 1st and 2nd {\it{normal}} charged unit-fields (particles) and the 3rd and 4th {\it{virtual}} unit-fields (particles). For more details of the interpretation based on the use of virtual particles see the comments to Eq. (\ref{eq199abc}). The {\it{virtual relativistic unit-fields (particles)}}, which {\it{carry}} the electromagnetic (Coulomb and Lorentz) forces and energy, logically to call the {\it{virtual photons}}. Since the physical properties of the interaction energy attributed to the energy of the 3rd and 4th {\it{virtual}} relativistic unit-fields (particles) are determined by the gauge (\ref{eq291abc}) with the eigen parameter ${\Gamma ^2 _{C}}={\Gamma _{C1}}{\Gamma _{C2}}=0$, which is interpreted in the total mass $(m_{01}m_{02} + {\Gamma _{C1}}{\Gamma _{C2}})^{1/2}=(m_{01}m_{02} )^{1/2}$ as the {\it{gauge-mass squared}} of the electrical AC, the virtual particles may be considered as the {\it{mass-less gauge bosons}} ({\it{virtual mass-less gauge photons}}). In the frame of the Heisenberg energy uncertainty relation and the perturbation approximation, the quantum field theories and SM have phenomenologically attributed the long range of the electromagnetic (Lorentz) interaction to absence of the rest mass of gravitons. The present model gives the microscopic explanation of the phenomenon. The long range of the electromagnetic interaction is attributed to the long-range geometrical factor $f_C(R)\sim 1/R$ (\ref{eq431abc}) of the unit-fields with the {\it{irregular}} {\it{Laplace-ACs}}, where the field strength of the electromagnetic interaction is determined by the electrical (Coulomb) constant $\gamma _C$ (\ref{eq434abc}). Here, one should not confuse the {\it{non-virtual}}, mass-less, one-component electromagnetic unit-fields (non-virtual photons) describing by Eq. (\ref{eq218abc}) with the rest mass $m_0=0$ and the index $I=C$,  which has the form $\phi_{Cn}\square {\tilde \psi} _{0n}  + {\tilde \psi} _{0n}\square \phi_{Cn} = 0$. The {\it{non-virtual}} electromagnetic unit-fields ({\it{non-virtual photons}}) that satisfy this equation have the {\it{structure-less}} form (\ref{eq53abc}) of the the time-harmonic plane waves $\psi_{0n} (\mathbf{r},t)= \tilde \psi_{0n} (\mathbf{r},t) { \phi} _{Cn} (\mathbf{r},t) = a_{0n} e^{i({\mathbf{k}_{0n}{\mathbf{r}}}-{\varepsilon}_{0n} t )}$ with the generator $\tilde \psi_{0n} (\mathbf{r},t)={\sqrt  {a_{0n}}} e^{i([{\mathbf{k}_{0n}/2]{\mathbf{r}}}-[{\varepsilon}_{0n}/2] t )}$ and associate-component $\phi_{Cn} (\mathbf{r},t)={\sqrt  {a_{0n}}} e^{i({[\mathbf{k}_{0n}/2]{\mathbf{r}}}-[{\varepsilon}_{0n}/2] t )}$ that are {\it{indistinguishable from each other}}. The {\it{non-virtual photons}} have the zero rest-mass ($m_0=0$) with the respective energy-mass relation ${\varepsilon}_{0n}^{2} =\mathbf{k}_{0n}^2$. In contrast to the massive unit-fields (particles), which obey the {\it{time-independent}} {\it{ACs}} ($\dot \phi_{Cn}=0$), the {\it{non-virtual}}, mass-less, one-component electromagnetic unit-fields ({\it{non-virtual photons}}) have the {\it{time-dependent}} {\it{ACs}} ($\dot \phi_{Cn} \neq 0$). In such a case the gauge (\ref{eq220abc}) provides the balance between the temporal and spatial variations of the non-virtual electromagnetic unit-field (non-virtual photon). Notice, in the case of the unit-field $\psi_{0n} (\mathbf{r},t)= \tilde \psi_{0n}{ \Phi}^a _{n}$ with the {\it{TAC}} $\Phi^a_{n}=1$, the structure-less unified unit-field ${\psi _{0n}}$ attributed to the $n$-th unified elementary particle {\it{is indistinguishable from the unit-field generator ${\tilde \psi} _{0n}$}} (for comparison, see Eq. (\ref{eq210abc}) and the relevant examples in Sec. I of the present study)
 
The model [Eqs. (\ref{eq426abc}) - (\ref{eq447abc})] describes interaction of the two ($N=2$) electrically-charged unit-fields (particles) with the moments $\mathbf{k}_{01}=\mathbf{k}_{02}=\mathbf{k}_{0}$. One can easily follow the model for an arbitrary number $N$ of the electrically-charged unit-fields (particles) having the de Broglie generators $\tilde \psi_{0n} (\mathbf{r},t)=a_0 e^{i({\mathbf{k}_{0n}{\mathbf{r}}}-{\varepsilon}_{0n} t )}$ with the moments $\mathbf{k}_{0n}=\pm \mathbf{k}_{0}$. The model [Eqs. (\ref{eq433abc}) - (\ref{eq443abc})] with $N>>1$ and $s_n=0$ describes the {\it{classical (non-quantum)}} electrostatic and magnetostatic fields and interactions that correspond to the case of the Lorentz ({\it{non-quantum}}) interaction. In other words, such a model corresponds to the {\it{non-quantum}} interactions of the classical electrostatics and magnetostatics of the electrically-charged unit-fields (particles). Among the electrically-charged unit-fields (particles) with the intrinsic magnetic quantum numbers $s_n = -1, 0, 1$ of the unit-field spin, {\it{the only  unit-field (particles) with the spin numbers $s_n \equiv m_n =0$ demonstrate the "spin-less" [${\cal E}_{12,21}(R,s_1,s_2 ) = {\cal E}_{12,21}(R)$] Lorentz classical behavior}} that satisfy the attractive ($q_{01}=-q_{02}$) and repulsive ($q_{01}=q_{02}$) electrostatic interactions $\mathbf{F}^C_{12}$ and the attractive ($q_{01}=q_{02}$, ${\mathbf{k}_{01}}={\mathbf{k}_{02}}$ or $q_{01}=-q_{02}$, ${\mathbf{k}_{01}}=-{\mathbf{k}_{02}}$ ) and repulsive ($q_{01}=q_{02}$, ${\mathbf{k}_{01}}= - {\mathbf{k}_{02}}$ or $q_{01}=- q_{02}$, ${\mathbf{k}_{01}}=  {\mathbf{k}_{02}}$) magnetostatic interactions $\mathbf{F}^M_{12}$ of the electrically-charged unit-fields (particles). The {\it{purely attractive}} spin-less (Lorentz-like) "non-quantum" interaction $\mathbf{F}^C_{12}+\mathbf{F}^M_{12}$ of the unit-field (particles) with the electric {\it{ACs}} having the spin numbers $s_n =0$, electric charges $q_{01}=-q_{02}$ and momentums ${\mathbf{k}_{01}}=-{\mathbf{k}_{02}}$ could be considered as the "{\it{non-quantum}}" classical attraction of charged unit-fields [{\it{boson-like}} ($q_{01}=-q_{02}$, ${\mathbf{k}_{01}}=-{\mathbf{k}_{02}}$) particles] that explains microscopically the Bose-Einstein condensation and statistics of such particles. Although the "spin-less" ($s_n=0$) electrically charged unit-fields  (particles) have not been yet detected, the model  [Eqs. (\ref{eq433abc}) - (\ref{eq443abc})] is good for the classical (Coulomb and Lorentz) description of the classical Coulomb and Lorentz forces. {\it{The spin-dependent quantum interaction between the two charged unit-fields, which is observed experimentally, is described by the Lorentz-like spin-dependent quantum interaction energy ${\cal E}_{12,21}(R,s_1,s_2 )$ determining by Eqs. (\ref{eq172abc}), (\ref{eq173abc}), (\ref{eq426abc}) - (\ref{eq432abc}) with the spin numbers ($s_1=1$, $s_2=1$), ($s_1=-1$, $s_2=-1$), ($s_1=1$, $s_2=-1$) or ($s_1=-1$, $s_2=1$)}}. That is to say that Eqs. (\ref{eq172abc}), (\ref{eq173abc}), (\ref{eq426abc}) - (\ref{eq432abc}) determine the electromagnetic properties of the electrically charged unit-field (particle) and their connections with the unit-field spin. A simple analysis of Eqs. (\ref{eq172abc}), (\ref{eq173abc}), (\ref{eq426abc}) - (\ref{eq432abc}) for the two {\it{identically charged}} unit-fields shows that the equations describe the {{\it{repulsive}} ($s_1=s_2, q_{01}=q_{02},{\mathbf{k}_{01}}=  {\mathbf{k}_{02}}$) and {\it{attractive}} ($s_1\neq s_2, q_{01}=q_{02},{\mathbf{k}_{01}}=  {\mathbf{k}_{02}}$) {\it{quantum}} forces associated with the {\it{Lorentz-like spin-dependent (quantum) interaction energy}} of the electrically charged fermion-like unit-fields (fermions). The Pauli exclusion principle states that no two identical ($s_1=s_2, q_{01}=q_{02},{\mathbf{k}_{01}}=  {\mathbf{k}_{02}}$) fermions may occupy the same quantum state simultaneously. The {\it{spin-dependent (quantum) repulsive or attractive interaction of the identically charged ($q_{01}=q_{02}$) unit-fields (fermions) having respectively the same ($s_1=s_2$) or different ($s_1 \neq s_2$) spin numbers could be considered as the physical origin  explaining microscopically the Pauli exclusion principle and the Fermi-Dirac statistics of the fermions, which in canonical quantum mechanics and SM have nature of the unexplained postulates}}. Unlike in the traditional quantum field theory and SM, which consider rather the fields of operators than the fields of particles, Eqs. (\ref{eq172abc}), (\ref{eq173abc}), (\ref{eq426abc}) - (\ref{eq432abc}) determine the electromagnetic properties of the electrically charged unit-field (particle) and their connections with the unit-field spin naturally, without introduction of any pure mathematical object, like the spin matrix  or operators. Moreover, Eqs. (\ref{eq172abc}), (\ref{eq173abc}), (\ref{eq426abc}) - (\ref{eq432abc}) {\it{explain, probably for the first time, why the spin $s_n=\pm 1/2$ of the electrically-charged electron (fermion) in the traditional quantum field theory and SM is artificially increased in two times up to the value $s_n=2(\pm 1/2)=\pm 1$ in order to correspond the experimentally observed value}}  
\begin{eqnarray} \label{eq448abc}
\mu _n = \mu _B 2 (\pm 1/2) = \pm \mu _B
\end{eqnarray} 
of the electron (fermion) magnetic moment $\mu _n \equiv \mu _B s_n $, where $\mu _B=q_{0n}/2m_0$ denotes the Bohr magneton. The present model gives the correct magnetic moment of the electrically-charged unit-field (electron) having the spin $s_n=\pm 1$ {\it{without any artificial adjustment}}: 
\begin{eqnarray} \label{eq449abc}
\mu _n = \mu _B (\pm 1) = \pm \mu _B.
\end{eqnarray} 
That also means the present model obeys the same (unified) g-factor $g=g_S=g_L=1$ for both the electron spin g-factor $g_S$ and the electron orbital g-factor $g_L$, which in in the traditional quantum field theory and SM have the different values $g_S=2$ and $g_L=1$. Notice, the g-factor $g=g_S=g_L=1$ corresponds to an electron placed into the absolute vacuum ("straight" empty space). In the case of the unit-field (electron) interacting (interfering) with the physical (non-absolute) vacuum associated with other spatially infinite or finite unit-fields, the value $g$ is slightly larger. The g-factor approaches the value corresponding to the anomalous magnetic moment of an electron in the traditional quantum electrodynamics. It should be also noted that in the case of the two electrically charged {\it{normal}} unit-fields (particles) having the spin numbers $s_1=s_2=\pm 1$ and charges $q_{01}=-q_{02}$, the unit-fields (particles) can annihilate. Indeed, the charged unit-fields are the solutions of the equations of motions of the present model for both the $A_{1s_n}$ and $-A_{1s_n}$ amplitudes. The {\it{Laplace-AC}} amplitude ${A_{1_ns_n}}$ can be an arbitrary constant [see, Eq. (\ref{eq317abc})]. The two charged unit-fields with $s_1=s_2$ can satisfy the annihilation condition in the case ${A_{1s_1}} = - {A_{1s_2}}$.  In Eqs. (\ref{eq426abc}) - (\ref{eq447abc}), the charged unit-field (particle) with the positive value of  ${A_{1s_1}}$ could be considered as a {\it{normal}} unit-field (particle), while the charged unit-field (particle) with the negative value of  ${A_{1m_2}}$ could be considered as an {\it{anti-unitfield}} ({\it{anti-particle}}). The annihilation of the two charged unit-fields with $s_1=s_2=0$ does not require adjustments of the unit-field spins. The easy annihilation of the charged unit-fields with $s_n=0$, if they really have been created under the cosmological Big Bang, could be considered as a physical  mechanism of the absence of such unit-fields (particles) in the present Universe. It should be noted again that Eqs. (\ref{eq426abc}) - (\ref{eq447abc}) are valid in the particular case of the {\it{de Broglie generators}} $\tilde \psi_{01} (\mathbf{r},t)=a_0 e^{i({\mathbf{k}_{01}{\mathbf{r}}}-{\varepsilon}_{01} t )}$ (\ref{eq384abc}) and $\tilde \psi_{02} (\mathbf{r},t)=a_0 e^{i({\mathbf{k}_{02}{\mathbf{r}}}-{\varepsilon}_{02} t )}$ (\ref{eq385abc}) . If the unit-field generators are not the de Broglie plane-waves, then the pure electromagnetic interaction of the unit-fields (particles) are determined rather by Eqs. (\ref{eq364abc}) - (\ref{eq374abc}) than Eqs. (\ref{eq426abc}) - (\ref{eq443abc}). The above-present model may be easily reformulated for any hypothetical unit-field (particle) with the spherically symmetric {\it{regular}} {\it{Laplace-AC}} (\ref{eq316abc}) if such a kind of elementary particles exists somewhere in the Universe. In such a case, the unit-fields with the spherically symmetric {\it{regular}} {\it{Laplace-ACs}} (\ref{eq316abc}) should have the finite external dimensions.

\subsection{7.3. The interference (interaction) of the unit-fields (particles) having the {\it{de Broglie generators}} $({\tilde \psi} _{0n})$ and the {\it{spherically symmetric, one-component Helmholtz-ACs}} (${\Phi^{a}_{n}}$): The non-quantum and quantum pure-{\it{week}} and pure-{\it{strong}} interactions}

\vspace{0.2cm}
1. {\it{The non-quantum and quantum pure-weak interactions of the unit-fields (particles)}}
\vspace{0.2cm}

For the sake of simplicity, I present here the equations for the {\it{two}} unit-fields, namely for the unit-fields $\psi _{01}={\tilde \psi _{01}}\Phi^{a}_{1}$ (\ref{eq356abc}) and $\psi _{02}={\tilde \psi _{02}}\Phi^{a}_{2}$ (\ref{eq357abc}) of the composite field ${\psi }={\tilde \psi _{01}}\Phi^{a}_{1}+{\tilde \psi _{02}}\Phi^{a}_{2}$ (\ref{eq358abc}), where the total associate-components ({\it{TACs}}) have the {\it{one-component forms}} $\Phi^{a}_{1} = {\phi_{i1}}$ and  $\Phi^{a}_{2} = {\phi_{i2}}$, and the relevant values are given by $\mathbf{k}_{01} = \mathbf{k}_{02}= \mathbf{k}_{0}$, $\varepsilon_{01} = \varepsilon_{02} =\varepsilon_{0} $, $m_{01}=m_{02}=m_{0}$, $\alpha_{1}=\alpha_{2}=\alpha _0$ and ${\Gamma_ {i1}}={\Gamma_ {i2}}={\Gamma_ {i}}$. A simple analysis of the unit-fields (particles) (\ref{eq356abc}) and (\ref{eq357abc}) having the de Broglie generators $\tilde \psi_{01} (\mathbf{r},t)=a_0 e^{i({\mathbf{k}_{01}{\mathbf{r}}}-{\varepsilon}_{01} t +\alpha _1 )}$ (\ref{eq384abc}) and $\tilde \psi_{02} (\mathbf{r},t)=a_0 e^{i({\mathbf{k}_{02}{\mathbf{r}}}-{\varepsilon}_{02} t +\alpha _2 )}$ (\ref{eq385abc}) and the spherically symmetric {\it{Helmholtz-ACs}} associated with the spherical Bessel functions of the first (\ref{eq326abc}) or second (\ref{eq328abc}) kind shows that only the {\it{one-component unit-fields}} (\ref{eq356abc}) related to the {\it{spherical Bessel functions}} (\ref{eq326abc}) of the first kind with the real (${\Gamma ^{a} _{in}} r = {\Gamma  _{Wn}}  r=|{\Gamma  _{Wn}} | r$) argument and the quantum numbers ($l_n = 0$, $s_n=0$) or ($l_n = 1$, $s_n=-1, 0, 1$) are suitable for description of the pure-weak interactions of the unit-fields (particles) with the short-range interaction energy. Moreover, among these unit-fields the only one-component unit-fields with the quantum numbers $l_n = 1$ and $s_n=\pm 1$ demonstrate the experimentally observed physical properties of the pure-weak interaction. Indeed, in the case of the first {\it{Helmholtz-AC}} [${\phi_{i1}}={\phi_{W1}}(\bf{r})$] centered at the origin and the second {\it{Helmholtz-AC}} [${\phi_{i2}}={\phi_{W2}}(\bf{r}-\bf{R})$] located at the distance $R=|\bf{R}|\equiv |\bf{R}_{12}|$ in the azimuthal direction, the relativistic interaction (cross-correlation) total term (\ref{eq389abc}) has the form
\begin{eqnarray} \label{eq450abc}
{\cal E}_{12,21}(R,l_1,l_2,s_1,s_2 ) = \mathbf{k}_{01} \mathbf{k}_{02} \left[  {{ {{A_{l_1s_1}A_{l_2s_2}}} a_0^2} \int_{\theta =0}^\pi d\theta sin\theta  \int_{\varphi  =0}^{2\pi} d \varphi  \int_{0}}^{\infty}r^2dr j_{l_1} ( {\Gamma  _{W1}} r)  j_{l_2} ( {\Gamma  _{W2}} | {\bf{r}}-{\bf{R}}| ) f  \right]  +   \nonumber  \\  +   \left[ m_{01}m_{02} + {\Gamma ^a_{I1}} {\Gamma ^a_{I2}} \right] \left[  {{ {{A_{l_1s_1}A_{l_2s_2}}} a_0^2} \int_{\theta =0}^\pi d\theta sin\theta  \int_{\varphi  =0}^{2\pi} d \varphi  \int_{0}}^{\infty}r^2dr j_{l_1} ({\Gamma  _{W1}} r)  j_{l_2} ({\Gamma  _{W2}} | {\bf{r}}-{\bf{R}}| ) f  \right] , 
\end{eqnarray}
where 
\begin{eqnarray} \label{eq451abc}
{\Gamma ^{a}_{i1}}= {\Gamma ^{a}_{i2}}=  {\Gamma  _{W1}}  = {\Gamma _{W2}}  = |{\Gamma  _{W1}} | = |{\Gamma _{W2}} | \neq 0 \nonumber  \\
l_1=l_2=1 .
\end{eqnarray}
Here, the angle factor $f=f(r,R,\theta , \varphi ,l_1,l_2, s_1, s_2 )$ associated with the spherical harmonics has the form
\begin{eqnarray} \label{eq452abc}
f(r,R,\theta , \varphi , l_1, l_2, s_1, s_2 )= {Y_{l_1}^{s_1}}^* (\theta , \varphi) Y_{l_2}^{s_2} (\theta , \varphi _2 ) + {Y_{l_1}^{s_1}} (\theta , \varphi) {Y_{l_2}^{s_2}}^* (\theta , \varphi _2 ), 
\end{eqnarray}
where the intrinsic "magnetic" quantum numbers are given by $s_1 = -1, 0, 1$ and $s_2 = -1, 0, 1$, and $ \varphi _2 =\varphi _2 (r, R, \varphi)$ denotes the azimuthal angle associated with the vector $\bf{r}-\bf{R}$ in the spherical coordinate system of the second {\it{Helmholtz-AC}}. Notice, a simple analysis also shows that Eqs. (\ref{eq390abc}) - (\ref{eq392abc}) derived by using the {\it{non-relativistic}} (Newton) approximation of the unit-field (particle) energy {\it{do not describe correctly the weak interaction of unit-fields (particles)}}. That means that the weakly interacting unit-fields (elementary particles) are {\it{relativistic objects}} describing rather by the Einstein particle energy than by the Newton particle energy. Also note that the relativistic interaction (cross-correlation) total term ${\cal E}_{12,21}(R,l_1,l_2,s_1,s_2)=0$ if the intrinsic orbital quantum numbers $l_1 \neq l_2$. That is to say that the two one-component unit-fields having the different intrinsic orbital quantum numbers do not interact weakly with each other. To exclude the singularities in the particle self-energy (\ref{eq228abc}) and the cross-correlation energy (\ref{eq450abc}) associated with the points $r=0$ and ${\bf{r}}-{\bf{R}}=0$, one should suppose that the weakly interacting {\it{Helmholtz-ACs}} ${\phi_{W1}}(\bf{r})$ and ${\phi_{W2}}(\bf{r}-\bf{R})$ are hollow around these points in the bolls of the radius $r_{0}^W$. A simple analysis shows that the infinite upper integration limit ($r^W_{up}=\infty$) in Eq. (\ref{eq450abc}) is inconsistent with the finite fields and energies of the weak interaction. Indeed, such a case corresponds to the infinite {\it{Helmholtz-ACs}} with infinite interaction energies. In order to satisfy the finite fields and energies of the weak interaction, one should suppose that the unit-field {\it{Helmholtz-AC}} of the weakly interacting unit-field is limited by the upper boundary of the radius $r_{up}^W<\infty$, where $r_{up}^W >r_{0}^W$. In other words, the unit-field {\it{Helmholtz-AC}} vanishes above the {\it{AC}} upper boundary (integration limit). The calibration of unit-fields by the "ultraviolet" and "infrared" cut-offs of the AC fields yielded the relativistic weak interaction (cross-correlation) term ${\cal E}_{12,21}={\cal E}_{12,21}(R,l_1,l_2,s_1,s_2 )$ with the modified upper and lower integration limits:
\begin{eqnarray} \label{eq453abc}
{\cal E}_{12,21}(R,l_1,l_2,s_1,s_2 ) = \mathbf{k}_{01} \mathbf{k}_{02} f_W  +   \left[ m_{01}m_{02} + {\Gamma _{W1}} {\Gamma _{W2}} \right] f_W, 
\end{eqnarray}
where the weak-interaction geometrical factor 
\begin{eqnarray} \label{eq454abc}
f_W= f_W(R,l_1,l_2,s_1,s_2 )  
\end{eqnarray}
is given by 
\begin{eqnarray} \label{eq455abc}
f_W(R,l_1,l_2,s_1,s_2 )  =   \nonumber  \\  =  {{{A_{l_1s_1}A_{l_2s_2}}} a_0^2} \int_{\theta =0}^\pi d\theta sin\theta  \int_{\varphi  =0}^{2\pi} d \varphi \int_{r_0^W}^{R-r_{0,low}^W/ 2} \int_{R+r_{0,up}^W/ 2}^{r_{up}^W}r^2 dr  j_{l_1} ({\Gamma _{W1}}  r)  j_{l_2} ({\Gamma  _{W2}} | {\bf{r}}-{\bf{R}}| ) f .
\end{eqnarray}
It should be stressed that the {\it{Helmholtz-AC}} amplitude ${A_{1_ns_n}}$ {\it{can be an arbitrary constant}} [see, Eq. (\ref{eq326abc})], which cold be called the {\it{weak isospin}}. If the $n$-th weakly interacting unit-field (particle) has the "charge-like" {\it{weak hypercharge}} $Q^h_n$, which is different from the electric (Coulomb) charge, then the amplitude ${A_{1_ns_n}}$ in Eq. (\ref{eq455abc}) could be presented in the form ${A_{1_ns_n}}(m_n/Q^h_n)$ that corresponds to the "electric-like" ("electromagnetic-like") weak interaction mediating by the {\it{hypercharge}} of the weakly interacting unit-field (particle) [for the comparison, see Eq. (\ref{eq434abc})]. Naturally, the {\it{Helmholtz-AC}} amplitude ({\it{weak isospins}} or {\it{weak hypercharges}}) of the unit-fields (particles) should be adjusted to the value corresponding to the experimentally observed parameters of the weak interaction. In contrast to the traditional quantum field theories, the {\it{Helmholtz-AC}} amplitude associated with the {\it{weak isospin}} or {\it{weak hypercharge}} of the weakly-interacting unit-field (particle) gives the {\it{physical (non-phenomenological) explanation}} of the {\it{weak isospin}} and {\it{weak hypercharge}} of an elementary particle. The radius
\begin{eqnarray} \label{eq456abc}
r = r_{up}^W
\end{eqnarray}
could be considered as the external radius of a weakly interacting unit-field (particle). The parameter $r = r_{0}^W$ logically to associate with the classical radius of a weakly interacting particle. The value $|{\bf{r}}-{\bf{R}}|$, which is calculated by using a simple geometrical analysis of the two vectors, is a function of the parameters $r$, $R$, $\theta$ and $\varphi$. Correspondingly, the integration limits $r_{0,low}^W$, $r_{0,up}^W$ and $r_{up}^W$ are calculated by using this geometrical analysis as functions of the parameters $r$, $R$, $\theta$, $\varphi$, $r_0^W$ and $r_{up}^W$. 

If the {\it{"non-quantum" (spin-independent) weak interaction}} does exist somewhere in the Universe, then the unit-fields (particles) with de Broglie generators and the spherically symmetric {\it{Helmholtz-ACs}} associated with the spherical Bessel functions of first kind and the quantum numbers $l_n = 0$, $s_n=0$ could be attributed to such an interaction. In the case of the unit-field {\it{Helmholtz-ACs}} with the quantum numbers $l_1=l_2=0$ and $s_1=s_2=0$, the spherical Bessel functions in Eq. (\ref{eq455abc}) are given simply by 
\begin{eqnarray} \label{eq457abc}
j_{l_1} ({\Gamma  _{W1}} r)=j_0 ({\Gamma _{W1}} r)  = \frac {sin({\Gamma  _{W1}} r)} {{\Gamma  _{W1}} r}
\end{eqnarray}
and 
\begin{eqnarray} \label{eq458abc}
j_{l_2} ({\Gamma  _{W2}} | {\bf{r}}-{\bf{R}}| ) = j_0 ({\Gamma  _{W2}}  | {\bf{r}}-{\bf{R}}| )  = \frac {sin({\Gamma  _{W2}} | {\bf{r}}-{\bf{R}}|)} {{\Gamma  _{W2}} |  {\bf{r}}-{\bf{R}}|},
\end{eqnarray}
and the angle factor $f=f(r,R,\theta , \varphi , l_1, l_2, s_1, s_2 )=f(r,R,\theta , \varphi , 0, 0, 0, 0 )$ has the spin-independent form
\begin{eqnarray} \label{eq459abc}
f(r,R,\theta , \varphi , 0, 0, 0, 0 )= {Y_{0}^{0}}^* (\theta , \varphi) Y_{0}^{0} (\theta , \varphi _2 ) + {Y_{0}^{0}} (\theta , \varphi) {Y_{0}^{0}}^* (\theta , \varphi _2 ) = 1/2\pi. 
\end{eqnarray}
The annihilation of the two weakly interacting unit-fields with the quantum numbers $l_1=l_2=0$ and $s_1=s_2=0$ does not require adjustments of the unit-field spins. The easy annihilation of the weakly interacting unit-fields with $l_1=l_2=0$ and $s_1=s_2=0$, if they really have been created under the cosmological Big Bang, could be considered as a physical  mechanism explaining the absence of such unit-fields (particles) at the present stage of Universe. 

The {\it{physical properties of quantum (spin-dependent) weak interactions}}, which are observed experimentally, could be attributed to the unit-fields (particles) having the de Broglie generators and the spherically symmetric {\it{Helmholtz-ACs}} associated with the spherical Bessel functions of the first kind and the quantum numbers $l_n = 1$, $s_n=\pm 1$. In the case of $l_1=l_2=1$ and $s_1=\pm 1$, $s_2=\pm 1$, the relativistic weak interaction (cross-correlation) term ${\cal E}_{12,21}={\cal E}_{12,21}(R,l_1,l_2,s_1,s_2 )$ is determine by Eqs. (\ref{eq453abc}) - (\ref{eq455abc}), where the spherical Bessel functions are given by 
\begin{eqnarray} \label{eq460abc}
j_{l_1} ({\Gamma  _{W1}} r)=j_{1} ({\Gamma  _{W1}} r) =  \frac {sin({\Gamma  _{W1}} r)} {({\Gamma  _{W1}} r)^2} - \frac {cos({\Gamma  _{W1}} r)} {{\Gamma  _{W1}} r}
\end{eqnarray}
and 
\begin{eqnarray} \label{eq461abc}
j_{l_2} (i {\Gamma  _{W2}} |{\bf{r}}-{\bf{R}}|)=j_{1} (i {\Gamma  _{W2}}  | {\bf{r}}-{\bf{R}}|) = \frac {sin({\Gamma  _{W2}} | {\bf{r}}-{\bf{R}}|)} {({\Gamma  _{W2}} |  {\bf{r}}-{\bf{R}}|)^2} - \frac {cos({\Gamma  _{W2}} | {\bf{r}}-{\bf{R}}|)} {({\Gamma  _{W2}} |  {\bf{r}}-{\bf{R}}|)},
\end{eqnarray}
and the angle factor $f=f(r,R,\theta , \varphi , l_1, l_2, s_1, s_2 )=f(r,R,\theta , \varphi , 1, 1, s_1, s_2 )$ has the spin-dependent form
\begin{eqnarray} \label{eq462abc}
f(r,R,\theta , \varphi , 1, 1, s_1, s_2 )= {Y_{1}^{s_1}}^* (\theta , \varphi) Y_{1}^{s_2} (\theta , \varphi _2 ) + {Y_{1}^{s_1}} (\theta , \varphi) {Y_{1}^{s_2}}^* (\theta , \varphi _2 ) .
\end{eqnarray}
Although the weakly interacting unit-fields (particles) with the quantum numbers $l_1=l_2=1$ and $s_1=s_2=0$ can exist in principle, the annihilation of such unit-fields does not require adjustments of the unit-field spins. The absence of the weakly interacting unit-fields (particles) with the quantum numbers $l_n=1$ and $s_n = 0$ in the present Universe, if they have been created after the cosmological Big Bang, could be considered as s a physical  mechanism explaining of the absence of such unit-fields (particles) at the present stage of Universe. It should be noted that the relativistic interaction (cross-correlation) total term ${\cal E}_{12,21}(R,l_1,l_2,s_1,s_2)=0$ if the intrinsic orbital quantum numbers $l_1 \neq l_2$. The "weak strength" of the weakly interacting unit-field is attributed to the properties of the weak-interaction geometrical factor (\ref{eq455abc}), while the short-range of the weak interaction is associated with the external radius $r_{up}^W$ of the {\it{Helmholtz-AC}}. 

The {\it{weak-relativistic}} (${\cal E}_{12,21}<<{\varepsilon}_{01}^{2} + {\varepsilon}_{02}^{2}$) interaction forces (\ref{eq172abc}) and (\ref{eq173abc}) acting upon the first and second unit-fields (particles) have respectively the forms
\begin{eqnarray} \label{eq463abc}
\mathbf{F}_{12}(\mathbf{R} ) = -{\frac 1  {2}}{\frac {\partial}  {\partial \mathbf{R}}} {\cal E}_{12,21}(R,l_1,l_2,s_1,s_2 ) 
\end{eqnarray} 
and  
\begin{eqnarray} \label{eq464abc}
\mathbf{F}_{21}(\mathbf{R} ) = {\frac 1  {2}}{\frac {\partial}  {\partial \mathbf{R}}} {\cal E}_{12,21}(R,l_1,l_2,s_1,s_2 ).
\end{eqnarray} 
Analysis of Eqs. (\ref{eq450abc}) - (\ref{eq464abc}) shows that the weakly interacting unit-fields $\psi _{01}={\tilde \psi _{01}}\phi_{W1}$ and $\psi _{02}={\tilde \psi _{02}}\phi_{W2}$ indeed have the finite interaction energies corresponding to the finite total interaction (cross-correlation) term ${\cal E}_{12,21} \neq \infty$ in the case of the {\it{Helmholtz-AC}} dimensions $r_{0}^W  \neq 0$ and $r_{up}^W  \neq \infty$. The interaction energy {\it{decreases}} with the increase of the value $R$ and approaches the zero value at  $R> r_{up}^W$. The dimensions $r_{0}^W  \neq 0$ and $r_{up}^W  \neq \infty$ should be adjusted to the values, which are in agreement with the physical properties of the weak interactions. The parameter $r_{0}^W$ could be interpreted as the classical radius of the particle, while $r_{up}^W$ as the external radius of the unit-field determining the short range of weak interactions {\it{Helmholtz-AC}}. 

Equations (\ref{eq450abc}) - (\ref{eq464abc}) are valid for the {\it{weak relativistic}} (${\cal E}_{12,21} << {\varepsilon}_{01}^{2} + {\varepsilon}_{02}^{2}$) interaction, which is the case of the most of experimental conditions. The other two interaction conditions, namely the {\it{strong relativistic}} and {\it{non-relativistic}} weak interactions, are also possible. Under the {\it{strong relativistic}} (${\cal E}_{12,21} \sim {\varepsilon}_{01}^{2} + {\varepsilon}_{02}^{2}$) weak interaction, which corresponds to the extremely small distance $R\sim r_0$, one should use the above-derived equations with the relations (\ref{eq165abc}), (\ref{eq168abc}) and (\ref{eq169abc}) instead of the relations (\ref{eq172abc}) and (\ref{eq173abc}). The {\it{strong-relativistic}} weak-interaction forces (\ref{eq168abc}) and (\ref{eq169abc}) acting upon the first and second unit-fields (particles) have respectively the forms
\begin{eqnarray} \label{eq465abc}
\mathbf{F}_{12}(\mathbf{R} ) = -{\frac 1  {2}}{\frac {\partial}  {\partial \mathbf{R}}} {\varepsilon }_{12,21}(R) 
\end{eqnarray} 
and  
\begin{eqnarray} \label{eq466abc}
\mathbf{F}_{21}(\mathbf{R} ) = {\frac 1  {2}}{\frac {\partial}  {\partial \mathbf{R}}} {\varepsilon }_{12,21}(R) ,
\end{eqnarray} 
where the weak-interaction energy ${\varepsilon }_{12}(R)$ is given by Eq. (\ref{eq165abc}) as 
\begin{eqnarray} \label{eq467abc}
{\varepsilon}_{12,21}=({\varepsilon}_{01}^{2} + {\varepsilon}_{02}^{2} + {\cal E}_{12,21})^{1/2}-({\varepsilon}_{01} + {\varepsilon}_{02}),
\end{eqnarray} 
with the relativistic cross-correlation (interaction) total term ${\cal E}_{12,21}$ given by Eqs. (\ref{eq450abc}) - (\ref{eq464abc}). It should be stressed that Eqs. (\ref{eq450abc}) - (\ref{eq464abc}) are valid in the particular case of the de Broglie generators of the unit-fields having the same moments. Although the distance $R\sim r_0$ probably can be realized in some very particular experimental conditions, the unit-field generators of such unit-fields (particles) are not the de Broglie waves. In other words, the {\it{strong relativistic}} weak-interaction of the real unit-fields (particles) should be described rather by the equations (\ref{eq422abc}) - (\ref{eq424abc}) than Eqs. (\ref{eq405abc}) - (\ref{eq421abc}). A simple analysis also shows that Eqs. (\ref{eq390abc}) - (\ref{eq392abc}) derived by using the {\it{non-relativistic}} (Newton) approximation of the unit-field (particle) energy {\it{do not describe correctly the weak interactions}}. That means that the weakly interacting unit-fields (elementary particles) are relativistic objects describing rather by the Einstein particle energy than by the Newton particle energy.  

The weak interaction between the two unit-fields determining by Eqs. (\ref{eq450abc}) - (\ref{eq464abc}) may be interpreted not only by using the "electric-like" and "magnetic-like" weak-interaction fields and potentials, which are somewhat similar to the respective Eqs. (\ref{eq415abc}) - (\ref{eq421abc}), but also in the terms of the virtual particles. The relativistic energy of the composite field (particle) composed from the weakly interacting unit-fields (particles) is given by Eq. (\ref{eq199abc}) as 
\begin{eqnarray} \label{eq468abc}
{\varepsilon}^{2} = {\varepsilon}_{01}^{2} + {\varepsilon}_{02}^{2} + {\cal E}_{12}+{\cal E}_{21} ={\varepsilon}_{01}^{2} + {\varepsilon}_{02}^{2} + {\cal E}_{12,21}. 
\end{eqnarray} 
The energies squared ${\varepsilon}_{01}^2$ and ${\varepsilon}_{02}^2$ could be attributed to the "moving" and "non-moving" {\it{masses}} of the 1st and 2nd relativistic {\it{normal}} unit-fields (elementary particles). The weak-interaction cross-correlation terms ${\cal E}_{12}={\cal E}_{12}(m_{01},\Gamma  _{W1}; m_{02},\Gamma  _{W2})$ and ${\cal E}_{21}={\cal E}_{21}(m_{02},\Gamma  _{W2}; m_{01},\Gamma  _{W1})$ could be attributed to the "moving" and "non-moving" {\it{masses}} $m_{01}$, $m_{02}$, $\Gamma  _{W1}$ and $\Gamma  _{W2}$ of the 1st and 2nd relativistic {\it{normal}} unit-fields. The weak-interaction cross-correlation terms ${\cal E}_{12}$ and ${\cal E}_{21}$ associated with the masses $m_{01}$, $m_{02}$, $\Gamma  _{W1}$ and $\Gamma  _{W2}$ and the weak interactions may be considered as interplay of the 3rd and 4th {\it{virtual}} relativistic unit-fields (elementary particles) attributed to the relativistic weak-interaction of the 1st and 2nd {\it{normal}} massive unit-fields (elementary particles). Thus the relativistic weak-interaction between the two {\it{normal}} relativistic unit-fields (elementary particles) could be considered (interpreted) as the interplay of the {\it{four}} unit-fields (particles), where the composite field (particle) is composed from the 1st and 2nd {\it{normal}} unit-fields (particles) and the 3rd and 4th {\it{virtual}} unit-fields (particles). For more details of the interpretation based on the use of virtual particles see the comments to Eq. (\ref{eq199abc}). The {\it{virtual relativistic unit-fields (elementary particles)}}, which {\it{carry}} the weak force and energy, logically to call the {\it{virtual W or Z bosons}} mediated by the {\it{weak isospin}} or the {\it{weak hypercharge}} of the 1st and 2nd {\it{normal}} unit-fields (particles). The names {\it{W and Z bosons}} correspond to the carriers of the weak interaction in the quantum field theories and SM. The weakly interacting normal unit-field (particle) with the mass $m_{0n}$ obey the relations $m_{0n} << \Gamma  _{Wn}$. In such a case  the weak-interaction cross-correlation terms are given by ${\cal E}_{12}(m_{01},\Gamma  _{W1}; m_{02},\Gamma  _{W2})={\cal E}_{12}(\Gamma  _{W1}, \Gamma  _{W2})$ and ${\cal E}_{21}(m_{02},\Gamma  _{W2}; m_{01},\Gamma  _{W1})={\cal E}_{21}(\Gamma  _{W2},\Gamma  _{W1})$, where the values $\Gamma  _{W1}$ and $\Gamma  _{W2}$ could be interpreted as the {\it{gauge masses}} of the {\it{virtual W or Z bosons}} associated with the 1st and 2nd weakly interacting normal unit-fields (particles). In the frame of the Heisenberg energy uncertainty relation and the perturbation approximation, the short range of the weak interaction is attributed phenomenologically (formally) by the quantum field theories and SM to the heaviness  of the virtual $W$ and $Z$ bosons. The present model {\it{gives the microscopic explanation of the phenomenon}}. The short-range of the weak interaction is attributed to the finite external radius $r_{up}^W$ of the heavy {\it{Helmholtz-AC}} of the weakly interacting unit-field (particle). It should be stressed that the parameter $\Gamma_{Wn} \neq 0$ could be attributed to the mass of the weakly interacting particles even in the case of the unit-fields (particles) that obey the zero rest-mass $m_{0}=0$ [see, Eq. (\ref{eq301abc})].

The model [Eqs. (\ref{eq450abc}) - (\ref{eq468abc})] describes interaction of the two ($N=2$) weakly interacting unit-fields (particles) with the moments $\mathbf{k}_{01}=\mathbf{k}_{02}=\mathbf{k}_{0}$. One can easily follow the model for an arbitrary number $N$ of the weakly interacting unit-fields (particles) having the de Broglie generators $\tilde \psi_{0n} (\mathbf{r},t)=a_0 e^{i({\mathbf{k}_{0n}{\mathbf{r}}}-{\varepsilon}_{0n} t )}$ with the moments $\mathbf{k}_{0n}=\pm \mathbf{k}_{0}$. The model [Eqs. (\ref{eq450abc}) - (\ref{eq468abc})] with $N>>1$ and $s_n=0$ describes the {\it{spin-independent (non-quantum)}} "electrostatic-like" and "magnetostatic-like" weak-interaction fields and weak interactions that correspond to the case of the Lorentz-like ({\it{non-quantum}}) spin-independent interaction. In other words, such a model corresponds to the {\it{non-quantum}} interactions of the "electrostatics" and "magnetostatics" of the weakly interacting unit-fields (particles). Among the weakly interacting unit-fields (particles) with the intrinsic magnetic quantum numbers $s_n = -1, 0, 1$ of the unit-field spin, {\it{the only  unit-field (particles) with the spin numbers $s_n \equiv m_n =0$ demonstrate the "spin-independent" [${\cal E}_{12,21}(R,s_1,s_2 ) = {\cal E}_{12,21}(R)$] Lorentz-like behavior}} that satisfy the attractive ($A_{l_1s_1}=-A_{l_2s_2}$) and repulsive ($A_{l_1s_1}=A_{l_2s_2}$) "electrostatic-like" weak interactions $\mathbf{F}^C_{W12}$ and the attractive ($A_{l_1s_1}=A_{l_2s_2}$, ${\mathbf{k}_{01}}={\mathbf{k}_{02}}$ or $A_{l_1s_1}=-A_{l_2s_2}$, ${\mathbf{k}_{01}}=-{\mathbf{k}_{02}}$ ) and repulsive ($A_{l_1s_1}=A_{l_2s_2}$, ${\mathbf{k}_{01}}= - {\mathbf{k}_{02}}$ or $A_{l_1s_1}=-A_{l_2s_2}$, ${\mathbf{k}_{01}}=  {\mathbf{k}_{02}}$) "magnetostatic-like" interactions $\mathbf{F}^M_{W12}$ of the weakly interacting unit-fields (particles). The {\it{purely attractive}} spin-independent (Lorentz-like) "non-quantum" interaction $\mathbf{F}^C_{W12}+\mathbf{F}^M_{W12}$ of the unit-field (particles) with the weak-interaction {\it{ACs}} having the spin numbers $s_n =0$, amplitudes $A_{l_1}=-A_{l_2}$ and momentums ${\mathbf{k}_{01}}=-{\mathbf{k}_{02}}$ could be considered as the "{\it{non-quantum}}" attraction of weakly interacting unit-fields [{\it{boson-like}} ($A_{l_1}=-A_{l_2}$, ${\mathbf{k}_{01}}=-{\mathbf{k}_{02}}$) particles] that explains microscopically the Bose-Einstein condensation and statistics of such particles. Although the "spin-less" ($s_n=0$) weakly interacting unit-fields (particles) have not been yet detected, the model  [Eqs. (\ref{eq450abc}) - (\ref{eq468abc})] is good for the (Coulomb-like and Lorentz-like) description of the weak-interaction forces. {\it{The spin-dependent (quantum) interaction between the weakly interacting unit-fields, which is observed experimentally, is described by the Lorentz-like spin-dependent (quantum) interaction energy ${\cal E}_{12,21}(R,s_1,s_2 )$ determining by Eqs. (\ref{eq172abc}), (\ref{eq173abc}), (\ref{eq450abc}) - (\ref{eq468abc}) with the spin numbers ($s_1=1$, $s_2=1$), ($s_1=-1$, $s_2=-1$), ($s_1=1$, $s_2=-1$) or ($s_1=-1$, $s_2=1$)}}. In other words, Eqs. (\ref{eq172abc}), (\ref{eq173abc}), (\ref{eq450abc}) - (\ref{eq468abc}) determine the interaction properties of the weakly interacting unit-field (particle) and their connections with the unit-field spin. A simple analysis of Eqs. (\ref{eq172abc}), (\ref{eq173abc}), (\ref{eq450abc}) - (\ref{eq468abc}) for the two unit-fields with the amplitudes $A_{l_1s_1}=A_{l_2s_2}$ shows that the equations describe the {{\it{repulsive}} ($s_1=s_2, A_{l_1s_1}=A_{l_2s_2},{\mathbf{k}_{01}}=  {\mathbf{k}_{02}}$) and {\it{attractive}} ($s_1\neq s_2, A_{l_1s_1}=A_{l_2s_2},{\mathbf{k}_{01}}=  {\mathbf{k}_{02}}$) {\it{quantum}} forces associated with the {\it{Lorentz-like spin-dependent (quantum) energy}} of the weakly interacting  fermion-like unit-fields (fermions). The Pauli exclusion principle states that no two identical ($s_1=s_2, A_{l_1s_1}=A_{l_2s_2},{\mathbf{k}_{01}}=  {\mathbf{k}_{02}}$) fermions may occupy the same quantum state simultaneously. The {\it{spin-dependent (quantum) repulsive or attractive weak-interaction of the unit-fields (fermions) with the amplitudes $A_{l_1s_1}=A_{l_2s_2}$ having respectively the same ($s_1=s_2$) or different ($s_1 \neq s_2$) spin numbers could be considered as the physical origin explaining microscopically the Pauli exclusion principle and the Fermi-Dirac statistics of the weakly interacting fermions, which in canonical quantum mechanics and SM have nature of the unexplained postulates}}. Unlike in the traditional quantum field theory and SM, which consider rather the fields of operators than the fields of particles, Eqs. (\ref{eq172abc}), (\ref{eq173abc}), (\ref{eq450abc}) - (\ref{eq468abc}) determine the "electromagnetic-like" properties of the weakly interacting unit-field (particle) and their connections with the unit-field spin naturally, without introduction of any pure mathematical object, like the spin matrix  or operators. 
It should be also noted that in the case of the two weakly interacting, {\it{normal}} unit-fields (particles) having the spin numbers $s_1=s_2=\pm 1$ and amplitudes $A_{l_1s_1}=-A_{l_2s_2}$, the unit-fields (particles) can annihilate. Indeed, the weakly interacting unit-fields are the solutions of the equations of motions of the present model for both the $A_{1s_n}$ and $-A_{1s_n}$ amplitudes. The {\it{Helmholtz-AC}} amplitude ${A_{1_ns_n}}$ can be an arbitrary constant [see, Eq. (\ref{eq326abc})]. The two weakly interacting unit-fields with $s_1=s_2$ can satisfy the annihilation condition in the case ${A_{1s_1}} = - {A_{1s_2}}$.  In Eqs. (\ref{eq450abc}) - (\ref{eq468abc}), the unit-field (particle) with the positive value of  ${A_{1s_1}}$ could be considered as a {\it{normal}} unit-field (particle), while the unit-field (particle) with the negative value of  ${A_{1m_2}}$ could be considered as an {\it{anti-unitfield}} ({\it{anti-particle}}). The annihilation of the two weakly interacting unit-fields with $s_1=s_2=0$ does not require adjustments of the unit-field spins. The easy annihilation of the weakly interacting unit-fields with $s_n=0$, if they really have been created under the cosmological Big Bang, could be considered as a physical  mechanism of the absence of such unit-fields (particles) in the present Universe. It should be noted again that Eqs. (\ref{eq450abc}) - (\ref{eq468abc}) are valid in the particular case of the {\it{de Broglie generators}} $\tilde \psi_{01} (\mathbf{r},t)=a_0 e^{i({\mathbf{k}_{01}{\mathbf{r}}}-{\varepsilon}_{01} t )}$ (\ref{eq384abc}) and $\tilde \psi_{02} (\mathbf{r},t)=a_0 e^{i({\mathbf{k}_{02}{\mathbf{r}}}-{\varepsilon}_{02} t )}$ (\ref{eq385abc}). If the unit-field generators are not the de Broglie time-harmonic plane-waves, then the pure-weak interaction of the unit-fields (particles) are determined rather by the general equations (\ref{eq364abc}) - (\ref{eq374abc}) than the particular equations (\ref{eq450abc}) - (\ref{eq468abc}). One can easily follow the present model for any hypothetical unit-field (particle) with the spherically symmetric {\it{Helmholtz-ACs}} associated with the spherical Bessel functions (\ref{eq328abc}) of the second kind, if such elementary particles exist somewhere in the Universe. In such a case, the spherically symmetric {\it{Helmholtz-ACs}} associated with the spherical Bessel functions (\ref{eq328abc}) of the second kind should have the finite external radius.

\vspace{0.2cm}
2. {\it{The non-quantum and quantum pure-strong interactions of the unit-fields (particles)}}
\vspace{0.2cm}

An analysis of the unit-fields (particles) (\ref{eq356abc}) and (\ref{eq357abc}) having the de Broglie generators (\ref{eq384abc}) and (\ref{eq385abc}) and the spherically symmetric {\it{Helmholtz-ACs}} associated with the {\it{modified}} spherical Bessel functions of the first (\ref{eq327abc}) or second (\ref{eq329abc}) kind, which is quite similar to the above-presented analysis of the weakly interacting unit-fields,  shows that only the {\it{one-component unit-fields}} (\ref{eq356abc}) related to the {\it{modified spherical Bessel functions}} (\ref{eq327abc}) of the first kind with the imaginary (${\Gamma ^{a} _{In}} r = i|{\Gamma  _{Sn}} | r$) argument and the quantum numbers ($l_n = 0$, $s_n=0$) or ($l_n = 1$, $s_n=-1, 0, 1$) may be considered as candidates for the pure-strong interactions of the unit-fields (particles) with the short-range interaction energy. Furthermore, the only one-component unit-fields with the quantum numbers $l_n = 1$ and $s_n=\pm 1$ obey the experimentally observed physical properties of the pure-strong interaction. In the case of the first {\it{Helmholtz-AC}} [${\phi_{i1}}={\phi_{S1}}(\bf{r})$] centered at the origin and the second {\it{Helmholtz-AC}} [${\phi_{i2}}={\phi_{W2}}(\bf{r}-\bf{R})$] located at the distance $R=|\bf{R}|\equiv |\bf{R}_{12}|$ in the azimuthal direction, the relativistic interaction (cross-correlation) total term (\ref{eq389abc}) has the form
\begin{eqnarray} \label{eq469abc}
{\cal E}_{12,21}(R,l_1,l_2,s_1,s_2 ) = \mathbf{k}_{01} \mathbf{k}_{02} \left[  {{ {{A_{l_1s_1}A_{l_2s_2}}} a_0^2} \int_{\theta =0}^\pi d\theta sin\theta  \int_{\varphi  =0}^{2\pi} d \varphi  \int_{0}}^{\infty}r^2dr j_{l_1} (i |{\Gamma  _{S1}} | r)  j_{l_2} (i |{\Gamma  _{S2}} | | {\bf{r}}-{\bf{R}}| ) f  \right]  +   \nonumber  \\  +   \left[ m_{01}m_{02} - |{\Gamma _{S1}}| |{\Gamma _{S2}}| \right] \left[  {{ {{A_{l_1s_1}A_{l_2s_2}}} a_0^2} \int_{\theta =0}^\pi d\theta sin\theta  \int_{\varphi  =0}^{2\pi} d \varphi  \int_{0}}^{\infty}r^2dr j_{l_1} (i |{\Gamma  _{S1}} | r)  j_{l_2} (i |{\Gamma  _{S2}} | | {\bf{r}}-{\bf{R}}| ) f  \right] , 
\end{eqnarray}
where 
\begin{eqnarray} \label{eq470abc}
{\Gamma ^{a}_{i1}}= i |{\Gamma  _{S1}} | = i|{\Gamma _{S2}} | \neq 0 \nonumber  \\
l_1=l_2=1 .
\end{eqnarray}
Here, the angle factor $f=f(r,R,\theta , \varphi ,l_1,l_2, s_1, s_2 )$ associated with the spherical harmonics is given by
\begin{eqnarray} \label{eq471abc}
f(r,R,\theta , \varphi , l_1, l_2, s_1, s_2 )= {Y_{l_1}^{s_1}}^* (\theta , \varphi) Y_{l_2}^{s_2} (\theta , \varphi _2 ) + {Y_{l_1}^{s_1}} (\theta , \varphi) {Y_{l_2}^{s_2}}^* (\theta , \varphi _2 ), 
\end{eqnarray}
where the intrinsic "magnetic" quantum numbers are given by $s_1 = -1, 0, 1$ and $s_2 = -1, 0, 1$, and $ \varphi _2 =\varphi _2 (r, R, \varphi)$ denotes the azimuthal angle associated with the vector $\bf{r}-\bf{R}$ in the spherical coordinate system of the second {\it{Helmholtz-AC}}. Notice, a simple analysis also shows that Eqs. (\ref{eq390abc}) - (\ref{eq392abc}) derived by using the {\it{non-relativistic}} (Newton) approximation of the unit-field (particle) energy {\it{do not describe correctly the strong interaction of unit-fields (particles)}}. That means that the strongly interacting unit-fields (elementary particles) are {\it{relativistic objects}} describing rather by the Einstein particle energy than by the Newton particle energy. The relativistic interaction (cross-correlation) total term ${\cal E}_{12,21}(R,l_1,l_2,s_1,s_2)=0$ if the intrinsic orbital quantum numbers $l_1 \neq l_2$. That is to say that the two one-component unit-fields having the different intrinsic orbital quantum numbers do not interact strongly with each other. In order to exclude the singularities in the self-correlation energy (\ref{eq228abc}) and the cross-correlation energy (\ref{eq469abc}) associated with the points $r=0$ and ${\bf{r}}-{\bf{R}}=0$, one should suppose that the strongly interacting {\it{Helmholtz-ACs}} ${\phi_{S1}}(\bf{r})$ and ${\phi_{S2}}(\bf{r}-\bf{R})$ are hollow around these points in the bolls of the radius $r_{0}^S$. A simple analysis shows that the infinite upper integration limit ($r_{up}^S=\infty$) in Eq. (\ref{eq450abc}) does not compare well with the finite fields and energies of the strong interaction. The case of $r_{up}^S=\infty$ corresponds to the infinite {\it{Helmholtz-ACs}} with infinite strong-interaction energies. In order to satisfy the finite fields and energies of the strong interaction, one should suppose that the unit-field {\it{Helmholtz-AC}} of the strongly interacting unit-field is limited by the upper boundary (integration limit) of the radius $r_{up}^S<\infty$. That means that the unit-field {\it{Helmholtz-AC}} vanishes above the {\it{AC}} upper boundary. The calibrations of unit-fields yielded the relativistic strong interaction (cross-correlation) term ${\cal E}_{12,21}={\cal E}_{12,21}(R,l_1,l_2,s_1,s_2 )$ with the modified integration boundaries:
\begin{eqnarray} \label{eq472abc}
{\cal E}_{12,21}(R,l_1,l_2,s_1,s_2 ) = \mathbf{k}_{01} \mathbf{k}_{02} f_S  +   \left[ m_{01}m_{02} - |{\Gamma _{S1}}| |{\Gamma _{S2}}| \right] f_S, 
\end{eqnarray}
where the strong-interaction geometrical factor $f_S= f_S(R,l_1,l_2,s_1,s_2 )$ is given by 
\begin{eqnarray} \label{eq473abc}
f_S(R,l_1,l_2,s_1,s_2 )  =   \nonumber  \\  =  {{{A_{l_1s_1}A_{l_2s_2}}} a_0^2} \int_{\theta =0}^\pi d\theta sin\theta  \int_{\varphi  =0}^{2\pi} d \varphi \int_{r_0^S}^{R-r_{0,low}^S/ 2} \int_{R+r_{0,up}^S/ 2}^{r_{up}^S}r^2 dr  j_{l_1} (i|{\Gamma _{S1}}|  r)  j_{l_2} (i|{\Gamma  _{S2}}| | {\bf{r}}-{\bf{R}}| ) f .
\end{eqnarray}
Notice, the {\it{Helmholtz-AC}} amplitude ${A_{1_ns_n}}$ {\it{can be an arbitrary constant}} [see, Eq. (\ref{eq327abc})]. If the $n$-th strongly interacting unit-field (particle) has the "{\it{colar charge}}" $Q^c_{0n}$, which is different from the electric (Coulomb) charge, then the amplitude ${A_{1_ns_n}}$ in Eq. (\ref{eq473abc}) could be presented in the form ${A_{1_ns_n}}(Q^c_{0n}/m_{0n})$ that corresponds to the "electric-like" ("electromagnetic-like") weak interaction mediating by the "hyper charges" $Q^c_{0n}$ [for the comparison, see Eqs. (\ref{eq434abc}) and (\ref{eq455abc})] . The {\it{Helmholtz-AC}} amplitudes {\it{color charges}} of the unit-fields (particles) should be adjusted to the values corresponding to the experimentally observed parameters of the strong interaction. Unlike in the conventional quantum field theories, the {\it{Helmholtz-AC}} amplitude associated with the {\it{hyper (color) charge}} of the strongly-interacting unit-field (particle) gives the {\it{physical (non-phenomenological) explanation}} of the {\it{color charge}}. The parameters 
\begin{eqnarray} \label{eq474abc}
r = r_{0}^S
\end{eqnarray}
and
\begin{eqnarray} \label{eq475abc}
r = r_{up}^S
\end{eqnarray}
could be considered respectively as the particle classical dimension and the external radius of a strongly interacting unit-field. The value $|{\bf{r}}-{\bf{R}}|$, which is calculated by using a simple geometrical analysis of the two vectors, is a function of the parameters $r$, $R$, $\theta$ and $\varphi$. Respectively, the integration limits $r_{0,low}^S$, $r_{0,up}^S$ and $r_{up}^S$ are calculated by using this geometrical analysis as functions of the parameters $r$, $R$, $\theta$, $\varphi$, $r_0^S$ and $r_{up}^S$. 

If the {\it{"non-quantum" (spin-independent) strong interaction}} does exist somewhere in the Universe, then the unit-fields (particles) with de Broglie generators and the spherically symmetric {\it{Helmholtz-ACs}} associated with the {\it{modified spherical Bessel functions}} of first kind and the quantum numbers $l_n = 0$, $s_n=0$ could be associated with such an interaction. In the case of the unit-field {\it{Helmholtz-ACs}} with the quantum numbers $l_1=l_2=0$ and $s_1=s_2=0$, the modified spherical Bessel functions in Eq. (\ref{eq473abc}) are given by 
\begin{eqnarray} \label{eq476abc}
j_{l_1} (i |{\Gamma  _{S1}} | r)=j_0 (i |{\Gamma  _{S1}} | r)  = \frac {(i)^{0+1/2}I_{1/2}(|{\Gamma  _{S1}} | r)} {(2\pi ^{-1}i|{\Gamma  _{S1}} | r)^{1/2}}=\frac {sinh(|{\Gamma  _{S1}} | r)} {|{\Gamma  _{S1}} | r}= \frac {1}{2}\left( \frac {e^{|{\Gamma  _{S1}} | r} }{|{\Gamma  _{S1}} | r}- \frac {e^{-|{\Gamma  _{S1}} | r} }{|{\Gamma  _{S1}} | r} \right)
\end{eqnarray}
and 
\begin{eqnarray} \label{eq477abc}
j_{l_2} (i |{\Gamma _{S2}} | | {\bf{r}}-{\bf{R}}| ) = j_0 (i |{\Gamma  _{S2}} | | {\bf{r}}-{\bf{R}}| )  = \frac {I_{1/2}(|{\Gamma  _{S2}} | | {\bf{r}}-{\bf{R}}|)} {(2\pi ^{-1}|{\Gamma _{S2}} | |  {\bf{r}}-{\bf{R}}|)^{1/2}} = \nonumber  \\ = \frac {sinh(|{\Gamma  _{S2}} | | {\bf{r}}-{\bf{R}}|)} {|{\Gamma  _{S2}} | | {\bf{r}}-{\bf{R}}|}= \frac {1}{2}\left( \frac {e^{|{\Gamma _{S2}} | |{\bf{r}}-{\bf{R}}|} }{|{\Gamma  _{S2}} | | {\bf{r}}-{\bf{R}}|}- \frac {e^{-|{\Gamma  _{S2}} | |{\bf{r}}-{\bf{R}}|} }{|{\Gamma _{S2}} ||{\bf{r}}-{\bf{R}}|} \right),
\end{eqnarray}
and the angle factor $f=f(r,R,\theta , \varphi , l_1, l_2, s_1, s_2 )=f(r,R,\theta , \varphi , 0, 0, 0, 0 )$ has the spin-independent form
\begin{eqnarray} \label{eq478abc}
f(r,R,\theta , \varphi , 0, 0, 0, 0 )= {Y_{0}^{0}}^* (\theta , \varphi) Y_{0}^{0} (\theta , \varphi _2 ) + {Y_{0}^{0}} (\theta , \varphi) {Y_{0}^{0}}^* (\theta , \varphi _2 ) = 1/2\pi. 
\end{eqnarray}
The annihilation of the two strongly interacting unit-fields (particles) with the quantum numbers $l_1=l_2=0$ and $s_1=s_2=0$ does not require adjustments of the unit-field spins. The easy annihilation of the strongly interacting unit-fields (particles) with the quantum numbers $l_1=l_2=0$ and $s_1=s_2=0$, if these unit-fields (particles) really have been created under the cosmological Big Bang, could be considered as a physical  mechanism explaining the absence of such unit-fields (particles) at the present stage of Universe. 

The {\it{physical properties of quantum (spin-dependent) strong interactions}}, which are observed experimentally, could be attributed to the unit-fields (particles) having the de Broglie generators and the spherically symmetric {\it{Helmholtz-ACs}} associated with the {\it{modified spherical Bessel functions}} of first kind and the quantum numbers $l_n = 1$, $s_n=\pm 1$. In the case of $l_1=l_2=1$ and $s_1=\pm 1$, $s_2=\pm 1$, the relativistic strong interaction (cross-correlation) term ${\cal E}_{12,21}={\cal E}_{12,21}(R,l_1,l_2,s_1,s_2 )$ has the form (\ref{eq472abc}), where the modified spherical Bessel functions are given by 
\begin{eqnarray} \label{eq479abc}
j_{l_1} (i |{\Gamma  _{S1}} | r)=j_{1} (i |{\Gamma  _{S1}} | r)  = \frac {(i)^{1+1/2}I_{1+1/2}(|{\Gamma _{S1}} | r)} {(2\pi ^{-1}i|{\Gamma  _{S1}} | r)^{1/2}}= \frac {i\left[ cosh(|{\Gamma  _{S1}} | r) -(|{\Gamma  _{S1}} | r)^{-1} sinh(|{\Gamma _{S1}} | r) \right]} {|{\Gamma  _{S1}} | r} = \nonumber  \\ = \frac {i}{2} \left[ \left( \frac {e^{|{\Gamma  _{S1}} | r} }{|{\Gamma _{S1}} | r} + \frac {e^{-|{\Gamma  _{S1}} | r} }{|{\Gamma  _{S1}} | r} \right) - \left( \frac {e^{|{\Gamma  _{S1}} | r} }{|{\Gamma  _{S1}}|^2 r^2} - \frac {e^{-|{\Gamma  _{S1}}| r} }{|{\Gamma  _{S1}}|^2 r^2} \right) \right]
\end{eqnarray}
and 
\begin{eqnarray} \label{eq480abc}
j_{l_2} (i |{\Gamma _{S2}} ||{\bf{r}}-{\bf{R}}|)=j_{1} (i |{\Gamma  _{S2}} | | {\bf{r}}-{\bf{R}}|) = \frac {(i)^{1+1/2}I_{1+1/2}(|{\Gamma  _{S2}} | | {\bf{r}}-{\bf{R}}| )} {(2\pi ^{-1}i|{\Gamma  _{S2}} | | {\bf{r}}-{\bf{R}}|)^{1/2}}= \nonumber  \\ = \frac {i\left[ cosh(|{\Gamma  _{S2}} || {\bf{r}}-{\bf{R}}|) -(|{\Gamma  _{S2}} | | {\bf{r}}-{\bf{R}}|)^{-1} sinh(|{\Gamma  _{S2}} || {\bf{r}}-{\bf{R}}|) \right]} {|{\Gamma _{S2}} | | {\bf{r}}-{\bf{R}}|} = \nonumber  \\ = \frac {i}{2} \left[ \left( \frac {e^{|{\Gamma _{S2}} ||{\bf{r}}-{\bf{R}}|} }{|{\Gamma _{S2}} ||{\bf{r}}-{\bf{R}}|} + \frac {e^{-|{\Gamma _{S2}} ||{\bf{r}}-{\bf{R}}|} }{|{\Gamma _{S2}} ||{\bf{r}}-{\bf{R}}|} \right) - \left( \frac {e^{|{\Gamma _{S2}} ||{\bf{r}}-{\bf{R}}|} }{|{\Gamma _{S2}}|^2 {|{\bf{r}}-{\bf{R}}|}^2} - \frac {e^{-|{\Gamma _{S2}}||{\bf{r}}-{\bf{R}}|} }{|{\Gamma _{S2}}|^2 {|{\bf{r}}-{\bf{R}}|}^2} \right) \right],
\end{eqnarray}
and the angle factor $f=f(r,R,\theta , \varphi , l_1, l_2, s_1, s_2 )=f(r,R,\theta , \varphi , 1, 1, s_1, s_2 )$ has the spin-dependent form
\begin{eqnarray} \label{eq481abc}
f(r,R,\theta , \varphi , 1, 1, s_1, s_2 )= {Y_{1}^{s_1}}^* (\theta , \varphi) Y_{1}^{s_2} (\theta , \varphi _2 ) + {Y_{1}^{s_1}} (\theta , \varphi) {Y_{1}^{s_2}}^* (\theta , \varphi _2 ) .
\end{eqnarray}
It should be noted that the relativistic strong-interaction (cross-correlation) total term ${\cal E}_{12,21}(R,l_1,l_2,s_1,s_2)=0$ if the intrinsic orbital quantum numbers $l_1 \neq l_2$. Although the strongly interacting unit-fields (particles) with the quantum numbers $l_1=l_2=1$ and $s_1=s_2=0$ can exist in principle, the annihilation of such unit-fields does not require adjustments of the unit-field spins. The absence of the strongly interacting unit-fields (particles) with the quantum numbers $l_n=1$ and $s_n = 0$ in the present Universe, if they have been created under the cosmological Big Bang, could be considered as s a physical  mechanism explaining of the absence of such unit-fields (particles) at the present stage of Universe. The "strong strength" of the strongly interacting unit-field is attributed to the properties of the strong-interaction geometrical factor (\ref{eq473abc}), while the short-range of the strong interaction is associated with the external radius $r_{up}^S$ of the {\it{Helmholtz-AC}}. Thus the parameter $r_{0}^S$ could be interpreted as the classical radius of the particle, while $r_{up}^S$ as the external dimension of the unit-field determining the short range of weak interactions {\it{Helmholtz-AC}}. 

The {\it{weak-relativistic}} (${\cal E}_{12,21}<<{\varepsilon}_{01}^{2} + {\varepsilon}_{02}^{2}$) interaction forces (\ref{eq172abc}) and (\ref{eq173abc}) acting upon the first and second unit-fields (particles) have respectively the forms
\begin{eqnarray} \label{eq482abc}
\mathbf{F}_{12}(\mathbf{R} ) = -{\frac 1  {2}}{\frac {\partial}  {\partial \mathbf{R}}} {\cal E}_{12,21}(R,l_1,l_2,s_1,s_2 ) 
\end{eqnarray} 
and  
\begin{eqnarray} \label{eq483abc}
\mathbf{F}_{21}(\mathbf{R} ) = {\frac 1  {2}}{\frac {\partial}  {\partial \mathbf{R}}} {\cal E}_{12,21}(R,l_1,l_2,s_1,s_2 ).
\end{eqnarray} 
Analysis of Eqs. (\ref{eq469abc}) - (\ref{eq481abc}) shows that the strongly interacting unit-fields $\psi _{01}={\tilde \psi _{01}}\phi_{S1}$ and $\psi _{02}={\tilde \psi _{02}}\phi_{S2}$ indeed have the finite interaction energies corresponding to the finite total interaction (cross-correlation) term ${\cal E}_{12,21} \neq \infty$ in the case of the {\it{Helmholtz-AC}} dimensions $r_{0}^S  \neq 0$ and $r_{up}^S  \neq \infty$. The dimensions $r_{0}^S  \neq 0$ and $r_{up}^S  \neq \infty$ should be adjusted to the values, which are in agreement with the physical properties of the strong interactions. The parameters $r_{0}^S$ and $r_{up}^S$ could be interpreted as the internal and internal dimensions of the unit-field having the strong-interaction {\it{Helmholtz-AC}}. The strong-interaction energy and force {\it{do increase exponentially}} with the increase of the distance $R$ between the strongly interacting unit-fields (particles) and approaches the zero value at  the distance $R>2 r_{up}^S$. Indeed, the volume of interaction associated with the two overlapping {\it{Helmholtz-ACs}} slowly decreases with increasing the distance $R$, while the energy of the interaction (cross-correlation) of the {\it{Helmholtz-ACs}} increases exponentially. On the short distance ($R<<r_{up}^S$), the strong force could hold the strongly interacting unit-fields (particles) together to form the composite field (hadron). On the longer distance ($R\sim r_{up}^S$), the volume of interaction (cross-correlation) associated with the two {\it{overlapping}} {\it{Helmholtz-ACs}} has the {\it{string-like shape}}. In other words, the {\it{lines of strong force could collimate into strings}} in such a case. In this way, the present model not only explains in microscopical details how the strongly interacting unit-fields (particles)  interact over short distances, but also the {\it{string-like behavior}}, which the virtual unit-fields (particles) manifest over longer distances. In the quantum field theories and SM, the {\it{color confinement}} is the physics phenomenon that color charged particles (such as quarks) cannot be isolated singularly, and therefore cannot be directly observed. In the present model, the above-considered unit-fields (particles) having the color charges do interact with each other by means of the strong force, whose value exponentially increases with the increase of distance between the unit-fields (particles). That gives the microscopic explanation of the {\it{color confinement phenomenon}}.   
 
The strong interaction between the two unit-fields determining by Eqs. (\ref{eq469abc}) - (\ref{eq483abc}) may be interpreted not only by using the "electric-like" and "magnetic-like" strong interaction fields and potentials, which are somewhat similar to the respective Eqs. (\ref{eq415abc}) - (\ref{eq421abc}), but also in the terms of the virtual particles. The relativistic energy of the composite field (particle) composed from the strongly interacting unit-fields (particles) is given by Eq. (\ref{eq199abc}) as 
\begin{eqnarray} \label{eq484abc}
{\varepsilon}^{2} = {\varepsilon}_{01}^{2} + {\varepsilon}_{02}^{2} + {\cal E}_{12}+{\cal E}_{21} ={\varepsilon}_{01}^{2} + {\varepsilon}_{02}^{2} + {\cal E}_{12,21}. 
\end{eqnarray} 
The energies squared ${\varepsilon}_{01}^2$ and ${\varepsilon}_{02}^2$ could be attributed to the "moving" and "non-moving" {\it{masses}} of the 1st and 2nd relativistic {\it{normal}} unit-fields (elementary particles). The strong-interaction cross-correlation terms ${\cal E}_{12}={\cal E}_{12}(m_{01},\Gamma  _{S1}; m_{02},\Gamma  _{S2})$ and ${\cal E}_{21}={\cal E}_{21}(m_{02},\Gamma  _{S2}; m_{01},\Gamma  _{S1})$ could be attributed to the "moving" and "non-moving" {\it{masses}} $m_{01}$, $m_{02}$, $\Gamma  _{S1}$ and $\Gamma  _{S2}$ of the 1st and 2nd relativistic {\it{normal}} unit-fields. The strong-interaction cross-correlation terms ${\cal E}_{12}$ and ${\cal E}_{21}$ associated with the masses $m_{01}$, $m_{02}$, $\Gamma  _{S1}$ and $\Gamma  _{S2}$ and the strong interactions may be considered as the interplay of the 3rd and 4th {\it{virtual}} relativistic unit-fields (elementary particles) attributed to the relativistic strong interaction of the 1st and 2nd {\it{normal}} massive unit-fields (elementary particles). Thus the relativistic strong-interaction between the two {\it{normal}} relativistic unit-fields (elementary particles) could be considered (interpreted) as the interplay of the {\it{four}} unit-fields (particles), where the composite field (particle) is composed from the 1st and 2nd {\it{normal}} unit-fields (particles) and the 3rd and 4th {\it{virtual}} unit-fields (particles). For more details of the interpretation based on the use of virtual particles see the comments to Eq. (\ref{eq199abc}). The {\it{virtual relativistic unit-fields (elementary particles)}}, which {\it{carry}} the strong force and energy, logically to call the {\it{virtual gluons}} induced by the {\it{color charges}} of the 1st and 2nd {\it{normal}} unit-fields (particles). The name {\it{gluon}} corresponds to the carrier of the strong interaction in the quantum field theories and SM. The strongly interacting normal unit-field (particle) with the mass $m_{0n}$ obey the relations $m_{0n} << \Gamma  _{Sn}$. In such a case  the strong-interaction cross-correlation terms are given by ${\cal E}_{12}(m_{01},\Gamma  _{S1}; m_{02},\Gamma  _{S2})={\cal E}_{12}(\Gamma  _{S1}, \Gamma  _{S2})$ and ${\cal E}_{21}(m_{02},\Gamma  _{S2}; m_{01},\Gamma  _{S1})={\cal E}_{21}(\Gamma  _{S2},\Gamma  _{S1})$, where the values $\Gamma  _{S1}$ and $\Gamma  _{S2}$ could be interpreted as the {\it{gauge masses}} of the {\it{virtual gluons}} associated with the 1st and 2nd strongly interacting normal unit-fields (particles). The parameter $\Gamma_{Wn} \neq 0$ could be attributed to the mass of the strongly interacting particles even in the case of the unit-fields (particles) that obey the zero rest-mass $m_{0}=0$ [see, Eq. (\ref{eq301abc})]. In the frame of the Heisenberg energy uncertainty relation and the perturbation approximation, the short range of the strong interaction is attributed phenomenologically (formally) by the quantum field theories and SM to the heaviness of the virtual gluons. The present model gives the microscopic explanation of the phenomenon. The short-range of the strong interaction is attributed to the finite external radius $r_{up}^S$ of the heavy {\it{Helmholtz-AC}} of the strongly interacting unit-field (particle). In the terms of the virtual particles (gluons) of the present model,  the strong force is assumed to be mediated by gluons, acting upon the strongly interacting unit-fields (particles). On the short distance, the strong force holds the strongly interacting unit-fields (particles) to form the composite field (particle). The lines of strong force of the gluons interacting with each other at long distances collimate into strings. In this way, the physical interpretation of the phenomena is similar to the traditional quantum field theories and SM.

Since the energy squared of the $n$-th unit-field is positive (${\varepsilon}_{0n}^{2}\geq 0$), the unit-fields (particles) with the parameters ${\Gamma _{Sn}}= i | {\Gamma _{Sn}}|$ under the condition $ | {\Gamma _{Sn}}|\geq m_0$ could exist only as the moving ($\left[ \mathbf{k}_{0n}^2+(m_{0}^2  - |\Gamma_{Sn}|^2 )  \right] > 0$) unit-fields (particles). The model [Eqs. (\ref{eq469abc}) - (\ref{eq484abc})] describes interaction of the two ($N=2$) strongly interacting unit-fields (particles) with the moments $\mathbf{k}_{01}=\mathbf{k}_{02}=\mathbf{k}_{0}$. One can easily follow the model for an arbitrary number $N$ of the weakly interacting unit-fields (particles) having the de Broglie generators $\tilde \psi_{0n} (\mathbf{r},t)=a_0 e^{i({\mathbf{k}_{0n}{\mathbf{r}}}-{\varepsilon}_{0n} t )}$ with the moments $\mathbf{k}_{0n}=\pm \mathbf{k}_{0}$. The model [Eqs. (\ref{eq469abc}) - (\ref{eq484abc})] with $N>>1$ and $s_n=0$ describes the {\it{spin-independent (non-quantum)}} "electrostatic-like" and "magnetostatic-like" strong-interaction fields and strong interactions that correspond to the case of the Lorentz-like ({\it{non-quantum}}) spin-independent interaction. In other words, such a model corresponds to the {\it{non-quantum}} interactions of the "electrostatics" and "magnetostatics" of the strongly interacting unit-fields (particles). Among the strongly interacting unit-fields (particles) with the intrinsic magnetic quantum numbers $s_n = -1, 0, 1$ of the unit-field spin, {\it{the only  unit-field (particles) with the spin numbers $s_n \equiv m_n =0$ demonstrate the "spin-independent" [${\cal E}_{12,21}(R,s_1,s_2 ) = {\cal E}_{12,21}(R)$] Lorentz-like behavior}} that satisfy the attractive ($A_{l_1s_1}=-A_{l_2s_2}$) and repulsive ($A_{l_1s_1}=A_{l_2s_2}$) "electrostatic-like" strong interactions $\mathbf{F}^C_{W12}$ and the attractive ($A_{l_1s_1}=A_{l_2s_2}$, ${\mathbf{k}_{01}}={\mathbf{k}_{02}}$ or $A_{l_1s_1}=-A_{l_2s_2}$, ${\mathbf{k}_{01}}=-{\mathbf{k}_{02}}$ ) and repulsive ($A_{l_1s_1}=A_{l_2s_2}$, ${\mathbf{k}_{01}}= - {\mathbf{k}_{02}}$ or $A_{l_1s_1}=-A_{l_2s_2}$, ${\mathbf{k}_{01}}=  {\mathbf{k}_{02}}$) "magnetostatic-like" interactions $\mathbf{F}^M_{W12}$ of the strongly interacting unit-fields (particles). The {\it{purely attractive}} spin-independent (Lorentz-like) "non-quantum" interaction $\mathbf{F}^C_{W12}+\mathbf{F}^M_{W12}$ of the unit-field (particles) with the strong-interaction {\it{ACs}} having the spin numbers $s_n =0$, amplitudes $A_{l_1}=-A_{l_2}$ and momentums ${\mathbf{k}_{01}}=-{\mathbf{k}_{02}}$ could be considered as the "{\it{non-quantum}}" attraction of strongly interacting unit-fields [{\it{boson-like}} ($A_{l_1}=-A_{l_2}$, ${\mathbf{k}_{01}}=-{\mathbf{k}_{02}}$) particles] that explains microscopically the Bose-Einstein condensation and statistics of such particles. Although the "spin-less" ($s_n=0$) strongly interacting unit-fields (particles) have not been yet detected, the model  [Eqs. (\ref{eq469abc}) - (\ref{eq484abc})] is good for the (Coulomb-like and Lorentz-like) description of the strong-interaction forces.{\it{The spin-dependent (quantum) interaction between the strongly interacting unit-fields, which is observed experimentally, is described by the Lorentz-like spin-dependent (quantum) interaction energy ${\cal E}_{12,21}(R,s_1,s_2 )$ determining by Eqs. (\ref{eq172abc}), (\ref{eq173abc}), (\ref{eq469abc}) - (\ref{eq484abc}) with the spin numbers ($s_1=1$, $s_2=1$), ($s_1=-1$, $s_2=-1$), ($s_1=1$, $s_2=-1$) or ($s_1=-1$, $s_2=1$)}}. In other words, Eqs. (\ref{eq172abc}), (\ref{eq173abc}), (\ref{eq469abc}) - (\ref{eq484abc}) determine the interaction properties of the strongly interacting unit-field (particle) and their connections with the unit-field spin. A simple analysis of Eqs. (\ref{eq172abc}), (\ref{eq173abc}), (\ref{eq469abc}) - (\ref{eq484abc}) for the two unit-fields with the amplitudes $A_{l_1s_1}=A_{l_2s_2}$ shows that the equations describe the {{\it{repulsive}} ($s_1=s_2, A_{l_1s_1}=A_{l_2s_2},{\mathbf{k}_{01}}=  {\mathbf{k}_{02}}$) and {\it{attractive}} ($s_1\neq s_2, A_{l_1s_1}=A_{l_2s_2},{\mathbf{k}_{01}}=  {\mathbf{k}_{02}}$) {\it{quantum}} forces associated with the {\it{Lorentz-like spin-dependent (quantum) energy}} of the strongly interacting  fermion-like unit-fields (fermions). The Pauli exclusion principle states that no two identical ($s_1=s_2, A_{l_1s_1}=A_{l_2s_2},{\mathbf{k}_{01}}=  {\mathbf{k}_{02}}$) fermions may occupy the same quantum state simultaneously. The {\it{spin-dependent (quantum) repulsive or attractive strong-interaction of the unit-fields (fermions) with the amplitudes $A_{l_1s_1}=A_{l_2s_2}$ having respectively the same ($s_1=s_2$) or different ($s_1 \neq s_2$) spin numbers could be considered as the physical origin explaining microscopically the Pauli exclusion principle and the Fermi-Dirac statistics of the strongly interacting fermions, which in canonical quantum mechanics and SM have nature of the unexplained postulates}}. Unlike in the traditional quantum field theory and SM, which consider rather the fields of operators than the fields of particles, Eqs. (\ref{eq172abc}), (\ref{eq173abc}), (\ref{eq469abc}) - (\ref{eq484abc}) determine the "electromagnetic-like" properties of the strongly interacting unit-field (particle) and their connections with the unit-field spin naturally, without introduction of any pure mathematical object, like the spin matrix  or operators. It should be also noted that in the case of the two strongly interacting, {\it{normal}} unit-fields (particles) having the spin numbers $s_1=s_2=\pm 1$ and amplitudes $A_{l_1s_1}=-A_{l_2s_2}$, the unit-fields (particles) can annihilate. Indeed, the strongly interacting unit-fields are the solutions of the equations of motions of the present model for both the $A_{1s_n}$ and $-A_{1s_n}$ amplitudes. The {\it{Helmholtz-AC}} amplitude ${A_{1_ns_n}}$ can be an arbitrary constant [see, Eq. (\ref{eq327abc})]. The two strongly interacting unit-fields with $s_1=s_2$ can satisfy the annihilation condition in the case ${A_{1s_1}} = - {A_{1s_2}}$.  In Eqs. (\ref{eq469abc}) - (\ref{eq484abc}), the unit-field (particle) with the positive value of  ${A_{1s_1}}$ could be considered as a {\it{normal}} unit-field (particle), while the unit-field (particle) with the negative value of  ${A_{1m_2}}$ could be considered as an {\it{anti-unitfield}} ({\it{anti-particle}}). The annihilation of the two strongly interacting unit-fields with $s_1=s_2=0$ does not require adjustments of the unit-field spins. The easy annihilation of the strongly interacting unit-fields with $s_n=0$, if they really have been created under the cosmological Big Bang, could be considered as a physical  mechanism of the absence of such unit-fields (particles) in the present Universe. It should be stressed again that Eqs. (\ref{eq469abc}) - (\ref{eq484abc}) are valid in the particular case of the {\it{de Broglie generators}} $\tilde \psi_{01} (\mathbf{r},t)=a_0 e^{i({\mathbf{k}_{01}{\mathbf{r}}}-{\varepsilon}_{01} t )}$ (\ref{eq384abc}) and $\tilde \psi_{02} (\mathbf{r},t)=a_0 e^{i({\mathbf{k}_{02}{\mathbf{r}}}-{\varepsilon}_{02} t )}$ (\ref{eq385abc}). If the unit-field generators are not the de Broglie plane-waves, then the pure-strong interaction of the unit-fields (particles) are determined rather by Eqs. (\ref{eq364abc}) - (\ref{eq374abc}) than Eqs. (\ref{eq469abc}) - (\ref{eq484abc}). The above-presented model may be easily reformulated for any hypothetical unit-field (particle) with the spherically symmetric {\it{Helmholtz-ACs}} associated with the spherical Bessel functions (\ref{eq329abc}) of the second kind, if such elementary particles exist somewhere in the Universe. Notice, the spherically symmetric {\it{Helmholtz-ACs}} associated with the spherical Bessel functions (\ref{eq329abc}) of the second kind should have the finite external radius.

\subsection{7.4. The interference (interaction) of the unit-fields having the {\it{de Broglie generators}} ${\tilde \psi} _{0n}$ and the {\it{TACs}} containing the {\it{spherically symmetric, Laplace-ACs and  Helmholtz-ACs}} ${\phi_{in}}$: The unit-fields corresponding to the {\it{experimentally observed particles}} obeying the different combinations of the {\it{gravitational}}, {\it{electromagnetic}}, {\it{weak}} and {\it{strong}} interactions}

In the standard model of particle physics (SM), the quarks, anti-quarks, leptons, anti-leptons and gauge bosons are {\it{experimentally observed}} elementary particles, the building blocks of the {\it{gravity-less}} physical matter. All other particles are made from these elementary particles. Quarks (up, down, charm, strange, top, bottom) and Leptons (electron neutrino, electron, muon neutrino, muon, tau neutrino, tau) and the respective anti-quarks and anti-leptons are fermions. If the particles have electric charges, weak isospins, weak hypercharges and color charges, then they interact with each other electromagnetically, weakly and strongly by exchanging the respective gauge virtual particles. The gauge virtual particles (photons, $W$-bosons, $Z$-bosons, gluons and Higgs particles) are bosons. If the bosons have the weak isospins or weak hypercharges, then they demonstrate the weak interactions between each other. The particles and interaction of SM are summarized in Tab. \ref{tab:t2}. 
\begin{table}[b]
\centering
\begin{tabular}{ll}
\hline\noalign{\smallskip}
\multicolumn{2}{c} {Standard model of elementary particles and interactions (SM)}   \\
\hline\noalign{\smallskip}
\multicolumn{2}{c}{\it{Elementary particles }}   \\
\noalign{\smallskip}\hline\noalign{\smallskip}
{\it{Matter:}} Quarks and Leptons &  \\
\noalign{\smallskip}\hline\noalign{\smallskip}
{\it{Antimatter:}} Antiquarks and antileptons & \\
\noalign{\smallskip}\hline\noalign{\smallskip}
{\it{Carriers of forces:}} Photons, W-bosons and Z-bosons, Gluons &\\
\noalign{\smallskip}\hline
{\it{Composite particles}}: Hadrons (Mesons and Barions), nuclei, atoms and molecules & \\
\hline\noalign{\smallskip}
\multicolumn{2}{c}{\it{Interactions}}   \\
\noalign{\smallskip}\hline
{\it{Electromagnetic Force}}&{\it{Force carriers:}} Photons \\
\noalign{\smallskip}\hline
{\it{Weak Force}}&{\it{Force carriers:}} W- and Z-Bosons \\
\noalign{\smallskip}\hline
{\it{Strong Force}}&{\it{Force carriers:}} Gluons\\
\noalign{\smallskip}\hline
\end{tabular}
\caption{The particles and interactions according to the standard model of elementary particles and interactions (SM). Some non-standard, candidate theories ({\it{Grand Unified Theories}} or {\it{Theories of Everything}}) do include into SM also the Gravitational Force by considering the exchange of virtual gravitons (hypothetical particles) or the curving of space by the massive particles of SM.}
\label{tab:t2}
\end{table}
In the present section (Sec. 7.4), the unit-fields having the {\it{de Broglie generators}} ${\tilde \psi} _{0n}$ and the different {\it{TACs}} containing the {\it{spherically symmetric, Laplace-ACs and  Helmholtz-ACs}} ${\phi_{in}}$ obeying the different combinations of the {\it{gravitational}}, {\it{electromagnetic}}, {\it{weak}} and {\it{strong}} interactions are identified as the {\it{elementary particles}}, namely the {\it{gravitons}} (mass-less), {\it{photons}} (mass-less), {\it{quarks}} (massive), {\it{anti-quarks}} (massive), {\it{leptons}} (massive), {\it{anti-leptons}} (massive), {\it{W-bosons}} (massive), {\it{Z-bosons}} (massive) and {\it{gluons}} (massive) as follows. 

\vspace{0.4cm}
1. {\it{The mass-less unit-fields corresponding to the non-virtual gravitons. Gravitational interaction of gravitons with each other. The gravitational interaction of massive unit-fields with each other by the virtual exchange of the mass-less unit-fields (gravitons) or the curving of space by the massive unit-fields (particles)}}
\vspace{0.2cm}

In Sec. 7.2., the {\it{massive}} gravitationally interacting unit-fields having the {\it{de Broglie generators}} ${\tilde \psi} _{0n}$, which do not carry electric charges, weak isospins, weak hypercharges and color charges have been attributed  to the one-component unit-fields with the {\it{one-component TACs}} containing the {\it{spherically symmetric gravitational Laplace-ACs}} ${\phi_{Gn}}$. The {\it{mass-less}} one-component unit-fields corresponding to the non-virtual or virtual {\it{gravitons}} are different from the {\it{massive}} gravitationally interacting unit-fields. The {\it{mass-less}} unit-fields, which do not have the rest masses, electric charges, weak isospins, weak hypercharges and color charges, are the simplest unit-fields. The non-virtual, one-component, gravitational unit-fields ({\it{non-virtual gravitons}}) $\psi_{0n} (\mathbf{r},t)= \tilde \psi_{0n} (\mathbf{r},t) { \phi} _{Gn} (\mathbf{r},t)$ are described by Eq. (\ref{eq218abc}) with the rest mass $m_0=0$ and the index $I=G$,  which has the form 
\begin{eqnarray} \label{eq485abc}
\phi_{Gn}\square {\tilde \psi} _{0n}  + {\tilde \psi} _{0n}\square \phi_{Gn} = 0.
\end{eqnarray} 
The solutions of Eq. (\ref{eq485abc}) are the {\it{non-virtual gravitons}} 
\begin{eqnarray} \label{eq486abc}
\psi_{0n} (\mathbf{r},t)= a_{0n} e^{i({\mathbf{k}_{0n}{\mathbf{r}}}-{\varepsilon}_{0n} t +\alpha_n )},
\end{eqnarray} 
which have the {\it{structure-less}} form (\ref{eq53abc}) of the the time-harmonic plane waves with the de Broglie generators 
\begin{eqnarray} \label{eq487abc}
\tilde \psi_{0n} (\mathbf{r},t)={\sqrt  {a_{0n}}} e^{i({\mathbf{k}_{0n}{\mathbf{r}}}-{\varepsilon}_{0n} t +\alpha_n )/2}
\end{eqnarray} 
and the de Broglie associate-components 
\begin{eqnarray} \label{eq488abc}
\phi_{Gn} (\mathbf{r},t)={\sqrt  {a_{0n}}} e^{i({\mathbf{k}_{0n}{\mathbf{r}}}-{\varepsilon}_{0n}t + \alpha_n )/2} 
\end{eqnarray} 
that are {\it{indistinguishable from each other}}. In contrast to the {\it{massive}} gravitationally interacting unit-fields (particles), which obey the {\it{time-independent}} {\it{ACs}} ($\dot \phi_{Gn}=0$), the non-virtual, {\it{mass-less}}, one-component gravitational unit-fields ({\it{non-virtual gravitons}}) have the {\it{time-dependent}} {\it{ACs}} ($\dot \phi_{Gn}\neq 0$). The {\it{non-virtual gravitons}} have the zero rest-mass ($m_0=0$) with the Einstein energy-mass relation 
\begin{eqnarray} \label{eq489abc}
{\varepsilon}_{0n}^{2} =\mathbf{k}_{0n}^2 
\end{eqnarray} 
for the normalization ${a_{0n}}=1/\sqrt{V}$. The present model predicts interaction of the identical or almost identical non-virtual gravitons with each other. The relativistic gravitational interaction (cross-correlation) term (\ref{eq389abc}) for the two interacting {\it{mass-less}} gravitons located at the distance $R$ from each other has the form 
\begin{eqnarray} \label{eq490abc}
{\cal E}_{12,21}=   \mathbf{k}_{01}\mathbf{k}_{02} {\int_{V} 2 a_0^2 cos({\Delta  \mathbf{k}_{12}{\mathbf{r}}}-{\Delta {\varepsilon}_{12} t  } +{ {\mathbf{k}}_{02}{\mathbf{R}}}+ \Delta \alpha _{12} ) d^3x}.   
\end{eqnarray}
For comparison see the relativistic gravitational interaction (cross-correlation) term (\ref{eq401abc}) for the interacting {\it{massive}} unit-fields (particles) separated by the distance $R$. In the case of the identical ($\mathbf{k}_{01}= \mathbf{k}_{02}= \mathbf{k}_{0}$, $\Delta \alpha _{12}= \alpha _{1} - \alpha _{2}=0$ and $R=0$) gravitons, the energy-mass relation for the graviton pair is given by Eq. (\ref{eq89abc}), (\ref{eq122abc}) and (\ref{eq143abc}) as
\begin{eqnarray} \label{eq491abc}
{\varepsilon}^{2} = {\varepsilon}_{01}^{2} + {\varepsilon}_{02}^{2} + {\cal E}_{12,21}= 4\mathbf{k}_{0}^2, 
\end{eqnarray} 
which corresponds the maximum 
\begin{eqnarray} \label{eq492abc}
{\varepsilon}_{max}= 2\mathbf{k}_{0}, 
\end{eqnarray} 
of the graviton-pair energy. 
In the case of the almost identical ($\mathbf{k}_{01}\sim   \mathbf{k}_{02}$, $\Delta \alpha _{12}= \alpha _{1} - \alpha _{2} \sim   0$ and $R \sim   0$) gravitons, the energy squired for the graviton pair could vary in the region
\begin{eqnarray} \label{eq493abc}
0<{\varepsilon}< 2\mathbf{k}_{0}. 
\end{eqnarray} 
The minimum energy 
\begin{eqnarray} \label{eq494abc}
{\varepsilon}_{min} \approx  0. 
\end{eqnarray} 
is attributed to the different  ($\mathbf{k}_{01}\neq \mathbf{k}_{02}$, $\Delta \alpha _{12}\neq 0$ and $R \neq 0$) gravitons. An analysis of Eqs. (\ref{eq490abc}) - (\ref{eq494abc}) predicts the bunching and anti-bunching of gravitons as well as the Bose-Einstein condensation of such unit-fields (particles). Notice, the similar coherent phenomena are known in the traditional quantum electrodynamics describing the other mass-less particles, namely the photons. 

The relativistic gravitational interaction (interference) between the {\it{massive}} unit-fields (particles) separated by the distance $R$, which is governed by the relativistic gravitational interaction (cross-correlation) term (\ref{eq401abc}), can be considered as the virtual exchange of the above-described gravitons. In other words, the gravitational interaction may be reinterpreted as the exchange of virtual gravitons at the every time moment $t$. For instance, in the case of the two one-component unit-fields $\psi_{01} (\mathbf{r},t)= \tilde \psi_{01} (\mathbf{r},t) { \phi} _{G1} (\mathbf{r},t)$ and $\psi_{02} (\mathbf{r},t)= \tilde \psi_{02} (\mathbf{r},t) { \phi} _{G2} (\mathbf{r},t)$ interacting {\it{only gravitationally}}, the relativistic gravitational interaction (cross-correlation) total term (\ref{eq386abc}) - (\ref{eq389abc}), (\ref{eq401abc}) has the form         
\begin{eqnarray} \label{eq495abc}
{\cal E}_{12,21}(R,s_1,s_2 ) = \mathbf{k}_{01} \mathbf{k}_{02}f_G +   m_{01} m_{02}f_G , 
\end{eqnarray}
where the gravitational geometrical factor $f_G$ determining by Eq. (\ref{eq403abc}) may be represented as 
\begin{eqnarray} \label{eq496abc}
f_G(R,s_1,s_2 ) =  {{{A_{1s_1}} {A_{1s_2}}} { a_0^2}} \int_{\theta =0}^\pi d\theta sin\theta  \int_{\varphi  =0}^{2\pi} d \varphi  \int_{r_{0}^G}^{R-r_{0,low}^G/ 2} \int_{R+r_{0,up}^G / 2}^{\infty} dr {\frac {f(r,R,\theta , \varphi , s_1, s_2 ) } {| {\bf{r}}-{\bf{R}}|^{2}}} \cdot  \nonumber  \\ \cdot [ e^{-i({\mathbf{k}_{01}{\mathbf{r}}}-{\varepsilon}_{01} t +\alpha_1 )}  e^{i({\mathbf{k}_{02}{\mathbf{r}}}-{\varepsilon}_{02} t +\alpha_2 )} +  e^{i({\mathbf{k}_{01}{\mathbf{r}}}-{\varepsilon}_{01} t +\alpha_1 )}  e^{-i({\mathbf{k}_{02}{\mathbf{r}}}-{\varepsilon}_{02} t +\alpha_2 )}]/2
\end{eqnarray}
with $\mathbf{k}_{01}=\mathbf{k}_{02}$, ${\varepsilon}_{01}={\varepsilon}_{02}$, $\alpha_1 = \alpha_2$ and $a_{01}=a_{02}$. In such a case, the interaction could be reinterpreted as the virtual exchange of real gravitons or simply as the {\it{exchange of virtual gravitons}}. The present model of the interference (interaction) of the unit-fields, which are not the point particles, does not really require the introduction of the virtual point-particles because such a reinterpretation does not give new inside the mechanism of gravitational interactions. Nevertheless, the use of the virtual unit-fields (gravitons) provides a background for comparison of the present model with the traditional quantum field theories, where the point-particles separated by the vacuum {\it{may interact with each other only by the exchange of virtual particles}}. In the terms of the Einstein general relativity, the gravitational interaction determining by Eqs. (\ref{eq495abc}) and (\ref{eq496abc}) may be reinterpreted as the consequence of the curving space by the massive unit-fields (particles). Indeed, the gravitational geometrical factor $f_G(R,s_1,s_2 )=const.$ could be attributed to the "non-curved" space with the freely moving particles, while the gravitational geometrical factor $f_G(R,s_1,s_2 )\neq const.$ could describe the gravitationally interacting unit-fields (particles). In the present model of the gravitational interaction of massive unit-fields, the virtual exchange of the mass-less unit-fields (gravitons) is equivalent to the curving of space by the massive unit-fields (particles). Also note the particles (gravitons) with the negative values of the unit-field amplitudes have been attributed in Sec. 7.2. to the ant-particles (anti-gravitons).

\vspace{0.2cm}
2. {\it{The mass-less unit-fields corresponding to the non-virtual photons. Electromagnetic interaction of photons with each other. The electromagnetic interaction of massive electrically charged unit-fields with each other by the virtual exchange of the mass-less unit-fields (photons)}}
\vspace{0.2cm}

The description of the mass-less unit-fields corresponding to the non-virtual and virtual photons is {\it{quite similar}} to the above-presented model of gravitons. Although the gravitons and photons have the unit-field amplitudes of different natures, the equations describing the photons are indistinguishable from the respective equations describing the gravitons. Indeed, the {\it{massive}} electromagnetically interacting unit-fields having the {\it{de Broglie generators}} ${\tilde \psi} _{0n}$, which do carry masses and electric charges but do not have weak isospins, weak hypercharges  and color charges, have been considered in Sec. 7.2. as the one-component unit-fields with the {\it{one-component TACs}} containing the {\it{spherically symmetric electrical Laplace-ACs}} ${\phi_{Cn}}$. The {\it{mass-less}} one-component unit-fields corresponding to the non-virtual or virtual {\it{photons}} are different from the {\it{massive}} electromagnetically interacting unit-fields. The {\it{mass-less}} unit-fields, which do not have the rest masses, electric charges, weak isospins, weak hypercharges and color charges, are the simplest unit-fields. The non-virtual, one-component, electromagnetic unit-fields ({\it{non-virtual photons}}) $\psi_{0n} (\mathbf{r},t)= \tilde \psi_{0n} (\mathbf{r},t) { \phi} _{Cn} (\mathbf{r},t)$ are described by Eq. (\ref{eq218abc}) with the rest mass $m_0=0$ and the index $I=C$,  which has the form 
\begin{eqnarray} \label{eq497abc}
\phi_{Cn}\square {\tilde \psi} _{0n}  + {\tilde \psi} _{0n}\square \phi_{Cn} = 0.
\end{eqnarray} 
The solutions of Eq. (\ref{eq497abc}) are the {\it{non-virtual photons}} 
\begin{eqnarray} \label{eq498abc}
\psi_{0n} (\mathbf{r},t)= a_{0n} e^{i({\mathbf{k}_{0n}{\mathbf{r}}}-{\varepsilon}_{0n} t +\alpha_n )},
\end{eqnarray} 
which have the {\it{structure-less}} form (\ref{eq53abc}) of the the time-harmonic plane waves with the de Broglie generators 
\begin{eqnarray} \label{eq499abc}
\tilde \psi_{0n} (\mathbf{r},t)={\sqrt  {a_{0n}}} e^{i({\mathbf{k}_{0n}{\mathbf{r}}}-{\varepsilon}_{0n} t +\alpha_n )/2}
\end{eqnarray} 
and the de Broglie associate-components 
\begin{eqnarray} \label{eq500abc}
\phi_{Cn} (\mathbf{r},t)={\sqrt  {a_{0n}}} e^{i({\mathbf{k}_{0n}{\mathbf{r}}}-{\varepsilon}_{0n}t + \alpha_n )/2} 
\end{eqnarray} 
that are {\it{indistinguishable from each other}}. Notice, that Eqs. (\ref{eq498abc}) - (\ref{eq500abc}) describing the photons are indistinguishable from Eqs. (\ref{eq486abc}) - (\ref{eq488abc}) describing the gravitons. In contrast to the {\it{massive}} electromagnetically interacting unit-fields (particles), which obey the {\it{time-independent}} {\it{ACs}} ($\dot \phi_{Cn}=0$), the non-virtual, {\it{mass-less}}, one-component electromagnetic unit-fields ({\it{non-virtual photons}}) have the {\it{time-dependent}} {\it{ACs}} ($\dot \phi_{Cn}\neq 0$). The {\it{non-virtual photons}} have the zero rest-mass ($m_0=0$) with the Einstein energy-mass relation 
\begin{eqnarray} \label{eq501abc}
{\varepsilon}_{0n}^{2} =\mathbf{k}_{0n}^2 
\end{eqnarray} 
for the normalization ${a_{0n}}=1/\sqrt{V}$. The present model predicts interaction of the identical or almost identical non-virtual photons with each other. The relativistic electromagnetic interaction (cross-correlation) term (\ref{eq389abc}) for the two interacting {\it{mass-less}} photons located at the distance $R$ from each other has the form 
\begin{eqnarray} \label{eq502abc}
{\cal E}_{12,21}=   \mathbf{k}_{01}\mathbf{k}_{02} {\int_{V} 2 a_0^2 cos({\Delta  \mathbf{k}_{12}{\mathbf{r}}}-{\Delta {\varepsilon}_{12} t  } +{ {\mathbf{k}}_{02}{\mathbf{R}}}+ \Delta \alpha _{12} ) d^3x}.   
\end{eqnarray}
For comparison see the relativistic electromagnetic interaction (cross-correlation) term (\ref{eq429abc}) for the interacting {\it{massive}} unit-fields (particles) separated by the distance $R$. In the case of the identical ($\mathbf{k}_{01}= \mathbf{k}_{02}= \mathbf{k}_{0}$, $\Delta \alpha _{12}= \alpha _{1} - \alpha _{2}=0$ and $R=0$) gravitons, the energy-mass relation for the photon pair is given by Eq. (\ref{eq89abc}), (\ref{eq122abc}) and (\ref{eq143abc}) as
\begin{eqnarray} \label{eq503abc}
{\varepsilon}^{2} = {\varepsilon}_{01}^{2} + {\varepsilon}_{02}^{2} + {\cal E}_{12,21}= 4\mathbf{k}_{0}^2, 
\end{eqnarray} 
which corresponds the maximum 
\begin{eqnarray} \label{eq504abc}
{\varepsilon}_{max}= 2\mathbf{k}_{0}, 
\end{eqnarray} 
of the photon-pair energy. 
In the case of the almost identical ($\mathbf{k}_{01}\sim   \mathbf{k}_{02}$, $\Delta \alpha _{12}= \alpha _{1} - \alpha _{2} \sim   0$ and $R \sim   0$) photons, the energy squired for the photon pair could vary in the region
\begin{eqnarray} \label{eq505abc}
0<{\varepsilon}< 2\mathbf{k}_{0}. 
\end{eqnarray} 
The minimum energy 
\begin{eqnarray} \label{eq506abc}
{\varepsilon}_{min} \approx  0. 
\end{eqnarray} 
is attributed to the different  ($\mathbf{k}_{01}\neq \mathbf{k}_{02}$, $\Delta \alpha _{12}\neq 0$ and $R \neq 0$) photons. A simple analysis of Eqs. (\ref{eq497abc}) - (\ref{eq506abc}) predicts the bunching and anti-bunching of photons as well as the Bose-Einstein condensation of such unit-fields (particles). The similar coherent phenomena are known in the traditional quantum electrodynamics describing the behavior of photons. 

The relativistic electromagnetic interaction (interference) between the {\it{electrically charged}} unit-fields (particles) separated by the distance $R$, which is governed by the relativistic gravitational interaction (cross-correlation) term (\ref{eq401abc}), can be considered as the virtual exchange of the above-described photons. That is to say that the electromagnetic interaction may be reinterpreted as the exchange of virtual photons at the every time moment $t$. For instance, in the case of the two one-component unit-fields $\psi_{01} (\mathbf{r},t)= \tilde \psi_{01} (\mathbf{r},t) { \phi} _{C1} (\mathbf{r},t)$ and $\psi_{02} (\mathbf{r},t)= \tilde \psi_{02} (\mathbf{r},t) { \phi} _{C2} (\mathbf{r},t)$ interacting {\it{only electromagnetically}}, the relativistic electromagnetic interaction (cross-correlation) total term (\ref{eq386abc}) - (\ref{eq389abc}), (\ref{eq429abc}) has the form         
\begin{eqnarray} \label{eq507abc}
{\cal E}_{12,21}(R,s_1,s_2 ) = \mathbf{k}_{01} \mathbf{k}_{02}f_C +   m_{01} m_{02}f_C , 
\end{eqnarray}
where the electromagnetic geometrical factor $f_C$ determining by Eq. (\ref{eq431abc}) may be represented as 
\begin{eqnarray} \label{eq508abc}
f_C(R,s_1,s_2 ) =  {{{A_{1s_1}} {A_{1s_2}}} { a_0^2}} \int_{\theta =0}^\pi d\theta sin\theta  \int_{\varphi  =0}^{2\pi} d \varphi  \int_{r_{0}^C}^{R-r_{0,low}^C/ 2} \int_{R+r_{0,up}^C / 2}^{\infty} dr {\frac {f(r,R,\theta , \varphi , s_1, s_2 ) } {| {\bf{r}}-{\bf{R}}|^{2}}}  \cdot  \nonumber  \\ \cdot [ e^{-i({\mathbf{k}_{01}{\mathbf{r}}}-{\varepsilon}_{01} t +\alpha_1 )}  e^{i({\mathbf{k}_{02}{\mathbf{r}}}-{\varepsilon}_{02} t +\alpha_2 )} +  e^{i({\mathbf{k}_{01}{\mathbf{r}}}-{\varepsilon}_{01} t +\alpha_1 )}  e^{-i({\mathbf{k}_{02}{\mathbf{r}}}-{\varepsilon}_{02} t +\alpha_2 )}]/2
\end{eqnarray}
with $\mathbf{k}_{01}=\mathbf{k}_{02}$, ${\varepsilon}_{01}={\varepsilon}_{02}$, $\alpha_1 = \alpha_2$ and $a_{01}=a_{02}$. The representation (\ref{eq508abc}) gives rise to the interpretation of the electromagnetic interaction as the virtual exchange of real photons or simply as the {\it{exchange of virtual photons}}. The present model of the interference (interaction) of the unit-fields, which are not the point particles, does not really require the introduction of the virtual point-particles because such a reinterpretation does not gives new inside the mechanism of electromagnetic  interactions. Nevertheless, the use of the virtual unit-fields (photons) provides a background for comparison of the present model with the traditional quantum electrodynamics, where the electrically charged point-particles separated by the vacuum {\it{may interact with each other only by the exchange of virtual photons}}. Also note the particles (photons) with the negative values of the unit-field amplitudes have been attributed in Sec. 7.2. to the ant-particles (anti-photons).

\vspace{0.2cm}
3. {\it{The massive unit-fields corresponding to the electrically charged leptons (electron, muon and  tau). Gravitational and electromagnetic interactions of the leptons with each other}}
\vspace{0.2cm}

The {\it{massive}} gravitationally and electromagnetically interacting unit-fields having the {\it{de Broglie generators}} ${\tilde \psi} _{0n}$, which do carry masses and electric charges but do not have weak isospins, weak hypercharges and color charges, should be attributed to the two-component unit-fields $\psi_{0n} (\mathbf{r},t)= \tilde \psi_{0n} (\mathbf{r},t) [{ \phi} _{Gn} (\mathbf{r},t) + { \phi} _{Cn} (\mathbf{r},t)]$ associated with the {\it{electrically charged leptons (electron, muon and  tau)}}. In the case of the two electrically charged leptons $\psi_{01} (\mathbf{r},t)= \tilde \psi_{01} (\mathbf{r},t) [{ \phi} _{G1} (\mathbf{r},t) + { \phi} _{Cn1} (\mathbf{r},t)]$ and $\psi_{02} (\mathbf{r},t)= \tilde \psi_{02} (\mathbf{r},t) [{ \phi} _{G2} (\mathbf{r},t) + { \phi} _{C2} (\mathbf{r},t)]$, the relativistic interaction (cross-correlation) total term is given by the field superposition principle as          
\begin{eqnarray} \label{eq509abc}
{\cal E}_{12,21}(R,s_1,s_2 ) = \mathbf{k}_{01} \mathbf{k}_{02}[f_G + f_C]+ m_{01} m_{02}[f_G + f_C], 
\end{eqnarray}
where the gravitational geometrical factor $f_G$ determining by Eq. (\ref{eq403abc}) may be represented as
\begin{eqnarray} \label{eq510abc}
f_G(R,s_1,s_2 ) =  {{{A_{1s_1}} {A_{1s_2}}} { a_0^2}} \int_{\theta =0}^\pi d\theta sin\theta  \int_{\varphi  =0}^{2\pi} d \varphi  \int_{r_{0}^G}^{R-r_{0,low}^G/ 2} \int_{R+r_{0,up}^G / 2}^{\infty} dr {\frac {f(r,R,\theta , \varphi , s_1, s_2 ) } {| {\bf{r}}-{\bf{R}}|^{2}}} \cdot  \nonumber  \\ \cdot [ e^{-i({\mathbf{k}_{01}{\mathbf{r}}}-{\varepsilon}_{01} t +\alpha_1 )}  e^{i({\mathbf{k}_{02}{\mathbf{r}}}-{\varepsilon}_{02} t +\alpha_2 )} +  e^{i({\mathbf{k}_{01}{\mathbf{r}}}-{\varepsilon}_{01} t +\alpha_1 )}  e^{-i({\mathbf{k}_{02}{\mathbf{r}}}-{\varepsilon}_{02} t +\alpha_2 )}]/2 
\end{eqnarray}
and the electric geometrical factor $f_C$ determining by Eq. (\ref{eq431abc}) may be represented as
\begin{eqnarray} \label{eq511abc}
f_C(R,s_1,s_2 ) =  {{{A_{1s_1}} {A_{1s_2}}} { a_0^2}} \int_{\theta =0}^\pi d\theta sin\theta  \int_{\varphi  =0}^{2\pi} d \varphi  \int_{r_{0}^C}^{R-r_{0,low}^C/ 2} \int_{R+r_{0,up}^C / 2}^{\infty} dr {\frac {f(r,R,\theta , \varphi , s_1, s_2 ) } {| {\bf{r}}-{\bf{R}}|^{2}}} \cdot  \nonumber  \\ \cdot [ e^{-i({\mathbf{k}_{01}{\mathbf{r}}}-{\varepsilon}_{01} t +\alpha_1 )}  e^{i({\mathbf{k}_{02}{\mathbf{r}}}-{\varepsilon}_{02} t +\alpha_2 )} +  e^{i({\mathbf{k}_{01}{\mathbf{r}}}-{\varepsilon}_{01} t +\alpha_1 )}  e^{-i({\mathbf{k}_{02}{\mathbf{r}}}-{\varepsilon}_{02} t +\alpha_2 )}]/2.
\end{eqnarray}
with $\mathbf{k}_{01}=\mathbf{k}_{02}$, ${\varepsilon}_{01}={\varepsilon}_{02}$, $\alpha_1 = \alpha_2$ and $a_{01}=a_{02}$. The representation (\ref{eq510abc}) gives rise to the reinterpretation of the gravitational interaction as the virtual exchange of real gravitons or simply as the {\it{exchange of virtual gravitons}}. Similarly, the representation (\ref{eq518abc}) may be reinterpreted as the electromagnetic interaction by the virtual exchange of real photons or simply as the {\it{exchange of virtual photons}}. The electrically charged leptons (electron, muon and  tau) are fermions ($s_n \neq 0$). Nevertheless, the case of the spins $s_1=s_2=0$ is instructive for understanding of the gravitational and electromagnetic interactions. If the spins $s_1=s_2=0$, then the relativistic interaction (cross-correlation) total term is given by Eqs. (\ref{eq405abc}), (\ref{eq433abc}) and (\ref{eq497abc}) as
\begin{eqnarray} \label{eq512abc}
{\cal E}_{12,21}(R) = {\frac {\mathbf{k}_{01}} {m_{01}} } {\frac {\mathbf{k}_{02}} {m_{02}} } \left[ 2\gamma _G{\frac {m_{01} m_{02}} {R} } \right]  +  \left[ 2\gamma _G{\frac {m_{01} m_{02}} {R} } \right]  + \nonumber  \\ + {\frac {\mathbf{k}_{01}} {m_{01}} } {\frac {\mathbf{k}_{02}} {m_{02}} } \left[ 2\gamma _C{\frac {q_{01} q_{02}} {R} } \right]  +     \left[ 2\gamma _C{\frac {q_{01} q_{02}} {R} } \right]. 
\end{eqnarray}
Formally, the relativistic interaction (cross-correlation) total term (\ref{eq512abc}) mediates the superposition of the Lorentz-like gravitational and electromagnetic interactions (forces ${\bf{F}}_{12}(R)=-{\bf{F}}_{21}(R)$), which compare well with the non-quantum gravitation and electromagnetism associated with the non-existing {\it{electrically charged particles}} having the spins $s_1=s_2=0$. The relativistic interaction (cross-correlation) total term (\ref{eq509abc}) mediates the superposition of the Lorentz-like gravitational [(\ref{eq407abc}), (\ref{eq408abc})] and electromagnetic [(\ref{eq435abc}), (\ref{eq436abc})] interactions (forces) (see, Sec. 7.2.) between the {\it{electrically charged leptons (electron, muon and  tau)}} carrying the masses and electric charges. Also note the particles (electrically charged leptons) with the negative values of the unit-field amplitudes have been attributed in Sec. 7.2. to the ant-particles (electrically charged antileptons).

\vspace{0.2cm}
4. {\it{The massive unit-fields corresponding to the leptons (electron neutrino, muon neutrino and tau neutrino) carrying the weak isospins or weak hypercharges. Gravitational and weak interactions of the leptons with each other}}
\vspace{0.2cm}

The {\it{massive}} gravitationally and weakly interacting unit-fields having the {\it{de Broglie generators}} ${\tilde \psi} _{0n}$, which do carry masses and weak isospins/hypercharges but do not have the electrical and color charges, should be attributed to the two-component unit-fields $\psi_{0n} (\mathbf{r},t)= \tilde \psi_{0n} (\mathbf{r},t) [{ \phi} _{Gn} (\mathbf{r},t) + { \phi} _{Wn} (\mathbf{r},t)]$ associated with the the electrically uncharged leptons (electron neutrino, muon neutrino and tau neutrino). In the case of the two leptons $\psi_{01} (\mathbf{r},t)= \tilde \psi_{01} (\mathbf{r},t) [{ \phi} _{G1} (\mathbf{r},t) + { \phi} _{Wn1} (\mathbf{r},t)]$ and $\psi_{02} (\mathbf{r},t)= \tilde \psi_{02} (\mathbf{r},t) [{ \phi} _{G2} (\mathbf{r},t) + { \phi} _{W2} (\mathbf{r},t)]$, the relativistic interaction (cross-correlation) total term is given by the field superposition principle as          
\begin{eqnarray} \label{eq513abc}
{\cal E}_{12,21}(R,s_1,s_2 ) = (\mathbf{k}_{01} \mathbf{k}_{02}f_G + m_{01} m_{02}f_G) + (\mathbf{k}_{01} \mathbf{k}_{02} f_W  +   \left[ m_{01}m_{02} + {\Gamma _{W1}} {\Gamma _{W2}} \right] f_W), 
\end{eqnarray}
where the gravitational geometrical factor $f_G$ determining by Eq. (\ref{eq403abc}) is presented as
\begin{eqnarray} \label{eq514abc}
f_G(R,s_1,s_2 ) =  {{{A_{1s_1}} {A_{1s_2}}} { a_0^2}} \int_{\theta =0}^\pi d\theta sin\theta  \int_{\varphi  =0}^{2\pi} d \varphi  \int_{r_{0}^G}^{R-r_{0,low}^G/ 2} \int_{R+r_{0,up}^G / 2}^{\infty} dr {\frac {f(r,R,\theta , \varphi , s_1, s_2 ) } {| {\bf{r}}-{\bf{R}}|^{2}}} \cdot  \nonumber  \\ \cdot [ e^{-i({\mathbf{k}_{01}{\mathbf{r}}}-{\varepsilon}_{01} t +\alpha_1 )}  e^{i({\mathbf{k}_{02}{\mathbf{r}}}-{\varepsilon}_{02} t +\alpha_2 )} +  e^{i({\mathbf{k}_{01}{\mathbf{r}}}-{\varepsilon}_{01} t +\alpha_1 )}  e^{-i({\mathbf{k}_{02}{\mathbf{r}}}-{\varepsilon}_{02} t +\alpha_2 )}]/2 
\end{eqnarray}
with $\mathbf{k}_{01}=\mathbf{k}_{02}$, ${\varepsilon}_{01}={\varepsilon}_{02}$, $\alpha_1 = \alpha_2$ and $a_{01}=a_{02}$, and the weak-interaction geometrical factor $f_W$ determining by Eq. (\ref{eq455abc}) has the form
\begin{eqnarray} \label{eq515abc}
f_W(R,s_1,s_2 ) =  {{{A_{l_1s_1}A_{l_2s_2}}} a_0^2} \int_{\theta =0}^\pi d\theta sin\theta  \int_{\varphi  =0}^{2\pi} d \varphi \int_{r_0^W}^{R-r_{0,low}^W/ 2} \int_{R+r_{0,up}^W/ 2}^{r_{up}^W}r^2 dr  j_{l_1} ({\Gamma _{W1}}  r)  j_{l_2} ({\Gamma  _{W2}} | {\bf{r}}-{\bf{R}}| ) f. 
\end{eqnarray}
The representation (\ref{eq514abc}) gives rise to the reinterpretation of the gravitational interaction as the virtual exchange of real {\it{mass-less}} particles (gravitons) or simply as the {\it{exchange of virtual gravitons}}. While the form (\ref{eq515abc}) could be reinterpreted according to Sec. 7.3 as the weak interaction by the virtual exchange of real {\it{massive}} particles ({\it{W and Z bosons}}). The relativistic interaction (cross-correlation) total term (\ref{eq513abc}) mediates the superposition of the Lorentz-like gravitational [(\ref{eq407abc}), (\ref{eq408abc})] and weak [(\ref{eq463abc}), (\ref{eq464abc})] interactions (forces ${\bf{F}}_{12}(R)=-{\bf{F}}_{21}(R)$) [for details, see Sec. 7.2.] between the electrically uncharged leptons (electron neutrino, muon neutrino and tau neutrino) carrying the masses and weak isospins/hypercharges. Also note the particles (electrically uncharged leptons) with the negative values of the unit-field amplitudes have been attributed in Sec. 7.3. to the ant-particles (electrically uncharged antileptons).

\vspace{0.2cm}
5. {\it{The massive unit-fields corresponding to the quarks (up, down, charm, strange, top and bottom) carrying the masses, electric charges, weak isospins, weak hypercharges and color charges. Gravitational, electromagnetic, weak and strong interactions of the quarks with each other}}
\vspace{0.2cm}

The {\it{massive}} gravitationally and weakly interacting unit-fields having the {\it{de Broglie generators}} ${\tilde \psi} _{0n}$, which do carry masses, electric charges, weak isospins, weak hypercharges and color charges, should be attributed to the two-component unit-fields $\psi_{0n} (\mathbf{r},t)= \tilde \psi_{0n} (\mathbf{r},t) [{ \phi} _{Gn} (\mathbf{r},t) + { \phi} _{Cn} (\mathbf{r},t) + { \phi} _{Wn} (\mathbf{r},t) + { \phi} _{Sn} (\mathbf{r},t) ]$ associated with the quarks (up, down, charm, strange, top and bottom). In the case of the two quarks $\psi_{01} (\mathbf{r},t)= \tilde \psi_{01} (\mathbf{r},t) [{ \phi} _{G1} (\mathbf{r},t) + { \phi} _{C1} (\mathbf{r},t) + { \phi} _{W1} (\mathbf{r},t) + { \phi} _{S1} (\mathbf{r},t)]$ and $\psi_{02} (\mathbf{r},t)= \tilde \psi_{02} (\mathbf{r},t) [{ \phi} _{Gn} (\mathbf{r},t) + { \phi} _{C2} (\mathbf{r},t) + { \phi} _{W2} (\mathbf{r},t) + { \phi} _{S2} (\mathbf{r},t)]$, the relativistic interaction (cross-correlation) total term is given by the field superposition principle as          
\begin{eqnarray} \label{eq516abc}
{\cal E}_{12,21}(R,s_1,s_2 ) = (\mathbf{k}_{01} \mathbf{k}_{02}f_G + m_{01} m_{02}f_G) + (\mathbf{k}_{01} \mathbf{k}_{02}f_C + m_{01} m_{02}f_C) +  \nonumber  \\ + (\mathbf{k}_{01} \mathbf{k}_{02} f_W  +   \left[ m_{01}m_{02} + {\Gamma _{W1}} {\Gamma _{W2}} \right] f_W) +  (\mathbf{k}_{01} \mathbf{k}_{02} f_S  +   \left[ m_{01}m_{02} + {\Gamma _{S1}} {\Gamma _{S2}} \right] f_S), 
\end{eqnarray}
where the gravitational geometrical factor $f_G$ determining by Eq. (\ref{eq403abc}) is represented as
\begin{eqnarray} \label{eq517abc}
f_G(R,s_1,s_2 ) =  {{{A_{1s_1}} {A_{1s_2}}} { a_0^2}} \int_{\theta =0}^\pi d\theta sin\theta  \int_{\varphi  =0}^{2\pi} d \varphi  \int_{r_{0}^G}^{R-r_{0,low}^G/ 2} \int_{R+r_{0,up}^G / 2}^{\infty} dr {\frac {f(r,R,\theta , \varphi , s_1, s_2 ) } {| {\bf{r}}-{\bf{R}}|^{2}}} \cdot  \nonumber  \\ \cdot [ e^{-i({\mathbf{k}_{01}{\mathbf{r}}}-{\varepsilon}_{01} t +\alpha_1 )}  e^{i({\mathbf{k}_{02}{\mathbf{r}}}-{\varepsilon}_{02} t +\alpha_2 )} +  e^{i({\mathbf{k}_{01}{\mathbf{r}}}-{\varepsilon}_{01} t +\alpha_1 )}  e^{-i({\mathbf{k}_{02}{\mathbf{r}}}-{\varepsilon}_{02} t +\alpha_2 )}]/2,
\end{eqnarray}
the electric geometrical factor $f_C$ determining by Eq. (\ref{eq431abc}) is represented as
\begin{eqnarray} \label{eq518abc}
f_C(R,s_1,s_2 ) =  {{{A_{1s_1}} {A_{1s_2}}} { a_0^2}} \int_{\theta =0}^\pi d\theta sin\theta  \int_{\varphi  =0}^{2\pi} d \varphi  \int_{r_{0}^C}^{R-r_{0,low}^C/ 2} \int_{R+r_{0,up}^C / 2}^{\infty} dr {\frac {f(r,R,\theta , \varphi , s_1, s_2 ) } {| {\bf{r}}-{\bf{R}}|^{2}}} \cdot  \nonumber  \\ \cdot [ e^{-i({\mathbf{k}_{01}{\mathbf{r}}}-{\varepsilon}_{01} t +\alpha_1 )}  e^{i({\mathbf{k}_{02}{\mathbf{r}}}-{\varepsilon}_{02} t +\alpha_2 )} +  e^{i({\mathbf{k}_{01}{\mathbf{r}}}-{\varepsilon}_{01} t +\alpha_1 )}  e^{-i({\mathbf{k}_{02}{\mathbf{r}}}-{\varepsilon}_{02} t +\alpha_2 )}]/2
\end{eqnarray}
with $\mathbf{k}_{01}=\mathbf{k}_{02}$, ${\varepsilon}_{01}={\varepsilon}_{02}$, $\alpha_1 = \alpha_2$ and $a_{01}=a_{02}$. The weak-interaction geometrical factor $f_W$ determining by Eq. (\ref{eq455abc}) has the form
\begin{eqnarray} \label{eq519abc}
f_W(R,s_1,s_2 ) =  {{{A_{l_1s_1}A_{l_2s_2}}} a_0^2} \int_{\theta =0}^\pi d\theta sin\theta  \int_{\varphi  =0}^{2\pi} d \varphi \int_{r_0^W}^{R-r_{0,low}^W/ 2} \int_{R+r_{0,up}^W/ 2}^{r_{up}^W}r^2 dr  j_{l_1} ({\Gamma _{W1}}  r)  j_{l_2} ({\Gamma  _{W2}} | {\bf{r}}-{\bf{R}}| ) f 
\end{eqnarray}
and the strong-interaction geometrical factor $f_W$ determining by Eq. (\ref{eq473abc}) has the form
\begin{eqnarray} \label{eq520abc}
f_S(R,l_1,l_2,s_1,s_2 )  =   \nonumber  \\  =  {{{A_{l_1s_1}A_{l_2s_2}}} a_0^2} \int_{\theta =0}^\pi d\theta sin\theta  \int_{\varphi  =0}^{2\pi} d \varphi \int_{r_0^S}^{R-r_{0,low}^S/ 2} \int_{R+r_{0,up}^S/ 2}^{r_{up}^S}r^2 dr  j_{l_1} (i|{\Gamma _{S1}}|  r)  j_{l_2} (i|{\Gamma  _{S2}}| | {\bf{r}}-{\bf{R}}| ) f .
\end{eqnarray}
The representations (\ref{eq517abc}) and (\ref{eq518abc}) give rise to the reinterpretation of the gravitational and  electromagnetic interaction as the exchange of virtual gravitons ({\it{mass-less gravitational particles)}} and the {exchange of virtual photons (\it{mass-less electromagnetic particles)}}, respectively. While the forms (\ref{eq519abc}) and (\ref{eq520abc}) could be reinterpreted according to Sec. 7.2 respectively as the weak interaction by the virtual exchange of the {\it{massive}}, weakly-interacting particles ({\it{W and Z bosons}}) and the strong  interaction by the virtual exchange of the {\it{massive}}, strongly-interacting particles ({\it{gluons}}). The virtual photons and gravitons are the mass-less particles associated with the gauge masses ${\Gamma _{G}}={\Gamma _{EM}}=0$, while the gauge masses of the W and Z bosons are associated with the eigen-parameters ${\Gamma _{W}}\neq 0$ and ${\Gamma _{S}}\neq 0$.  The relativistic interaction (cross-correlation) total term (\ref{eq516abc}) mediates the superposition of the Lorentz-like gravitational [(\ref{eq407abc}), (\ref{eq408abc})], electromagnetic [(\ref{eq435abc}), (\ref{eq436abc})], weak [(\ref{eq463abc}), (\ref{eq464abc})] and strong [(\ref{eq482abc}), (\ref{eq483abc})] interactions (forces ${\bf{F}}_{12}(R)=-{\bf{F}}_{21}(R)$) (see, Sec. 7.3.) between the quarks (up, down, charm, strange, top and bottom) carrying the masses, electric charges, weak isospins, weak hypercharges and color charges. Also note the particles (quarks) with the negative values of the unit-field amplitudes have been attributed in Sec. 7.3. to the ant-particles (antiquarks). The comparison of the gravitational ($f_G$), electromagnetic ($f_{EM}$), weak ($f_W$) and  strong ($f_S$) geometric factors presents a physical explanation why the field strength of the strong interaction is a few orders of magnitude higher than that of the weak force, electromagnetism and gravitation. The different dependences of the geometric factors on the distance $R$ give a natural explanation of the long range of the gravitation and electromagnetism and the short range of the weak and strong interactions.

\section{8. The particles, interactions and forces according to the present model unifying the all-known elementary particles and interactions}

Unification of the all-known particles, interactions and forces by the present model is summarized in Tab. \ref{tab:t3}. The present model considers an {\it{elementary particle}} as a {\it{material unit-field of mass-energy}}. The {\it{unified interaction}} between unit-fields (particles) is considered to be mediated by the {\it{cross-correlation energy}} connected with the {\it{interference}} between the unit-fields. The {\it{unified force}} between unit-fields (particles) is attributed to the {\it{gradient}} of the cross-correlation energy.
A unit-field ${\psi _{0n}}$ obeys the internal structure associated with the generator ${\tilde \psi _{0n}}$ and the total associate-component  $\Phi^{a}_{n}$ of the unit-field, ${\psi _{0n}}={\tilde \psi _{0n}}\Phi^{a}_{n}$. In other words, the unit-fields are distinguished from each other by the generators ${\tilde \psi _{0n}}$ and the total associate-components  $\Phi^{a}_{n}$. The free unit-fields with different generators have the different values of the unit-field momentum. The differences in the associate components yields the different masses, electric charges, weak isospins, weak hypercharges and color charges of the unit-fields. That leads to the differences in the interactions and forces between the unit-fields (particles). The Gravitational Force is mediated by the interference of massive unit-fields, which may be reinterpreted as the exchange of virtual gravitons (hypothetical mass-less particles) or as the relativistic curving of empty space by the massive particles. 

If the unit-field generators have the configuration of the de Broglie time-harmonic plane-waves, then the gravitational, electromagnetic, weak and strong interactions of the unit-fields (particles) are described by Eqs. (\ref{eq384abc}) - (\ref{eq520abc}). The behavior of unit-fields (elementary particles) having the arbitrary generators ${\tilde \psi _{0n}}$ and the multi-component {\it{TACs}} $\Phi^{a}_{1}$ with the calibrations (gauges) (\ref{eq287abc}), (\ref{eq288abc}) and (\ref{eq291abc}) is described by Eqs. (\ref{eq364abc}) - (\ref{eq374abc}) of Sec. 6. {\it{In the most general case, one should use the all above-presented equations without the calibrations (gauges) (\ref{eq287abc}), (\ref{eq288abc}) and (\ref{eq291abc})}}. For an example, the system of motion equations 
\begin{eqnarray} \label{eq521abc}
\sum_{n=1}^{N } {\Phi^{a}_{n}} \left( \square {\tilde \psi} _{0n} + m_{0}^2 {\tilde \psi} _{0n} + {\tilde \psi} _{0n} (\Phi^{a}_{n})^{-1} \nabla ^2 \Phi^{a}_{n} \right)   = 0
\end{eqnarray}
describing the interfering (interacting) unit-fields (particles) under the calibrations (gauges) (\ref{eq287abc}), (\ref{eq288abc}) and (\ref{eq291abc})}}, in the most general case [see, Eq. (\ref{eq218abc})] has the {\it{general form}} of the $N$ coupled equations    
\begin{eqnarray} \label{eq522abc}
\sum_{n=1}^{N } \Phi ^a_{n}(\square {\tilde \psi} _{0n} + m_{0}^2 {\tilde \psi} _{0n})  + {\tilde \psi} _{0n}\square \Phi ^a_{n}+2\dot {\tilde \psi} _{0n} \dot {\Phi}^a_{n}-2\nabla {\tilde \psi} _{0n} \nabla {\Phi}^a_{n}=0, 
\end{eqnarray}
which is more complicated, but is suitable for the comparison of formulas of Part II based on the direct generalization of the Einstein energy-mass relation for a structured unit-field (particle) with  the Standard Model of Particle Physics (SM) and/or with Part I based on the insertion of interference between particles into the traditional interaction-free (interference-free) Hamiltonians of the traditional quantum field theories. The detailed relations of the present model with the Standard Model of Particle Physics will be illustrated in the following study. Notice, the transition from the state describing by the unit-field generator ${\tilde \psi _{0n}}$ to the state characterizing by the generator ${\tilde  \psi '_{0n}}$ could be described simply by application of the {\it{perturbation theory}} to Eqs. (\ref{eq364abc}) - (\ref{eq374abc}) based on the calibrations (gauges) (\ref{eq287abc}), (\ref{eq288abc}) and (\ref{eq291abc})}} or to the above-presented equations that do not use these calibrations. In the present model, the transition from one kind of a unit-field (particles) to another is accompanied by the change ("spontaneous or non-spontaneous breaking") of symmetry of the total associate component ({\it{TAC}}) of the unit-field (see, Secs. 7.2 and 7.3.). It should be also noted that the total relativistic cross-correlation terms (\ref{eq502abc}), (\ref{eq507abc}), (\ref{eq509abc}), (\ref{eq513abc}) and (\ref{eq516abc}) associated with the $m$-th and $n$-th unit-fields {\it{do depend on the phases}} $\alpha_m$ and $\alpha_n$ of the interfering (interacting) unit-fields (particles). In the case of unit-fields (particles) having the probabilistic (for instance, thermal) distributions of the values $\mathbf{k}_n$, $\mathbf{k}_m$, $\varepsilon_{n}$, $\varepsilon_{m}$, $\alpha_{n}$ and $\alpha_{m}$ in macroscopic bodies (macroscopic composite "particles"), the energies and total relativistic cross-correlation terms {\it{should be averaged}} using these probabilistic distributions. That is important for description of the well-known macroscopic coherent quantum phenomena, such as the Bose-Einstein condensation, superfluidity, superconductivity, supermagnetism, super-radiation, Bosenova effect, and quantum anomalous and fractional Hall effects. For the elementary particles of the present model, the macroscopic quantum phenomena will be presented in the next study. 

\begin{table}[t]
\centering
\begin{tabular}{ll}
\hline\noalign{\smallskip}
\multicolumn{2}{c} {The unified model of elementary particles and interactions}   \\
\hline\noalign{\smallskip}
\multicolumn{2}{c}{\it{Elementary particles $\equiv$ Unit-fields}}   \\
\noalign{\smallskip}\hline\noalign{\smallskip}
{\it{Matter:}} Quarks and Leptons &  \\
\noalign{\smallskip}\hline\noalign{\smallskip}
{\it{Antimatter:}} Antiquarks and antileptons & \\
\noalign{\smallskip}\hline\noalign{\smallskip}
{\it{Carriers of forces:}} Gravitons (Curved Space), Photons, W-bosons and Z-bosons, Gluons &\\
\noalign{\smallskip}\hline
{\it{Composite particles}}: Hadrons (Mesons and Barions), nuclei, atoms and molecules & \\
\hline\noalign{\smallskip}
\multicolumn{2}{c}{\it{ Interaction $\equiv$ Interference (cross-correlation)}}   \\
\noalign{\smallskip}\hline
{\it{Gravitational Force}} by interference of massive unit-fields &  Force carriers: Gravitons   \\
\noalign{\smallskip}\hline
{\it{Electromagnetic Force}} by interference of electrically charged unit-fields & Force carriers: Photons \\
\noalign{\smallskip}\hline
{\it{Weak Force}} by interference of unit-fields having isospins and hypercharges & Force carriers: W- and Z-Bosons \\
\noalign{\smallskip}\hline
{\it{Strong Force}} by interference of unit-fields having color charges & Force carriers: Gluons\\
\noalign{\smallskip}\hline
\end{tabular}
\caption{The particles and interactions according to the present model, which unifies the all-known particles and interactions. Similarly to the some non-standard, candidate theories ({\it{Grand Unified Theories}} and {\it{Theories of Everything}}), the present model do contain the Gravitational Force by considering the interference of massive unit-fields, which may be reinterpreted as the exchange of virtual gravitons (hypothetical mass-less particles) or as the curving of space by the massive particles.}
\label{tab:t3}
\end{table}

In conclusion, Part I of the present study has developed the theoretical background for a unified description of the classical and quantum fields and interactions in terms of the interference between elementary particles (indivisible unit-fields) and the respective cross-correlation energy, which do not exist from the point of view of the canonical quantum mechanics, quantum field theories, SM and string theories. The Hamiltonians that describe the cross-correlation energy mediating by interference of the basic classical and quantum material fields composed from the 3-dimensional (in space) unit-fields of physical matter (mass-energy) have been derived by the generalization of the traditional Hamiltonians of the classical and quantum field theories for the superposition of interfering unit-fields associated with the interacting particles. It has been shown that the gradient of the cross-correlation energy induced by the interference between particles (unit-fields) mediates the attractive or repulsive forces, which could be attributed to all known classical and quantum fields (for details, see Part I). The present study (Part II) has used this theoretical background for unified description of the fundamental (electromagnetic, weak, strong and gravitational) fields and interactions. However, unlike in Part I, the unification was performed rather by the generalization of the basic (energy-mass) relation of the Einstein special relativity than the traditional Hamiltonians of the classical and quantum field theories. The model unified the all-known fields, particles and interactions by the straightforward generalization of the Einstein relativistic energy-mass relation ${\varepsilon}^{2} ={\mathbf{k}}^2+m^2$ for the interacting particles and bodies, which are composed from the interfering, indivisible unit-fields associated with the elementary particles. The unit-fields have simultaneously properties of the point particles and waves of quantum mechanics (particle-wave duality), the point particles of SM, the strings of theories of strings and the 3-D (in space) waves of theories of classical fields. Therefore the unification of the fundamental fields and interactions could be considered as the further development and generalization of the canonical quantum mechanics, classical and quantum field theories, SM and string theories.

{\begin{center} \bf \large ENERGY MEDIATED BY INTERFERENCE OF PARTICLES (Parts I-IV): The Way to Unified Classical and Quantum Fields and Interactions \\ 
\vspace{0.2cm}
{\it{Part III. New Approach for Constructing a Unified Quantum Model of Gravitational and Electromagnetic Fields and Interactions}} \end{center}}

{\begin{center} S. V. Kukhlevsky \end{center}} 
{\begin{center}{\it Department of Physics, Faculty of Natural Sciences,\\ University of Pecs, Ifjusag u. 6, H-7624 Pecs, Hungary} \end{center}}

\begin{quote}
\small Part I of the present study has developed the theoretical background for unified description of the all-known classical and quantum fields in terms of the interference between particles and the respective cross-correlation energy, which do not exist from the point of view of quantum mechanics and SM. Part II developed this background for unification of the electromagnetic, weak, strong and gravitational fields and interactions. However, unlike in Part I, the unification was performed rather by the generalization of the basic (energy-mass) relation of the Einstein special relativity than the traditional Hamiltonians of the classical and quantum field theories. Part III presents the new approach for constructing a unified quantum model of gravitational and electromagnetic fields and interactions is presented. The approach is based on a concept of interfering (cross-correlating) material unit-fields. Starting from the Einstein mass-energy relation and using this concept, the quantum equation for united gravitation and electromagnetism is derived. The unified equation yields all known solutions to the Dirac equation, for example, the fine and hyperfine structure of the atom spectrum. Furthermore, the unified model suggests explanations of the Pauli exclusion principle and the physical nature of spin and anomalous gyromagnetic factor of an electron. For weak potentials, in the classical limit, the model simplifies to the Lorentz-Maxwell electromagnetism and the so-called gravitoelectromagnetic approximation of Einstein's general relativity. In case of the strong potentials, the model yields new predictions. For instance, the cross-correlation of gravitational and electric strong potentials predicts the "anti-gravity force". The model suggests explanations also of "dark matter" and "dark energy". 
\end{quote}

\section{1. Introduction}
\label{sec1y}
Unification of the Lorentz-Maxwell electromagnetism with the Einstein general relativity is a long-standing problem in physics. Today we know that the electromagnetism is part of a larger gauge group described by the Standard Model of Particle Physics (SM). Unfortunatelly, SM does not include gravity. The main difficulty in unification of the Lorentz-Maxwell and Einstein theories is the principal difference between the models. The model of \textit{material} electromagnetic fields is based on the scalar and vector potentials mediated by electrical charges, whereas the geometrical framework of general relativity deals with gravitational fields attributed to the immaterial spacetime around massive particles. In spite of the difference, there are many physical and mathematical connections between the two theories. The early studies~\cite{maxw,heav,thir} have revealed the gravitational analogues to Maxwell's equations, called the "gravitoelectromagnetic equations". The study~\cite{lens} has derived the Lorentz-like force law for a massive particle moving in gravitational field. In 1916, A. Einstein noticed that the Maxwell equations can be formulated in a form independent of the metric gravitational potential~\cite{eins}. This was followed by further important studies~\cite{weyl,murn,kott,cart}. From a huge amount of early studies one should mention the famous article~\cite{eins1} of A. Einstein and N. Rosen, which tried to extend the geometric framework of general relativity to also include electromagnetism. Despite the fact that the unification is challenging, a lot of other physical and mathematical connections between the two models were uncovered during the last 100 years, e.g. string theory and quantum gravity (see, for instance, the modern reviews~\cite{hehl,lind,hube} and references therein). Analysis of the connections, especially the gravitational analogues to Maxwell's and Lorentz's equations~\cite{maxw,heav,thir,lens}, supports rather the old idea that objects with mass attract each other through a \textit{material} gravitational field ("Laplace's fluid") than the modern hypothesis of immaterial spacetime as bendable. Although the Newton, Laplace and Poisson equations didn't explain the \textit{material} nature and mass of gravitational field, the field mass could be connected, at least formally, with the field energy by using the Einstein mass-energy relation.

The basic hypothesis of present study is similar in spirit to both the aforementioned idea of a \textit{material} gravitational field and the concept of \textit{material} gravitons of quantum gravity. The new approach for constructing a quantum model of united gravitation and electromagnetism is based on the concept of \textit{material} one-particle fields~\cite{kukh}, i.e. the gravito-electromagnetic unit-fields (quantum particles) that make up the local and global gravitoelectromagnetic quantum fields. The interaction of quantum fields is determined by interference (cross-correlation) of the unit-fields. The mass-energy of quantum gravitoelectromagnetic field is given by the Einstein mass-energy relation modified by the interference of \textit{material} unit-fields. The fields are modelled to be representations of the Lorentz group in order to provide consistency of the model with the Einstein relativity, Lorentz-Maxwell electromagnetism and quantum mechanics (QM). 

\section{2. The idea of a single free unit-field with gravito-electric dress}
\label{sec2y}

According to the concept~\cite{kukh}, the present model assumes that any field ${\Psi (\mathbf{r},t)}$ of electrically-charged material \textit{quantum particles} consists of the gravitoelectromagnetic (GEM) material unit-fields ${\Psi _{0i}(\mathbf{r},t)}$ attributed to the particles: 
\begin{eqnarray} 
\Psi (\mathbf{r},t) =\sum_{i=1}^N{\Psi _{0i}(\mathbf{r},t)}, 
\label{eq1y}
\end{eqnarray} 
where $N$ is the number  of particles. The superposition of unit-fields ${\Psi _{0i}(\mathbf{r},t)}$, which are one-particle fields, yields the local (beams) and global quantum GEM fields $\Psi (\mathbf{r},t)$. In the global fields, $N\rightarrow \infty$. The pure gravitational (G) and pure electromagnetic (EM) fields present different aspects of the unified GEM-field $\Psi (\mathbf{r},t)$. The interaction of unit-fields ${\Psi _{0i}(\mathbf{r},t)}$ with each other is determined by interference (cross-correlation) of these unit-fields (see, Sect. 10). The mass-energy of field $\Psi (\mathbf{r},t)$ is found by using the Einstein mass-energy relation modified by the interference of unit-fields ${\Psi _{0i}(\mathbf{r},t)}$. For conformity of the model with the special relativity, electromagnetism and QM, the fields are modelled to be representations of the Lorentz group. 

Before presenting the model of united gravitation and electromagnetism (gravitoelectromagnetism), let me explain relationship of the GEM unit-field to the special relativity and QM. The concept of GEM unit-fields is based on the de Broglie idea of a \textit{quantum particle} (material wave or, in other words, material \textit{unit-field} $\psi _{0}(\mathbf{r},t)$) connected to the Einstein mass-energy relation for a point-like particle. In the special relativity, the mass-energy relation for a single free $\textit{point-particle}$ located at the spacetime point $(\mathbf{r},t)$ reads     
\begin{eqnarray} 
{\varepsilon}_0^{2} = m_{0}^2 c^4 + \mathbf{p}_{0}^2 c^2, 
\label{eq2y}
\end{eqnarray} 
where ${\varepsilon}_{0}$, $m_{0}$ and $\mathbf{p}_{0}$ are the particle energy, rest mass and momentum. Using of Eq.~(\ref{eq2y}) for a single free {\it{unit-field}} $\psi _{0}(\mathbf{r},t)$ yields the relativistic quantum (Klein-Gordon) mass-energy relation:
\begin{eqnarray} 
\langle \psi_0 | {\hat{\varepsilon}^{2}} | \psi_0 \rangle = \langle \psi_0 | m_0^2c^4+{\hat{\mathbf{p}}^{2}c^2}|\psi_0 \rangle ,
\label{eq3y}
\end{eqnarray}
where the squared energy and squared momentum are determined by the operators ${\hat{\varepsilon}^{2}}={-\hbar}^{2}{\frac {{\partial}^2} {\partial {t}^2}}$ and ${\hat{\mathbf{p}}^{2}}={-\hbar}^{2}{\nabla}^{2}$. In the non-relativistic case of Eqs. (\ref{eq2y}) and (\ref{eq3y}), the immaterial psi-wave described by the psi-wavefunction ${\psi} _{0}(\mathbf{r},t)$ obeys rather the Copenhagen (statistical) meaning of QM than the field interpretation. In order to introduce the GEM interactions into Eq.~(\ref{eq3y}), the state $ |{\psi_{0}}\rangle$ of the unit-field ${\psi_{0}} = {\psi_{0}}(\mathbf{r},t)$ is replaced by  
\begin{eqnarray} 
| {\Psi_{0}} (\mathbf{r},t) \rangle  =  \Phi_{0ge}|{\psi} _{0}(\mathbf{r},t) \rangle .
\label{eq4y}
\end{eqnarray} 
The operator $\Phi_{0ge}$ is interpreted as the "operator of gravitoelectric (GE) dress" that yields the dress mass. The bare unit-field ${\psi_{0}}$ could be thought as a carrier of massive GE dress (field). It is well known how to construct the operator $\Phi=\Phi_{0e}$ for electromagnetism by using the Dirac equation and so-called minimal substitution assumption (${\varepsilon} \rightarrow {\varepsilon}- q\varphi $, $\mathbf{p} \rightarrow \mathbf{p}+(q/c)\mathbf{A}$)~\cite{dira}. However, it is not clear how to do that for gravity ($\Phi=\Phi_{0g}$) and gravitoelectromagnetism ($\Phi=\Phi_{0ge}$). The construction of the operator $\Phi_{0ge}$ for the gravitoelectrically dressed unit-field ${\Psi_{0}}$ is impossible without additional assumptions.

\section{3. Model of a single free unit-field with GE dress}
\label{sec3y}

In the frame of the Euler-Lagrange formalism, the gravitoelectrically dressed unit-field is described by the Lagrangian 
\begin{eqnarray} 
L = {\int_{V}} {\cal L}d^3x,
\label{eq5y}
\end{eqnarray}
where
\begin{eqnarray} 
{\cal L} =  {\hbar}^{2} \dot{\psi_0 ^*} \dot{\psi_0} - {\hbar}^{2}\Phi_{0ge}^{2} \nabla \psi_{0}^*  \nabla \psi_{0} - m_0^2 c^4\Phi_{0ge}^{2}\psi_{0} ^* \psi_{0}
\label{eq6y}
\end{eqnarray}
is the Lagrangian density, the psi-wavefunction ${\psi} _{0}(\mathbf{r},t)$ describes the unit-field (de Broglie's psi-wave), $\Phi_{0ge}$ denotes the operator of GE dress (field), and $\dot{\psi_0} = {\frac {\partial} {\partial {t}}}{\psi}_0$. The equation of motion of the unit-field $\psi_{0}$ is found by varying the value ${\psi} _{0}$ under the fixed dress-parameter $\Phi_{0ge}$. The least action principle for the action $S={\int}{\cal L}d^4x$ yields the Euler-Lagrange equation 
\begin{eqnarray} 
 \frac {{\delta }{\cal L}} {\delta  {\psi}^*}-{\partial}_{\mu}\frac {{\delta }{\cal L} } {\delta {\partial}_{\mu}{\psi}^*} =0
\label{eq7y}
\end{eqnarray}
in the form of quantum equation for the united gravitation and electromagnetism:    
\begin{eqnarray} 
{-\hbar}^{2} {\frac {{\partial}^2} {\partial {t}^2}}{\psi}_0  = m_0^2 c^4 \Phi_{0ge}^{2}\psi_{0} - {\hbar}^{2}\Phi_{0ge}^{2}\nabla ^2{\psi_{0}}.
\label{eq8y}
\end{eqnarray}
The unified mass-energy relation~(\ref{eq8y}) reduces to the Klein-Gordon form~(\ref{eq3y}) in case of the bare unit-field ($\Phi_{0ge}=1$). In case of the time-dependent dress $\Phi_{0ge}=\Phi_{0ge}(\mathbf{r},t)$, Eq.~(\ref{eq8y}) can be presented as
\begin{eqnarray} 
{-\hbar}^{2} {\frac {{\partial}^2} {\partial {t}^2}}|\psi_0 ({\mathbf{r}},t)\rangle  = {\hat{\mathbf{H}}^{2}}({\mathbf{r}},t) |\psi_0 (\mathbf{r},t)\rangle,
\label{eq9y}
\end{eqnarray}
where the nonstationary squared-energy ${\varepsilon}_0^{2}={\varepsilon}_0^{2}(t)$ of transient unit-field ${\psi} _{0}(\mathbf{r},t)$ is determined by the time-dependent operator 
\begin{eqnarray} 
{\hat{\mathbf{H}}^{2}}(\mathbf{r},t) = \Phi_{0ge}^{2}(\mathbf{r},t)\left(  m_{0}^2 c^4 +  c^2 {\hat{\mathbf{p}}^{2}}  \right), 
\label{eq10y}
\end{eqnarray}
which is identified with the squared Hamiltonian operator. For the stationary state $(| {\psi} _{0}(\mathbf{r},t) \rangle = \exp\left[\pm{{\frac {i} {{\hbar}}}} {\varepsilon}_{0}t \right] |\psi_{0}(\mathbf{r})\rangle$), which is provided by the time-independent dress $\Phi_{0ge} =\Phi_{0ge}(\mathbf{r})$, Eq.~(\ref{eq8y}) reads
\begin{eqnarray} 
{\varepsilon}_0^{2} | \psi_0 (\mathbf{r})\rangle =  \Phi_{0ge}^{2}(\mathbf{r}) \left(m_{0}^2 c^4 +  c^2 {\hat{\mathbf{p}}^{2}}  \right) | \psi_0 (\mathbf{r})\rangle. 
\label{eq11y}
\end{eqnarray} 
The time-independent operator  
\begin{eqnarray} 
{\hat{\mathbf{H}}^{2}}(\mathbf{r}) = \Phi_{0ge}^{2}(\mathbf{r})\left(  m_{0}^2 c^4 +  c^2 {\hat{\mathbf{p}}^{2}}  \right) 
\label{eq12y}
\end{eqnarray} 
determines the stationary squared-energy ${\varepsilon}_0^{2}\neq{\varepsilon}_0^{2}(t)$ of the eigen-state $|\psi_0 (\mathbf{r})\rangle$ with the normalization $\langle \psi_{0}(\mathbf{r}) | \psi_{0}(\mathbf{r}) \rangle =1$. 

The unit-field $\psi_0$ could be described also in the frame of the Hamilton-Jacobi formalism. The Lagrangian density (\ref{eq6y}) and the canonical relations ${\cal H}=\Pi \dot{\psi_0} -{\cal L}$ and $\Pi = \partial {\cal L}/\partial \dot{\psi_0}$ yielded the Hamiltonian density 
\begin{eqnarray} 
{\cal H}= m_0^2 c^4\Phi_{0ge}^{2}\psi_{0} ^* \psi_{0} + {\hbar}^{2}\Phi_{0ge}^{2} \nabla \psi_{0}^*  \nabla \psi_{0}.
\label{eq13y}
\end{eqnarray}
Thus, the Hamiltonian $H = {\int_{V}}{\cal H}d^3x$ reads 
\begin{eqnarray} 
H = {\varepsilon}_0^{2} = {\int_{V}}\psi_{0}^* {\hat{\mathbf{H}}^{2}} \psi_{0} d^3x,
\label{eq14y}
\end{eqnarray}
where the squared Hamiltonian operator ${\hat{\mathbf{H}}^{2}}$ is given by Eqs.~(\ref{eq10y}) and (\ref{eq12y}) for the transient ($\psi_{0} = \psi_{0}(\mathbf{r},t))$ and stationary ($\psi_{0} = \exp\left[\pm{{\frac {i} {{\hbar}}}} {\varepsilon}_{0}t \right] \psi_{0}(\mathbf{r}))$ unit-fields, respectively. Here, I used the relation ${\int_{V}} \nabla \psi_{0}^* \nabla \psi_{0} d^3x = - {\int_{V}} \psi_{0}^* \nabla^2{\psi}_0 d^3x$, because the parameter $\Phi_{0ge}$ was fixed under variation of the value ${\psi}_{0}$. Note that the Hamiltonian-energy relation $H = {\varepsilon}_0^{2}$ differs from the relation $H = {\varepsilon}_0$ of QM and SM.

\section{4. Explicit expressions for $\Phi_{0ge}$ and rest-masses of G and E dresses} 
\label{sec4y}

We now assume that the quantum particle (unit-field) at rest satisfies the Einstein mass-energy relation for the total rest-mass. The assumption, which is consistent with the special relativity, helps us to determine an explicit expression for $\Phi_{0ge}$. The concrete form and physical meaning of $\Phi_{0ge}$ are uncovered by analysing the steady state $|\psi_{0}(\mathbf{r},t)\rangle=\exp\left[\pm{{\frac {i} {{\hbar}}}} {\varepsilon}_{0}t \right]| \psi_{0}(\mathbf{r})\rangle$. In this state, the squared energy ${\varepsilon}_0^{2}$ is given by Eq.~(\ref{eq11y}), where the dress is time-independent ($\Phi_{0ge}=\Phi_{0ge}(\mathbf{r})$). The kinetic energy of a classic (point-like) particle at rest vanishes. The kinetic energy of the quantum particle at rest ($\psi_{0}(\mathbf{r})= s_{0}(\mathbf{r})$), which is determined by the term $\Phi_{0ge}^{2}(\mathbf{r}) c^2 {\hat{\mathbf{p}}^{2}} |\psi_0 (\mathbf{r})\rangle$ in Eq.~(\ref{eq11y}), also should be zero. The relation $\Phi_{0ge}^{2}(\mathbf{r}) c^2 {\hat{\mathbf{p}}^{2}} |s_{0} (\mathbf{r})\rangle =0$ implies
\begin{eqnarray} 
{\hat{\mathbf{p}}}^{2} | s_{0}(\mathbf{r}) \rangle  =  0.  
\label{eq15y}
\end{eqnarray}
Respectively, Eq.~(\ref{eq11y}) yields
\begin{eqnarray} 
{\varepsilon}_0  = m_{0}\langle s_{0}(\mathbf{r}) | \Phi_{0ge}(\mathbf{r}) | s_{0}(\mathbf{r}) \rangle c^2 . 
\label{eq16y}
\end{eqnarray} 

The rest masses of the bare unit-field $s_{0}(\mathbf{r})$ and the G and E dresses (fields) are assumed to be additive. Therefore, the total rest mass $M_0$ of the gravitoelectrically dressed unit-field is given by 
\begin{eqnarray} 
M_0=m_{0} + \langle m_{0g}\rangle + \langle m_{0e}\rangle, 
\label{eq17y}
\end{eqnarray}
where $m_{0}$, $\langle m_{0g}\rangle$ and $\langle m_{0e}\rangle$ denote the rest masses of the bare unit-field $\psi_{0}(\mathbf{r})= s_{0}(\mathbf{r})$ and the G and E dresses, respectively. Thus, the Einstein mass-energy relation for the total rest-mass reads
\begin{eqnarray} 
{\varepsilon}_{0} = \left(m_{0} + \langle m_{0g}\rangle + \langle m_{0e}\rangle  \right) c^2,
\label{eq18y}
\end{eqnarray} 
The G and E rest-masses are connected to the gravitostatic and electrostatic potential energies of G and E fields by using Eq. (\ref{eq18y}) for the potential energies:
\begin{eqnarray} 
\langle m_{0g}\rangle = \langle s_{0}(\mathbf{r}) | U_{0g}(\mathbf{r})| s_{0}(\mathbf{r}) \rangle c^{-2},\nonumber \\ \langle m_{0e}\rangle = \langle s_{0}(\mathbf{r}) | U_{0e}(\mathbf{r})| s_{0}(\mathbf{r}) \rangle c^{-2},
\label{eq19y}
\end{eqnarray} 
where $\langle s_{0}(\mathbf{r}) | U_{0g}(\mathbf{r})| s_{0}(\mathbf{r}) \rangle$ and $\langle s_{0}(\mathbf{r}) | U_{0e}(\mathbf{r})| s_{0}(\mathbf{r}) \rangle$ denote the gravitostatic and electrostatic potential energies determined by the G and E static potentials $\varphi_{0g} (\mathbf{r})$ and $\varphi_{0e} (\mathbf{r})$:  
\begin{eqnarray}
 U_{0g}(\mathbf{r}) =  m_0 \varphi_{0g}(\mathbf{r}), \nonumber \\ U_{0e}(\mathbf{r}) = q_0 \varphi_{0e} (\mathbf{r}). 
\label{eq20y}
\end{eqnarray}
In order to provide consistency of the G and E potentials with the Newton gravitation and the Lorentz-Maxwell electromagnetism, we assume that the static potentials $\varphi_{0g} (\mathbf{r})$ and $\varphi_{0e} (\mathbf{r})$ satisfy the Poisson equation: 
\begin{eqnarray}
{\nabla}^{2} \varphi_{0g}=\pm 4\pi \gamma_{N}m_0 \delta (\mathbf{r}),\nonumber \\ {\nabla}^{2} \varphi_{0e}(\mathbf{r})=\pm 4\pi \gamma_{C}q_0 \delta (\mathbf{r}), 
\label{eq21y}
\end{eqnarray}
where $\delta (\mathbf{r})$ is the Dirac delta. Equations~(\ref{eq21y}) yielded the Newton-like and Coulomb-like potentials
\begin{eqnarray}
\varphi_{0g} (\mathbf{r}) =\pm  \gamma_{N}{\frac {m_0} {r}},\nonumber \\ \varphi_{0e} (\mathbf{r}) =\pm \gamma_{C}{\frac {q_0} {r}},
\label{eq22y}
\end{eqnarray}
where $q_{0}$ denotes the unit-field charge, and $\gamma_{N}$ and $\gamma_{C}$ are the Newton and Coulomb constants. The signs ($\pm$) are explained in Sects. 10 and 12.2. 

The comparison of Eq.~(\ref{eq16y}) with (\ref{eq18y}) and the use of the rest-masses $\langle m_{0g}\rangle$ and $\langle m_{0e}\rangle$ determined by Eqs.~(\ref{eq19y})-(\ref{eq22y}) imply the explicit expression      
\begin{eqnarray} 
\Phi_{0ge}(\mathbf{r}) =  1 + {\frac {U_{0ge}(\mathbf{r})} {m_{0} c^2}}  
\label{eq23y}
\end{eqnarray}
for $\Phi_{0ge}(\mathbf{r})$. In other words, the explicit expressions for $\Phi_{0ge}(\mathbf{r})$, $\langle m_{0g}\rangle$ and $\langle m_{0e}\rangle$ yielded the Einstein mass-energy relation ${\varepsilon}_{0} = M_{0}c^2$ for the total rest-mass $M_0=m_{0} + \langle m_{0g}\rangle + \langle m_{0e}\rangle$. Notice, the additivity ($U_{0ge}(\mathbf{r})=U_{0g}(\mathbf{r})+U_{0e}(\mathbf{r})$) of the fields $U_{0g}(\mathbf{r})$ and $U_{0e}(\mathbf{r})$ is in agreement with the additivity ($\langle m_{0ge}\rangle = \langle m_{0g}\rangle +\langle m_{0e}\rangle$) of the rest-masses $\langle m_{0g}\rangle$ and $\langle m_{0e}\rangle$. Although the relation~(\ref{eq17y}) looks like a naive model of the total rest-mass, the model explains the two long-standing problems in physics, namely the nature of an electron spin and the gravitostatic and electrostatic self-energies of an electron.

\section{5. The spin and gravitostatic and electrostatic self-energies of unit-field}
\label{sec5y}

\subsection{5.1 The particle spin attributed to the internal state of unit-field}
\label{sec5ay}

The internal state $|s_{0}(\mathbf{r})\rangle$ of the unit-field at rest ($\psi_{0}(\mathbf{r})\equiv s_{0}(\mathbf{r})$) satisfies Eq.~(\ref{eq15y}). For the spherically symmetric unit-field, the equation reads
\begin{eqnarray}
{\hbar}^{2} \nabla^2 s_{0}(\mathbf{r})  =    \left(     \frac {{\hbar}^{2}} {r}    \frac {\partial ^2} {\partial r^2} r - \frac {\hat{\mathbf{L}}^2} { r^2} \right)  s_{0}(\mathbf{r}) =0,
\label{eq24y}
\end{eqnarray}
where ${\hat{\mathbf{L}}}^2$ is the square of the angular momentum operator ${\hat{\mathbf{L}}} = \mathbf{r}\times {\hat{\mathbf{p}}} =i{\hbar}(\mathbf{r}\times \nabla)$, ${\hat{\mathbf{p}}}$ is the momentum operator, and $s_{0}(\mathbf{r})$ is the spin s-wavefunction with the normalization  
\begin{eqnarray}
\langle s_{0}(\mathbf{r}) | s_{0}(\mathbf{r}) \rangle = \int_{\theta =0}^\pi   \int_{\varphi  =0}^{2\pi} d\Omega {\int_{0}^{r_{0}}} s_{0}^{\ast}(\mathbf{r}) s_{0}(\mathbf{r})r^{2} dr = 1.
\label{eq25y}
\end{eqnarray}
Equation~(\ref{eq25y}) means that $s_{0}(\mathbf{r})=0$ for $r>r_{0}$. The value $r_0$ could be interpreted as the internal radius of unit-field.
We are looking for the solutions of Eq.~(\ref{eq24y}) that do not diverge at $r\rightarrow 0$. They are given by the regular solid harmonics
\begin{eqnarray}
s_{0}(\mathbf{r})  = B_{lm} r^{l} Y_l^m (\theta, \varphi).
\label{eq26y}
\end{eqnarray}
Here, $Y_l^m (\theta, \varphi  )$ is the joint eigenfunction of the operator ${\hat{\mathbf{L}}}^2$ and the generator ${\hat{L}}_{z} \equiv {\hat{s}}_{z}$ of rotations around the azimuthal axis:
\begin{eqnarray} 
{\hat{\mathbf{L}}}^2  Y_l^m (\theta, \varphi  ) = {\hbar}^{2} l(l+1) Y_l^m (\theta, \varphi ), \nonumber \\{\hat{L}_{z}} s_{0}(\mathbf{r})  =  \hbar m  s_{0}(\mathbf{r}). 
\label{eq27y}
\end{eqnarray}
For the Laplace spherical harmonics $Y_l^m (\theta, \varphi)$ normalized by $\langle Y_l^m | Y_l^m \rangle  = \int_{\theta =0}^\pi   \int_{\varphi  =0}^{2\pi} Y_l^{m\ast} Y_l^m d\Omega =1$, the constant $(B_{lm})^{2}= (2l+3)r_0^{-(2l+3)}$ provides the normalization~(\ref{eq25y}). The different species of quantum particles (unit-fields) should correspond to the different values of $l$. For an orbital quantum number $l$, there are $2l+1$ independent spin wavefunctions (\ref{eq26y}), one for each magnetic quantum number $m$ with $-l \leq  m \leq  l$. Although the squared internal momentum $\langle {{\mathbf{p}}}^{2}_{0s}\rangle$ of the unit-field vanishes ($\langle {{\mathbf{p}}}^{2}_{0s}\rangle=\langle s_{0}(\mathbf{r})  |{\hat{\mathbf{p}}}^{2} | s_{0}(\mathbf{r}) \rangle  =  0$) due to Eq.~(\ref{eq15y}), the non-zero internal momentum $\langle {{\mathbf{p}}}_{0s} \rangle = \langle s_{0}(\mathbf{r}) |{\hat{\mathbf{p}}} | s_{0}(\mathbf{r})\rangle \neq 0$ for $m\neq 0$ yields the non-zero internal angular momentum (spin) $\langle {{s}_{z}}\rangle =\langle  s_{0}(\mathbf{r})|  {\hat{L}_{z}} | s_{0}(\mathbf{r}) \rangle =  \hbar m$. Such behaviour is different from the classical mechanics of point particles, where $\mathbf{p}= 0$ if $\mathbf{p}^{2}= 0$. 

The above-described state $|s_{0}(\mathbf{r})\rangle$ is the internal state of unit-field $\psi_{0}(\mathbf{r})$. The internal state $|s_{0}(\mathbf{r})\rangle$ and the internal momentum $\langle {{\mathbf{p}}}_{0s} \rangle = \langle s_{0} (\mathbf{r})|{\hat{\mathbf{p}}} | s_{0}(\mathbf{r})\rangle$ could be extracted from the total state $| \psi_{0}(\mathbf{r}) \rangle$ and the total momentum $\langle {{\mathbf{p}}}\rangle = \langle \psi_{0}(\mathbf{r}) | {\hat{\mathbf{p}}} | \psi_{0} (\mathbf{r})\rangle$ as follows: 
\begin{eqnarray} 
| \psi_{0}(\mathbf{r}) \rangle = | s_{0}(\mathbf{r})\rangle | \tilde \psi_0 (\mathbf{r})\rangle,
\label{eq28y}
\end{eqnarray} 
\begin{eqnarray} 
\langle {{\mathbf{p}}} \rangle = \langle {{\mathbf{p}}}_{0s} \rangle + \langle \tilde {{\mathbf{p}}} \rangle = \langle \tilde \psi_0 (\mathbf{r})|\langle s_{0}(\mathbf{r})| {\hat{\mathbf{p}}}| s_{0}(\mathbf{r}) \rangle | \tilde \psi_0 (\mathbf{r})\rangle.
\label{eq29y}
\end{eqnarray} 
Here, the external momentum $\langle \tilde {{\mathbf{p}}} \rangle = \langle \tilde \psi_0 (\mathbf{r}) | {\hat{\mathbf{p}}} | \tilde \psi_0  (\mathbf{r})\rangle$ is determined by the external state $| \tilde \psi_0 (\mathbf{r}) \rangle$ of the external unit-field $\tilde \psi_0 (\mathbf{r})$. The Hilbert state~(\ref{eq28y}) means that the internal ($s_{0}(\mathbf{r})$) and external ($ \tilde \psi_0 (\mathbf{r})$) unit-fields cannot be separated from each other. Since the internal unit-field $s_{0}(\mathbf{r})$ is local ($r\in [0,r_0]$), the nonlocality of state $| s_{0}(\mathbf{r}) \rangle | \tilde \psi_0 (\mathbf{r})\rangle$ is provided by the nonlocality ($r\in [0,\infty]$) of $\tilde \psi_0 (\mathbf{r})$. The external unit-field $\tilde \psi_0 (\mathbf{r})$ is assumed to be undividable for agreement with QM. For the total unit-field $\psi_{0}(\mathbf{r})$ at rest, Eq.~(\ref{eq15y}) reads 
\begin{eqnarray} 
{\hat{\mathbf{p}}}^{2}|s_{0}(\mathbf{r})\rangle | \tilde \psi_0 (\mathbf{r}) \rangle =0.
\label{eq30y}
\end{eqnarray} 
The equations (\ref{eq15y}) and (\ref{eq30y}) should have the same solution $| \psi_{0}(\mathbf{r}) \rangle = |s_{0}(\mathbf{r})\rangle $. That is ensured by the solution $\tilde \psi_0 (\mathbf{r})$ =  $\exp\left[\pm{{\frac {i} {{\hbar}}}} (0\cdot\mathbf{r}) \right]$ = $1$ to the equation ${\hat{\mathbf{p}}}^{2} | \tilde \psi_0 (\mathbf{r}) \rangle =0$ and the \textit{transverse gauge} 
\begin{eqnarray} 
\langle {{\mathbf{p}}}_{0s} \rangle \langle \tilde {{\mathbf{p}}} \rangle =\langle s_{0} (\mathbf{r})|{\hat{\mathbf{p}}} | s_{0} (\mathbf{r})\rangle \langle \tilde \psi_0 (\mathbf{r}) | {\hat{\mathbf{p}}} | \tilde \psi_0  (\mathbf{r})\rangle  =  0
\label{eq31y}
\end{eqnarray}
imposed on the unit-field state (\ref{eq28y}), where $\langle \tilde {{\mathbf{p}}} \rangle \neq 0$ or $\langle \tilde {{\mathbf{p}}} \rangle =0$. Notice, the solution $| \psi_{0}(\mathbf{r}) \rangle = |s_{0}(\mathbf{r})\rangle $ also means that the total rest-mass $ M_0 = \langle \psi_{0}(\mathbf{r}) | \left(m_{0} + U_{0g}(\mathbf{r})c^{-2} +  U_{0e}(\mathbf{r})c^{-2}  \right) | \psi_{0}(\mathbf{r}) \rangle$ doesn't depend on $\tilde \psi_0 (\mathbf{r})$.  

Let me now explain the electron spin. A simple analysis of Eqs.~(\ref{eq24y})-(\ref{eq27y}) shows that the spin state $|s_{0}(\mathbf{r})\rangle$ of the internal unit-field $s_{0}(\mathbf{r})$ should obey the quantum numbers $l=1$ and $m=\pm1$. The numbers yielded the spin states $|s_{0,+}\rangle\equiv| s_{0,l=1,m=+1} \rangle $ and $|s_{0,-}\rangle \equiv| s_{0,l=1,m=-1} \rangle $ with the spins $\langle s_{0,+}| {\hat{L}_{z}} | s_{0,+} \rangle = \hbar$ and $\langle s_{0,-} | {\hat{L}_{z}} | s_{0,-} \rangle = - \hbar$. It is interesting that the state $| s_{0,l=0,m=0} \rangle $ doesn't satisfy the transverse gauge~(\ref{eq31y}). Indeed, the spin wavefunction $s_{0,l=0,m=0}(\mathbf{r})= B_{00}r\sqrt{\frac{1}{4\pi}}$ (see, Eq,~(\ref{eq26y})) yields 
\begin{eqnarray} 
\langle {{\mathbf{p}}}_{0s} \rangle \langle \tilde {{\mathbf{p}}} \rangle = \textbf{e}_r\langle p_{0sr} \rangle (\textbf{e}_r\langle \tilde{p}_r \rangle +\textbf{e}_\Theta\langle {\tilde{p}}_\Theta \rangle +\textbf{e}_\varphi\langle \tilde {p}_\varphi \rangle ) = \langle p_{0sr} \rangle \langle \tilde{p}_r \rangle \neq  0
\label{eq32y}
\end{eqnarray} 
for $\langle \tilde{p}_r  \rangle \neq 0$. Notice, the assumption of $l=1$ and $m=\pm1$ corresponds to the postulate of SU(2) symmetry in QM. The difference between the value $\langle s_{0,\pm} | {\hat{L}_{z}} | s_{0,\pm}\rangle = \pm \hbar$ and the electron spin $\langle s_{0,\pm} | {\hat{L}_{z}} | s_{0,\pm}\rangle = \pm \hbar /2$ in QM is clarified in Sect. 11.   

\subsection{5.2 The gravitostatic and electrostatic self-energies of an electron}
\label{sec5by}

The model explains the gravitostatic ($U^{s}_{0g} $) and electrostatic ($U^{s}_{0e} $) self-energies of an electron. Indeed, the classic values $U^{s}_{0g} =\pm  \gamma_{N}{\frac {m_0^{2}} {2r}}$ and $U^{s}_{0e} =\pm \gamma_{C}{\frac {q_0^{2}} {2r}}$ of a point-like electron placed at $r=0$ are infinite~\cite{lan}. While, the gravitostatic ($U^{s}_{0g}=\langle \psi_{0}(\mathbf{r}) | U_{0g}(\mathbf{r})| \psi_{0}(\mathbf{r}) \rangle$) and electrostatic ($U^{s}_{0e}=\langle \psi_{0}(\mathbf{r}) | U_{0e}(\mathbf{r})| \psi_{0}(\mathbf{r}) \rangle$) self-energies of the electron unit-field at rest ($\tilde \psi_0 (\mathbf{r}) =  1, \psi_{0}(\mathbf{r})= s_{0}(\mathbf{r})$) obey the finite values: 
\begin{eqnarray}
U^{s}_{0g} =   \int_{\theta =0}^\pi   \int_{\varphi  =0}^{2\pi} d\Omega {\int_{0}^{r_{0}}} s_{0}^{\ast}(\mathbf{r}) \left(\pm  \gamma_{N}{\frac {m_0^{2}} {r}}\right) s_{0}(\mathbf{r})r^{2} dr =  \pm {\frac {5\gamma_{N}m_0^{2}} {4r_0}},
\label{eq33y}
\end{eqnarray}
\begin{eqnarray}
U^{s}_{0e} = \int_{\theta =0}^\pi   \int_{\varphi  =0}^{2\pi} d\Omega {\int_{0}^{r_{0}}} s_{0}^{\ast}(\mathbf{r}) \left(\pm  \gamma_{C}{\frac {q_0^{2}} {r}}\right) s_{0}(\mathbf{r})r^{2} dr = \pm {\frac {5\gamma_{C}q_0^{2}} {4r_0}},
\label{eq34y}
\end{eqnarray}
where the spin wave-function is given by $s_{0}(\mathbf{r})= B_{l=1,m=\pm1} r Y_1^{\pm 1} (\theta, \varphi)$, and the value $r_{0}$ is determined in Sect. 12.2. The gravitostatic ($\langle m_{0g}\rangle$) and electrostatic ($\langle m_{0e}\rangle$) rest-masses of G ($U_{0g}(\mathbf{r})$) and E ($U_{0e}(\mathbf{r})$) dresses (fields), which are determined by Eqs.~(\ref{eq19y}), (\ref{eq33y}) and (\ref{eq34y}), don't depend on the external unit-field $\tilde \psi_0 (\mathbf{r})$. This means that 
\begin{eqnarray}
U^{s}_{0g}= \langle s_{0}(\mathbf{r}) |U_{0g}(\mathbf{r})| s_{0}(\mathbf{r}) \rangle = \nonumber \\ \langle \tilde \psi_{0}(\mathbf{r}) | \langle s_{0}(\mathbf{r}) |U_{0g}(\mathbf{r})| s_{0}(\mathbf{r}) \rangle | \tilde \psi_{0}(\mathbf{r}) \rangle =   \langle s_{0}(\mathbf{r}) | \langle \tilde \psi_{0}(\mathbf{r}) |U_{0g}(\mathbf{r})| \tilde \psi_{0}(\mathbf{r}) \rangle| s_{0}(\mathbf{r}) \rangle, 
\label{eq35y}
\end{eqnarray}
\begin{eqnarray}
U^{s}_{0e}= \langle s_{0}(\mathbf{r}) |U_{0e}(\mathbf{r})| s_{0}(\mathbf{r}) \rangle =\nonumber \\ \langle \tilde \psi_{0}(\mathbf{r}) | \langle s_{0}(\mathbf{r}) |U_{0e}(\mathbf{r})| s_{0}(\mathbf{r}) \rangle | \tilde \psi_{0}(\mathbf{r}) \rangle =   \langle s_{0}(\mathbf{r}) | \langle \tilde \psi_{0}(\mathbf{r}) |U_{0e}(\mathbf{r})| \tilde \psi_{0}(\mathbf{r}) \rangle| s_{0}(\mathbf{r}) \rangle 
\label{eq36y}
\end{eqnarray}
for $\tilde \psi_0 (\mathbf{r}) \neq  1$ ($\psi_{0}(\mathbf{r})\neq s_{0}(\mathbf{r})$). Note that the values $U^{s}_{0g}$, $U^{s}_{0e}$, $\langle m_{0g}\rangle=U^{s}_{0g}c^{-2}$ and $\langle m_{0e}\rangle=U^{s}_{0e}c^{-2}$ don't depend on the magnetic quantum number $m=\pm1$.

\section{6. The squared eigen-energy ${\varepsilon}_0^{2}$ of unit-field by ${\hat{\mathbf{H}}^{2}}$}
\label{sec6y}

According to Eqs.~(\ref{eq11y}), (\ref{eq12y}) and (\ref{eq23y}), the squared eigen-energy ${\varepsilon}_0^{2}$ of the stationary unit-field is determined by the equation
\begin{eqnarray} 
{\varepsilon}_0^{2}| s_{0} \rangle | \tilde \psi_0 \rangle = {\hat{\mathbf{H}}^{2}}| s_{0} \rangle | \tilde \psi_0 \rangle,  
\label{eq37y}
\end{eqnarray} 
where the squared Hamiltonian-operator ${\hat{\mathbf{H}}^{2}}={\hat{\mathbf{H}}^{2}}(\mathbf{r})$ is given by
\begin{eqnarray} 
{\hat{\mathbf{H}}^{2}} =  m^{2}_{0} c^4 (1+x)
\label{eq38y}
\end{eqnarray} 
with the operator
\begin{eqnarray} 
x =  {\frac {\hat{\mathbf{p}}^{2}} {m_{0}^{2} c^2}} +  {\frac {2U_{0ge}} {m_{0} c^2}} + {\frac {U_{0ge}^{2} } {m_{0}^{2} c^4}}  +   {\frac {2{U_{0ge}}\hat{\mathbf{p}}^{2} } {m_{0}^{3} c^4}} + {\frac {{{U_{0ge}^{2}}\hat{\mathbf{p}}^{2}} } {m_{0}^{4} c^6}}
\label{eq39y}
\end{eqnarray} 
determined by the time-independent ($U_{0ge}=U_{0g}(\mathbf{r})+U_{0e}(\mathbf{r})$) GE dress.

\section{7. The eigen-energy ${\varepsilon}_0$ of unit-field by ${\hat{\mathbf{H}}}$}
\label{sec7y}

The Lorentz-Maxwell electromagnetism and Einstein general relativity look for the particle energy rather than the energy squared. In order to find the eigen-energy ${\varepsilon}_0$ of quantum particle (unit-field), I presented Eq.~(\ref{eq37y}) in the form
\begin{eqnarray} 
{\varepsilon}_0 {\varepsilon}_0 | s_{0} \rangle | \tilde \psi_0 \rangle = {\hat{\mathbf{H}}} {\hat{\mathbf{H}}}| s_{0} \rangle | \tilde \psi_0 \rangle. 
\label{eq40y}
\end{eqnarray} 
Thus, Eq.~(\ref{eq11y}), (\ref{eq12y}) and (\ref{eq40y}) yielded the unit-field self-energy as the eigen-solution of the equation   
\begin{eqnarray} 
{\varepsilon}_{0} | s_{0} \rangle | \tilde \psi_0 \rangle = {\hat{\mathbf{H}}}| s_{0} \rangle | \tilde \psi_0 \rangle ,  
\label{eq41y}
\end{eqnarray} 
where the Hamiltonian operator ${\hat{\mathbf{H}}}$ reads
\begin{eqnarray} 
{\hat{\mathbf{H}}} = \left( m_{0} c^2 + U_{0ge}\right)\left( 1 + {\frac {\hat{\mathbf{p}}^{2}} {m_{0}^{2} c^2}} \right)^{1/2}.
\label{eq42y}
\end{eqnarray} 
Using of the notation $y = {\frac {\hat{\mathbf{p}}^{2}} {m_{0}^{2} c^2}}$ in Eq.~(\ref{eq42y}) yields the operator 
\begin{eqnarray} 
{\hat{\mathbf{H}}} = \left( m_{0} c^2 + U_{0ge}\right)(1+y)^{1/2} =  \left( m_{0} c^2 + U_{0ge}\right) \left( 1 + {\frac {1} {2}}y -{\frac {1} {8}}y^{2} + {\frac {1} {16}}y^{3} - \ldots \right),
\label{eq43y}
\end{eqnarray} 
which ensures nondivergence of the eigen-energy 
\begin{eqnarray} 
{\varepsilon}_{0} = \langle \tilde \psi_0 |\langle s_{0} |{\hat{\mathbf{H}}}| s_{0} \rangle | \tilde \psi_0 \rangle
\label{eq44y}
\end{eqnarray} 
only if the dimensionless parameter $\langle y \rangle = \langle \tilde \psi_0 |\langle s |y| s \rangle | \tilde \psi_0 \rangle $ satisfies the condition $\langle y \rangle<1$. The Hamiltonian operator~(\ref{eq42y}) could be presented also in the equivalent form 
\begin{eqnarray} 
{\hat{\mathbf{H}}} = m_{0} c^2 (1+x)^{1/2} =  m_{0} c^2 \left( 1 + {\frac {1} {2}}x -{\frac {1} {8}}x^{2} + {\frac {1} {16}}x^{3} - \ldots \right), 
\label{eq45y}
\end{eqnarray} 
where $x$ is given by Eq.~(\ref{eq39y}). The energy determined by Eqs.~(\ref{eq44y}) and (\ref{eq45y}) does not diverge only if the dimensionless parameter $\langle x \rangle = \langle \tilde \psi_0 |\langle s_{0} |x| s_{0} \rangle | \tilde \psi_0 \rangle <1$. If the dimensionless parameters $\langle y \rangle$ and $\langle x \rangle$ in Eqs.~(\ref{eq43y}) and (\ref{eq45y}) are of order one or larger, then the \textit{non-perturbation} model based on Eqs.~(\ref{eq37y})-(\ref{eq39y}) should be used to describe the  interactions.

\section{8. The operators ${\hat{\mathbf{H}}}$ and ${\hat{\mathbf{H}}^{2}}$ by using the scalar-potential ($\varphi_{0}$) and vector-potential ($\hat{\mathbf{A}}$) operators}
\label{sec8y}

The squared Hamiltonian operator (\ref{eq38y}) and the Hamiltonian operator (\ref{eq45y}) are determined by the operator (\ref{eq39y}) that includes the G ($U_{0g}$) and E ($U_{0e}$) static dresses (fields). The dresses are determined by Eq.~(\ref{eq20y}), where the G and E internal \textit{scalar} potentials (self-potentials) $\varphi_{0g}$ and $\varphi_{0e}$ are given by Eq. (\ref{eq22y}). 
Thus, the static dresses $U_{0g}=U_{0g}(\mathbf{r})$ and $U_{0e}=U_{0e}(\mathbf{r})$ and the static potentials $\varphi_{0g}=\varphi_{0g}(\mathbf{r})$ and $\varphi_{0e}=\varphi_{0e}(\mathbf{r})$ could be considered as the time-independent operators in the operators (\ref{eq38y}) and (\ref{eq45y}). The definition  
\begin{eqnarray} 
\hat{\mathbf{A}}_{0g}= \frac {\varphi_{0g} \hat{\mathbf{p}}} {m_{0}c},\nonumber \\ \hat{\mathbf{A}}_{0e} =  \frac {\varphi_{0e} \hat{\mathbf{p}}} {m_{0}c},
\label{eq48y}
\end{eqnarray}
of the operators of G and M internal \textit{vector} potentials (self-potentials) yielded the operator (\ref{eq39y}) in the form
\begin{eqnarray} 
x =  {\frac {\hat{\mathbf{p}}^{2}} {m_{0}^{2} c^2}}  + \nonumber \\ {\frac {2(m_0 \varphi_{0g} + q_0\varphi_{0e})} {m_{0} c^2}} + {\frac {(m_0 \varphi_{0g} + q_0\varphi_{0e})^{2}} {m_{0}^{2} c^4}} +  {\frac {2(m_0 \hat{\mathbf{A}}_{0g} + q_0\hat{\mathbf{A}}_{0e})\hat{\mathbf{p}} } {m_{0}^{2} c^3}} + {\frac {(m_0 \hat{\mathbf{A}}_{0g} + q_0\hat{\mathbf{A}}_{0e})^{2} } {m_{0}^{2} c^4}}.
\label{eq49y}
\end{eqnarray} 
Notice, if a unit-field (quantum particle) is interpreted as a string, then the above-presented model could be considered as a model of GEM string.

\section{9. Model of a unit-field in external GEM static-field}
\label{sec9y}

Let me extend the model of a single unit-field dressed into \textit{internal} GE static-field $U_{0ge}(\mathbf{r})=U_{0g}(\mathbf{r})+U_{0e}(\mathbf{r})$ to the unit-field placed in an \textit{external} GE static-field $U_{ge}(\mathbf{r})=U_{g}(\mathbf{r})+U_{e}(\mathbf{r})$. Because of the principle of additivity applied to  the \textit{internal} and \textit{external} fields (dresses) $U_{0ge}(\mathbf{r})$ and $U_{ge}(\mathbf{r})$, the operator $U_{0ge}(\mathbf{r})=U_{0g}(\mathbf{r})+U_{0e}(\mathbf{r})$ in Eqs.~(\ref{eq38y}), (\ref{eq43y}) and (\ref{eq45y}) is replaced by the operator of \textit{total} potential energy
\begin{eqnarray}
U^{\star}_{ge}(\mathbf{r}) = U^{\star}_{g}(\mathbf{r}) + U^{\star}_{e}(\mathbf{r}),
\label{eq50y}
\end{eqnarray}
where
\begin{eqnarray}
U^{\star}_{g}(\mathbf{r})= U_{0g}(\mathbf{r}) + U_{g}(\mathbf{r}),
\label{eq51y}
\end{eqnarray}
\begin{eqnarray}
U^{\star}_{e}(\mathbf{r})= U_{0e}(\mathbf{r}) + U_{e}(\mathbf{r}).
\label{eq52y}
\end{eqnarray}
The fields $U_{0ge}(\mathbf{r})$ and $U_{ge}(\mathbf{r})$ determine the \textit{internal} potential energy (self-energy) $\langle U_{0ge}\rangle= \langle \tilde \psi_0 (\mathbf{r})|\langle s_{0} (\mathbf{r})| U_{0ge}(\mathbf{r})| s_{0}(\mathbf{r}) \rangle | \tilde \psi_0 (\mathbf{r})\rangle$ and the \textit{external} potential energy $\langle U_{ge}\rangle= \langle s_{0} (\mathbf{r})| \langle \tilde \psi_0 (\mathbf{r})| U_{ge}(\mathbf{r}) | \tilde \psi_0 (\mathbf{r})\rangle | s_{0}(\mathbf{r}) \rangle$, respectively. Using of Eqs.~(\ref{eq35y}) and (\ref{eq36y}) yielded the relation 
\begin{eqnarray}
\langle U_{0ge}\rangle= \langle \tilde \psi_0 (\mathbf{r})|\langle s_{0} (\mathbf{r})| U_{0ge}(\mathbf{r})| s_{0}(\mathbf{r}) \rangle | \tilde \psi_0 (\mathbf{r})\rangle =  \langle s_{0} (\mathbf{r})| \langle \tilde \psi_0 (\mathbf{r})| U_{0ge}(\mathbf{r})| \tilde \psi_0 (\mathbf{r})\rangle | s_{0}(\mathbf{r}) \rangle. 
\label{eq53y}
\end{eqnarray}
The independence of the \textit{external} potential energy $\langle U_{ge}\rangle$ from the internal state $| s_{0}(\mathbf{r}) \rangle $ implies the relation 
\begin{eqnarray}
\langle U_{ge}\rangle= \langle s_{0} (\mathbf{r})| \langle \tilde \psi_0 (\mathbf{r})| U_{ge}(\mathbf{r})| \tilde \psi_0 (\mathbf{r})\rangle | s_{0}(\mathbf{r}) \rangle = \langle \tilde \psi_0 (\mathbf{r})|\langle s_{0} (\mathbf{r})| U_{ge}(\mathbf{r})| s_{0}(\mathbf{r}) \rangle | \tilde \psi_0 (\mathbf{r})\rangle.
\label{eq54y}
\end{eqnarray}

For the the operators~(\ref{eq38y}) and (\ref{eq45y}), the principle of additivity for the \textit{internal} and \textit{external} scalar and vector potentials (dresses) yields the parameter (\ref{eq49y}) in the form 
\begin{eqnarray} 
x = {\frac {\hat{\mathbf{p}}^{2}} {m_{0}^{2} c^2}} + \nonumber \\ {\frac {2(m_0 \varphi^{\star}_{g} + q_0\varphi^{\star}_{e})} {m_{0} c^2}}  + {\frac {(m_0 \varphi^{\star}_{g} + q_0\varphi^{\star}_{e})^{2}} {m_{0}^{2} c^4}}+ {\frac {2(m_0 \hat{\mathbf{A}}^{\star}_{g} + q_0\hat{\mathbf{A}}^{\star}_{e})\hat{\mathbf{p}} } {m_{0}^{2} c^3}} +{\frac {(m_0 \hat{\mathbf{A}}^{\star}_{g}+q_0\hat{\mathbf{A}}^{\star}_{e})^{2} } {m_{0}^{2} c^4}}.
\label{eq55y}
\end{eqnarray} 
This means that the operators (\ref{eq51y}) and (\ref{eq52y}) in Eq.~(\ref{eq55y}) are given by 
\begin{eqnarray}
U^{\star}_{g}= m_0 \varphi^{\star}_{g},
\label{eq56y}
\end{eqnarray}
\begin{eqnarray}
U^{\star}_{e}= q_0 \varphi^{\star}_{e},
\label{eq57y}
\end{eqnarray}
where the operators of \textit{total} scalar-potentials read  
\begin{eqnarray}
\varphi^{\star}_{g}  = \varphi_{0g} (\mathbf{r}) + \varphi_{g} (\mathbf{r}),\nonumber \\ \varphi^{\star}_{e} = \varphi_{0e} (\mathbf{r}) + \varphi_{e} (\mathbf{r}).
\label{eq58y}
\end{eqnarray}
Respectively, the operators of \textit{total} vector-potentials are given by
\begin{eqnarray} 
\hat{\mathbf{A}}^{{\star}}_{g} = \hat{\mathbf{A}}_{0g}(\mathbf{r}) + \hat{\mathbf{A}}_{g}(\mathbf{r}), \nonumber \\ \hat{\mathbf{A}}^{{\star}}_{e} = \hat{\mathbf{A}}_{0e}(\mathbf{r}) + \hat{\mathbf{A}}_{e}(\mathbf{r}),
\label{eq59y}
\end{eqnarray}
where $\hat{\mathbf{A}}_{0g}(\mathbf{r})= \frac {\varphi_{0g}(\mathbf{r}) \hat{\mathbf{p}}} {m_{0}c}$ and $\hat{\mathbf{A}}_{0e}(\mathbf{r}) =  \frac {\varphi_{0e}(\mathbf{r}) \hat{\mathbf{p}}} {m_{0}c}$ are the operators of \textit{internal} vector-potentials, and $\hat{\mathbf{A}}_{g}(\mathbf{r})$ and $\hat{\mathbf{A}}_{e}(\mathbf{r})$ denote the operators of \textit{external} vector-potentials. In case of the non-quantum external fields, the operators $\hat{\mathbf{A}}_{g}(\mathbf{r})$ and $\hat{\mathbf{A}}_{e}(\mathbf{r})$ obey the forms of G and M non-quantum vector-potentials ${\mathbf{A}}_{g}(\mathbf{r})$ and ${\mathbf{A}}_{e}(\mathbf{r})$. The pure-G and pure-EM interactions are described by the operator~(\ref{eq55y}) with $\varphi^{\star}_{e}(\mathbf{r}) =\hat{\mathbf{A}}^{{\star}}_{e}(\mathbf{r})=0$ and $\varphi^{\star}_{g}(\mathbf{r})=\hat{\mathbf{A}}^{{\star}}_{g}(\mathbf{r})=0$, respectively. 

The relations (\ref{eq37y}), (\ref{eq38y}), (\ref{eq41y}), (\ref{eq45y}) and (\ref{eq55y}) predict new physical phenomena for \textit{strong} potentials. For instance, the united GE interaction is described by the operators $\varphi^{\star}_{g} \varphi^{\star}_{e}$ and $\hat{\mathbf{A}}^{\star}_{g}\hat{\mathbf{A}}^{\star}_{e}$ of Eq.~(\ref{eq55y}). The physical picture of interactions based on the Taylor expansion of ${\hat{\mathbf{H}}}(x)$ in Eq.~(\ref{eq45y}) depends on the number of terms in the expansion. The terms of more than second order in the potentials describe the \textit{combined} interactions, which are the high-order combinations of G, EM and GEM quantum interactions. In case of the \textit{weak} scalar ($\langle\varphi^{\star}_{g}\rangle$, $\langle\varphi^{\star}_{e}\rangle$) and/or vector ($\langle{\mathbf{A}}^{\star}_{g}\rangle$,  $\langle{\mathbf{A}}^{\star}_{e}\rangle$) potentials, the combined interactions could be neglected.

\section{10. Model of the stationary multi-unitfield. The Pauli exclusion principle}
\label{sec10y}

We now consider the stationary multi-unitfield $\psi =\sum_{i=1}^N{\psi _{0i}(\mathbf{r}_{i})}$ that consists of the \textit{N} unit-fields ${\psi _{0i}}={\psi _{0i}(\mathbf{r}_{i})}$. The \textit{n-th} unit-field  ${\psi _{0n}(\mathbf{r}_{n})}$, which is surrounded by the ($N-1$) \textit{external} unit-fields, satisfies Eqs. (\ref{eq37y}) and (\ref{eq41y}), where $|s_{0} \rangle | \tilde \psi_{0} \rangle$, ${\hat{\mathbf{H}}^{2}}(x)$, ${\hat{\mathbf{H}}}(x)$ and ${\hat{\mathbf{H}}}(y)$ are replaced by $|s_{0n}\rangle | \tilde \psi_{0n} \rangle$, ${\hat{\mathbf{H}}_{n}^{2}}(x_{n})$, ${\hat{\mathbf{H}}_{n}}(x_{n})$ and ${\hat{\mathbf{H}}_{n}}(y_{n})$ with $y_{n} = {\frac {\hat{\mathbf{p}}_{n}^{2}} {m_{0n}^{2} c^2}}$ and  
\begin{eqnarray} 
x_{n} = {\frac {\hat{\mathbf{p}}_{n}^{2}} {m_{0n}^{2} c^2}}  +  {\frac {2(m_{0n} \varphi^{\star}_{gn} + q_{0n}\varphi^{\star}_{en})} {m_{0n} c^2}}  + {\frac {(m_{0n} \varphi^{\star}_{gn} + q_{0n}\varphi^{\star}_{en})^{2}} {m_{0n}^{2} c^4}} + \nonumber \\{\frac {2(m_{0n} \hat{\mathbf{A}}^{\star}_{gn} + q_{0n}\hat{\mathbf{A}}^{\star}_{en})\hat{\mathbf{p}}_{n} } {m_{0n}^{2} c^3}} + {\frac {(m_{0n} \hat{\mathbf{A}}^{\star}_{gn} + q_{0n}\hat{\mathbf{A}}^{\star}_{en})^{2} } {m_{0n}^{2} c^4}}.
\label{eq60y}
\end{eqnarray} 
The operators of scalar and vector \textit{total}-potentials for the \textit{n-th} unit-field are given by
\begin{eqnarray}
\varphi^{\star}_{gn} = \varphi_{0gn} (\mathbf{r}_{n}) + \sum_{m\neq n}^{N-1}\varphi_{gm} (\mathbf{r}_{n},\mathbf{r}_{m}),\nonumber  \\ \hat{\mathbf{A}}^{{\star}}_{gn} = \frac {\varphi_{0gn}(\mathbf{r}_{n}) \hat{\mathbf{p}}_{n}} {m_{0n}c} + \sum_{m\neq n}^{N-1}\frac {\varphi_{gm}(\mathbf{r}_{m}) \hat{\mathbf{p}}_{m}} {m_{0m}c},
\label{eq61y}
\end{eqnarray}
\begin{eqnarray}
\varphi^{\star}_{en} = \varphi_{0en} (\mathbf{r}_{n}) + \sum_{m\neq n}^{N-1}\varphi_{em}(\mathbf{r}_{n},\mathbf{r}_{m}),\nonumber  \\\hat{\mathbf{A}}^{{\star}}_{en} = \frac {\varphi_{0en}(\mathbf{r}_{n}) \hat{\mathbf{p}}_{n}} {m_{0n}c} + \sum_{m\neq n}^{N-1}\frac {\varphi_{em}(\mathbf{r}_{m}) \hat{\mathbf{p}}_{m}} {m_{0m}c},
\label{eq62y}
\end{eqnarray}
where $\varphi_{0gn} (\mathbf{r}_{n})$, $\varphi_{0en}(\mathbf{r}_{n})$, $\frac {\varphi_{0gn}(\mathbf{r}_{n}) \hat{\mathbf{p}}_{n}} {m_{0n}c}$ and $\frac {\varphi_{0en}(\mathbf{r}_{n}) \hat{\mathbf{p}}_{n}} {m_{0n}c}$ are the internal potentials of ${\psi _{0n}(\mathbf{r}_{n})}$; $\sum_{m\neq n}^{N-1}\varphi_{gm} (\mathbf{r}_{n},\mathbf{r}_{m}) =\varphi_{g} (\mathbf{r}_{n})$ and $\sum_{m\neq n}^{N-1}\varphi_{em}(\mathbf{r}_{n},\mathbf{r}_{m}) = \varphi_{e} (\mathbf{r}_{n})$ are the operators of \textit{external} scalar potentials, and $\sum_{m\neq n}^{N-1}\frac {\varphi_{gm}(\mathbf{r}_{m}) \hat{\mathbf{p}}_{m}} {m_{0m}c} = \hat{\mathbf{A}}_{g}$ and $\sum_{m\neq n}^{N-1}\frac {\varphi_{em}(\mathbf{r}_{m}) \hat{\mathbf{p}}_{m}} {m_{0m}c}=\hat{\mathbf{A}}_{e}$ denote the operators of \textit{external} vector potentials. The explicit expressions for the scalar potentials are given by 
\begin{eqnarray}
\varphi_{0gn}(\mathbf{r}_{n}) = -\gamma_{N}{\frac {m_{0n}} {|\mathbf{r}_{n}|}} ,\nonumber \\ \varphi_{0en}(\mathbf{r}_{n}) =- \gamma_{C}{\frac {q_{0n}} {|\mathbf{r}_{n}|}},
\label{eq63y}
\end{eqnarray}
\begin{eqnarray}
\varphi_{gm}(\mathbf{r}_{m}) = -\gamma_{N}{\frac {m_{0m}} {|\mathbf{r}_{n}-\mathbf{r}_{m}|}} ,\nonumber \\ \varphi_{em}(\mathbf{r}_{m}) =+ \gamma_{C}{\frac {q_{0m}} {|\mathbf{r}_{n}-\mathbf{r}_{m}|}}.
\label{eq64y}
\end{eqnarray}
For the comparison, see Eqs.~(\ref{eq22y}), (\ref{eq58y}) and (\ref{eq59y}). The sign ($-$) in Eqs.~(\ref{eq22y}) and (\ref{eq63y}) is explained in Sect. 12.2. Note that $x_{n}=x$ for $N=1$.

For the multi-unitfield $\psi =\sum_{i=1}^N{\psi _{0i}(\mathbf{r}_{i})}$, the use of Eqs.~(\ref{eq37y}), (\ref{eq41y}), (\ref{eq60y})-(\ref{eq64y}) and the principle of additivity for the squared energies and energies of unit-fields $\psi _{0i}(\mathbf{r}_{i})$ yielded the \textit{unified mass-energy relations} 
\begin{eqnarray} 
\langle{\psi}|  \left(\sum_{i=1}^N {\varepsilon}^{2}_{0i} \right) | {\psi}\rangle  = \langle{\psi}|  \left(\sum_{i=1}^N {\hat{\mathbf{H}}}^{2}_{i}(x_{i}) \right) |{\psi}\rangle,  
\label{eq65y}
\end{eqnarray} 
\begin{eqnarray} 
\langle{\psi}|  \left(\sum_{i=1}^N {\varepsilon}_{0i} \right) | {\psi}\rangle  = \langle{\psi}|  \left(\sum_{i=1}^N {\hat{\mathbf{H}}}_{i}(x_{i}) \right) |{\psi}\rangle,
\label{eq66y}
\end{eqnarray} 
where $| {\psi}_{i}(\mathbf{r}_{i})\rangle=|s_{0i}(\mathbf{r}_{i})\rangle | \tilde \psi_{0i}(\mathbf{r}_{i}) \rangle$, $\langle {\psi}_{i} |{\psi}_{j} \rangle=\delta_{ij}$, and $\delta_{ij}$ denotes the Kronecker symbol. Replacement of the field state $|\psi\rangle  =\sum_{i=1}^N|{\psi _{0i}(\mathbf{r}_{i})}\rangle$ by the QM state $|\psi\rangle =\prod ^N _{i=1}| {\psi}_{0i}(\mathbf{r}_{i})\rangle$ in Eqs.~(\ref{eq65y}) and (\ref{eq66y}) solves the long-standing problem in Quantum Field Theory (QFT), namely the transition from QFT to QM. 
The dependence $x_{i}=x_{i}(\mathbf{r}_{i},\mathbf{r}_{j})$ in the multi-unitfield energy $\langle{\psi}|  \left(\sum_{i=1}^N {\hat{\mathbf{H}}}_{i}(x_{i}) \right) |{\psi}\rangle$ causes the cross-correlation (interference) of unit-fields ${\psi}_{0i}(\mathbf{r}_{i})$ and ${\psi}_{0j}(\mathbf{r}_{j})$. That \textit{explains} what is the interference of material unit-fields mentioned in Sect. 2. The multi-unitfield $\psi =\sum_{i=1}^N{\psi _{0i}(\mathbf{r}_{i})}$ consists of the \textit{bare} unit-fields ${\psi _{0i}}={\psi _{0i}(\mathbf{r}_{i})}$. The stationary multi-unitfield $\Psi =\sum_{i=1}^N{\Psi_{0i}(\mathbf{r}_{i})}$ discussed in Sect. 2 is a superposition of the \textit{dressed} unit-fields $\Psi_{0i}(\mathbf{r}_{i})=\Phi_{0i,ge}(\mathbf{r}_{i}){\psi}_{0i}(\mathbf{r}_{i})$, which should be representations of the Lorentz group. The unit-fields $\Psi_{0i}(\mathbf{r}_{i})$ are representations of the Lorentz group, because the operator $\hat{\mathbf{p}}$ and the potentials~(\ref{eq61y}) and (\ref{eq62y}) were modelled to satisfy the Lorentz transformations. Notice, the relations~(\ref{eq65y}) and (\ref{eq66y}) indicate the orbit-orbit, spin-orbit and spin-spin (exchange) interactions of the unit-fields $ {\psi}_{i}(\mathbf{r}_{i})$ (see, Sects.~11 and 12). 

The \textit{Pauli exclusion principle} is a postulate in QM. In the present model of the two (i=1,2) electrons in the different internal states and identical external states ($|\psi_{01}(\mathbf{r}_{1}) \rangle=| s_{0,\pm}(\mathbf{r}) \rangle | \tilde \psi_{0}(\mathbf{r}) \rangle$, $|\psi_{02}(\mathbf{r}_{2}) \rangle=| s_{0,\mp}(\mathbf{r}) \rangle | \tilde \psi_{0}(\mathbf{r}) \rangle$, ${\hat{\mathbf{H}}}_{1}(\mathbf{r}_{1})={\hat{\mathbf{H}}}_{2}(\mathbf{r}_{2})={\hat{\mathbf{H}}}(\mathbf{r})$), the gauge (\ref{eq31y}), Eq.~(\ref{eq66y}) and the condition
\begin{eqnarray} 
\langle \tilde \psi_{0} |\langle s_{0,\pm}| {{\hat{\mathbf{H}}}} | s_{0,\pm}\rangle | \tilde \psi_{0} \rangle + \langle \tilde \psi_{0} | \langle s_{0,\mp}|{{\hat{\mathbf{H}}}} | s_{0,\mp}\rangle | \tilde \psi_{0} \rangle  < \nonumber  \\   \langle \tilde \psi_{0} | \langle s_{0,\pm}| {{\hat{\mathbf{H}}}} | s_{0,\pm}\rangle | \tilde \psi_{0} \rangle +  \langle \tilde \psi_{0} |\langle s_{0,\pm}| {{\hat{\mathbf{H}}}} | s_{0,\pm}\rangle | \tilde \psi_{0} \rangle
\label{eq67y}
\end{eqnarray} 
indicate the minimum of energy only if no two identical electrons occupy the same quantum state $| \tilde \psi_{0} \rangle$. The energy minimum explains the physical nature of \textit{Pauli's principle}, which states that two identical electrons cannot occupy the same quantum state, simultaneously.

\section{11. Comparison of the basic Hamiltonian operators of QM with the present model}
\label{sec11y}
 
Let me derive the Hamiltonian operators for an electron unit-field (charged quantum particle) placed in a E or EM static field, which are the low-order approximations to the unified Hamiltonian (\ref{eq45y}) in the parameter (\ref{eq60y}). For the comparison with QM, the G-field $U^{\star}_{g}= U_{0g}+U_{g}$ and E self-field $U_{0e}$ are neglected, i.e. $U^{\star}_{ge}= U_{0g}+U_{0e} + U_{g}+ U_{e}\approx U_{e}$.  

In the E or EM static field, the electron unit-field obeys the stationary state $| {\psi_{0}} (\mathbf{r},t)\rangle  =\exp\left(\pm{{\frac {i} {{\hbar}}}} {\varepsilon}_{0}t \right) $ $| s_{\pm}(\mathbf{r})\rangle  |\tilde \psi_0(\mathbf{r}) \rangle $. For the E field, keeping the first two terms in Eqs.~(\ref{eq45y}) and (\ref{eq55y}) yields the unified quantum equation~(\ref{eq8y}) in the general form
\begin{eqnarray} 
\langle \tilde \psi_0|\langle s_{\pm}|({\varepsilon}_{0} -  m_{0} c^2 ) | s_{\pm} \rangle  | \tilde \psi_0 \rangle = \langle \tilde \psi_0|\langle s_{\pm}|\hat{\mathbf{H}}| s_{\pm} \rangle  | \tilde \psi_0 \rangle= \langle s_{\pm}|\langle \tilde \psi_0|\hat{\mathbf{H}}| \tilde \psi_0 \rangle| s_{\pm} \rangle,  
\label{eq68y}
\end{eqnarray} 
which can be presented in the Shr\"odinger form
\begin{eqnarray} 
\langle\tilde\psi_0| ({\varepsilon}_{0} -  m_{0} c^2 ) |\tilde\psi_0 \rangle = \langle\tilde\psi_0| \hat{\mathbf{H}}|\tilde\psi_0 \rangle,  
\label{eq69y}
\end{eqnarray}
where the Hamiltonian operator $\hat{\mathbf{H}}$ reads 
\begin{eqnarray} 
\hat{\mathbf{H}}={\frac {\hat{\mathbf{p}}^{2}} {2m_{0}}}  +  U_{e}.  
\label{eq70y}
\end{eqnarray}
Here, $|s_{\pm}(\mathbf{r})\rangle\equiv |s_{0,\pm}(\mathbf{r})\rangle$, $U_{e}=q_{0}\varphi_{e}(\mathbf{r})$ is the E field mediated by the external scalar potential $\varphi_{e}(\mathbf{r})$, and $q_{0}$ denotes the electron charge. Notice, in Eqs.~(\ref{eq68y}) and (\ref{eq69y}), I also use the relations~(\ref{eq25y}), (\ref{eq30y}), (\ref{eq31y}), (\ref{eq35y}), (\ref{eq36y}), (\ref{eq53y}) and (\ref{eq54y}).

For the electron unit-field placed in the EM field, taking into account the first two terms in Eq.~(\ref{eq45y}) and the 1st, 2nd, 4th and 5th terms in Eq.~(\ref{eq55y}) yields Eq.~(\ref{eq8y}) in the form~(\ref{eq68y}), where   
\begin{eqnarray} 
\hat{\mathbf{H}}={\frac {1} {2m_{0}}} \left( \hat{\mathbf{p}} + {\frac {q_{0}} {c}}\hat{\mathbf{A}}_{e} \right)^{2}+ q_{0}\varphi_{e}.  
\label{eq71y}
\end{eqnarray} 
Here, the operators $\varphi_{e}=\varphi_{e}(\mathbf{r})$ and $\hat{\mathbf{A}}_{e}=\hat{\mathbf{A}}_{e}(\mathbf{r})$ correspond to the external scalar and vector (M) potentials. The Hamiltonian~(\ref{eq71y}) is well-known in QM \cite{land,ber}. For the non-quantum M-field ${\mathbf{B}}_{e}(\mathbf{r})$ determined by the relations $\mathbf{A}_{e}=(1/2){\mathbf{B}}_{e}\times {\mathbf{r}}$ and $\nabla {\mathbf{A}}_{e} =0$~\cite{lan}, Eq. (\ref{eq71y}) yields the equation of QM~\cite{land,ber}: 
\begin{eqnarray} 
\langle \tilde \psi_0|\langle s_{\pm}|({\varepsilon}_{0} -  m_{0} c^2 ) | s_{\pm} \rangle  | \tilde \psi_0 \rangle = \nonumber \\ \langle s_{\pm}| s_{\pm} \rangle  \langle \tilde \psi_0|\left({\frac {\hat{\mathbf{p}}^{2}} {2m_{0}}} + q_{0}\varphi_{e} \right)| \tilde \psi_0 \rangle +  \langle s_{\pm}| s_{\pm} \rangle {\frac { q_{0}g_{L}} {2m_{0}c}} {\mathbf{B}}_{e} \langle \tilde \psi_0|\hat{\mathbf{L}} |\tilde \psi_0 \rangle  + \nonumber \\\langle \tilde \psi_0|{\frac {q_{0}g_{s}} {2m_{0}c}} {\mathbf{B}}_{e}\langle s_{\pm}| \hat{\mathbf{L}} | s_{\pm} \rangle \tilde |\psi_0 \rangle   +   \langle \tilde \psi_0|\langle s_{\pm}|  {\frac {q_{0}^{2}} {8m_{0}c^{2}}} (\mathbf{B}_{e}\times {\mathbf{r}} )^{2} |s_{\pm} \rangle  | \tilde \psi_0 \rangle.  
\label{eq72y}
\end{eqnarray} 
The constants $g_{L}\equiv1$ and $g_{s}\equiv1$ were introduced into Eq.~(\ref{eq72y}) for the comparison with QM. In Eq.~(\ref{eq72y}), the orbital momentum $\langle {{L}_{z}}\rangle$ and the spin $\langle {{s}_{z}}\rangle$ are given by $\langle {{L}_{z}}\rangle =\langle  \tilde \psi_0|{\hat{L}_{z}} |\tilde \psi_0 \rangle = \pm \hbar m$ and $\langle {{s}_{z}}\rangle = \langle  s_{\pm}| {\hat{L}_{z}} | s_{\pm}\rangle = \pm \hbar$. In QM, the respective equation~\cite{land,ber} obeys the form~(\ref{eq72y}), where the orbital momentum $\langle {{L}_{z}}\rangle = \pm \hbar m$, the spin $\langle {{s}_{z}}\rangle = \pm \hbar/2$, and the non-unified gyromagnetic factors are given by $g_{L} = 1$ and $g_{s} = 2$. The comparison of Eq.~(\ref{eq72y}) with QM reveals the unified gyromagnetic factor $g = g_{L}=g_{s}=1$ in the present model. That explains difference between the spin $\langle s_{\pm} | {\hat{L}_{z}} | s_{\pm}\rangle = \pm \hbar$ in  Eq.~(\ref{eq72y}) and the value $\langle s_{\pm} | {\hat{s}_{z}} | s_{\pm}\rangle = \pm \hbar /2$ in QM.  

\section{12. Comparison of the unified quantum equation with the Dirac equation. The electron anomalous gyromagnetic factor}
\label{subsec12y}

\subsection{12.1 The unified quantum equation versus the Dirac equation}
\label{subsec12a}

The quantum equation~(\ref{eq8y}) for the united gravitation and electromagnetism yields all known solutions to the Dirac equation~\cite{dira}. As an example, I derive the \textit{fine} and \textit{hyperfine} structure of the atom spectrum and compare that with the Dirac model. For the electron unit-field of hydrogen atom in the state $| {\psi_{0}} \rangle$, the Hamiltonian operator~(\ref{eq43y}) reads 
\begin{eqnarray} 
\hat{\mathbf{H}} =  m_{0} c^2 \left( 1 + \frac{\hat{\mathbf{p}}^2}{m_{0}^{2}c^{2}} \right)^{1/2}+ U^{\star}_{ge}\left( 1 + \frac{\hat{\mathbf{p}}^2}{m_{0}^{2}c^{2}} \right)^{1/2}\approx \nonumber \\   m_{0} c^2 + \frac{\hat{\mathbf{p}}^2}{2m_{0}} + U^{\star}_{ge} - \frac{\hat{\mathbf{p}}^4}{8m_{0}^{3}c^{2}} + U^{\star}_{ge}\frac{\hat{\mathbf{p}}^2}{2m_{0}^{2}c^{2}}. 
\label{eq73y}
\end{eqnarray} 
Using of the approximation $U^{\star}_{ge}= U_{0g}+U_{0e} + U_{g}+ U_{e} \approx U_{e}$ yields the operator~(\ref{eq73y}) in the form  
\begin{eqnarray} 
\hat{\mathbf{H}} = m_{0} c^2 +  \hat{\mathbf{H}}_{0} + \hat{\mathbf{H}}^{1}_{fine}, 
\label{eq74y}
\end{eqnarray} 
where 
\begin{eqnarray} 
\hat{\mathbf{H}}_{0} = \frac{\hat{\mathbf{p}}^2}{2m_{0}} + U_{e}
\label{eq75y}
\end{eqnarray}
is the unperturbed Hamiltonian operator, and 
\begin{eqnarray} 
\hat{\mathbf{H}}^{1}_{fine} = -\frac{\hat{\mathbf{p}}^4}{8m_{0}^{3}c^{2}} + U_{e}\frac{\hat{\mathbf{p}}^2} {2m_{0}^{2}c^{2}} 
\label{eq76y}
\end{eqnarray} 
is the perturbation operator. For $\hat{\mathbf{H}}_{0} | s_{\pm} \rangle | \tilde{\psi_0} \rangle = {\varepsilon}_0 | s_{\pm} \rangle | \tilde{\psi_0} \rangle $ and $| \psi_{0} \rangle \approx | s_{\pm} \rangle | \tilde{\psi_0} \rangle$, the first-order perturbation energy  $ \langle \tilde{\psi_0} | \langle s_{\pm} | -\frac{\hat{\mathbf{p}}^4}{8m_{0}^{3}c^{2}} | s_{\pm} \rangle | \tilde{\psi_0} \rangle$ yields the \textit{fine} structure~\cite{land,ber} of the atom spectrum attributed to the relativistic movement in the Dirac model. While, the perturbation energy $\langle \tilde\psi_0| \langle s_{\pm}|U_{e}\frac{\hat{\mathbf{p}}^2}{2m_{0}^{2}c^{2}}|s_{\pm} \rangle |\tilde\psi_0 \rangle$ causes the \textit{fine} structure attributed to the spin-orbit and Darwin interactions. Indeed, for the M field described by the operator $\hat{\mathbf{B}}_{e}$ (see, Eq.~(\ref{eq72y})) and $U_{e} \frac{\hat{\mathbf{p}}^2}{2m_{0}^{2}c^{2}} \equiv \frac{q_{0}\hat{\mathbf{A}}_{e} \hat{\mathbf{p}}}{2m_{0}c}$, we have the perturbation energy 
\begin{eqnarray} 
\langle \tilde{\psi_0} | \langle s_{\pm} | \left( U_{e} \frac{\hat{\mathbf{p}}^2}{2m_{0}^{2}c^{2}} \right) | s_{\pm} \rangle | \tilde{\psi_0} \rangle =  \langle \tilde{\psi_0} | \frac {q_{0}}{4m_{0}c} \hat{\mathbf{B}}_{e} | \tilde{\psi_0} \rangle \langle s_{\pm} | \hat{\mathbf{L}} | s_{\pm} \rangle. 
\label{eq77y}
\end{eqnarray} 
According to electromagnetism, the electron orbital M-moment ${\mathbf{m}}_{e}=\frac{q_{0}g}{2m_{0}c}{\mathbf{L}}$ generates the M-field ${\mathbf{B}}_{e}$ with the operator
\begin{eqnarray} 
\hat{\mathbf{B}}_{e}= \frac{1}{4\pi\epsilon_{0}c^{2}r^{3}} [3(\hat{\mathbf{m}}_{e} \mathbf{e}_{r})\mathbf{e}_{r}-\hat{\mathbf{m}}_{e}] +\frac{2}{3\epsilon_{0}c^{2}}\hat{\mathbf{m}}_{e}\delta(\mathbf{r}),
\label{eq78y}
\end{eqnarray} 
which yields the perturbation energy~(\ref{eq77y}) in the form
\begin{eqnarray} 
\langle \tilde{\psi_0} | \frac{q_{0}}{4m_{0}c} \hat{\mathbf{B}}_{e} | \tilde{\psi_0} \rangle \langle s_{\pm} | \hat{\mathbf{L}} | s_{\pm} \rangle =  -\frac{q_{0}^{2}g}{16\pi \epsilon_{0}c^{4}m_{0}^{2}r^{3}} \langle \tilde{\psi_0} | \hat{\mathbf{L}} | \tilde{\psi_0} \rangle \langle s_{\pm} | \frac{\hat{\mathbf{L}}}{2} | s_{\pm} \rangle + \nonumber  \\ \frac{q_{0}^{2}g}{6 \epsilon_{0}c^{4}m_{0}^{2}}\langle \tilde{\psi_0} | \hat{\mathbf{L}} | \tilde{\psi_0} \rangle \langle s_{\pm} | \frac{\hat{\mathbf{L}}}{2} | s_{\pm} \rangle \delta (\mathbf{r}). 
\label{eq79y}
\end{eqnarray} 
Here, $g\equiv 1$, $\delta(\mathbf{r})$ is the Dirac delta, and the Coulomb constant reads $\gamma_C=1/4\pi\epsilon_0$. The first term on the right-hand side of Eq.~(\ref{eq79y}) yields the atom \textit{fine} structure mediated by the spin-orbit interaction, whereas the second term is the Darwin energy in the \textit{fine} structure~\cite{land,ber}. Thus, the fine structure determined by the operator~(\ref{eq76y}) is in agreement with the Dirac equation.

The proton in a hydrogen atom possesses a M-moment $\langle{\mathbf{m}}_{p}\rangle =\langle s_{p} | \frac{q_{0p}g_p}{2m_{0p}c}{\mathbf{L}}| s_{p} \rangle$, where $q_{0p}$, $m_{0p}$,  $g_p$ and $|s_{p} \rangle$ are the proton charge, mass, gyromagnetic factor and state. The M field $\langle{\mathbf{B}}_{p}\rangle$ induced by the M-moment $\langle{\mathbf{m}}_{p}\rangle$ yields the atom \textit{hyperfine} structure attributed to the spin-spin interaction. Indeed, replacement of the term $\langle \tilde{\psi_0} | {\frac{q_{0}}{4m_{0}c}} \hat{\mathbf{B}}_{e} | \tilde{\psi_0} \rangle$ of Eq.~(\ref{eq77y}) by $\langle \tilde{\psi_0} |\langle s_{p} | {\frac{q_{0}}{4m_{0}c}} \hat{\mathbf{B}}_{p}| s_{p} \rangle | \tilde{\psi_0} \rangle$ yields the \textit{hyperfine} energy~\cite{land,ber}
\begin{eqnarray} 
\langle \tilde{\psi_0} |\langle s_{p} | {\frac{q_{0}}{4m_{0}c}} \hat{\mathbf{B}}_{p}| s_{p} \rangle | \tilde{\psi_0} \rangle \langle s_{\pm} | \hat{\mathbf{L}} | s_{\pm} \rangle =  \frac{3q_{0}q_{0p}g_p}{8\pi\epsilon_{0}c^{4}m_{0}m_{0p}r^{3}} \langle s_{p} | \frac{\mathbf{e}_{r}\hat{\mathbf{L}}}{2} | s_{p} \rangle \langle s_{\pm} | \frac{\mathbf{e}_{r} \hat{\mathbf{L}}}{2} | s_{\pm} \rangle - \nonumber  \\ \frac{q_{0}q_{0p}g_p}{8\pi\epsilon_{0}c^{4}m_{0}m_{0p}r^{3}} \langle s_{p} | \frac{\hat{\mathbf{L}}}{2} | s_{p} \rangle \langle s_{\pm} | \frac{\hat{\mathbf{L}}}{2} | s_{\pm} \rangle +  \frac{q_{0}q_{0p}g_p}{3\epsilon_{0}c^{4}m_{0}m_{0p}} \langle s_{p} | \frac{\hat{\mathbf{L}}}{2} | s_{p} \rangle \langle s_{\pm} | \frac{\hat{\mathbf{L}}}{2} | s_{\pm} \rangle\delta(\mathbf{r}),
\label{eq80y}
\end{eqnarray} 
which compares well with the Dirac model. For the ground-state of hydrogen, which is spherically symmetric, the first two terms on the right-hand side of Eq.~(\ref{eq80y}) vanish because of symmetry. The comparison of Eqs.~(\ref{eq79y}) and (\ref{eq80y}) with QM reveals the unified gyromagnetic factor $g = g_{L}=g_{s}=1$ in the present model. For the \textit{strong} potentials, the unified equation~(\ref{eq8y}) predicts the new phenomena compared to the Dirac equation, e.g. the \textit{quantum gravitoelectromagnetism} ($U^{\star}_{ge}= U_{0g}+U_{0e} + U_{g}+ U_{e}$) and \textit{quantum gravitation} ($U^{\star}_{ge}= U_{0g}+U_{g}$).

\subsection{12.2 The electron anomalous gyromagnetic factor}
\label{subsec12by}

Sections~11 and 12.1 revealed the unified gyromagnetic factor $g = g_{L}=g_{s}=1$ under the approximation $U^{\star}_{ge}= U_{0g}+U_{0e} + U_{g}+ U_{e}\approx U_{e}$. In quantum electrodynamics (QE), calculations of the anomalous gyromagnetic factor $g^{a}_{s}$ agree with the experimental value $g^{a}_{s} = 2\times 1.00115965218$ with more than ten digits. For instance, the QE calculus~\cite{schw,weis} of one loop contribution to anomalous part of the magnetic moment yields $g^{a}_{s} = 2\times 1.00116$. 

In the present model, the approximation $U^{\star}_{ge}\approx U_{0e} + U_{e}$ in Eq.~(\ref{eq73y}) yields the value of $g_s=g^{a}_{s}/2 $. Under the approximation, Eq.~(\ref{eq73y}) reads 
\begin{eqnarray} 
\hat{\mathbf{H}} = \left(m_{0}c^2+ U_{0e} \right)  + \hat{\mathbf{H}}^{\prime}_{0} + \hat{\mathbf{H}}^{1}_{fine},
\label{eq81y}
\end{eqnarray} 
where 
\begin{eqnarray} 
 \hat{\mathbf{H}}^{\prime}_{0}=\left( \frac{1}{2m_{0}}+\frac{U_{0e}}{2m_{0}^{2}c^{2}} \right)\hat{\mathbf{p}}^2 + U_{e} 
\label{eq82y}
\end{eqnarray} 
is the unperturbed Hamiltonian operator, and the perturbation operator $\hat{\mathbf{H}}^{1}_{fine}$ is given by Eq.~(\ref{eq76y}). In the case of $\hat{\mathbf{H}}^{\prime}_{0}|s_{\pm} \rangle |\tilde\psi^{\prime}_0 \rangle={\varepsilon}_0 |s_{\pm} \rangle |\tilde\psi^{\prime}_0 \rangle$ and $|{\psi_{0}}\rangle \approx |s_{\pm} \rangle |\tilde\psi^{\prime}_0 \rangle$, the operators~(\ref{eq81y}) and (\ref{eq82y}) read $\hat{\mathbf{H}} = \left(m_{0}c^2+ \langle m_{0e} \rangle c^2  \right)  + \hat{\mathbf{H}}^{\prime}_{0} + \hat{\mathbf{H}}^{1}_{fine}$ 
and 
\begin{eqnarray} 
\hat{\mathbf{H}^{\prime}_0}= \frac{\hat{\mathbf{p}}^2}{2m^{\prime}_{0}} + U_{e}, 
\label{eq83y}
\end{eqnarray} 
where $\langle m_{0e} \rangle c^{2} = \langle s_{\pm}|U_{0e}|s_{\pm} \rangle $ and $m^{\prime}_{0} = m_{0}\left(1+\frac{\langle m_{0e} \rangle}{m_{0}}\right)^{-1}$ are the electrostatic self-energy and the modified mass. The comparison of Eq.~(\ref{eq75y}) with Eq.~(\ref{eq83y}) indicates the change 
\begin{eqnarray} 
\tilde\psi_0(m_0) \rightarrow \tilde\psi^{\prime}_0(m^{\prime}_0)=  \tilde\psi_0(m^{\prime}_0)
\label{eq84y}
\end{eqnarray} 
in Eqs.~(\ref{eq77y}) and (\ref{eq79y}). Using of the well-known explicit expression of $\tilde\psi_0(m_0)$ for a hydrogen atom~\cite{land} and the change~(\ref{eq84y}) in Eq.~(\ref{eq77y}) yields 
\begin{eqnarray} 
\langle \tilde{\psi}^{\prime}_0(m^{\prime}_0) | \frac{q_{0}}{4m_{0}c} \hat{\mathbf{B}}_{e} | \tilde{\psi}^{\prime}_0(m^{\prime}_0)\rangle\langle s_{\pm} | \hat{\mathbf{L}} | s_{\pm} \rangle = \nonumber  \\-\frac{q_{0}^{2}g^{a}_{s}}{16\pi\epsilon_{0}c^{4}m_{0}^{2}r^{3}} \langle \tilde{\psi}_0|\hat{\mathbf{L}} | \tilde{\psi_0} \rangle \langle s_{\pm} | \frac{\hat{\mathbf{L}}}{2} | s_{\pm} \rangle +\frac{q_{0}^{2}g^{a}_{s}}{6\epsilon_{0}c^{4}m_{0}^{2}}\langle \tilde{\psi_0} | \hat{\mathbf{L}} | \tilde {\psi_0} \rangle \langle s_{\pm} | \frac{\hat{\mathbf{L}}}{2} | s_{\pm} \rangle\delta(\mathbf{r}), 
\label{eq85y}
\end{eqnarray} 
where the anomalous gytomagnetic factor $g^{a}_{s}=m^{\prime}_{0}/ m_{0} $ reads
\begin{eqnarray} 
g^{a}_{s} = \left(1+\frac{\langle m_{0e} \rangle}{m_{0}}\right)^{-1}  
\label{eq86y}
\end{eqnarray} 
Equation~(\ref{eq85y}) indicates the change in the electron M-moment, $\langle s_{\pm} | \hat{\mathbf{m}}_{e} | s_{\pm} \rangle = \langle s_{\pm} | \frac{q_{0}g}{2m_{0}c}\hat{\mathbf{L}} | s_{\pm} \rangle \rightarrow \langle s_{\pm} | \hat{\mathbf{m}}^{a}_{e} | s_{\pm} \rangle = \langle s_{\pm} | \frac{q_{0}g^{a}_{s}}{2m_{0}c}\hat{\mathbf{L}} | s_{\pm} \rangle$. The same result is obtained for Eq.~(\ref{eq80y}). The value~(\ref{eq86y}) is equal to 1.00116 under the condition
\begin{eqnarray} 
\frac{\vert \langle m_{0e} \rangle \vert}{m_{0}} = \frac {\alpha}{2\pi}, 
\label{eq87y}
\end{eqnarray} 
which compares well with Ref.~\cite{schw}. Here, $\langle m_{0e}\rangle=U^{s}_{0e}c^{-2}  =-|\langle s_{\pm}  |U_{0e}(\mathbf{r})| s_{\pm} \rangle| c^{-2} = -\vert \langle s_{\pm}| q_{0}\varphi_{0e} |s_{\pm} \rangle \vert c^{-2}$ is the unit-field electric rest-mass determined by Eq.~(\ref{eq34y}), and $\alpha=e^{2}/4\pi\epsilon_0\hbar c$ is the fine structure constant. The relation $\langle m_{0e}\rangle  = -\vert \langle s_{\pm}| q_{0}\varphi_{0e} |s_{\pm} \rangle \vert c^{-2}$ explains the sign ($-$) in Eqs.~(\ref{eq22y}) and (\ref{eq63y}). Notice, Eqs.~(\ref{eq34y}) and (\ref{eq87y}) yielded $r_{0}=\frac {5\pi}{2\alpha}\left(\frac {\gamma_{C}q^{2}_{0}}{m_{0}c^{2}}  \right)$. The anomalous gytomagnetic factor $g^{a}_{s}=1.00116$ was obtained by using the perturbation model of first-order. Taking into account the perturbation terms of higher order yields the more accurate value of $g^{a}_{s}$. 

The weak ($|\langle x \rangle|\ll 1$) interplay between the electric ($\langle m_{0e} \rangle$) and mechanical ($m_{0}$) masses under the transition $\tilde\psi_0(m_0) \rightarrow \tilde\psi_0(m^{\prime}_0)$ explains the anomalous magnetic moment $\langle s_{\pm}|\hat{\mathbf{m}}^{a}_{e}| s_{\pm} \rangle$ and gyromagnetic factor $g^{a}_{s}$. The explanation is in agreement with QE in principle. Indeed, the QE model~\cite{schw,weis} is based on the hypothesis of a weak interaction between the mechanical and electromagnetic masses (matter and radiation) under the zero-point oscillation of the EM field and the current fluctuations induced in the Dirac vacuum. 

\section{13. The non-stationary unit-field}
\label{sec13yabc}

In case of the non-stationary unit-field ($|{\psi_{0}}(\mathbf{r},t)\rangle = |s(\mathbf{r},t)\rangle | \tilde \psi_{0}(\mathbf{r},t) \rangle$), Eqs.~(\ref{eq8y}), (\ref{eq50y})-(\ref{eq55y}) and (\ref{eq60y})) yielded the time-dependent mass-energy relation
\begin{eqnarray} 
{-\hbar}^{2} {\frac {{\partial}^2} {\partial {t}^2}} |{\psi_{0}}(\mathbf{r},t)\rangle   = {{\hat{\mathbf{H}}}^{2}}(x) |{\psi_{0}}(\mathbf{r},t)\rangle,  
\label{eq88y}
\end{eqnarray} 
where the squared energy ${\varepsilon}_0^{2}(t)$ is determined by the operator $x = x(\mathbf{r},t)=x(\varphi^{\star}_{g} (\mathbf{r},t), \varphi^{\star}_{e} (\mathbf{r},t), \hat{\mathbf{A}}^{\star}_{g} (\mathbf{r},t), \hat{\mathbf{A}}^{\star}_{e} (\mathbf{r},t))$ in ${\hat{\mathbf{H}}}^{2}(x)$. The operators of scalar and vector potentials are found by using the quantum wave-equations  
\begin{eqnarray}
\left(\bigtriangledown^{2}-\frac{1}{c^{2}} \frac{\partial^{2}}{\partial t^{2}}\right) \varphi^{\star} (\mathbf{r},t) =\pm4\pi \Gamma
\delta (t{\mp}r/c), 
\label{eq89y}
\end{eqnarray}
\begin{eqnarray}
\left(\bigtriangledown^{2}-\frac{1}{c^{2}} \frac{\partial^{2}}{\partial t^{2}}\right) \hat{\mathbf{A}}^{\star}({\mathbf{r}},t) =\pm4\pi\Gamma \delta (t{\mp}r/c) \frac{\hat{\mathbf{p}}}{m_{0}c} , 
\label{eq90y}
\end{eqnarray}
coupled to the matter equation~(\ref{eq88y}), where $\delta (t{\mp}r/c)$ is the Dirac delta, and $\Gamma = \gamma_{N} m_0$ and $\Gamma = \gamma_{C} q_0$ correspond to the G and E potentials. The relations (\ref{eq89y}) and (\ref{eq90y}) for the E potentials are equivalent to the Maxwell equations under the Lorenz gauge, $\hat{\mathbf{A}}^{\star}_{e} (\mathbf{r},t)=\frac{\varphi^{\star} (\mathbf{r},t) \hat{\mathbf{p}}}{m_{0}c} \rightarrow{\mathbf{A}}_{e} (\mathbf{r},t)=\frac{\varphi (\mathbf{r},t)\mathbf{p}}{m_{0}c}$ and ${\mathbf{p}} = m_{0}{\mathbf{v}}$ (see, Sect. 14.2.). For the static fields, Eqs.~(\ref{eq89y}) and (\ref{eq90y}) imply the potentials that compare well with the static ones of Sects. 4, 10 and 12.2. For instance, the time-dependent solution $\varphi_{0n}(\mathbf{r}_n,t) =-\Gamma \frac{\delta (t{\mp}r_n /c)}{r_n} $ of Eq.~(\ref{eq89y}) for the \textit{n}-th unit-field in stationary state simplifies to the form~(\ref{eq63y}). The relations~(\ref{eq88y}), (\ref{eq89y}) and (\ref{eq90y}) are consistent with Lorentz's transformations. In terms of the energy ${\varepsilon}_0(t)$, the relation~(\ref{eq88y}) reads
\begin{eqnarray} 
{\pm i\hbar} {\frac {{\partial}} {\partial {t}}} |{\psi_{0}}(\mathbf{r},t)\rangle   = {{\hat{\mathbf{H}}}}(x)|{\psi_{0}}(\mathbf{r},t)\rangle .  
\label{eq91y}
\end{eqnarray} 
By analogy with the very simple derivations of formulas in Sects. 11 and 12, one can easily show that the time-dependent perturbation models of transient physical phenomena used in QM and QE, for instance, the models of scattering, emission and absorption of EM waves, are the low-order approximations to Eq.~(\ref{eq91y}) in the parameter $x=x(\mathbf{r},t)$.

\section{14. The classical limit of quantum gravitoelectromagnetism}
\label{sec14y}

\subsection{14.1 The Hamiltonian formalism}
\label{subsec14ay}

In the classical limit, the physical parameters of unit-fields should obey the non-quantum values of classical (spin-less) point particles. Therefore, for the unit-field parameter $f$ determined by the operator ${\hat{f}}$, the classical limit yields
\begin{eqnarray} 
f(\mathbf{r},t) =  {\int^{\infty}_{-\infty}} {\int^{\infty}_{-\infty}} \psi_{0}^{\ast}(\mathbf{r}, \mathbf{r}';t,t') \hat f(\mathbf{\mathbf{r'}},t') \psi_{0}(\mathbf{r}, \mathbf{r}';t,t') d\mathbf{r}'dt' =\nonumber \\ {\int^{\infty}_{-\infty}} {\int^{\infty}_{-\infty}} \rho (\mathbf{r}, \mathbf{r}';t,t') f(\mathbf{\mathbf{r'}},t')d\mathbf{r}'dt', 
\label{eq92y}
\end{eqnarray} 
where $\rho (\mathbf{r}, \mathbf{r}';t,t')= \psi_{0}^{\ast}(\mathbf{r}, \mathbf{r}';t,t')\psi_{0}(\mathbf{r}, \mathbf{r}';t,t') = \delta(\mathbf{r}-\mathbf{r}')\delta(t-t')$, and $\delta(\mathbf{r}-\mathbf{r}')$ and $\delta(t-t')$ denote the Dirac deltas. For the stationary unit-fields, Eq.~(\ref{eq92y}) reads  
\begin{eqnarray} 
f(\mathbf{r}) = \langle {f(\mathbf{r})} \rangle =  {\int^{\infty}_{-\infty}} \psi_{0}^{\ast}(\mathbf{r-r'}) \hat f(\mathbf{\mathbf{r'}}) \psi_{0}(\mathbf{\mathbf{r-r'}}) d\mathbf{r}' = {\int^{\infty}_{-\infty}} \rho(\mathbf{\mathbf{r-r'}}) f(\mathbf{\mathbf{r'}})  d\mathbf{r}', 
\label{eq92yy}
\end{eqnarray} 
where $\rho (\mathbf{r} - \mathbf{r}') = \psi_{0}^{\ast}(\mathbf{r}-\mathbf{r}')\psi_{0}(\mathbf{r}-\mathbf{r}')= \delta(\mathbf{r}-\mathbf{r}')$. Using of Eqs.~(\ref{eq10y}), (\ref{eq14y}) and (\ref{eq92y}) implies   
\begin{eqnarray} 
H_{ge} = {\varepsilon}_{0ge}^{2} = \left( 1 + {\frac {U^{\star}_{ge} } {m_{0} c^2}} \right)^{2} \left( m_{0}^2 c^4 + \mathbf{p}^2 c^2 \right),
\label{eq93y}
\end{eqnarray}
\begin{eqnarray} 
\varepsilon_{0ge} = \left( 1 + {\frac {U^{\star}_{ge} } {m_{0} c^2}} \right) \left( m_{0}^2 c^4 + \mathbf{p}^2 c^2 \right)^{1/2},
\label{eq94y}
\end{eqnarray} 
where
\begin{eqnarray}
U^{\star}_{ge} = U_{g} (\mathbf{r},t)+U_{e} (\mathbf{r},t), \nonumber  \\ U_{g} (\mathbf{r},t)= m_{0}\varphi_{g} (\mathbf{r},t), \nonumber \\ U_{e} (\mathbf{r},t)=q_{0}\varphi_{e} (\mathbf{r},t), 
\label{eq95y}
\end{eqnarray}
\begin{eqnarray} 
{\mathbf{A}}^{\star}_{g}(\mathbf{r},t)=\frac {\varphi_{g}(\mathbf{r},t) {\mathbf{p}}} {m_{0}c}, \nonumber \\{\mathbf{A}}^{\star}_{e}(\mathbf{r},t)=  \frac {\varphi_{e}(\mathbf{r},t) {\mathbf{p}}} {m_{0}c},
\label{eq96y}
\end{eqnarray}
\begin{eqnarray} 
{\mathbf{p}} = \frac{m_{0}{\mathbf{v}}}{{\sqrt{1-\frac{{\mathbf{v}}^{2}}{c^{2}}}}}.
\label{eq97y}
\end{eqnarray}
For the stationary unit-fields, $U^{\star}_{ge} = U_{g} (\mathbf{r})+U_{e} (\mathbf{r})$, $U_{g} (\mathbf{r})= m_{0}\varphi_{g} (\mathbf{r})$, $U_{e} (\mathbf{r})=q_{0}\varphi_{e} (\mathbf{r})$, ${\mathbf{A}}^{\star}_{g}(\mathbf{r})=\frac {\varphi_{g}(\mathbf{r}) {\mathbf{p}}} {m_{0}c}$ and ${\mathbf{A}}^{\star}_{e}(\mathbf{r})=  \frac {\varphi_{e}(\mathbf{r}) {\mathbf{p}}} {m_{0}c}$. In case of the pure-EM point particles ($U^{\star}_{ge} = U_{0e}+U_{e}=U_{e} $) and weak ($|q_{0}{\varphi_{e} } | \ll 2m_{0} c^2$) potentials, Eq.~(\ref{eq94y}) yields the canonical Hamiltonian
\begin{eqnarray} 
H^{L}_{e} = \varepsilon_{0e} = \sqrt{m_{0}^{2} c^4 + c^{2}\left(\mathbf{p} +\frac{q_{0}}{c}{{\mathbf{A}}_{e}} \right)^{2}} + q_{0}{\varphi_{e} }
\label{eq98y}
\end{eqnarray} 
of the Lorentz electromagnetism, which is usually derived~\cite{lan} by using the minimal substitution assumption (${\varepsilon}_{0} \rightarrow {\varepsilon}_{0}- q_{0}\varphi_{e}  $, $\mathbf{p}_{0} \rightarrow \mathbf{p}_{0}+(q_{0}/c)\mathbf{A}_{e}$) in the Einstein mass-energy relation~(\ref{eq2y}).
 
\subsection{14.2 The Lagrangian formalism}
\label{subsec14by}

Using of Eqs.~(\ref{eq5y}) and (\ref{eq92y}) yields the non-quantum Lagrangian 
\begin{eqnarray} 
L^{NQ}_{ge} =   \left(-{m_{0} c^2} - {U_{ge} } \right)^{2}\left(1-\frac{{\mathbf{p}}^{2}}{m_{0}^{2}c^{2}}\right)
\label{eq99y}
\end{eqnarray} 
that contains the squared energies. In terms of the energy (see, Eqs.~(\ref{eq93y}), (\ref{eq94y}) and (\ref{eq99y})), the respective Lagrangian reads
\begin{eqnarray} 
L_{ge} = -  \left({m_{0} c^2} + {U_{ge} } \right)\left(1-\frac{{\mathbf{p}}^{2}}{m_{0}^{2}c^{2}}\right)^{1/2}.
\label{eq100y}
\end{eqnarray} 
The unified non-quantum equation of motion 
\begin{eqnarray} 
\frac{d}{dt} \frac{\partial L_{ge}}{\partial {\mathbf{v}}} = \frac{\partial L_{ge}}{\partial {\mathbf{r}}} 
\label{eq101y}
\end{eqnarray} 
is determined by the GEM force ${\mathbf{F}}_{ge}=\frac{\partial L_{ge}}{\partial {\mathbf{r}}}$. The conditions $U_{ge}=U_{g}$ and $U_{ge}=U_{e}$ yielded the pure-G (${\mathbf{F}}_{g}=\frac{\partial L_{g}}{\partial {\mathbf{r}}}$) and pure-EM (${\mathbf{F}}_{e}=\frac{\partial L_{e}}{\partial {\mathbf{r}}}$) forces in the equations of motion $\frac{d}{dt} \frac{\partial L_{g}}{\partial {\mathbf{v}}} = \frac{\partial L_{g}}{\partial {\mathbf{r}}}$ and $\frac{d}{dt} \frac{\partial L_{e}}{\partial {\mathbf{v}}} = \frac{\partial L_{e}}{\partial {\mathbf{r}}}$, respectively. Thus, the pure-G and pure-EM forces are different aspects of a single GEM-interaction described by $U_{ge}=U_{g}+U_{e}$. The unified (GEM) force ${\mathbf{F}}_{ge}=\frac{\partial L_{ge}}{\partial {\mathbf{r}}}$ predicts the "anti-gravitation force" acting in opposite direction to the G attraction.

In case of the pure-EM particles ($U_{ge} = U_{e}$) with the weak ($|q_{0}{\varphi_{e} } | \ll2m_{0} c^2$) potentials and ${\mathbf{p}}= m_{0}{\mathbf{v}}$, Eq.~(\ref{eq100y}) simplifies to the Lagrangian~\cite{lan} 
\begin{eqnarray} 
L^{L}_{e} = -m_{0} c^2{\sqrt{1-\frac{{\mathbf{v}}^{2}}{c^{2}}}} + \frac{q_{0}}{c}{{\mathbf{A}}_{e}}{{\mathbf{v}}} - q_{0}\varphi_{e}
\label{eq102y}
\end{eqnarray} 
of Lorentz's electromagnetism, which yields the Lorentz EM force ${\mathbf{F}}^{L}_{e}=\frac{\partial L_{Le}}{\partial {\mathbf{r}}}$ and the Maxwell first two equations~\cite{lan}. Whereas, the formal addition of the Lagrangian $L_{EM}= \frac{1}{4\pi}{\int} ({{\mathbf{E}}^{2}_{e}}-{{\mathbf{B}}^{2}_{e}})d^3x$ of the free EM-field with $\mathbf{E}=-\bigtriangledown \varphi_{e} -c^{-1}\dot{\mathbf{A}}_{e}$ and $\mathbf{B}=\bigtriangledown \times {\mathbf{A}}_{e}$ to Eq.~(\ref{eq102y}) and the variations $\delta\varphi_{e}$ and $\delta{{\mathbf{A}}_{e}}$ yielded the Maxwell two second equations~\cite{lan}. In the present model, the quantum relations~(\ref{eq89y}) and (\ref{eq90y}) for the pure-EM unit-fields imply the non-quantum wave-equations
$\left(\bigtriangledown^{2}-\frac{1}{c^{2}} \frac{\partial^{2}}{\partial t^{2}}\right) \varphi_{e} (\mathbf{r},t) =-4\pi \gamma_{C} q_0 \delta (t{\mp}r/c)$ and $\left(\bigtriangledown^{2}-\frac{1}{c^{2}} \frac{\partial^{2}}{\partial t^{2}}\right) {\mathbf{A}}_{e}({\mathbf{r}},t) =-4\pi \gamma_{C} q_0 \frac{\mathbf{p}}{m_{0}c} \delta (t{\mp}r/c)$ coupled to the matter equations~(\ref{eq100y}) and (\ref{eq102y}). For Eqs.~(\ref{eq100y}) and (\ref{eq102y}), the scalar ($\varphi_{e}$) and vector (${\mathbf{A}}_{e}$) potentials are found by using the aforementioned wave-equations, which are equivalent to Maxwell's equations under the Lorenz gauge and ${\mathbf{p}} = \frac{m_{0}{\mathbf{v}}}{{\sqrt{1-\frac{{\mathbf{v}}^{2}}{c^{2}}}}}\approx m_{0}{\mathbf{v}}$. 

For the weak pure-gravitation ($ U_{ge} = U_{g}, |m_{0}{\varphi_{g} } | \ll2m_{0} c^2$), the matter equation~(\ref{eq100y}) coupled to the generalized wave-equations based on the non-quantum versions of Eqs.~(\ref{eq89y}) and (\ref{eq90y}) for the transient pure-G unit-fields yielded the so-called GEM approximation of general relativity, i.e. the G analogues~\cite{maxw,heav,thir,lens} to the Lagrangian (\ref{eq102y}), Lorentz force, Maxwell equations and EM waves. 

The Einstein special and general relativities are described by the metric 
\begin{eqnarray} 
ds^{2} = g_{ik}dx^{i}dx^{k},
\label{eq103y}
\end{eqnarray} 
where $g_{ik}$ is the metric tensor and $x^{i}=(ct,\mathbf{r})$. For the flat spacetime, $g_{ik}=g^{M}_{ik}$. Here, I use the Minkowski tensor $g^{M}_{ik}$ with the signature ($+---$)~\cite{lan}. In case of the pure-G particle ($U_{ge} = U_{g}$), the Lagrangian~(\ref{eq100y}) and the canonical relation $L_{g}=-m_{0}cds/dt$~\cite{lan} yielded
\begin{eqnarray} 
ds^{2} =  \left(1+ \frac{\varphi_{g}({\mathbf{r}})}{c^{2}} \right)^{2}c^{2}dt^{2} - \left(1+ \frac{\varphi_{g}({\mathbf{r}})}{c^{2}} \right)^{2} \frac{d{\mathbf{r}}^{2}}{1-\frac{{\mathbf{v}}^{2}}{c^{2}}}.
\label{eq104y}
\end{eqnarray} 
For the geodesic $\left( m_0c^{2} (1-\frac{{\mathbf{v}}^{2}}{c^{2}})^{-1/2}-m_0c^{2}=m_0(-\varphi_{g})\right)$ movement, Eq.~(\ref{eq104y}) implies the relation 
\begin{eqnarray} 
ds^{2} =   \left(1+ \frac{\varphi_{g}({\mathbf{r}})}{c^{2}} \right)^{2}c^{2}dt^{2} - \left(1- \frac{\varphi_{g}({\mathbf{r}})}{c^{2}} \right)^{2} d{\mathbf{r}}^{2} - \nonumber \\\left(1- \frac{\varphi_{g}({\mathbf{r}})}{c^{2}} \right)^{2}\left(\frac{2\varphi_{g}({\mathbf{r}})}{c^{2}}+\frac{\varphi_{g}({\mathbf{r}})^{2}}{c^{4}} \right)d{\mathbf{r}}^{2},
\label{eq105y}
\end{eqnarray} 
which compares well with the special relativity ($ds^{2} = c^{2}dt^{2} - d{\mathbf{r}}^{2}$) in the limit $\frac{\varphi_{g}({\mathbf{r}})}{c^{2}} \rightarrow 0$. The Schwarzschild metric is the low-order approximation to the 1st and 2nd terms on the right-hand side of Eq.~(\ref{eq105y}) in case of the weak $([1\pm {\varphi_{g}({\mathbf{r}})}/{c^{2}} ]^{2}\approx 1\pm {2\varphi_{g}({\mathbf{r}})}/{c^{2}})$ potentials, whereas the Lense-Thirring gravitomagnetic term~\cite{lens} is the low-order approximation to 3rd term. 
This means that Eqs.~(\ref{eq100y}) and (\ref{eq105y}) coupled to the non-quantum wave-equations based on the relations~(\ref{eq89y}) and (\ref{eq90y}) for the transient pure-G unit-fields yielded the aforementioned gravitational analogues to the Lagrangian (\ref{eq102y}), Lorentz force, Maxwell equations and EM waves. They are in agreement with all known solutions of the special and general relativities proved by experiments. 

The model suggests explanations of the so-called dark matter and dark energy. Indeed, in principle, a GE unit-field may lose its electric dress ($U_{0ge} \rightarrow U_{0g}$) in transient (scattering) process by radiation of the photon energy $\hbar\omega=\langle m_{0e}\rangle c^{2}$ (Sects. 3 and 13). The pure-G particle is the dark matter, because the particle doesn't interact through EM forces. The G-potential is determined by the equation ${\nabla}^{2} \varphi_{0g}=4\pi \gamma_{N}m_0 \delta (\mathbf{r})$ with the general solution $\varphi_{0g} (\mathbf{r}) =-\gamma_{N}{\frac {m_0} {r}} \pm m_0|B_{lm}| (\frac{r}{R_{0}})^{l} Y_l^m (\theta, \varphi)$ (see, Eqs.~(\ref{eq21y}), (\ref{eq22y}), (\ref{eq24y}) and (\ref{eq63y})). If the potential $-m_0|B_{lm}|(\frac{r}{R_{0}})^{l}Y_l^m (\theta, \varphi)$ does exist in the nature, then this term explains the dark energy. Indeed, for $r\gg R_0$, the potential induces the repulsive force ${\mathbf{F}}_{g}=\frac{\partial L_{g}}{\partial {\mathbf{r}}}$ that increases with the increase of $r$. The repulsive force could be induced also by the attractive G-force of dark particles, because they are invisible in EM experiments.  

\section{15. Summary and Conclusions}

The new approach for constructing a unified quantum model of G and EM fields and interactions has been presented. The approach is based on a concept of interfering (cross-correlating) material unit-fields. Starting from the Einstein mass-energy relation and using this concept, the quantum equation for united gravitation and electromagnetism was derived. The unified equation yields all known solutions to the Dirac equation. As an example, the fine and hyperfine structure of the atom spectrum was derived. Furthermore, the unified model suggests explanations of the Pauli exclusion principle, the physical nature of spin and anomalous gyromagnetic factor of an electron and the electron gravitostatic and electrostatic self-energies. Replacement of the field state $|\psi\rangle  =\sum_{i=1}^N|{\psi _{0i}(\mathbf{r}_{i})}\rangle$ by the QM state $|\psi\rangle =\prod ^N _{i=1}| {\psi}_{0i}(\mathbf{r}_{i})\rangle$ in Eqs.~(\ref{eq65y}) and (\ref{eq66y}) explains how QFT transits to QM and vice versa. In case of the transient unit-fields, the quantum equation for united gravitation and electromagnetism is coupled to the generalised wave-equations for G and E scalar and vector potentials. The coupled equations describe, for instance, the scattering, emission and absorption of G and EM waves. For weak potentials, in the classical limit, the model simplifies to the Lorentz-Maxwell electromagnetism and the so-called GEM approximation of Einstein's general relativity. This means that the model is in agreement with all known experimental tests of the special and general relativities. In case of the strong potentials, the model yielded new predictions. For instance, the cross-correlation of G and E strong potentials predicts the "anti-gravity force". If the G and E potentials don't correlate with each other, then the model simplifies to the pure-G and pure-EM fields and interactions. Finally, the model suggested explanations also of "dark matter" and "dark energy". 

The concept of interfering material unit-fields doesn't require any metaphysical interpretation. Nevertheless, one could connect the concept to the most popular philosophic explanation of evolution (\textit{E}) of material and/or immaterial substances~\cite{Bir}. The evolution is described by the formula $E = Q4P$, were $Q4P$ denotes the quest for possibility. The present model yields the formula $E = I^{3}$, which shows that the evolution is caused by the interference-induced interaction ($I^{3}$) mediated by interference (cross-correlation) of the substances. 

The unified quantum equation could be important for analysis of more particular physical problems, such as the cosmology of black holes and the near field (subwavelength) optics and photonics that deal with the strong potentials and relativistic particles. It is well known that EM interactions are unified with weak interactions into electroweak theory. The GEM and nuclear fields and interactions will be united in the next article.

%

{\begin{center} \bf \large ENERGY MEDIATED BY INTERFERENCE OF PARTICLES (Parts I-IV): The Way to Unified Classical and Quantum Fields and Interactions \\ 
\vspace{0.2cm}
{\it{Part IV. New Approach for Constructing a Unified Quantum Model of Gravitational, Electromagnetic, Weak and Strong Fields and Interactions}} \end{center}}

{\begin{center} S. V. Kukhlevsky \end{center}} 
{\begin{center}{\it Department of Physics, Faculty of Natural Sciences,\\ University of Pecs, Ifjusag u. 6, H-7624 Pecs, Hungary} \end{center}}

\begin{quote}
\small Part I of the present study has developed the theoretical background for unified description of the all-known classical and quantum fields in terms of the interference between particles and the respective cross-correlation energy. Part II developed this background for unification of the electromagnetic, weak, strong and gravitational fields and interactions. However, unlike in Part I, the unification has been performed rather by the generalization of the basic (energy-mass) relation of the Einstein special relativity than the traditional Hamiltonians of the classical and quantum field theories. Part III presented a unified quantum model of gravitation and electromagnetism. The model is based on a concept of cross-correlating material unit-fields. Starting from the energy-mass relation of Einstein's special relativity and using this concept, the unified equations for gravitation and electromagnetism are derived. For strong fields and relativistic particles, the unified model yields new predictions compared to the Einstein gravity and Maxwell-Lorentz electromagnetism. The model explains physical nature of the spin and anomalous gyromagnetic factor of an electron, as well as the Pauli exclusion principle. The cross-correlation of gravitational and electric potentials predicts the existence of "antigravitational" forces. Keeping terms of no more than second order in the potential yields the Hamiltonians of canonical quantum mechanics. Part IV extends the model to weak-nuclear and strong-nuclear fields and interactions. Using the energy-mass relation of Einstein's special relativity and the concept of unit-fields carrying the gravitational, electric and strong-nuclear "dressings", the unified quantum equations and their classical limits for gravitational, electromagnetic, weak-nuclear and strong-nuclear fields and interactions are derived. The results of unified model are in agreement with the Standard Model. Furthermore, the cross-correlation of electric and strong-nuclear potentials of the electric and strong-nuclear "dressings" explains the physical nature of weak-nuclear force. While the cross-correlation of gravitational and strong-nuclear potentials predicts the new kind of interactions, namely the gravito-strongnuclear forces.  
\end{quote}

\section{I. Introduction}	
\label{sec1}

Unification of the gravitational, electromagnetic, weak-nuclear and strong-nuclear fields and interactions is a long-standing problem in physics. The generally accepted unified description of electromagnetic, weak, and strong interactions, which does not include gravity, is provided by the Standard Model of Particle Physics (SM) in terms of the gauge and broken-gauge symmetries. SM is considered as the final outcome of quantum field theory combining the basic physical conceptions of quantum mechanics with special relativity~\cite{4Planck,4Einst1,4Bohr,4Einst2,4Noeth,4Brog,4Born,4Schr,4Ferm,4Heis1,4Dir1,4Wign,4Weyl,4Heis2,4Amb,4Paul,4Dir2,4Tomo,4Beth,4Schw,4Feyn,4Dyso}. In its generally accepted form, SM has been developed mainly by S. Glashow, A. Salam, S. Weinberg, F. Englert, R. Brout, and P.W. Higgs (see, for example, the studies \cite{4Yang,4Namb,4Gold,4Glash,4Engl,4Higg,4Wein1,4Sal} and references therein). SM is based on the three independent gauge symmetries, interactions and coupling constants, which are attributed to the electromagnetic, weak-nuclear and strong-nuclear interactions of particles. There are several similar candidate models, for instance, Grand Unified Theories (GUTs), in which at high energy, the three gauge interactions of SM are merged into one single interaction characterized by one larger gauge symmetry and thus one unified coupling constant. Some models, which exhibit similar properties, do not unify all interactions using one simple Lie group as the gauge symmetry, but do so using semi-simple groups or other super-symmetries. GUTs are considered as an intermediate step towards a Theory of Everything (TE) that would unify gravity with electromagnetic, weak-nuclear, and strong-nuclear interactions. String Theory is currently considered as a candidate for TE. The description of candidate theories and generally accepted models can be found in research articles and textbooks published in the last 100 years. 

Although today we know that the electromagnetism is part of a larger gauge group described by the Standard Model (SM), SM does not include gravity. In this regard, one could mention the famous article of A. Einstein and N. Rosen that tried to unify the Maxwell-Lorentz electromagnetism with the Einstein theory of gravity~\cite{4eins1}. The new approach for constructing a unified quantum model of gravitation and electromagnetism has been presented in the study~\cite{4kukh}. This approach has been inspired by the de Broglie idea of material waves (quantum particles). The approach is based on a concept of cross-correlating material unit-fields carrying the gravitational and electric associate components ("dressings"). Connecting the mass-energy relation of Einstein's special relativity with this concept, the unified quantum equations for gravitation and electromagnetism were derived. For strong fields and relativistic particles, the unified model yielded new predictions in comparison to the Einstein gravity and Maxwell-Lorentz electromagnetism. The model explains the spin and anomalous gyromagnetic factor of an electron, as well as the Pauli exclusion principle. The cross-correlation of gravitational and electric potentials of the gravitational and strong-nuclear "dressings" predicts the existence of "antigravitational" interactions, i.e., the gravitoelectric repulsive forces acting in opposite direction to the gravitational attraction. Keeping terms of no more than second order in the potential yields the Hamiltonians of canonical quantum mechanics. For weak fields, the model simplifies to the Maxwell-Lorentz electromagnetism and the so-called gravitoelectromagnetic approximation of the Einstein gravity. 

The aforementioned model~\cite{4kukh} is based on a concept of cross-correlating material unit-fields (quantum particles) that only have gravitational and electric "dressings". The present study extends this model to the unit-fields carrying strong-nuclear "dressing". Using the unit-fields with gravitational, electric, and strong-nuclear "dressings" yields a unified quantum model of gravitational, electromagnetic, weak-nuclear, and strong-nuclear fields and interactions. 

\section{II. Unified Quantum Model of a Single Unit-field (Particle)}
\label{sec2}

According to the concept of cross-correlating material unit-fields~\cite{4kukh}, any field ${\Psi (\mathbf{r},t)}$ of quantum particles consists of the cross-correlating (interfering), material unit-fields ${\Psi _{0n}(\mathbf{r},t)}$ attributed to the particles: 
\begin{eqnarray} 
\Psi (\mathbf{r},t) =\sum_{n=1}^N{\Psi _{0n}(\mathbf{r},t)}, 
\label{eq01}
\end{eqnarray} 
where $N$ is the number of particles. Unification of gravitational, electromagnetic, weak-nuclear, and strong-nuclear fields and interactions is performed by generalization of the Einstein energy-mass relation for the unit-fields carrying the gravitational, electric, and strong-nuclear "dressings".  

\subsection{A. Generalized energy-mass relation}
\label{sec2.1}

Let us connect the energy-mass relation of Einstein's special relativity of a classical point-like particle ($N=1$) with the de Broglie idea of material wave attributed to a particle. The energy-mass relation for a $\textit{point particle}$ located at the spacetime point $(\mathbf{r},t)$ is given by Einstein's special relativity,     
\begin{eqnarray} 
{\varepsilon}_0^{2} = m_{0}^2 c^4 + \mathbf{p}_{0}^2 c^2, 
\label{eq02}
\end{eqnarray} 
where ${\varepsilon}_{0}$, $m_{0}$ and $\mathbf{p}_{0}$ are particle energy, rest mass and momentum. Connecting the concept~(\ref{eq01}) for a {\it{material unit-field}} (quantum particle) with the relation~(\ref{eq02}) yields the quantum energy-mass relation in the Klein-Gordon form, which in the Dirac notations reads  
\begin{eqnarray} 
\langle \Psi_0 | {\hat{\varepsilon}^{2}} | \Psi_0 \rangle = \langle \Psi_0 | m_0^2c^4 + c^2{\hat{\mathbf{p}}^{2}} | \Psi_0 \rangle,
\label{eq03}
\end{eqnarray}
where, ${\Psi _{01}(\mathbf{r},t)}\equiv{\Psi _{0}(\mathbf{r},t)} $, and ${\hat{\mathbf{p}}^{2}}={-\hbar}^{2}{\nabla}^{2}$ and ${\hat{\varepsilon}^{2}}={-\hbar}^{2}{\frac {{\partial}^2} {\partial {t}^2}}$ denote the operators of the squared momentum and energy, respectively. 

In order to describe the gravitational, electromagnetic, weak-nuclear and strong-nuclear fields and interactions, the  unit-field ${\Psi_{0}}$ is assumed to be structured as 
\begin{eqnarray} 
| {\Psi_{0}} (\mathbf{r},t) \rangle  =  \Phi_{0ges}(\mathbf{r})  |{\psi} _{0}(\mathbf{r},t) \rangle ,
\label{eq04}
\end{eqnarray} 
where $| {\Psi_{0}} (\mathbf{r},t) \rangle$ and $|{\psi} _{0}(\mathbf{r},t) \rangle$ denote the states of the unit-field ${\Psi_{0}} (\mathbf{r},t)$ and the unit-field generator ${\psi} _{0}(\mathbf{r},t)$. The intrinsic structure indicates indivisible connection of the generator ${\psi} _{0}$ with the associate-component $ \Phi_{0ges}$. We assume that the unit-field ${\Psi_{0}}$ is mediated by the generator $\psi_0$, because ${\Psi_{0}}=0$ for $\psi_0=0$. The associate-component is interpreted as the gravitoelectic-strongnuclear (\textit{ges}) "dressing" of the unit-field generator, which can be thought as a "carrier of the gravitational (\textit{g}), electric (\textit{e}) and strong-nuclear (\textit{s})  "dressings".

The generalization of Eq.~(\ref{eq03}) for the "gravitoelectic-strongnuclearly dressed" unit-field is impossible without additional assumptions. The derivation of the generalized energy-mass relation for the structured unit-field~(\ref{eq04}) is based on the following assumptions. The stationary value of the squared energy~(\ref{eq03}) is given by the stationary form   
\begin{eqnarray} 
|{\Psi} _{0}(\mathbf{r},t) \rangle  =  exp\left(\pm{{\frac {i} {{\hbar}}}} {\varepsilon}_{0}t \right) \Phi_{0ges}(\mathbf{r}) | \psi_{0}(\mathbf{r}) \rangle 
\label{eq05}
\end{eqnarray} 
of the unit-field state (\ref{eq04}), which indicates that the energy operator could be presented as
\begin{eqnarray} 
{\hat{\varepsilon}^{2}}={-\hbar}^{2}{\frac {1} {\Phi_{0ges}^{*}\Phi_{0ges}}} {\frac {{\partial}^2} {\partial {t}^2}}.
\label{eq06}
\end{eqnarray} 
In such case, Eq.~(\ref{eq03}) yields the stationary squared energy
\begin{eqnarray} 
\langle \psi_0 | {\varepsilon}_0^{2} |\psi_0 \rangle = \nonumber  \\ \langle \psi_0 |\Phi_{0ges}^{*}\Phi_{0ges} m_0^2c^4 + \Phi_{0ges}^{*}c^2{\hat{\mathbf{p}}^{2}} \Phi_{0ges}| \psi_0 \rangle,
\label{eq07}
\end{eqnarray}
The mass-energy relation~(\ref{eq07}) should reduce to the Klein-Gordon form~(\ref{eq03}) if the unit-field ${\Psi_{0}}(\mathbf{r},t)$ looses the \emph{ges} "dressing". That indicates that the associate component $\Phi_{0ges}(\mathbf{r})$ has the form    
\begin{eqnarray} 
\Phi_{0ges}(\mathbf{r}) =  1 + {\frac {U_{0ges}(\mathbf{r}) } {m_{0} c^2}} .  
\label{eq08}
\end{eqnarray}
We assume the additivity of gravitational ($U_{0g}(\mathbf{r})$), electric ($U_{0g}(\mathbf{r})$) and strong-nuclear ($U_{0s}(\mathbf{r})$) components:
\begin{eqnarray}
U_{0ges}(\mathbf{r}) = U_{0g}(\mathbf{r}) + U_{0e}(\mathbf{r}) + U_{0s}(\mathbf{r}).
\label{eq09}
\end{eqnarray}
The Laplace and Helmholtz calibrations  
\begin{eqnarray} 
{\nabla}^{2} U_{0ge}(\mathbf{r})  =  0  
\label{eq010}
\end{eqnarray}
\begin{eqnarray} 
{\nabla}^{2} U_{0s}(\mathbf{r})  =  \vert\Gamma_{s}\vert ^{2} U_{0s}(\mathbf{r})  
\label{eq011}
\end{eqnarray}
are imposed respectively on the terms $U_{0ge}(\mathbf{r})=U_{0g}(\mathbf{r})+U_{0e}(\mathbf{r})$ and $U_{0s}(\mathbf{r})$. The transverse gauge 
\begin{eqnarray} 
{\nabla}U_{0ge}(\mathbf{r}) {\nabla} \psi_0 (\mathbf{r})  =  0.  
\label{eq012}
\end{eqnarray}
is used for the term $U_{0ge}(\mathbf{r})$ and generator $ \psi_{0}(\mathbf{r})$, whereas the gauge 
\begin{eqnarray} 
{\nabla}U_{0s}(\mathbf{r}) {\nabla} \psi_0 (\mathbf{r}) + {\vert\Gamma_{s}\vert ^{2} } U_{0s}(\mathbf{r})\psi_0 (\mathbf{r}) =  0.  
\label{eq013}
\end{eqnarray}
is used for $U_{0s}(\mathbf{r})$ and $ \psi_{0}(\mathbf{r})$. Notice, Eqs.~(\ref{eq08})-(\ref{eq013}) yielded the relation $\hat{\mathbf{p}}^{2}(\Phi_{0ges}\psi_{0})=\Phi_{0ges} \hat{\mathbf{p}}^{2} \psi_{0}$. In addition, we impose the conditions 
\begin{eqnarray} 
 \Phi_{0ges} = \Phi_{0ges}^{*}
\label{eq014}
\end{eqnarray} 
\begin{eqnarray} 
\langle \psi_{0}(\mathbf{r}) | \psi_{0}(\mathbf{r}) \rangle =1 
\label{eq015},
\end{eqnarray} 
which provide the real value of $\Phi_{0ges}$ and the appropriate normalization of $\psi_{0}(\mathbf{r})$. The meaning of "ad hoc" assumptions~(\ref{eq04})-(\ref{eq015}) becomes evident in Sects. II.B.3 and II.C (See Eqs. (\ref{eq046})-(\ref{eq056}) and (\ref{eq069}) that support the assumptions). One can compare these assumptions with the so-called "minimal substitution" assumptions $\varepsilon \rightarrow \varepsilon - q \varphi$ and $\mathbf{p} \rightarrow\mathbf{p} +(q/c)\mathbf{A}$ of Field Theory.
Using the relations~(\ref{eq08})-(\ref{eq015}) in Eq.~(\ref{eq07}) yields the $\textit{generalized}$ energy-mass relation
\begin{eqnarray} 
\langle \psi_0 | {\varepsilon}_0^{2} |\psi_0 \rangle =  \langle \psi_0 | \left( 1 + {\frac {U_{0ges}(\mathbf{r}) } {m_{0} c^2}} \right)^{2} m_0^2c^4 | \psi_0 \rangle  \nonumber  \\ + \langle \psi_0 | \left( 1 + {\frac {U_{0ges}(\mathbf{r}) } {m_{0} c^2}} \right)^{2} c^2{\hat{\mathbf{p}}^{2}} | \psi_0 \rangle . 
\label{eq016}
\end{eqnarray} 
In the differential form, the integral relation (\ref{eq016}) reads  
\begin{eqnarray} 
{\varepsilon}_0^{2} | \psi_0 \rangle = \left( 1 + {\frac {U_{0ges}(\mathbf{r}) } {m_{0} c^2}} \right)^{2} \left( m_0^2c^4 + c^2{\hat{\mathbf{p}}^{2}} \right) | \psi_0 \rangle, 
\label{eq017}
\end{eqnarray} 
where the operator  
\begin{eqnarray} 
{\hat{\mathbf{H}}^{2}} = \left( 1 + {\frac {U_{0ges}(\mathbf{r}) } {m_{0} c^2}} \right)^{2} \left( m_0^2c^4 + c^2{\hat{\mathbf{p}}^{2}} \right) 
\label{eq018}
\end{eqnarray} 
is the squared Hamiltonian operator having the eigen-generator state $|\psi_0 \rangle$ and the eigen-value ${\varepsilon}_0^{2}$. Notice, the relation (\ref{eq016}) simplifies to the Klein-Gordon form (\ref{eq03}) in the case of $U_{0ges}(\mathbf{r}) = 0$.


\subsubsection{1. Identification of the gravitational $(U_{0g}(\mathbf{r}))$, electric $(U_{0e}(\mathbf{r}))$ and strong-nuclear $(U_{0s}(\mathbf{r}))$ components}
\label{subsec2.1.1}

Spherically symmetric solutions of Eq.~(\ref{eq010}) that do not diverge at $r \rightarrow \infty$ are given by the irregular solid harmonics  
\begin{eqnarray} 
U_{0g}(\mathbf{r}) = A^{g}_{lm} r^{-l-1} Y_l^m (\theta, \varphi)
\label{eq019}
\end{eqnarray}
\begin{eqnarray} 
U_{0e}(\mathbf{r}) = A^{e}_{l'm'} r^{-l'-1} Y_{l'}^{m'} (\theta, \varphi) ,
\label{eq020}
\end{eqnarray}
where $Y_l^m (\theta,\varphi)$ is the spherical harmonic of degree $l$ and order $m$, $A^{g}_{lm}$ and $A^{e}_{l'm'}$ are the constants, and $\mathbf{r} = (r, \theta\varphi)$. For a given non-negative integer $l$, there are $2l+1$ independent configurations (\ref{eq019}), one for each integer $m$ with $-l \leq  m \leq  l$. The divergent solutions, which are based on the regular solid harmonics, are described in Sec. VI.A. The gravitational ~(\ref{eq019}) and electric (\ref{eq020}) components with the lowest degree ($l=l'=0$) and the respective order ($m=m'=0$), which satisfy the condition (\ref{eq014}), uncover the physical meaning of the components:
\begin{eqnarray}
U_{0g}(\mathbf{r}) = \pm  \gamma_{g} {\frac {m_0^{2}} {r}}
\label{eq021}
\end{eqnarray}
\begin{eqnarray}
U_{0e}(\mathbf{r}) = \pm \gamma_{e} {\frac {q_0^{2}} {r}} ,
\label{eq022}
\end{eqnarray}
where $\gamma_{g}$ and $\gamma_{e}$ are respectively the gravitational (Newton) and electric (Coulomb) constants, and $q_{0}$ denotes the electrical charge of unit-field. The terms (\ref{eq021}) and (\ref{eq022}) are identified with the gravitational and electrical potential self-energies of the unit-field. The positive mass ($m_{0}\geq0$) excludes the meaningless negative values of the Einstein energy ${\varepsilon}_0= m_{0} c^2 $ of a particle at rest. The sign ($-$) in Eqs.~(\ref{eq021}) and (\ref{eq022}) yields the attractive gravitational and electric self-forces, whereas the repulsive self-forces are explained by the sign ($+$). To provide the attractive forces, we will use the sign ($-$). 

Spherically symmetric solutions of Eq.~(\ref{eq011}) that could be associated with the strong-nuclear potential self-energy  $U_{0s}(\mathbf{r})$ are given by  
\begin{eqnarray} 
U_{0s}(\mathbf{r}) = A^{s}_{lm} j_{l} \left(i \vert\Gamma_{s}\vert r \right) Y_l^m (\theta, \varphi),
\label{eq023}
\end{eqnarray}
where $A^{s}_{lm}$ is the constant, $j_l (i |{\Gamma_{s}} | r)$ is the Bessel function of first kind with the imaginary argument $i |{\Gamma_{s}} | r$. The Bessel functions of first kind with the real argument $|{\Gamma_{s}} | r$ and/or the Bessel functions of second kind, which are also solutions of Eq.~(\ref{eq011}) have been considered Chapter II.
The harmonic degree $l=0$ and the respective order $m=0$ in Eq.~(\ref{eq023}), which satisfy the condition (\ref{eq014}), 
yielded  
\begin{eqnarray} 
U_{0s}(\mathbf{r}) = a \frac {e^{|{\Gamma_{s}} | r} }{r} - b \frac {e^{-|{\Gamma_{s}} | r} }{r} ,
\label{eq024}
\end{eqnarray}
where $a$ and $b$ are the constants. For instance, in the case of $a=0$ and $b=-\gamma_{s}Q_{0}^{2}$, the potential self-energy $U_{0s}(\mathbf{r})$ obeys the Yukawa form
\begin{eqnarray} 
U_{0s}(\mathbf{r}) =  \gamma_{s}Q_{0}^{2} \frac {e^{-|{\Gamma_{s}} | r} }{r},
\label{eq025}
\end{eqnarray}
where $\gamma_{s}Q_{0} \frac {e^{-|{\Gamma_{s}} | r} }{r}$ is the Yukawa potential, $\gamma_{s}$ is the strong-nuclear constant, and $Q_{0}$ is the color charge with the value $Q_{0}\geq 0$. The self-energy~(\ref{eq025}) vanishes at $r \gg |{\Gamma_{s}} |^{-1}$ and becomes infinite ($U_{0s}(\mathbf{r}) \rightarrow \infty$) at $r \rightarrow 0$. The potential self-energy~(\ref{eq024}) with $a=-b=-\gamma_{s}Q_{0}^{2}$ reads
\begin{eqnarray} 
U_{0s}(\mathbf{r}) = -\gamma_{s}Q_{0}^{2} \left(\frac {e^{|{\Gamma_{s}} | r} }{r} + \frac {e^{-|{\Gamma_{s}} | r} }{r}\right) .
\label{eq026}
\end{eqnarray}
The potential self-energy~(\ref{eq026}) becomes infinite ($U_{0s}(\mathbf{r}) \rightarrow -\infty$) at $r \rightarrow 0$ and $r \rightarrow \infty$. 

We will see that the strong-nuclear fields and interactions correspond to Eq.~(\ref{eq026}) rather than Eq.~(\ref{eq025}). Notice, Eqs.~(\ref{eq010}) and (\ref{eq011}) have the same solutions as the respective equations ${\nabla}^{2} U_{0g}(\mathbf{r})=4\pi\gamma_{g} m_0^{2} \delta (\mathbf{r}) $, ${\nabla}^{2} U_{0e}(\mathbf{r})  =  4\pi\gamma_{e} q_0^{2} \delta (\mathbf{r}) $ and ${\nabla}^{2} U_{0s}(\mathbf{r})  -  \vert\Gamma_{s}\vert ^{2} U_{0s}(\mathbf{r}) = 4\pi\gamma_{s} Q^{2} \delta (\mathbf{r})$, where $\delta (\mathbf{r})$ is the Dirac delta. It could be mentioned that the solutions of Eqs.~(\ref{eq010}) and (\ref{eq011}) having the non-spherical symmetries are discussed in Ref.~\cite{4kukh}.

\subsubsection{2. The spin of a unit-field}
\label{subsec2.1.2}

The second term on the right-hand side of Eq.~(\ref{eq016}) is associated with the unit-field kinetic energy. The kinetic energy of a particle (unit-field) at rest vanishes. Therefore, the generator state $|\psi_{0}(\mathbf{r})\rangle \equiv |s(\mathbf{r})\rangle$ of the unit-field at rest satisfies the equation
\begin{eqnarray} 
{\hat{\mathbf{p}}}^{2} | s(\mathbf{r}) \rangle  =  0.  
\label{eq027}
\end{eqnarray}
For the spherically symmetric generator, Eq.~(\ref{eq027}) reads
\begin{eqnarray}
{\hbar}^{2} \nabla^2  s(\mathbf{r})  =    \left[     \frac {{\hbar}^{2}} {r}    \frac {\partial ^2} {\partial r^2} r - \frac {\hat{\mathbf{L}}^2} { r^2} \right]  s(\mathbf{r}) =0,
\label{eq028}
\end{eqnarray}
where ${\hat{\mathbf{L}}}^2$ is the square of the angular momentum operator ${\hat{\mathbf{L}}} = \mathbf{r}\times {\hat{\mathbf{p}}} =i{\hbar}(\mathbf{r}\times \nabla)$ and ${\hat{\mathbf{p}}}$ is the momentum operator. Therefore the generator $\psi_{0}(\mathbf{r})=s(\mathbf{r})$ is called the spin psi-wavefunction. Similarly to Eq.~(\ref{eq010}), the solutions of Eq.~(\ref{eq028}) are given by  in the regular 
\begin{eqnarray}
s(\mathbf{r})  = B_{lm} r^{l} Y_l^m (\theta, \varphi) 
\label{eq029}
\end{eqnarray}
and/or irregular 
\begin{eqnarray}
s(\mathbf{r})  = C_{lm} r^{-l-1} Y_l^m (\theta, \varphi) 
\label{eq030}
\end{eqnarray}
solid harmonics, where $B_{lm}$ and $C_{lm}$ are constants; $Y_l^m (\theta, \varphi  )$ are the joint eigenfunctions of the operator ${\hat{\mathbf{L}}}^2$ and the generator ${\hat{L}}_{z}$ of rotations around the azimuthal axis:
\begin{eqnarray} 
{\hat{\mathbf{L}}}^2  Y_l^m (\theta, \varphi  ) = {\hbar}^{2} l(l+1) Y_l^m (\theta, \varphi )
\label{eq031}
\end{eqnarray}
\begin{eqnarray} 
{\hat{L}_{z}}  s(\mathbf{r}) =  \hbar m  s(\mathbf{r}). 
\label{eq032}
\end{eqnarray}
The different species of particles (unit-fields) correspond to the different values of $l$ and $m$. For an orbital quantum number $l$, there are $2l+1$ independent spin wavefunctions (\ref{eq029}), one for each magnetic quantum number $m$ with $-l \leq  m \leq  l$. Although the squared intrinsic quantum momentum of the unit-field vanishes ($\langle {{\mathbf{p}}}^{2}_{0s}\rangle=\langle s(\mathbf{r})  |{\hat{\mathbf{p}}}^{2} | s(\mathbf{r}) \rangle  =  0$), the intrinsic quantum momentum $\langle {{\mathbf{p}}}_{0s} \rangle = \langle s(\mathbf{r}) |{\hat{\mathbf{p}}} | s(\mathbf{r})\rangle \neq 0$ yields the non-vanishing intrinsic angular momentum (spin) $\langle {{s}_{z}}\rangle =\langle  s(\mathbf{r})|  {\hat{L}_{z}} | s(\mathbf{r}) \rangle =  \hbar m$ for $m\neq 0$. Such behaviour is different from the classic mechanics of a point particle, where $\mathbf{p}= 0$ if $\mathbf{p}^{2}= 0$. 

\subsection{B. Generalized energy-mass relation for a unit-field with spin}
\label{sec2.2}

The intrinsic momentum $\langle {{\mathbf{p}}}_{0s} \rangle = \langle s (\mathbf{r})|{\hat{\mathbf{p}}} | s (\mathbf{r})\rangle$ and the intrinsic generator $|s (\mathbf{r})\rangle$ can be extracted from the total momentum $\langle {{\mathbf{p}}}\rangle = \langle \psi_{0}(\mathbf{r}) | {\hat{\mathbf{p}}} | \psi_{0} (\mathbf{r})\rangle$ and the total generator $| \psi_{0}(\mathbf{r}) \rangle$: 
\begin{eqnarray} 
\langle {{\mathbf{p}}} \rangle = \langle {{\mathbf{p}}}_{0s} \rangle + \langle \tilde {{\mathbf{p}}} \rangle = \langle \tilde \psi_0 (\mathbf{r})|\langle s (\mathbf{r})| {\hat{\mathbf{p}}}| s(\mathbf{r}) \rangle | \tilde \psi_0 (\mathbf{r})\rangle
\label{eq033}
\end{eqnarray} 
\begin{eqnarray} 
| \psi_{0}(\mathbf{r}) \rangle = | s (\mathbf{r})\rangle | \tilde \psi_0 (\mathbf{r})\rangle,
\label{eq034}
\end{eqnarray} 
where $\langle \tilde {{\mathbf{p}}} \rangle = \langle \tilde \psi_0 (\mathbf{r}) | {\hat{\mathbf{p}}} | \tilde \psi_0  (\mathbf{r})\rangle$ is the external momentum corresponding to the external generator $| \tilde \psi_0 (\mathbf{r}) \rangle$, and the intrinsic generator $|s (\mathbf{r})\rangle$ denotes the spin state. 

Using the representation (\ref{eq034}) yields the energy-mass relation (\ref{eq016}) in the new form:      
\begin{eqnarray} 
\langle \tilde \psi_0 |\langle s | {\varepsilon}_0^{2} | s \rangle | \tilde \psi_0 \rangle =  \langle \tilde \psi_0 |\langle s | {\hat{\mathbf{H}}^{2}} | s \rangle | \tilde \psi_0 \rangle.
\label{eq035}
\end{eqnarray} 
Taking into account Eqs.~(\ref{eq015}), (\ref{eq018}) and (\ref{eq027}), the energy-mass relation~(\ref{eq035}) yields    
\begin{eqnarray} 
{\varepsilon}_0^{2} =  \langle \tilde \psi_0 |\langle s | \left( 1 + {\frac {U_{0ges}} {m_{0} c^2}}   \right)^{2}| s \rangle | \tilde \psi_0 \rangle m_{0}^2 c^4 + \nonumber  \\ \langle \tilde \psi_0 |\langle s | \left( 1 + {\frac {U_{0ges}} {m_{0} c^2}} \right)^{2}\left( {2\hat{\mathbf{p}}} | s \rangle {\hat{\mathbf{p}}}| \tilde \psi_0 \rangle + {| s \rangle \hat{\mathbf{p}}^{2}}  | \tilde \psi_0 \rangle \right) c^2.  
\label{eq036}
\end{eqnarray} 
In the case of $U_{0ges}(\mathbf{r}) = 0$, Eqs.~(\ref{eq015}) and (\ref{eq036}) for a unit-field at rest ($| \tilde \psi_0 \rangle = 1$) yielded the Einstein energy-mass relation
\begin{eqnarray} 
{\varepsilon}_0^{2} = \langle s | s \rangle m_{0}^2 c^4 = m_{0}^2 c^4  
\label{eq037}
\end{eqnarray} 
under the condition $\langle s | s \rangle =1$. For the orbital quantum number $l\geq1$, the condition is provided by the normalizations
\begin{eqnarray} 
\langle s | s \rangle = \int_{\theta =0}^\pi   \int_{\varphi  =0}^{2\pi} d\Omega {\int_{0}^{r_{0}\neq\infty}} s^{\ast}(\mathbf{r}) s(\mathbf{r})r^{2} dr = 1.  
\label{eq038}
\end{eqnarray} 
\begin{eqnarray} 
\langle s | s \rangle =\int_{\theta =0}^\pi   \int_{\varphi  =0}^{2\pi} d\Omega {\int_{r_{0}\neq\infty}^{R_{0}=\infty}} s^{\ast}(\mathbf{r}) s(\mathbf{r}) r^{2}dr = 1.  
\label{eq039}
\end{eqnarray} 
for the regular~(\ref{eq029}) and irregular~(\ref{eq030}) solid harmonics, respectively. The relation~(\ref{eq038}) indicates that the intrinsic (spin) wavefunctions corresponding to the regular solid harmonics obey the finite ($r_{0}\neq\infty$) dimension, i.e., the spin wavefunctions are local. While, Eq.~(\ref{eq039}) reveals the non-locality of the spin wavefunctions based on the irregular solid harmonics with infinite ($R_{0}=\infty$) radius. 
If $l=0$, respectively $m=0$, then Eq.~(\ref{eq032}) yields the zero spin ($\langle {{s}_{z}}\rangle =\langle  s(\mathbf{r})|  {\hat{L}_{z}} | s(\mathbf{r}) \rangle =  0$). For irregular harmonics with $l=0$, Eqs.~(\ref{eq038}) and (\ref{eq039}) yielded $\langle s | s \rangle = \infty$. In the case of the unit-fields with zero spins, the condition $\langle s | s \rangle =1$ is provided by the regular harmonics and Eq.~(\ref{eq038}). Thus the unit-fields with non-zero spins are described by the regular (local) or irregular (non-local) harmonics with the orbital quantum number $l\geq1$, whereas the zero-spin unitfields correspond to the regular harmonics with $l=0$. Notice, the  relation~(\ref{eq038}) means that the spin wavefunction vanishes ($s(\mathbf{r})=0$) for $r>r_{0}$. In the case of Eq.~(\ref{eq039}), $s(\mathbf{r})=0$ for $r<r_{0}$. 


\subsubsection{1. The pure gravitational, pure electric, pure strong-nuclear, gravitoelectric, gravito-strongnuclear and electro-strongnuclear terms in the squared energy}
\label{subsec2.2.1}

The additivity~(\ref{eq09}) of gravitational ($U_{0g}(\mathbf{r})$), electric ($U_{0g}(\mathbf{r})$) and strong-nuclear ($U_{0s}(\mathbf{r})$) potential self-energies in Eq.~(\ref{eq035}) yields
\begin{eqnarray} 
{\varepsilon}_0^{2} =  \langle \tilde \psi_0 |\langle s | {\hat{\mathbf{H}}^{2}} | s \rangle | \tilde \psi_0 \rangle, 
\label{eq040}
\end{eqnarray} 
where
\begin{eqnarray} 
{\hat{\mathbf{H}}^{2}} = \left( 1 + {\frac {U_{0ges}} {m_{0} c^2}}   \right)^{2} \left( m_0^2c^4 + c^2{\hat{\mathbf{p}}^{2}} \right) 
\label{eq041}
\end{eqnarray} 
and $ \left( 1 + {\frac {U_{0ges}} {m_{0} c^2}}   \right)^{2} \equiv X^{2}$ is given by
\begin{eqnarray} 
X^{2} =  1 + \left( {\frac {2U_{0g}} {m_{0} c^2}} + {\frac {U^{2}_{0g}} {m_{0}^{2} c^4}} \right) + \left( {\frac {2U_{0e}} {m_{0} c^2}} + {\frac {U^{2}_{0e}} {m_{0}^{2} c^4}} \right) \nonumber  \\ +\left( {\frac {2U_{0s}} {m_{0} c^2}} + {\frac {U^{2}_{0s}} {m_{0}^{2} c^4}} \right) + \left( {\frac {2U_{0g}U_{0e}} {m_{0}^{2} c^4}} \right) \nonumber  \\ + \left( {\frac {2U_{0g}U_{0s}} {m_{0}^{2} c^4}} \right)+ \left( {\frac {2U_{0e}U_{0s}} {m_{0}^{2} c^4}} \right).
\label{eq042}
\end{eqnarray} 
The generalized energy-mass relation~(\ref{eq040}) uncovers physical meanings of the \textit{pure gravitational}, \textit{pure electric}, \textit{pure strong-nuclear}, \textit{gravitoelectric}, \textit{gravito-strongnuclear} and \textit{electro-strongnuclear (electro-weaknuclear)} terms in the squared energy. The second, third and forth terms on the right-hand side of Eq.~(\ref{eq042}), which describe the \textit{pure gravitational}, \textit{pure electric}, \textit{pure strong-nuclear} contributions to the squared energy, are mediated by the self-correlation of gravitational, electric and strong-nuclear potential self-energies. The fifth, sixth and seventh terms correspond to the \textit{gravitoelectric}, \textit{gravito-strongnuclear} and \textit{electro-strongnuclear} contributions induced by the cross-correlation of the respective potential self-energies.

\subsubsection{2. The squared eigen-energy of unit-field determined by the squared Hamiltonian ${\hat{\mathbf{H}}^{2}}$}
\label{subsec2.2.2}

According to the relations (\ref{eq017}) and (\ref{eq034}), the squared eigen-energy ${\varepsilon}_0^{2}$ of unit-field (quantum particle) is determined by the equation
\begin{eqnarray} 
{\varepsilon}_0^{2}| s \rangle | \tilde \psi_0 \rangle = {\hat{\mathbf{H}}^{2}}| s \rangle | \tilde \psi_0 \rangle .  
\label{eq043}
\end{eqnarray} 
Using Eq.~(\ref{eq018}) yields the squared Hamiltonian operator
\begin{eqnarray} 
{\hat{\mathbf{H}}^{2}} =  m^{2}_{0} c^4 (1+x),
\label{eq044}
\end{eqnarray} 
where
\begin{eqnarray} 
x = {\frac {\hat{\mathbf{p}}^{2}} {m_{0}^{2} c^2}}  +  {\frac {2U_{0ges}} {m_{0} c^2}} + {\frac {U_{0ges}^{2} } {m_{0}^{2} c^4}} \nonumber  \\ +   {\frac {2{U_{0ges}}\hat{\mathbf{p}}^{2} } {m_{0}^{3} c^4}} + {\frac {{{U_{0ges}^{2}}\hat{\mathbf{p}}^{2}} } {m_{0}^{4} c^6}}.
\label{eq045}
\end{eqnarray}

\subsubsection{3. The eigen-energy of unit-field determined by the Hamiltonian ${\hat{\mathbf{H}}}$}
\label{subsec2.2.3}

The classical mechanics, quantum mechanics and SM consider the particle energy rather than the squared energy. For the comparison with these models, we find the quantum energy of unit-field using the representation of Eq.~(\ref{eq043}) in the form
\begin{eqnarray} 
{\varepsilon}_0 {\varepsilon}_0 | s \rangle | \tilde \psi_0 \rangle = {\hat{\mathbf{H}}} {\hat{\mathbf{H}}}| s \rangle | \tilde \psi_0 \rangle. 
\label{eq046}
\end{eqnarray} 
Equations~(\ref{eq046}) and (\ref{eq044}) yielded the unit-field self-energy as the eigen-solution of the equation   
\begin{eqnarray} 
{\varepsilon}_{0} | s \rangle | \tilde \psi_0 \rangle = {\hat{\mathbf{H}}}| s \rangle | \tilde \psi_0 \rangle ,  
\label{eq047}
\end{eqnarray} 
where the Hamiltonian operator 
is given by  
\begin{eqnarray} 
{\hat{\mathbf{H}}} = \left( m_{0} c^2 + U_{0ges}\right)\left( 1 + {\frac {\hat{\mathbf{p}}^{2}} {m_{0}^{2} c^2}} \right)^{1/2}.
\label{eq048}
\end{eqnarray} 
Equation~(\ref{eq048}) with the notation $y = {\frac {\hat{\mathbf{p}}^{2}} {m_{0}^{2} c^2}}$ has the form 
\begin{eqnarray} 
{\hat{\mathbf{H}}} = \left( m_{0} c^2 + U_{0ges}\right)(1+y)^{1/2} = \nonumber  \\  \left( m_{0} c^2 + U_{0ges}\right) \left( 1 + {\frac {1} {2}}y -{\frac {1} {8}}y^{2} + {\frac {1} {16}}y^{3} - \ldots \right),
\label{eq049}
\end{eqnarray} 
which yields the non-divergent eigen self-energy 
\begin{eqnarray} 
{\varepsilon}_{0} = \langle \tilde \psi_0 |\langle s |{\hat{\mathbf{H}}}| s \rangle | \tilde \psi_0 \rangle
\label{eq050}
\end{eqnarray} 
only if the dimensionless parameter $\langle y \rangle \equiv \langle \tilde \psi_0 |\langle s |y| s \rangle | \tilde \psi_0 \rangle $ satisfies the condition $\langle y \rangle<1$. For the unit-field at rest ($| \tilde \psi_0 \rangle =1$), Eq.~(\ref{eq050}) yields 
\begin{eqnarray} 
{\varepsilon}_{0} = \left( m_{0} + \langle m_{0g}\rangle + \langle m_{0e}\rangle + \langle m_{0s}\rangle \right)c^2,
\label{eq051}
\end{eqnarray} 
where
\begin{eqnarray} 
\langle m_{0g}\rangle=\langle s | U_{0g}c^{-2}| s \rangle 
\label{eq052}
\end{eqnarray} 
\begin{eqnarray} 
\langle m_{0e}\rangle=\langle s | U_{0e}c^{-2}| s \rangle 
\label{eq053}
\end{eqnarray} 
\begin{eqnarray} 
\langle m_{0s}\rangle=\langle s | U_{0s}c^{-2}| s \rangle 
\label{eq054}
\end{eqnarray} 
are identified with the gravitational, electric and strong-nuclear rest-masses of the unit-field (particle), which satisfy the relations
\begin{eqnarray} 
\langle m_{0g}\rangle \ll \langle m_{0e}\rangle \ll \langle m_{0}\rangle \ll \langle m_{0s}\rangle. 
\label{eq055}
\end{eqnarray} 

The Hamiltonian operator~(\ref{eq048}) can be presented also in the equivalent form 
\begin{eqnarray} 
{\hat{\mathbf{H}}} = m_{0} c^2 (1+x)^{1/2} = \nonumber  \\ m_{0} c^2 \left( 1 + {\frac {1} {2}}x -{\frac {1} {8}}x^{2} + {\frac {1} {16}}x^{3} - \ldots \right), 
\label{eq056}
\end{eqnarray} 
where $x$ is given by Eq.~(\ref{eq045}). The eigen self-energy determined by Eqs.~(\ref{eq050}) and (\ref{eq056}) does not diverge only if the dimensionless parameter $\langle x \rangle \equiv \langle \tilde \psi_0 |\langle s |x| s \rangle | \tilde \psi_0 \rangle <1$.  If $\langle y \rangle <1$, then the conditions $\langle x \rangle <1$ is satisfied by the \textit{pure gravitational} ($U_{0ges}=U_{0g}$), \textit{pure electric} ($U_{0ges}=U_{0e}$) and \textit{gravitoelectric} ($U_{0ge}=U_{0g}+U_{0e}$) unit-fields because of the relation~(\ref{eq055}). 

In the case of $\langle x \rangle<1$ or $\langle y \rangle<1$, the exact values of eigen self-energies are provided by the infinite number of terms of the Taylor expansions in Eqs.~(\ref{eq049}) and (\ref{eq056}). Surprisingly, the divergence problem does not exist in the case of Eq.~(\ref{eq043}), which yields the exact eigen-values of squared self-energies by using the six terms in Eq.~(\ref{eq044}).

\subsubsection{4. The gravitational, electric and strong-nuclear rest-masses}
\label{subsec2.2.4}

The rest-masses of the gravitational $(U_{0g}(\mathbf{r}))$, electric $(U_{0e}(\mathbf{r}))$ and strong-nuclear $(U_{0s}(\mathbf{r}))$ fields of the unit-field (particle) depend on the internal (spin) wavefunction $s(\mathbf{r})$. The spherical symmetry of the non-local spin wave-functions based on the irregular solid harmonics (\ref{eq030}) could be affected by a border placed at infinite distance from the particle. The rest-masses determined by such exotic (non-local) spin wavefunctions are calculated in the Appendix. Here, we consider the unit-fields with local spin wavefunctions based on the regular solid harmonics (\ref{eq029}). For the unit-fields with the spins ($l\geq0$), Eqs.~(\ref{eq052}), (\ref{eq038}), (\ref{eq029}) and (\ref{eq021}) and Eqs.~(\ref{eq053}), (\ref{eq038}), (\ref{eq029}) and (\ref{eq022}) yielded the gravitational and electric rest-masses  
\begin{eqnarray} 
\langle m_{0g}\rangle=-{\frac {\gamma_{g} m_0^{2}(2l+3)} {c^{2}r_{0g}(2l+2)}} 
\label{eq057}
\end{eqnarray} 
\begin{eqnarray} 
\langle m_{0e}\rangle= - {\frac {\gamma_{e} q_0^{2}(2l+3)} {c^{2}r_{0e}(2l+2)}}, 
\label{eq058}
\end{eqnarray} 
whereas the strong-nuclear rest-mass is given by Eqs.~(\ref{eq054}), (\ref{eq038}), (\ref{eq029}) and (\ref{eq026}) as  
\begin{eqnarray} 
\langle m_{0s}\rangle=-{\frac {\gamma_{s} Q_0^{2}(2l+3)} {c^{2}r_{0s}^{2l+2}|{\Gamma_{s}} |}}  \left({e^{|{\Gamma_{s}} | r_{0}} } +  {e^{-|{\Gamma_{s}} | r_{0}} }\right) .
\label{eq059}
\end{eqnarray} 
Notice, $r_{0g}$, $r_{0e}$ and $r_{0s}$ are the gravitational, electric and strong-nuclear dimensions of the spin wavefunction $s(\mathbf{r})$.

\subsection{C. The Hamiltonian ${\hat{\mathbf{H}}}$ and squared Hamiltonian ${\hat{\mathbf{H}}^{2}}$ in terms of the scalar ($\varphi_{0}$) and vector ($\hat{\mathbf{A}}_{0}$) potentials}
\label{sec2.3}

The squared Hamiltonian (\ref{eq044}) and Hamiltonians (\ref{eq049}) and (\ref{eq056}) are determined by the parameter $x$ (see, Eq.~(\ref{eq045})) that contains the gravitational (\ref{eq021}), electric (\ref{eq022}) and strong-nuclear (\ref{eq026}) potential self-energies of the unit-field. These energies can be presented as
\begin{eqnarray}
U_{0g}(\mathbf{r}) = m_0 \varphi_{0g} (\mathbf{r})
\label{eq060}
\end{eqnarray}
\begin{eqnarray}
U_{0e}(\mathbf{r}) = q_0\varphi_{0e}(\mathbf{r})
\label{eq061}
\end{eqnarray}
\begin{eqnarray}
U_{0s}(\mathbf{r}) = Q_0\varphi_{0s}(\mathbf{r})
\label{eq062}
\end{eqnarray}
by introducing the gravitational, electric and strong-nuclear scalar potentials
\begin{eqnarray}
\varphi_{0g} (\mathbf{r}) = - \gamma_{g}{\frac {m_0} {r}}
\label{eq063}
\end{eqnarray}
\begin{eqnarray}
\varphi_{0e} (\mathbf{r}) = -\gamma_{e}{\frac {q_0} {r}}
\label{eq064}
\end{eqnarray}
\begin{eqnarray}
\varphi_{0s} (\mathbf{r}) = -\gamma_{s}Q_{0}\left(\frac {e^{|{\Gamma_{s}} | r} }{r} + \frac {e^{-|{\Gamma_{s}} | r} }{r}\right).
\label{eq065}
\end{eqnarray}
Introduction of the scalar potentials and the respective gravitational, electric (magnetic) and strong-nuclear vector potentials 
\begin{eqnarray} 
\hat{\mathbf{A}}_{0g} = \frac {\varphi_{0g} \hat{\mathbf{p}}} {m_{0}c}
\label{eq066}
\end{eqnarray} 
\begin{eqnarray} 
\hat{\mathbf{A}}_{0e} = \frac {\varphi_{0e} \hat{\mathbf{p}}} {m_{0}c},
\label{eq067}
\end{eqnarray}
\begin{eqnarray} 
\hat{\mathbf{A}}_{0s} = \frac {\varphi_{0s} \hat{\mathbf{p}}} {m_{0}c},
\label{eq068}
\end{eqnarray}
yielded the parameter $x$ in the form
\begin{eqnarray} 
x = {\frac {\hat{\mathbf{p}}^{2}} {m_{0}^{2} c^2}}  +  {\frac {2(m_0 \varphi_{0g} + q_0\varphi_{0e}+ Q_0\varphi_{0s})} {m_{0} c^2}} \nonumber  \\ + {\frac {(m_0 \varphi_{0g} + q_0\varphi_{0e}+ Q_0\varphi_{0s})^{2}} {m_{0}^{2} c^4}} +  \nonumber  \\ {\frac {2(m_0 \hat{\mathbf{A}}_{0g} + q_0\hat{\mathbf{A}}_{0e}+ Q_0\hat{\mathbf{A}}_{0s})\hat{\mathbf{p}} } {m_{0}^{2} c^3}}\nonumber  \\ + {\frac {(m_0 \hat{\mathbf{A}}_{0g} + q_0\hat{\mathbf{A}}_{0e}+ Q_0\hat{\mathbf{A}}_{0s})^{2} } {m_{0}^{2} c^4}}.
\label{eq069}
\end{eqnarray}

\subsubsection{1. The pure gravitational, pure electric, pure strong-nuclear, gravitoelectric, gravito-strongnuclear, electro-strongnuclear and gravito-electric-strongnuclear unit-fields}
\label{subsec2.3.1}

The species of unit-fields (particles) are determined by the parameter $x$ in the squared Hamiltonian (\ref{eq044}) and the Hamiltonians (\ref{eq049}) and (\ref{eq056}). 

For the \textit{pure gravitational}, \textit{pure electric (electromagnetic)} and \textit{pure strong-nuclear} unit-fields, the parameter $x$ is given by Eq.~(\ref{eq072}) in the forms
\begin{eqnarray} 
x = {\frac {\hat{\mathbf{p}}^{2}} {m_{0}^{2} c^2}}  +  {2 \frac {\varphi_{0g} } { c^2}} + {\frac {\varphi_{0g}^{2}} {c^4}} +  {2\frac {\hat{\mathbf{A}}_{0g} \hat{\mathbf{p}} } {m_{0} c^3}} + {\frac {\hat{\mathbf{A}}_{0g}^{2}} { c^4}} 
\label{eq070}
\end{eqnarray} 
\begin{eqnarray} 
x = {\frac {\hat{\mathbf{p}}^{2}} {m_{0}^{2} c^2}}  +  {2 \frac {q_{0}\varphi_{0e} } {m_{0} c^2}} + {\frac {q_{0}^{2}\varphi_{0e}^{2}} {m_{0}^{2}c^4}} +  {2\frac {q_{0}\hat{\mathbf{A}}_{0e} \hat{\mathbf{p}} } {m_{0}^{2} c^3}} + {\frac {q_{0}^{2}\hat{\mathbf{A}}_{0e}^{2}} {m_{0}^{2} c^4}}
\label{eq071}
\end{eqnarray} 
\begin{eqnarray} 
x = {\frac {\hat{\mathbf{p}}^{2}} {m_{0}^{2} c^2}}  +  {2 \frac {Q_{0}\varphi_{0s} } {m_{0} c^2}} + {\frac {Q_{0}^{2}\varphi_{0s}^{2}} {m_{0}^{2}c^4}} \nonumber  \\+  {2\frac {Q_{0}\hat{\mathbf{A}}_{0s} \hat{\mathbf{p}} } {m_{0}^{2} c^3}} + {\frac {Q_{0}^{2}\hat{\mathbf{A}}_{0s}^{2}} {m_{0}^{2} c^4}},
\label{eq072}
\end{eqnarray} 
respectively. Notice, in the case of $-\gamma_{g}m_{0}= \gamma_{e}q_{0} = 1$, the pure gravitation (Eq.~(\ref{eq070})) and pure electromagnetism (Eq.~((\ref{eq071})) of the unit-field become undistinguishable from each other. Imposing the commutation relation $[\hat{\mathbf{p}},\hat{\mathbf{A}_{0}}]=0$, Eqs.~(\ref{eq070})-(\ref{eq072}) can be represented as 
\begin{eqnarray} 
x = {\frac {1} {m_{0}^{2} c^2}}\left({\hat{\mathbf{p}}}+\frac {m_{0}} {c}\hat{\mathbf{A}}_{0g} \right)^{2}  +  {2 \frac {\varphi_{0g} } { c^2}} + {\frac {\varphi_{0g}^{2}} {c^4}} 
\label{eq073}
\end{eqnarray} 
\begin{eqnarray} 
x = {\frac {1} {m_{0}^{2} c^2}}\left({\hat{\mathbf{p}}}+\frac {q_{0}} {c}\hat{\mathbf{A}}_{0e} \right)^{2}  +  {2 \frac {q_{0}\varphi_{0e} } {m_{0} c^2}} + {\frac {q_{0}^{2}\varphi_{0e}^{2}} {m_{0}^{2}c^4}} 
\label{eq074}
\end{eqnarray} 
\begin{eqnarray} 
x = {\frac {1} {m_{0}^{2} c^2}}\left({\hat{\mathbf{p}}}+\frac {Q_{0}} {c}\hat{\mathbf{A}}_{0s} \right)^{2}  +  {2 \frac {Q_{0}\varphi_{0s} } {m_{0} c^2}} + {\frac {Q_{0}^{2}\varphi_{0s}^{2}} {m_{0}^{2}c^4}}. 
\label{eq075}
\end{eqnarray} 

The \textit{gravitoelectric} (\textit{gravitoelectromagnetic}), \textit{gravito-strongnuclear} and \textit{electro-strongnuclear} unit-fields are determined respectively by 
\begin{eqnarray} 
x = {\frac {\hat{\mathbf{p}}^{2}} {m_{0}^{2} c^2}}  +  {\frac {2(m_0 \varphi_{0g} + q_0\varphi_{0e})} {m_{0} c^2}}\nonumber  \\ + {\frac {(m_0 \varphi_{0g} + q_0\varphi_{0e})^{2}} {m_{0}^{2} c^4}} \nonumber  \\ +{\frac {2(m_0 \hat{\mathbf{A}}_{0g} + q_0\hat{\mathbf{A}}_{0e})\hat{\mathbf{p}} } {m_{0}^{2} c^3}}\nonumber  \\ + {\frac {(m_0 \hat{\mathbf{A}}_{0g} + q_0\hat{\mathbf{A}}_{0e})^{2} } {m_{0}^{2} c^4}}
\label{eq076}
\end{eqnarray} 
\begin{eqnarray} 
x = {\frac {\hat{\mathbf{p}}^{2}} {m_{0}^{2} c^2}}  +  {\frac {2(m_0 \varphi_{0g} + Q_0\varphi_{0s})} {m_{0} c^2}} \nonumber  \\ + {\frac {(m_0 \varphi_{0g} + Q_0\varphi_{0s})^{2}} {m_{0}^{2} c^4}} +  \nonumber  \\ {\frac {2(m_0 \hat{\mathbf{A}}_{0g} + Q_0\hat{\mathbf{A}}_{0s})\hat{\mathbf{p}} } {m_{0}^{2} c^3}}\nonumber  \\ + {\frac {(m_0 \hat{\mathbf{A}}_{0g} + Q_0\hat{\mathbf{A}}_{0s})^{2} } {m_{0}^{2} c^4}}
\label{eq077}
\end{eqnarray} 
\begin{eqnarray} 
x = {\frac {\hat{\mathbf{p}}^{2}} {m_{0}^{2} c^2}}  +  {\frac {2(q_0 \varphi_{0e} + Q_0\varphi_{0s})} {m_{0} c^2}} \nonumber  \\ + {\frac {(q_0 \varphi_{0e} + Q_0\varphi_{0s})^{2}} {m_{0}^{2} c^4}} +  \nonumber  \\ {\frac {2(q_0 \hat{\mathbf{A}}_{0e} + Q_0\hat{\mathbf{A}}_{0s})\hat{\mathbf{p}} } {m_{0}^{2} c^3}}\nonumber  \\ + {\frac {(q_0 \hat{\mathbf{A}}_{0e} + Q_0\hat{\mathbf{A}}_{0s})^{2} } {m_{0}^{2} c^4}}
\label{eq078}
\end{eqnarray} 
The \textit{pure gravitational} (Eqs.~(\ref{eq070}) and (\ref{eq073})), \textit{pure electromagnetic} (Eqs.~(\ref{eq071} and (\ref{eq074})) and \textit{gravitoelectromagnetic} (Eq.~(\ref{eq076})) unit-fields have been discussed in Ref.~\cite{4kukh}. The \textit{gravitoelectromagnetic} properties of unit-fields are determined by Eq.~(\ref{eq076}), where the intrinsic cross-correlation (interference) of gravitational and electric potentials is given by the operators $\varphi_{0g} \varphi_{0e}$ and $\hat{\mathbf{A}}_{0g} \hat{\mathbf{A}}_{0e}$. The cross-correlation of gravitational and strong-nuclear potentials inside the \textit{pure gravito-strongnuclear} unit-fields is described by the operators $\varphi_{0g} \varphi_{0s}$ and $\hat{\mathbf{A}}_{0g} \hat{\mathbf{A}}_{0s}$ in Eq.~(\ref{eq077}), whereas the operators $\varphi_{0e} \varphi_{0s}$ and $\hat{\mathbf{A}}_{0e} \hat{\mathbf{A}}_{0s}$ in equation (\ref{eq078}) describe the cross-correlation of electric and strong-nuclear intrinsic potentials of the \textit{pure electro-strongnuclear} unit-fields.   

The physical properties of \textit{gravitoelectromagnetic-strongnuclear} unit-fields are determined by the parameter $x$ in the forms~(\ref{eq069})-(\ref{eq078}). If a unit-field is called \textit{string}, then the above-presented description of the unit-field could be considered as a model of the gravitoelectromagnetic-strongnuclear \textit{string}.

\section{III. Unified Quantum Model of a Unit-Field (Particle) in the External Gravitoelectromagnetic-strongnuclear Field}
\label{sec3}

In order to extend the unified quantum model of a single gravitoelectromagnetic-strongnuclear unit-field to the unit-field placed into \textit{external} gravitoelectromagnetic-strongnuclear field, we use 
the principle of additivity for potential energies and potentials. The potential \textit{self-energy} $U_{0ges}(\mathbf{r})$ of a single unit-field is replaced in Eqs.~(\ref{eq044}), (\ref{eq049}) and (\ref{eq056}) by the \textit{total} potential energy $U^{\star}_{ges}(\mathbf{r})$ of the unit-field:
\begin{eqnarray}
U^{\star}_{ges}(\mathbf{r}) = U^{\star}_{g}(\mathbf{r}) + U^{\star}_{e}(\mathbf{r}) +U^{\star}_{s}(\mathbf{r}),
\label{eq079}
\end{eqnarray}
with 
\begin{eqnarray}
U^{\star}_{g}(\mathbf{r})= U_{0g}(\mathbf{r}) + U_{g}(\mathbf{r}) = m_0 \varphi^{\star}_{g} (\mathbf{r})
\label{eq080}
\end{eqnarray}
\begin{eqnarray}
U^{\star}_{e}(\mathbf{r})= U_{0e}(\mathbf{r}) + U_{e}(\mathbf{r}) = q_0 \varphi^{\star}_{e} (\mathbf{r})
\label{eq081}
\end{eqnarray}
\begin{eqnarray}
U^{\star}_{s}(\mathbf{r})= U_{0s}(\mathbf{r}) + U_{s}(\mathbf{r}) = Q_0 \varphi^{\star}_{s} (\mathbf{r}).
\label{eq082}
\end{eqnarray}
The components $U_{g}(\mathbf{r}) = m_0 \varphi_{g} (\mathbf{r})$, $U_{e}(\mathbf{r}) = q_0 \varphi_{e} (\mathbf{r})$ and $U_{s}(\mathbf{r}) = Q_0 \varphi_{s} (\mathbf{r})$ denote respectively the gravitational, electric and strong potential energies attributed to the gravitational ($\varphi_{g} (\mathbf{r})$), electric ($\varphi_{e} (\mathbf{r})$) and strong-nuclear ($\varphi_{s} (\mathbf{r})$) \textit{external} potentials. The \textit{total} scalar and \textit{total}vector potentials are given by
\begin{eqnarray}
\varphi^{\star}_{g} (\mathbf{r}) = \varphi_{0g} (\mathbf{r}) + \varphi_{g} (\mathbf{r})
\label{eq083}
\end{eqnarray}
\begin{eqnarray}
\varphi^{\star}_{e} (\mathbf{r}) = \varphi_{0e} (\mathbf{r}) + \varphi_{e} (\mathbf{r})
\label{eq084}
\end{eqnarray}
\begin{eqnarray}
\varphi^{\star}_{s} (\mathbf{r}) = \varphi_{0s} (\mathbf{r}) + \varphi_{s} (\mathbf{r})
\label{eq085}
\end{eqnarray}
\begin{eqnarray} 
\hat{\mathbf{A}}^{{\star}}_{g} = \frac {\varphi^{\star}_{g} \hat{\mathbf{p}}} {m_{0}c}
\label{eq086}
\end{eqnarray} 
\begin{eqnarray} 
\hat{\mathbf{A}}^{\star}_{e} = \frac {\varphi^{\star}_{e} \hat{\mathbf{p}}} {m_{0}c}
\label{eq087}
\end{eqnarray}
\begin{eqnarray} 
\hat{\mathbf{A}}^{\star}_{s} = \frac {\varphi^{\ast}_{s} \hat{\mathbf{p}}} {m_{0}c}.
\label{eq088}
\end{eqnarray}
Notice, the parameters $\varphi_{0g} (\mathbf{r})$, $\varphi_{0e} (\mathbf{r})$ and $\varphi_{0s} (\mathbf{r})$ are \textit{internal} potentials. Thus the \textit{gravitoelectromagnetic-strongnuclear} unit-field placed into the external gravitoelectromagnetic-strongnuclear field $U_{ges}(\mathbf{r})= U_{g}(\mathbf{r}) + U_{e}(\mathbf{r})+ U_{s}(\mathbf{r})$ satisfies Eqs. (\ref{eq043}) and (\ref{eq047}), where the squared Hamiltonian (\ref{eq044}) and the Hamiltonians (\ref{eq049}) and (\ref{eq056}) are described by the parameter $x$ in the form
\begin{eqnarray} 
x = {\frac {\hat{\mathbf{p}}^{2}} {m_{0}^{2} c^2}}  +  {\frac {2(m_0 \varphi^{\star}_{g} + q_0\varphi^{\star}_{e}+ Q_0\varphi^{\star}_{s})} {m_{0} c^2}} \nonumber  \\ + {\frac {(m_0 \varphi^{\star}_{g} + q_0\varphi^{\star}_{e}+ Q_0\varphi^{\star}_{s})^{2}} {m_{0}^{2} c^4}} +  \nonumber  \\ {\frac {2(m_0 \hat{\mathbf{A}}^{\star}_{g} + q_0\hat{\mathbf{A}}^{\star}_{e}+ Q_0\hat{\mathbf{A}}^{\star}_{s})\hat{\mathbf{p}} } {m_{0}^{2} c^3}}\nonumber  \\ + {\frac {(m_0 \hat{\mathbf{A}}^{\star}_{g} + q_0\hat{\mathbf{A}}^{\star}_{e}+ Q_0\hat{\mathbf{A}}^{\star}_{s})^{2} } {m_{0}^{2} c^4}}.
\label{eq089}
\end{eqnarray} 
Thus the \textit{gravitoelectromagnetic-strongnuclear interaction} of the unit-field with the external field is determined by the parameter~(\ref{eq089}). The \textit{pure gravitational}, \textit{pure electric (electromagnetic)} and \textit{pure strong-nuclear} unit-fields and interactions correspond to the parameter $x$, which is given by Eq.~(\ref{eq089}) with $\varphi^{\star}_{e} =\varphi^{\star}_{s}=0$, $\varphi^{\star}_{g} =\varphi^{\star}_{s}=0$ and $\varphi^{\star}_{g} =\varphi^{\star}_{e}=0$, respectively. The \textit{gravitoelectric} (\textit{gravitoelectromagnetic}), \textit{gravito-strongnuclear} and \textit{electro-strongnuclear} unit-fields and interactions are determined by the parameter~(\ref{eq089}) with $\varphi^{\star}_{s}=0$, $\varphi^{\star}_{e}=0$ and $\varphi^{\star}_{g}=0$, respectively. 

The relations (\ref{eq043}), (\ref{eq047}), (\ref{eq044}), (\ref{eq049}), (\ref{eq056}) and (\ref{eq089}) predict the \textit{new physical phenomena} attributed to \textit{interaction} (cross-correlation) of the unit-field with the external gravitoelectromagnetic-strongnuclear field. The operators $\varphi_{0g} \varphi_{g}$ and $\hat{\mathbf{A}}_{0g} \hat{\mathbf{A}}_{g}$ uncover the interaction mediated by means of the cross-correlation of intrinsic end external gravitational potentials. The terms $\varphi_{0e} \varphi_{e}$ and $\hat{\mathbf{A}}_{0e} \hat{\mathbf{A}}_{e}$ reveal the interaction induced by the intrinsic and external electromagnetic potentials, whereas the operators $\varphi_{0s} \varphi_{s}$ and $\hat{\mathbf{A}}_{0s} \hat{\mathbf{A}}_{s}$ predict the interaction by the cross-correlation of intrinsic and external strong-nuclear potentials. The \textit{gravito-electric} \textit{interaction} is predicted by the operators $\varphi^{\star}_{g} \varphi^{\star}_{e}$ and $\hat{\mathbf{A}}^{\star}_{g}\hat{\mathbf{A}}^{\star}_{e}$. The \textit{gravito-strongnuclear} \textit{interaction} is revealed by the operators $\varphi^{\star}_{g} \varphi^{\star}_{s}$ and $\hat{\mathbf{A}}^{\star}_{g}\hat{\mathbf{A}}^{\star}_{s}$, whereas the \textit{electro-strongnuclear} (\textit{electro-weak}) \textit{interaction} correspond to the operators $\varphi^{\star}_{e} \varphi^{\star}_{s}$ and $\hat{\mathbf{A}}^{\star}_{e}\hat{\mathbf{A}}^{\star}_{s}$. The second order term ${{(m_{0} \varphi^{\star}_{g} + q_{0n}\varphi^{\star}_{e})^{2}} {m_{0}^{-2} c^{-4}}}$, which does not exist in the general relativity and electromagnetism could play a key role in the interactions mediated by the \textit{strong} gravitational and electromagnetic potentials.

It is surprising that the physical picture of the interactions based on the approximation~(\ref{eq056}) of ${\hat{\mathbf{H}}}(x)$ under the condition $\langle x \rangle = \langle \tilde \psi_0 |\langle s | x | s \rangle | \tilde \psi_0 \rangle<1$ depends on the number of terms in the Taylor expansion. The terms of more than second order in the potentials yielded the \textit{combined interactions}, which are the \textit{high-order combinations} of \textit{gravitational}, \textit{electromagnetic}, \textit{strong-nuclear}, \textit{gravito-electromagnetic}, \textit{gravito-strongnuclear} and \textit{electro-strongnuclear} interactions. The interaction models based on Eqs.~(\ref{eq056}) and (\ref{eq049}) are perturbative theories, i.e., they are described by expansions in powers of $\langle x \rangle$ or $\langle y\rangle = \langle \tilde \psi_0 |\langle s | y | s \rangle | \tilde \psi_0 \rangle $. In the case of the relativistic unit-fields ($\langle {\mathbf{p}} \rangle \approx {\mathbf{p}}$), the condition $\langle y \rangle <1$ is satisfied by the \textit{gravitoelectro-strongnuclear} unit-fields if $v<c/\sqrt{2}$, where ${\mathbf{p}} = {m_{0}{\mathbf{v}}} /{{\sqrt{1-\frac{{\mathbf{v}}^{2}}{c^{2}}}}}$ is the particle momentum. If the dimensionless parameters $\langle y \rangle$ or $\langle x \rangle$ are of order one or larger, then the non-perturbative theory based on Eq.~(\ref{eq044}) have to be used to describe the interactions. An example of the latter is the strongnuclear and electro-strongnuclear interactions, where the parameter $\langle x \rangle \gg 1$ due to Eq.~(\ref{eq055}). 
 

\section{IV. Unified Quantum Model of Multi-Unitfield System}
\label{sec4}

We now extend the single-unitfield ($N=1$) model developed in the previous sections to the multi-unitfield ($N\geq2$) system. The multi-unitfield system (\ref{eq01}), which consists of the cross-correlating gravito-electromagnetic-strongnuclear unit-fields ${\Psi _{0n}(\mathbf{r},t)}$, is described by the field  
\begin{eqnarray} 
\Psi (\mathbf{r},t) =\sum_{n=1}^{N\geq 2}{\Psi _{0n}(\mathbf{r},t)}. 
\label{eq090}
\end{eqnarray} 

The \textit{n-th} unit-field, which is surrounded by the ($N-1$) \textit{external} unit-fields, does satisfy Eqs. (\ref{eq043}) and (\ref{eq047}), where $|s \rangle | \tilde \psi_{0} \rangle$, ${\hat{\mathbf{H}}^{2}}(x)$, ${\hat{\mathbf{H}}}(x)$ and ${\hat{\mathbf{H}}}(y)$ are replaced respectively by $| s_{n} \rangle | \tilde \psi_{0n} \rangle$, ${\hat{\mathbf{H}}_{n}^{2}}(x_{n})$, ${\hat{\mathbf{H}}_{n}}(x_{n})$ and ${\hat{\mathbf{H}}_{n}}(y_{n})$ with $y_{n} = {\frac {\hat{\mathbf{p}}_{n}^{2}} {m_{0n}^{2} c^2}}$ and 
\begin{eqnarray} 
x_{n} = {\frac {\hat{\mathbf{p}}_{n}^{2}} {m_{0n}^{2} c^2}}  +  {\frac {2(m_{0n} \varphi^{\star}_{gn} + q_{0n}\varphi^{\star}_{en}+ Q_{0n}\varphi^{\star}_{sn})} {m_{0n} c^2}} \nonumber  \\ + {\frac {(m_{0n} \varphi^{\star}_{gn} + q_{0n}\varphi^{\star}_{en}+ Q_{0n}\varphi^{\star}_{sn})^{2}} {m_{0n}^{2} c^4}} +  \nonumber  \\ {\frac {2(m_{0n} \hat{\mathbf{A}}^{\star}_{gn} + q_{0n}\hat{\mathbf{A}}^{\star}_{en}+ Q_{0n}\hat{\mathbf{A}}^{\star}_{sn})\hat{\mathbf{p}} } {m_{0n}^{2} c^3}}\nonumber  \\ + {\frac {(m_{0n} \hat{\mathbf{A}}^{\star}_{gn} + q_{0n}\hat{\mathbf{A}}^{\star}_{en}+ Q_{0n}\hat{\mathbf{A}}^{\star}_{sn})^{2} } {m_{0n}^{2} c^4}}.
\label{eq091}
\end{eqnarray} 
(for comparison, see Eq.~(\ref{eq089})).
The respective \textit{total} scalar and \textit{total} vector potentials are given by
\begin{eqnarray}
\varphi^{\star}_{gn} (\mathbf{r}) = \varphi_{0gn} (\mathbf{r}) + \sum_{m\neq n}^{N-1}\varphi_{0gmn} (\mathbf{r})
\label{eq092}
\end{eqnarray}
\begin{eqnarray}
\varphi^{\star}_{en} (\mathbf{r}) = \varphi_{0en} (\mathbf{r}) + \sum_{m\neq n}^{N-1}\varphi_{0emn}(\mathbf{r})
\label{eq093}
\end{eqnarray}
\begin{eqnarray}
\varphi^{\star}_{sn} (\mathbf{r}) = \varphi_{0sn} (\mathbf{r}) + \sum_{m\neq n}^{N-1}\varphi_{0smn} (\mathbf{r})
\label{eq094}
\end{eqnarray}
\begin{eqnarray} 
\hat{\mathbf{A}}^{{\star}}_{gn} = \frac {\varphi^{\star}_{gn} \hat{\mathbf{p}}_{n}} {m_{0n}c}
\label{eq095}
\end{eqnarray} 
\begin{eqnarray} 
\hat{\mathbf{A}}^{\star}_{en} = \frac {\varphi^{\star}_{en} \hat{\mathbf{p}}_{n}} {m_{0n}c}
\label{eq096}
\end{eqnarray}
\begin{eqnarray} 
\hat{\mathbf{A}}^{\star}_{sn} = \frac {\varphi^{\star}_{sn} \hat{\mathbf{p}}_{n}} {m_{0n}c},
\label{eq097}
\end{eqnarray}
where $\varphi_{gn} (\mathbf{r})= \sum_{m\neq n}^{N-1}\varphi_{0gmn} (\mathbf{r})$, $\varphi_{en} (\mathbf{r})=\sum_{m\neq n}^{N-1}\varphi_{0emn}(\mathbf{r})$ and $\varphi_{sn} (\mathbf{r}) = \sum_{m\neq n}^{N-1}\varphi_{0smn} (\mathbf{r})$ denote the gravitational, electric and strong-nuclear \textit{external} potentials (for comparison, see Eq.~(\ref{eq083})-(\ref{eq088})). Using the principle of additivity for the squared energies and energies yields the unified mass-energy relations
\begin{eqnarray} 
\sum_{n=1}^N {\varepsilon}^{2}_{0n}  | s_{n} \rangle | \tilde \psi_{0n} \rangle =\sum_{n=1}^N {\hat{\mathbf{H}}}^{2}_{n} | s_{n} \rangle | \tilde \psi_{0n} \rangle  
\label{eq098}
\end{eqnarray} 
\begin{eqnarray} 
\sum_{n=1}^N {\varepsilon}_{0n}  | s_{n} \rangle | \tilde \psi_{0n} \rangle =\sum_{n=1}^N {\hat{\mathbf{H}}}_{n} | s_{n} \rangle | \tilde \psi_{0n} \rangle. 
\label{eq099}
\end{eqnarray} 
Imposing the relations ${\hat{\mathbf{H}}}_{n} | s_{m} \rangle | \tilde \psi_{0m} \rangle = \delta_{nm} {\hat{\mathbf{H}}}_{n} | s_{m} \rangle | \tilde \psi_{0m} \rangle$ and $ \langle \tilde \psi_{0n} |\langle  s_{n}  | s_{m} \rangle | \tilde \psi_{0m} \rangle=\delta_{nm}$, where $\delta_{nm}$ is the Kronecker symbol, Eqs.~(\ref{eq098}) and (\ref{eq099}) read
\begin{eqnarray} 
\left(\sum_{n=1}^N {\varepsilon}^{2}_{0n} \right) \prod ^N _{n=1}| s_{n} \rangle | \tilde \psi_{0n} \rangle =\nonumber  \\ \left(\sum_{n=1}^N {\hat{\mathbf{H}}}^{2}_{n} \right) \prod ^N _{n=1}| s_{n} \rangle | \tilde \psi_{0n} \rangle  
\label{eq0100}
\end{eqnarray} 
\begin{eqnarray} 
\left(\sum_{n=1}^N {\varepsilon}_{0n} \right) \prod ^N _{n=1}| s_{n} \rangle | \tilde \psi_{0n} \rangle =\nonumber  \\\left(\sum_{n=1}^N {\hat{\mathbf{H}}}_{n} \right) \prod ^N _{n=1}| s_{n} \rangle | \tilde \psi_{0n} \rangle,  
\label{eq0101}
\end{eqnarray} 
where $|\Psi \rangle =\prod ^N _{n=1}| s_{n} \rangle | \tilde \psi_{0n} \rangle$ denotes the state of the field $\Psi (\mathbf{r},t)$ of the multi-unitfield system. 

It has been shown in Ref.~\cite{4kukh} that the Hamiltonians of canonical quantum mechanics~\cite{4land},\cite{4ber} of a system of electrically charged particles are the low-order approximations of the unified Hamiltonian of the system of pure electromagnetic unit-fields. The unified energy-mass relation (\ref{eq0101}) provides the minimum energy of the system only if no two identical electrons occupy the same quantum state.~\cite{4kukh} The model explained physically the \textit{Pauli exclusion principle}. In the quantum mechanics, the principle was introduced as a postulate. 

\section{V. The Unified Energy-Mass Relation for Non-stationary Unit-Fields}
\label{sec5}

According to equations (\ref{eq04})-(\ref{eq06}) and (\ref{eq034}), the unified energy-mass relation for the time-dependent energies of the $n$-th non-stationary unit-field 
\begin{eqnarray} 
|\psi_{0n} (\mathbf{r},t)\rangle = | s_{n} (\mathbf{r},t)\rangle | \tilde \psi_{0n} (\mathbf{r},t)\rangle \neq \nonumber  \\ exp\left(\pm{{\frac {i} {{\hbar}}}} {\varepsilon}_{0n}t \right) | s_{n} (\mathbf{r})\rangle | \tilde \psi_{0n} (\mathbf{r})\rangle    
\label{eq0102}
\end{eqnarray} 
has the form
\begin{eqnarray} 
{-\hbar}^{2} {\frac {{\partial}^2} {\partial {t}^2}} | s_{n} (\mathbf{r},t)\rangle | \tilde \psi_{0n} (\mathbf{r},t)\rangle  = \nonumber  \\{{\hat{\mathbf{H}}}_{n}^{2}}| s_{n} (\mathbf{r},t)\rangle | \tilde \psi_{0n} (\mathbf{r},t)\rangle  ,  
\label{eq0103}
\end{eqnarray} 
In terms of the energy, the relation (\ref{eq0103}) can be presented as
\begin{eqnarray} 
{\pm i\hbar} {\frac {{\partial}} {\partial {t}}} | s_{n} (\mathbf{r},t)\rangle | \tilde \psi_{0n} (\mathbf{r},t)\rangle  = \nonumber  \\{{\hat{\mathbf{H}}}_{n}}| s_{n} (\mathbf{r},t)\rangle | \tilde \psi_{0n} (\mathbf{r},t)\rangle  ,  
\label{eq0104}
\end{eqnarray} 
where the operators ${\hat{\mathbf{H}}}_{n}^{2}$ and ${\hat{\mathbf{H}}}_{n}$ have been presented in the previous sections. Eqs. (\ref{eq0102})-(\ref{eq104}) describe the transient physical processes, for instance, scattering, emission and absorption of a unit-field (quantum particle) by a system of unit-fields. 

\section{VI. The Classical Limit of Unified Quantum Model}
\label{sec6}

The physical parameters of quantum particles (unit-fields) obey the non-quantum values of classical (spin-less) particles in the classical limit. That means that a physical value $f$ determined by the respective operator ${\hat{f}}$ is given by
\begin{eqnarray} 
f(\mathbf{r}) = \langle {f(\mathbf{r})} \rangle = {\int^{\infty}_{-\infty}} \psi_{0}^{\ast}(\mathbf{r'}) \hat f(\mathbf{r'}) \psi_{0}(\mathbf{r'}) d^3x' = \nonumber  \\ {\int^{\infty}_{-\infty}} \rho(\mathbf{r'}) f(\mathbf{r'})  d^3x', 
\label{eq0105}
\end{eqnarray} 
where $\rho(\mathbf{r'}) = \delta(\mathbf{r-r'})$ denotes the Dirac delta. Therefore, in the classical limit, the gravitoelectro-strongnuclear unit-fields are described by the unified non-quantum model of classical particles:   
\begin{eqnarray} 
H_{ges}^{2} = \left( 1 + {\frac {U^{\star}_{ges} } {m_{0} c^2}} \right)^{2} \left( m_{0}^2 c^4 + \mathbf{p}^2 c^2 \right) = \nonumber  \\   \left( 1 +{\frac {\langle m_{0ges}\rangle } {m_{0} }}+ {\frac {U_{ges} } {m_{0} c^2}} \right)^{2} \left( m_{0}^2 c^4 + \mathbf{p}^2 c^2 \right)  
\label{eq0106}
\end{eqnarray} 
\begin{eqnarray} 
H_{ges} = \left( 1 + {\frac {U^{\star}_{ges} } {m_{0} c^2}} \right) \left( m_{0}^2 c^4 + \mathbf{p}^2 c^2 \right)^{1/2} = \nonumber  \\\left( 1 +{\frac {\langle m_{0ges}\rangle } {m_{0} }}+ {\frac {U_{ges} } {m_{0} c^2}} \right) \left( m_{0}^2 c^4 + \mathbf{p}^2 c^2 \right)^{1/2},
\label{eq0107}
\end{eqnarray} 
where
\begin{eqnarray}
U^{\star}_{ges}(\mathbf{r}) = U^{\star}_{g}(\mathbf{r}) + U^{\star}_{e}(\mathbf{r})+ U^{\star}_{e}(\mathbf{r})
\label{eq0108}
\end{eqnarray}
\begin{eqnarray}
U^{\star}_{g} (\mathbf{r}) = \langle m_{0g}\rangle c^{2} +U_{g} (\mathbf{r}) 
\label{eq0109}
\end{eqnarray}
\begin{eqnarray}
U^{\star}_{e} (\mathbf{r}) = \langle m_{0e}\rangle c^{2} + U_{e} (\mathbf{r})
\label{eq0110}
\end{eqnarray}
\begin{eqnarray}
U^{\star}_{s} (\mathbf{r}) = \langle m_{0s}\rangle c^{2} + U_{s} (\mathbf{r})
\label{eq0111}
\end{eqnarray}
\begin{eqnarray}
\langle m_{0ges}\rangle = \langle m_{0g}\rangle +\langle m_{0e}\rangle +\langle m_{0s}\rangle =  \nonumber  \\ \langle s | U_{0g}c^{-2}| s \rangle + \langle s | U_{0e}c^{-2}| s \rangle  \nonumber  \\+\langle s | U_{0s}c^{-2}| s \rangle 
\label{eq0112}
\end{eqnarray}
\begin{eqnarray}
U_{ges}(\mathbf{r}) = U_{g}(\mathbf{r}) + U_{e}(\mathbf{r})+ U_{e}(\mathbf{r})
\label{eq0113}
\end{eqnarray}
\begin{eqnarray}
U_{g} (\mathbf{r}) = m_{0}\varphi_{g} (\mathbf{r})
\label{eq0114}
\end{eqnarray}
\begin{eqnarray}
U_{e} (\mathbf{r}) =  q_{0}\varphi_{e} (\mathbf{r})
\label{eq0115}
\end{eqnarray}
\begin{eqnarray}
U_{s} (\mathbf{r}) = Q_{0}\varphi_{s} (\mathbf{r})
\label{eq0116}
\end{eqnarray}
\begin{eqnarray} 
{\mathbf{A}}_{g} = \frac {\varphi_{g} {\mathbf{p}}} {m_{0}c}
\label{eq0117}
\end{eqnarray} 
\begin{eqnarray} 
{\mathbf{A}}_{e} =  \frac {\varphi_{e} {\mathbf{p}}} {m_{0}c}
\label{eq0118}
\end{eqnarray}
\begin{eqnarray} 
{\mathbf{A}}_{s} = \frac {\varphi_{s} {\mathbf{p}}} {m_{0}c}
\label{eq0119}
\end{eqnarray}
\begin{eqnarray} 
{\mathbf{p}} = \frac{m_{0}{\mathbf{v}}}{{\sqrt{1-\frac{{\mathbf{v}}^{2}}{c^{2}}}}}.
\label{eq0120}
\end{eqnarray}
In other words, in the non-quantum model, the quantum parameter $x$ determined by Eqs.~(\ref{eq045}), (\ref{eq069}), (\ref{eq0107}), (\ref{eq089}), and (\ref{eq091}) is replaced by the respective non-quantum value, while the quantum parameter $y = {\frac {\hat{\mathbf{p}}^{2}} {m_{0}^{2} c^2}}$ should have the non-quantum form $y = {\frac {{\mathbf{p}}^{2}} {m_{0}^{2} c^2}}$. Notice, the energy-mass relation~(\ref{eq0107}) reduces to the Einstein form (\ref{eq02}) in the case of $U^{\star}_{ges} = 0$. In the case of pure electromagnetic particles ($U^{\star}_{ges} = U^{\star}_{e}$, $\langle m_{0e}\rangle \ll m_{0}$) and weak external scalar potentials [$(q_{0}{\varphi^{\star}_{e} } )^{2} \ll2m_{0} c^2|q_{0}{\varphi^{\star}_{e} } |$], Eqs.~(\ref{eq0107}), (\ref{eq0110}), (\ref{eq0115}) and (\ref{eq0118}) yielded the canonical Hamiltonian~\cite{4lan}
\begin{eqnarray} 
H_{e} = \sqrt{m_{0}^{2} c^4 + c^{2}\left(\mathbf{p} +\frac{q_{0}}{c}{{\mathbf{A}}_{e}} \right)^{2}} + q_{0}{\varphi_{e} }
\label{eq0121}
\end{eqnarray} 
of the Lorentz electromagnetism.~\cite{4kukh}

\subsection{A. The unified model of non-quantum gravitoelectromagnetic-strongnuclear particles by using the Lagrange formalism}
\label{subsec6.1.}

The classical fields and interactions, for instance, the Einstein relativity and Maxwell-Lorentz electromagnetism are usually formulated in frames of the Lagrangian formalism. For the comparison, we reformulate the above-presented model using this formalism. The non-quantum Lagrangian that corresponds to equations (\ref{eq0106}) and (\ref{eq0107}) is given by 
\begin{eqnarray} 
L_{ges} = -\left( m_{0} c^2 + {U^{\star}_{ges}}({\mathbf{r}}) \right) \sqrt{1-\frac{{\mathbf{v}}^{2}}{c^{2}}} = \nonumber  \\ - \sqrt{ \left({m_{0} c^2} +{\langle m_{0ges}\rangle }c^{2}+ {U_{ges} } \right)^{2}\left(1-\frac{{\mathbf{v}}^{2}}{c^{2}}\right)}.
\label{eq0122}
\end{eqnarray} 
The unified equations of motion 
\begin{eqnarray} 
\frac{d}{dt} \frac{\partial L_{ges}}{\partial {\mathbf{v}}} = \frac{\partial L_{ges}}{\partial {\mathbf{r}}}
\label{eq0123}
\end{eqnarray} 
\begin{eqnarray} 
\frac{d}{dt} \frac{\partial L_{ge}}{\partial {\mathbf{v}}} = \frac{\partial L_{ge}}{\partial {\mathbf{r}}}
\label{eq0124}
\end{eqnarray} 
\begin{eqnarray} 
\frac{d}{dt} \frac{\partial L_{gs}}{\partial {\mathbf{v}}} = \frac{\partial L_{gs}}{\partial {\mathbf{r}}}
\label{eq0125}
\end{eqnarray} 
\begin{eqnarray} 
\frac{d}{dt} \frac{\partial L_{es}}{\partial {\mathbf{v}}} = \frac{\partial L_{es}}{\partial {\mathbf{r}}}
\label{eq0126}
\end{eqnarray} 
\begin{eqnarray} 
\frac{d}{dt} \frac{\partial L_{g}}{\partial {\mathbf{v}}} = \frac{\partial L_{g}}{\partial {\mathbf{r}}}
\label{eq0127}
\end{eqnarray} 
\begin{eqnarray} 
\frac{d}{dt} \frac{\partial L_{e}}{\partial {\mathbf{v}}} = \frac{\partial L_{e}}{\partial {\mathbf{r}}}
\label{eq0128}
\end{eqnarray} 
\begin{eqnarray} 
\frac{d}{dt} \frac{\partial L_{s}}{\partial {\mathbf{v}}} = \frac{\partial L_{s}}{\partial {\mathbf{r}}},
\label{eq0129}
\end{eqnarray} 
yielded the respective forces, namely the non-quantum \textit{gravitoelectromagnetic-strongnuclear}, \textit{gravitoelectric} (\textit{gravitoelectromagnetic}), \textit{gravito-strongnuclear},  \textit{electro-strongnuclear} (\textit{electro-weak}), \textit{pure gravitational}, \textit{pure electromagnetic} and \textit{pure strongnuclear} \textit{forces}. Notice, the Lagrangian~(\ref{eq0122}) describes these forces as different aspects of a single gravitoelectromagnetic-strongnuclear interaction. In other words, at very high energies, the gravitational, electromagnetic, strong-nuclear and weak-nuclear forces are combined into one. The \textit{gravitoelectromagnetic} (${\mathbf{F}}_{ge}=\frac{\partial L_{ge}}{\partial {\mathbf{r}}}$) and \textit{gravito-strongnuclear} (${\mathbf{F}}_{gs}=\frac{\partial L_{gs}}{\partial {\mathbf{r}}}$) \textit{forces} are the new kinds of forces, compared to SM. The \textit{fundamental forces} correspond to Eqs.~(\ref{eq0127})-(\ref{eq0129}), whereas the \textit{combined forces} are described by Eqs.~(\ref{eq0123})-(\ref{eq0126}).  Notice, the Tailor expansion in the Lagrangian~(\ref{eq0122}) changes the physical picture of interactions. Indeed, the cross-correlation terms of more than second order in the potentials yielded other combinations of gravitational, electromagnetic, strong-nuclear, gravito-electromagnetic, gravito-strongnuclear and electro-strongnuclear forces. 

In the case of \textit{pure electromagnetic} ($U^{\star}_{ges} = U^{\star}_{e}$, $\langle m_{0e}\rangle \ll m_{0}$) particles with the weak [$(q_{0}{\varphi^{\star}_{e} } )^{2} \ll2m_{0} c^2|q_{0}{\varphi^{\star}_{e} } |$] external potentials $\varphi_{e}$ and ${\mathbf{A}}_{e} = \frac {\varphi_{e} {\mathbf{v}}} {2c}$, Eqs.~(\ref{eq0108})-(\ref{eq0120}), (\ref{eq0122}) and (\ref{eq0128}) yielded the canonical Lagrangian~\cite{4lan} 
\begin{eqnarray} 
L_{e} = -m_{0} c^2{\sqrt{1-\frac{{\mathbf{v}}^{2}}{c^{2}}}} + \frac{q_{0}}{c}{{\mathbf{A}}_{e}}{{\mathbf{v}}} - q_{0}\varphi_{e}
\label{eq0130}
\end{eqnarray} 
of Lorentz's electromagnetism, the Lorentz electromagnetic force ${\mathbf{F}}_{e}=\frac{\partial L_{e}}{\partial {\mathbf{r}}}$ and the Maxwell equations.~\cite{4kukh} 

The Einstein relativity is described by the spacetime metric 
\begin{eqnarray} 
ds^{2} = g_{ik}dx^{i}dx^{k},
\label{eq0131}
\end{eqnarray} 
where $g_{ik}$ is the metric tensor and $x^{i}=(ct,\mathbf{r})$. For the flat spacetime, $g_{ik}=g^{M}_{ik}$. Here, I use the Minkowski tensor $g^{M}_{ik}$ with the signature ($+---$).~\cite{4lan} For the \textit{pure gravitational particles} ($U^{\star}_{ges} = U_{g}$, $\langle m_{0g}\rangle \ll m_{0}$), the Lagrangian (\ref{eq0122}) and the canonical relation $L_{g}=-m_{0}cds/dt$ yielded~\cite{4kukh}
\begin{eqnarray} 
ds^{2} = \left(1+ \frac{\varphi_{g}({\mathbf{r}})}{c^{2}} \right)^{2}c^{2}dt^{2} - \left(1+ \frac{\varphi_{g}({\mathbf{r}})}{c^{2}} \right)^{2}d{\mathbf{r}}^{2}
\label{eq0132}
\end{eqnarray} 
with the respective unified metric tensor 
\begin{eqnarray} 
g^{u}_{ik} = \left(1+ \frac{\varphi_{g} ({\mathbf{r}})}{c^{2}} \right)^{2} g^{M}_{ik}, 
\label{eq0133}
\end{eqnarray} 
which is similar to the Schwarzschild tensor. The general relativity and the unified non-quantum pure gravitation becomes undistinguishable from each other in the case of $g_{ik} = g^{u}_{ik}$. It can be mentioned that the divergent solutions to Eq.~(\ref{eq010}), which are based on the regular solid harmonics (see, Sec. II.A.1), yielded a repulsive gravitational force which increases infinitely with the increase of $r$. The other properties of forces ${\mathbf{F}}_{g}=\frac{\partial L_{g}}{\partial {\mathbf{r}}}$, ${\mathbf{F}}_{e}=\frac{\partial L_{e}}{\partial {\mathbf{r}}}$ and ${\mathbf{F}}_{ge}=\frac{\partial L_{ge}}{\partial {\mathbf{r}}}$ have been considered in Ref.~\cite{4kukh}. 

\section{VII. Discussions and Conclusions}
\label{sec7}

Starting from the mass-energy relation of Einstein's special relativity and using the concept for the unit-fields with gravitational, electric, and strong-nuclear "dressings", the unified equations for gravitational, electromagnetic, strong-nuclear, gravito-electric, gravito-strongnuclear and electro-strongnuclear (weak-nuclear) fields and interactions were derived. The cross-correlation of the gravitational ($U^{\star}_{g}$), electric ($U^{\star}_{e}$) and strong-nuclear ($U^{\star}_{s}$) fields ("dressings") or the respective potentials explain the three fundamental kinds of interactions, namely the gravitational, electric and strong-nuclear forces. The gravito-electric ($U^{\star}_{ge}=U^{\star}_{g}+U^{\star}_{e}$), gravito-strongnuclear ($U^{\star}_{gs}=U^{\star}_{g}+U^{\star}_{s}$) and electro-strongnuclear ($U^{\star}_{es}=U^{\star}_{e}+U^{\star}_{s}$) fields are the combined "dressings" of the unit-field, which are responsible for the gravito-electric, gravito-strongnuclear and electro-strongnuclear (weak-nuclear) combined forces. The gravitational ($U^{\star}_{g} = U_{0g} + U_{g}$), electric ($U^{\star}_{e} = U_{0e} + U_{e}$) and strong-nuclear ($U^{\star}_{s} = U_{0s} + U_{s}$) fields ("dressings") consist of the gravitational ($U_{0g}$), electric ($U_{0e}$), and strong-nuclear ($U_{0s}$) intrinsic fields ("intrinsic dressings") and the gravitational ($U_{g}$), electric ($U_{e}$), and strong-nuclear ($U_{s}$) external fields ("external dressings"). The total rest-mass $\langle m_{0ges}\rangle $ of the gravitoelectric-strongnuclear "intrinsic dressing" $U_{0ges}$ of the unit-field, which is given by \begin{eqnarray}
\langle m_{0ges}\rangle = \langle m_{0g}\rangle +\langle m_{0e}\rangle +\langle m_{0s}\rangle =  
\nonumber  \\ \langle s | (U_{0g}c^{-2})| s \rangle + \langle s | (U_{0e}c^{-2})| s \rangle 
 \nonumber  \\+\langle s | (U_{0s}c^{-2})| s \rangle ,
\label{eq0134}
\end{eqnarray}
consists of the gravitational ($\langle m_{0g}\rangle$), electric ($\langle m_{0e}\rangle$) and strong-nuclear ($\langle m_{0s}\rangle$) masses of the gravitational ($U_{0g}$), electric ($U_{0g}$) and strong-nuclear ($U_{0g}$)  "intrinsic dressings" (intrinsic fields). In terms of SM, the gravitoelectric-strongnuclear unit-field carries the \textit{ intrinsic gauge-bosons} associated with the respective "intrinsic dressings". The gravitonal, electric, strong-nuclear, gravito-electric, gravito-strongnuclear and electro-strongnuclear (electroweak) intrinsic gouge-bosons have the masses $\langle m_{0g}\rangle$, $\langle m_{0e}\rangle$, $\langle m_{0s}\rangle$, $\langle m_{0ge}\rangle=\langle m_{0g}\rangle + \langle m_{0e}\rangle$, $\langle m_{0gs}\rangle=\langle m_{0g}\rangle + \langle m_{0s}\rangle$ and $\langle m_{0es}\rangle=\langle m_{0s}\rangle + \langle m_{0e}\rangle$, respectively. 

The physical picture of quantum interactions is determined by the unit-field state $| s \rangle | \tilde \psi \rangle$, the squared Hamiltonian ${\hat{\mathbf{H}}}^{2}(x)$ and the Hamiltonians ${\hat{\mathbf{H}}}(x)$ and ${\hat{\mathbf{H}}}(y)$. The model predicts gravitational, electromagnetic, strong-nuclear, gravito-electromagnetic, gravito-strongnuclear and electro-strongnuclear (weak-nuclear) quantum interactions described by the dimensionless quantum parameters~(\ref{eq070})-(\ref{eq072}) and (\ref{eq076})-(\ref{eq078}). If the dimensionless parameter $\langle y \rangle = \langle \tilde \psi_0 |\langle s | y | s \rangle | \tilde \psi_0 \rangle$ or $\langle x \rangle = \langle \tilde \psi_0 |\langle s | x | s \rangle | \tilde \psi_0 \rangle$ is of order one or larger, then the quantum interactions are described by the non-perturbative theory based on Eq.~(\ref{eq044}). The quantum models based on Eqs.~(\ref{eq049}) and (\ref{eq056}) are perturbative theories, i.e., they are described by expansions in powers of $\langle y \rangle$ or $\langle x \rangle$. The physical picture of quantum interactions based on the approximation~(\ref{eq056}) of ${\hat{\mathbf{H}}}(x)$ under the condition $\langle x \rangle<1$ depends on the number of terms in the Taylor expansion. The terms of more than second order in the potentials yielded the combined quantum interactions, which are the high-order combinations of gravitational, electromagnetic, strong-nuclear, gravito-electromagnetic, gravito-strongnuclear and electro-strongnuclear interactions. In terms of SM, the interaction of unit-fields is provided by the virtual ($U^{\star}=U^{\star}(\mathbf{r}$)) or real ($U^{\star}=U^{\star}(\mathbf{r},t$)) exchange of the virtual or real gauge-bosons, respectively. For instance, the real gravitons, real photons, electro-strong (electroweak) real bosons and strong-nuclear real bosons (real gluons) are the real gauge-bosons associated with the gravitational ($U^{\star}_{g}(\mathbf{r},t)$), electric ($U^{\star}_{g}(\mathbf{r},t)$), strong-nuclear ($U^{\star}_{g}(\mathbf{r},t)$), gravito-electric ($U^{\star}_{ge}(\mathbf{r},t)=U^{\star}_{g}(\mathbf{r},t)+U^{\star}_{e}(\mathbf{r},t)$), gravito-strongnuclear ($U^{\star}_{gs}(\mathbf{r},t)=U^{\star}_{g}(\mathbf{r},t)+U^{\star}_{s}(\mathbf{r},t)$) and electro-strongnuclear ($U^{\star}_{es}(\mathbf{r},t)=U^{\star}_{e}(\mathbf{r},t)+U^{\star}_{s}(\mathbf{r},t)$) intrinsic and external fields ("intrinsic and external dressings"). In other words, the interaction of unit-fields is provided by exchange of the "intrinsic dressings" (virtual or real intrinsic gauge-bosons) or "external dressings" (virtual or real external gauge-bosons).

In the standard model of particle physics, the relative strengths of the four fundamental (gravitational, electromagnetic, strong-nuclear and weak-nuclear) forces are quoted in terms of the respective dimensionless coupling constants $\alpha_{g}$, $\alpha_{e}$, $\alpha_{s}$ and $\alpha_{w}$. The coupling constants $\alpha_{g}$ and $\alpha_{e}$ are given by $\alpha_{e}= \frac {\gamma_{e}q_0^{2}} {\hslash c}$, $\alpha_{g}=\alpha_{e}\frac {{\mathbf{F}}_{g}}{{\mathbf{F}}_{e}} = \alpha_{e} \frac {\gamma_{g}m_0^{2}} {\gamma_{e}q_0^{2}}$. The experimental data describing the strong-nuclear (color) force between nucleons is consistent with the value $\alpha_{s}\sim1$. SM sees the color force as the force between the constituent quarks. The quantum chromodynamics of SM describing the color force gives the well-known expression for the strong-nuclear coupling constant:
\begin{eqnarray}
\alpha_{s}({\varepsilon}) \approx \frac {12\pi}{(33-2n_{q})ln \left( \frac{\varepsilon^{2} }{\Lambda^{2}} \right) }, 
\label{eq0135}
\end{eqnarray}
where $n_{q}$ is the number of quarks, $\varepsilon$ denotes the energy, and $\Lambda\approx 0.2 GeV$ is experimentally determined parameter. Equation~(\ref{eq0135}) shows that the strong force coupling constant decreases to zero ($\alpha_{s}({\varepsilon})\rightarrow 0$) at small distances ($r\rightarrow 0$, respectively $\varepsilon \rightarrow \infty$), i.e., the quarks approach a state where they can move without resistance in the tiny volume of the hadron. In other words, the color force diminishes inside the nucleons. The phenomenon is called the "asymptotic freedom" of quarks. In the quantum picture of SM, the nuclear strong force between quarks is based on exchange of gluons. The color force acts like a spring between quarks of different colors, becoming stronger as they get farther apart. That explained why quarks cannot be isolated singularly. According to SM, the weak-nuclear coupling constant $\alpha_{w}$ is related to the strong-nuclear $\alpha_{s}$ coupling constant as $\alpha_{w}/\alpha_{s}\approx 10^{-7}$. SM describes the electromagnetic force and the weak-nuclear force as two different aspects of a single electroweak interaction. At very high energies, SM combines the electromagnetic and weak forces into one.

In the unified model, the counterparts of the coupling constants of SM are the unified dimensionless coupling parameters (see, Eqs.~(\ref{eq044}), (\ref{eq049})(\ref{eq056}), (\ref{eq0106}), (\ref{eq0107}), and (\ref{eq0122})):   
\begin{eqnarray}
\alpha_{g} = \frac {\langle m_{0g}\rangle}{m_{0}} = \frac {\langle s | U_{0g}c^{-2}| s \rangle}{m_{0}}
\label{eq0136}
\end{eqnarray}
\begin{eqnarray}
\alpha_{e} = \frac {\langle m_{0e}\rangle}{m_{0}} = \frac {\langle s | U_{0e}c^{-2}| s \rangle}{m_{0}}
\label{eq0137}
\end{eqnarray}
\begin{eqnarray}
\alpha_{s} = \frac {\langle m_{0s}\rangle}{m_{0}} = \frac {\langle s | U_{0s}c^{-2}| s \rangle}{m_{0}}
\label{eq0138}
\end{eqnarray}
\begin{eqnarray}
\alpha_{w} = \frac {\langle m_{0w}\rangle}{m_{0}} = \frac {\langle s | (U_{0e}U_{0s})^{-1/2}c^{-2}| s \rangle}{m_{0}}.
\label{eq0139}
\end{eqnarray}
The unified dimensionless coupling parameters are consistent with the respective coupling constants of SM. For instance, the strong-nuclear force between two quarks with the color charges $Q_{1}$ and $Q_{2}$, which is described by Eq.~(\ref{eq026}), (\ref{eq0122}) and (\ref{eq0129}), reads 
\begin{eqnarray}
{\mathbf{F}}_{s}=\frac{\partial L_{s}}{\partial {\mathbf{r}}} = -\frac{\partial U_{s}}{\partial {\mathbf{r}}} = -\gamma_{s}Q_{1}Q_{2} \nonumber  \\ \left[ \left(\frac {|{\Gamma_{s}} | }{r} + \frac {1}{r^{2}} \right) {e^{-|{\Gamma_{s}} | r} } - \left(\frac {|{\Gamma_{s}} | }{r} - \frac {1}{r^{2}} \right) {e^{|{\Gamma_{s}} | r} }\right].
\label{eq0140}
\end{eqnarray}
Here, we considered the quarks at rest ($\mathbf{v}=0$). The strong-nuclear force (\ref{eq0140}) diminishes at the distance $r\approx\sqrt{3}|{\Gamma_{s}}|^{-1}$ between the quarks independently from the color charge $Q$. That means that the quarks move freely (${\mathbf{F}}_{s}\sim 0$) within the hadrons, whose dimensions are given by $r\sim \sqrt{3}|{\Gamma_{s}}|^{-1}$. Like in SM, the quark color charges have three kinds. Antiquarks come with the anti-color charges. At large distance between quarks, Eq.~(\ref{eq0140}) yields the force ${\mathbf{F}}_{s}\approx \frac {\gamma_{s}Q_{1}Q_{2} |{\Gamma_{s}} | }{r}  {e^{|{\Gamma_{s}} | r} }$ which increases infinitely with the increase of $r$. In case of $Q^{i}Q^{j}=(-1+2\delta_{ij}) \vert Q^{i} \vert \vert Q^{j} \vert$, the quarks of different colors attract each other, and the quarks of the same color repel each other. That explains why the attractive quarks are apparently inseparable, though they can move freely when they are close together. The attractive quarks are elementary particles from which nucleons (protons and neutrons) are composed. Like in SM, a neutron, proton or any other three-quark particle have one of each color to exist. In the present model, the electro-weak (electro-strongnuclear) force between two unit-fields (particles) with the electric charges $q_{1}$ and $q_{2}$ and the color charges $Q_{1}$ and $Q_{2}$, which is described by Eq.~(\ref{eq022}), (\ref{eq026}), (\ref{eq0126}) and (\ref{eq0129}), reads 
\begin{eqnarray}
{\mathbf{F}}_{ew}=\frac{\partial L_{es}}{\partial {\mathbf{r}}} = \nonumber  \\-\frac{\partial }{\partial {\mathbf{r}}} \sqrt{ \left({m_{0} c^2} +{\langle m_{0es}\rangle }c^{2}+ {U_{es}({\mathbf{r}}) } \right)^{2}\left(1-\frac{{\mathbf{v}}^{2}}{c^{2}}\right)}.
\label{eq0141}
\end{eqnarray}
The parameters $U^{2}_{es}({\mathbf{r}})= U^{2}_{e}({\mathbf{r}})+2U_{e}({\mathbf{r}})U_{s}({\mathbf{r}})+ U^{2}_{s}({\mathbf{r}})$ and $U_{e}({\mathbf{r}})U_{s}({\mathbf{r}})$ in Eq.~(\ref{eq0141}) explain the electro-weak (electro-strongnuclear) force. Notice, Eq.~(\ref{eq0141}) describes the electromagnetic force and the weak-nuclear force as two different aspects of a single electroweak interaction. Indeed, at very high energies, the electromagnetic (${\mathbf{F}}_{e}=\frac{\partial L_{e}}{\partial {\mathbf{r}}}$) and strong ${\mathbf{F}}_{s}=\frac{\partial L_{s}}{\partial {\mathbf{r}}}$ forces are combined into the electro-weak force ${\mathbf{F}}_{ew}=\frac{\partial L_{es}}{\partial {\mathbf{r}}}$. 

The role of the unit-field spin wave-functions in the interactions is described rather by the quantum equations of Sects. I-V than the non-quantum equations of Sect. VI. For an orbital quantum number $l$, there are $2l+1$ independent spin wavefunctions (\ref{eq029}), one for each magnetic quantum number $m$ with $-l \leq  m \leq  l$ (see, Sec. II.A.2). It was shown that the electron corresponds to the unit-field with the orbital and magnetic quantum numbers $l=1$ and $m=\pm1$, which yielded the well-known normal and anomalous values of the electron magnetic moment.~\cite{4kukh}. For other particles, the values of $l$ and $m$ are determined by the experimental data and SM. Notice, the spin wave-functions $s(\mathbf{r})$ and $-s(\mathbf{r})$ of a particle and antiparticle provide their annihilation.


The present model has explained the weak-nuclear interaction (force) by the cross-correlation of electric and strong-nuclear potentials of the electric and strong-nuclear "dressings". The cross-correlations of gravitational potential with the electric and strong-nuclear potentials predict the existence of new forces, namely gravito-electromagnetic and gravito-strongnuclear interactions, which are unknown in SM. Keeping terms of more than second order in the potentials of perturbative model (\ref{eq056}) uncovers other combinations of gravitational, electromagnetic, weak-nuclear and strong-nuclear interactions. The unified model is surprisingly simple in comparison to the Maxwell-Lorentz electrodynamics, Einstein relativity, quantum mechanics, and SM. The unified model may be useful in other branches of physics, especially in the case of strong gravitational, electric and strong-nuclear potentials. As an example, the new aspects of the physics of cosmological black-holes, solid-state physics, near-field optics and photonics based on the unified model will be presented in the next paper.

\end{document}